\documentclass[a4paper,10pt,epjc3,twoside]{svjour3}
\ifdefined\pdfpxdimen
   \let\sphinxpxdimen\pdfpxdimen\else\newdimen\sphinxpxdimen
\fi \sphinxpxdimen=.75bp\relax
\ifdefined\pdfimageresolution
    \pdfimageresolution= \numexpr \dimexpr1in\relax/\sphinxpxdimen\relax
\fi

\PassOptionsToPackage{bookmarksdepth=5}{hyperref}

\PassOptionsToPackage{warn}{textcomp}
\usepackage[utf8]{inputenc}

\ifdefined\DeclareUnicodeCharacter
  \ifdefined\DeclareUnicodeCharacterAsOptional
    \def\sphinxDUC#1{\DeclareUnicodeCharacter{"#1}}
  \else
    \let\sphinxDUC\DeclareUnicodeCharacter
  \fi
  \sphinxDUC{00A0}{\nobreakspace}
  \sphinxDUC{2500}{\sphinxunichar{2500}}
  \sphinxDUC{2502}{\sphinxunichar{2502}}
  \sphinxDUC{2514}{\sphinxunichar{2514}}
  \sphinxDUC{251C}{\sphinxunichar{251C}}
  \sphinxDUC{2572}{\textbackslash}
\fi
\usepackage{cmap}
\usepackage[T1]{fontenc}
\usepackage{fix-cm}
\usepackage{amsmath,amssymb,amstext}

\usepackage{bm}
\usepackage{babel}

\usepackage{tgtermes}
\usepackage{tgheros}

\usepackage[,numfigreset=2,mathnumfig,mathnumsep={.}]{sphinx}
\sphinxsetup{HeaderFamily=\bfseries, TitleColor=black}
\fvset{fontsize=auto}
\usepackage{geometry}

\usepackage{subcaption}

\usepackage{hyperref}
\usepackage{hypcap}% it must be loaded after hyperref.
\usepackage{tocvsec2}

\usepackage{sphinxmessages}
\counterwithin*{footnote}{section}

    \preprintnumbers{CERN-TH-2025-252/IPPP-25-57/HERWIG-2025-01/KA-TP-36-2025/MCNET-25-31}
    \authorrunning{The Herwig collaboration}
    \usepackage{etoolbox}
    \AtBeginEnvironment{figure}{\renewcommand{\phantomsection}{}}
    \AtBeginEnvironment{figure}{\pretocmd{\hyperlink}{\protect}{}{}}
    \AtBeginEnvironment{table}{\pretocmd{\hyperlink}{\protect}{}{}}
    \AtBeginEnvironment{savenotes}{\pretocmd{\hyperlink}{\protect}{}{}}
    \institute{Department of Astronomy and Theoretical Physics, Lund University \label{addr1}
 \and Institute for Particle Physics Phenomenology, Department of Physics, University of Durham \label{addr2}
 \and Dipartimento di Fisica, Universit\'a di Torino, and INFN, Sezione di Torino \label{addr3} %Via P. Giuria 1, I-10125 Torino, Italy 
 \and Institute for Theoretical Physics, Karlsruhe Institute of Technology \label{addr4}
 \and Department of Informatics, University of Bergen \label{addr5}
 \and Higgs Centre for Theoretical Physics, University of Edinburgh \label{addr6}
 \and Department of Physics, Kennesaw State University \label{addr7}
 \and Theoretical Physics, NAWI Graz, University of Graz \label{addr8}
 \and Department of Physics and Astronomy, University of Victoria \label{addr9}
 \and Marietta Blau Institute for Particle Physics, Austrian Academy of Sciences \label{addr10}
 \and Jagiellonian University, Kraków \label{addr11}
 \and Particle Physics Group, Department of Physics and Astronomy, University of Manchester \label{addr13}
 \and Theoretical Physics Department, CERN \label{addr14}
 \and DESY, Hamburg \label{addr15}
 }
 
\usepackage{cite}
      \newcommand{\PS}{\mathcal{PS}}

      \newcommand{\dd}{{\mathrm{d}}}

      \newcommand{\tPS}{\mathcal{\widetilde{PS}}}

      \newcommand{\tPSV}{\mathcal{\widetilde{PSV}}}

      \newcommand{\MT}{M_{T}}

\newcommand\splusminus{{\mathchoice
{\vplusminus\displaystyle}
{\vplusminus\scriptstyle}
{\vplusminus\scriptscriptstyle}
{\vplusminus\scriptscriptstyle}}}
\newcommand\sminusplus{{\mathchoice
{\vminusplus\displaystyle}
{\vminusplus\scriptstyle}
{\vminusplus\scriptscriptstyle}
{\vminusplus\scriptscriptstyle}}}

      \newcommand{\db}{\mathrm{d}^2\vect b \ }

      \newcommand{\dpt}{\mathrm{d} p^2_t \ }

      \newcommand{\vect}[1]{{\bf #1}}

      \newcommand{\br}[1] {\frac{1}{#1}}

      \newcommand{\abs}[1]{\left| #1 \right|}

      \newcommand{\myexp}[1]{\: e^{#1} \: }

      \newcommand{\di}[2]{{\rm d}^{#2}#1}

      \newcommand{\mydiff}[2]{{ \frac{ \di{#1}{}  }{ \di{#2}{} }  }}

      \newcommand{\stot}{{\sigma_{\rm tot}}}

      \newcommand{\sela}{{\sigma_{\rm el}}}

      \newcommand{\sinel}{{\sigma_{\rm inel}}}

      \newcommand{\sigmasoft}{{\sigma^{\rm inc}_{\rm soft}}}

      \newcommand{\sigmahard}{{\sigma^{\rm inc}_{\rm hard}}}

      \newcommand{\slope}{{b_{\rm el}}}

      \newcommand{\avgN}{\langle n(\vect b, s) \rangle}

      \newcommand{\pt}{p_t}

      \newcommand{\ptmin}{p_t^{\rm min}}

      \newcommand{\ptmino}{p_{t,0}^{\rm min}}

      \newcommand{\eik}[1]{\chi_{\rm #1}(\vect{b}, s)}

      \newcommand{\Pcal}{\mathcal{P}}

      \newcommand{\dSig}{\mathrm{d\sigma}}

      \newcommand{\dPhi}{\mathrm{d\phi}}

      \newcommand{\rmPhi}{\mathrm{\phi}}

      \newcommand{\rmPhiTilde}{\tilde{\mathrm{\phi}}}

      \newcommand{\tsum}{\textstyle\sum}

      \newcommand{\obu}{u}

      \newcommand{\df}{\mathrm{d}f}

      \newcommand{\hardProcScale}{\mu_\mathrm{H}}                                      

      \newcommand{\vetoScale}{Q_\perp}                                                 

      \newcommand{\showerScale}{\mu_\mathrm{S}}

      \newcommand{\atantwo}{\operatorname{atan2}}

\newdimen\hbigcirc
\newdimen\wbigcirc
\newcommand\vplusminus[1]{%
\settoheight{\hbigcirc}{$\scriptstyle\bigcirc$}%
\settowidth{\wbigcirc}{$\scriptstyle\bigcirc$}%
\makebox[\wbigcirc]{%
\makebox[0pt]{$\scriptstyle\pm$}%
\makebox[0pt]{$\scriptstyle\bigcirc$}}%
}
\newcommand\vminusplus[1]{%
\settoheight{\hbigcirc}{$\scriptstyle\bigcirc$}%
\settowidth{\wbigcirc}{$\scriptstyle\bigcirc$}%
\makebox[\wbigcirc]{%
\makebox[0pt]{$\scriptstyle\mp$}%
\makebox[0pt]{$\scriptstyle\bigcirc$}}%
}

\title{The Physics of Herwig 7}
\date{\today}
\release{7.3}
\author{J.~Bellm\thanksref{addr1} \and G.~Bewick\thanksref{addr2} \and S.~Ferrario Ravasio\thanksref{addr3} \and S.~Gieseke\thanksref{addr4} \and D.~Grellscheid\thanksref{addr5} \and S.~Kiebacher\thanksref{addr4} \and P.~Kirchgaeßer\thanksref{addr4} \and F.~Loshaj\thanksref{addr4} \and M.R.~Masouminia\thanksref{addr2} \and G.~Nail\thanksref{addr6} \and A.~Papaefstathiou\thanksref{addr7} \and S.~Plätzer\thanksref{addr8} \and M.~Rauch\thanksref{addr4} \and P.~Reimitz\thanksref{addr9} \and C.~Reuschle\thanksref{addr1} \and P.~Richardson\thanksref{addr2} \and D.~Samitz\thanksref{addr10} 
\and P.~Sarmah\thanksref{addr11} \and P.~Schichtel\thanksref{addr2} \and M.H.~Seymour\thanksref{addr13} \and A.~Siódmok\thanksref{addr11,addr14} \and D.~Stafford\thanksref{addr15} \and C.B.~Strange\thanksref{addr4} \and S.~Sule\thanksref{addr13} \and S.~Webster\thanksref{addr2} \and J.~Whitehead\thanksref{addr11}}

\renewcommand{\releasename}{Release}
\makeindex

\journalname{Eur. Phys. J. C}

\makeatletter
\def\@mkboth#1#2{}
\newlength\appendixwidth
\preto\appendix{\addtocontents{toc}{\protect\patchl@section}}
\newcommand{\patchl@section}{%
  \settowidth{\appendixwidth}{\textbf{Appendix }}%
  \addtolength{\appendixwidth}{1.5em}%
  \patchcmd{\l@section}{1.5em}{\appendixwidth}{}{}%
}

\makeatother

\makeatletter
\def\sphinxattablestart{\par\vskip\dimexpr\sphinxtablepre\relax
  \spx@inframedtrue\vbox\bgroup}
\def\sphinxattableend{\par\egroup\sphinxatlongtableend}
\appto\appendix{%
  \renewcommand\thetable{\@Alph\c@section.\@arabic\c@table}%
  \renewcommand\thefigure{\@Alph\c@section.\@arabic\c@figure}%
}
\makeatother

\usepackage[switch]{lineno}
\begin{document}
\ifdefined\shorthandoff
  \ifnum\catcode`\=\string=\active\shorthandoff{=}\fi
  \ifnum\catcode`\"=\active\shorthandoff{"}\fi
\fi

\maketitle
\begin{abstract}
We present the physics foundations and recent developments of Herwig 7, the modern C++ successor of the original Fortran-based HERWIG and the Herwig++~2.x series. Herwig 7 provides a flexible and systematically improvable framework for the simulation of high-energy lepton-lepton, lepton-hadron, and hadron-hadron collisions, with particular emphasis on  QCD and electroweak (EW) effects. Hard scattering processes are generated within the automated \emph{Matchbox} framework, which integrates external amplitude providers, supports tree-level, next-to-leading-order (NLO), and loop-induced matrix elements, and implements subtraction schemes, multi-channel phase-space sampling, dynamic scale choices, and both POWHEG- and MC@NLO-type matching algorithms. Consistent multijet merging at LO and NLO is provided, enabling precise predictions across a wide range of Standard Model processes. Parton radiation is simulated using two complementary showers: an angular-ordered shower incorporating QCD coherence and the heavy-quark dead-cone effect, and a dipole shower optimised for NLO matching and multijet merging. Higher-order corrections are included through matrix-element corrections and dedicated reweighting techniques, while QED and EW radiation are treated using a Yennie--Frautschi--Suura formalism and EW showering algorithms. The modelling of non-perturbative physics employs an advanced cluster hadronization framework with improved cluster formation, fission and decay, as well as colour reconnection models, heavy-quark effects, and interfaces to alternative hadronization schemes. The simulation of the underlying event is achieved with an extended eikonal multiple-partonic-scattering model, incorporating both semi-hard and soft components together with diffractive interactions, thereby enabling realistic descriptions of minimum-bias and underlying-event data. Herwig 7 thus represents a versatile event generator, providing a coherent, modular, and extensible platform for Standard Model and beyond-the-Standard-Model collider phenomenology at current and future facilities.
\end{abstract}

\clearpage\tableofcontents
\phantomsection\label{\detokenize{review/index::doc}}

\section{Preface}
\label{\detokenize{review/intro:preface}}\label{\detokenize{review/intro::doc}}

The Monte Carlo event generator Herwig 7 is the result of the collective effort of a large number of people who have contributed to the project over the years. The development began in 2001 as a complete rewrite of the original Fortran code Herwig, which was then available as Herwig 6.5 \cite{Marchesini:1984bm, Webber:1983if, Marchesini:1987cf, Corcella:2000bw, Corcella:2002jc}. The author list of this article comprises all contributors who have made significant contributions to the development of Herwig 7 since the publication of the last write-up for the Herwig++ 2 series \cite{Bahr:2008pv}. Over the years, many authors have joined the collaboration and later moved on, after making valuable contributions to the code. We would like to express our gratitude for their efforts by acknowledging their contributions to the various releases. These are, in alphabetical order:
Ken Arnold (2.6),
Manuel Bähr (M, 2.1, 2.2, 2.3),
Luca d’Errico (2.6),
Nadine Fischer (7.0),
Martyn Gigg (M, 2.1, 2.2, 2.3),
Keith Hamilton (M, 2.0, 2.1, 2.2, 2.3, 2.5, 2.6),
Marco Harrendorf (7.0),
Seyi Latunde-Dada (M, 2.1, 2.2), 
Radek Podskubka (7.1),
Daniel Rauch (7.0), 
Alberto Ribon (1.0, 2.0 $\beta$, 2.0),
Christian Röhr (2.5, 2.6, 2.7),
Pavel Růžička (2.5),
Alex Schofield (2.6),
Thomas Schuh (2.7), 
Alexander Sherstnev (M, 2.1, 2.2),
Phil Stephens (1.0, 2.0 $\beta$, 2.0),
Martin Stoll (2.6),
Louise Suter (2.5),
Jon Tully (M, 2.3),
Alexandra Wilcock (2.7, 7.0),
David Winn (2.5, 2.6),
Benedikt Zimmermann (2.7).
In parentheses we list their contributions to the previous release notes for the respective Herwig++ versions 1.0 \cite{Gieseke:2003hm}, 2.0 $\beta$ \cite{Gieseke:2006rr}, 2.0 \cite{Gieseke:2006ga}, 2.1 \cite{Bahr:2007ni}, 2.2 \cite{Bahr:2008tx}, 2.3 \cite{Bahr:2008tf}, 2.5 \cite{Gieseke:2011na}, 2.6 \cite{Arnold:2012fq}, 2.7 \cite{Bellm:2013hwb}, 7.0 \cite{Bellm:2015jjp}, 7.1 \cite{Bellm:2017bvx}, 7.2 \cite{Bellm:2019zci}, and 7.3 \cite{Bewick:2023tfi}, as well as authorship of the previous write-up (M) \cite{Bahr:2008pv}. At this point the name change from Herwig++ 2.x to Herwig 7 should be explained. As of version 7, we consider the program the true successor of the previous Fortran series, as all parts of the simulation are at least on an equal, and in many cases even superior, level of sophistication and detail.

A very special thank you goes to Bryan Webber, one of the principal authors of the original Herwig program and founders of the Herwig++ project (M, 1.0, 2.0 $\beta$, 2.0, 2.1, 2.2), whose guidance and enthusiasm have made him a constant source of inspiration and a true {\it spiritus rector\/} for our collaboration. His continuous engagement and generous advice, from the early days of the project to the present, have been invaluable.  
We would also like to acknowledge Pino Marchesini, whose pioneering contributions laid essential foundations for our work.  
Our gratitude extends to Leif L\"onnblad for his close collaboration and steadfast support of the ThePEG framework, which underpins much of our development.
Finally, we thank Torbj\"orn Sj\"ostrand and Frank Krauss
for many stimulating discussions, which have greatly enriched this project.

\clearpage

\section{Introduction}
\label{\detokenize{review/index:introduction}}\label{\detokenize{review/index:sec-intro}}

Event generators are indispensable tools for high energy physics. General-purpose event generators describe high energy reactions across vastly different energy scales, multiplicity regimes, and at all levels of detail thereby enabling direct comparison to measurements carried out at colliders. Within the LHC era, there are three multi-purpose event generators, which have extensively been improved as compared to their older versions. The \textsc{Pythia} event generator is now available as \textsc{Pythia~8}~\cite{Sjostrand:2007gs,Bierlich:2022pfr}, \textsc{Sherpa}~\cite{Gleisberg:2003xi,Sherpa:2024mfk} has been developed as an independent project, and the \textsc{ThePEG} framework~\cite{Lonnblad:2006pt} for multi-purpose event generators has emerged from the \textsc{Pythia~7} development.
In the present document we describe the \textsc{Herwig~7} event generator, based on \textsc{ThePEG}, which has originated from the \textsc{Herwig++} development and continues in the spirit of the old \textsc{HERWIG}~6 program. 

Multi-purpose event generators like \textsc{Herwig} describe high energy particle collisions in several stages, expecting factorization due to the disparate energy and length scales:
\begin{enumerate}
\item \emph{Elementary hard subprocess.} The incoming particles
  interact to produce the primary outgoing states. In lepton
  collisions these are the fundamental incoming leptons; in hadron
  collisions the incoming partons are extracted from the hadrons.
  The hard interaction is generally calculated at LO or NLO in
  perturbation theory. The energy scale of the hard scattering, together with
  its flow of colour, electric and weak charge, sets the initial conditions
  for subsequent QCD and EW parton showers.
\item \emph{Initial- and final-state parton showers.} Coloured
  particles are perturbatively evolved from the hard scale down to an
  infrared cutoff. QCD coherence can be implemented through angular ordering, and dipole showers provide an alternative to describe QCD parton showers in the large-$N_c$ limit. Corrections beyond the large-$N_c$ limit, and the quest for logarithmic accuracy 
  pose additional constraints.
  QED and electroweak emission can also be incorporated in these showers.
\item \emph{Decay of heavy objects.} Massive particles such as the top
  quark, EW bosons, Higgs bosons, and many BSM states decay
  on timescales comparable to parton showering. Radiation in both
  production and decay needs to be simulated consistently, ideally with full spin correlations preserved between stages.
\item \emph{Multiple scattering.} At high energies, multiple partonic
  interactions within a single hadron-hadron collision are
  significant. Eikonal multiple interaction models are central to the \textsc{Herwig} program, and can be extended to include diffractive contributions.
\item \emph{Hadronization.} After perturbative evolution, coloured
  partons are projected into hadrons. This can be done using the string model, or through projections on colour-singlet clusters
  \cite{Webber:1983if}, which decay into hadrons and resonances. Colour reconnection plays an important role in dense environments with many coloured particles.
\item \emph{Hadron decays.} Hadronic decays are typically modelled with
  matrix-element techniques, including off-shell effects and spin
  correlations, fully consistent with those in perturbative
  production and decay.
\end{enumerate}

Herwig is a general-purpose event generator for the simulation of
high-energy lepton-lepton, lepton-hadron and hadron-hadron collisions, built around the above paradigm. Particular emphasis is put on the accurate simulation of QCD radiation. It builds upon the heritage of the
\textsc{Herwig} Fortran program
\cite{Marchesini:1984bm,Webber:1983if,Marchesini:1987cf,Corcella:2000bw,Corcella:2002jc},
while providing a much more flexible modular structure for future
development. In comparison to the final Fortran version, \textsc{Herwig~7}
already incorporates a number of substantially more advanced features.
It provides a full simulation of high-energy collisions with the
following special components:

\begin{itemize}
  \item \textbf{Initial- and final-state QCD parton showers:}  
  Herwig provides two complementary shower algorithms. The traditional
  angular-ordered shower incorporates the coherence of soft-gluon
  emission through angular ordering, correctly reproducing the
  logarithmic structure of QCD radiation \cite{Gieseke:2003rz}. A dipole-based shower
  \cite{Platzer:2009jq,Platzer:2011bc} is also available, designed for
  improved treatment of colour coherence and for systematic NLO
  matching and multijet merging. Developments of logarithmic accuracy
  and initial-state radiation in the angular-ordered shower are
  discussed in~\cite{Bewick:2019rbu, Bewick:2021nhc}, the accuracy in the presence of massive quarks has been addressed in \cite{Hoang:2018zrp}, while studies of
  quark/gluon jet properties are presented in~\cite{Reichelt:2017hts}.
  Mass effects in both showers are extensively discussed in \cite{Cormier:2018tog}, and spin correlations in shower branchings are detailed
  in~\cite{Richardson:2018pvo}. Parton-shower uncertainties, along
  with their evaluation by reweighting, are analysed
  in~\cite{Bellm:2016rhh,Bellm:2016voq,Cormier:2018tog}. All Herwig showers dynamically reproduce heavy quark contributions and hence the
  \emph{dead-cone effect}~\cite{Marchesini:1989yk} by using quasi-collinear splitting functions. The sensitivity to the parton shower cutoff and the logarithmic accuracy for heavy quark contributions have been studied in
  detail in~\cite{Hoang:2024nqi}.

  \item \textbf{Electroweak showers:}  
  The simulation of the emission of quasi-collinear $W$ and $Z$ bosons
  has been developed to provide a consistent description of mixed QCD-EW
  radiation in high-energy collisions. The implementation of the
  EW parton shower in Herwig is described
  in~\cite{Masouminia:2021kne}. Electroweak corrections within hard processes have also been considered in~\cite{Gieseke:2014gka}.

  \item \textbf{Matching and merging to higher-order calculations:}  
  NLO QCD calculations are facilitated by the \textsc{Matchbox} framework \cite{Platzer:2011bc}. \textsc{Matchbox} automates NLO fixed-order and matched calculations, to either of the two Herwig shower algorithms, with both the MC@NLO and POWHEG matching methods. Interfaces to external matrix-element libraries allow a wide variety of processes to be calculated.
  Systematic NLO multi-jet merging \cite{Platzer:2012bs,Bellm:2017ktr} is also implemented in \textsc{Matchbox}, for the dipole shower.
  An alternative matching method, the KrkNLO method, 
  is described in \cite{Jadach:2016qti,Sarmah:2024hdk}.

  \item \textbf{Colour matrix element corrections} provide full-colour real emission corrections to QCD radiation in the dipole shower. Their implementation is discussed in detail in \cite{Platzer:2012np,Platzer:2018pmd}

  \item \textbf{Hadronization and non-perturbative dynamics:}  
  Jet hadronization is modelled using the cluster approach originally
  proposed in~\cite{Webber:1983if}. The model has been substantially
  extended in Herwig~7 to describe baryon production
  \cite{Gieseke:2017clv}, enhanced strangeness production
  \cite{Duncan:2018gfk}, and colour reconnection effects based on
  soft-gluon evolution~\cite{Gieseke:2018gff}. The interplay of the
  perturbative cutoff with hadronization is analysed
  in~\cite{Hoang:2024nqi}, where first steps towards a new hadronization model are proposed. Spin effects in the hadronization of heavy hadrons have been described in~\cite{Masouminia:2023zhb}.
  Colour-sextet diquarks are incorporated as discussed
  in~\cite{Richardson:2011df}.

  \item \textbf{Decays of partons and hadrons:}  
  A comprehensive framework is provided for the decays of hadrons and
  heavy particles. The treatment of multiple soft QED radiation is
  based on the Yennie--Frautschi--Suura (YFS) formalism~\cite{Hamilton:2006xz}.
  Radiation in top-quark decays is described in~\cite{Hamilton:2006ms},
  while general spin correlations in hadronic decays are implemented
  following~\cite{Richardson:2001df}. The simulation of hard radiation
  in BSM decays is given in~\cite{Richardson:2013nfo}, $\tau$-lepton
  decays are detailed in~\cite{Grellscheid:2007tt}, and a general
  framework for many-body BSM decays is provided
  in~\cite{Gigg:2007cr}.

  \item \textbf{BSM physics:}  
  Beyond-the-Standard-Model scenarios are supported through flexible
  interfaces to new models encoded via Feynman rules. Specific studies
  include parton-shower simulations of $R$-parity-violating
  supersymmetry, as well as collider
  phenomenology of split supersymmetry~\cite{Kilian:2005kr}.

  \item \textbf{Underlying event and soft physics:}  
  The description of the underlying event relies on an eikonal multiple
  parton-interaction model~\cite{Bahr:2008dy}, rooted in the earlier
  \textsc{Jimmy} approach~\cite{Butterworth:1996zw} and supplemented by
  non-perturbative extensions~\cite{Bahr:2008wk}. Subsequent developments
  include colour-reconnection models~\cite{Gieseke:2012ft} and a
  dedicated treatment of soft and diffractive scattering based on the
  cluster model~\cite{Gieseke:2016fpz}.
\end{itemize}

Some of these features were already present in the first release of
Herwig++~\cite{Gieseke:2003hm} 
However, there have been extensive improvements to both the physics
content and the internal structure of the simulation since that first
release, and Herwig~7 has added and consolidated many features on top of the first generations of Herwig++, ultimately replacing both the HERWIG 6 and Herwig++ developments. Given the significant
differences between the current version of the program, 7.3, and that
described in~\cite{Gieseke:2003hm}, we present here a comprehensive
description of the physics implemented in the code.
Updates to this manual will cover the novel features developed for the 7.4 series and beyond.

The program and additional technical documentation are available from

\begin{center}
   \url{https://herwig.hepforge.org} 
\end{center}

Herwig is distributed under the GNU General Public License (GPL),
version~3. This ensures that the source code remains available to
users, who may use and modify it under the conditions of the licence.
As Herwig is developed within an academic research collaboration and
represents the result of many years of work, it is accompanied by
guidelines~\footnote{These guidelines are contained in the
\texttt{GUIDELINES} file distributed with the release and are also
available from
\url{http://www.montecarlonet.org/publications_guidelines}.}
agreed within the MCnet collaboration. These set out expectations of
responsible use, in particular the citation of relevant physics
publications. The designated reference for each Herwig release (this
manual for versions~7.0 onwards) should always be cited, along with
the original literature on which a given study relies. To facilitate
this, Herwig automatically produces a \LaTeX\ file listing the
appropriate primary citations for the modules used in any given run.
The authors are happy to assist users in determining the correct
citations for their studies.

The structure of this manual closely follows the general workflow of an event generator. In Sec.~3 we discuss calculations of the hard process, including the interfaces to external matrix element libraries which can provide input for matrix elements. Sec.~4 focuses on the two available parton showers, the angular ordered shower and the dipole shower. Sec.~5 presents how parton showers and hard matrix elements at leading and next-to-leading order can be consistently combined with each other using matching and merging algorithms. Sec.~6 is devoted to the simulation of physics beyond the standard model, while Sec.~7 discusses hadronization and colour reconnection. Sec.~8 describes how \textsc{Herwig} models multiple parton interactions and the beam remnants. Finally, hadronic decays are covered in Sec.~9, before we briefly summarize what has been presented in this manual, and give
pointers to
further resources. Each section discusses the physics assumptions and models, as well as the user flags and some of the code structure. We devote appendices to tuning, and some of the core algorithms used.

\clearpage

\section{Hard processes}
\label{\detokenize{review/index:hard-processes}}\label{\detokenize{review/index:sect-hard}}

\subsection{Overview}
\label{\detokenize{review/hardprocess/general:hard-processes-overview}}\label{\detokenize{review/hardprocess/general:sect-hard-process}}\label{\detokenize{review/hardprocess/general::doc}}

\subsubsection{Philosophy}
\label{\detokenize{review/hardprocess/general:philosophy}}

The improved physics implementation of the hard process, specifically the
approaches that include higher-order corrections and the production of additional
jets, require a very detailed control over the hard process input, typically
beyond the level that can be achieved by using event files. Within Herwig
7 the philosophy for generating the hard process has hence been one of
automation, making use of external libraries to calculate scattering
amplitudes, while all of the other methods required to assemble full NLO QCD
matched and/or merged cross sections are provided within the program
itself.

The core module responsible for the set up of hard cross sections along with
the subtractions required for matching, and the further handling of
contributions to form a multijet-merged simulation, is called
Matchbox. Matchbox contains modules for process bookkeeping, diagram
generation and mapping, the creation of subtraction terms, phase-space
sampling, scale choices, generation cuts and parton shower matching to both
available shower modules. Amplitudes for a small number of processes are
available through in-house calculations shipped with a Herwig release, while
Matchbox in general requires external amplitude libraries to work in a truly
automated fashion.

In previous Herwig releases, hard processes have been implemented in a small
library of built-in matrix elements for Herwig++, some of them at
NLO. These can be used for the simulation of dedicated processes and are
documented in the section on \hyperref[\detokenize{review/hardprocess/me:builtin-matrix-elements}]{Section \ref{\detokenize{review/hardprocess/me:builtin-matrix-elements}}}.

\subsubsection{General components and workflow}
\label{\detokenize{review/hardprocess/general:general-components-and-workflow}}

The Matchbox module is centred around a Factory object responsible for
assembling hard process cross sections using external input at the level of
amplitudes, where interfaces are inherited from the MatchboxAmplitude base
class. Details of the code are documented in
\hyperref[\detokenize{review/hardprocess/code-structure:matchbox-code-structure}]{Section \ref{\detokenize{review/hardprocess/code-structure:matchbox-code-structure}}}.

Once a process at a certain coupling order has been requested,
Matchbox determines the possibly contributing partonic
subprocesses based on typical restrictions such as charge conservation
(these restrictions can be lifted if required for physics simulation
beyond the Standard Model), scans the repository for MatchboxAmplitude
objects which claim to be capable of calculating the desired
process either at tree level, loop level or both and will then
determine Feynman diagram topologies; a similar treatment is applied
to the real emission process. The diagram information is used to
determine both the subtraction terms required to render the individual
cross section contributions finite, as well as to set up a
multi-channel phase-space generator to perform efficient sampling of
events, see \hyperref[\detokenize{review/hardprocess/tree-level:phase-space-generation}]{Section \ref{\detokenize{review/hardprocess/tree-level:phase-space-generation}}}.

Since each amplitude can represent more than one process through the use of
crossing, amplitudes are hooked into the actual (subtracted) matrix element
objects which represent a contribution to the total cross section. If a
matching to a parton shower is requested, Matchbox will also supplement the
subtraction terms by the matching subtractions and re-arrange the cross
section into Born/virtual, integrated subtraction, and real/shower subtracted
contributions. Details of the matching algorithm are discussed in
\hyperref[\detokenize{review/matching/general:matching-merging}]{Section \ref{\detokenize{review/matching/general:matching-merging}}}.

Additional components are used to perform the jet finding and cuts on
various objects in the final state of the hard process; also a number
of (dynamical) scale choices is provided within the module. The
merging of cross sections for the production of different jet
multiplicities is described in \hyperref[\detokenize{review/matching/merging:multijet-merging}]{Section \ref{\detokenize{review/matching/merging:multijet-merging}}}.

Matchbox has been used and is used to perform fixed-order calculations
for state-of-the-art processes. One of the earliest examples where it
has been used for challenging processes is the full EW Higgs
boson plus three jet production at Next-to-Leading Order (NLO) QCD
\cite{Campanario:2013fsa, Campanario:2018ppz}, from which example
results are shown is shown in \hyperref[\detokenize{review/hardprocess/general:h3jets-results}]{Fig.\@ \ref{\detokenize{review/hardprocess/general:h3jets-results}}}.

\begin{figure}[tp]
\centering
\capstart
\begin{subfigure}{0.49\textwidth}
\centering
\noindent\includegraphics[width=\linewidth]{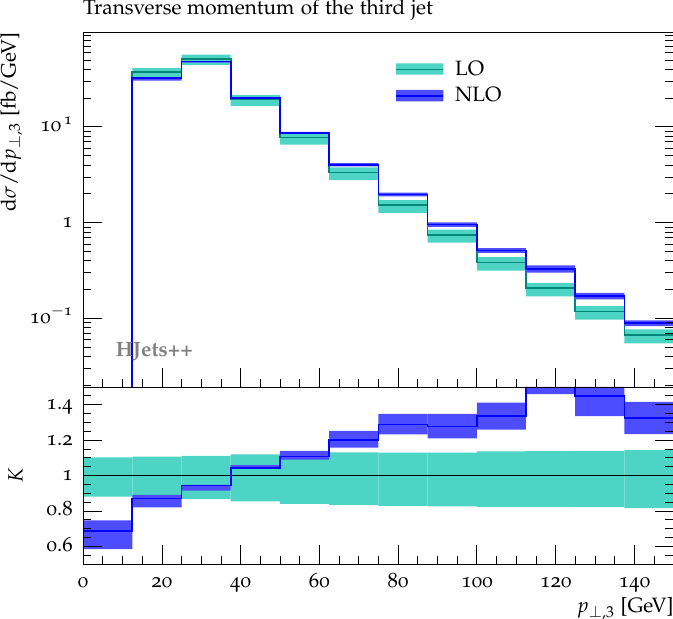}
\end{subfigure}
\begin{subfigure}{0.49\textwidth}
\centering
\noindent\includegraphics[width=\linewidth]{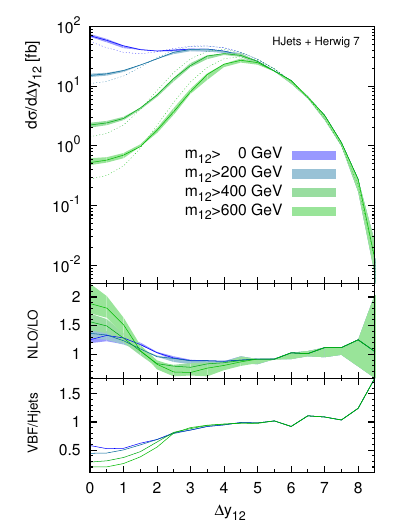}
\end{subfigure}
\caption{\label{\detokenize{review/hardprocess/general:id5}}\label{\detokenize{review/hardprocess/general:h3jets-results}}Sample results from fixed-order calculations of VBF processes obtained with Matchbox,
using the HJets plugin described in Section \ref{\detokenize{review/hardprocess/amplitude-providers:hjets}}. The left panel shows the spectrum of the third jet in Higgs-plus-three-jet events at NLO QCD using the full EW process. The right panel shows a comparison of the rapidity difference of the tagged jets, comparing the VBF approximation to the full calculations in events with one Higgs and three jets.}\end{figure}

\subsubsection{Input file steering}
\label{\detokenize{review/hardprocess/general:input-file-steering}}

In preparation for the Herwig 7 release the input files were simplified in
order to reflect the more generic handling of the hard process. Several
sequences of logically interrelated commands have been grouped into small
building block input files called snippets and should serve as the main source
of steering the hard process generation and the selection of matching paradigm
and hard process accuracy within Herwig. A small number of changes to the
example input files are required to set up a hard process, notably the
\begin{itemize}
\item {} 

incoming and outgoing particles of the hard process, also referring to
groups of particles such as jets,

\item {} 

desired coupling orders,

\item {} 

scale choice and generation cuts, and the

\item {} 

matching algorithm and hard process accuracy.

\end{itemize}

As such, a typical Matchbox-based collider simulation is set up with
an input file enabling the Matchbox module and making adjustments for
a particular collider type before setting the couplings and defining
the process of interest. More details of these input files are covered
in the \href{https://herwig.hepforge.org/tutorials/index.html}{herwig-tutorials}. Tree-level and loop-induced
leading-order cross sections are both supported, and phase-space generation can
be adapted to the latter, see \hyperref[\detokenize{review/hardprocess/tree-level:phase-space-generation}]{Section \ref{\detokenize{review/hardprocess/tree-level:phase-space-generation}}}.

\subsection{Leading-Order computations}
\label{\detokenize{review/hardprocess/tree-level:leading-order-computations}}\label{\detokenize{review/hardprocess/tree-level:leading-order}}\label{\detokenize{review/hardprocess/tree-level::doc}}

The starting point for all cross-section calculations are cross sections at the
Leading Order (LO), mostly mediated through tree-level scattering amplitudes.%
\footnote{
If matrix elements for loop-induced processes are provided, these can be
used in leading-order computations too. The details for the kinematic mapping
in the phase-space generation differ slightly in this case, as described in
\hyperref[\detokenize{review/hardprocess/tree-level:phase-space-generation}]{Section \ref{\detokenize{review/hardprocess/tree-level:phase-space-generation}}}.
}
In this section we outline the basic definitions and set some notation, before
presenting the algorithmic techniques used for the generation of leading-order
cross sections.
Advanced topics relevant for NLO computations, as well as matching and merging
are covered in \hyperref[\detokenize{review/hardprocess/nlo:nlo-subtraction}]{Section \ref{\detokenize{review/hardprocess/nlo:nlo-subtraction}}} as well as \hyperref[\detokenize{review/matching/general:matching-merging}]{Section \ref{\detokenize{review/matching/general:matching-merging}}}
respectively.

\subsubsection{LO cross section}
\label{\detokenize{review/hardprocess/tree-level:lo-cross-section}}\label{\detokenize{review/hardprocess/tree-level:cross-section}}

The LO cross section, $\sigma^{\mathrm{LO}}[\obu]$,
is given by integrals over longitudinal momentum fractions,
$x_a,x_b$,
of two (incoming) initial state partons,
and the Lorentz invariant phase-space,
$\phi_n$,
of $n$ (outgoing) final-state partons.
In a notation in anticipation of \hyperref[\detokenize{review/hardprocess/nlo:nlo-subtraction}]{Section \ref{\detokenize{review/hardprocess/nlo:nlo-subtraction}}},
we write the LO cross section for the scattering of two incoming hadrons,
$A,B$,
weighted by an observable $u$,
as
\begin{equation*}
\begin{split}&\sigma^{\mathrm{LO}}[\obu]
=\int\df\,\dSig^{\mathrm{B}}(\rmPhi_n)\,\obu(\rmPhi_n)\\
&=\sum\limits_{a,b}\sum\limits_{\{n|ab\}}\int
\df_{ab}(x_a,x_b)\,
\dSig^{\mathrm{B}}_{ab}\!\left(\rmPhi_{n|ab}|p_a(x_a),p_b(x_b)\right)
\obu\!\left(\rmPhi_{n|ab}|p_a(x_a),p_b(x_b)\right)\,,\end{split}
\end{equation*}

where we sum over all possible partonic configurations, also referred
to as the hard processes or subprocesses, $a,b\rightarrow
f_n=\{n|ab\}$, that can occur in the hadronic scattering.  The
longitudinal momentum fractions define the partonic initial state
momenta, $p_a,p_b$, in terms of the incoming hadron momenta,
$P_A,P_B$, through $p_a=p_a(x_a)=x_aP_A$ and
$p_b=p_b(x_b)=x_bP_B$ respectively.  Momentum conservation in
the partonic centre-of-mass frame,
$p_a(x_a)+p_b(x_b)=Q=p_1+\dots+p_n$, with a total momentum
$Q$, tells us that the (Lorentz invariant) partonic final-state
phase-space depends implicitly on the longitudinal momentum fractions,
i.e.\ $\rmPhi_n=\{\rmPhi_{n|ab}|p_a(x_a),p_b(x_b)\}=\{\rmPhi_{n|ab}|x_aP_A,x_bP_B\}$.
More details on phase-space generation, and the mapping of
$\phi_n$ to a unit hypercube, will be covered in
\hyperref[\detokenize{review/hardprocess/tree-level:phase-space-generation}]{Section \ref{\detokenize{review/hardprocess/tree-level:phase-space-generation}}}.

The differential cross section for the subprocess
$a,b\rightarrow f_n=\{n|ab\}$ is given by the Born matrix
element (or amplitude) squared, weighted by the parton distribution
functions for the incoming partons and the flux factor (plus
additional factors for the unit conversion),
\begin{equation*}
\begin{split}\frac{\df_{ab}\,\dSig^{\mathrm{B}}_{ab}}{\dPhi_{n|ab}\,\dd x_a\,\dd x_b}
&=\frac{\hbar^2c^2}{\mathcal{F}\!\left(p_a(x_a)p_b(x_b)\right)}
\,f_{a/A}(x_a,\mu_F)\,f_{b/B}(x_b,\mu_F)\times\\
&\times\mathcal{B}_{n|ab}\!\left(\rmPhi_{n|ab}|p_a(x_a),p_b(x_b)\right)
\theta_c\!\left(\rmPhi_{n|ab}|p_a(x_a),p_b(x_b)\right)\,,\end{split}
\end{equation*}

and where the Born matrix element squared is essentially evaluated as
the square of a vector in spin and colour space
\footnote{
Here we assume the leading-order (or Born) cross section for
$n$ final-state partons to be defined through tree-level
matrix elements ${\cal M}^{(0)}_n$. This is different for
loop-induced processes, for which the lowest order matrix elements
contain loops. If the corresponding matrix elements are provided,
these can be used in leading-order computations too. The details
for the kinematic mapping in the phase-space generation differ
slightly in this case, as described in
\hyperref[\detokenize{review/hardprocess/tree-level:phase-space-generation}]{Section \ref{\detokenize{review/hardprocess/tree-level:phase-space-generation}}}.
}
$\langle {\cal
M}^{(0)}|{\cal M}^{(0)}\rangle$,
\begin{equation*}
\begin{split}&\mathcal{B}_{n|ab}\!\left(\rmPhi_{n|ab}|p_a(x_a),p_b(x_b)\right)\\
&=N_{n|ab}
\Big\langle{\cal M}^{(0)}_{n|ab}\!\left(\rmPhi_{n|ab}|p_a(x_a),p_b(x_b)\right)
\Big|\,{\cal M}^{(0)}_{n|ab}\!\left(\rmPhi_{n|ab}|p_a(x_a),p_b(x_b)\right)\Big\rangle\,,\end{split}
\end{equation*}

with the additional weight
$N_{n|ab}=1 / (S_{n|ab}\,n_s(a,b)\,n_c(a,b))$ containing
symmetry factors for identical final-state partons and averaging
factors for initial state spin and colour degrees of freedom.  The
default convention for amplitudes within the Matchbox framework is
that all physical processes are calculated by crossing an amplitude
with all legs outgoing to the reaction under consideration, though
this can be overwritten within the amplitude interfaces to be
described in \hyperref[\detokenize{review/hardprocess/amplitude-providers:amplitude-providers}]{Section \ref{\detokenize{review/hardprocess/amplitude-providers:amplitude-providers}}}. The amplitude interfaces also
provide the possibility to return results for squared matrix elements
directly. The choice of a basis for colour space and the evaluation of
the matrix element squared will be discussed in detail in
\hyperref[\detokenize{review/hardprocess/tree-level:colour-basis}]{Section \ref{\detokenize{review/hardprocess/tree-level:colour-basis}}}. The spin degrees of freedom are currently summed
over, though some amplitude providers can provide spin density
matrices.

The flux factor is $\mathcal{F}(p_ap_b)=4p_ap_b=2x_ax_bS$ and
$S$ denotes the hadronic centre-of-mass energy.  The definition
of the total cross section requires a choice for renormalization and
factorization scales $\vec{\mu}=(\mu_R,\mu_F,\mu^{\text{QED}})$,
to be discussed in detail in \hyperref[\detokenize{review/hardprocess/scales:scale-choices}]{Section \ref{\detokenize{review/hardprocess/scales:scale-choices}}}, and a set of
acceptance criteria, or phase-space cuts $\theta_c(\rmPhi_n)$
applied to the kinematic configuration (on the Monte Carlo generation
level), which will be covered in \hyperref[\detokenize{review/hardprocess/cuts-and-jets:cuts-and-jet-definitions}]{Section \ref{\detokenize{review/hardprocess/cuts-and-jets:cuts-and-jet-definitions}}}.  The
cross section $\sigma^{\mathrm{LO}}[\obu]$ is a functional of
the infrared-safe observable $\obu(\rmPhi_n)$, also defined over
the phase-space configuration $\rmPhi_n$. In order for it to be
calculated correctly, the phase-space cuts $\theta_c(\rmPhi_n)$
must not be stronger than those implied by $\obu(\rmPhi_n)$
(that is, there must be no region of $\rmPhi_n$ in which
$\theta_c$ is zero but $\obu$ is non-zero).  Events are
distributed according to the differential cross section, normalized to
the total cross section.

\subsubsection{Process setup}
\label{\detokenize{review/hardprocess/tree-level:process-setup}}

Subprocesses are typically specified by groups of allowed incoming partons,
and desired outgoing partons, as well as a coupling power which here refers to
the cross section at the lowest order in perturbation theory. At the moment
only powers of the strong and electromagnetic coupling are considered, with
the assumption that Yukawa interactions are mediated by an electromagnetic
coupling. For loop-induced processes, the proper powers of couplings required
to make the loop graphs need to be chosen.

Groups of particles can also be provided as the initial- and final-states of a
process, e.g.\ to supply a list of jet constituents, charged leptons, or the
like. For all possible combinations the MatchboxFactory class then
determines if the subprocess satisfies charge conservation, colour
conservation, conservation of lepton and quark number, and possibly the
compatibility with flavour-diagonal interactions if desired. These
restrictions help reducing the complexity of further bookkeeping tasks but can
be relaxed individually.

Once a set of subprocesses compatible with the restrictions is identified, the
list of available MatchboxAmplitude objects is considered to provide the
actual matrix element calculations and further steps required for the assembly
of a complete cross section. Notably, MatchboxAmplitude objects will be
queried if they are able to provide matrix elements for the process considered
and will then be used to build full matrix element objects, as well as
instances of the XComb objects responsible for caching information associated with
a given phase-space point. In the case that several amplitude objects claim
responsibility for a given subprocess, the user can choose to prioritize some
implementations amongst others.

A typical matrix element implementation will then invoke a
Tree2toNGenerator to determine the diagrams contributing to the
process, given the coupling powers considered as well as a set of
vertices to be considered. The diagrams are determined by a clustering
procedure on the basis of the list of vertices, however their use is only
in aiding some phase-space generator implementations, to determine a
set of subtraction terms contributing to a given process at NLO, and
to provide (unphysical) information when assigning mother-daughter
relations to the partons contributing to a hard subprocess. For an
overview of the core components and code structure see
\hyperref[\detokenize{review/hardprocess/code-structure:matchbox-code-structure}]{Section \ref{\detokenize{review/hardprocess/code-structure:matchbox-code-structure}}}.

\subsubsection{Low-level amplitude interfaces}
\label{\detokenize{review/hardprocess/tree-level:low-level-amplitude-interfaces}}

While Matchbox fully supports interfaces at the level of matrix elements
squared, the default choice is to work at the level of colour and helicity
dependent amplitudes. Once a colour basis has been fixed and the contributing
colour structures have been enumerated, a matrix element implementation needs
to be able to supply the amplitude multiplying a given colour structure, for a
fixed set of spins of the external legs.
$2s+1$ combinations are
considered for a spin $s$ particle, however after a number of
evaluations of the matrix element only those combinations are kept for which
the implementation has not returned a zero amplitude, i.e.\ we assume that such
combinations vanish for all phase-space points. A similar treatment is
foreseen for colour structures where up to now we rely on the colour basis
object (see \hyperref[\detokenize{review/hardprocess/tree-level:colour-basis}]{Section \ref{\detokenize{review/hardprocess/tree-level:colour-basis}}}) supplied with the matrix element implementation to
be aware of the colour structures actually contributing to a given process.

The amplitude values obtained at a given phase-space point are cached together
in a MatchboxXComb object, and the colour basis object will then be
responsible for evaluating squared matrix elements, and subsequently colour and
spin correlated matrix elements required in NLO calculations, see
\hyperref[\detokenize{review/hardprocess/nlo:nlo-subtraction}]{Section \ref{\detokenize{review/hardprocess/nlo:nlo-subtraction}}}. Using the spin dependent amplitudes, spin density
matrices can also be evaluated and be supplied to the spin correlation
algorithms in our decay and parton shower modules. Full flexibility is
provided as far as the inclusion of fixed reference or running couplings,
averaging factors, and other conventions relevant to virtual corrections are
concerned.

\subsubsection{BLHA-type interfaces}
\label{\detokenize{review/hardprocess/tree-level:blha-type-interfaces}}

BLHA-type interfaces are an interface level for evaluating cross sections at
the level of amplitudes squared, in the spirit of directly implementing
interfaces of the Binoth Les Houches Accord (BLHA) \cite{Alioli:2013nda}. This is
the preferred mechanism to obtain one-loop virtual corrections for NLO
calculations, and we hence typically refer to amplitude
interfaces using this paradigm as ‘One-Loop Providers (OLPs)’. OLP interfaces
do inherit from the base MatchboxAmplitude object but are limited to the case
of matrix elements squared, interferences of one-loop with tree level
amplitudes, and the direct evaluation of colour and spin correlated
amplitudes. They can hence fully be used for fixed-order computations, however
if a physically sensible assignment of colour flows to the hard process for
subsequent showering is desired, amplitude information is required. If not
available, colour flows will be randomly assigned without any specific
weighting.

\subsubsection{Handling of colour bases}
\label{\detokenize{review/hardprocess/tree-level:handling-of-colour-bases}}\label{\detokenize{review/hardprocess/tree-level:colour-basis}}

For a fixed colour basis a general (QCD) amplitude $|{\cal M}\rangle$
can be decomposed in terms of a finite set (dimension $d_c$) of colour
structures $|\alpha\rangle$ as
\begin{equation*}
\begin{split}|{\cal M}\rangle = \sum_{\alpha=1}^{d_c}{\cal M}_\alpha |\alpha\rangle \ .\end{split}
\end{equation*}

Examples for such bases are the so-called trace basis or colour flow basis,
both available in the Matchbox module.  Suppressing spin indices for
readability, we consider colour amplitudes to fill a complex vector
${\cal M}_\alpha$ and information on the colour basis is handled through
linear algebra in this complex vector space.

In general colour bases are not orthogonal, thus calculating a squared matrix element
or similarly for NLO calculations a Born/one-loop interference requires
knowledge of a scalar product matrix $S_{\alpha\beta}=\langle \alpha|\beta\rangle$,
in terms of which a colour-summed, squared matrix element is calculated
as $|{\cal M}|^2 ={\cal M}^*_\alpha S_{\alpha\beta} {\cal M}_\beta$.

Similarly, colour-correlated matrix elements -- needed for NLO corrections --
can be expressed as
\begin{equation*}
\begin{split}\langle {\cal M}| {\mathbf T}_i\cdot {\mathbf T}_j |{\cal M}\rangle =
{\cal M}^*_\alpha \ T^\dagger_{i,\alpha\gamma}\ S'_{\gamma\delta}\ T_{j,\delta\beta}\ {\cal M}_\beta\ ,\end{split}
\end{equation*}

where $S'$ is the scalar product matrix for a final state with
an additional parton and the $T_{i,\alpha\gamma}$ are
appropriate representations of the colour charge operators, see
e.g.~\cite{Platzer:2011bc}. Once $T^{*}_{i,\alpha\gamma}\
S'_{\gamma\delta}\ T_{j,\delta\beta}$ is determined, the
problem reduces to linear algebra, which we perform with the boost
library \cite{boost}.

While setting up a process calculation, the amplitudes used initialize
the colour basis object, which calculates the scalar product matrices
and colour charge representations required. The results are then
stored in an external file for further reference. Currently, some
built-in colour bases for low multiplicities are available, and we also
release a copy of the ColorFull package \cite{Sjodahl:2014opa}, as
well as some components of the CVolver library
\cite{Platzer:2013fha, Martinez:2018ffw} with appropriate interfaces
to Matchbox amplitudes.

In order to generate colour flow information for the hard process,
the ColourBasis objects provide information on colour connections and
the respective partial subamplitudes to evaluate the weights by which
a certain colour flow is selected. In all cases we use, a single
colour flow can be assigned from a given colour structure and we
choose those with weights
\begin{equation*}
\begin{split}w_\alpha = \frac{|{\cal M}_\alpha|^2}{\sum_\beta |{\cal M}_\beta|^2} \ .\end{split}
\end{equation*}

\subsubsection{Phase-space generation}
\label{\detokenize{review/hardprocess/tree-level:phase-space-generation}}\label{\detokenize{review/hardprocess/tree-level:id11}}

The Lorentz invariant phase-space for $n$ outgoing particles with
momenta $p_i$ and masses $m_i$, and of total four-momentum $Q$ is
given by
\begin{equation*}
\begin{split}{\rm d}\phi_n (p_1,m_1,...,p_n,m_n|Q) = (2\pi)^4 \; \delta\left(\sum_{i=1}^n p_i-Q\right)
\prod_{i=1}^n \frac{{\rm d}^4 p_i}{(2\pi)^3} \; \delta(p_i^2-m_i^2) \; \theta(p_i^0)\end{split}
\end{equation*}

A phase-space generator maps random numbers to physical momenta,
\begin{equation*}
\begin{split}\phi_n = \{p_1,...,p_n\}= \Phi_n(\vec{r})\end{split}
\end{equation*}

where the mapping from random numbers $\vec{r}$ to the actual momenta is
in general not unique, and more random numbers might be required then there
are actual degrees of freedom. An importance sampling phase-space generator
determines a mapping in such a way that the resulting Jacobian factor broadly
resembles the inverse of the behaviour of the cross section to integrate in
order to ensure a fast convergence of the Monte Carlo integration.

Phase-space generators within the Matchbox module inherit from a general
interface, and are also responsible for assigning a tree-structure to a hard
process that represents the diagram that most probably contributes
the bulk of the cross section at this point. This is, by default, done based
on the propagator and coupling structure of the topologies encountered,
evaluated with the momenta which have just been generated. Indeed, if
$D_i=p_i^2-M_i^2+i M_i\Gamma_i$ is the propagator for the $i$-th
internal line, and $g_k$ is the (canonical) coupling of vertex
$k$, we associate a weight
\begin{equation*}
\begin{split}w_{(E,V)} = \prod_{\text{lines }i\in E} \prod_{\text{vertices }k\in V}
\frac{|g_k |^2}{(|D_i|^2)^\xi} f(|p_i^2|)\end{split}
\end{equation*}

where $\xi$ and $f$ are additional quantities depending on
the particle species propagating along the given line, and a topology
is selected with a probability proportional to this weight.

The phase-space generators can, if desired, generate the incoming legs’
momentum fraction either directly or indirectly through generating the other
momenta, though the default way of operation is that the incoming momenta have
been fixed by the responsible PartonExtractor instance and the phase-space
generators solely need to determine a set of final-state momenta.

\paragraph{The TreePhasespace generator}
\label{\detokenize{review/hardprocess/tree-level:the-treephasespace-generator}}

The TreePhasespace generator uses the diagram information supplied previously
to determine several different mappings according to decomposing the phase-space
measure as
\begin{equation*}
\begin{split}{\rm d}\phi_n = \frac{1}{\sum_{(E',V')} w_{(E',V')}} \sum_{(E,V)} w_{(E,V)}
\left. {\rm d}\phi_n\right|_{\phi_n = \Phi_{E,V}(\vec{r})}\end{split}
\end{equation*}

The tree mappings are organized as $1\to 2$ splittings of either
incoming to incoming and outgoing (‘spacelike’) or outgoing to two outgoing
(‘timelike’) particles and are iteratively used to build up the tree diagram
identified by $V,E$.

The factorization underlying timelike splittings is the standard phase-space
factorization for decays,
\begin{equation*}
\begin{split}{\rm d}\phi_n(p_1,m_1,...,p_i,m_i,...,p_j,m_j,...,p_n,m_n|Q) =\\ {\rm
d}m_{ij}^2\,\theta(m_{ij} -m_i-m_j)\times\\ {\rm
d}\phi_{n-1}(p_1,m_1,...,p_{ij},m_{ij},...,p_n,m_n|Q)\,{\rm
d}\phi_2(p_i,p_j|p_{ij})\end{split}
\end{equation*}

For spacelike splittings the situation is more complicated. Here we use a
composition similar to the one above, however we consider scattering of a
possibly off-shell and on-shell leg, and use a light-cone type decomposition
for the outgoing momentum considered. This way, rungs on a $t$-channel
ladder are subsequently removed until a core two to two scattering can be
generated. As the procedure has a preferred direction, we randomly select one
of the incoming legs to start the algorithm. Note that we could have chosen to randomize over splittings performed on either side of the ladder, however the structure of the algorithm is simplified if only one off-shell incoming leg needs to be considered. In particular, we can use the lightlike momentum of the on-shell incoming leg in a decomposition of the momenta to be generated. Once all space-like splittings
have been performed, the time-like ones are initiated using the standard decay
factorization.

In both cases the virtualities of the intermediate line $ij$ or
$ia$ are determined in a way to cancel the behaviour of the propagator
factor $D_{ij}$ or $D_{ia}$ such that the Jacobian is adapting to
the structure of the weight for this topology, with the expectation that the
normalization of the topology weights will resemble the leading behaviour of
the (tree level) cross section. The procedure of successive splitting proceeds
until a 2-to-2 (sub-)diagram is encountered for which the final momenta are
then determined to cancel either the $s$- or $t$-channel we are
left with. Since we do introduce a preferred direction by applying splittings
from one incoming leg towards the other one, the choice of direction the
algorithm proceeds in is randomized.

\paragraph{Phase-space generation for loop-induced processes}
\label{\detokenize{review/hardprocess/tree-level:phase-space-generation-for-loop-induced-processes}}\label{\detokenize{review/hardprocess/tree-level:loopinduced-phasespace}}

For setting up more suitable kinematic mappings in the case of loop
induced processes we introduce fictitious particles, and the
corresponding vertices and couplings, to emulate the kinematics
transmitted through loops.
The TreePhasespace generator will find associated fictitious
diagrammatic information in that case, and assemble the corresponding
weights in the kinematic mappings.
Currently implemented are versions of such fictitious particles
corresponding to colour and charge flowing through loops
($ccLP-$ and $ccLP+$), only colour ($cLP$), or none
of the above ($nLP$), more specifically the ones necessary for
loop-induced $W^+W^-$ production at the LHC, which means only
colour octets and integer charges.
For example for the loop-induced process $gg\rightarrow W^+W^-$,
we need such configurations as the vertices $(g,g,nLP)$, with
coupling $\alpha_s^2\alpha_e^0$, and $(nLP,W^+,W^-)$, with
coupling $\alpha_s^0\alpha_e^2$, with an intermediate
$nLP$ particle in an $s$-channel configuration, or as the vertices
$(ccLP-,g,W^+)$, with coupling $\alpha_s^1\alpha_e^1$, and
$(ccLP+,g,W^-)$, with coupling $\alpha_s^1\alpha_e^1$,
with an intermediate $ccLP$ particle in a $t$-channel
configuration.

\paragraph{The RAMBO generator}
\label{\detokenize{review/hardprocess/tree-level:the-rambo-generator}}

The RAMBO phase-space generator performs flat phase-space sampling as detailed
in \cite{Kleiss:1985gy, Platzer:2013esa}; it is used for cross checks and phenomenological
studies and not recommended as a production algorithm.

\paragraph{The FlatInvertible generator}
\label{\detokenize{review/hardprocess/tree-level:the-flatinvertible-generator}}

The FlatInvertible phase-space generator performs flat phase-space sampling as
detailed in \cite{Platzer:2013esa}; this is the only phase-space generator
that provides a unique and invertible mapping from $3n-4$ random
numbers to $3n-4$ physical degrees of freedom in the phase-space
point. It is used for pre-sampling matrix element correction factors and not
recommended as a production algorithm.

\paragraph{Two-to-one processes}
\label{\detokenize{review/hardprocess/tree-level:two-to-one-processes}}

The phase-space generators described above do not apply for two-to-one
processes and are overwritten in the Matchbox framework, by a routine that
generates a rapidity for the final-state system, and determines the incoming
momentum fractions from this rapidity and the system’s mass.

\subsection{Next-to-Leading Order computations}
\label{\detokenize{review/hardprocess/nlo:next-to-leading-order-computations}}\label{\detokenize{review/hardprocess/nlo:nlo-subtraction}}\label{\detokenize{review/hardprocess/nlo::doc}}

\subsubsection{NLO cross section}
\label{\detokenize{review/hardprocess/nlo:nlo-cross-section}}\label{\detokenize{review/hardprocess/nlo:id1}}

For NLO computations, we follow the paradigm of NLO subtraction and
matching, further described in \hyperref[\detokenize{review/matching/general:matching-merging}]{Section \ref{\detokenize{review/matching/general:matching-merging}}}, as outlined in
\cite{Bellm:2015jjp, Platzer:2011bc} and further in
\cite{Cormier:2018tog}, and as detailed in
\cite{Catani:1996vz, Catani:2002hc}.  Within the subtraction paradigm
the NLO cross section, e.g.\ for a process with $n$ final-state
partons at the Born level, is written as
\begin{equation*}
\begin{split}\sigma^{\mathrm{NLO}}[\obu]
=\sigma^{\mathrm{LO}}[\obu]
+\sigma^{\mathrm{R-A}}[\obu]
+\sigma^{\mathrm{V+A+C}}[\obu]\,,\end{split}
\end{equation*}

where $\sigma^{\mathrm{LO}}$ denotes the LO or Born cross
section, $\sigma^{\mathrm{R-A}}$ the NLO subtracted
real-emission cross section, and $\sigma^{\mathrm{V+A+C}}$ the
combination of NLO virtual cross section, analytically integrated
subtraction terms and collinear counterterms (PDF counterterms) for
initial state partons.  The LO or Born cross section was already
discussed in \hyperref[\detokenize{review/hardprocess/tree-level:cross-section}]{Section \ref{\detokenize{review/hardprocess/tree-level:cross-section}}}. Below we will briefly outline how
these cross sections are calculated, focusing on the real-subtracted
cross section which is the most relevant for matching to parton
showers.

\paragraph{NLO subtracted real-emission cross section}
\label{\detokenize{review/hardprocess/nlo:nlo-subtracted-real-emission-cross-section}}\label{\detokenize{review/hardprocess/nlo:nlo-real-cross-section}}

The NLO subtracted real-emission cross section is written as
\begin{equation*}
\begin{split}\sigma^{\mathrm{R-A}}[\obu]
=\int\df&\Big[\dSig^{\mathrm{R}}(\rmPhi_{n+1})\,\obu(\rmPhi_{n+1})
-\tsum\limits_{i}\dSig^A_{(i)}(\rmPhi_{n+1})\,\obu\!\left(\rmPhiTilde_{n|(i)}(\rmPhi_{n+1})\right)\Big]\\
=\sum\limits_{a,b}\sum\limits_{\{n+1|ab\}}\int
\df_{ab}(x_a,x_b)
&\Big[\dSig^{\mathrm{R}}_{ab}\!\left(\rmPhi_{n+1|ab}|p_a(x_a),p_b(x_b)\right)
\obu\!\left(\rmPhi_{n+1|ab}|p_a(x_a),p_b(x_b)\right)\\
&-\tsum\limits_{i}\dSig^A_{(i)}\!\left(\rmPhi_{n+1|ab}|p_a(x_a),p_b(x_b)\right)
\obu\!\left(\rmPhiTilde_{n|(i)}\!\left(\rmPhi_{n+1|ab}|p_a(x_a),p_b(x_b)\right)\right)\Big]\,,\end{split}
\end{equation*}

where
\begin{equation*}
\begin{split}\frac{\df_{ab}\,\dSig^{\mathrm{R}}_{ab}}{\dPhi_{n+1|ab}\,\dd x_a\,\dd x_b}
&=\frac{\hbar^2c^2}{\mathcal{F}\!\left(p_a(x_a)p_b(x_b)\right)}
\,f_{a/A}(x_a,\mu_F)\,f_{b/B}(x_b,\mu_F)\times\\
&\times\mathcal{R}_{n+1|ab}\!\left(\rmPhi_{n+1|ab}\Big|\,p_a(x_a),p_b(x_b)\right)
\theta_c\!\left(\rmPhi_{n+1|ab}|p_a(x_a),p_b(x_b)\right)\,,\end{split}
\end{equation*}

with
\begin{equation*}
\begin{split}&\mathcal{R}_{n+1|ab}\!\left(\rmPhi_{n+1|ab}|\,p_a(x_a),p_b(x_b)\right)\\
&=N_{n+1|ab} \;
\Big\langle{\cal M}^{(0)}_{n+1|ab}\!\left(\rmPhi_{n+1|ab}|\,p_a(x_a),p_b(x_b)\right)
\Big|\,{\cal M}^{(0)}_{n+1|ab}\!\left(\rmPhi_{n+1|ab}|p_a(x_a),p_b(x_b)\right)\Big\rangle\,,\end{split}
\end{equation*}

and where
\begin{equation*}
\begin{split}\frac{\df_{ab}\,\dSig^{\mathrm{A}}_{(i)}}{\dPhi_{n+1|ab}\,\dd x_a\,\dd x_b}
&=\frac{\hbar^2c^2}{\mathcal{F}\!\left(p_a(x_a)p_b(x_b)\right)}
\,f_{a/A}(x_a,\mu_F)\,f_{b/B}(x_b,\mu_F)\times\\
&\times\mathcal{A}_{(i)}\!\left(\rmPhi_{n+1|ab}|p_a(x_a),p_b(x_b)\right)
\theta_c\!\left(\rmPhiTilde_{n|(i)}\!\left(\rmPhi_{n+1|ab}|p_a(x_a),p_b(x_b)\right)\right)\,,\end{split}
\end{equation*}

with
\begin{equation*}
\mathcal{A}_{(i)}\!\left(\rmPhi_{n+1|ab}|p_a(x_a),p_b(x_b)\right)
=
N_{n+1|ab}\; n_c(a,b) \;
\,\mathcal{D}_{(i)}\!\left(\rmPhi_{n+1|ab}|p_a(x_a),p_b(x_b)\right)\,.
\end{equation*}

$\{\rmPhi_{n+1|ab}|p_a,p_b\}$ denotes the phase-space of the hard process $a,b\rightarrow f_{n+1}=\{n+1|ab\}$.
The corresponding differential phase-space is $\dPhi_{n+1|ab}=\dPhi_{n+1|ab}(\rmPhi_{n+1|ab}|p_a,p_b)$.
Here we have again $p_a=p_a(x_a)=x_aP_A$, $p_b=p_b(x_b)=x_bP_B$, and $\mathcal{F}(p_ap_b)=4p_ap_b=2x_ax_bS$.
$N_{n+1|ab}=\frac{1}{S_{n+1|ab}\,n_s(a,b)\,n_c(a,b)}$ accounts for symmetry factors for identical final-state partons,
as well as for the averaging over spin and colour degrees of freedom of the initial state partons, which are denoted by $n_s(a,b)$
and $n_c(a,b)$ respectively.

The sum over $i$ is taken over all possible dipole configurations,
where the $\mathcal{D}_{(i)}\in\{\mathcal{D}_{ij,k},\mathcal{D}_{ij}^a,\mathcal{D}^{ai}_k,\mathcal{D}^{ai,b}\}$ denote the dipole subtraction terms,
as defined in \cite{Catani:1996vz, Catani:2002hc}.
The dipole subtraction terms depend on spin- and colour-correlated Born matrix elements,
further discussed in \hyperref[\detokenize{review/hardprocess/nlo:correlated-me}]{Section \ref{\detokenize{review/hardprocess/nlo:correlated-me}}}.
The $\rmPhiTilde_{n|(i)}(\rmPhi_{n+1|ab}|p_a,p_b)$ denote the corresponding so-called reduced tilde kinematic mappings,
and for later reference we further define the inverse tilde kinematic mappings $\rmPhiTilde_{n+1|(i)}$,
such that
\begin{equation*}
\begin{split}\rmPhiTilde_{n+1|(i)}\!\left(\rmPhiTilde_{n|(i)}(\rmPhi_{n+1|ab}|p_a,p_b),R_{(i)}(\rmPhi_{n+1|ab}|p_a,p_b)\right)
=\{\rmPhi_{n+1|ab}|p_a,p_b\}\,,\end{split}
\end{equation*}

or in short
\begin{equation*}
\begin{split}\rmPhiTilde_{n+1|(i)}\!\left(\rmPhiTilde_{n|(i)}(\rmPhi_{n+1}),R_{(i)}(\rmPhi_{n+1})\right)
=\rmPhi_{n+1}\,,\end{split}
\end{equation*}

and where $R_{(i)}$ denotes the set of additional emission variables --
an emission scale, a momentum fraction and an azimuthal variable --
corresponding to the dipole configuration $i$.

\paragraph{NLO virtual cross section and integrated subtraction terms}
\label{\detokenize{review/hardprocess/nlo:nlo-virtual-cross-section-and-integrated-subtraction-terms}}\label{\detokenize{review/hardprocess/nlo:nlo-virtual-cross-section}}

The combination of NLO virtual cross section,
analytically integrated subtraction terms and collinear counterterms (PDF counterterms) for initial state partons,
is cast in terms of the insertion operators $\mathbf{I}$, $\mathbf{P}$ and $\mathbf{K}$,
such that
\begin{equation*}
\begin{split}\sigma^{\mathrm{V+A+C}}[\obu]
=\sigma^{\mathrm{V+I}}[\obu]
+\sigma^{\mathrm{P+K}}[\obu]\,,\end{split}
\end{equation*}

The explicit poles in dimensional regularisation cancel between the (UV renormalised) NLO virtual cross section,
the integrated subtraction terms and the PDF counterterms,
such that $\sigma^{\mathrm{V+A+C}}$ can be integrated numerically in four
dimensions.
The $\mathbf{I}$ operator contains all the explicit poles in dimensional regularisation,
from the integrated subtraction terms and the PDF counterterms,
that are necessary to cancel the IR poles in the NLO virtual cross section.
The $\mathbf{P}$ and $\mathbf{K}$ operators contain finite remainders,
which are left after the factorisation of initial state collinear singularities into the PDFs. The $\mathrm{V+I}$ contributions have identical kinematics to the leading order contribution, while the $\mathrm{P+K}$ contribution contains an additional convolution over an incoming momentum fraction, as will be discussed in more detail below.

By default,
all contributions that are considered to be virtual corrections are evaluated together with the leading order contribution.
For efficiency reasons,
however,
the contributions from the virtual corrections together with the ${\mathbf I}$ operator,
i.e.\ $\sigma^{\mathrm{V+I}}$,
and the contributions from the ${\mathbf P}$ and ${\mathbf K}$ operators,
i.e.\ $\sigma^{\mathrm{P+K}}$,
can be treated as independent contributions to the same subprocess and will in this case be integrated separately.
The partons which are summed in the insertion operator contributions are all inferred from the content of the particle groups defining the proton and jet content,
including the possibility of massive jet constituents.

Several internal cross-checks can be performed for the cancellation of $\epsilon$ terms,
and the independence of phase-space restrictions on the subtraction terms and other finite-term ambiguities,
see \hyperref[\detokenize{review/hardprocess/internal-checks:internal-checks}]{Section \ref{\detokenize{review/hardprocess/internal-checks:internal-checks}}}

\paragraph{The V+I contribution}
\label{\detokenize{review/hardprocess/nlo:the-v-i-contribution}}\label{\detokenize{review/hardprocess/nlo:vi}}

The V+I contribution is written as
\begin{equation*}
\begin{split}\sigma^{\mathrm{V+I}}[\obu]
=\int\df&\Big[\dSig^{\mathrm{V}}(\rmPhi_n)+\dSig^{\mathrm{B}}(\rmPhi_n)\otimes\mathbf{I}(\epsilon)\Big]
\obu(\rmPhi_n)\\
=\sum\limits_{a,b}\sum\limits_{\{n|ab\}}\int
\df_{ab}(x_a,x_b)
&\Big[\dSig^{\mathrm{V}}_{ab}\!\left(\rmPhi_{n|ab}|p_a(x_a),p_b(x_b)\right)\\
&+\dSig^{\mathrm{B}}_{ab}\!\left(\rmPhi_{n|ab}|p_a(x_a),p_b(x_b)\right)
\otimes\mathbf{I}(\epsilon)\Big]
\obu\!\left(\rmPhi_{n|ab}|p_a(x_a)p_b(x_b)\right)\,,\end{split}
\end{equation*}

where
\begin{equation*}
\begin{split}\frac{\df_{ab}\,\dSig^{\mathrm{V}}_{ab}}{\dPhi_{n|ab}\,\dd x_a\,\dd x_b}
&=\frac{\hbar^2c^2}{\mathcal{F}\!\left((p_a(x_a)p_b(x_b)\right)}
\,f_{a/A}(x_a,\mu_F)\,f_{b/B}(x_b,\mu_F)\times\\
&\times \mathcal{V}_{n|ab}\!\left(\rmPhi_{n|ab}|p_a(x_a),p_b(x_b)\right)
\theta_c\!\left(\rmPhi_{n|ab}|p_a(x_a),p_b(x_b)\right)\,,\end{split}
\end{equation*}

with
\begin{equation*}
\begin{split}&\mathcal{V}_{n|ab}\!\left(\rmPhi_{n|ab}|p_a(x_a),p_b(x_b)\right)\\
&=N_{n|ab}\,
2\mathrm{Re} \Big\langle{\cal M}^{(0)}_{n|ab}\!\left(\rmPhi_{n|ab}|p_a(x_a),p_b(x_b)\right)
\Big|\,{\cal M}^{(1)}_{n|ab}\!\left(\rmPhi_{n|ab}|p_a(x_a),p_b(x_b)\right)\Big\rangle\,,\end{split}
\end{equation*}

and where
\begin{equation*}
\begin{split}\frac{\df_{ab}(\dSig^{\mathrm{B}}_{ab}\otimes\mathbf{I})}{\dPhi_{n|ab}\,\dd x_a\,\dd x_b}
&=\frac{\hbar^2c^2}{\mathcal{F}\!\left(p_a(x_a)p_b(x_b)\right)}
\,f_{a/A}(x_a,\mu_F)\,f_{b/B}(x_b,\mu_F)\times\\
&\times \left(\mathcal{B}_{n|ab}\!\left(\rmPhi_{n|ab}|p_a(x_a),p_b(x_b)\right)\otimes\mathbf{I}(\epsilon)\right)
\theta_c\!\left(\rmPhi_{n|ab}|p_a(x_a),p_b(x_b)\right)\,,\end{split}
\end{equation*}

with
\begin{equation*}
\begin{split}&\mathcal{B}_{n|ab}\!\left(\rmPhi_{n|ab}|p_a(x_a),p_b(x_b)\right)\otimes\mathbf{I}(\epsilon)\\
&=N_{n|ab}\,n_c(a,b)\,
\Big\langle{\cal M}^{(0)}_{n|ab}\!\left(\rmPhi_{n|ab}|p_a(x_a),p_b(x_b)\right)
\Big|\,\mathbf{I}(\epsilon)\,\Big|\,{\cal M}^{(0)}_{n|ab}\!\left((\rmPhi_{n|ab}|p_a(x_a),p_b(x_b)\right)\Big\rangle\,,\end{split}
\end{equation*}

containing the colour-correlated Born matrix elements,
with the colour correlations induced by the $\mathbf{I}$ operator,
as defined in \cite{Catani:1996vz, Catani:2002hc}.

The V+I contribution contains $\dSig^{\mathrm{V}}_{ab}(\rmPhi_{n|ab}|x_aP_A,x_bP_B)$ and $\mathbf{I}(\epsilon)\otimes\dSig^{\mathrm{B}}_{ab}(\rmPhi_{n|ab}|x_aP_A,x_bP_B)$,
both evaluated at the same partonic phase-space point $\{\rmPhi_{n|ab}|x_aP_A,x_bP_B\}$ with which also the LO contribution as well as scales, cuts and observables are calculated.
While the (UV-renormalized) virtual one-loop matrix elements $|\mathcal{M}^{(1)}_n(\rmPhi_n)\rangle$ with $n$ final-state partons,
contained in $\dSig^{\mathrm{V}}_{ab}(\rmPhi_{n|ab}|x_aP_A,x_bP_B)$,
are received from the one-loop amplitude providers, see \hyperref[\detokenize{review/hardprocess/amplitude-providers:amplitude-providers}]{Section \ref{\detokenize{review/hardprocess/amplitude-providers:amplitude-providers}}},
$\mathbf{I}(\epsilon)\otimes\dSig^{\mathrm{B}}_{ab}(\rmPhi_{n|ab}|x_aP_A,x_bP_B)$ is implemented as described in \cite{Catani:1996vz, Catani:2002hc}.

\paragraph{The P+K contribution}
\label{\detokenize{review/hardprocess/nlo:the-p-k-contribution}}\label{\detokenize{review/hardprocess/nlo:pk}}

The implementation of the P+K contribution is a bit more involved.
The $\mathbf{P}$ and $\mathbf{K}$ operators contain finite remainders,
which are left after the factorization of initial state collinear singularities into the PDFs.
This involves an integration over an additional initial state momentum fraction $z$ and,
in line with the definitions in \cite{Catani:1996vz} and the main text of \cite{Catani:2002hc},
a continuous reevaluation of the underlying Born matrix elements,
in the convolution of the initial state dependent $\mathbf{P}$ and $\mathbf{K}$ operators,
at the corresponding underlying Born phase-space points $\{\rmPhi_{n|a'b}|zx_aP_A,x_bP_B\}$ and $\{\rmPhi_{n|ab'}|x_aP_A,zx_bP_B\}$.

Continuously reevaluating matrix elements, even Born matrix elements, can be computationally expensive,
and is thus not ideal for an implementation in Monte Carlo event generators.
A change of variables $x_a\rightarrow\frac{x_a}{z}$ or $x_b\rightarrow\frac{x_b}{z}$,
together with a rearrangement of integrations and summations,
can be performed,
such that the corresponding correlations are shifted from the underlying Born matrix elements to correspondingly redefined PDFs.
In this parameterization the P+K contribution is written as
\begin{equation*}
\begin{split}\hspace{-25ex}
\sigma^{\mathrm{P+K}}[\obu]\\
=
\sum\limits_{a',b}\sum\limits_{\{n|a'b\}}\int
\df_{a'b}\!\left(\tfrac{x_a}{z},x_b\right)
\otimes&\Big[\big(\mathbf{P}(z,x_aP_A,\mu_F)+\mathbf{K}(z)\big)_{a'}
\otimes\dSig^{\mathrm{B}}_{a'b}\!\left(\rmPhi_{n|a'b}|x_aP_A,x_bP_B\right)\Big]\times\\
\times&\obu\!\left(\rmPhi_{n|a'b}|x_aP_A,x_bP_B\right)\\
+
\sum\limits_{a,b'}\sum\limits_{\{n|ab'\}}\int
\df_{ab'}\!\left(x_a,\tfrac{x_b}{z}\right)
\otimes&\Big[\big(\mathbf{P}(z,x_bP_B,\mu_F)+\mathbf{K}(z)\big)_{b'}
\otimes\dSig^{\mathrm{B}}_{ab'}\!\left(\rmPhi_{n|ab'}|x_aP_A,x_bP_B\right)\Big]\times\\
\times&\obu\!\left(\rmPhi_{n|ab'}|x_aP_A,x_bP_B\right)\,,\end{split}
\end{equation*}

where the differential P+K contribution for the first incoming parton is written as
\begin{equation*}
\begin{split}&\frac{\df_{a'b}\otimes((\mathbf{P}\!+\!\mathbf{K})_{a'}\otimes\dSig^{\mathrm{B}}_{a'b})}{\dPhi_{n|a'b}\,\dd z\,\dd x_a \,\dd x_b}\\
&=\frac{\hbar^2c^2}{\mathcal{F}(x_aP_Ax_bP_B)}\,f_{b/B}(x_b,\mu_F)\,N_{n|a'b}n_c(a',b)\times\\
&\times{\big\langle{\cal M}^{(0)}_{n|a'b}\Big|\,
\tsum\limits_{a}\theta(z-x_a)f_{a/A}\!\left(\tfrac{x_a}{z},\mu_F\right)\!\tfrac{1}{z}\big(\mathbf{P}_{aa'}\!\left(z,x_aP_A,\mu_F\right)+\mathbf{K}_{aa'}(z)\big)
\Big|\,{\cal M}^{(0)}_{n|a'b}\big\rangle}\times\\
&\times\theta_c\!\left(\rmPhi_{n|a'b}|x_aP_A,x_bP_B\right)\,,\end{split}
\end{equation*}

with ${\cal M}^{(0)}_{n|a'b}={\cal M}^{(0)}_{n|a'b}(\rmPhi_{n|a'b}|x_aP_A,x_bP_B)$,
and similarly for the second incoming parton.

The advantage of the above parameterization is that the matrix
elements of the hard underlying Born process are evaluated at the
phase-space points $\{\rmPhi_{n|a'b}|x_aP_A,x_bP_B\}$ and
$\{\rmPhi_{n|ab'}|x_aP_A,x_bP_B\}$, which are kept fixed during
$z$-integration and match the $x$-dependencies of those in
the LO and V+I contributions, at the expense of evaluating the PDFs at
the correspondingly reparameterized $x$-values
$\frac{x_a}{z}$ or $\frac{x_b}{z}$.

This can also be imagined as follows: If a splitting occurs with a
momentum fraction $z$ transferred to the child parton $a'$
of the splitting, which participates in the hard underlying Born
process, in order to keep $p_{a'}$ fixed the parent parton
$a$ is extracted from the proton $A$ with a
correspondingly higher momentum.

This reparameterization has further consequences, due to different quantities being fixed during $z$-integration.
Before the reparameterization: $p_a=x_aP_A$ is kept fixed, whereas $p_{a'}=zp_a=zx_aP_A$ is not.
After the reparameterization: $p_{a'}=x_aP_A$ is kept fixed, whereas $p_a=\frac{x_a}{z}P_A$ is not.
The prescription to match the integrated endpoint contributions is not readily applicable after reparameterization and finite terms have to be rearranged.
As a consequence,
compared to \cite{Catani:1996vz} and the main text of \cite{Catani:2002hc},
the $\mathbf{P}$ and $\mathbf{K}$ operators need to be modified.
This has no effect in the massless case and in general also not for the $\mathbf{P}$ operator,
but only for the $\mathbf{K}$ operator in the massive case,
for which it is necessary to make use of the formulation in Appendix B of \cite{Catani:2002hc}.

Matchbox sums over all required contributions of all those initial state partons $a$ or $b$ that might,
upon splitting,
give rise to the partons $a'$ or $b'$ participating in the hard underlying Born process.
It generates a list of hard underlying Born processes and for each the action of $(\mathbf{P}+\mathbf{K})$ is evaluated.
The folding with the rescaled PDFs is thereby included in the definition of the various terms contributing to $(\mathbf{P}+\mathbf{K})$.
The sums over the correlated products of rescaled PDFs times $(\mathbf{P}+\mathbf{K})$ are implemented in \sphinxtitleref{Dipole*PKOperator::sumParton(id)},
inheriting from \sphinxtitleref{MatchboxInsertionOperator},
where \sphinxtitleref{id} = $a$ or $b$.
For further details see \hyperref[\detokenize{review/hardprocess/code-structure:matchbox-code-structure}]{Section \ref{\detokenize{review/hardprocess/code-structure:matchbox-code-structure}}}.

\subsubsection{Process setup and analyses}
\label{\detokenize{review/hardprocess/nlo:process-setup-and-analyses}}

After a leading order process has been determined for the setup of an
NLO computation the amplitude instances claiming responsibility for
the leading order process are also queried for virtual corrections. A
number of conventions need to be communicated by the interface chosen,
which we discuss in detail below.

For the real emission contribution, the leading order process is
considered with an additional outgoing jet, at an increased order in
the strong coupling, and the partons entered in the jet particle group
determine what real emission processes are considered. The lookup
of these processes is subject to the same constraints as have already
been applied by searching for contributing leading order processes. An
enhanced bookkeeping that allows for mixed expansions and QED
corrections is currently under development.

While the Born and virtual contributions can be analysed just as plain
leading order phase-space points, care needs to be taken for the
subtracted real emission contribution. In an unmatched calculation
(which is determined by the MatchboxFactory object through the absence
of a ShowerApproximation object), real emission phase-space points are
generated, and the contributing ‘tilde’ kinematics are determined,
together with the weights governed by the individual subtraction
terms. We then form a so-called event group, consisting of a ‘head’
subprocess, along with a set of dependent phase-space points
reflecting the subtraction contributions. Such a composite needs to be
analysed in a correlated way, i.e.\ the individual subprocesses in the
group can in general not be viewed as independent events. If the
results of the calculation are to be written to HepMC events, Herwig
indeed writes out a sequence of independent events, however the fact
that they are correlated is signaled through an identical event
number for all of them.

A set of built-in analyses is available for fixed order calculations,
as well, which directly takes into account the correlations between
real emission and subtraction term contributions. These analyses also
apply a fuzzy bin definition in order to aid the convergence of
distributions and reduce the impact of mis-binning effects in a way
consistent with recovering the underlying integrable divergence in the
limit of zero smearing. A similar approach is applied to generation
cuts on the hard process where we choose a linear smearing,
corresponding to replacing the event contribution by a box
approximation to a $\delta$-function.

\subsubsection{One-loop amplitude interfaces}
\label{\detokenize{review/hardprocess/nlo:one-loop-amplitude-interfaces}}

Interfaces to one-loop amplitudes, or interferences of tree-level and one-loop
amplitudes, are implemented much along the lines of the tree level amplitude
providers; in contrast to the tree-level contributions to a NLO QCD cross
section, one-loop providers can also only exclusively provide input at the
cross-section level.

In any case, a convention needs to be adopted to define the finite part by
fixing what quantities are expanded in the dimensional regularization
parameter $\varepsilon$ prior to removing the poles of the virtual
contribution. We support one scheme where everything but a global prefactor of
$(4\pi)^\varepsilon/\Gamma(1-\varepsilon)$ is expanded, one where the
amplitudes are cast in a form similar to the one reported in
\cite{Bern:1993mq}, and one where they are cast into the form of the
insertion operators presented in \cite{Catani:1996vz, Catani:2002hc}.

Further to this information, the interfaces of the one-loop providers need to
communicate about the precise type of regularization (conventional dimensional
regularization or dimensional reduction), and their normalization scheme
($\overline{\text{MS}}$ or $\overline{\text{DR}}$). The insertion
operators will be adjusted accordingly. The philosophy of the virtual
corrections used inside Matchbox is that the t’Hooft mass parameter and the
actual renormalization scale are kept different, and the value of the former
will also be reported by the one loop interface. This allows for strong cross
checks as the t’Hooft mass $\mu$ needs to exactly cancel between the
integrated subtraction terms and the one-loop amplitudes, while the
renormalization scale dependence will only cancel out to ${\cal
O}(\alpha_s^2)$. In view of mixed expansions, however, we also allow the
amplitude providers, and the one-loop providers in particular, to use their
own running coupling in case of which the insertion operators will not contain
the logarithms relating the t’Hooft mass and renormalization scale as dictated
by the one-loop UV counter term.

\subsubsection{Generation of subtraction terms}
\label{\detokenize{review/hardprocess/nlo:generation-of-subtraction-terms}}

At the core of the framework for an NLO calculation in the subtraction
formalism is the identification of the subtraction terms that contribute to a
given process, along with the projection of the phase-space point to an
underlying Born process.

Within the Matchbox module, we perform the setup of the subtraction terms on a
diagrammatic basis as follows: For each contributing real emission diagram, two
external legs are clustered together, i.e.\ the vertex connecting exactly these
two external legs is removed from the diagram. The resulting partonic
subprocess is checked against the list of Born processes contributing to the
computation at leading order. If no such process has been found, the
clustering is not considered, while if there is a process the diagram obtained
after clustering is compared against the list of diagrams contributing to the
Born process. This allows dipoles to be ruled out that, e.g.~are not compatible
with an approximation leaving out a specific set of diagrams, as is the case in
VBF processes.

While the clustering determines how the momenta of an emission and emitter are
combined into the underlying Born phase-space point, the assignment of all
other momenta is not unique and we therefore compare the diagrams on the basis of
their topology to finally obtain a dictionary of how the other momenta need to
be assigned to the hard process. Once such a diagrammatic comparison and mapping
has been successfully obtained, a list of SubtractionDipole objects is checked
for a contributing subtraction term (indeed, these objects are queried at
several stages of the dipole identification process to rule out
non-contributing clusterings and spectator assignments as early as
possible).

All of the SubtractionDipole objects are then considered as functions
of the real emission phase-space point, and the subtraction cross
sections generically take the form described in \hyperref[\detokenize{review/hardprocess/nlo:nlo-real-cross-section}]{Section \ref{\detokenize{review/hardprocess/nlo:nlo-real-cross-section}}}.

\subsubsection{Spin- and colour-correlated Born matrix elements}
\label{\detokenize{review/hardprocess/nlo:spin-and-colour-correlated-born-matrix-elements}}\label{\detokenize{review/hardprocess/nlo:correlated-me}}

The subtraction dipoles
$\mathcal{D}_{(i)}\in\{\mathcal{D}_{ij,k},\mathcal{D}_{ij}^a,\mathcal{D}^{ai}_k,\mathcal{D}^{ai,b}\}$,
as defined in \hyperref[\detokenize{review/hardprocess/nlo:nlo-cross-section}]{Section \ref{\detokenize{review/hardprocess/nlo:nlo-cross-section}}} and in detail in \cite{Catani:1996vz, Catani:2002hc},
require colour-correlated and in the case of external state gluons at Born level
also spin-correlated Born-type matrix elements.

For example, in the case of gluons splitting into gluons or quarks we have
\begin{equation*}
\begin{split}\mathcal{D}_{\tilde{g},k}
\propto \left|{\cal M}_{a,b\to f_n}^{\tilde{g},k}\right|^2\end{split}
\end{equation*}

where
\begin{equation*}
\begin{split}\left|{\cal M}_{a,b\to f_n}^{\tilde{g},k}\right|^2
= -\Big\langle {\cal M}_{a,b\to f_n}^{\mu} \Big|
{\cal C}_{\mu\nu} \frac{{\mathbf T}_{\tilde{g}}\cdot {\mathbf T}_k}{C_A}
\Big| {\cal M}_{a,b\to f_n}^\nu \Big\rangle\end{split}
\end{equation*}

is a spin- and colour-correlated matrix element in the standard
notation using an open Lorentz index to denote the gluon’s polarization
degree of freedom. While an implementation at this level is possible
by representing the amplitude as an explicit, complex four-vector and
the spin correlation tensor as a four-by-four matrix, the information
content is significantly less in this case. In fact, by inserting a
polarization sum for the contractions of the amplitude and conjugate
amplitude with the spin correlation tensor, we find that, for
\begin{equation*}
\begin{split}{\cal C}_{\mu\nu} = {\cal A}\ \eta_{\mu\nu} +{\cal B} n_\mu n_\nu\end{split}
\end{equation*}

the correlated matrix element can as well be expressed as
\begin{equation*}
\begin{split}\left|{\cal M}_{a,b\to f_n}^{\tilde{g},k}\right|^2
&= -\left({\cal A}-{\cal B}|\epsilon_+\cdot n|^2\right)
\Big\langle {\cal M}_{a,b\to f_n} \Big|
\frac{{\mathbf T}_{\tilde{g}}\cdot {\mathbf T}_k}{C_A}
\Big| {\cal M}_{a,b\to f_n} \Big\rangle\\
&\hspace{2cm}+ 2{\rm Re} \left({\cal B}\ (n\cdot \epsilon_+)^2 \,
\Big\langle {\cal M}_{a,b\to f_n}^{(+)} \Big|
\frac{{\mathbf T}_{\tilde{g}}\cdot {\mathbf T}_k}{C_A}
\Big| {\cal M}_{a,b\to f_n}^{(-)} \Big\rangle \right)\end{split}
\end{equation*}

where instead of an open Lorentz index the spin information is now
encoded in a colour-correlated matrix element interfering a positive
and negative helicity amplitude for the gluon in question. In the case of
gluon emission from a quark only colour correlations exist, for which
we have
\begin{equation*}
\begin{split}\left|{\cal M}_{a,b\to f_n}^{\tilde{q},k}\right|^2
= -\Big\langle {\cal M}_{a,b\to f_n} \Big|
\frac{{\mathbf T}_{\tilde{q}}\cdot {\mathbf T}_k}{C_F}
\Big| {\cal M}_{a,b\to f_n} \Big\rangle\end{split}
\end{equation*}

The correlated matrix elements can be determined at various levels of
interfaces, either directly for the BLHA-type handling, or as a net
result from the colour basis data and the colour and helicity
subamplitudes. Our treatment of the spin correlation has also served
as the reasoning behind extensions of the BLHA standard \cite{Alioli:2013nda}
and proof of concept of an implementation thereof, see \cite{Andersen:2014efa}.
The spin and colour-correlated matrix
elements in each of the cases are stored in caches associated with the
phase-space point considered, and downstream use of these does not
require knowledge of the level of interface they have been obtained
with. Note, however, that in the case of an amplitude level interface
an agreement and possibly a method to return, a polarization vector
for the splitting gluon is required to use a consistent phase
definition.

\subsection{Cuts and jet definitions}
\label{\detokenize{review/hardprocess/cuts-and-jets:cuts-and-jet-definitions}}\label{\detokenize{review/hardprocess/cuts-and-jets:id1}}\label{\detokenize{review/hardprocess/cuts-and-jets::doc}}

At the level of the hard process of interest, acceptance cuts might need to be
imposed for a finite answer, if the underlying matrix element is singular in soft
or collinear limits.
They may also be imposed to increase the efficiency of the event generation.
The cuts discussed in this section are applied at the level of the hard process,
that is, to the partonic objects that subsequently undergo showering.

The successive emissions occurring within a parton shower involve momentum mappings
which distribute the recoil from each generated emission among the remaining particles.
This allows events that are below some cut threshold at one stage of the evolution to
migrate above it.
As such, it is necessary to carefully ensure that these cuts are compatible with the desired analysis criteria,
such that the exclusion of events due to the parton-level cuts has no effect
(or a negligible effect) on the cross section that passes the fiducial cuts
after showering and hadronisation.
A common heuristic is to choose (parton-level) generation cuts
that are significantly less restrictive than the analysis cuts
(for example, a factor of two in a transverse momentum cut),
and to test sensitivity for each analysis empirically by varying the cut.
The most restrictive generation cuts, for which no residual effect on the analysis cross sections
and histograms can be detected, will be the most efficient;
if an effect can be detected, that result depends on the (unphysical) generation cuts used,
indicating that they must be loosened to give physical predictions.%
\footnote{Note that these considerations do not apply to fixed-order generation,
for which there is no subsequent recoil and the generation and analysis cuts are applied to the same objects/phase-space(s).}

Within the process of implementing NLO QCD contributions in Herwig, the
framework for cuts needed to be revisited considerably as most of the cuts had
parton level origin and were hence problematic or impossible to be used in
the context of an NLO calculation. They are still provided in a legacy manner
and we document them towards the end of this section. The new cut
implementations have seen a twofold improvement:
\begin{itemize}
\item {} 

It is now possible that a jet clustering can take place prior to the actual
evaluation of a final state definition. This jet clustering is steered
through a Matcher object, which contains the information on the jet
constituents. Definitions involving, e.g.\ a democratic clustering algorithm
for photons are hence straightforward to implement.

\item {} 

Cuts on the hard process, in general, are not anymore implemented as
sharp theta-function constraints on the partonic cross section but can
actually reflect a definition involving some smearing function. This is
highly important in the context of convergence of NLO calculation, where
such flexibility allows for a (one-sided) bin smearing in the definition of
the cross section. Currently a linear smearing is provided, consistent with
approximating the delta function to represent a single event by a box
approximation.

\end{itemize}

\subsubsection{Jet finding}
\label{\detokenize{review/hardprocess/cuts-and-jets:jet-finding}}\label{\detokenize{review/hardprocess/cuts-and-jets:id2}}

ThePEG provides a general interface to jet finding algorithms through the
JetFinder class. This class, if present, will be considered to act on the
final state presented to the Cut objects to perform jet clustering prior to
the actual cut evaluation. Currently fastjet \cite{Cacciari:2011ma} is used
as the only working implementation behind this interface, with nearly all
options available to the jet finding algorithm. Jet finders make use of
Matcher objects, which classify which of the final state particles of the hard
process should be considered for jet clustering, and can also be synchronized
with ParticleGroups set up for the MatchboxFactory object.  The \href{https://thepeg.hepforge.org/doxygen/classThePEG\_1\_1FastJetFinder.html}{FastJetFinder} class
instantiates the more general \href{https://thepeg.hepforge.org/doxygen/classThePEG\_1\_1JetFinder.html}{JetFinder}
interface for using the fastjet package and currently supports the Kt, CA and
AntiKt algorithms, as well as their spherical counterparts to be used in the
context of electron-positron collisions. All algorithms can be operated with
two different recombination schemes.

After jets have been formed by the clustering algorithm, a definite jet final
state needs to be required. These constraints are collected in a \href{https://thepeg.hepforge.org/doxygen/classThePEG\_1\_1JetCuts.html}{JetCuts} object,
which assembles single or pairwise constraints on the jet final state, subject
to an ordering of jets. Ordering of jets can be applied through the
\sphinxtitleref{JetCuts:Ordering} interface and can be either set to ordering in
transverse momentum or in rapidity. Individual phase-space requirements on the
final state jets are encoded in JetRegion, JetPairRegion and MultiJetRegion
objects, see \hyperref[\detokenize{review/hardprocess/cuts-and-jets:jet-regions}]{Section \ref{\detokenize{review/hardprocess/cuts-and-jets:jet-regions}}} for more details. A \texttt{JetCuts} object
then assembles an entire jet final state definition composed of a Matcher
object set through \sphinxtitleref{JetCuts:UnresolvedMatcher}, an ordering choice, and
several jet regions, jet veto regions, jet pair regions as well as multi-jet
regions.

On top of this a simple jet multiplicity cut, possibly used together with an
exclusive resolution for the jet finder object is provided for typical jet
definitions in $e^+e^-$ collisions using the \href{https://thepeg.hepforge.org/doxygen/classThePEG\_1\_1NJetsCut.html}{NJetsCut} class.

\subsubsection{Jet and jet pair regions}
\label{\detokenize{review/hardprocess/cuts-and-jets:jet-and-jet-pair-regions}}\label{\detokenize{review/hardprocess/cuts-and-jets:jet-regions}}

Each JetRegion object represents a slice in phase-space composed out of an
interval in transverse momentum
$[p_{\perp,\text{min}},p_{\perp,\text{max}}]$ and a set of (mutually
disjoint) rapidity intervals
$([y_{1,-},y_{1,+}],....,`[y_{n,-},y_{n,+}])$ in which a jet is accepted
(no restrictions are applied in azimuthal angle in the transverse plane). Each
jet region can accept (or match) any, or only certain entries in the ordered
list of reconstructed jets, however only one jet per region is
allowed. Multi-jet final states can then be composed through combining several
JetRegion objects into a JetCuts object. Each jet region might also serve as a
jet veto region in case of which no jets are allowed in certain phase-space
slices, i.e.\ events for which such a region matches will not be accepted.

JetPairRegion objects can be composed out of two JetRegion objects and will
impose additional constraints on the pair of matched jets, such as distances in
rapidity, azimuth, lego-plot separation or invariant mass of the pair.
Finally, MultiJetRegion objects contain an arbitrary number of JetRegion
objects onto which, once matched, pairwise constraints on typical inter-jet
separations can be applied.

\subsubsection{Default cuts}
\label{\detokenize{review/hardprocess/cuts-and-jets:default-cuts}}

A number of cuts are available by default, specifically to be used
within NLO calculations and the Matchbox framework. Apart from jet
cuts the cuts in Herwig are defined either on individual, pair or
multi particle level (ThePEG’s OneCut, TwoCut or MultiCut).  Here the
property of the objects affected by the cuts are defined via
matchers. Internally the cut object will then check for this property
and apply the restrictions on the phase-space point.  For numerical
stability the cuts in Herwig can be and are by default defined with a
fuzzy cut instead of sharp cuts, i.e.\ we replace
\begin{equation*}
\begin{split}\theta(x-x_c) \to \theta_\lambda (x-x_c)\end{split}
\end{equation*}

where $\theta_\lambda (x-x_c)$ is a turn-on function, in
practice chosen to rise linearly from zero to one along an interval of
width $\lambda$, centred around $x_c$ in units of the cut
quantity considered. This width can be changed in the FuzzyTheta
object.

Several cut classes, which can be supplemented by Matcher objects to
identify the final state particles to which they will be applied, are
provided as a default. To be precise, Herwig provides the following
classes:
\begin{itemize}
\item {} 

IdentifiedParticleCut, which cuts on individual particle properties
like transverse momentum or rapidity;

\item {} 

InvariantMassCut, which checks for pairwise invariant masses;

\item {} 

MatchboxDeltaRCut which checks for pairwise lego plot separation;

\item {} 

MissingPtCut, cutting on the sum of invisible momenta such as
neutrinos; and

\item {} 

FrixionePhotonSeparationCut, providing photon isolation according to
\cite{Frixione:1998jh}.

\item {} 

JetCuts are instances of JetRegion objects outlined in
\hyperref[\detokenize{review/hardprocess/cuts-and-jets:jet-regions}]{Section \ref{\detokenize{review/hardprocess/cuts-and-jets:jet-regions}}}, they default to accept jets with $|y_j|<5$
and $p_{\perp,j}> 20 {\rm GeV}$.

\end{itemize}

Further details on using cut objects can be found in the \href{https://herwig.hepforge.org/tutorials/hardprocess/matchbox.html\#cuts}{tutorials}. This
library of typically used cuts however is nowhere
comprehensive. Custom constraints on the final state at the hard
process level can be implemented rather easily by inheriting from
ThePEG’s OneCut, TwoCut or MultiCut base classes or the JetRegion or
JetPairRegion classes. Even a jet clustering algorithm not provided by
our fastjet interface can be used within Herwig through a class
deriving from ThePEG’s JetFinder class. Any such implementation does
not require change or other interaction with an existing Herwig
installation but can be compiled into a shared library readily
available for a custom run.

\subsection{Scale choices}
\label{\detokenize{review/hardprocess/scales:scale-choices}}\label{\detokenize{review/hardprocess/scales:id1}}\label{\detokenize{review/hardprocess/scales::doc}}

Matchbox provides full flexibility regarding (dynamic) scale choices
for the hard process, and all dynamical scales inherit from the
MatchboxScaleChoice class. In particular, the QCD renormalization and
factorization scales $\mu_R$ and $\mu_F$, as well as the
QED renormalization scale $\mu_{R,QED}$ can be implemented
depending on the phase-space point considered. By default, if only an
implementation of $\mu_R$ is provided, we put
$\mu_F=\mu_R$, and $\mu_{R,QED}= M_Z$. In order to
guarantee a consistent subtraction calculation of the NLO QCD cross
sections, dynamic scales need to obey infrared safety in the sense
that their functional form depending on the real emission phase-space
point approaches the functional form defined on the Born phase-space
point in any of the unresolved limits:
\begin{equation*}
\begin{split}\mu_R(\phi_{n+1}|a b) \to \mu_R(\tilde{\phi}_{n|(i)}(\phi_{n|ab}))\end{split}
\end{equation*}

if the unresolved limit is associated to the dipole configuration
$(i)$. Any scale based on jets or jet-like objects fulfills this
property. Notice that this also applies to a dynamic QED scale even
in absence of NLO EW corrections.

In the case of NLO plus parton shower matched calculations, the hard
veto scale of the parton shower evolution can also be determined in
the scale setting object. This scale determines the maximally allowed
hardness of emissions generated in the parton shower, which is needed
to avoid a double counting of hard emissions in particular when the
hard process already involves additional jets. Note that a simple
prescription of taking the minimum transverse momentum at the hard
process to set the limit on parton shower emissions is not applicable
in the NLO matched context anymore. More details about this can be
found in the section on matching parton showers to NLO QCD
corrections. By default, the hard shower scale, which we refer to
as $\mu_Q$, is set equal to the implementation of the QCD
factorization scale choice, though a scale based on the
average transverse momentum of the hard objects is available in several
scale setting implementations and might become the default in a future
version.

The tutorials provide instructions on how custom scale choices can be
implemented by inheriting from the MatchboxScaleChoice base class,
while a range of implementations are available by default, among them
\begin{itemize}
\item {} 

Fixed Scale:

$\mu= \mu_{\rm fixed}$

The base class “MatchboxScaleChoice” implements a fixed scale choice that can be modified by the interface “FixedScale”.
This choice is for testing. Actual physics simulations should make use of a variable scale choice.

\item {} 

MatchboxHtScale:

$\mu= \sum_i p^i_T + M^{\text{non-jets}}_T$

This scale choice sums the scalar transverse momenta of the jets above a given scale, defined by the interface ‘JetPtCut’.
Furthermore, if the interface ‘IncludeMT’ is set to ‘Yes’, the $\MT$ of the vector resulting when adding all non jet
particles is added to the scale.

\item {} 

MatchboxParticlePtScale:

$\mu= \sum_i p^i_T \delta_{{\rm flav}(i) {\rm flav}(X) }$

This scale needs a ‘Matcher’ $X$, see “Matchers”, and then sums the scalar transverse momenta of all objects
that match the definition of $X$.

\end{itemize}

as well as a range of scales more suited to top quark production, and processes involving multiple gauge bosons.

\subsection{Cross-section integrators and event sampling}
\label{\detokenize{review/hardprocess/integrators:cross-section-integrators-and-event-sampling}}\label{\detokenize{review/hardprocess/integrators:sampling}}\label{\detokenize{review/hardprocess/integrators::doc}}

\subsubsection{Overview}
\label{\detokenize{review/hardprocess/integrators:overview}}

When calculating integrated cross sections, the basic problem is the
calculation of a many-dimensional integral $\sigma = \int {\rm d}^d x \
f(\vec{x})$, where $f(\vec{x})$ contains the matrix element, the
phase-space and any cuts, the latter parameterized as, possibly
smeared, step functions. Typically, the parameterization of the phase-space
is written such that it takes a vector of $d$ real numbers $\vec{x}$
ranging between $0$ and $1$. Then the integration is
performed over the $d$-dimensional unit hypercube $[0,1]^d$.

For a finite
number of sampled points $n$, the integral can be approximated by its
discrete sum, $\sigma_n = \frac1n \sum_{i=1}^n f(\vec{x}_i)$.
In the limit of $n\to\infty$, the law of large numbers ensures
that the discrete sum converges to the value of the integral. The
corresponding error can be approximated by an estimate for the variance,
so that we obtain
$\Delta\sigma_n = \frac1{\sqrt{n(n-1)}} \sqrt{\sum_i(f(\vec{x}_i) - \sigma_n)^2}$.

The contribution of each individual point to the sum can differ vastly.
From the estimate of the integration error we see that for a given
number of points this becomes smaller when the contribution of the
points is more uniform. In the ideal case each point would give the same
contribution, and a single point would hence be sufficient to exactly
determine the integral. As a step towards this goal, two important
techniques exist: 
\begin{itemize}
    \item Stratified sampling divides the integration region
into subparts, performs a Monte Carlo integration in each part, and then
sums the results;
    \item Importance sampling introduces a weight function
$p(x)$ into the integral, $\sigma = \int {\rm d}^d x \; p(\vec{x})
\ \frac{f(\vec{x})}{p(\vec{x})},$
which is positive-valued and
integrates to unity on the hypercube; 
then the integral can be computed by
sampling according to ${\rm d}^d \, P(\vec{x}) = p(\vec{x}) \;
{\rm d}^d x$.
Typically this requires the inverse function of the integral of
$p(\vec{x})$ to be available, either numerically or analytically.
If the weight function
approximates the true integrand reasonably well, the integration error
is significantly reduced.
\end{itemize}
Both techniques are also helpful when generating unweighted events, as
the generated events should ideally already have comparable weights.

The \sphinxtitleref{GeneralSampler} class manages the cross section integration
and is responsible for selecting among the different subprocesses
contributing to the cross section. Each subprocess is integrated using
a \sphinxtitleref{BinSampler}, which is a base class to the different adaptive
methods we offer (flat integration is of course available as a cross
check). The \sphinxtitleref{GeneralSampler} class also controls the final unweighting
step. The main switches used to control the sampling independent of
the adaptive algorithm chosen are:
\begin{itemize}
\item {} 

\sphinxtitleref{MaxEnhancement}

The actual maximum weight used for unweighting should be chosen
slightly higher than the largest weight found during the integration
step, so there is an additional safety margin for larger events
appearing in the run phase. This option specifies the factor by which
the weight is increased.

\item {} 

\sphinxtitleref{MinSelection}

Individual subprocesses are usually sampled according to their
relative weights. This option sets a minimal probability for how often a
subprocess must be chosen nevertheless.

\item {} 

\sphinxtitleref{AlmostUnweighted}

When using unweighted events and an individual event has a weight
larger than the guessed maximum weight, the event is still written out
with a unit weight. Hence, this introduces a difference on the cross
section. If this option is switched on, these events are output with
the correct weight, which is larger than one.

\item {} 

\sphinxtitleref{ParallelIntegration}

Setting this option allows the integration to be performed in parallel
using multiple jobs. The number of jobs can be controlled with the
following two options:
\begin{itemize}
\item {} 

\sphinxtitleref{IntegratePerJob}

This sets the number of subprocesses per integration job. The number
of jobs necessary is then calculated from the total number of
subprocesses.

\item {} 

\sphinxtitleref{IntegrationJobs}

This sets the number of integration jobs. The subprocesses are then
distributed equally between the different jobs.

\end{itemize}

\end{itemize}

The typical setup, which we follow in Herwig, is that for
all samplers at first a certain number of points with the starting
setup, a flat distribution on the hypercube of random numbers, is
performed. Then these results are used for the optimization procedure of
the sampler, i.e.\ splitting the integration space for stratified
sampling and calculating the appropriate weight function for importance
sampling. Then a new iteration is started, where again we collect points
and afterwards use the total results to improve the sampling. This
iterative procedure will then usually be performed a few times until the
integration space splitting or weight function is sufficiently well
adapted.

The following options are common to all adaptive algorithms inheriting
from the \sphinxtitleref{BinSampler} class and control the behaviour of the
integration routine, as defined in the \sphinxtitleref{BinSampler} class which gives
the common framework.
\begin{itemize}
\item {} 

\sphinxtitleref{InitialPoints}

This sets the number of points used in the first iteration of the
adaption.

\item {} 

\sphinxtitleref{NIterations}

The number of iterations is controlled with this setting. Note that
if results from a previous integration run exist already, only a
single iteration is performed, independently of the setting of this
variable, as we assume that the previous integration has already given
a reasonably well converging adaptation and only additional
improvements are necessary. If build parameters like masses or
phase-space cuts have been altered and a new integration starting from
scratch becomes necessary, the old integration results in
\sphinxtitleref{Herwig-scratch/} must be deleted manually. Herwig will also issue a
warning reminding you about this if previous integration data is
found.

\item {} 

\sphinxtitleref{EnhancementFactor}

Calculating the first iterations with fewer points and then
subsequently increasing the points with the following iterations can
be helpful to lower the time needed for integration, as the coarse
adaptation during the first few iterations can be done with fewer
points and the refinement later on is performed with higher
statistics. This parameter is the factor by which the number of
points increases from iteration to iteration.

\end{itemize}

For some samplers (specifically the \sphinxtitleref{CellGridSampler} and the
\sphinxtitleref{FlatSampler}), besides the normal behaviour of the algorithm, a
‘remapping’ can be applied to individual random number dimensions, in
which case projections of the integrand are recorded along these
axes and an importance sampling deduced from the histogram. This
particularly improves the sampling of those dimensions that
correspond to the incoming parton’s luminosity, though other options
can be chosen, see the \href{https://herwig.hepforge.org/tutorials/advanced/index.html\#integrators}{tutorials} for more details.

By default, two different samplers are available in Herwig, which are
described below. Additionally, for simple checks and debugging, an
identity mapping, meaning that the input random number vector is returned
as output unmodified, is implemented as \sphinxtitleref{FlatSampler}.

\subsubsection{CellGridSampler}
\label{\detokenize{review/hardprocess/integrators:cellgridsampler}}

The \sphinxtitleref{CellGridSampler} is based on the \sphinxtitleref{ExSample} library
\cite{Platzer:2011dr}. The integration hypercube is represented by a
tree of so-called cells, where the top cell represents the full unit
hypercube. This one and each intermediate cell contains two children
cells, which are derived from the parent cell by splitting it
along one of the integration dimensions.
The measure to evaluate possible splits of a dimension $k$ is
called the gain, and is defined as
\begin{align*}
\mathrm{gain}_k(r_k) = \frac{\left| \int_{\vec{r}_{-}}^{\vec{r}_{+,k}} {\rm d}\vec{r} \ f(\vec{r}) - \int^{\vec{r}_{+}}_{\vec{r}_{-,k}} {\rm d}\vec{r} \ f(\vec{r}) \right| }{ \int_{\vec{r}_{-}}^{\vec{r}_{+}} {\rm d}\vec{r} \ f(\vec{r}) },
\end{align*}
where $\vec{r}_{\pm}$ is the upper or lower bound of the
cell, respectively, and $\vec{r}_{\pm,k}$ denotes $r_k$ for
the random number in dimension $k$ and again the upper or lower
bound for all other dimensions. This means that the two integrals in the
numerator denote the two half cells which have been split at $r_k$
for dimension $k$ and span the whole cell space for the other
dimensions, whereas the denominator is simply the integral over the
whole cell, i.e.\ the gain is the normalized difference between the lower
left and upper right half of the cell. A cell split happens along the
dimension where the gain is largest. In the current version of the
\sphinxtitleref{CellGridSampler}, for performance reasons only splits along the middle
of the cell, i.e.\  $r_k=\frac{r_{+,k}-r_{-,k}}2$ are considered.
After a cell has been split, the new child cell which does not
contain the largest weight is immediately sampled again, so that we get
a good estimate of the maximum weight also for this half.

The usual splitting algorithm described so far can give
non-satisfactory results. One example would be an integrand which is
symmetric in one dimension, but strongly peaked at $\frac12$.
Then, the splitting algorithm would see a gain of 0 and thus never
consider this dimension for splitting, while in practice for example
an initial split into three equally sized pieces would have been
highly beneficial. Usually, these symmetries are relevant only at the
level of the full unit hypercube, though particular integrands,
possibly defined as step-wise functions, might exhibit this behaviour
also for sub-cells. Therefore, it is possible to generate some forced
splittings of the cells already before any sampling has taken place.
This then further refines the set of cells with which the algorithm
starts. The main parameters to be considered for this algorithm are
\begin{itemize}
\item {} 

\sphinxtitleref{ExplorationPoints}

To get an initial estimate of the size of the contributions, each cell
needs to be sampled with a certain number of points, before possible
splits can be evaluated. \sphinxtitleref{ExplorationPoints} defines how many points
are used for this.

\item {} 

\sphinxtitleref{Epsilon}

This variable is a cell-efficiency threshold applied to the ratio of the average
integration point weight over the maximum weight. If this ratio
exceeds the parameter \sphinxtitleref{Epsilon}, i.e.\ the distribution of weights is
sufficiently flat, the cell is no longer considered for splitting.

\item {} 

\sphinxtitleref{Gain}

Correspondingly, this variable defines the minimum value the gain must
reach before splitting the cell is considered.

\item {} 

\sphinxtitleref{MinimumSelection}

This variable enforces a minimum selection probability for each cell
that the algorithm guarantees, even if the integral value of the cell
would yield only a smaller probability.

\end{itemize}

\subsubsection{MonacoSampler}
\label{\detokenize{review/hardprocess/integrators:monacosampler}}

The \sphinxtitleref{MonacoSampler} is an implementation of the importance sampling
method \sphinxtitleref{Vegas} \cite{Lepage:1977sw, Lepage:1980dq}. It is derived from
the VBFNLO implementation of the same name.

The basic idea is a combination of importance sampling and stratified
sampling. Each integration dimension is split into a number of bins,
where the bin boundaries are chosen such that the integral over each bin
has the same value.
Over the full unit hypercube, this leads to a $d$-dimensional
grid of smaller hyperrectangles. The random points are then distributed
such that each hyperrectangle receives the same number of random points
and the distribution is uniformly inside each hyperrectangle, and the
weight function factorizes over the different dimensions,
$p(\vec{r}) = \prod_{i=1}^d p_i(r_i)$.
To calculate the bin boundaries, the integrand is first sampled with
\sphinxtitleref{InitialPoints} points using equally spaced bins and hence uniformly
distributed random points over the whole hypercube. This information is
then used to generate the new bin boundaries. A damping factor protects
against over-optimization, which could destabilize the grid optimization
and lead to strong fluctuations in the grid adaptation.

In the Herwig implementation, in addition to the general parameters
discussed beforehand, two parameters specific to this algorithm can be
set:
\begin{itemize}
\item {} 

\sphinxtitleref{GridDivisions}

This specifies how many bins are used for each integration dimension.

\item {} 

\sphinxtitleref{Alpha}

The damping behaviour is controlled with this parameter. Instead of
the actual relative contribution $c_i$ of each bin, the
modified expression

$\left(\frac{c_i}{(1-c_i)\ln(c_i)}\right)^\alpha$

is used to calculated the new bins. Typical values of the exponent are
between 0.2 and 2, where larger values result in less damping. When
setting \sphinxtitleref{Alpha} to zero, no grid adaptation occurs and the algorithm
reduces to the \sphinxtitleref{FlatBinSampler}.

\end{itemize}

\subsection{Amplitude providers}
\label{\detokenize{review/hardprocess/amplitude-providers:amplitude-providers}}\label{\detokenize{review/hardprocess/amplitude-providers:id1}}\label{\detokenize{review/hardprocess/amplitude-providers::doc}}

The Matchbox module can use several libraries to calculate scattering
amplitudes and/or amplitudes squared to build up cross sections. The
interfaces are set up within a broad range, from run-time interfaces
providing low level amplitude information, up to programs which
generate code at the squared amplitude level in correspondence to the
BLHA2 interface standard \cite{Alioli:2013nda, Binoth:2010xt}, which is
then loaded as a dynamical library by the Matchbox framework. In this
section we summarize the available libraries.

While some libraries provide the entire list of quantities required to set up
NLO calculations, some of them are only capable of calculating cross section
level quantities and thus need to be combined with a native amplitude provider
in a hybrid setup to obtain meaningful selection of colour flows or the
filling of spin density matrices for the hard process.

\subsubsection{MatchboxCurrents}
\label{\detokenize{review/hardprocess/amplitude-providers:matchboxcurrents}}

Several matrix elements in Matchbox are based on a library of leptonic
and hadronic currents \cite{Platzer:SpinorHelicity} which have been
set up using spinor helicity techniques and caching of sub-amplitudes,
together with a small collection of colour structures for low jet
multiplicities. These amplitude building blocks are also used for the
amplitudes available in the {\hyperref[\detokenize{review/hardprocess/amplitude-providers:hjets}]{\sphinxcrossref{\DUrole{std}{\DUrole{std-ref}{HJets}}}}} library. Herwig interfaces to
MatchboxCurrents at the amplitude level, and all quantities relevant
to NLO predictions for the processes provided can be obtained from the
single library.

The MatchboxCurrents library is directly shipped with the Herwig release and
can be used to generate any process involving a lepton pair, and two or three
partons at leading and next-to-leading order. This includes $e^+e^-\to
2,3\ \text{jets}$, DIS $0+1$ and $1+1$ jet production, as well as
Drell-Yan and $V+\text{jet}$ processes, all at NLO QCD with the real
emission additional jet production also available to leading order multi-jet
merging.

\subsubsection{GoSam}
\label{\detokenize{review/hardprocess/amplitude-providers:gosam}}\label{\detokenize{review/hardprocess/amplitude-providers:id4}}

GoSam \cite{Cullen:2011ac, Cullen:2014yla,Braun:2025afl} provides functionality to generate
code for scattering amplitudes at the tree and one-loop level, and interfaces to
Herwig via the BLHA2 interface standard at the level of the squared amplitudes.
Process code is generated as requested and the corresponding libraries compiled
and linked dynamically at run time.
Spin- and colour-correlated Born matrix elements can be provided by GoSam, as
described in \cite{Alioli:2013nda, Andersen:2014efa} (see \hyperref[\detokenize{review/hardprocess/nlo:correlated-me}]{Section \ref{\detokenize{review/hardprocess/nlo:correlated-me}}}
for more details), as well as squared one-loop amplitudes for loop induced
processes at the leading order.

Per default Herwig’s GoSam interface assumes GoSam’s Standard Model
implementation, with a diagonal CKM matrix, but it is possible to switch to a
Higgs effective model as well. All Standard Model parameters as well as the
EW scheme are automatically communicated from Herwig to GoSam, and the
complex-mass scheme can be used. Per BLHA the light lepton masses are set to
zero by default though (GoSam then assumes its own hard-coded default for the
tau mass). The interface assumes GoSam’s matrix elements to be provided in
conventional dimensional regularization, but they may be provided in dimensional
reduction if required. Further, settings related to the one-loop reduction in
GoSam might be accessed through the interface as well, such as the one-loop
reduction algorithm that is used in GoSam or the target accuracy of the one-loop
matrix elements.

GoSam uses a setup file, in which model assumptions, diagram filters, one-loop
reduction settings, etc. are specified beyond its hard-coded defaults. Herwig
provides a template GoSam setup file, which is used per default, modified
according to the settings that are chosen through Herwig’s GoSam interface.
However, a custom GoSam setup file may also be provided, e.g.\ to implement
custom diagram filters or to specify physics models beyond what can be specified
through the interface (as for instance used in \cite{Bellm:2016cks} for
anomalous $ggW^+W^-$ couplings, originating from dimension-8 operators).

More details on the various interface options can be found in the \href{https://herwig.hepforge.org/tutorials/hardprocess/matchbox.html\#using-external-amplitude-providers}{tutorials}.
More information on GoSam specific settings can be found in the GoSam manual.

\subsubsection{HJets}
\label{\detokenize{review/hardprocess/amplitude-providers:hjets}}\label{\detokenize{review/hardprocess/amplitude-providers:id8}}

HJets \cite{Campanario:2013fsa} builds on the MatchboxCurrents
library, the ColorFull package \cite{Sjodahl:2014opa}, and the
methods developed in \cite{Hagiwara:1988pp} to provide a library
which can calculate EW (VBF) Higgs production in association
with two or three jets, including all contributing diagrams (VBF-type
and VH-type) at the orders ${\cal O}(\alpha^3\alpha_s)$ and
${\cal O}(\alpha^3\alpha_s^2)$ relevant to NLO QCD
corrections. Herwig interfaces to HJets at the amplitude level.

\subsubsection{NJet}
\label{\detokenize{review/hardprocess/amplitude-providers:njet}}

NJet \cite{Badger:2012pg} is a library dedicated to QCD jet production at
hadron colliders, and provides tree and one-loop amplitudes for up to
$pp\to 4\ \text{jet}$ production. NJet is interfaced to Herwig at the
level of squared amplitudes, and purely as a dynamic library such that no code
generation is happening while the processes are set up.

\subsubsection{MadGraph5\_aMC@NLO}
\label{\detokenize{review/hardprocess/amplitude-providers:madgraph5-amc-nlo}}\label{\detokenize{review/hardprocess/amplitude-providers:madgraph}}

MadGraph5\_aMC@NLO \cite{Alwall:2014hca} provides functionality to generate code
for scattering amplitudes at the tree and one-loop level, and interfaces to
Herwig at the amplitude level as well as on the level of the squared amplitude.
Process code is generated as requested and the corresponding libraries compiled
and linked dynamically at run time.
At the amplitude level, partial subamplitudes, in either the trace or the color
flow basis, can be provided (see \hyperref[\detokenize{review/hardprocess/tree-level:colour-basis}]{Section \ref{\detokenize{review/hardprocess/tree-level:colour-basis}}} for more information),
particularly also in the large-N limit, for use within the parton showers. At
the level of the squared amplitude, spin- and colour-correlated Born matrix
elements can be provided (see \hyperref[\detokenize{review/hardprocess/nlo:correlated-me}]{Section \ref{\detokenize{review/hardprocess/nlo:correlated-me}}} for more information).

Per default Herwig’s MadGraph interface assumes MadGraph’s Standard Model
implementation, but it is possible to switch to a Higgs effective model as well.
All Standard Model parameters are automatically communicated from Herwig to
MadGraph, and the complex-mass scheme can be used.

More details on the various interface options can be found in the \href{https://herwig.hepforge.org/tutorials/hardprocess/matchbox.html\#using-external-amplitude-providers}{tutorials}.

\subsubsection{OpenLoops}
\label{\detokenize{review/hardprocess/amplitude-providers:openloops}}

OpenLoops \cite{Buccioni:2019sur} implements the Open Loops algorithm for the
fast numerical evaluation of scattering amplitudes at the tree and one-loop level \cite{Cascioli:2011va},
and interfaces to Herwig via the BLHA2 interface standard at the level of the
squared amplitudes, and purely as a dynamic library such that no code generation
is happening while the processes are set up. OpenLoops also provides spin- and
colour-correlated Born matrix elements (see \hyperref[\detokenize{review/hardprocess/nlo:correlated-me}]{Section \ref{\detokenize{review/hardprocess/nlo:correlated-me}}} for more
information).

By default, Herwig’s OpenLoops interface assumes OpenLoops’ Standard Model
implementation, but it is possible to switch to a Higgs effective model as well.
All Standard Model parameters are automatically communicated from Herwig to
OpenLoops, and the complex-mass scheme can be used. For the one-loop reduction
OpenLoops uses the Collier library \cite{Denner:2016kdg} per default.

More details on the various interface options can be found in the \href{https://herwig.hepforge.org/tutorials/hardprocess/matchbox.html\#using-external-amplitude-providers}{tutorials}.

\subsubsection{VBFNLO}
\label{\detokenize{review/hardprocess/amplitude-providers:vbfnlo}}\label{\detokenize{review/hardprocess/amplitude-providers:id17}}

VBFNLO \cite{Arnold:2008rz, Arnold:2011wj, Baglio:2014uba, Baglio:2024gyp} is a flexible
parton-level Monte Carlo program for processes with EW bosons. It
provides amplitudes at the tree and one-loop level, and interfaces to Herwig via
the BLHA2 interface standard at the level of the squared amplitudes.
VBFNLO also provides spin- and colour-correlated Born
matrix elements (see \hyperref[\detokenize{review/hardprocess/nlo:correlated-me}]{Section \ref{\detokenize{review/hardprocess/nlo:correlated-me}}} for more information). These and the
standard Born matrix elements can in particularly also be provided in the
large-N limit, for use within the parton showers.

All Standard Model parameters are automatically communicated from Herwig to
VBFNLO. In addition anomalous couplings of the EW bosons, originating
from dimension-6 and possibly dimension-8 operators, can be enabled.

In order to improve the convergence of the Monte Carlo integration, Herwig’s
VBFNLO interface has the possibility to use the VBFNLO internal phase-space
generator, which is particularly tailored to the processes available through
VBFNLO and which for more complex final states, as the gains can be highly
significant, is in general recommended.

Further, there is the possibility to sample the helicities of the leptons and
photons in the final state randomly instead of summing them; the computation
time spent per phase-space point will be reduced at the expense of a larger
variation, but overall a slightly improved integration accuracy is expected.

More details on the various interface options can be found in the \href{https://herwig.hepforge.org/tutorials/hardprocess/matchbox.html\#using-external-amplitude-providers}{tutorials}.
More information on VBFNLO specific settings can be found in the VBFNLO manual
\cite{Arnold:2011wj}.

\subsection{Internal cross-checks}
\label{\detokenize{review/hardprocess/internal-checks:internal-cross-checks}}\label{\detokenize{review/hardprocess/internal-checks:internal-checks}}\label{\detokenize{review/hardprocess/internal-checks::doc}}

For calculations beyond LO, the different pieces of virtual corrections
and real emissions are individually divergent. Only when looking at
their sum, and well-defined, infrared-safe observables, the KLN theorem
\cite{Kinoshita:1962ur, Lee:1964is} guarantees that the sum of all
contributions is finite. Subtraction schemes like Catani--Seymour
\cite{Catani:1996vz} or FKS \cite{Frixione:1995ms} make use of that by
constructing approximations to the real-emission process from the
corresponding Born plus a term describing the extra emission, such that
it equals the full matrix element in the soft or collinear divergent
region, and the difference of the two can be integrated numerically in
four dimensions. Integrating analytically over the extra-emission term
in $d=4-2\epsilon$ dimensions generates poles in $1/\epsilon^2$ and
$1/\epsilon$, which then exactly cancel the corresponding poles of the
virtual amplitude.

Herwig generates the Catani--Seymour dipoles themselves, using the
Born, colour-correlated and spin-colour-correlated Born processes from
the matrix-element provider as an input. Therefore, comparing these
expressions numerically to the real-emission process in the singular
limits and the $\epsilon$ poles of the Born-virtual interference,
respectively, provides a powerful check of the consistency of the
calculation.

\subsubsection{Singularity cancellation}
\label{\detokenize{review/hardprocess/internal-checks:singularity-cancellation}}

For the real-emission part, we can check that the approximation from the
dipole subtraction term approaches the real-emission matrix element when
coloured final-state particles become soft or a pair of them becomes
collinear. This check can be switched on by setting \sphinxtitleref{SubtractionData} to
a string and creating a directory with the name of this string in the
run directory. The exact type of output can be controlled by two variables.
In all cases a histogram for each singular limit is generated. The
x-axis value contains the energy of the gluon in the soft limit, and the
virtuality $q=\sqrt{2 p_i \cdot p_j}$ of the two collinear
particles \sphinxtitleref{i} and \sphinxtitleref{j} in the collinear limit. The corresponding y value
is defined by the setting of
\begin{description}
\sphinxlineitem{\sphinxtitleref{SubtractionPlotType}:}\begin{itemize}
\item {} 

\sphinxtitleref{LinRatio}
The result is the ratio of all dipole contributions over the real-emission
matrix element, $\left|\frac{\sum
\mathcal{D}}{|\mathcal{M}_\text{real-emission}|^2}\right|$, i.e.\ should go to
one in the soft or collinear limit. The scaling of the y axis is linear.

\item {} 

\sphinxtitleref{LogRelDiff}
The result is the sum of dipole and real-emission matrix element normalized
over the real-emission matrix element, $\left|\frac{\sum
\mathcal{D}+|\mathcal{M}_\text{real-emission}|^2}{|\mathcal{M}_\text{real-emission}|^2}\right|$,
i.e.\ should go to zero in the soft or collinear limit. The scaling of the y
axis is logarithmic.

\end{itemize}

\sphinxlineitem{\sphinxtitleref{SubtractionScatterPlot}}

Setting this to \sphinxtitleref{Yes} instead of the default \sphinxtitleref{No} additionally
generates the data file for a scatter plot from the same data points as
used for the histograms.

\end{description}

\subsubsection{Pole cancellation}
\label{\detokenize{review/hardprocess/internal-checks:pole-cancellation}}

For the virtual corrections we can verify explicitly that the poles of
the Born-virtual interference are cancelled by the corresponding
contributions from the dipole subtraction terms, namely the
$I$-operator. To activate this check one should set
\sphinxtitleref{PoleData} to a string and create a directory of the
same name in the run directory, where the histogram output will then be
written. For each subprocess two histogram files are produced:
\sphinxtitleref{epsilonSquarePoles}, containing the contributions
proportional to $1/\epsilon^{2}$, and \sphinxtitleref{epsilonPoles},
containing those proportional to $1/\epsilon$. The entries are binned
according to
\[
  \log_{10}\!\left(1-\left|\frac{|\mathcal{M}_{\text{Dipole}}|^{2}}{2\,\mathrm{Re}\!\left(\mathcal{M}_{\text{Born}}^{*}\,\mathcal{M}_{\text{Virtual}}\right)}\right|\right),
\]
with the ordinate giving the number of sampled phase-space points. The
distribution should be sharply peaked near $-16$, reflecting the fact
that the cancellation holds up to the level of double-precision
numerical accuracy (i.e.\ relative differences of order $10^{-16}$).

\subsubsection{Independence of finite terms}
\label{\detokenize{review/hardprocess/internal-checks:independence-of-finite-terms}}

In Ref. \cite{Nagy:1998bb, Nagy:2003tz}, a modification of the
Catani--Seymour dipole subtraction terms has been introduced, which
restricts the contribution of the dipoles to a region around the
singular poles. This is realized by multiplying the dipole expressions
with an additional step function $\Theta(\alpha-y_{ij,k})$, where
$y_{ij,k}$ is the dimensionless variable of the Catani--Seymour
algorithm that is used to calculate the momenta of spectator and emitter
in the tilde kinematic, and $\alpha$ is a dimensionless parameter,
$\alpha \in (0,1]$. Setting $\alpha$ to 1 restores the
original Catani--Seymour version, while using smaller values subsequently
restricts the dipole phase-space closer to the pole. In Herwig, this
parameter can be set via \sphinxtitleref{AlphaParameter}. The integrated dipoles, which
enter in the calculation of the virtual corrections, then receive
corresponding terms also depending on $\alpha$ such that in the
full NLO cross section everything drops out again.

The use of this parameter is two-fold. Restricting the contribution of
the dipoles to a smaller region than the full real-emission phase-space
means that they do not need to be calculated there. This will therefore
reduce the computation time spent on the phase-space point. In practice,
the gain turns out to be rather low only. Additionally, varying the
parameter allows one to check that the total contribution of the dipoles
on the cross section drops out, as the cross section should be
independent of the value of $\alpha$. In particular, this also checks
that the values of the coupling constants and the cuts are the same for
the virtual corrections and the real-emission contributions.

\subsection{Built-in matrix elements}
\label{\detokenize{review/hardprocess/me:built-in-matrix-elements}}\label{\detokenize{review/hardprocess/me:builtin-matrix-elements}}\label{\detokenize{review/hardprocess/me::doc}}

In Herwig 7 the library of matrix elements for QCD and
EW processes is similar to that available in its FORTRAN predecessor
\cite{Corcella:2000bw, Corcella:2002jc}. While the library of Standard
Model processes provides a core of important
processes with which to test the program it is our intention
that, in general, users should study most processes of interest via
the Matchbox module, or external programs using the Les Houches Accord.

Nevertheless, there are still some cases for which it is useful to have
Herwig handle all stages of the event generation process. This is
particularly true for processes in which spin correlations between the
production and decay stages are significant, \textit{e.g.} those involving top
quarks or tau leptons. Such correlation effects are hard to treat
correctly if different programs handle different steps of the simulation
process.

In order to facilitate the process of adding new matrix elements, where
needed, and to enable us to generate the spin correlation effects
\cite{Richardson:2001df, Knowles:1988vs, Collins:1987cp},
we have based all matrix element calculations on the helicity libraries
of ThePEG. As well as providing a native library of Standard Model
processes and an interface to parton-level generators, Herwig also
includes matrix elements for hard $2\to2$ collisions and
$1\to2$ and $1\to3$ decays, arising in various models of new
physics (see \hyperref[\detokenize{review/index:sect-bsm}]{Section \ref{\detokenize{review/index:sect-bsm}}}).

We also include a number of next-to-leading-order (NLO) matrix
elements in the POsitive Weight Hardest Emission Generator (POWHEG)
scheme of Refs. \cite{Nason:2004rx, Frixione:2007vw}.

\subsubsection{Leading-order matrix elements}
\label{\detokenize{review/hardprocess/me:leading-order-matrix-elements}}\label{\detokenize{review/hardprocess/me:sec-specificmes}}

For $e^+e^-$ colliders only a small number of processes are included natively:
\begin{itemize}
\item {} 

Quark-antiquark production, via interfering photon and $Z^0$
bosons, is implemented in the
\href{https://herwig.hepforge.org/doxygen/classHerwig\_1\_1MEee2gZ2qq.html}{MEee2gZ2qq}
class. No approximation is made regarding the masses of the
particles. This process is essential for us to validate the program
using QCD analyses of LEP data.

\item {} 

Dilepton pair production, via interfering photon and $Z^0$
bosons, is implemented in the
\href{https://herwig.hepforge.org/doxygen/classHerwig\_1\_1MEee2gZ2ll.html}{MEee2gZ2ll}
class. No approximation is made regarding the masses of the
particles %
\begin{footnote}\sphinxAtStartFootnote
$t$-channel photon and $Z^0$ boson exchange are not
included.
\end{footnote}. This process is used to check the implementation of
spin correlations in $\tau$ decays.

\item {} 

The Bjorken process, $Z^0h^0$ production, which is implemented
in the
\href{https://herwig.hepforge.org/doxygen/classHerwig\_1\_1MEee2ZH.html}{MEee2ZH}
class. This process is included as it is very similar to the
production of $Z^0h^0$ and $W^\pm h^0$ in hadron-hadron
collisions and uses the same base class for most of the calculation.

\item {} 

The vector-boson fusion (VBF) processes, $e^+e^-\to e^+e^-h^0$
and $e^+e^-\to \nu_e\bar{\nu}_eh^0$, are implemented in the
\href{https://herwig.hepforge.org/doxygen/classHerwig\_1\_1MEee2HiggsVBF.html}{MEee2HiggsVBF}
class.

\item {} 

The production of a pair of EW gauge bosons, $W^+W^-$
and $Z^0Z^0$, is simulated using the
\href{https://herwig.hepforge.org/doxygen/classHerwig\_1\_1MEee2VV.html}{MEee2VV}
class. The decays of the gauge bosons are not included in the matrix
element but the spin correlations for the decays are correctly
described (\hyperref[\detokenize{review/showers/perturbative_decays:sec-perturbative-decays}]{Section \ref{\detokenize{review/showers/perturbative_decays:sec-perturbative-decays}}}).

\item {} 

The resonant production of the Higgs boson followed by its decay
to Standard Model particles is simulated using the
\href{https://herwig.hepforge.org/doxygen/classHerwig\_1\_1MEee2Higgs2SM.html}{MEee2Higgs2SM}
class. This process has a minuscule production rate and is only included to allow the
easy simulation of gluon jets and the decay $h^0\to\tau^+\tau^-$

\item {} 

The resonant production of vector mesons is simulated using the
\href{https://herwig.hepforge.org/doxygen/classHerwig\_1\_1MEee2VectorMeson.html}{MEee2VectorMeson}
class. This process is included to allow the simulation of the production of
charmonium and bottomonium resonances, primarily the $\Upsilon(4S)$, to test
the hadron decay model.

\end{itemize}

For deep inelastic scattering (DIS) only two processes are included.
Neutral and charged current processes are implemented in the
\href{https://herwig.hepforge.org/doxygen/classHerwig\_1\_1MENeutralCurrentDIS.html}{MENeutralCurrentDIS}
and
\href{https://herwig.hepforge.org/doxygen/classHerwig\_1\_1MEChargedCurrentDIS.html}{MEChargedCurrentDIS}
classes, respectively. In neutral current processes both the incoming
and outgoing partons are considered to be massless, whereas in the
charged current process the masses of the outgoing partons are included.
For neutral current scattering both photon and $Z^0$ boson
exchange are included.

A much wider range of matrix elements is included for the simulation of
events in hadron colliders:
\begin{itemize}
\item {} 

Difermion production via $s$-channel EW gauge bosons.
The matrix elements for the production of fermion-antifermion pairs
through $W^\pm$ bosons, or interfering photons and $Z^0$
bosons, are implemented in the
\href{https://herwig.hepforge.org/doxygen/classHerwig\_1\_1MEqq2W2ff.html}{MEqq2W2ff}
and
\href{https://herwig.hepforge.org/doxygen/classHerwig\_1\_1MEqq2gZ2ff.html}{MEqq2gZ2ff}
classes respectively. Only $s$-channel EW gauge boson
diagrams are included for the hadronic modes.

\item {} 

The production of a $Z^0$ or $W^\pm$ boson in association
with a hard jet is simulated using the
\href{https://herwig.hepforge.org/doxygen/classHerwig\_1\_1MEPP2ZJet.html}{MEPP2ZJet}
or
\href{https://herwig.hepforge.org/doxygen/classHerwig\_1\_1MEPP2WJet.html}{MEPP2WJet}
class respectively. The decay products of the bosons are included in
the $2\to3$ matrix element and the option of including the
photon for $Z^0$ production is supported.

\item {} 

The production of a pair of EW gauge bosons, $W^+W^-$,
$W^\pm Z^0$ and $Z^0Z^0$ is implemented in the
\href{https://herwig.hepforge.org/doxygen/classHerwig\_1\_1MEPP2VV.html}{MEPP2VV}
class. The decays of the gauge bosons are not included in the matrix
element, although the spin correlations for the decay products are
correctly treated using the algorithm described in \hyperref[\detokenize{review/showers/perturbative_decays:sec-perturbative-decays}]{Section \ref{\detokenize{review/showers/perturbative_decays:sec-perturbative-decays}}}.

\item {} 

The production of a single top quark is implemented in the
\href{https://herwig.hepforge.org/doxygen/classHerwig\_1\_1MEPP2SingleTop.html}{MEPP2SingleTop}
class. This process proceeds via either $t$-channel $W^\pm$ exchange,
$s$-channel $W^\pm$ exchange, or in association with a $W^\pm$ boson.

\item {} 

The production of an EW gauge boson, $W^\pm$ or
$Z^0$, in association with a hard photon is simulated using the
\href{https://herwig.hepforge.org/doxygen/classHerwig\_1\_1MEPP2VGamma.html}{MEPP2VGamma}
class. As with EW gauge boson pair production the decays of
the gauge bosons are not included in the matrix element, although the
spin correlations for the decay products are correctly treated as
described in \hyperref[\detokenize{review/showers/perturbative_decays:sec-perturbative-decays}]{Section \ref{\detokenize{review/showers/perturbative_decays:sec-perturbative-decays}}}.

\item {} 

The $2\to2$ QCD scattering processes are implemented in the
\href{https://herwig.hepforge.org/doxygen/classHerwig\_1\_1MEQCD2to2.html}{MEQCD2to2}
class. Currently all the particles are treated as massless in these
processes.

\item {} 

The matrix element for the production of a heavy quark-antiquark pair
(top, bottom or charm quark pairs), is coded in the
\href{https://herwig.hepforge.org/doxygen/classHerwig\_1\_1MEPP2QQ.html}{MEPP2QQ}
class. No approximations are made regarding the masses of the
outgoing $q\bar{q}$ pair.

\item {} 

The
\href{https://herwig.hepforge.org/doxygen/classHerwig\_1\_1MEPP2GammaGamma.html}{MEPP2GammaGamma}
class implements the matrix element for the production of prompt
photon pairs. In addition to the tree-level
$q\bar{q}\to\gamma\gamma$ process the loop-mediated
$gg\to\gamma\gamma$ process is included.

\item {} 

Direct photon production in association with a jet is simulated using
the
\href{https://herwig.hepforge.org/doxygen/classHerwig\_1\_1MEPP2GammaJet.html}{MEPP2GammaJet}
class. As with the QCD $2\to2$ process all of the particles are
treated as massless in these processes.

\item {} 

The production of an $s$-channel Higgs boson via both
$gg\to h^0$ and $q\bar{q}\to h^0$ is simulated using the
\href{https://herwig.hepforge.org/doxygen/classHerwig\_1\_1MEPP2Higgs.html}{MEPP2Higgs}
class.

\item {} 

The production of a Higgs boson in association with the $Z^0$
or $W^\pm$ bosons is simulated using the
\href{https://herwig.hepforge.org/doxygen/classHerwig\_1\_1MEPP2ZH.html}{MEPP2ZH}
or
\href{https://herwig.hepforge.org/doxygen/classHerwig\_1\_1MEPP2WH.html}{MEPP2WH}
class respectively.

\item {} 

The production of the Higgs boson in association with a hard jet is
simulated using the
\href{https://herwig.hepforge.org/doxygen/classHerwig\_1\_1MEPP2HiggsJet.html}{MEPP2HiggsJet}
class.

\item {} 

The production of the Higgs boson via the vector-boson fusion
process is implemented in the
\href{https://herwig.hepforge.org/doxygen/classHerwig\_1\_1MEPP2HiggsVBF.html}{MEPP2HiggsVBF}
class.

\item {} 

The production of the Higgs boson in association with either a heavy
quark-antiquark pair, either top or bottom, is implemented in the
\href{https://herwig.hepforge.org/doxygen/classHerwig\_1\_1MEPP2QQHiggs.html}{MEPP2QQHiggs}
class.

\end{itemize}

In addition we have a matrix element class,
\href{https://herwig.hepforge.org/doxygen/classHerwig\_1\_1MEQCD2to2Fast.html}{MEQCD2to2Fast},
that uses hard-coded formulae for the QCD $2\to2$ scattering
matrix elements rather than the helicity libraries of ThePEG. This class
is significantly faster than the default
\texttt{MEQCD2to2}
class, although it does not implement spin correlations. It is intended
to be used in the generation of the multiple parton-parton scatterings
for the underlying event where the spin correlations are not important
but due to the number of additional scatterings that must be generated
the speed of the calculation can significantly affect the run time of
the event generator. There are also the:
\begin{itemize}
\item {} 

\href{https://herwig.hepforge.org/doxygen/classHerwig\_1\_1MEMinBias.html}{MEMinBias}
class which is only used to simulate soft scattering processes as part
of the underlying event model;

\item {} 

\href{https://herwig.hepforge.org/doxygen/classHerwig\_1\_1MEDiffraction.html}{MEDiffraction}
class which is only used to simulate diffractive processes.

\end{itemize}

There are a small range of matrix elements for processes initiated by a
real photon. The
\href{https://herwig.hepforge.org/doxygen/classHerwig\_1\_1MEGammaGamma2ff.html}{MEGammaGamma2ff}
class implements the matrix elements for the production of a
fermion-antifermion pair in photon--photon collisions,
$\gamma\gamma\to f \bar{f}$. The
\href{https://herwig.hepforge.org/doxygen/classHerwig\_1\_1MEGammaGamma2WW.html}{MEGammaGamma2WW}
simulates the production of a $W^+W^-$ pair in photon--photon
collisions. The
\href{https://herwig.hepforge.org/doxygen/classHerwig\_1\_1MEGammaP2Jets.html}{MEGammaP2Jets}
simulates jet production in photon--hadron collisions via the
subprocesses $q\gamma\to qg$, $\bar{q}\gamma\to qg$ and
$g\gamma\to q\bar{q}$.

\subsubsection{Next-to-leading-order matrix elements}
\label{\detokenize{review/hardprocess/me:next-to-leading-order-matrix-elements}}\label{\detokenize{review/hardprocess/me:sect-powheg-me}}

A small number of hard-coded processes \cite{Hamilton:2008pd, Hamilton:2010mb, DErrico:2011cgc, Hamilton:2009za, DErrico:2011wfa} in the POWHEG scheme
are implemented in Herwig together with a full implementation of
the truncated shower. These processes are implemented in the following
way:
\begin{itemize}
\item {} 

the matrix elements are calculated with NLO accuracy and a Born
configuration supplied in the same way as for the leading-order
matrix elements;

\item {} 

the matrix element class generates a real emission configuration
according to the POWHEG Sudakov form-factor;

\item {} 

the event is then showered, including the truncated shower.

\end{itemize}

Currently the following processes are implemented:
\begin{itemize}
\item {} 

$e^+e^-\to q\bar{q}$ via photon and  $Z^0$
boson exchange is simulated using the
\href{https://herwig.hepforge.org/doxygen/classHerwig\_1\_1MEee2gZ2qqPowheg.html}{MEee2gZ2qqPowheg}
class and includes both the QCD and FSR QED corrections.

\item {} 

$e^+e^-\to l^+l^-$ via photon and  $Z^0$
boson exchange is simulated using the
\href{https://herwig.hepforge.org/doxygen/classHerwig\_1\_1MEee2gZ2llPowheg.html}{MEee2gZ2llPowheg}
class and includes FSR QED corrections.

\item {} 

neutral- and charged-current DIS \cite{DErrico:2011wfa} are implemented in the
\texttt{MENeutralCurrentDIS}
and
\texttt{MEChargedCurrentDIS}
classes, respectively which also implement the leading-order processes;

\item {} 

the Drell Yan production \cite{Hamilton:2008pd} of neutral vector bosons $\gamma/Z^0$
is simulated using the
\href{https://herwig.hepforge.org/doxygen/classHerwig\_1\_1MEqq2gZ2ffPowheg.html}{MEqq2gZ2ffPowheg}
class;

\item {} 

the Drell Yan production \cite{Hamilton:2008pd} of charged vector bosons,
\textit{i.e.} $W^\pm$, is implemented in the
\href{https://herwig.hepforge.org/doxygen/classHerwig\_1\_1MEqq2W2ffPowheg.html}{MEqq2W2ffPowheg}
class;

\item {} 

the production of the Higgs boson via the gluon-gluon fusion process \cite{Hamilton:2009za}
is simulated using the
\href{https://herwig.hepforge.org/doxygen/classHerwig\_1\_1MEPP2HiggsPowheg.html}{MEPP2HiggsPowheg}
class;

\item {} 

the production of the Higgs boson in association with the
$W^\pm$ boson \cite{Hamilton:2008pd} is implemented in the
\href{https://herwig.hepforge.org/doxygen/classHerwig\_1\_1MEPP2WHPowheg.html}{MEPP2WHPowheg}
class;

\item {} 

the production of the Higgs boson in association with the $Z^0$
boson \cite{Hamilton:2008pd} is simulated using the
\href{https://herwig.hepforge.org/doxygen/classHerwig\_1\_1MEPP2ZHPowheg.html}{MEPP2ZHPowheg} class.

\item {} 

the production of pairs of EW gauge bosons $W^+W^-$, $W^\pm Z^0$ and
$Z^0Z^0$ \cite{Hamilton:2010mb} is simulated using the
\href{https://herwig.hepforge.org/doxygen/classHerwig\_1\_1MEPP2VVPowheg.html}{MEPP2VVPowheg} class.

\item {} 

the production of the Higgs boson via the vector boson fusion process \cite{DErrico:2011wfa} is implemented in the
\href{https://herwig.hepforge.org/doxygen/classHerwig\_1\_1MEPP2HiggsVBFPowheg.html}{MEPP2HiggsVBFPowheg} class.

\item {} 

the production of photon pairs \cite{DErrico:2011cgc} is implemented in the
\href{https://herwig.hepforge.org/doxygen/classHerwig\_1\_1MEPP2GammaGammaPowheg.html}{MEPP2GammaGammaPowheg} class.

\end{itemize}

More details
of the simulation of QCD radiation can be found in Ref. \cite{Hamilton:2008pd}.

\subsection{Les Houches interface}
\label{\detokenize{review/hardprocess/leshouches:les-houches-interface}}\label{\detokenize{review/hardprocess/leshouches:sect-leshoucheseg}}\label{\detokenize{review/hardprocess/leshouches::doc}}

The
\href{https://thepeg.hepforge.org/doxygen/classThePEG\_1\_1LesHouchesEventHandler.html}{LesHouchesEventHandler}
class inherits from the EventHandler class of ThePEG. The class has a
list of
\href{https://thepeg.hepforge.org/doxygen/classThePEG\_1\_1LesHouchesReader.html}{LesHouchesReader}
objects that are normally connected to files with event data produced by
an external matrix element generator program, although it could in
principle include a direct run-time link to the matrix element generator
or read events ‘on the fly’ from the output of a matrix element
generator connected to a pipe.

When an event is requested by
\texttt{LesHouchesEventHandler},
one of the readers is chosen according to the cross section of the
process for which events are supplied by that reader. An event is read
in and subsequently handled in the same way as for an internally
generated process.

There are a number of matrix element generators available that can
generate parton-level events using either the original Les Houches
Accord \cite{Boos:2001cv} or the subsequent extension
\cite{Alwall:2006yp}, which specified a file format for the transfer of
the information between the matrix element generator and a
general-purpose event generator, such as Herwig, rather than the
original FORTRAN COMMON block.

In addition to the internal mechanism for the generation of hard
processes, ThePEG provides a general
\texttt{LesHouchesEventHandler}
class, which generates the hard process using the Les Houches Accord. In
principle a run-time interface could be used to directly transfer the
information between the matrix element generator and Herwig, however
we expect that the majority of such interfaces will be via data files
containing the event information using the format specified in Ref.
\cite{Alwall:2006yp}.

We also support option of reading the particle data for BSM models from the event file,
as described in \cite{Alwall:2007mw}, and controlled by the \href{https://thepeg.hepforge.org/doxygen/LesHouchesFileReaderInterfaces.html\#QNumbers}{QNumbers} switch. We also support the reading of the event file
as a SUSY Les Houches Accord (SLHA) \cite{Skands:2003cj, Allanach:2008qq} file for internal supersymmetric models,
see \hyperref[\detokenize{review/BSM:sect-susy-models}]{Section \ref{\detokenize{review/BSM:sect-susy-models}}} for more details.

It is important that the settings for the parton shower are consistent with
the matching scheme used by the external package generating the events.
In all cases where NLO matching is included we advise that all the kinematics
generation cuts are applied in program which generates the hard process.
The two major packages supplying external events are:
\begin{description}
\sphinxlineitem{POWHEG BOX \cite{Alioli:2010xd}}

In the case of hard configurations generated using the POWHEG approach
the parton shower should not generate any radiation with transverse momentum
harder than the hardest emission supplied by the external program, \textit{i.e.}
the settings \href{https://herwig.hepforge.org/doxygen/ShowerHandlerInterfaces.html\#MaxPtIsMuF}{MaxPtIsMuF=Yes}
and \href{https://herwig.hepforge.org/doxygen/ShowerHandlerInterfaces.html\#RestrictPhasespace}{RestrictPhasespace=Yes}
should be used.

\sphinxlineitem{mg5\_aMC@NLO \cite{Alwall:2014hca}}

For hard configurations generated using the \texttt{{mg5\_aMC@NLO}} program the settings of the parton
shower must be consistent with those used to calculate the subtraction terms in the MC@NLO approach.
This means that only the angular-ordered parton shower can be used and the settings no longer
correspond to our current defaults. The options
\href{https://herwig.hepforge.org/doxygen/PartnerFinderInterfaces.html\#PartnerMethod}{PartnerMethod=Random},
\href{https://herwig.hepforge.org/doxygen/PartnerFinderInterfaces.html\#ScaleChoice}{ScaleChoice=Partner},
\href{https://herwig.hepforge.org/doxygen/QTildeReconstructorInterfaces.html\#InitialInitialBoostOption}{InitialInitialBoostOption=LongTransBoost},
\href{https://herwig.hepforge.org/doxygen/QTildeReconstructorInterfaces.html\#ReconstructionOption}{ReconstructionOption=General},
\href{https://herwig.hepforge.org/doxygen/QTildeReconstructorInterfaces.html\#FinalStateReconOption}{FinalStateReconOption=Default},
\href{https://herwig.hepforge.org/doxygen/QTildeReconstructorInterfaces.html\#InitialStateReconOption}{InitialStateReconOption=Rapidity},
must be used.

\end{description}

\subsection{`Blob' Matrix Elements}
\label{\detokenize{review/hardprocess/blobme:blobme}}\label{\detokenize{review/hardprocess/blobme::doc}}
\begin{figure}[tp]
\centering
\capstart
\noindent\includegraphics[width=0.500\linewidth]{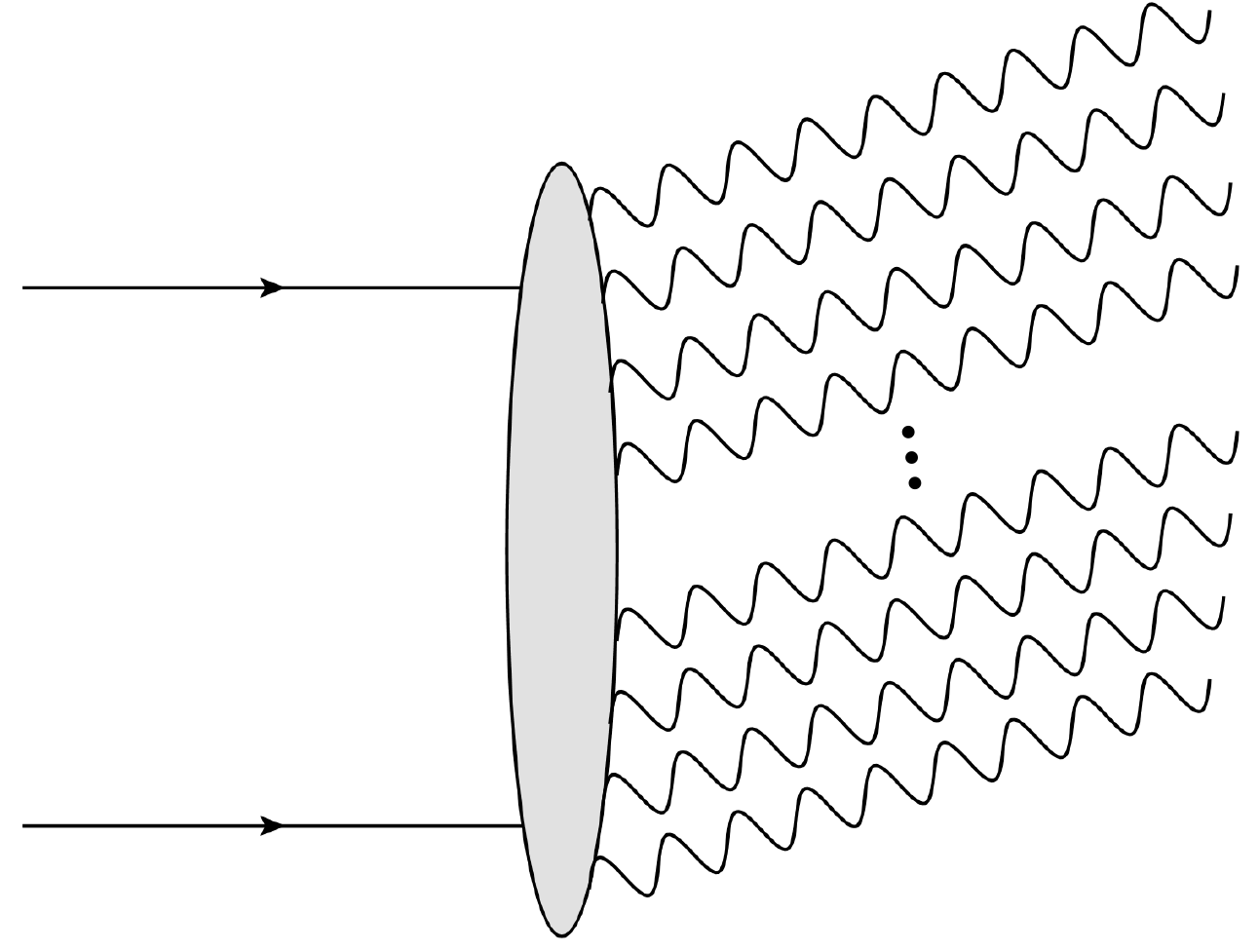}
\caption{A schematic diagram of a $2 \rightarrow n$ processes that can be simulated via the “blob” matrix elements.}\label{fig:blobme}\end{figure}

The default method for constructing the diagrams contributing to a
given hard process is through the \href{https://herwig.hepforge.org/doxygen/classHerwig_1_1Tree2toNGenerator.html}{Tree2toNGenerator} class, which only allows $1\rightarrow 2$ interactions. However, for several models containing new types of interactions, such as those involving two incoming particles and $n$ outgoing, $2\rightarrow n$, where $n\gg 2$, this procedure is not ideal. This is in particular the case for processes for which only the $2\rightarrow n$ matrix element is known, such as sphaleron~\cite{Papaefstathiou:2019djz} or instanton-like processes, shown schematically in Fig.~\ref{fig:blobme}. The \href{https://herwig.hepforge.org/doxygen/classHerwig_1_1BlobME.html}{BlobME} class provides the necessary functionality, replacing the Tree2toNGenerator class, where the two incoming and $n$ outgoing particles are now specified by the \texttt{processes} class function, which is of the type \texttt{multimap<tcPDPair,tcPDVector>}. The matrix element squared then needs to be specified through the \texttt{me2} function. If the matrix element requires a variable number of final-state particles $n$, that number can be accessed via the \texttt{mePartonData} function.

\subsection{Processes with incoming photons}
\label{\detokenize{review/hardprocess/photons:processes-with-incoming-photons}}\label{\detokenize{review/hardprocess/photons::doc}}

It is possible to have hard scattering processes with incoming photons
in hadron-hadron collisions, for example in the higher-order QED
corrections to the Drell-Yan production of $W^\pm$ or
$Z^0$ bosons. Some of these are available through the built-in
matrix elements, but photon induced processes can also be generated
within the Matchbox framework. While these can now be directly
showered by the angular-ordered parton shower we also provide
\href{https://herwig.hepforge.org/doxygen/classHerwig\_1\_1IncomingPhotonEvolver.html}{IncomingPhotonEvolver}
class which can be used as one of the \href{https://thepeg.hepforge.org/doxygen/EventHandlerInterfaces.html\#PreCascadeHandlers}{PreCascadeHandlers}
in these processes to perform the backward evolution of the photon to
a quark or antiquark which can then be evolved by the parton shower.

This performs one backward branching evolving in transverse momentum
from a starting scale $p_{T{\rm start}}$ given by the $p_T$
of the softest particle in the event, or a minimum scale
\href{https://herwig.hepforge.org/doxygen/IncomingPhotonEvolverInterfaces.html\#minpT}{minpT}
if the scale is below the minimum allowed value. This is performed using
a Sudakov form factor,
\begin{equation*}
\begin{split}\Delta(p_T) = \exp\left\{ -\int^{p^2_{T{\rm start}}}_{p^2_T}\frac{{\rm d}p^{\prime2}_T}{p^{\prime2}_T}\frac{\alpha}{2\pi}\int^1_x{\rm d}z \; P(z)\sum_{i}e_i^2
            \frac{\frac{x}{z}f_i\left(\frac{x}{z},p'_T\right)}{xf_\gamma\left(x,p'_T\right)}
\right\},\end{split}
\end{equation*}

where $p_T$ is the transverse momentum of the branching,
$\alpha$ is the fine structure constant, $x$ is the momentum
fraction of the photon, $e_i$ is the electric charge of the
particle produced in the backward evolution and the sum over $i$
runs over all the quarks and antiquarks. The splitting function is
\begin{equation*}
\begin{split}P(z) = \frac{1+(1-z)^2}{z},\end{split}
\end{equation*}

where $z$ is the fraction of the momentum of incoming parton
produced in the backward evolution given to the photon. The $p_T$
and momentum fraction of the branching are generated in the same way as
those in the parton shower, as described in \hyperref[\detokenize{review/showers/qtilde:sect-angular-shower}]{Section \ref{\detokenize{review/showers/qtilde:sect-angular-shower}}}.
The momenta of the particles, including the new branching are then
reconstructed as described in \hyperref[\detokenize{review/showers/qtilde:sub-initial-state-radiation}]{Section \ref{\detokenize{review/showers/qtilde:sub-initial-state-radiation}}}.

\subsection{Code structure}
\label{\detokenize{review/hardprocess/code-structure:code-structure}}\label{\detokenize{review/hardprocess/code-structure:matchbox-code-structure}}\label{\detokenize{review/hardprocess/code-structure::doc}}

In ThePEG the generation of the hard process is the responsibility of
the
\href{https://thepeg.hepforge.org/doxygen/classThePEG\_1\_1EventHandler.html}{EventHandler}.
The base
\texttt{EventHandler}
class only provides the abstract interfaces for the generation of the
hard process with the actual generation of the kinematics being the
responsibility of inheriting classes. There are two such classes
provided in ThePEG: the
\href{https://thepeg.hepforge.org/doxygen/classThePEG\_1\_1StandardEventHandler.html}{StandardEventHandler},
which implements the internal mechanism of ThePEG for the generation of
the hard process; and the
\texttt{LesHouchesEventHandler},
which allows events to be read from data files.

The \texttt{StandardEventHandler}
provides the high-level interface of both built-in Herwig 7 matrix elements as
described in \hyperref[\detokenize{review/hardprocess/me:builtin-matrix-elements}]{Section \ref{\detokenize{review/hardprocess/me:builtin-matrix-elements}}}, as well as
matrix elements provided by the Matchbox module. The handling of event files
is covered in \hyperref[\detokenize{review/hardprocess/leshouches:sect-leshoucheseg}]{Section \ref{\detokenize{review/hardprocess/leshouches:sect-leshoucheseg}}}.

The \texttt{StandardEventHandler}
uses a \href{https://thepeg.hepforge.org/doxygen/classThePEG\_1\_1SubProcessHandler.html}{SubProcessHandler}
to generate the kinematics of the particles involved in the hard process. In
turn the \texttt{SubProcessHandler}
makes use of a number of \href{https://thepeg.hepforge.org/doxygen/classThePEG\_1\_1MEBase.html}{MEBase}
and \href{https://thepeg.hepforge.org/doxygen/classThePEG\_1\_1MEBase.html}{MEGroup}
objects to calculate the matrix element and generate the kinematics for
specific processes. For Matchbox-generated processes
additional subclasses are used, most of which we describe here.
Due to the inheritance structure, the common interfaces
for these matrix elements also resemble the high-level structures
used by ThePEG.
In particular, they need to
\begin{itemize}
\item {} 

define the particles that interact in a given process, by
specifying a number of
\href{https://thepeg.hepforge.org/doxygen/classThePEG\_1\_1DiagramBase.html}{DiagramBase}
objects; one
\texttt{DiagramBase}
is specified per flavour combination, and

\item {} 

return the differential partonic cross section
\begin{equation*}
\begin{split}\frac{{\rm d}\sigma}{{\rm d}r_1 ... {\rm d}r_n},\end{split}
\end{equation*}

when supplied with the partonic centre-of-mass energy of the
collision and $n$ random numbers between $0$ and
$1$, and to

\item {} \begin{description}
\sphinxlineitem{create a \href{https://herwig.hepforge.org/doxygen/classHerwig\_1\_1HardVertex.html}{HardVertex}}
object describing the interaction that occurred, including the
spin-unaveraged matrix element (where available) to allow spin correlation
effects to be generated.

\end{description}

\end{itemize}

One \texttt{MEBase}
object is generally used to describe one physical process, possibly with
different partonic flavours, though the Matchbox framework chooses to
represent individual subprocesses by individual MEBase objects. The selection
of flavours within each subprocess is carried out internally by the
\texttt{EventHandler},
through the SamplerBase object in charge, cf \hyperref[\detokenize{review/hardprocess/integrators:sampling}]{Section \ref{\detokenize{review/hardprocess/integrators:sampling}}}. The resulting
cross sections can be output with varying levels of detail, controlled by the
\href{https://thepeg.hepforge.org/doxygen/EventHandlerInterfaces.html\#StatLevel}{StatLevel}
switch; by default they are only broken down by \texttt{MEBase}
objects.

\subsubsection{Cross-section assembly within Matchbox}
\label{\detokenize{review/hardprocess/code-structure:cross-section-assembly-within-matchbox}}

Within the Matchbox framework a central object of the class
MatchboxFactory is administrating the assembly of all cross sections,
both at leading and next-to-leading order and possibly matched to the
showers or used within a multijet merging approach. This class is
deriving from a standard ThePEG SubProcessHandler and will, for a
given process and desired order of cross section, generate MEBase and
MEGroup objects to represent various contributions to the cross
sections. While these classes form the high-level interface to the
hard process handling of ThePEG, Matchbox’s internal class structure
is in close correspondence to the physical quantities required for a
cross-section calculation, accompanied by book-keeping algorithms and
data-base functionality required for the handling of different
contributions. This core layer, which facilitates the calculation of
the hard cross sections, is interfaced to amplitudes for specific
processes, or external libraries delivering a number of different
subprocesses; several other interfaces are provided for the choice of
scales in the hard process, the input of matching subtractions, or a
hook for multijet merging. The backbone structure of the Matchbox
module is shown in \hyperref[\detokenize{review/hardprocess/code-structure:matchbox-classes}]{Fig.\@ \ref{\detokenize{review/hardprocess/code-structure:matchbox-classes}}}. In the following we
list the different classes with a short description of their main
functionality, and grouped with respect to the different layers.

\begin{figure}[htp]
\centering
\capstart

\noindent\includegraphics[width=\linewidth]{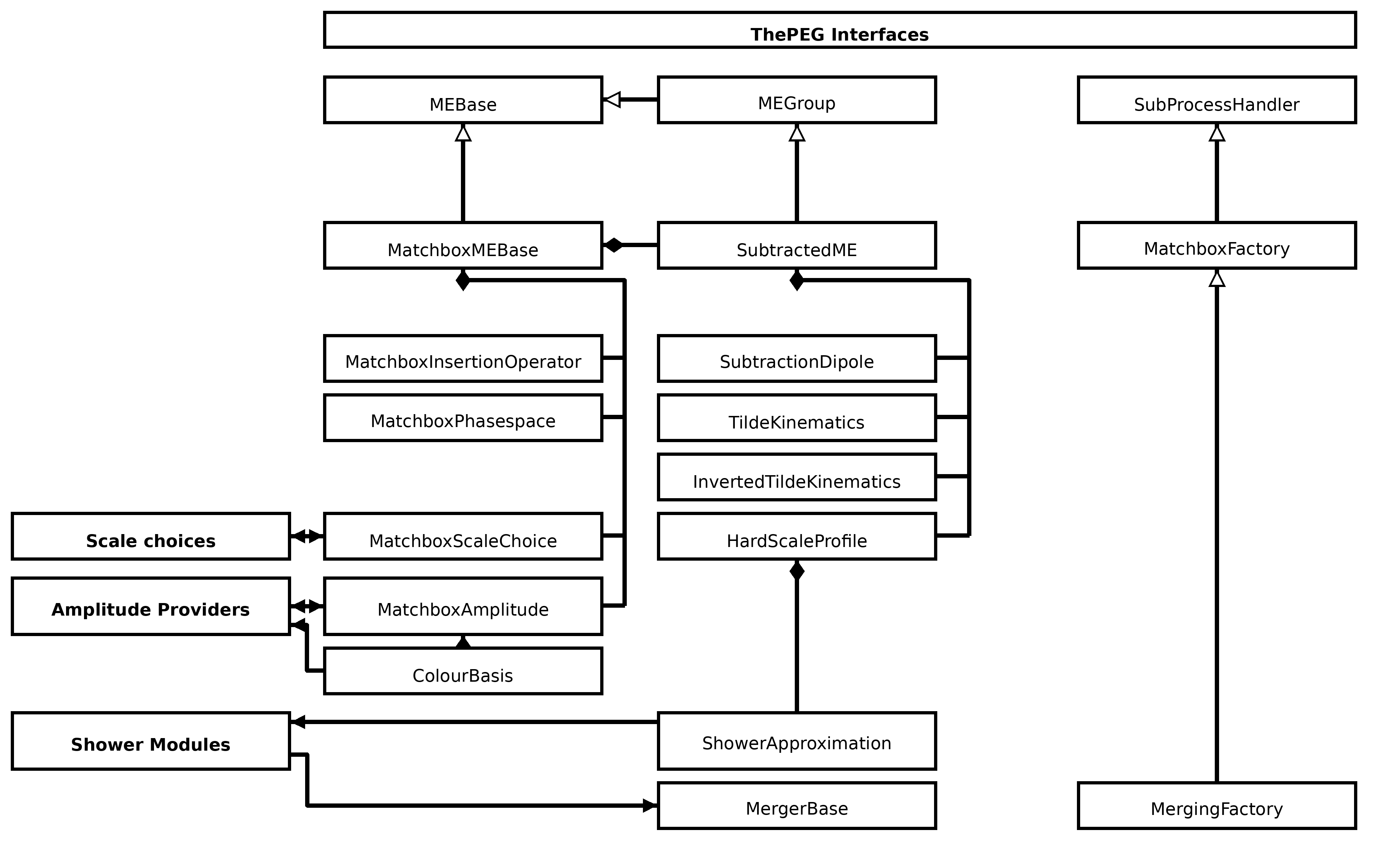}
\caption{An overview of the classes in Matchbox and their interaction.
We use standard UML notation in this diagram: Open triangular arrow heads denote direct programmatic inheritance, pointing from a derived class to its base; filled diamond arrows denote configured components, pointing from a component to its container and configurer. Black filled arrows denote interaction or interfaces.}
\label{\detokenize{review/hardprocess/code-structure:id4}}\label{\detokenize{review/hardprocess/code-structure:matchbox-classes}}\end{figure}

The MatchboxFactory and the deriving MergingFactory classes inherit from
ThePEG’s subprocess handler and are the main classes to identify the required
contributions to a given process of interest. They will determine leading
order, virtual as well as subtraction terms required and create and configure
MatchboxMEBase and SubtractedME objects to calculate the actual cross-section
contributions. Once ready, these will be inserted into the ME vector of the
SubProcessHandler which does not require access by a user in this case
anymore.

\paragraph{Core cross-section contributions}
\label{\detokenize{review/hardprocess/code-structure:core-cross-section-contributions}}

All of the cross sections that are calculated by Matchbox are handled by
ThePEG through the standard mechanism sketched above; on top of these, some
extensions to NLO have been introduced specifically to represent groups of
events, typically a real emission contribution with accompanying subtraction
terms. Since much more detailed communication in between different matrix
element classes is required by Matchbox matrix elements to use colour and spin
correlated matrix elements, ThePEG’s XComb objects have been extended to
contain more per-phase-space information, in particular the caching of colour
and helicity amplitudes, as well as resulting correlated matrix elements. The
inner core of the Matchbox framework closely follows the layout of its
original development as discussed in \cite{Platzer:2011bc, Platzer:2010ppa}.

Cross-section contributions to an observable at one fixed phase-space point,
i.e.\ all contributions that are of leading order, virtual or integrated
subtraction term type are handled by a MatchboxMEBase object which uses a
MatchboxAmplitude object to calculate matrix elements for a particular
subprocess. MatchboxMEBase will also incorporate insertion operators deriving
from the MatchboxInsertionOperator class to be combined with virtual
contributions provided by the amplitude object. The diagrams contributing to a
given hard process are generated by the Tree2toNGenerator class starting from
a given coupling order and a set of vertex objects to be considered. A
physical phase-space point is generated through a MatchboxPhaseSpace object,
which is also responsible to select a diagram topology regarded as the most
significant contribution to the cross section at the selected phase-space
point. This information is used to subsequently set up mother/daughter
relations in the event record. The ColourBasis object contains information
about the colour basis which has been used to express the amplitudes and is
responsible of selecting a colour flow to be assigned to the event at the hard
process level. The choice of renormalization and factorization scales, as well
as the hard shower scale are contained in classes deriving from
MatchboxScaleChoice. In order to optimize memory consumption to store
information shared among several objects, such as diagrams contributing to a
given subprocess or the tree structures describing hierarchic phase-space
generation within a multi-channel approach, the ProcessData object holds such
data in a quasi-static way.

NLO cross sections within Matchbox are strictly calculated in the subtraction
paradigm for which we assume that an organization of the subtraction terms in
one or the other incarnation of a dipole-type kernel is in place. While the
default subtraction method is dipole subtraction along the lines of
\cite{Catani:1996vz, Catani:2002hc} other approaches of this kind are
present. The subtraction terms all derive from the SubtractionDipole class,
which are accompanied by respective kinematic mappings deriving from the
TildeKinematics class, and possibly their inverse in InvertedTildeKinematics
to be used for a different kind of phase-space sampling or the generation of
matrix element corrections. The subtraction term classes register with the
DipoleRepository class to be available to the dipole finding algorithm, which
is mainly contained in the MatchboxMEBase class and works using the diagram
information supplied for the real emission and leading order processes in
order to determine the dipole mappings. An object of class SubtractedME,
inheriting from ThePEG’s MEGroup and operating on an StdXCombGroup to
represent the collection of correlated phase-space points, then holds the
various subtraction terms along with the real emission matrix element
considered. SubtractedME is also capable of projecting a contribution
calculated from a real emission phase-space point down to an underlying Born
phase-space point derived from one of the subtraction terms. This mechanism is
heavily used within matching to a parton shower or the multijet merging
algorithm. SubtractionDipole objects also contain a reference to a shower
kernel to be used in setting up the subtraction required in NLO plus parton
shower matching. All shower subtractions follow the matching paradigm outlined
at the beginning of this chapter and their implementation inherits from the
ShowerApproximation class. The MatchboxMEBase class is also using objects of
the MergerBase class for hooks into the multijet merging algorithm.

\paragraph{Other helper classes and algorithms}
\label{\detokenize{review/hardprocess/code-structure:other-helper-classes-and-algorithms}}

A number of additional helper algorithms have been incorporated into
the Matchbox framework; this specifically includes a versatile set of
template-based sampling of one-dimensional densities, spinor helicity
support, caching of intermediate amplitudes, determination of diagrams
to only name a few. Most of these can also be used
standalone outside of the Herwig framework and support is available from
the authors on request.

\subsubsection{Code structure of Herwig 7 built-in matrix elements}
\label{\detokenize{review/hardprocess/code-structure:code-structure-of-herwig-built-in-matrix-elements}}

For the built-in matrix elements which are supplied with Herwig 7,
typically $2\to2$ scattering processes, the generation of the
kinematics, including
off-shell effects if required, and other technical steps is handled by the
\href{https://herwig.hepforge.org/doxygen/classHerwig\_1\_1HwMEBase.html}{HwMEBase} class
which in turn inherits from
the \texttt{MEBase}
class of ThePEG.
The actual calculation of the matrix elements is implemented
using the Helicity classes of ThePEG%
\begin{footnote}\sphinxAtStartFootnote
The only exception is the
\texttt{MEQCD2to2Fast}
class, which is ‘hand-written’ for speed.
\end{footnote}.
This allows us to include spin correlations between the production, decay
and parton showering of the particles, as described in \hyperref[\detokenize{review/showers/perturbative_decays:sec-perturbative-decays}]{Section \ref{\detokenize{review/showers/perturbative_decays:sec-perturbative-decays}}}.
The helicity amplitudes for the production process are stored in the
\href{https://herwig.hepforge.org/doxygen/classHerwig\_1\_1ProductionMatrixElement.html}{ProductionMatrixElement}
which is used by the \texttt{HardVertex}
class to calculate the $\rho$ and $D$ matrices required to generate the spin correlations.

The main switch for the generation of the hard process is
the \href{https://thepeg.hepforge.org/doxygen/SubProcessHandlerInterfaces.html\#MatrixElements}{MatrixElements}
interface, which allows the
\texttt{MEBase}
objects to be specified and hence determines which hard scattering
processes are generated. In addition, each class inheriting from
\texttt{MEBase}
in Herwig 7 has a number of switches that control the incoming,
outgoing and intermediate particles in a specific process. These are
controlled by Interfaces in the specific matrix element classes. A
number of different partonic subprocesses can be handled at the same
time by simply specifying several
\texttt{MEBase}
objects.

\clearpage

\section{Parton showers}
\label{\detokenize{review/index:parton-showers}}

\subsection{Overview}
\label{\detokenize{review/showers/general:overview}}\label{\detokenize{review/showers/general::doc}}

\subsubsection{Philosophy}
\label{\detokenize{review/showers/general:philosophy}}

A major success of the original HERWIG program was its treatment of soft gluon
interference effects, in particular the phenomenon of \textit{colour coherence}, via
the angular ordering of emissions in the parton shower
\cite{Marchesini:1984bm, Bassetto:1982ma, Bassetto:1984ik, Catani:1983bz, Ciafaloni:1980pz, Ciafaloni:1981bp, Dokshitzer:1988bq, Mueller:1981ex, Ermolaev:1981cm, Dokshitzer:1982fh}.
This emphasis on the accurate treatment of perturbative QCD radiation
continues in Herwig 7 with two parton-shower algorithms:
\begin{itemize}
\item {}

the default angular-ordered algorithm \cite{Gieseke:2003rz}, which generalizes
that used in the original HERWIG program
\cite{Marchesini:1984bm, Webber:1983if, Marchesini:1987cf}; and

\item {}

the alternative dipole-type algorithm \cite{Platzer:2009jq, Platzer:2011bc},
which was added in a later version of the Herwig++ program.

\end{itemize}

\begin{figure}[H]
    \centering
    \includegraphics[width=\linewidth]{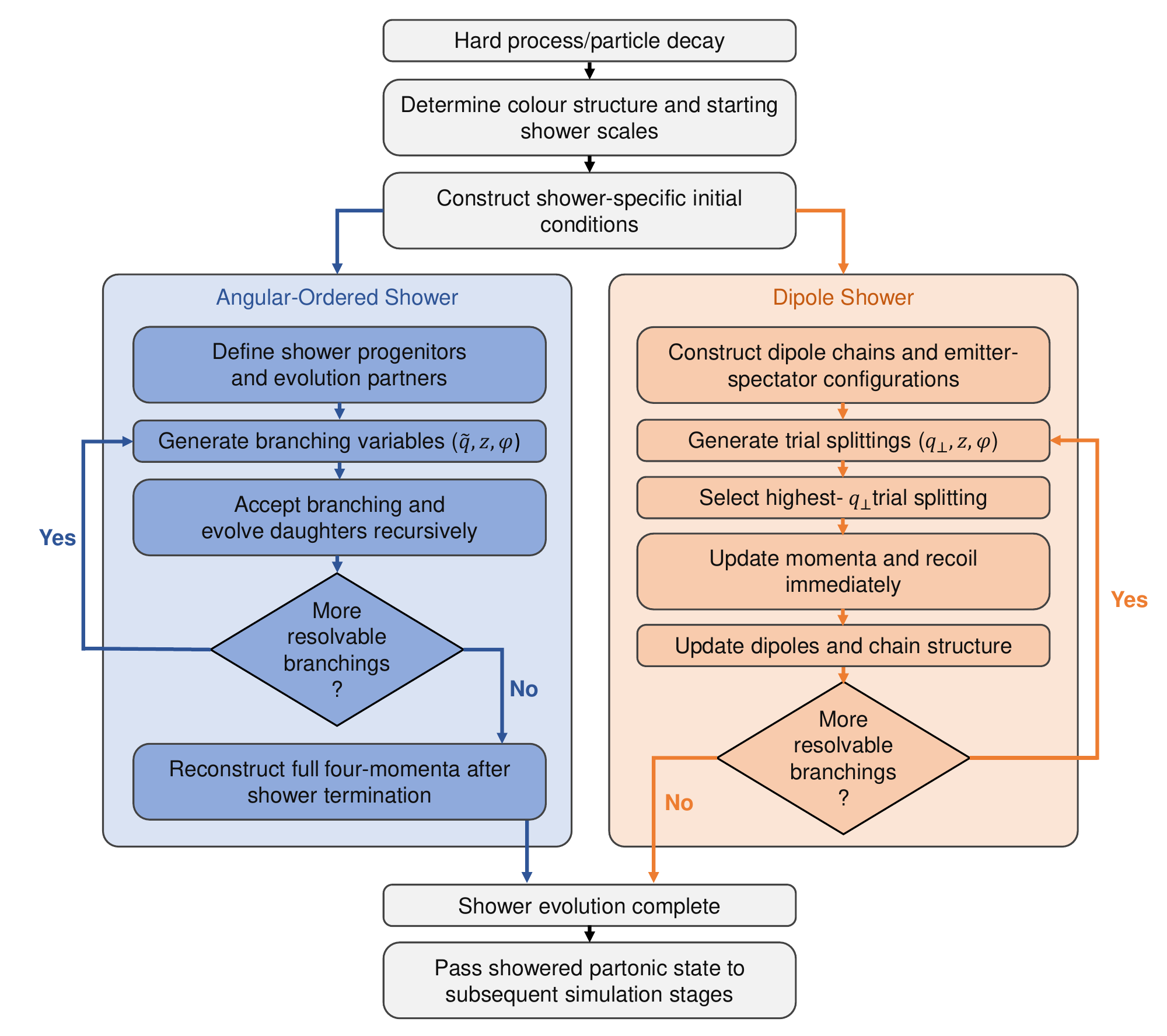}
    \caption{Schematic structure of the two parton-shower algorithms available in \textsc{Herwig}~7.}
    \label{fig_shower_flowchart}
\end{figure}

The overall structures of the two algorithms are compared schematically in
Fig.~\ref{fig_shower_flowchart}. In the angular-ordered shower, the branching
history is first generated in terms of shower variables, and the physical
four-momenta are reconstructed after the shower evolution has terminated. In
the dipole shower, the event kinematics are instead updated at each accepted
branching, with the recoil assigned according to the relevant dipole mapping.

Both shower modules are intended to provide the same level of physics
predictions in order to be able to assess uncertainties, in so far as is possible,
by cross-comparisons when first-principle theoretical prescriptions are not
available. Using the approach described in Ref.~\cite{Bellm:2016rhh}, a number of
studies have been performed for specific processes
\cite{Cormier:2018tog, Rauch:2016upa}. A number of efficient techniques are
provided to evaluate parton-shower variations \cite{Bellm:2016voq}.

In this section, following a brief review of the core concepts common to both
parton-shower algorithms, we describe the two algorithms in detail. This is
followed by a discussion of issues common to both algorithms. The section then
concludes with details of the C++ code structure.

\subsubsection{Basics}
\label{\detokenize{review/showers/general:basics}}\label{\detokenize{review/showers/general:sect-shower-basics}}

Both parton-shower algorithms in Herwig, and in general all modern parton-shower simulations \cite{Buckley:2011ms}, have their outset from the basic properties of QCD radiation.

All parton showers start from the approximation that, in the \textit{(quasi-)collinear} %
\begin{footnote}\sphinxAtStartFootnote
While older parton-shower programs generally just used the collinear limit, \textit{i.e.} assuming all the
partons are massless, most modern programs including both shower algorithms in Herwig use the
quasi-collinear limit to improve the treatment of radiation from massive particles.
\end{footnote}
limit \cite{Catani:2000ef}, in which the transverse
momentum of the branching and the masses of the particles are small
compared to the other scales, the matrix element for a process with an additional parton, $\mathcal{M}_{n+1}$,
is approximately
\begin{equation}\label{equation:review/showers/general:eq:me_factorization}
\begin{split}\left|\mathcal{M}_{n+1}\right|^{2}=\frac{8\pi\alpha_{s}}{q_{\widetilde{ij}}w^{2}-m_{\widetilde{ij}}^{2}}\,
P_{\widetilde{ij}\to ij}(z,q_{\widetilde{ij}}^2)\,\left|\mathcal{M}_{n}\right|^{2},\end{split}
\end{equation}

where $\alpha_{s}$ is the strong coupling constant, $q_{\widetilde{ij}}^2$ is the virtuality of the branching
parton $\widetilde{ij}$ %
\begin{footnote}\sphinxAtStartFootnote
We reserve the tilde notation $\widetilde{ij}$ exclusively to
denote the parent parton, which decays into daughters $i$ and
$j$.
\end{footnote}, $m_{\widetilde{ij}}^2$ is the physical mass of the branching particle,
$z$ is the fraction of the momentum of $\widetilde{ij}$ carried by the parton $i$,
$P_{\widetilde{ij}\to ij}\left(z,q_{\widetilde{ij}}^2\right)$ is the quasi-collinear splitting function and
$\mathcal{M}_{n}$ is the matrix element for the process without the additional collinear parton.
As the phase-space also factorizes in this limit, we can write the
probability of the $\widetilde{ij}$ branching to produce partons $i$ and
$j$ in the \textit{quasi-collinear limit} as
\begin{equation}\label{equation:review/showers/general:eq:branchquasi}
\begin{split}\mathrm{d}\mathcal{P}_{\widetilde{ij}\to ij}=\frac{\alpha_{s}}{2\pi}\,\frac{\mathrm{d}q^2}{q^2-m_{\widetilde{ij}}^2}\,\mathrm{d}z\, P_{\widetilde{ij}\to ij}\left(z,q^2\right).\end{split}
\end{equation}

The splitting functions with soft singularities, i.e.\ where the emitted particle $j$ is a gluon%
\begin{footnote}\sphinxAtStartFootnote
i.e.\ $P_{q\to qg}$, $P_{\tilde{q}\to\tilde{q}g}$, $P_{g\to gg}$, and
$P_{\tilde{g}\to\tilde{g}g}$.
\end{footnote}, all behave as
\begin{equation}\label{equation:review/showers/general:eq:soft_splitting_fns}
\begin{split}\lim_{z\rightarrow1}P_{\widetilde{ij}\to ij}=\frac{2C_{\widetilde{ij}}}{1-z}\left(1-\frac{m_{i}^{2}}{z(1-z)(q^2-m^2_{\widetilde{ij}})}\right),\end{split}
\end{equation}

in the soft $z\rightarrow1$ limit, where
$C_{\widetilde{ij}}$ equals $C_{F}$ for $P_{q\to qg}$
and $P_{\tilde{q}\to\tilde{q}g}$, $\frac12C_{A}$%
\begin{footnote}\sphinxAtStartFootnote
Note that for $g\to gg$, there is also a soft singularity at
$z\to0$ with the same strength, so that the \textit{total} emission
strength for soft gluons from particles of all types in a given
representation is the same: $C_F$ in the fundamental
representation and $C_A$ in the adjoint.
\end{footnote} for
$P_{g\to gg}$, and $C_{A}$ for
$P_{\tilde{g}\to\tilde{g}g}$.

Similarly, if we consider the emission of a soft gluon, the amplitude
for the process including an extra soft gluon with momentum $q_j$ is
\begin{equation*}
\begin{split}\mathcal{M}_{n+1}  = g_s \mathcal{M}_n {\bf J}(q_j),\end{split}
\end{equation*}

where ${\bf J}(q_j)$ is the non-Abelian semi-classical ‘eikonal’
current for the emission of the soft gluon
from the hard partons. In general the eikonal current is
\begin{equation*}
\begin{split}{\bf J}(q_j) = \varepsilon^*_\mu
    \sum_{\mathrm {\stackrel{\scriptstyle external}{partons}}}
    C^b_\alpha P^{ab}_\alpha
    \left(\frac{p_{\mathrm{parton}}}{p_{\mathrm{parton}}\cdot q_j}\right)^\mu,\end{split}
\end{equation*}

for the emission of a soft gluon with momentum $q_j$ and polarization vector $\varepsilon$.
The colour structure of the leading-order process without the soft gluon is
given by $C^b_\alpha$, while $P^{ab}_\alpha$ is
the colour matrix for the emission of a gluon with
colour $a$. If we consider the simplest case of two partons, $i,k$, with four-momenta
$q_i,q_k$ respectively, which form a colour-singlet system, in the soft limit
\begin{equation}\label{equation:review/showers/general:eq:eikonal_shower}
\begin{split}\left|\mathcal{M}_{n+1}\right|^{2} = -4\pi\alpha_{s}C_{\widetilde{ij}}\left(\frac{q_k}{q_k\cdot
q_j}-\frac{q_i}{q_i\cdot q_j}\right)^{2}\left|\mathcal{M}_{n}\right|^{2}.\end{split}
\end{equation}

In order to correctly treat both the soft and collinear limits for gluon emission, and describe
multiple emissions we need to define an appropriate emission probability
\begin{equation*}
\begin{split}\mathrm{d}\mathcal{P}_{\mathrm{branching}} = \mathrm{d} \kappa \frac{\alpha_{s}}{2\pi} \mathcal{K}_{\mathrm{branching}}(\kappa),\end{split}
\end{equation*}

with an evolution variable $\kappa$ and kernel $\mathcal{K}_{\mathrm{branching}}(\kappa)$.

In Herwig the two parton-shower algorithms make different choices :
\begin{itemize}
\item {} 

the angular-ordered parton shower uses an evolution variable which reduces
to the angle of emission in the collinear limit, together with the quasi-collinear branching probability, Eq. \eqref{equation:review/showers/general:eq:branchquasi},
and a disjoint phase-space for emission from different partons which correctly treats both soft and collinear emission
as discussed in \hyperref[\detokenize{review/showers/qtilde:sect-angular-shower}]{Section \ref{\detokenize{review/showers/qtilde:sect-angular-shower}}};

\item {} 

the dipole shower uses the transverse momentum for the evolution variable together with a
dipole phase-space including both an emitter and a spectator which correctly treats both soft and collinear
emission using a kernel which reduces to Eqs. \eqref{equation:review/showers/general:eq:branchquasi} and \eqref{equation:review/showers/general:eq:eikonal_shower} in the appropriate
limits.

\end{itemize}

In general, the emission probability for the radiation of gluons is
infinite in the soft and collinear
limits. Physically these divergences would be cancelled by virtual
corrections, which we do not explicitly calculate but rather include
through unitarity. We therefore impose a physical cutoff
and call radiation above this limit resolvable. The
cutoff ensures that the contribution from resolvable radiation is
finite. Equally the uncalculated virtual corrections ensure that the
contribution of the virtual and unresolvable emission below the cutoff
is also finite. Imposing unitarity
\begin{equation*}
\begin{split}\mathcal{P}\left(\mathrm{resolved}\right)+\mathcal{P}\left(\mathrm{unresolved}\right)=1\end{split}
\end{equation*}

gives the probability of no branching in an infinitesimal increment of
the evolution variable $\mathrm{d}\kappa$ as
\begin{equation*}
\begin{split}1-\sum_{\mathrm{branchings}}\mathrm{d}\mathcal{P}_{\mathrm{branching}},\end{split}
\end{equation*}

where the sum runs over all possible branchings of the emitting particles
$\widetilde{ij}$ in the angular-ordered shower, and
all possible branchings and spectators in the dipole shower.
The probability that a parton does not branch
between two scales is given by the product of the probabilities that it
did not branch in any of the small increments
$\mathrm{d}\kappa$ between the two scales. Hence, in the limit
$\mathrm{d}\kappa\to0$ the probability of no branching
exponentiates, giving the \textit{Sudakov form factor}
\begin{equation}\label{equation:review/showers/general:eq:product_of_sudakovs}
\begin{split}\Delta\left(\kappa,\kappa_{h}\right)=\prod_{\mathrm{branchings}}\Delta_{\mathrm{branching}}\left(\kappa,\kappa_{h}\right),\end{split}
\end{equation}

which is the probability of evolving between the scale
$\kappa_{h}$ and $\kappa$ without resolvable emission. The Sudakov form factor for
an individual branching is given by
\begin{equation}\label{equation:review/showers/general:eq:gen_sudakov}
\begin{split}\Delta_{\mathrm{branching}}\left(\kappa,\kappa_{h}\right) = \exp\left(-\int \mathrm{d}\mathcal{P}_{\mathrm{branching}}\right)
 = \exp\left(-\int^{\kappa_h}_{\kappa}\mathrm{d} \kappa'  \frac{\alpha_{s}}{2\pi} \mathcal{K}_{\mathrm{branching}}(\kappa')\right).\end{split}
\end{equation}

The formalism discussed above allows us, if given a starting scale
$\kappa_{h}$, to evolve a parton down in scale and generate the
next branching of this particle at a lower scale. The no-emission
probability encoded in the Sudakov form factor is used to generate
the evolution variable and any other variables, \textit{e.g.} the momentum fraction $z$, for this branching.
This procedure can then be iterated to generate subsequent branchings of the particles
produced until no further emission occurs above the cutoff. The procedures used solve
\begin{equation}\label{equation:review/showers/general:eq:sudakov_master}
\begin{split}\Delta\left(\kappa,\kappa_{h}\right)=\mathcal{R},\end{split}
\end{equation}

where $\mathcal{R}$ is a random number uniformly distributed between 0 and 1. In solving Eq.~\eqref{equation:review/showers/general:eq:sudakov_master}, the variables describing the emission are generated as discussed in \hyperref[\detokenize{review/appendix/sudakov:sect-sudakov-solution}]{\ref{\detokenize{review/appendix/sudakov:sect-sudakov-solution}}}.

The treatment of the strong coupling constant, and its argument, are described in detail in \hyperref[\detokenize{review/appendix/alphaS:sub-the-running-coupling}]{\ref{\detokenize{review/appendix/alphaS:sub-the-running-coupling}}}.

\subsubsection{Structure of the evolution and showering of scattering processes}
\label{\detokenize{review/showers/general:structure-of-the-evolution-and-showering-of-scattering-processes}}

Both parton shower algorithms can be used to shower both
$2\to n$ scattering processes and $1\to n$ decay processes,
where $n$ is the number of outgoing particles.
In the case of electron-positron collisions this is simply the hard subprocess and
any decay processes, however in a hadron-hadron collision, secondary collisions,
or multiple parton interactions, also require showering in addition to the
primary hard subprocess. The details of the simulation of the multiple and soft scattering processes
are discussed in \hyperref[\detokenize{review/index:sect-ue}]{Section \ref{\detokenize{review/index:sect-ue}}}.
In this section we will explicitly discuss the showering of
the hard subprocess and decay processes, however the procedure also
applies to secondary processes.

In the first stage of these processes, the particles from the hard scattering
processes are separated into the core hard scattering processes and the subsequent decays
of any unstable particles%
\begin{footnote}\sphinxAtStartFootnote
This behaviour is controlled using the  \href{https://herwig.hepforge.org/doxygen/ShowerHandlerInterfaces.html\#SplitHardProcess}{SplitHardProcess} switch which by default separates the hard and decay processes. This is necessary in order to correctly shower any almost on-shell intermediates and preserve their masses when the decays of the particles have been generated as part of the hard scattering processes. It is not usually required in processes where Herwig 7 generates the hard processes as we handle the decays as part of the showering process but can be required for external events supplied in the Les Houches Event format.
\end{footnote}. By default both parton shower
algorithms work in the  narrow width approximation,
therefore the hard process and any decay processes (i.e.\ an unstable particle
and its decay products) are showered separately.

The hard process sets the initial conditions for
the parton-shower evolution. There are two initial conditions: the first, and most important,
is the large-$N_C$ colour structure of the hard process, which determines:
\begin{itemize}
  \item the direction of the colour partner for each shower progenitor
  (see Section~\ref{\detokenize{review/showers/qtilde:sub-shower-kinematics}}) and hence the maximum
  evolution scale in the angular-ordered parton shower;
  \item which dipoles are present in the dipole shower.
\end{itemize}

In addition to the constraints given by the colour structure, we impose that the maximum transverse
momentum for the radiation in the parton shower is below a veto scale determined from the factorization
scale for the hard process. This scale is interpreted differently in the two parton-shower
modules:
\begin{itemize}
\item {} 

in the angular-ordered parton shower any emissions with transverse momenta relative to the jet axis
above this scale are vetoed, as the shower emissions are not ordered in transverse momentum;

\item {} 

in the dipole shower, which is ordered in transverse momentum, the scale simply sets the maximum emission
scale in the parton shower.

\end{itemize}

These scales and their variations, in order to assess uncertainties, are discussed in more detailed in \hyperref[\detokenize{review/showers/variations:shower-scale-variations}]{Section \ref{\detokenize{review/showers/variations:shower-scale-variations}}}.

The two parton showers also differ in their treatment of mass effects. The dipole shower requires a fixed flavour
number scheme so that the masses are consistently defined in both the hard process and subsequent parton showering, for
example if working in a five-flavour scheme then the bottom mass will consistently be set to zero.
The angular-ordered parton shower uses a more pragmatic approach where the masses of the partons in the hard process can be zero
but they are put on mass-shell in the parton shower in order to ensure the correct radiation at small scales.

\subsubsection{Outline}
\label{\detokenize{review/showers/general:outline}}

We will now discuss the two parton-shower algorithms in detail.
This is followed by discussions of issues common to both
parton-shower algorithms:
\begin{itemize}
\item {}

perturbative decays and spin correlations;

\item {}

intrinsic transverse momentum;

\item {}

forced splitting in initial-state radiation;

\item {}

the simulation of QED radiation using the YFS formalism
\cite{Yennie:1961ad};

\item {}

shower variations and reweighting.

\end{itemize}

\subsection{Angular-Ordered parton shower}
\label{\detokenize{review/showers/qtilde:angular-ordered-parton-shower}}\label{\detokenize{review/showers/qtilde:sect-angular-shower}}\label{\detokenize{review/showers/qtilde::doc}}

The default parton shower algorithm is the \textit{coherent branching algorithm}
\cite{Gieseke:2003rz}, which extends the formalism originally developed
for the HERWIG program \cite{Marchesini:1984bm, Webber:1983if, Marchesini:1987cf}.
The new algorithm:
\begin{itemize}
\item {} 

retains angular ordering as a central feature;

\item {} 

uses a covariant formulation of the showering algorithm, which is
invariant under boosts along the jet axis;

\item {} 

improves the treatment of heavy quark fragmentation through the use of
mass-dependent splitting functions \cite{Catani:2000ef} and
kinematics, providing a complete description of the so-called
\textit{dead-cone} region;

\item {} 

includes azimuthal correlations between the hard scattering process,
perturbative decays and all shower emissions using the approach of Refs.
\cite{Richardson:2001df, Knowles:1988vs, Collins:1987cp, Richardson:2018pvo};

\item {} 

includes interleaved QCD, QED and EW radiations, with the option of choosing/omitting interactions.

\item {} 

supports multiple choices for the control of the evolution scale $\tilde{q}$, i.e.\
the transverse momentum-preserving \cite{Gieseke:2003hm, Gieseke:2003rz},
virtuality-preserving \cite{Reichelt:2017hts} and dot-product-preserving
\cite{Bewick:2019rbu, Bewick:2021nhc} schemes.

\end{itemize}

In this section we give a full description of the angular-ordered parton shower model.
We begin by introducing the
fundamental kinematics and dynamics underlying the shower algorithm.
This is followed by descriptions of the initial conditions and the Monte
Carlo algorithms used to generate the showers.

\subsubsection{Shower kinematics}
\label{\detokenize{review/showers/qtilde:shower-kinematics}}\label{\detokenize{review/showers/qtilde:sub-shower-kinematics}}

Each of the colour-, weak- or electromagnetically-charged legs of the hard sub-process is considered to be a
\textit{shower progenitor}. The angular-ordered shower first generates a sequence of splittings in terms of the shower variables, while the corresponding physical four-momenta are reconstructed only after the shower evolution has terminated, as described for final-state and initial-state radiation in Sections~\ref{\detokenize{review/showers/qtilde:sect-finalrecon}} and~\ref{\detokenize{review/showers/qtilde:sect-initialrecon}}, respectively. We associate a set of basis vectors to each
progenitor, in terms of which we can express the momentum
$\left(q_{i}\right)$ of each particle in the resulting shower as
\begin{equation}\label{equation:review/showers/qtilde:eq:sudbasis}
\begin{split}q_{i}=\alpha_{i}p+\beta_{i}n+q_{\perp i},\end{split}
\end{equation}
that is the well-known \textit{Sudakov basis}. The vector $p$ is equal to
the momentum of the shower progenitor generated by the prior simulation
of the hard scattering process, \textit{i.e.} $p^{2}=m^{2}$, where
$m$ is the on-shell mass of the progenitor. The \textit{reference vector}
$n$ is a light-like vector that satisfies $n\cdot p>m^{2}$.
In practice $n$ is chosen anticollinear to $p$ in the frame
where the shower is generated, maximizing $n\cdot p$. Since we
almost always generate the shower in the rest frame of the progenitor
and an object with which it shares a colour line%
\begin{footnote}\sphinxAtStartFootnote
In the case that the particle is both colour and electromagnetically charged we use the
colour partner for the evolution, however if the particle is only electromagnetically
charged we pick a partner from the other charged particles using the
product of the charges of the particles as required for coherence of the QED radiation.
\end{footnote}, $n$ is therefore
collinear with this \textit{evolution partner} object. The $q_{\perp i}$
vector gives the remaining components of the momentum, transverse to
$p$ and $n$.

Our basis vectors satisfy the following relations:
\begin{equation*}
\begin{aligned}
q_{\perp i}\cdot p &= 0, &\qquad p^{2} &= m^{2}, &\qquad q_{\perp i}^{2} &= -\mathbf{q}_{\perp i}^{2},\\
q_{\perp i}\cdot n &= 0, &\qquad n^{2} &= 0,     &\qquad n\cdot p &> m^{2},
\end{aligned}
\end{equation*}
where $\mathbf{q}_{\perp i}$ is the spatial component of
$q_{\perp i}$ in the frame where the shower is generated
$\left(\mathbf{q}_{\perp i}^{2}\ge0\right)$. Given these
definitions, calculating $q_{i}^{2}$, one finds that
$\beta_{i}$ may be conveniently expressed in terms of the mass and
transverse momentum of particle $i$ as
\begin{equation}\label{equation:review/showers/qtilde:eq:beta_i}
\begin{split}\beta_{i}=\frac{q_{i}^{2}-\alpha_{i}^{2}m^{2}-q_{\perp
    i}^{2}}{2\alpha_{i}n\cdot p}.\end{split}
\end{equation}

The shower algorithm does not generate the momenta or Sudakov parameters
directly. In practice what is generated first is a set, each element of
which consists of three \textit{shower variables}, which fully parameterize
each parton branching. One of these variables parameterizes the scale of
each branching, the so-called \textit{evolution scale}, which we shall discuss
in more detail below. Typically this evolution scale starts at a high
value, characteristic of the process that produces the progenitors, and
continually reduces as the shower develops, via the radiation of
particles. When the evolution scale has reduced to the point where there
is insufficient phase-space to produce any more branchings, the
resulting partons are considered to be on-shell, and the reconstruction
of the momenta from the shower variables may begin in full.

The first shower variable  is the
\textit{light-cone momentum fraction}
$z$. Given a branching,
$\widetilde{ij}\rightarrow i+j$, this parameterizes how the
momentum component of the parent parton, $\widetilde{ij}$, in the
direction of the shower progenitor\footnote{Here, the \textit{progenitor} denotes the original external leg that initiates
the shower, whereas the \textit{parent} denotes the particle undergoing a
particular branching within that shower.}, is divided between its two
daughter partons, $i$ and $j$. We define $z$ as
\begin{equation}\label{equation:review/showers/qtilde:eq:z_defn}
\begin{split}z=\frac{\alpha_{i}}{\alpha_{\widetilde{ij}}}=\frac{n\cdot q_{i}}{n\cdot q_{\widetilde{ij}}}.\end{split}
\end{equation}

For particles in the final state we use a forward evolution algorithm
where the parton shower consists of a sequence of branchings
$\widetilde{ij}\rightarrow i+j$, ordered in the evolution scale.
For incoming particles we use a backward evolution algorithm where we
start at the large evolution scale of the scattering process and evolve
the incoming particles backwards toward the incoming hadron to give the
mother $\widetilde{ij}$ and the sister parton $j$, again
with a decreasing evolution scale. We use the definition of $z$ in
Eq. \eqref{equation:review/showers/qtilde:eq:z_defn} both for forward and backward parton
shower algorithms.

The second variable used to parameterize a branching is the azimuthal
angle, $\phi$, of the relative transverse momentum of each
branching $p_{\perp}$, measured with respect to the direction of the shower progenitor momentum $p$. The relative transverse momentum $p_{\perp}$ is
\textit{defined} to be
\begin{equation}\label{equation:review/showers/qtilde:eq:pt_fwd}
\begin{split}p_{\perp}=q_{\perp i}-zq_{\perp\widetilde{ij}}.\end{split}
\end{equation}

As with the definition of $z$, this definition of the relative
transverse momentum is the same for both forward and backward
parton-shower evolution algorithms.

The last, and most important, of the shower variables defining a
branching is the evolution scale. Parton shower algorithms may be
formulated as an evolution in the virtualities of the branching partons,
or as an evolution in the transverse momentum of the branching products.
However, a careful treatment of colour coherence effects
\cite{Marchesini:1984bm, Bassetto:1982ma, Bassetto:1984ik, Catani:1983bz, Ciafaloni:1980pz, Ciafaloni:1981bp, Dokshitzer:1988bq, Mueller:1981ex, Ermolaev:1981cm, Dokshitzer:1982fh}
reveals that branchings involving soft gluons should be ordered in the
angle between the branching products.

The key finding of these studies is that, when soft gluon emissions are
considered, branchings that are not angular-ordered do not give any
leading-logarithmic contributions. This is a dynamical effect whereby
radiation from the emitting partons, with smaller angular separations,
interferes destructively in these non-ordered regions. Some intuitive
understanding of the effect may be gained by considering that a soft
gluon, emitted at a large angle from a jet-like configuration of
partons, does not have sufficient transverse resolving power to probe
the internal jet structure. As a result, it is only sensitive to the
\textit{coherent sum} of the collinear singular contributions associated with
the constituents, resulting in a contribution equivalent to that from
the original progenitor parton. Destructive interference in the
non-ordered region effectively decreases the available phase-space for
each branching, from the virtuality-ordered region to the
angular-ordered region.

It may be shown that the contributions that angular ordering misses are
purely soft and suppressed by at least one power of $1/N_C^2$,
where $N_{C}=3$, the number of colours in QCD. Formally then,
omitting such contributions amounts to neglecting terms of
next-to-leading-logarithmic accuracy that are \textit{also} strongly colour suppressed.
We stress however, that whereas angular ordering leads to an \textit{omission}
of these suppressed higher order terms, other forms of ordering must
prove that they do not overestimate leading-logarithmic contributions.

For the forward evolution of partons with time-like virtualities, the
variable used to achieve such ordering, $\tilde{q}^{2}$, is
defined according to
\begin{equation}\label{equation:review/showers/qtilde:eq:qtilde_timelike}
\begin{split}z\left(1-z\right)\tilde{q}^{2}=-m_{\widetilde{ij}}^{2}+\frac{m_{i}^{2}}{z}+\frac{m_{j}^{2}}{1-z}-\frac{p_{\perp}^{2}}{z\left(1-z\right)},\end{split}
\end{equation}

where $m_{i}$ is the on-shell mass of particle $i$ \textit{etc.}
This definition is arrived at by generalizing the FORTRAN HERWIG angular-evolution variable, $\tilde{q}^{2}=q_{i}^{2}/\left(z(1-z)\right)$,
to include the effects of the mass of the emitting
parton. This may be seen by writing
$q_{\widetilde{ij}}=q_{i}+q_{j}$, and calculating
$q_{\widetilde{ij}}^{2}\left(z,p_{\perp}^{2},q_{i}^{2},q_{j}^{2}\right)$,
which shows
\begin{equation}\label{equation:review/showers/qtilde:eq:qtilde_timelike_explicit}
\begin{split}\tilde{q}^{2}=\left.\frac{q_{\widetilde{ij}}^{2}-m_{\widetilde{ij}}^{2}}{z\left(1-z\right)}\right|_{q_{i}^{2}=m_{i}^{2},\mbox{ }q_{j}^{2}=m_{j}^{2}}.\end{split}
\end{equation}

For showers involving the evolution of partons with space-like
virtualities, the evolution variable is instead defined by
\begin{equation}\label{equation:review/showers/qtilde:eq:qtilde_spacelike}
\begin{split}\left(1-z\right)\tilde{q}^{2}=-zm_{\tilde{ij}}^{2}+m_{i}^{2}+\frac{zm_{j}^{2}}{1-z}-\frac{p_{\perp}^{2}}{1-z}.\end{split}
\end{equation}

Once again this definition of the evolution variable is a generalization
of the analogous FORTRAN HERWIG angular-evolution variable used for
initial-state radiation:
$\tilde{q}^{2}=-q_{i}^{2}/\left(1-z\right)$. Using momentum
conservation, $q_{\widetilde{ij}}=q_{i}+q_{j}$, we may calculate
$q_{i}^{2}\left(z,p_{\perp}^{2},q_{\widetilde{ij}}^{2},q_{j}^{2}\right)$,
giving
\begin{equation*}
\begin{split}\tilde{q}^{2}=\left.\frac{m_{i}^{2}-q_{i}^{2}}{1-z}\right|_{q_{\widetilde{ij}}^{2}=m_{\widetilde{ij}}^{2},\mbox{ }q_{j}^{2}=m_{j}^{2}}.\end{split}
\end{equation*}
To see how these variables relate to the angle between the branching
products, consider that the parton shower is generated in the frame
where the light-like basis vector $n$ is anticollinear to the
progenitor. For forward evolving partons with small time-like
virtualities, expanding $z$ and $q_{\widetilde{ij}}^{2}$ in
component form,
\begin{equation*}
\begin{split}\tilde{q}^{2}=\frac{2E_{\widetilde{ij}}^{2}\left(1-\cos\theta_{ij}\right)\left(1+\cos\theta_{\widetilde{ij}}\right)^{2}}{\left(1+\cos\theta_{i}\right)\left(1+\cos\theta_{j}\right)},\end{split}
\end{equation*}
where $\theta_{i}$ and $\theta_{j}$ are the angles between
the daughter particles $i$, $j$ and the progenitor,
$\theta_{\tilde{ij}}$ is the angle between the parent and the
progenitor, and $\theta_{ij}$ is the angle between the two
daughters, and $E_{\tilde{ij}}$ is the energy of the parent. This
expression for the time-like evolution variable in terms of angles is
more complicated than the analogous FORTRAN HERWIG formula:
$\tilde{q}^{2}=2E_{\widetilde{ij}}^{2}\left(1-\cos\theta_{ij}\right)$.
This is because in FORTRAN HERWIG $z$ was defined to
be the energy fraction $E_{i}/E_{\widetilde{ij}}$, instead of the
light-cone momentum fraction as given in Eq. \eqref{equation:review/showers/qtilde:eq:z_defn}. Nevertheless, for small angles we find that
the Herwig 7 and FORTRAN HERWIG evolution variables are both given by
\begin{equation}\label{equation:review/showers/qtilde:eq:qtilde_timelike_small_angles}
\begin{split}\tilde{q}=E_{\widetilde{ij}}\theta_{ij}\left(1-\mathcal{O}\left(\theta_{x}^{2}\right)\right).\end{split}
\end{equation}

When a branching occurs, the daughter partons $i$ and $j$,
with momentum fractions $z$ and $1-z$, have their starting
evolution scales set to $z\tilde{q}$ and
$\left(1-z\right)\tilde{q}$ respectively, where
$z\tilde{q}\approx E_{i}\theta_{ij}$ and
$\left(1-z\right)\tilde{q}\approx E_{j}\theta_{ij}$. In this way
the maximum opening angle of any subsequent branching is
$\theta_{ij}$, thereby implementing angular ordering.

For initial-state showers the same QCD coherence argument applies, so in
evolving backwards, away from the hard process, the angle between the
mother of the branching and its final-state daughter parton must
decrease. Writing the space-like evolution variable (Eq.
\eqref{equation:review/showers/qtilde:eq:qtilde_spacelike}) in terms of angles, neglecting
parton virtualities, one finds the same form as for the time-like
variable in Eq. \eqref{equation:review/showers/qtilde:eq:qtilde_timelike_small_angles}.
This means that once a branching has occurred in the course of the
backward evolution, the mother of the branching evolves backwards from
scale $\tilde{q}$, and the daughter evolves forwards from scale
$\left(1-z\right)\tilde{q}$, as in the time-like case.

Even at leading-logarithmic accuracy there are correlations between subsequent emissions in the
parton-shower which affect the azimuthal angles of the partons. There can also be correlations
between emissions from different progenitors which depend on the nature of the
hard scattering process. As in FORTRAN HERWIG, Herwig 7 includes these correlations
using a generalisation \cite{Richardson:2001df, Richardson:2018pvo} of the spin correlation algorithm
\cite{Knowles:1988vs, Collins:1987cp}
used in FORTRAN HERWIG. This approach allows the emissions to be generated as a Markov
chain while including the correlations using the spin density formalism.
The approach is described in detail in \hyperref[\detokenize{review/showers/perturbative_decays:sec-perturbative-decays}]{Section \ref{\detokenize{review/showers/perturbative_decays:sec-perturbative-decays}}} for the simulation
of perturbative decays. In addition to the collinear spin correlations,
the correlations for soft emission which are present in the full eikonal factor
are also included multiplicatively.

When the evolution in terms of the shower variables has
run its course, \textit{i.e.} there is no more phase-space available for
further emissions, the external particles are taken as being on-shell
and the reconstruction in terms of the physical momenta can start.

In performing this reconstruction we must make a number of formally
subleading choices which can however significantly affect the results.
In particular in the case of subsequent emission from the daughter partons $i$ and/or $j$
we must decide which properties of the originally generated kinematics to preserve once the
masses of $i$ and/or $j$ in Eqs.~\eqref{equation:review/showers/qtilde:eq:qtilde_timelike}--\eqref{equation:review/showers/qtilde:eq:qtilde_spacelike}
are no longer the infrared cut-off masses but the virtualities generated by any subsequent emissions.
The choice of which kinematic quantities to preserve affects how we proceed with the reconstruction
of the momenta.

First, all of the $\alpha$ coefficients in the Sudakov
decomposition of each momentum are calculated. This is done by first
setting $\alpha$ equal to one for final-state progenitors and to
the associated PDF light-cone momentum fraction $x$, generated in
the preceding simulation of the hard process, for initial-state
progenitors. Using the relation defining $z$, Eq. \eqref{equation:review/showers/qtilde:eq:z_defn}, together with the momentum conservation
relation $\alpha_{\widetilde{ij}}=\alpha_{i}+\alpha_{j}$, one can
iteratively calculate all $\alpha$ values, starting from the hard
process and working outward to the external legs.

To proceed with the reconstruction, we need to decide which quantity
is preserved after multiple emissions. We now illustrate the
alternative options supported by Herwig.
\begin{enumerate}
\sphinxsetlistlabels{\arabic}{enumi}{enumii}{}{.}%
\item {} 

The simplest choice we can then make for the reconstruction is to
preserve the transverse momentum $p_T$. Therefore, we always
calculate the transverse momenta of the branchings using
Eqs.~\eqref{equation:review/showers/qtilde:eq:qtilde_timelike}--\eqref{equation:review/showers/qtilde:eq:qtilde_spacelike} using the
cutoff values of the masses. This was the originally implemented
option. However, when generating final-state radiation the
virtuality of the emitting parton is increased and too much
radiation can be generated into the non-angular-ordered region
after multiple emissions, motivating implementation of other
options \cite{Bewick:2019rbu, Bewick:2021nhc}.

\item {} 

Alternatively, we can choose to preserve
the virtuality of the branching parton. The transverse momentum of
the branching is then computed using the virtual masses the
daughter partons develop due to subsequent emissions. This means
that if the daughter partons develop large virtual masses the
transverse momentum of the branching is reduced. This can lead to
situations where there is no solution of
Eq.~\eqref{equation:review/showers/qtilde:eq:qtilde_timelike} and the emission would have to be
vetoed, which inhibits further soft emission and significantly
changes the evolution, leading to incorrect evolution of
observables. Thus, instead of vetoing the emission, we set $p_T=0$ in
Eq.~\eqref{equation:review/showers/qtilde:eq:qtilde_timelike} and allow the virtuality to increase. Although
this alternative reconstruction retains in general better agreement
with the data in the hard region of the spectrum, it has been shown
that this can formally violate the logarithmic accuracy of the
parton shower \cite{Bewick:2019rbu}.

\item {} 

The last option that has been introduced,
which is the current default, preserves the dot product
of the emitted partons \cite{Bewick:2019rbu, Bewick:2021nhc}.  The ordering
variable can be rewritten as
\begin{equation}\label{equation:review/showers/qtilde:eq:qtilde_dot_fsr}
\begin{split}\tilde{q}^{2}=\frac{2 q_i \cdot q_j + m^2_j + m^2_i-m^2_{\tilde{ij}}}{z(1-z)},\end{split}
\end{equation}
and it is easy to check that if just one emission takes place,
Eqs.~\eqref{equation:review/showers/qtilde:eq:qtilde_dot_fsr} and \eqref{equation:review/showers/qtilde:eq:qtilde_timelike_explicit} are
identical. With this definition the transverse momentum is thus
given by
\begin{equation*}
\begin{split}p_T^2 = z^2(1-z)^2 \tilde{q}^2 -q^2_i (1-z)^2 -q^2_j z ^2 + z(1-z) \left[ m^2_{\tilde{ij}} -m^2_i -m^2_j \right].\end{split}
\end{equation*}
This can still lead to a negative $p_T^2$ if one of the
particles is massive, but the fraction of events where this happens is
significantly smaller than the one obtained preserving the
virtuality. In particular, if an emission is soft, the variation it
introduces in the transverse momentum of the previous emission is
small, while this happens to be the case in the virtuality-preserving
scheme only if the emission is softer than the previous one.  The
default behaviour is thus to preserve the dot product and in cases in which a
solution with negative transverse momentum is encountered, $p_T$
is set to 0 and the dot product increases.

\end{enumerate}

In all the approaches the magnitude of the relative
transverse momentum
$\left|\mathbf{p}_{\perp}\right|=\sqrt{-p_{\perp}^{2}}$ is
calculated in terms of the evolution variables $z$ and
$\tilde{q}^{2}$ once the evolution of the daughter partons has terminated.

For final-state showers the $q_{\perp}$ components of each
momentum may then be calculated simultaneously. Final-state showering cannot
change the direction of the progenitor since the transverse momentum
must be conserved at each branching, hence the $q_{\perp}$
component of the progenitor is zero.

The $q_{\perp}$ components of
the branching products are iteratively computed by adding the relative
transverse momentum,
\begin{equation}\label{equation:review/showers/qtilde:eq:pt_components}
\begin{split}p_{\perp}=\left(\left|\mathbf{p}_{\perp}\right|\cos\phi,\left|\mathbf{p}_{\perp}\right|\sin\phi,0;0\right),\end{split}
\end{equation}

to $z$ times the transverse momentum of the mother,
$q_{\perp\widetilde{ij}}$, to give $q_{\perp i}$ according
to Eq. \eqref{equation:review/showers/qtilde:eq:pt_fwd};
$q_{\perp j}=q_{\perp\widetilde{ij}}-q_{\perp i}$ immediately
follows by momentum conservation.

The only remaining Sudakov parameters to be determined are the
$\beta$ values. These can be obtained once the evolution in terms
of the shower variables is complete, by using the fact that the external
partons are on-shell, in order to compute their $\beta$
coefficients from Eq. \eqref{equation:review/showers/qtilde:eq:beta_i}. The coefficients of
their parent momenta may then be computed using momentum conservation:
$\beta_{\widetilde{ij}}=\beta_{i}+\beta_{j}$. The latter step may
be iterated until the progenitor is reached, yielding all $\beta$
coefficients.

The reconstruction of the initial-state parton showers is slightly
different but it follows essentially the same reasoning. Our aim here
has been to simply sketch how the reconstruction occurs. More detailed
presentations of these procedures will be given later in
\hyperref[\detokenize{review/showers/qtilde:sub-final-state-radiation}]{Section \ref{\detokenize{review/showers/qtilde:sub-final-state-radiation}}}, \hyperref[\detokenize{review/showers/qtilde:sub-initial-state-radiation}]{Section \ref{\detokenize{review/showers/qtilde:sub-initial-state-radiation}}} and
\hyperref[\detokenize{review/showers/qtilde:sub-radiation-in-particle}]{Section \ref{\detokenize{review/showers/qtilde:sub-radiation-in-particle}}}.

\subsubsection{Shower dynamics}
\label{\detokenize{review/showers/qtilde:shower-dynamics}}\label{\detokenize{review/showers/qtilde:sub-shower-dynamics}}

With the kinematics defined, we now consider the dynamics governing the
parton branchings. Each parton branching is approximated by the
\textit{quasi-collinear limit} \cite{Catani:2000ef}, in which the transverse
momentum squared, $\mathbf{p}_{\perp}^{2}$, and the mass squared
of the particles involved are small (compared to $p\cdot n$) but
$\mathbf{p}_{\perp}^{2}/m^{2}$ is not necessarily small.
In this
limit the probability of the branching
$\widetilde{ij}\rightarrow i+j$ and with our choice of
evolution variables Eq. \eqref{equation:review/showers/general:eq:branchquasi} can be written as
\begin{equation}\label{equation:review/showers/qtilde:eq:branchprob}
\begin{split}\mathrm{d}\mathcal{P}_{\widetilde{ij}\to ij}=\frac{\alpha_{s}}{2\pi}\,\frac{\mathrm{d}\tilde{q}^{2}}{\tilde{q}^{2}}\,\mathrm{d}z\, P_{\widetilde{ij}\to ij}\left(z,\tilde{q}\right),\end{split}
\end{equation}

where $P_{\widetilde{ij}\to ij}\left(z,\tilde{q}\right)$ are the
quasi-collinear splitting functions derived in \cite{Catani:2000ef}. In
terms of our light-cone momentum fraction and (time-like) evolution
variable the quasi-collinear splitting functions are%
\begin{footnote}\sphinxAtStartFootnote
The $P_{g\to gg}$ splitting presented here is for final-state
branching where the outgoing gluons are not identified and therefore
it lacks a factor of two due to the identical particle symmetry
factor. For initial-state branching one of the gluons is identified
as being space-like and one as time-like and therefore an additional
factor of 2 is required.
\end{footnote}
\begin{equation}\label{equation:review/showers/qtilde:eq:AP}
\left.
\begin{aligned}P_{q\to qg} & =\frac{C_{F}}{1-z}\left[1+z^{2}-\frac{2m_{q}^{2}}{z\tilde{q}^{2}}\right],\\
P_{g\to gg} & =C_{A}\left[\frac{z}{1-z}+\frac{1-z}{z}+z\left(1-z\right)\right],\\
P_{g\to q\bar{q}} & =T_{R}\left[1-2z\left(1-z\right)+\frac{2m_{q}^{2}}{z\left(1-z\right)\tilde{q}^{2}}\right],\\
P_{\tilde{g}\to\tilde{g}g} & =\frac{C_{A}}{1-z}\left[1+z^{2}-\frac{2m_{\tilde{g}}^{2}}{z\tilde{q}^{2}}\right],\\
P_{\tilde{q}\to\tilde{q}g} & =\frac{2C_{F}}{1-z}\left[z-\frac{m_{\tilde{q}}^2}{z\tilde{q}^{2}}\right],\end{aligned}
\qquad\right\}
\end{equation}

for QCD and singular SUSY QCD branchings%
\begin{footnote}\sphinxAtStartFootnote
The splitting functions for photon emission from fermions and scalars
can be obtained by replacing the colour charge $C_F$ with
the square of the charge of the radiating particle in the $P_{q\to qg}$
and $P_{\tilde{q}\to\tilde{q}g}$ splitting functions.
Similarly the splitting function for $\gamma\to f\bar{f}$ can
be obtained from that for $P_{g\to q\bar{q}}$ by replacing
the colour factor $T_R$ with the square of the charge of the fermion.
The splitting functions for EW emissions are, however, more involved and
require treatment beyond simple charge substitutions.
These are discussed in detail in \hyperref[\detokenize{review/showers/qtilde:sect-ewsplittings}]{Section \ref{\detokenize{review/showers/qtilde:sect-ewsplittings}}}.
\end{footnote}. These splitting functions
give a correct physical description of the dead-cone region
$\mathbf{p}_{\perp}\lesssim m$, where the collinear singular limit
of the matrix element is screened by the mass $m$ of the emitting
parton.

The soft limit of the splitting functions is also important.
As described in \hyperref[\detokenize{review/showers/general:sect-shower-basics}]{Section \ref{\detokenize{review/showers/general:sect-shower-basics}}}, in the soft
limits the probability for emission of a soft gluon is
universal, Eq. \eqref{equation:review/showers/general:eq:soft_splitting_fns}.
Using the definitions of our shower variables, Eq. \eqref{equation:review/showers/qtilde:eq:z_defn},
and making the soft emission
approximations $q_{\widetilde{ij}}\approx q_{i}\approx p$,
$q_{i}^{2}\approx m_{i}^{2}=m_{\widetilde{ij}}^{2}$ in
Eqs. \eqref{equation:review/showers/general:eq:soft_splitting_fns}, \eqref{equation:review/showers/general:eq:me_factorization} we find \cite{Hamilton:2006ms}
\begin{equation}\label{equation:review/showers/qtilde:eq:eikonal_dipole_fn}
\begin{split}\lim_{z\rightarrow1}\,\frac{8\pi\alpha_{s}}{q_{\widetilde{ij}}^{2}-m_{\widetilde{ij}}^{2}}\,
P_{\widetilde{ij}\to ij}=-4\pi\alpha_{s}C_{\widetilde{ij}}\left(\frac{n}{n\cdot
q_{j}}-\frac{p}{p\cdot q_{j}}\right)^{2}.\end{split}
\end{equation}

Recalling that we choose our Sudakov basis vector $n$ to point in
the direction of the colour partner of the gluon emitter
($\widetilde{ij}/i$), Eq. \eqref{equation:review/showers/qtilde:eq:eikonal_dipole_fn} is then just the usual
soft eikonal dipole function, Eq. \eqref{equation:review/showers/general:eq:eikonal_shower}, describing soft gluon radiation by a colour
dipole \cite{Ellis:1991qj}, at least for the majority of cases where
the colour partner is massless or nearly massless. In practice, the
majority of processes we intend to simulate involve massless or light
partons, or partons that are light enough that $n$ reproduces the
colour partner momentum to high accuracy%
\begin{footnote}\sphinxAtStartFootnote
Even when the colour partner has a large mass, as in
$e^{+}e^{-}\rightarrow t\bar{t}$, the fact that each shower
evolves into the forward hemisphere, in the opposite direction to the
colour partner, means that the difference between Eq. \eqref{equation:review/showers/qtilde:eq:eikonal_dipole_fn}
and the exact dipole
function is rather small in practice.
\end{footnote}.

For the case that the underlying process with matrix element
$\mathcal{M}_{n}$ is comprised of a single colour dipole (as is
the case for a number of important processes), the parton shower
approximation to the matrix element $\mathcal{M}_{n+1}$, Eq. \eqref{equation:review/showers/general:eq:me_factorization}, then becomes exact in the soft
limit as well as, and independently of, the collinear limit. This leads
to a better description of soft wide angle radiation, at least for the
first emission, which is of course the widest angle emission in the
angular-ordered parton shower. Should the underlying hard process
consist of a quark anti-quark pair, this exponentiation of the full
eikonal current, Eq. \eqref{equation:review/showers/qtilde:eq:eikonal_dipole_fn}, combined with a careful treatment of
the running coupling (\hyperref[\detokenize{review/appendix/alphaS:sub-the-running-coupling}]{\ref{\detokenize{review/appendix/alphaS:sub-the-running-coupling}}}), will
resum all leading and next-to-leading logarithmic corrections
\cite{Frixione:2007vw, Catani:1990rr, Bonciani:2003nt, Cacciari:2002re}.
In the event that there is more than one colour dipole in the underlying
process, the situation is more complicated due to the ambiguity in
choosing the colour partner of the gluon, and the presence of non-planar
colour topologies.

With this choice of emission kernel and evolution variable the
no-emission probability for a given type of radiation, Eq. \eqref{equation:review/showers/general:eq:gen_sudakov}, is
\begin{equation}\label{equation:review/showers/qtilde:eq:sudakovmaster}
\begin{split}\Delta_{\widetilde{ij}\to ij}\left(\tilde{q},\tilde{q}_{h}\right)=\exp\left\{ -\int_{\tilde{q}}^{\tilde{q}_{h}}\frac{\mathrm{d}\tilde{q}^{\prime2}}{\tilde{q}^{\prime2}}\int\mathrm{d}z\mbox{ }\frac{\alpha_{s}\left(z,\tilde{q}^{\prime}\right)}{2\pi}\mbox{ }P_{\widetilde{ij}\to ij}\left(z,\tilde{q}^{\prime}\right)\Theta\left(\mathbf{p}_{\perp}^{2}>0\right)\right\} .\end{split}
\end{equation}

In practice we use the transverse momentum of the branching as the scale, \textit{i.e.} we use
\begin{equation*}
\begin{split}\alpha_{s}\left(z^{2}\left(1-z\right)^{2}\tilde{q}^{2}\right),\end{split}
\end{equation*}
as argument of the strong coupling (\hyperref[\detokenize{review/appendix/alphaS:sub-the-running-coupling}]{\ref{\detokenize{review/appendix/alphaS:sub-the-running-coupling}}}).

We currently support two different choices for the cutoff, controlled by the
\href{https://herwig.hepforge.org/doxygen/SudakovFormFactorInterfaces.html\#CutOffOption}{CutOffOption} switch:
\begin{itemize}
\item {} 

Our default option (\href{https://herwig.hepforge.org/doxygen/SudakovFormFactorInterfaces.html\#CutOffOption}{CutOffOption=pT}) is to impose a cutoff on the transverse momentum of
the branching, ensuring that $p_\perp>p_\perp^{\min}$ as this gives
a more physical behaviour and better agreement with experimental data,
see Ref. \cite{Reichelt:2017hts} and \hyperref[\detokenize{review/appendix/tuning:sect-tuning}]{\ref{\detokenize{review/appendix/tuning:sect-tuning}}} for more details. In
this case the $p_\perp^{\min}$ (\href{https://herwig.hepforge.org/doxygen/SudakovFormFactorInterfaces.html\#pTmin}{pTmin}) is one of the main parameters which is tuned to data.

\item {} 

Alternatively, the option named
\href{https://herwig.hepforge.org/doxygen/SudakovFormFactorInterfaces.html\#CutOffOption}{CutOffOption=default},
which corresponds to the default behaviour of earlier versions, imposes
a physical cutoff on the gluon and light-quark virtualities.
The allowed phase-space for each branching is obtained by requiring that
the relative transverse momentum is real, or
$\mathbf{p}_{\perp}^{2}>0$. For a general time-like branching
$\widetilde{ij}\rightarrow i+j$ this gives
\begin{equation*}
\begin{split}z^{2}\left(1-z\right)^{2}\tilde{q}^{2}-\left(1-z\right)m_{i}^{2}-zm_{j}^{2}+z\left(1-z\right)m_{\widetilde{ij}}^{2}>0,\end{split}
\end{equation*}
from Eq. \eqref{equation:review/showers/qtilde:eq:qtilde_timelike}.

In practice rather than using the physical masses for the light quarks
and gluon we impose a cutoff to ensure that the emission probability is
finite. We use a cutoff, $Q_{g}$, for the gluon mass, and we take
the masses of the other partons to be $\mu=\mbox{max}\left(m,Q_{g}\right)$, \textit{i.e.} $Q_{g}$ is the
lowest mass allowed for any particle.

The cutoff parameter, $Q_{g}$, is the minimum virtuality of the
gluon. However, if we consider the phase-space that is available to the
parton shower we would expect a natural threshold of order
$m+Q_{g}$ for gluon emission from a quark of mass $m$. In
practice for the radiation of a gluon from a quark, Eq. \eqref{equation:review/showers/qtilde:eq:qtilde_timelike}
gives a threshold that behaves as $Q_{{\rm thr}}\simeq1.15\left(m_{q}+2Q_{g}\right)$.
This means that the phase-space threshold is significantly higher than the naive expectation, particularly for heavy quarks.

There is no reason why $Q_{g}$ should be the same for all quark
flavours. Therefore, we have chosen to parameterize the threshold for
different flavours as
\begin{equation*}
\begin{split}Q_{g}=\max\left(\frac{\delta-am_{q}}{b},c\right),\end{split}
\end{equation*}

where $a$ (\href{https://herwig.hepforge.org/doxygen/SudakovFormFactorInterfaces.html\#aParameter}{aParameter=0.3}) and $b$ (\href{https://herwig.hepforge.org/doxygen/SudakovFormFactorInterfaces.html\#bParameter}{bParameter=2.3}) are parameters chosen to give a threshold $Q_{{\rm thr}}= \beta m_{q}+\delta$,
with $\beta=0.85$, in order to slightly reduce the threshold
distance for heavier quarks. As a result, the threshold for radiation
from heavy quarks is closer to its physical limit. The parameter
$\delta$ is tuned to data and, only relevant for partons
heavier than the bottom quark, the parameter $c$ (\href{https://herwig.hepforge.org/doxygen/SudakovFormFactorInterfaces.html\#cParameter}{cParameter=0.3}) is chosen to
prevent the cutoff becoming too small.

\end{itemize}

We can combine the calculation of the limits on $z$ in both
approaches using
\begin{equation}\label{equation:review/showers/qtilde:eq:zlimits:all}
\begin{split}z^{2}\left(1-z\right)^{2}\tilde{q}^{2}-\left(1-z\right)m_{i}^{2}-zm_{j}^{2}+z\left(1-z\right)m_{\widetilde{ij}}^{2}>{p_\perp^{\min}}^2,\end{split}
\end{equation}

There are two important special cases.
\begin{enumerate}
\sphinxsetlistlabels{\arabic}{enumi}{enumii}{}{.}%
\item {} 

$q\to qg$, the radiation of a gluon from a quark, or indeed any
massive particle, with mass $\mu=m_{\widetilde{ij}}$. In this case Eq.~\eqref{equation:review/showers/qtilde:eq:zlimits:all} simplifies
to
\begin{equation*}
\begin{split}z^2(1-z)^2\tilde{q}^2> {p_\perp^{\min}}^2   +\left(1-z\right)^{2}\mu^{2}+zQ_{g}^{2},\end{split}
\end{equation*}

which gives a complicated boundary in the
$\left(\tilde{q},z\right)$ plane. However as
\begin{equation*}
{p_\perp^{\min}}^2+
\left(1-z\right)^{2}\mu^{2}+zQ_{g}^{2}
>
\max\left[
\left(1-z\right)^{2}\left({p_\perp^{\min}}^2+\mu^2\right),
z^{2}\left({p_\perp^{\min}}^2+Q_{g}^2\right)
\right].
\label{equation:review/showers/qtilde:eq:zlimits-bound}
\end{equation*}
the phase-space lies inside the region
\begin{equation}\label{equation:review/showers/qtilde:eq:zlimits2}
\begin{split}\frac{\sqrt{\mu^2+{p_\perp^{\min}}^2}}{\tilde{q}}<z<1-\frac{\sqrt{Q_{g}^2+{p_\perp^{\min}}^2}}{\tilde{q}}\end{split}
\end{equation}

and approaches these limits for large values of $\tilde{q}$. In
this case the relative transverse momentum of the branching can be
determined from the evolution scale as
\begin{equation}\label{equation:review/showers/qtilde:eq:quarkpT}
\begin{split}\mathbf{p}_{\perp}=\sqrt{\left(1-z\right)^{2}\left(z^2\tilde{q}^{2}-\mu^{2}\right)-zQ_{g}^{2}}.\end{split}
\end{equation}
\item {} 

$g\to gg$ and $g\to q\bar{q}$, or the branching of a
gluon into any pair of particles with the same mass. In this case the
limits on $z$ are
\begin{equation*}
\begin{split}z_- < z < z_+,  z_\pm = \frac12 \left(1 \pm \sqrt{1-\frac{4\sqrt{\mu^2+{p_\perp^{\min}}^2}}{\tilde{q}}}\right) \mbox{and}\ \tilde{q} >4\sqrt{\mu^2 + {p_\perp^{\min}}^2},\end{split}
\end{equation*}

where $\mu=m_i=m_j$.
Therefore analogously to Eq. \eqref{equation:review/showers/qtilde:eq:zlimits2} the phase-space
lies within the range
\begin{equation}\label{equation:review/showers/qtilde:eq:zlimits3}
\begin{split}\frac{\sqrt{\mu^2+{p_\perp^{\min}}^2}}{\tilde{q}}<z<1-\frac{\sqrt{\mu^2+{p_\perp^{\min}}^2}}{\tilde{q}}\end{split}
\end{equation}

In this case the relative transverse momentum of the branching can be
determined from the evolution scale as
\begin{equation}\label{equation:review/showers/qtilde:eq:gluonpT}
\begin{split}\mathbf{p}_{\perp}=\sqrt{z^{2}\left(1-z\right)^{2}\tilde{q}^{2}-\mu^{2}}.\end{split}
\end{equation}
\end{enumerate}

These two special cases are sufficient for all the branchings
currently included in the simulation, although the general case of
three unequal masses for the particles in the branching is supported.

The formalism discussed above allows us, if given a starting scale
$\tilde{q}_{h}$, to evolve a parton down in scale and generate the
next branching of this particle at a lower scale. The no-emission
probability encoded in the Sudakov form factor is used to generate
$\left(\tilde{q},z\right)$ for this branching. This procedure can
then be iterated to generate subsequent branchings of the particles
produced until no further emission occurs above the cutoff.

\subsubsection{Initial conditions}
\label{\detokenize{review/showers/qtilde:initial-conditions}}\label{\detokenize{review/showers/qtilde:sect-showerinitial}}

Before simulating radiation from a hard process, we must first determine the initial conditions, \textit{i.e.}, the scale $\tilde{q}_{h}$ from which the evolution begins. The maximal kinematically allowed scale cannot in general be assigned independently to every shower progenitor, since the starting scales of colour-connected particles are correlated by the requirement of soft coherence. The available soft-radiation phase space must therefore be partitioned between the evolution partners, rather than assigning its full extent to both. This issue already arises for a two-body colour-connected system and is not specific to decays, although decay and more complicated hard-process topologies require different treatments of the corresponding phase-space constraints. The initial conditions for QCD radiation in the parton shower are set by the colour flow in the hard process \cite{Marchesini:1987cf}. For each particle involved, a colour partner is assigned. In the case of particles in the fundamental representation of the $\mathrm{SU}(3)$ gauge group, this partner is uniquely defined, at least in processes that conserve baryon number. For gluons, a uniform random choice is made between the two possible partners. In baryon-number-violating processes, the colour partner is selected with equal probability from the available colour-connected particles \cite{Dreiner:1999qz, Gibbs:1995bt}. In the angular-ordered parton shower, the direction of the chosen colour partner determines the maximum angle for QCD radiation from the particle.

If a particle carries both colour and EW charge, the colour
partner is used to determine the direction of radiation in the parton shower.
This ensures that QCD radiation dominates the coherence structure. If the
particle carries only EW charges, the evolution partner is selected from among
the other particles carrying the relevant electroweak charge in the event. For
QED radiation, the partner is chosen based on the product of the electric
charges of the particle and its partner, to preserve the correct coherence
pattern. For weak radiation, the selection depends on the relevant
$\mathrm{SU}(2)$ interactions and takes into account weak isospin and particle
identity.

Following the choice of the \textit{evolution partner}, the maximum scale for
radiation from the particle must be calculated, as must the choice of
the $p$ and $n$ reference vectors defined in Eq. \eqref{equation:review/showers/qtilde:eq:sudbasis}.
We always take the choice of $p$ along the
direction of the radiating particle but the choice of $n$ is
related to the direction of the evolution partner.

\paragraph{Final-final evolution}
\label{\detokenize{review/showers/qtilde:final-final-evolution}}\label{\detokenize{review/showers/qtilde:sect-final-final-evolution}}

Consider the evolution of a final-state particle, $b$, with a final-state
evolution partner, $c$. In their centre-of-mass frame, the momenta are
\begin{equation*}
\begin{split}p_{b} &= \frac{1}{2}Q\left(\mathbf{0},\lambda;1+b-c\right), \\
p_{c} &= \frac{1}{2}Q\left(\mathbf{0},-\lambda;1-b+c\right),\end{split}
\end{equation*}

where $Q^{2} = (p_{b} + p_{c})^{2}$, $b = m_{b}^{2}/Q^{2}$,
$c = m_{c}^{2}/Q^{2}$, and
\begin{equation*}
\begin{split}\lambda = \lambda(1,b,c) = \sqrt{1 + b^{2} + c^{2} - 2b - 2c - 2bc},\end{split}
\end{equation*}

is the Källén function.

As described in \hyperref[\detokenize{review/showers/qtilde:sub-shower-kinematics}]{Section \ref{\detokenize{review/showers/qtilde:sub-shower-kinematics}}}, the $p$ basis vector
(Eq. \eqref{equation:review/showers/qtilde:eq:sudbasis}) corresponds to the progenitor momentum generated in the
hard process. The light-like basis vector $n$ is aligned with the
partner’s momentum in the rest frame of the pair. For radiation from $b$,
we define
\begin{equation}\label{equation:review/showers/qtilde:eq:final_final}
\begin{split}n=\frac{1}{2}Q\left(\mathbf{0},-\lambda;\lambda\right).\end{split}
\end{equation}

To simulate parton showering from $c$, the spatial components of
$n$ in Eq. \eqref{equation:review/showers/qtilde:eq:final_final} are reversed.

To ensure soft coherence, the initial evolution scales $\tilde{q}_{h\,b}$
and $\tilde{q}_{h\,c}$ are constrained via
\begin{equation*}
\begin{split}(\tilde{\kappa}_{b} - b)(\tilde{\kappa}_{c} - c)
= \frac{1}{4}(1 - b - c + \lambda)^{2},\end{split}
\end{equation*}

where $\tilde{\kappa}_{b} = \tilde{q}_{h\,b}^{2}/Q^{2}$ and
$\tilde{\kappa}_{c} = \tilde{q}_{h\,c}^{2}/Q^{2}$ \cite{Gieseke:2003rz}.
In Herwig 7, the choice of initial conditions is implemented via the
\texttt{{PartnerFinder}} class. The method \texttt{{calculateFinalFinalScales}}
assigns evolution scales to each particle based on an internal key.
Available strategies include:
\begin{itemize}
\item {} 

\sphinxtitleref{Symmetric}: assigns equal phase-space coverage,
\begin{equation*}
\begin{split}\tilde{\kappa}_{b} &= \frac{1}{2}(1 + b - c + \lambda), \\
\tilde{\kappa}_{c} &= \frac{1}{2}(1 - b + c + \lambda).\end{split}
\end{equation*}
\item {} 

\sphinxtitleref{Maximal}: maximises the radiation scale for one leg,
\begin{equation*}
\begin{split}\tilde{\kappa}_{b} = 4(1 - 2\sqrt{b} - b + c),\end{split}
\end{equation*}
with the converse applied to $\tilde{\kappa}_{c}$.

\end{itemize}

In Ref. \cite{Bewick:2019rbu} it has been observed that the phase
space for the production of $n$ additional partons from a 2-body
final state, that forms a colour singlet, requires an additional
Jacobian factor
\begin{equation*}
\begin{split}J = \frac{\lambda(b^\prime,c^\prime)}{\lambda(b,c)},\end{split}
\end{equation*}

where $b^\prime=q_b^2/Q^2$ and $c^\prime=q_c^2/Q^2$ are
calculated with the two final-jet momenta. To take into account this factor, at the end of the
shower evolution we accept the event with a probability given by
$J$.  This prevents an overpopulation of the
non-angular-ordered region after multiple emissions. Currently, this
behaviour is only available when the initial state is colourless and
we have only two shower progenitors, which thus form a colour singlet.
In decay topologies, an internal flag \texttt{{isDecayCase}} enables an alternate
strategy for assigning evolution scales, accounting for the different phase-space
structure in decaying systems. This ensures appropriate soft coverage and coherence
in configurations such as heavy-particle decays.

\paragraph{Initial-initial evolution}
\label{\detokenize{review/showers/qtilde:initial-initial-evolution}}\label{\detokenize{review/showers/qtilde:sub-initial-initial-colour-connection}}

Again we opt to work in the rest frame of the evolution partners, so
that the momenta of the particles are:
\begin{equation*}
\begin{split}p_{b} & =\frac{1}{2}Q\left(\mathbf{0},1;1\right); \\
p_{c} & =\frac{1}{2}Q\left(\mathbf{0},-1;1\right);\end{split}
\end{equation*}

where $Q$ is the partonic centre-of-mass energy of the collision.

In this case, as we assume that the incoming particles are massless, we
can simply take the $p$ reference vector to be the momentum of the
beam particle from which the emitting parton was extracted and the
$n$ reference vector to be the momentum of the beam particle from
which its colour partner was extracted. The fact that $p$ is
parallel to the momentum of the emitting parton makes it easier to
reconstruct the momenta of the shower particles in terms of the fraction
of the beam momentum they carry.

Defining the $p$ and $n$ vectors as being equal to
the beam momenta rather than the actual parton momenta does not affect
our earlier assertions relating to the soft limit of the splitting
functions, since Eq. \eqref{equation:review/showers/qtilde:eq:eikonal_dipole_fn} is
clearly invariant under overall rescalings of the dipole momenta
$n$ and $p$.

In this case the requirement that the soft region of phase-space is
smoothly covered is simply
\begin{equation*}
\begin{split}\tilde{\kappa}_{b}\tilde{\kappa}_{c}=1.\end{split}
\end{equation*}
Contrary to the case of the final-final evolution, there
is no upper bound on the values of $\tilde{\kappa}_{b}$ or
$\tilde{\kappa}_{c}$, \textit{i.e.} there is no choice that maximizes the
phase-space available to one parton relative to the other (at least none
that might reasonably be expected to give sensible results). Currently
only the most symmetric choice,
\textit{i.e.} $\tilde{\kappa}_{b}=\tilde{\kappa}_{c}=1$, is implemented.

\paragraph{Initial-final evolution in the hard process}
\label{\detokenize{review/showers/qtilde:initial-final-evolution-in-the-hard-process}}\label{\detokenize{review/showers/qtilde:sect-initialfinalhard}}

We consider the initial-final-state evolution in the context of a
process $a+b\to c$, where $a$ is a colour-singlet system and
$b$ and $c$ are evolution partners, \textit{e.g.} deep inelastic
scattering. As in the last two cases we work in the rest frame of the
evolution partners, in this case the Breit frame, where we may write:
\begin{equation*}
\begin{split}p_{b} & =\frac{1}{2}Q\left(\mathbf{0},1+c;1+c\right); \\
p_{c} & =\frac{1}{2}Q\left(\mathbf{0},-1+c;1+c\right);\end{split}
\end{equation*}

with $Q^{2}=-p_{a}^{2}$.

For emission from the final-state particle, the $p$ vector is
taken to be the momentum of the radiating particle and the $n$
reference vector is set equal to the momentum of the beam particle from
which the initial-state evolution partner was extracted. For emission from
the initial-state particle the $p$ vector is defined to be the
momentum of the beam particle from which the radiating parton was
extracted and
\begin{equation*}
\begin{split}n=\frac{1}{2}Q\left(\mathbf{0},-1-c;1+c\right),\end{split}
\end{equation*}

in the Breit frame. As discussed in \hyperref[\detokenize{review/showers/qtilde:sub-initial-initial-colour-connection}]{Section \ref{\detokenize{review/showers/qtilde:sub-initial-initial-colour-connection}}},
the normalization of $n$ and/or
$p$, does not affect the eikonal dipole limit of the splitting
functions Eq. \eqref{equation:review/showers/qtilde:eq:eikonal_dipole_fn}.

Achieving a smooth matching of the phase-space for the first emission
from parton $b$’s shower with that of parton $c$’s
shower, at wide angles, requires the initial evolution scales
$\left(\tilde{q}_{h\, b},\,\tilde{q}_{h\, c}\right)$ obey
\begin{equation}\label{equation:review/showers/qtilde:eq:initialfinalcondition}
\begin{split}\tilde{\kappa}_{b}\left(\tilde{\kappa}_{c}-c\right)=\left(1+c\right)^{2}.\end{split}
\end{equation}

In practice, we opt to assign roughly the same phase-space volume
to each shower, \textit{i.e.} we use the most symmetric choice:
$\tilde{\kappa}_{b}=1+c$, $\tilde{\kappa}_{c}=1+2c$. Of
course, a larger or smaller combination that satisfies
Eq. \eqref{equation:review/showers/qtilde:eq:initialfinalcondition} is also allowed.

\paragraph{Initial-final evolution in decays}
\label{\detokenize{review/showers/qtilde:initial-final-evolution-in-decays}}\label{\detokenize{review/showers/qtilde:sub-initial-final-colour-connection-in-decays}}

The Herwig 7 angular-ordered shower differs from other approaches in including
initial-state radiation from a decaying particle, as well as
final-state radiation from the decay products. This is required
in order to ensure that the full soft region of phase-space is filled by
radiation from the parton shower \cite{Hamilton:2006ms, Gieseke:2003rz}.

Consider the decay $b \to a\,c$, where $b$ and $c$ are
evolution partners and $a$ is a colour-singlet system, in the rest
frame of the decaying particle. In this frame, the momenta of $b$
and its evolution partner $c$ are:
\begin{equation*}
\begin{split}p_{b} &= m_{b}\left(\mathbf{0}, 0; 1\right); \\
p_{c} &= \frac{1}{2}m_{b}\left(\mathbf{0}, \lambda; 1 - a + c\right),\end{split}
\end{equation*}

where $c = m_{c}^{2}/m_{b}^{2}$, $a = m_{a}^{2}/m_{b}^{2}$ and
$\lambda = \lambda(1, a, c)$.

For radiation from the decaying particle, $p$ is chosen to be the
momentum of the decaying particle and
\begin{equation*}
\begin{split}n = \frac{1}{2}m_{b}\left(\mathbf{0}, 1; 1\right),\end{split}
\end{equation*}

in its rest frame, \textit{i.e.} $n$ is aligned with the evolution
partner.

In the case of radiation from the final-state particle, $p$ is set
equal to its momentum as generated in the hard decay process. However,
there is no obvious choice of $n$ related to the evolution partner
since we are working in its rest frame. We therefore choose $n$
such that it is in the opposite direction to the radiating particle in
this frame, \textit{i.e.}
\begin{equation*}
\begin{split}n = \frac{1}{2}\left(\mathbf{0}, -\lambda; \lambda\right).\end{split}
\end{equation*}

A more rigorous approach to this problem was carried out in
\cite{Hamilton:2006ms}, using a more generalised splitting function
derived assuming a massive gauge vector $n$. This feature is not
implemented in the standard released code, since any related deficiency
in the shower is completely avoided by using the associated matrix
element correction (\hyperref[\detokenize{review/matching/matrix-element-corrections:matrix-element-corrections}]{Section \ref{\detokenize{review/matching/matrix-element-corrections:matrix-element-corrections}}}).

In this case, the requirement that the full soft region of phase-space is
filled by radiation from the parton shower gives
\begin{equation*}
\begin{split}(\tilde{\kappa}_{b} - 1)(\tilde{\kappa}_{c} - c) = \frac{1}{4}(1 - a + c + \lambda)^{2}.\end{split}
\end{equation*}
While there is no limit on the value of $\tilde{\kappa}_{b}$, as
with the final--final evolution, the maximum value of
$\tilde{\kappa}_{c}$ is
\begin{equation*}
\begin{split}\tilde{\kappa}_{c} = 4(1 - \sqrt{a})^{2} - 4c.\end{split}
\end{equation*}
Herwig 7 currently uses a fixed, symmetric configuration of the initial
evolution scales, given by
\begin{equation*}
\begin{split}\tilde{\kappa}_{b} &= \frac{1}{2}(3 - a + c + \lambda), \\
\tilde{\kappa}_{c} &= \frac{1}{2}(1 - a + 3c + \lambda),\end{split}
\end{equation*}

which ensures approximately equal phase-space coverage from both the
decaying particle and its colour-connected final-state partner.

\subsubsection{Final-state radiation}
\label{\detokenize{review/showers/qtilde:final-state-radiation}}\label{\detokenize{review/showers/qtilde:sub-final-state-radiation}}

\paragraph{Evolution}
\label{\detokenize{review/showers/qtilde:evolution}}\label{\detokenize{review/showers/qtilde:sub-finalfinalevolution}}

The parton-shower algorithm generates the radiation from each progenitor
independently, \textit{modulo} the prior determination of the initial evolution
scale and the $n$ and $p$ basis vectors. Consider the
evolution of a given final-state progenitor, downward from its initial
evolution scale $\tilde{q}_{h}$. Given that
$\Delta\left(\tilde{q},\tilde{q}_{h}\right)$ gives the
\textit{probability} that this parton evolves from scale $\tilde{q}_{h}$
to $\tilde{q}$ without any resolvable branchings, we may generate
the scale of this first branching $\left(\tilde{q}\right)$ by
solving
\begin{equation}\label{equation:review/showers/qtilde:eq:sudakov_equals_random}
\begin{split}\Delta\left(\tilde{q},\tilde{q}_{h}\right)=\mathcal{R},\end{split}
\end{equation}

where $\mathcal{R}$ is a random number uniformly distributed
between 0 and 1.

The details of the solution of this equation, the generation of
the type of branching and variables describing the evolution
are described in \hyperref[\detokenize{review/appendix/sudakov:sect-sudakov-solution}]{\ref{\detokenize{review/appendix/sudakov:sect-sudakov-solution}}}.

The relative transverse momentum for the branching $p_{\perp}$
(i.e.\ Eq. \eqref{equation:review/showers/qtilde:eq:pt_fwd}) is then calculated, using Eq.
\eqref{equation:review/showers/qtilde:eq:quarkpT} or Eq. \eqref{equation:review/showers/qtilde:eq:gluonpT} depending on the type
of branching.
The azimuthal angle of $p_{\perp}$ is generated including spin and soft correlations, as
described in Section~\ref{\detokenize{review/showers/perturbative_decays::doc}} and
Refs.~\cite{Richardson:2001df, Knowles:1988vs, Collins:1987cp,
Richardson:2018pvo}.

The requirement of angular ordering, as discussed in
\hyperref[\detokenize{review/showers/qtilde:sub-shower-kinematics}]{Section \ref{\detokenize{review/showers/qtilde:sub-shower-kinematics}}}, defines the initial scales for the
daughter particles, $\tilde{q}_{h\, i}$ and
$\tilde{q}_{h\, j}$, produced in each branching,
$\widetilde{ij}\rightarrow i+j$, to be
\begin{equation*}
\begin{split}\begin{aligned}
\tilde{q}_{h\, i} & =z\tilde{q}, &
\tilde{q}_{h\, j} & =\left(1-z\right)\tilde{q},\end{aligned}\end{split}
\end{equation*}
where $\tilde{q}$ and $z$, are the evolution scale and
light-cone momentum fraction of the branching. By imposing these upper
bounds on the evolution scale of the emitted partons, subsequent
branchings will have a nesting of the angular separation of the
resulting daughters, where each one is smaller than the one preceding
it.

All of the steps above, required to generate the shower variables
associated with this initial branching, may then be repeated for the
daughter partons, and their daughter partons, should they also branch.
All showering terminates when the evolution scale
$\left(\tilde{q}\right)$ for each final-state parton falls below
its minimum value, when there is no phase-space for any more resolvable
emissions. The resulting partons, at the end of each shower, are deemed
to be on mass-shell, as defined in
\hyperref[\detokenize{review/index:sec-hadronization}]{Section \ref{\detokenize{review/index:sec-hadronization}}}, at which point the perturbative parton-shower
evolution is no longer sensible, since hadronization effects dominate at
these scales. By default, the parton shower puts partons onto their
\textit{constituent} parton mass-shell, thereby including some part of the
non-perturbative hadronization effects into the termination of the
parton shower. As an alternative, from Herwig 7.3, it is possible to
keep partons on their current mass-shells (massless for gluons) and
momentum is ‘shuffled’ as the first stage of hadronization to put them
onto their constituent mass-shells, as discussed in
\hyperref[\detokenize{review/index:sec-hadronization}]{Section \ref{\detokenize{review/index:sec-hadronization}}}.

\paragraph{Kinematic reconstruction}
\label{\detokenize{review/showers/qtilde:kinematic-reconstruction}}\label{\detokenize{review/showers/qtilde:sect-finalrecon}}

At this point we have a set of partons produced in the parton shower
from each of the progenitor partons, the scales $\tilde{q}$ at
which they are produced, the momentum fractions $z$ and azimuthal
angles $\phi$ of the branchings. Mapping these kinematic variables
into physical momenta is what we call \textit{kinematic reconstruction}. We
will now describe this procedure for showers generated by final-state
progenitors. First, the kinematics of the individual showers are
reconstructed by putting the external masses on their
mass-shell and working back through the shower, as described in
\hyperref[\detokenize{review/showers/qtilde:sub-shower-kinematics}]{Section \ref{\detokenize{review/showers/qtilde:sub-shower-kinematics}}}.

The shower evolution causes all progenitor partons, $J$, produced
in the hard process to gain a virtual mass, \textit{i.e.} the progenitor
partons, which initiated the jets, are no longer on mass shell,
$q_{J}^{2}\neq m_{J}^{2}$. We want to preserve the total energy of
the system in the centre-of-mass frame of the hard collision. If the
momenta of the progenitor partons before the shower evolution were
$p_{J}=\left(\mathbf{p}_{J};\sqrt{\mathbf{p}_{J}^{2}+m_{J}^{2}}\right)$
in this frame, then
\begin{equation*}
\begin{split}\sum_{J=1}^{n}\sqrt{\mathbf{p}_{J}^{2}+m_{J}^{2}}=\sqrt{s},\end{split}
\end{equation*}

while the sum of the spatial momenta is zero. As the jet parents have
momenta $q_{J}=\left(\mathbf{q}_{J};\sqrt{\mathbf{q}_{J}^{2}+q_{J}^{2}}\right)$
after the parton showering, we need to restore momentum conservation in
a way that disturbs the internal structure of the jet as little as
possible. The easiest way to achieve this is by boosting each jet along
its axis so that their momenta are rescaled by a common factor $k$
determined from
\begin{equation*}
\begin{split}\sum_{J=1}^{n}\sqrt{k^{2}\mathbf{p}_{J}^{2}+q_{J}^{2}}=\sqrt{s},\end{split}
\end{equation*}

which can be solved analytically for two jets or numerically for higher
multiplicities. For every jet a Lorentz boost is applied such that
\begin{equation*}
\begin{split}q_{J}=\left(\mathbf{q}_{J};\sqrt{\mathbf{q}_{J}^{2}+q_{J}^{2}}\right)\stackrel{{\rm boost}}{\longrightarrow}q_{J}^{\prime}=\left(k\mathbf{p}_{J};\sqrt{k^{2}\mathbf{p}_{J}^{2}+q_{J}^{2}}\right).\end{split}
\end{equation*}

Applying these boosts to each of the jets, in the centre-of-mass frame
of the collision, ensures global energy-momentum conservation. Typically
the rescaling parameters $k$ are close to unity, hence the
resulting boosts and rotations are small.

\subsubsection{Initial-state radiation}
\label{\detokenize{review/showers/qtilde:initial-state-radiation}}\label{\detokenize{review/showers/qtilde:sub-initial-state-radiation}}

\paragraph{Evolution}
\label{\detokenize{review/showers/qtilde:sub-isr-evolution}}\label{\detokenize{review/showers/qtilde:id31}}

As stated in  \hyperref[\detokenize{review/showers/qtilde:sub-shower-kinematics}]{Section \ref{\detokenize{review/showers/qtilde:sub-shower-kinematics}}}, in generating the
initial-state radiation we use a backward evolution algorithm, starting
with the space-like daughter parton that initiates the hard scattering
process, $i$, and evolving it backward to give its space-like
parent, $\widetilde{ij}$, and time-like sister parton $j$.
This evolution algorithm therefore proceeds from the high scale of the
hard process to the low scale of the external hadrons. Such a procedure
is significantly more efficient than the alternative forward evolution
algorithm, which would start from the incoming beam partons and evolve
them to the scale of the hard collision. This is because the forward
evolution cannot be constrained to end on the $x$ and
$Q^{2}$ values associated to the hard process, which in turn makes
it impossible to perform importance sampling of any significant resonant
contributions.

While forward evolution would dynamically generate the parton
distribution functions (PDFs), backward evolution uses the measured PDFs
to guide the evolution. As with the final-state shower, the initial
conditions for the initial-state shower are determined by the evolution
partners of the incoming particles (\hyperref[\detokenize{review/showers/qtilde:sub-initial-initial-colour-connection}]{Section \ref{\detokenize{review/showers/qtilde:sub-initial-initial-colour-connection}}}).

The angular-evolution variable $\tilde{q}^{2}$ for space-like
showers was defined in Eq. \eqref{equation:review/showers/qtilde:eq:qtilde_spacelike}. We
shall work exclusively with light initial-state partons so we take
$m_{\widetilde{ij}}=m_{i}=0$, and $m_{j}=\mu$ if $j$
is a quark and $m_{j}=Q_g$ if $j$ is a gluon, to regulate
the infrared divergent regions, hence Eq.
\eqref{equation:review/showers/qtilde:eq:qtilde_spacelike} simplifies to
\begin{equation}\label{equation:review/showers/qtilde:eq:backwardevolutionvariable}
\begin{split}\tilde{q}^{2}=\frac{zm_{j}^{2}-p_{\perp}^{2}}{\left(1-z\right)^{2}},\end{split}
\end{equation}

where $p_{\perp}^{2}=-\mathbf{p}_{\perp}^{2}$ (Eqs.~\eqref{equation:review/showers/qtilde:eq:pt_fwd} and \eqref{equation:review/showers/qtilde:eq:pt_components}).

The requirement that $\mathbf{p}_{\perp}^{2}\geq0$, Eq.
\eqref{equation:review/showers/qtilde:eq:backwardevolutionvariable} implies an upper limit on
$z$,
\begin{equation*}
\begin{split}z\le z_{+}=1+\frac{Q_{g}^{2}}{2\tilde{q}^{2}}-\sqrt{\left(1+\frac{Q_{g}^{2}}{2\tilde{q}^{2}}\right)^{2}-1}.\end{split}
\end{equation*}

In addition, if the light-cone momentum fraction of parton $i$ is
$x$, we must have $z\ge x$ to prevent the initial-state
branching simulation evolving backward into a parent with $x>1$.

In this case the Sudakov form factor for backward evolution is
\cite{Marchesini:1987cf, Sjostrand:1985xi}
\begin{equation*}
\begin{split}\Delta\left(x,\tilde{q},\tilde{q}_{h}\right)=\prod_{\widetilde{ij},j}\Delta_{\widetilde{ij}\to ij}\left(x,\tilde{q},\tilde{q}_{h}\right),\end{split}
\end{equation*}

where the Sudakov form factor for the backward evolution of a given
parton $i$ is
\begin{equation}\label{equation:review/showers/qtilde:eq:sudakovbackward}
\begin{split} \Delta_{\widetilde{ij}\to ij}\left(x,\tilde{q},\tilde{q}_{h}\right)=\exp\left\{ -\int_{\tilde{q}}^{\tilde{q}_{h}}\frac{\mathrm{d}\tilde{q}^{\prime2}}{\tilde{q}^{\prime2}}\int_{x}^{z_{+}}\mathrm{d}z\mbox{ }\frac{\alpha_{s}\left(z,\tilde{q}^{\prime}\right)}{2\pi}\mbox{ }P_{\widetilde{ij}\to ij}\left(z,\tilde{q}^{\prime}\right)\mbox{ }\frac{\frac{x}{z}f_{\widetilde{ij}}\left(\frac{x}{z},\tilde{q}^{\prime}\right)}{xf_{i}\left(x,\tilde{q}^{\prime}\right)}\Theta\left(\mathbf{p}_{\perp}^{2}>0\right)\right\},\end{split}
\end{equation}

and the product runs over all possible branchings
$\widetilde{ij}\rightarrow i+j$ capable of producing a
parton of type $i$. This is similar to the form factor used for
final-state radiation, Eq. \eqref{equation:review/showers/qtilde:eq:sudakovmaster}, with the
addition of the PDF factor, which guides the backward evolution.
The solution of Eq. \eqref{equation:review/showers/qtilde:eq:sudakovbackward} and the generation of
the variables describing the emission is discussed in \hyperref[\detokenize{review/appendix/sudakov:sect-sudakov-solution}]{Section \ref{\detokenize{review/appendix/sudakov:sect-sudakov-solution}}}.

When a branching is generated, the relative transverse momentum
$p_{\perp}$ (Eqs.~\eqref{equation:review/showers/qtilde:eq:pt_fwd} and \eqref{equation:review/showers/qtilde:eq:pt_components}) is calculated according to Eq.
\eqref{equation:review/showers/qtilde:eq:backwardevolutionvariable}. The azimuthal angle
is generated
including spin and soft correlations as described in
\cite{Richardson:2001df, Knowles:1988vs, Collins:1987cp, Richardson:2018pvo}.
In the case of backward evolution the angular-ordering requirement is
satisfied by simply continuing the backward evolution downward in
$\tilde{q}$, starting from the value generated in the previous
generated branching.

As stated above, when the evolution scale has reduced to the point where
there is no phase-space for further resolvable branchings, the
backward evolution ends. The incoming particle produced in the last
backward branching, assumed to be on-shell (massless), has no transverse
momentum, since this is measured with respect to the beam axis%
\begin{footnote}\sphinxAtStartFootnote
Herwig 7 supports the option of including a non-perturbative
intrinsic transverse momentum for the partons inside the incoming
hadron, as described in \hyperref[\detokenize{review/showers/intrinsic:shower-intrinsic}]{Section \ref{\detokenize{review/showers/intrinsic:shower-intrinsic}}} and
\hyperref[\detokenize{review/appendix/tuning:sect-tuning}]{\ref{\detokenize{review/appendix/tuning:sect-tuning}}}, which can give the initial incoming parton a
transverse momentum.
\end{footnote}.
This final parton also has a light-cone momentum fraction
$x/\prod_{i}z_{i}$, with respect to the incoming hadron’s
momentum, where $x$ is the light-cone momentum fraction generated
in the initial simulation of the hard process, and the product is
comprised of all $z$ values generated in the backward evolution.

Before any momentum reconstruction can begin, we must simulate the
effects of final-state showers from each time-like daughter parton
$j$, generated from the backward evolution of each space-like
parton $i$, in branchings $\widetilde{ij}\rightarrow i+j$.
As discussed in \hyperref[\detokenize{review/showers/qtilde:sub-shower-kinematics}]{Section \ref{\detokenize{review/showers/qtilde:sub-shower-kinematics}}}, for such a
branching occurring at scale $\tilde{q}$ with light-cone momentum
fraction $z$, angular ordering is achieved by evolving $j$
down from an initial scale
$\tilde{q}_{h}=\left(1-z\right)\tilde{q}$. This initial condition
ensures that for each parton $j$, the angular separation of any of
$j$’s subsequent branching products is less than the angle
between $j$ and $j$’s sister $i$.

This algorithm is all that is needed to generate the values of the
scales, momentum fractions and azimuthal angles, for the evolution of
both the incoming particles and the time-like particles emitted in their
backward evolution. These values are sufficient for us to determine the
momenta of all of the particles in the associated showers, to perform
the kinematic reconstruction.

\paragraph{Kinematic reconstruction}
\label{\detokenize{review/showers/qtilde:sect-initialrecon}}\label{\detokenize{review/showers/qtilde:id35}}

The kinematic reconstruction begins by finding the last initial-state
particle produced in the backward evolution of each of the beam
particles. This parton momentum is calculated as described in the
previous section. The momentum of the final-state time-like jet that it
radiates is then reconstructed in the same way as for the final-state
shower. Knowing the momenta of the former light-like parent parton and
the latter final-state, time-like, daughter parton, the momentum of the
initial-state, space-like, daughter, follows by momentum conservation.
This process is iterated for each initial-state branching, eventually
giving the momentum of the space-like progenitor parton, colliding in
the hard process.

The reconstructed momentum of the colliding parton incident from the
$+z$ direction is denoted $q_{\oplus}$, and that of the
colliding parton incident from the $-z$ direction is denoted
$q_{\ominus}$.

The final reshuffling of the momentum then proceeds in different ways
depending on whether the colour partner is an initial- or final-state
parton.
Final-state radiation is reconstructed first, including the global recoil
from the parton shower, which rescales the jet momenta to restore
overall energy-momentum conservation in the centre-of-mass frame.
Initial-state radiation is then reconstructed by backward evolution,
starting from the incoming partons participating in the hard process
and proceeding down to the hadron scale, using the final-state kinematics
as fixed input.

\subparagraph{Initial-state partner}
\label{\detokenize{review/showers/qtilde:initial-state-partner}}

As discussed in \hyperref[\detokenize{review/showers/qtilde:sub-initial-initial-colour-connection}]{Section \ref{\detokenize{review/showers/qtilde:sub-initial-initial-colour-connection}}} the
hadronic beam momenta, $p_{\oplus}$ and $p_{\ominus}$,
define the Sudakov basis for the initial-state shower algorithms, in
terms of which we can write the momenta of the original colliding
partons at the end of the showering phase
\begin{equation*}
\begin{split}q_{\splusminus} = \alpha_{\splusminus}p_{\splusminus} +\beta_{\splusminus}p_{\sminusplus}+q_{\perp\splusminus}.\end{split}
\end{equation*}

Here $q_{\splusminus}$ refers to the parton momentum at the end of the
backward evolution, before the final reshuffling.
The Sudakov coefficients may be calculated using the fact that
$p_{\oplus}$ and $p_{\ominus}$ are light-like and orthogonal
to the $q_{\perp}$ component:
\begin{equation*}
\begin{split}\alpha_{\splusminus} & =2p_{\sminusplus}\cdot q_{\splusminus}/s; \\
\beta_{\splusminus}  & =2p_{\splusminus}\cdot q_{\splusminus}/s;\end{split}
\end{equation*}

where $s=2p_{\oplus}\cdot p_{\ominus}$ is the hadronic
centre-of-mass energy squared. The $q_{\perp}$ components follow
by subtracting
$\alpha_{\splusminus}p_{\splusminus}+\beta_{\splusminus}p_{\splusminus}$
from the reconstructed momentum
$q_{\splusminus}$.

Through the emission of initial-state radiation, the colliding partons
acquire both space-like virtualities and transverse momenta, of which
they had neither in the initial simulation of the hard process.
Consequently, whereas momentum conservation in the prior simulation of
the hard process implies that the total initial- and final-state
momentum were equal to
$p_{\mathrm{cms}}=x_{\oplus}p_{\oplus}+x_{\ominus}p_{\ominus}$,
we now have a momentum imbalance between the two:
$q_{\oplus}+q_{\ominus}\ne
x_{\oplus}p_{\oplus}+x_{\ominus}p_{\ominus}$, where
$x_{\oplus}p_{\oplus}+x_{\ominus}p_{\ominus}$ now corresponds to
the sum of the momenta of the final-state shower progenitors,
\textit{i.e.} the momenta after final-state shower reconstruction, including
the global boost applied to restore energy--momentum conservation.

We first impose that the invariant mass of the system comprising the
two colliding partons is equal to the one of the system comprising the
final-state progenitors: this guarantees the existence of a Lorentz boost that we can
apply to the final state to achieve full momentum conservation.

To achieve this task, we begin by rescaling the energies and
longitudinal momenta of the colliding initial-state partons, in such a
way that their invariant mass is left unchanged.  Because of this, we
can then calculate a Lorentz boost (for each colliding parton) that
produces the same effect.  This boost is then applied to all elements
of the initial-state shower, including the final-state jets they emit.
We label with $k_{\oplus}$ and $k_{\ominus}$ the two
rescaling factors, and we introduce the \textit{shuffled momenta}
$q_{\oplus}^{\prime}$ and $q_{\ominus}^{\prime}$
\begin{equation*}
\begin{split}q_{\splusminus}^{\prime}=\alpha_{\splusminus}\, k_{\splusminus}\, p_{\splusminus}+\frac{\beta_{\splusminus}}{k_{\splusminus}}\, p_{\sminusplus}+q_{\perp\splusminus}.\end{split}
\end{equation*}

It is trivial to check that $q_{\splusminus}^{\prime 2} = q_{\splusminus}^2$.

In simulating the hard process the momentum of the partonic
centre-of-mass system was
\begin{equation}\label{equation:review/showers/qtilde:eq:isr_recon_1}
\begin{split}p_{\mathrm{cms}}=x_{\oplus}p_{\oplus}+x_{\ominus}p_{\ominus},\end{split}
\end{equation}

and in terms of the shuffled momenta it is
\begin{equation}\label{equation:review/showers/qtilde:eq:isr_recon_2}
\begin{split}q_{\mathrm{cms}}^{\prime}=\left(\alpha_{\oplus}k_{\oplus}+\frac{\beta_{\ominus}}{k_{\ominus}}\right)p_{\oplus}+\left(\alpha_{\ominus}k_{\ominus}+\frac{\beta_{\oplus}}{k_{\oplus}}\right)p_{\ominus}+q_{\perp\oplus}+q_{\perp\ominus}.\end{split}
\end{equation}

Imposing that the centre-of-mass energy generated in the simulation of
the hard process is preserved,
$q_{\mathrm{cms}}^{\prime2}=p_{\mathrm{cms}}^{2}$ and using the Sudakov
decompositions of Eqs.~\eqref{equation:review/showers/qtilde:eq:isr_recon_1} and \eqref{equation:review/showers/qtilde:eq:isr_recon_2} imply that the product of the rescalings
$k_{\oplus}$ and $k_{\ominus}$ obeys the relation
\begin{equation*}
\begin{split}\alpha_{\oplus}\alpha_{\ominus}s\, k_{\oplus\ominus}^{2}+\left(\left(\alpha_{\oplus}\beta_{\oplus}+\alpha_{\ominus}\beta_{\ominus}-x_{\oplus}x_{\ominus}\right)s+\left(q_{\perp\oplus}+q_{\perp\ominus}\right)^{2}\right)k_{\oplus\ominus}+\beta_{\oplus}\beta_{\ominus}s=0,\end{split}
\end{equation*}
where we used the short notation $k_{\oplus\ominus} = k_{\oplus}k_{\ominus}$.
In order to fully determine the rescaling factors we need a second constraint.
We provide a number of options, which are controlled by the
\href{https://herwig.hepforge.org/doxygen/QTildeReconstructorInterfaces.html\#InitialStateReconOption}{InitialStateReconOption} switch:
\begin{itemize}
\item {} 

Our default choice
(\href{https://herwig.hepforge.org/doxygen/QTildeReconstructorInterfaces.html\#InitialStateReconOption}{InitialStateReconOption=SofterFraction})
is to set $k_{\oplus\ominus}$ equal to the rescaling factor of the parton that had the emission
with the largest transverse momentum, and to set the other rescaling factor to 1.
This choice exactly reproduces the kinematics of the Catani--Seymour dipoles \cite{Catani:1996vz},
which makes matching to higher-order matrix elements simpler.

\item {} 

The option to preserve the rapidity of the partonic centre-of-mass
(\texttt{InitialStateReconOption=Rapidity})
requires that the ratio of the $p_{\oplus}$ coefficient to the
$p_{\ominus}$ Sudakov coefficient in
$q_{\mathrm{cms}}^{\prime}$ equal that in $p_{\mathrm{cms}}$.
This implies a second constraint on $k_{\oplus}$ and $k_{\ominus}$:
\begin{equation*}
\begin{split}k_{\oplus}^{2} = k_{\oplus\ominus} \cdot \frac{x_{\oplus}}{x_{\ominus}}
\cdot \frac{\beta_{\oplus} + \alpha_{\ominus}k_{\oplus\ominus}}{\alpha_{\oplus}k_{\oplus\ominus} + \beta_{\ominus}}\end{split}
\end{equation*}
This is more physically motivated than the first choice and was the default
in Herwig++ and FORTRAN HERWIG.

\item {} 

Another option is to preserve the longitudinal momentum of the system
(\href{https://herwig.hepforge.org/doxygen/QTildeReconstructorInterfaces.html\#InitialStateReconOption}{InitialStateReconOption=Longitudinal}).
In this case, $k_{\oplus}$ is the solution of:
\begin{equation*}
\begin{split}\left(\alpha_{\oplus} + \frac{\beta_{\ominus}}{k_{\oplus\ominus}}\right) k_{\oplus}^2
+ (x_\oplus - x_\ominus)k_{\oplus} - (\alpha_\ominus k_{\oplus\ominus} + \beta_\oplus) = 0\end{split}
\end{equation*}
\end{itemize}

In Herwig 7, the hardest emission is defined as the one with the largest
transverse momentum $p_T$ \textit{at the time of emission}. However, because
momentum reshuffling is applied to ensure global energy and momentum
conservation, the final $q_{\perp}$ of this emission may not correspond
exactly to the originally generated $p_T$. Among the available reshuffling
strategies, the \sphinxtitleref{Rapidity} option, which preserves the centre-of-mass energy
and the rapidity of the hard system, best maintains the kinematics of the
hardest emission. This makes it the preferred choice when a consistent definition
of the hardest emission is needed, such as in matching to fixed-order calculations.

The two relations above fully determine the
$k_{\oplus}$ and $k_{\ominus}$ rescaling factors. Having
solved these equations for $k_{\oplus}$ and $k_{\ominus}$, we
go on to determine a longitudinal boost for each initial-state jet such
that
\begin{equation*}
\begin{split}q_{\splusminus}\stackrel{\mathrm{boost}}{\longrightarrow}q_{\splusminus}^{\prime}.\end{split}
\end{equation*}
This boost may then be applied to all elements of the initial-state
shower, including any final-state partons or jets they emit.

Since the initial and final state now have the same invariant mass, a
second boost can be defined to ensure overall momentum conservation:
\begin{equation*}
\begin{split}x_{\oplus} p_{\oplus} + x_{\ominus} p_{\ominus}
\stackrel{\mathrm{boost}^\prime}{\longrightarrow}
q_{\mathrm{cms}}^{\prime} =
\left(\alpha_{\oplus}k_{\oplus}+\frac{\beta_{\ominus}}{k_{\ominus}}\right)p_{\oplus}
+ \left(\alpha_{\ominus}k_{\ominus}+\frac{\beta_{\oplus}}{k_{\oplus}}\right)p_{\ominus}
+ q_{\perp\oplus} + q_{\perp\ominus}\end{split}
\end{equation*}

This second boost is applied to the entire final-state system, including
all partons resulting from the evolution of the final-state shower progenitors,
to ensure that the total momentum of the event is conserved.
This procedure is sufficient for the production of colour-singlet
systems, such as EW gauge bosons in the Drell-Yan process.

\subparagraph{Final-state partner}
\label{\detokenize{review/showers/qtilde:final-state-partner}}

For systems that have an initial-state parton that is colour connected
to a final-state parton the reconstruction is performed in their Breit
frame in order to preserve the $Q^2$ of the system in, for
example, DIS processes.

The momenta of the initial- and final-state jets are first
reconstructed as described above for initial-state jets and in
\hyperref[\detokenize{review/showers/qtilde:sect-finalrecon}]{Section \ref{\detokenize{review/showers/qtilde:sect-finalrecon}}} for final-state jets. The momenta of the jet
progenitors, which are now off-shell, are then boosted to the
Breit-frame of the original system before the radiation. We take
$p_b$ to be the momentum of the original (on-shell) incoming
parton and $p_c$ the one of the original (on-shell) outgoing
parton. We thus define the space-like momentum $p_a=p_c-p_b$
that in the Breit frame is simply given by
\begin{equation*}
\begin{split}p_a = Q(0,0,-1; 0).\end{split}
\end{equation*}

We construct a set of basis vectors, similar to the Sudakov
basis defined in \hyperref[\detokenize{review/showers/qtilde:sect-initialfinalhard}]{Section \ref{\detokenize{review/showers/qtilde:sect-initialfinalhard}}} for the
initial-final colour connection,
\begin{equation*}
\begin{split}n_1 & = Q(0,0,+1;1), \\
n_2 & = Q(0,0,-1;1).\end{split}
\end{equation*}

In this basis the original momentum $p_a$ is simply given by
\begin{equation*}
\begin{split}p_a = \frac{1}{2}(n_1 - n_2),\end{split}
\end{equation*}

while the momentum of the incoming jet is decomposed as
\begin{equation*}
\begin{split}q_{\rm in} = \alpha_{\rm in} n_1 + \beta_{\rm in} n_2 + q_{\perp},\end{split}
\end{equation*}

where $\alpha_{\rm in} = \frac{n_2 \cdot q_{\rm in}}{n_1 \cdot n_2}$,
$\beta_{\rm in} = \frac{n_1 \cdot q_{\rm in}}{n_1 \cdot n_2}$, and
$q_\perp = q_{\rm in} - \alpha_{\rm in} n_1 - \beta_{\rm in} n_2$.
Note that the incoming jet momentum $q_{\rm in}$ includes a transverse
component $q_\perp$ due to the emission of radiation, whereas the
original momentum $p_a$ is defined from on-shell partons and lies
purely in the longitudinal direction.

In order to reconstruct the final-state momentum, we first apply a
rotation to align the frame such that the momentum of the outgoing jet can be
expressed as
\begin{equation*}
\begin{split}q_{\rm out} = \alpha_{\rm out} n_1 + \beta_{\rm out} n_2 + q_{\perp},\end{split}
\end{equation*}

where $q_{\perp}$ is the same transverse component as in the incoming
jet. The rotation ensures that the total momentum $p_a$ lies entirely
in the longitudinal plane, i.e.~it has no transverse component.

We take $\beta_{\rm out}$ to be approximately one in the limit where
the recoil is small and $n_1 \sim Q$. More precisely, if $n_1$
and $n_2$ are defined with $Q(0,0,\pm1;1)$, then $\beta_{\rm out}$
is approximately $\tfrac{1}{2}$. The correct value is determined from
the requirement that the virtual mass is preserved, which yields
\begin{equation*}
\begin{split}\alpha_{\rm out} = \frac{q^2_{\rm out} + p^2_{\perp}}{2 n_1 \cdot n_2},\end{split}
\end{equation*}

where $q^2_{\perp} = -p^2_{\perp}$.

We now rescale the momenta of the jets to implement momentum conservation.
We introduce the rescaling factors $k_{\rm in,out}$ and define the
\textit{shuffled momenta} as
\begin{equation*}
\begin{split}q'_{\rm in,out} = \alpha_{\rm in,out} k_{\rm in,out} n_1 +
\frac{\beta_{\rm in,out}}{k_{\rm in,out}} n_2 + q_\perp,\end{split}
\end{equation*}

which have the same virtual mass as the original jets. The requirement that
\begin{equation*}
\begin{split}p_a = \frac{1}{2}(n_1 - n_2) \equiv q'_{\rm out} - q'_{\rm in}
= (\alpha_{\rm in} k_{\rm in} - \alpha_{\rm out} k_{\rm out}) n_1
+ \left(\frac{\beta_{\rm in}}{k_{\rm in}} - \frac{\beta_{\rm out}}{k_{\rm out}} \right) n_2,\end{split}
\end{equation*}

leads to the equations
\begin{equation*}
\begin{split}\alpha_{\rm in} k_{\rm in} - \alpha_{\rm out} k_{\rm out} &= \frac{1}{2}, \\
\frac{\beta_{\rm in}}{k_{\rm in}} - \frac{\beta_{\rm out}}{k_{\rm out}} &= -\frac{1}{2}.\end{split}
\end{equation*}

As with the initial-initial case, once the rescaling factors have been determined,
the jets are transformed using a longitudinal boost such that
\begin{equation*}
\begin{split}q_{\rm in,out} \stackrel{\mathrm{boost}}{\longrightarrow} q_{\rm in,out}^{\prime}.\end{split}
\end{equation*}

This boost is defined in the rotated Breit frame. After the boost, a reverse
rotation is applied to bring the momenta back into the original frame.
Any net transverse momentum carried by the incoming jet prior to reshuffling is
absorbed by the final-state system via this boost--rotate--boost sequence,
ensuring momentum conservation across the full event.

\subsubsection{Reconstruction options}
\label{\detokenize{review/showers/qtilde:reconstruction-options}}

The procedures described above are sufficient for simple cases such as
the Drell--Yan production of vector bosons in hadron-hadron collisions or
deep inelastic scattering. In general however the colour structure of
the event, particularly in hadron collisions, requires a more general
procedure.

We currently support a number of procedures which have developed and become
more sophisticated during the evolution of the program.
The options are controlled by the
\href{https://herwig.hepforge.org/doxygen/QTildeReconstructorInterfaces.html\#ReconstructionOption}{ReconstructionOption} switch. In order of
the sophistication and our recommendation the options are:
\begin{enumerate}
\sphinxsetlistlabels{\arabic}{enumi}{enumii}{}{.}%
\item {} 

In Herwig 7 our default approach (\href{https://herwig.hepforge.org/doxygen/QTildeReconstructorInterfaces.html\#ReconstructionOption}{ReconstructionOption=Colour3}) attempts to use as much information as possible
on the colour structure of the hard process when performing the reconstruction. In order to
achieve this we consider all the partons in the hard process and commence the
reconstruction with the parton which had the hardest, \textit{i.e.} the highest $p_\perp$, emission in the parton
shower. The system formed by this parton and its colour partner is then reconstructed, with
either a full reconstruction of the jet produced by the evolution partner (\texttt{ReconstructionOption=Colour3}), the default, or optionally
just using the partner to absorb the recoil leaving it on its partonic mass shell (\href{https://herwig.hepforge.org/doxygen/QTildeReconstructorInterfaces.html\#ReconstructionOption}{ReconstructionOption=Colour4})
without a reconstruction of the full jet, where the jet denotes the complete
set of particles generated by the shower evolution of the corresponding
progenitor. This procedure is then repeated for the parton with
the hardest shower emission which has not been reconstructed until all the kinematics of
all the jets have been reconstructed. Together with an additional option of preserving the momentum
fraction of the softer incoming parton in the hard process for systems with colour connections
between initial-state partons this means that for a single emission the kinematics
reduce to those of the Catani--Seymour dipoles making matching in the MC@NLO approach simpler.

\item {} 

Another option using the colour structure is to first construct colour singlet systems
from the jet progenitors (\href{https://herwig.hepforge.org/doxygen/QTildeReconstructorInterfaces.html\#ReconstructionOption}{ReconstructionOption=Colour}). This was the default approach in
Herwig++ from version 2.3. Depending on the result different approaches
are used.
\begin{itemize}
\item {} 

If the incoming particles are colour neutral then any final-state
colour singlet systems are reconstructed as described in
\hyperref[\detokenize{review/showers/qtilde:sect-finalrecon}]{Section \ref{\detokenize{review/showers/qtilde:sect-finalrecon}}}, for example in $e^+e^-\to q\bar{q}$.

\item {} 

If there is a colour-singlet system consisting of the incoming
particles together with a number of final-state colour singlet
systems, \textit{e.g.} Drell-Yan vector boson production, then the
kinematics are reconstructed as described above for the
initial-initial system. The final-state systems are then
reconstructed in their rest frames as described in
\hyperref[\detokenize{review/showers/qtilde:sect-finalrecon}]{Section \ref{\detokenize{review/showers/qtilde:sect-finalrecon}}} and boosts applied to ensure the recoil from
the initial-state radiation is absorbed by the final-state systems.

\item {} 

If the system consists of colour-neutral particles and an
initial-final state colour connected system, e.g.\ deep inelastic
scattering, then the kinematics are reconstructed as described above
for an initial-final system.

\item {} 

If the system consists of two separate initial-final state colour
connected systems together with a number of colour-singlet
final-state systems, for example Higgs boson production via vector
boson fusion or $q\bar{q}\to t\bar{t}$, then the colour-singlet
initial-final systems are reconstructed as described above and the
final-state systems as described in \hyperref[\detokenize{review/showers/qtilde:sect-finalrecon}]{Section \ref{\detokenize{review/showers/qtilde:sect-finalrecon}}}.

\item {} 

In general in hadron-collisions the hard process cannot be decomposed
into colour singlet systems and a general procedure which preserves
the rapidity and mass of the hard collision is used. The
initial-state jets are reconstructed as discussed above for the
initial-initial connection. The final-state jets are then
reconstructed in the partonic centre-of-mass frame of the original
hard scattering process as described in
\hyperref[\detokenize{review/showers/qtilde:sect-finalrecon}]{Section \ref{\detokenize{review/showers/qtilde:sect-finalrecon}}}. This is effectively the same as
reconstructing them in the $q_{\mathrm{cms}}^{\prime}$ rest
frame, since the kinematic reconstruction for initial-initial
connection preserves the invariant mass of the hard process. In the
end, the jets originating from the final-state particles in the hard
process are boosted back to the lab frame, where they have a total
momentum $q_{\mathrm{cms}}^{\prime}$.

\end{itemize}

This procedure uses the underlying colour flow in the hard process to
determine how global energy and momentum conservation is enforced where
possible and resorts to the general approach used before Herwig++ 2.3
when this is not possible.

\item {} 

The general procedure described above is used in all cases (\texttt{ReconstructionOption=General}). This
was the default in FORTRAN HERWIG and Herwig++ versions prior to 2.3.

\end{enumerate}
The choice of reconstruction procedure can have a phenomenologically relevant
impact, particularly for observables sensitive to recoil and the detailed
kinematics of the showered jets, while its effect on sufficiently inclusive
observables is generally smaller. The \texttt{Colour3} option is therefore our
recommended default.

It is still possible to use the general procedure, which ignores the colour
flow for all processes, rather than the default option, which uses the colour
structure where available. The colour information supplied to the shower is a
specific colour-flow assignment for the hard event, generally in the
leading-colour approximation, and therefore does not in general represent the
complete finite-$N_C$ colour structure of the full QCD matrix element
\begin{footnote}\sphinxAtStartFootnote
When showering partonic configurations calculated in the MC@NLO scheme by MadGraph5/aMC@NLO \cite{Alwall:2014hca} program
a specific set of reconstruction options have to be selected due to the calculation of the shower
subtraction terms inside the MadGraph5/aMC@NLO program. In particular
\texttt{ReconstructionOption=General},
\texttt{InitialInitialBoostOption=LongTransBoost}, and
\texttt{InitialStateReconOption=Longitudinal}.
\end{footnote}.

\subsubsection{Radiation in particle decays}
\label{\detokenize{review/showers/qtilde:radiation-in-particle-decays}}\label{\detokenize{review/showers/qtilde:sub-radiation-in-particle}}

In general, the hard processes simulated by Herwig 7 consist of
$2 \rightarrow n$ scatterings. These are generated by first using
the relevant matrix elements to produce an initial configuration, and
then initiating parton showers from the external legs of the matrix element.
After this showering phase, the final state consists of a set of partons with
effective (constituent) masses assigned during the shower.

For processes involving the production and decay of unstable particles,
including decay chains, rather than attempting to calculate high-multiplicity
matrix elements, the simulation is simplified by appealing to the
\textit{narrow width approximation}—that is, by treating the production and decay
processes according to separate matrix elements, assuming no interference
between them. Unstable coloured particles are therefore produced in hard
processes or as the decays of other unstable particles, and are showered
like any other final-state coloured particle. In this case, the showering
process preserves the mass assigned during the production stage, rather than
assigning a new constituent mass.

For very high-mass coloured particles, \textit{e.g.} the top quark, the available
phase-space for decay can be so large that the decay occurs before any
hadronization can take place. Consequently, as well as undergoing time-like
showers $(q^2 > m^2)$ during their production phase, these partons
also undergo additional \textit{space-like} showering $(q^2 < m^2)$
of QCD radiation prior to their decay. Additionally, due to colour
conservation, the decay products themselves will also initiate time-like showers.

Since, under the narrow width approximation, the matrix element factorizes
into a component for the production and another for the decay, we may regard
these as two independent hard processes. This is the interpretation used in
Herwig’s simulation of the associated parton showers. Within this framework,
the time-like showers from coloured decay products follow an \textit{identical}
evolution to that used for final-state radiation in the production process.
Only the initial conditions for the evolution differ, although these are still
determined by examining the colour flow in the underlying hard decay
(see \hyperref[\detokenize{review/showers/qtilde:sub-initial-final-colour-connection-in-decays}]{Section \ref{\detokenize{review/showers/qtilde:sub-initial-final-colour-connection-in-decays}}}).

In contrast, the initial-state space-like shower created by a decaying particle
differs significantly from that of an incoming parton from the production
process (\hyperref[\detokenize{review/showers/qtilde:sub-initial-state-radiation}]{Section \ref{\detokenize{review/showers/qtilde:sub-initial-state-radiation}}}). In particular, it involves no
parton distribution functions (PDFs), since the heavy parton originates from
a hard scattering, not a hadron. In hard production processes, the initial-state
partons are evolved backward from the hard interaction to the incoming hadrons
to efficiently sample any resonant structure in the matrix element. In decays,
however, the emission of radiation that reduces the invariant mass of the
decaying particle does not impact the efficiency with which any resonant
structures are sampled. Therefore, it is natural to evolve the space-like
shower forward from the heavy unstable particle produced in the hard process
towards its eventual decay.

\paragraph{Evolution}
\label{\detokenize{review/showers/qtilde:id38}}

As in our discussion of the other showering algorithms, the description
here uses the Sudakov decomposition of the momenta given in Eq.
\eqref{equation:review/showers/qtilde:eq:sudbasis}. In space-like decay showers, the decaying particle
$\widetilde{ij}$ undergoes branchings
$\widetilde{ij}\rightarrow i+j$, where $j$ is a final-state
time-like parton and $i$ is the same decaying particle with an
increased space-like virtuality:
$q_{i}^{2}<q_{\widetilde{ij}}^{2}\le m_{\widetilde{ij}}^{2}$. In
this process the original particle acquires a space-like virtuality,
\begin{equation*}
\begin{split}q_{i}^{2}=zq_{\widetilde{ij}}^{2}+\frac{p_{\perp}^{2}-zq_{j}^{2}}{1-z},\end{split}
\end{equation*}

where $z=\alpha_{i}/\alpha_{\widetilde{ij}}$,
$\mathbf{p}_{\perp}^{2}=-p_{\perp}^{2}\ge0$, and
$p_{\perp}=q_{\perp i}-zq_{\perp\widetilde{ij}}$. Since, in the decay shower,
the invariant mass of the decaying particle remains unchanged,
i.e.~$m_i = m_{\widetilde{ij}}$, the space-like evolution variable from
Eq.~\eqref{equation:review/showers/qtilde:eq:qtilde_spacelike} simplifies to
\begin{equation*}
\begin{split}\tilde{q}^{2}=m_{i}^{2}+\frac{zm_{j}^{2}-p_{\perp}^{2}}{\left(1-z\right)^{2}}.\end{split}
\end{equation*}
Unlike in the final- and initial-state showers, the evolution variable
in this forward-evolving decay shower increases toward the decay scale.
Requiring the transverse momentum of the branching to be real,
$\mathbf{p}_{\perp}^{2}\geq0$, imposes an upper limit,
$z_+$, on $z$ where
\begin{equation*}
\begin{split}\begin{aligned}
z_+ & = & 1-\frac{m_{j}^{2}}{2\left(\tilde{q}^{2}-m_{i}^{2}\right)}\left(\sqrt{1+4\left(\tilde{q}^{2}-m_{i}^{2}\right)/m_{j}^{2}}-1\right).\end{aligned}\end{split}
\end{equation*}

For the space-like decay shower we have the further constraint that the
parton showering cannot degrade the invariant mass of the decaying
object below the threshold required for the decay process, which imposes
a lower limit on $z$.

Since no PDF is involved in this forward parton-shower evolution
algorithm, the Sudakov form factor has exactly the same form as that
used for final-state radiation in Eqs.~\eqref{equation:review/showers/general:eq:product_of_sudakovs} and \eqref{equation:review/showers/qtilde:eq:sudakovmaster}. Consequently the forward evolution can be
performed using the veto algorithm in almost exactly the same way as was
done for the final-state showers, see \hyperref[\detokenize{review/appendix/sudakov:sect-sudakov-solution}]{\ref{\detokenize{review/appendix/sudakov:sect-sudakov-solution}}}.
The main difference is in the implementation of the angular-ordering
bounds for subsequent branchings. For final-state radiation involving
branchings $\widetilde{ij}\rightarrow i+j$, where $i$ has a
light-cone momentum fraction $z$, we evolved $i$ and
$j$ \textit{downward} from $\tilde{q}_{h\, i}=z\tilde{q}$ and
$\tilde{q}_{h\, j}=\left(1-z\right)\tilde{q}$ respectively, where
$\tilde{q}$ was the scale of the $\widetilde{ij}$ branching.
Since the decay shower is really a forward-evolving initial-state
shower, we evolve $i$ \textit{upward} from
$\tilde{q}_{h\, i}=\tilde{q}$ and $j$ \textit{downward} from
$\tilde{q}_{h\, j}=\left(1-z\right)\tilde{q}$. This procedure is
iterated until the scale $\tilde{q}$ approaches the minimum
compatible with the threshold for the underlying decay process.

\paragraph{Kinematic reconstruction}
\label{\detokenize{review/showers/qtilde:id39}}

In the approach of \cite{Gieseke:2003rz}, for the simulation of QCD
radiation in particle decays, the recoil due to the radiation emitted
from the decaying particle is absorbed by its final-state colour
partner. The reconstruction described in \cite{Hamilton:2006ms}, valid
for the decay of a coloured particle to a colour connected final-state
particle and a colour-singlet system, was designed to preserve the mass
of the colour-singlet system. In the case of top decay this amounts to
preserving the mass of the W boson, and the momentum of the decaying
particle. More complicated colour structures, involving more coloured
particles in the final-state, \textit{e.g.} gluino decays, require a
generalization of this momentum reconstruction procedure.

Consider the decay of a coloured particle with momentum $p$, to
$n+1$ particles. We denote the momentum of the colour partner of
the decaying particle $\bar{p}$, and the momenta of the remaining
primary decay products are denoted $p_{i=1,n}$. Prior to
simulating the effects of QCD radiation,
\begin{equation*}
\begin{split}p=\bar{p}+\sum_{i=1}^{n}\, p_{i}.\end{split}
\end{equation*}

After simulating parton-shower radiation in the decay, the original
momenta of the decay products must be shifted and rescaled to
accommodate the additional \textit{initial}-\textit{state} radiation. We require the
sum of the new momenta of the colour partner, $\bar{q}$, the other
primary decay products, $q_{i}$, \textit{and} the radiation emitted prior
to the decay, $q_{{\rm ISR}}$, to equal that of the decaying particle:
\begin{equation}
\begin{split}
p=\bar{q}+q_{{\rm ISR}}+\sum_{i=1}^{n}\, q_{i}.
\end{split}
\label{equation:review/showers/qtilde:eq:decay-momentum-balance}
\end{equation}
To satisfy Eq.~\eqref{equation:review/showers/qtilde:eq:decay-momentum-balance},
we rescale by a common factor $k_{1}$ the three-momenta $p_i$ of the
remaining primary decay products, and by a separate factor $k_{2}$ the
three-momentum $\bar{p}$ of the colour partner. The component of the momentum of the emitted radiation
transverse to the colour partner is absorbed by the colour partner. In
the rest frame of the decaying particle these rescalings and shifts are:
\begin{equation*}
\begin{split}p & =  \left(\mathbf{0};m\right);\\
q_{i} & =  \left(k_{1}\mathbf{p}_{i};\sqrt{k_{1}^{2}\left|\mathbf{p}_{i}\right|^{2}+p_{i}^{2}}\right);\\
\bar{q} & =  \left(k_{2}\mathbf{\bar{p}}-\mathbf{q}_{\perp {\rm ISR}};\sqrt{k_{2}^{2}\left|\bar{\mathbf{p}}\right|^{2}+\left|\mathbf{q}_{\perp {\rm ISR}}\right|^{2}+\bar{p}^{2}}\right),\end{split}
\end{equation*}

where $m$ is the mass of the decaying particle and
$\mathbf{q}_{\perp {\rm ISR}}$ is the component of the three-momentum of
the initial-state radiation perpendicular to $\bar{p}$.

The rescaling factors $k_{1,2}$ allow for the
conservation of energy and of momentum in the longitudinal direction.
Three-momentum conservation in the longitudinal,
$\mathbf{\bar{p}}$, direction requires that
\begin{equation}\label{equation:review/showers/qtilde:eq:decay_recon_factors_1}
\begin{split}k_{2}\mathbf{\bar{p}}+k_{1}\sum_{i=1}^{n}\,\mathbf{p}_{i}+\mathbf{q}_{\parallel {\rm ISR}}=0.\end{split}
\end{equation}

The momentum of the initial-state radiation perpendicular to the
direction of the colour partner, $\mathbf{q}_{\perp {\rm ISR}}$, can be
projected out, leaving the parallel component
$\mathbf{q}_{\parallel {\rm ISR}}$, by taking the dot product with the
spatial component of the $n$ basis vector (aligned with
$\bar{p}$), \textit{i.e.}
\begin{equation}\label{equation:review/showers/qtilde:eq:decay_recon_factors_2}
\begin{split}k_{1}=k_{2}+\frac{{\mathbf{q}_{{\rm ISR}}}\cdot\mathbf{n}}{\mathbf{\bar{p}}\cdot\mathbf{n}}.\end{split}
\end{equation}

Finally, from the conservation of energy we have
\begin{equation}\label{equation:review/showers/qtilde:eq:decay_recon_factors_3}
\begin{split}\sum_{i=1}^{n}\sqrt{k_{1}^{2}{\left|\mathbf{p}_{i}\right|}^{2}+p_{i}^{2}}+\sqrt{k_{2}^{2}\left|\mathbf{\bar{p}}\right|^{2}+\left|\mathbf{q}_{\perp {\rm ISR}}\right|^{2}+\bar{p}^{2}}+E_{{\rm ISR}}=m,\end{split}
\end{equation}

where $E_{{\rm ISR}}$ is the energy of the initial-state
radiation. This system of equations, Eqs.~
\eqref{equation:review/showers/qtilde:eq:decay_recon_factors_1},
\eqref{equation:review/showers/qtilde:eq:decay_recon_factors_2},
\eqref{equation:review/showers/qtilde:eq:decay_recon_factors_3}) for the
rescaling factors can be solved analytically for two-body decays, or
numerically, using the Newton-Raphson method, for higher multiplicities.

\subsubsection{Shower interactions}
\label{\detokenize{review/showers/qtilde:shower-interactions}}\label{\detokenize{review/showers/qtilde:sect-showerinteractions}}

The conventional \textit{QCD+QED} schemes for generating collinear parton showers effectively describe experimental data up to current LHC energies (e.g.~\cite{ATLAS:2019lpk,ATLAS:2019lbg,ATLAS:2019old,ATLAS:2019gkg}). However, in higher CoM frameworks, contributions from pure EW radiations become significant. At these energies, heavy particles like EW gauge bosons, Higgs bosons, and top quarks may appear as part of jets and behave as massless partons as $\widetilde{q}$ increases beyond their masses. This expectation is supported by LHC observations of Higgs boson production via vector-boson fusion \cite{CMS:2019kqw,ATLAS:2018jvf} and recent studies \cite{Dawson:2014pea,Han:2014nja,Bellm:2016cks,Darvishi:2019uzp,Darvishi:2020paz}. Furthermore, excluding EW emissions from high-energy processes can cause an imbalance due to large, negative virtual corrections \cite{Beenakker:2000kb}. This justifies introducing a process-independent EW parton shower to enhance the production rate of underlying events and upgrade the conventional parton shower to a \textit{QCD+QED+EW} scheme, as outlined in \cite{Masouminia:2021kne}. Several theoretical studies have addressed parts of the EW parton shower \cite{Ciafaloni:2000rp,Ciafaloni:2000gm,Ciafaloni:2005fm,Baur:2006sn}, with more complete studies on EW splitting functions in \cite{Chen:2016wkt}. Attempts to incorporate EW parton showers in event generators include \cite{Chiesa:2013yma, Christiansen:2014kba, Krauss:2014yaa, Mangano:2002ea, Kleiss:2020rcg}, and most recently \cite{Brooks:2021kji}.

The implementation of the \textit{QCD+QED+EW} AO shower includes the calculation of initial-state (IS) and final-state (FS) quasi-collinear EW splittings in their spin-unaveraged forms for both massless and massive cases, including FFV and FFS splittings, where $F$, $V$, and $S$ denote fermion, vector, and scalar particles, respectively:
\begin{equation*}
\begin{split}
q \to q'W^{\pm}, \quad
q \to qZ^{0}, \quad
q \to qH,
\end{split}
\end{equation*}
in addition to VVV and VVS splittings:
\begin{equation*}
\begin{split}
W^{\pm} \to W^{\pm}Z^{0}, \quad
W^{\pm} \to W^{\pm}\gamma, \quad
Z^{0} \to W^{+}W^{-}, \quad
\gamma \to W^{+}W^{-},
\end{split}
\end{equation*}
\begin{equation*}
\begin{split}
W^{\pm} \to W^{\pm}H, \quad
Z^{0} \to Z^{0}H.
\end{split}
\end{equation*}
Thus, FFV, FFS, VVV, and VVS denote fermion-fermion-vector, fermion-fermion-scalar, three-vector, and vector-vector-scalar vertices, respectively. These, along with the existing SVV and VFF decay modes, corresponding to scalar-vector-vector and vector-fermion-fermion vertices, provide a comprehensive framework for IS and FS EW radiation in simulated events. Among these splittings, only FFV and FFS contribute to both ISR and FSR. Implementing VVV and VVS for ISR would require incorporating EW PDFs into the backward evolution of the corresponding progenitors. However, the required calculations are computationally expensive, and their phenomenological impact on initial-state radiation is negligible. Additionally, the available EW PDFs are neither sufficiently reliable nor accurate for inclusion in a general-purpose event generator \cite{Kane:1984bb, Dawson:1984gx}.

In the following discussions, we outline the method for deriving the explicit analytic forms of the EW splitting functions and analyse their numerical efficiency.

\paragraph{EW splitting functions}
\label{\detokenize{review/showers/qtilde:ew-splitting-functions}}\label{\detokenize{review/showers/qtilde:sect-ewsplittings}}

Following Herwig 7’s convention, we define the helicity-dependent splitting function of a generic $\widetilde{ij} \to i + j$ splitting as
\begin{equation}\label{equation:review/showers/qtilde:eq:splitFunc}
\begin{split}P_{\widetilde{ij} \to i + j}(z,\tilde{q}) =
\sum_{\lambda_i,i=0}^{2} \left| H_{\widetilde{ij} \to i + j}(z,\tilde{q};\lambda_0,\lambda_1,\lambda_2) \right|^2,\end{split}
\end{equation}

with a helicity amplitude $H_{\widetilde{ij} \to i + j}(z,\tilde{q};\lambda_0,\lambda_1,\lambda_2)$ defined as
\begin{equation}\label{equation:review/showers/qtilde:eq:heliAmp}
\begin{split}H_{\widetilde{ij} \to i + j}(z,\tilde{q};\lambda_0,\lambda_1,\lambda_2) =
g F_{\lambda_0,\lambda_1,\lambda_2}^{\widetilde{ij} \to i + j},\end{split}
\end{equation}

where $\lambda_i$ are the helicity states of the progenitor ($i=0$) and the children ($i=1,2$), and $F_{\lambda_0,\lambda_1,\lambda_2}^{\widetilde{ij} \to i + j}$ is the corresponding vertex function, which is determined through the Feynman rules.

One notes that most QED branchings can be derived from the equivalent QCD splittings by substituting $\alpha_{s}$ with $\alpha_{em}$ and replacing the colour factor with the charge squared of the fermion or scalar boson involved. However, this approach does not apply to EW branchings. EW branchings are more complex due to the mass of the gauge bosons and the presence of additional longitudinal polarisation states. Moreover, all particles involved in these processes have non-zero masses. In the following, we will outline these computations case-by-case, following the footsteps of Ref. \cite{Masouminia:2021kne}.

\subparagraph{FFV splittings}
\label{\detokenize{review/showers/qtilde:ffv-splittings}}\label{\detokenize{review/showers/qtilde:sect-ffv}}

In a $FFV$ branching with $V=W^{\pm}, \; Z^0$, the
transverse polarisation vectors ($\lambda_2=\pm 1$) of the
vector boson are the same as for gluonic radiation from a quark
splitting, i.e.
\begin{equation*}
\begin{split}\epsilon^\mu_{\lambda_2=\pm 1}(q_2) = \left[ 0;
-\frac{\lambda_2}{\sqrt{2}}\left(1-\frac {p_{\perp}^2 \lambda^2{\rm e}^{i\lambda_2\phi} \cos\phi}{2p^2\left(1-z\right)^2}\right),
-\frac{i}{\sqrt{2}} +\frac {\lambda_2p_\perp^2\lambda^2{\rm e}^{i\phi} \sin\phi}{2\sqrt {2}p^2 \left(1-z\right)^2},
-\frac { \lambda_2p_\perp\lambda{\rm e}^{i\lambda_2\phi} }{\sqrt {2} \left(1-z\right) p}\right],\end{split}
\end{equation*}
while the longitudinal polarisation vector ($\lambda_2=0$) is
\begin{equation}\label{equation:review/showers/qtilde:outPol2T}
\begin{split}\epsilon^\mu_0(q_2) &= \left[ \frac {p \left(1-z\right)}{\lambda m_2}
+\frac{p_\perp^2+m_0^2(1-z)^2-m_2^2}{4p \left(1-z\right) m_2}\lambda;
\cos\phi\left( \frac {p_\perp}{m_2}-\frac {m_2 p_\perp\lambda^2}{2p^2\left(1-z\right)^2}\right),
\right. \\
& \left. -\sin\phi\left( \frac {p_\perp}{m_2}
-\frac{m_2 p_\perp\lambda^2}{2p^2\left(1-z\right)^2}\right),
\frac {p\left(1-z\right)}{\lambda m_2}
-{\frac {p_\perp^2-m_0^2(1-z)^2-m_2^2}{4p \left(1-z\right)m_2}}\lambda
\right].\end{split}
\end{equation}

In these and subsequent equations, $\lambda$ is a power-counting
parameter that helps to identify the quasi-collinear limit, which is
the leading term of the small-$\lambda$ limit, and $p$ is
the magnitude of the 3-momentum of the incoming parton,
$p=|\mathbf{p}_0|$.

The spinors for the incoming and outgoing fermion are given by
\begin{equation}\label{equation:review/showers/qtilde:eq:inSpinors}
\begin{split}u_{\frac12}(p) = \left(\begin{array}{c} \frac {m_0}{\sqrt{2p}}\lambda\\ 0 \\ \sqrt {2p}\left(1+\frac{m_0^2\lambda^2}{8p^2}\right) \\ 0
\end{array}\right), \quad u_{-\frac12}(p) = \left(\begin{array}{c} 0\\ \sqrt {2p}\left(1+\frac {m_0^2\lambda^2}{8p^2}\right)\\ 0 \\ \frac{m_0}{\sqrt{2p}}\lambda
\end{array}\right),\end{split}
\end{equation}\begin{equation}\label{equation:review/showers/qtilde:eq:outSpinor1}
\begin{split}\bar{u}_{\frac12}(q_1) = \left[
\sqrt {2zp}\left(1+\frac{m_0^2\lambda^2}{8p^2}\right),
\frac {{\rm e}^{-i\phi} p_{\perp} \lambda}{\sqrt {2zp}},
\frac {m_1}{\sqrt {2zp}}\lambda,
\frac {{\rm e}^{-i\phi}p_\perp m_1{\lambda}^2}{\left[2zp\right]^{3/2}}
\right],\end{split}
\end{equation}\begin{equation}\label{equation:review/showers/qtilde:eq:outSpinor2}
\begin{split}\bar{u}_{-\frac12}(q_1) = \left[
-\frac { {\rm e}^{i\phi} p_{\perp} m_1\lambda^2}{\left[2zp\right]^{3/2}},
\frac{m_1}{\sqrt {2zp}}\lambda,
-\frac { {\rm e}^{i\phi} p_{\perp} \lambda}{\sqrt{2zp}},
\sqrt {2zp}\left(1+\frac {m_0^2\lambda^2}{8zp}\right)
\right].\end{split}
\end{equation}

Consequently, the vertex function for a FFV branching can be written as
\begin{equation}\label{equation:review/showers/qtilde:eq:FFVVF}
\begin{split}F_{\lambda_0, \lambda_1, \lambda_2}^{FFV} =
\sqrt{\frac{1}{2 (\tilde{q}^2_0 - m_0^2)}} \bar{u}_{\lambda_1}(q_1) \left(g_L P_L + g_R P_R\right)
\gamma \cdot \epsilon_{\lambda_2} \, u_{\lambda_0}(q_0),\end{split}
\end{equation}

with separate couplings to the left- and right-handed helicities, $g_L$ and $g_R$. Combining Eqs. \eqref{equation:review/showers/qtilde:eq:FFVVF},  \eqref{equation:review/showers/qtilde:eq:heliAmp} and \eqref{equation:review/showers/qtilde:eq:splitFunc} would readily result in the spin-unaveraged splitting function for the generic FFV branching. One however, immediately realizes that due to the presence of $p_\perp/m$ type terms in the longitudinal polarisation vector of the vector particle, the longitudinal part of the splitting function would become divergent in the infinite-momentum frame $\tilde{q}^2 \gg m^2$ as \cite{Masouminia:2021kne}
\begin{equation*}
\begin{split}P^{\rm L}_{FFV}(z,\tilde{q}) \stackrel{p_\perp\gg m_2}{\longrightarrow}
\frac{1}{2} \left( {g_L}^{2} \rho_{{-1,-1}}+{g_R}^{2}\rho_{{1,1}} \right)
{\frac {{\tilde{q}}^{2}{z}^{2} \left( 1-z \right) }{{m_2}^{2}}},\end{split}
\end{equation*}

where $\rho$ is the spin-density matrix of the emitter particle
$0$.

To address this divergent behaviour, we adopt Dawson’s approach \cite{Dawson:1984gx}, where the component of the longitudinal polarisation vector proportional to its momentum is subtracted. This yields:
\begin{equation}\label{equation:review/showers/qtilde:eq:outSpinor2reduced}
\begin{split}\epsilon^\mu_{0^*}(q_2) = \frac{\lambda m_2}{2p(1-z)}
\left[-1;
\frac{\lambda \cos\phi p_\perp}{p(1-z)},
\frac{\lambda \sin\phi p_\perp}{p(1-z)},
1\right],\end{split}
\end{equation}

which vanishes as $m_2 \to 0$. Using this approach, we derive the longitudinal polarisation as:
\begin{equation}\label{equation:review/showers/qtilde:eq:FFVL}
\begin{split}P^{\rm L}_{FFV}(z, \tilde{q}) &= \sum_{\lambda_0, \lambda_1= \pm \frac{1}{2}; \lambda_2 =0}
\left| H_{FFV}(z,\tilde{q};\lambda_0,\lambda_1,\lambda_2) \right|^2
\\ &=
\left(g_L^2 \rho_{-1,-1} + g_R^2 \rho_{1,1}\right)
\frac{2m_2^2}{\tilde{q}^2 (1-z)^3}.\end{split}
\end{equation}

Additionally, the transverse parts of the spin-averaged FFV splitting function can be calculated as
\begin{equation}\label{equation:review/showers/qtilde:eq:FFVT}
\begin{split}P^{\rm T}_{FFV}(z,\tilde{q}) &= \sum_{\lambda_0, \lambda_1 = \pm \frac{1}{2}; \lambda_2 = \pm 1}
\left| H_{FFV}(z,\tilde{q};\lambda_0,\lambda_1,\lambda_2) \right|^2
\\
&=
\frac {1}{1-z} \Bigg(\frac{\left( {g_L}^2+{g_R}^2\right)}
2\left[1+z^2+\frac {\left(1-z^2\right)\left({m_0}^2-{m_1}^2\right)-\left(1+ z^2\right)
{m_2}^2}{z \left( 1-z \right)^2 \tilde{q}^2}\right] \\
& -2g_Lg_R\frac {m_0m_1}{z {\tilde{q}}^2} \Bigg)\end{split}
\end{equation}

The total FFV splitting function can be simply written as the summation of Eqs.~\eqref{equation:review/showers/qtilde:eq:FFVL} and \eqref{equation:review/showers/qtilde:eq:FFVT}, but a more useful format would be to decompose it into massless and massive parts:
\begin{equation*}
\begin{split}P_{FFV}^{\rm massless}(z, \tilde{q}) = \left( g_L^2 \rho_{-1, -1} + g_R^2 \rho_{1, 1} \right) \frac{1 + z^2}{1 - z},\end{split}
\end{equation*}
\begin{equation*}
\begin{split}P_{FFV}^{\rm massive}(z, \tilde{q}) &= \frac{1}{1 - z}
\left[ \left( g_L^2 \rho_{-1, -1} + g_R^2 \rho_{1, 1} \right)
\left\{ \frac{m_0^2 (1 + z^2)}{\tilde{q}^2 z (1 - z)} -
\frac{m_1^2 (1 + z)}{z \tilde{q}^2 (1 - z)} - \frac{m_2^2}{z \tilde{q}^2}
\right\} \right.
\\ &+ \left.
\frac{m_0^2}{\tilde{q}^2} \left( g_L^2 \rho_{1, 1} + g_R^2 \rho_{-1, -1} \right) -
\frac{2 m_0 m_1 g_L g_R}{z \tilde{q}^2} \left( \rho_{1, 1} + \rho_{-1, -1} \right) \right].\end{split}
\end{equation*}
It is easy to see that in the massless limit, the EW FFV splitting function reduces to its QCD counterpart.

\subparagraph{FFS splittings}
\label{\detokenize{review/showers/qtilde:ffs-splittings}}\label{\detokenize{review/showers/qtilde:sect-ffs}}

For the case of Higgs boson radiation from a parent quark, the spinors of the incoming and outgoing quarks are the same as in Eqs. \eqref{equation:review/showers/qtilde:eq:inSpinors}, \eqref{equation:review/showers/qtilde:eq:outSpinor1}, and \eqref{equation:review/showers/qtilde:eq:outSpinor2}. The corresponding vertex function can be written as:
\begin{equation*}
\begin{split}F_{\lambda_0, \lambda_1}^{FFS} = \sqrt{\frac{1}{2(\tilde{q}^2_0 - m_0^2)}} \bar{u}_{\lambda_1}(q_1) u_{\lambda_0}(q_0),\end{split}
\end{equation*}
resulting in the following explicit forms:
\begin{equation*}
\begin{split}F_{+, +}^{FFS} &= \frac{m_0 (1 + z)}{\sqrt{2 z (\tilde{q}^2 - m_0^2)}}, \qquad
F_{+, -}^{FFS} = -\frac{p_{\perp} e^{i\phi}}{\sqrt{2 z (\tilde{q}^2 - m_0^2)}}, \\
F_{-, +}^{FFS} &= \frac{p_{\perp}}{\sqrt{2 z (\tilde{q}^2 - m_0^2)}}, \qquad
F_{-, -}^{FFS} = \frac{m_0 (1 + z) e^{-i\phi}}{\sqrt{2 z (\tilde{q}^2 - m_0^2)}}.\end{split}
\end{equation*}

These can be combined to form the $P_{FFS}$ splitting function,
\begin{equation}\label{equation:review/showers/qtilde:eq:FFSSF}
\begin{split}P_{FFS}(z, \tilde{q}) &= \sum_{\lambda_0, \lambda_1 = \pm \frac{1}{2}}
\left| H_{FFS}(z, \tilde{q}; \lambda_0, \lambda_1) \right|^2 \\
&= g^2 \left( \frac{m_0}{m_W} \right)^2 \left[ (1 - z) + \frac{ (m_0 + m_1)^2 - m_2^2}{\tilde{q}^2 (1 - z) z} \right],\end{split}
\end{equation}

where $m_i$, for ( i=0,1,2 ), are the running masses of the progenitor and the children. While working within the Standard Model (SM), one can always further simplify Eq. \eqref{equation:review/showers/qtilde:eq:FFSSF} by setting $m_0 = m_1$.

\subparagraph{VVV splittings}
\label{\detokenize{review/showers/qtilde:vvv-splittings}}\label{\detokenize{review/showers/qtilde:sect-vvv}}

In the most general case of a VVV EW splitting, both the parent and the children are considered to be massive gauge vector bosons with both transverse and longitudinal polarisation vectors. For the parent gauge boson, we can write the transverse polarisation vectors as:
\begin{equation}\label{equation:review/showers/qtilde:inPol0T}
\begin{split}\epsilon^\mu_{\lambda_0=\pm 1}(p) = \left[ 0,-\frac{\lambda_0}{\sqrt{2}},-\frac{i}{\sqrt{2}},0 \right],\end{split}
\end{equation}

The longitudinal polarisation vector is given by:
\begin{equation*}
\begin{split}\epsilon^\mu_{0}(p) = \left[ \frac{p}{\lambda m_0}, 0, 0, \frac{\sqrt{\lambda^2 m_0^2 + p^2}}{\lambda m_0} \right],\end{split}
\end{equation*}
To avoid the singularities that would emerge from the longitudinal polarisation vector as $m_0 \to 0$, we employ Dawson’s approach and rewrite this vector as:
\begin{equation}\label{equation:review/showers/qtilde:inPol0Ls}
\begin{split}\epsilon^\mu_{0^*}(p) = \left[ -\frac{\lambda m_0}{p + \sqrt{\lambda^2 m_0^2 + p^2}}, 0, 0, \frac{\lambda m_0}{p + \sqrt{\lambda^2 m_0^2 + p^2}} \right],\end{split}
\end{equation}

Furthermore, we use the polarisation vectors from Eqs. \eqref{equation:review/showers/qtilde:eq:outSpinor1} and \eqref{equation:review/showers/qtilde:eq:outSpinor2reduced} for the second child, while applying the transformation $z \to (1-z)$ will reproduce the polarisation vectors of the first child. Therefore, we can write
\begin{equation*}
\begin{split}H_{VVV}(z, \tilde{q}; \lambda_0, \lambda_1, \lambda_2) = i g F_{\lambda_0, \lambda_1, \lambda_2}^{VVV},\end{split}
\end{equation*}

where $g = e \tan \theta_W$ when $V', V'' = W^{\pm}, Z^0$ and $g = e \tan \theta_W$ when either $V$ or $V''$ is a photon. For these splittings, the vertex functions take on the form:
\begin{equation*}
\begin{split}F_{\lambda_0, \lambda_1, \lambda_2}^{VVV} = \sqrt{\frac{1}{2 (\tilde{q}^2_0 - m_0^2)}} \left[
(q_1 \cdot \epsilon_{\lambda_2}^*) (\epsilon_{\lambda_0} \cdot \epsilon_{\lambda_1}^*)
+ (q_2 \cdot \epsilon_{\lambda_0}^*) (\epsilon_{\lambda_1} \cdot \epsilon_{\lambda_2}^*)
- (q_2 \cdot \epsilon_{\lambda_1}^*) (\epsilon_{\lambda_0} \cdot \epsilon_{\lambda_2}^*)
\right].\end{split}
\end{equation*}

In Ref. \cite{Masouminia:2021kne} this vertex function is tabulated for different transverse/longitudinal configurations while considering Dawson’s approach to carry out the $1/m_i$ terms. It is shown that the massless and massive terms can be summarised as
\begin{equation*}
\begin{split}P_{VVV}^{\rm massless}(z,\tilde{q}) = 2(\rho_{-1,-1}+\rho_{1,1})
\frac{(1-(1-z) z)^2}{(1-z) z},\end{split}
\end{equation*}
\begin{equation*}
\begin{split}P_{VVV}^{\rm massive}(z,\tilde{q}) &=
\frac{1}{(1-z) z} \biggl[ (\rho_{-1,-1}+\rho_{1,1})
\Bigl\{
2 m_{0,t}^2 (1-(1-z) z)^2
- 2 m_{1,t}^2 \left(1-(1-z) z^2\right)
\\
&- 2 m_{2,t}^2 \left(1-(1-z)^2 z\right) \Bigr\}
+4 \rho_{0,0} \; m_{0,t}^2 \; z (1-z)^3 \biggr],\end{split}
\end{equation*}
with $m_{i,t}^2 = m_i^2/(\tilde{q}^2z(1-z))$.

\subparagraph{VVS splittings}
\label{\detokenize{review/showers/qtilde:vvs-splittings}}\label{\detokenize{review/showers/qtilde:sect-vvs}}

Similarly, for the generic case of VVS EW branching, one can write a vertex function
\begin{equation*}
\begin{split}F_{\lambda_0,\lambda_1}^{VVS} = m_0 \sqrt{\frac1{2(\tilde{q}^2_0-m_0^2)}}
\left( \epsilon_{\lambda_0} \cdot \epsilon_{\lambda_1}^* \right),\end{split}
\end{equation*}

to be used in the VVS helicity amplitude
\begin{equation*}
\begin{split}H_{VVS}(z,\tilde{q};\lambda_0,\lambda_1) = g \; F_{\lambda_0,\lambda_1}^{VVS},\end{split}
\end{equation*}

where $g = e/\sin \theta_W$ for $W^{\pm}$ bosons and $g = e/(\sin \theta_W \cos \theta_W)$ for $Z^{0}$ bosons, while the dynamics of the particles will remain the same as the previous cases, i.e.\ Eqs. \eqref{equation:review/showers/qtilde:inPol0T} and \eqref{equation:review/showers/qtilde:inPol0Ls} for the progenitor and Eqs. \eqref{equation:review/showers/qtilde:outPol2T} and \eqref{equation:review/showers/qtilde:eq:outSpinor2reduced} for the children. Accordingly, different helicity configurations of the splitting function can be individually derived \cite{Masouminia:2021kne}, resulting in
\begin{equation*}
\begin{split}P_{VVS}^{\rm Massless}(z,\tilde{q}) = \frac{1-z}{4 z}
\left[ z^2 (\rho_{-1,-1} + \rho_{1,1}) + 2 \rho_{0,0} \right],\end{split}
\end{equation*}
\begin{equation*}
\begin{split}P_{VVS}^{\rm Massive}(z,\tilde{q}) &= -\frac{m_{H,t}^2 }{4 z}
\left[ z^2 (\rho_{-1,-1} + \rho_{1,1}) + 2 \rho_{0,0} \right]
- \frac{m_{0,t}^2 }{4 z^2} \left[ \left( 2 z^2-4 z+2\right) \rho_{0,0} \right.
\\ &+ \left.
\left(z^4-2 z^3-z^2 \right) (\rho_{-1,-1}+\rho_{1,1}) \right].\end{split}
\end{equation*}

\paragraph{Validation of EW shower}
\label{\detokenize{review/showers/qtilde:validation-of-ew-shower}}\label{\detokenize{review/showers/qtilde:sect-ewvalidation}}

The validation of the EW parton shower implementation has been an essential aspect of ensuring the accuracy and reliability of the simulation results, particularly due to the use of Dawson’s approach in dealing with longitudinal divergences in the polarisation vectors of the involving massive vector bosons. In Ref. \cite{Masouminia:2021kne}, all four pure EW splitting classes - FFV, FFH, VVV and VVS - have been subjected to rigorous performance tests. For this purpose, we initially chose a suitable fixed order (FO) process that one can meaningfully split into a smaller hard process plus a single resummed splitting (RS). The aforementioned splitting then can be replaced by a single-radiation EW shower, providing a test bed for investigating the performance of individual splitting functions and their kinematic reconstruction within our implementation. In this context, special care was paid to VVV and VVS cases, where, in order to keep gauge invariance of the FO analysis intact, additional channels had to be included, some of which could not be seen as an interpretation of an RS case, i.e.\ where a child is not emitted directly from the target parent. In these scenarios, one can make use of the assumption that if the parent and child are connected, one expects to observe a small distance between them in the pseudorapidity-azimuthal angle plane, compared to the non-RS-related cases. Therefore, applying a $\Delta R$ cut to ensure a smaller separation between the prospective children could suppress non-RS-related channels \cite{Masouminia:2021kne}.
\begin{figure}[htp]
\centering
\capstart
\noindent\includegraphics[width=1.000\linewidth]{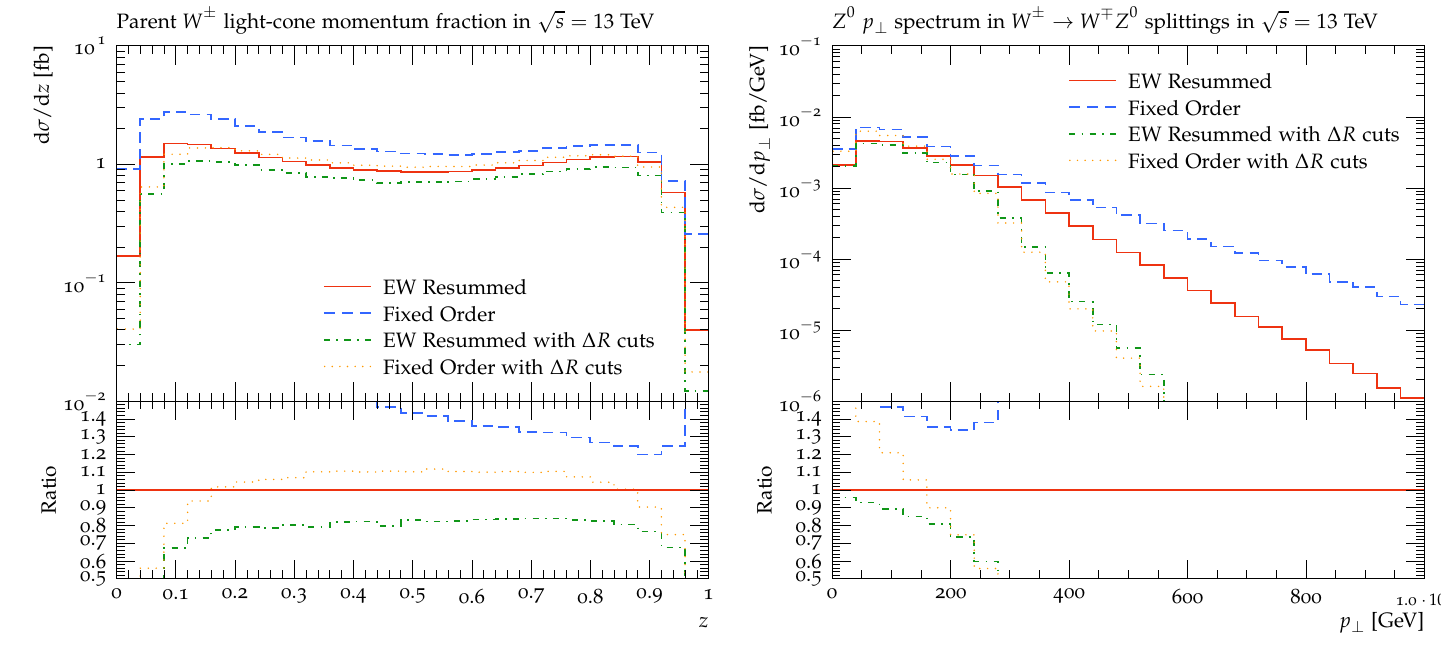}
\caption{Performance test for $W^{\pm} \to W^{\pm} Z^0$ EW branching.
In order to suppress the non-RS-related contributions, we have imposed the following cuts: $\Delta R_{W^{\pm},V} > 1$,
$\Delta R_{W^{\pm},jet} < 1$ and $\Delta R_{V,jet}< 1$.}\label{\detokenize{review/showers/qtilde:id66}}\label{\detokenize{review/showers/qtilde:fig-ewvalivation}}\end{figure}
The comparison in Fig.~\ref{\detokenize{review/showers/qtilde:fig-ewvalivation}}
tests how well the single-emission shower approximation reproduces the
relevant behaviour of the exact matrix element. The RS prediction is expected
to lie below the FO result, since the latter also contains contributions from
channels without an RS counterpart, while the longitudinal components in the
shower are treated using the Dawson prescription. The imposed kinematic cuts
suppress the non-RS-related contributions, after which the RS result closely
follows the behaviour of the FO prediction, indicating that the splitting
kernels and kinematic reconstruction perform as expected.

Secondly, it is essential to examine the effects of incorporating the entire tree of interleaved EW shower on the resulting predictions. To this end, in Ref. \cite{Masouminia:2021kne}, we looked at the production of $W^\pm$ bosons in the presence of a high-$p_\perp$ leading jet from the LHC’s first run, as detailed in \cite{ATLAS:2016jbu}. Here, the explicit production of prompt $W^\pm$ bosons with one/two jets and inclusively was compared to a case where a pure QCD $2 \to 2$ hard process (with no explicit $W^\pm$ boson) is showered with the EW shower, allowing the EW boson to be sourced solely from the shower. Comparing the results showed an interesting resemblance between the explicit ME($pp \to W^\pm j j$) $\oplus$ QCD+QED shower and ME($pp \to j j$) $\oplus$ QCD+QED+EW shower. Further phenomenological studies were also conducted, explicitly targeting EW-sensitive observables in Refs. \cite{Darvishi:2021het} and \cite{Darvishi:2020paz}, both of which outlined the positive effects of incorporating the EW shower in increasing the precision of the predictions.
\begin{figure}[htp]
\centering
\capstart
\noindent\includegraphics[width=1.000\linewidth]{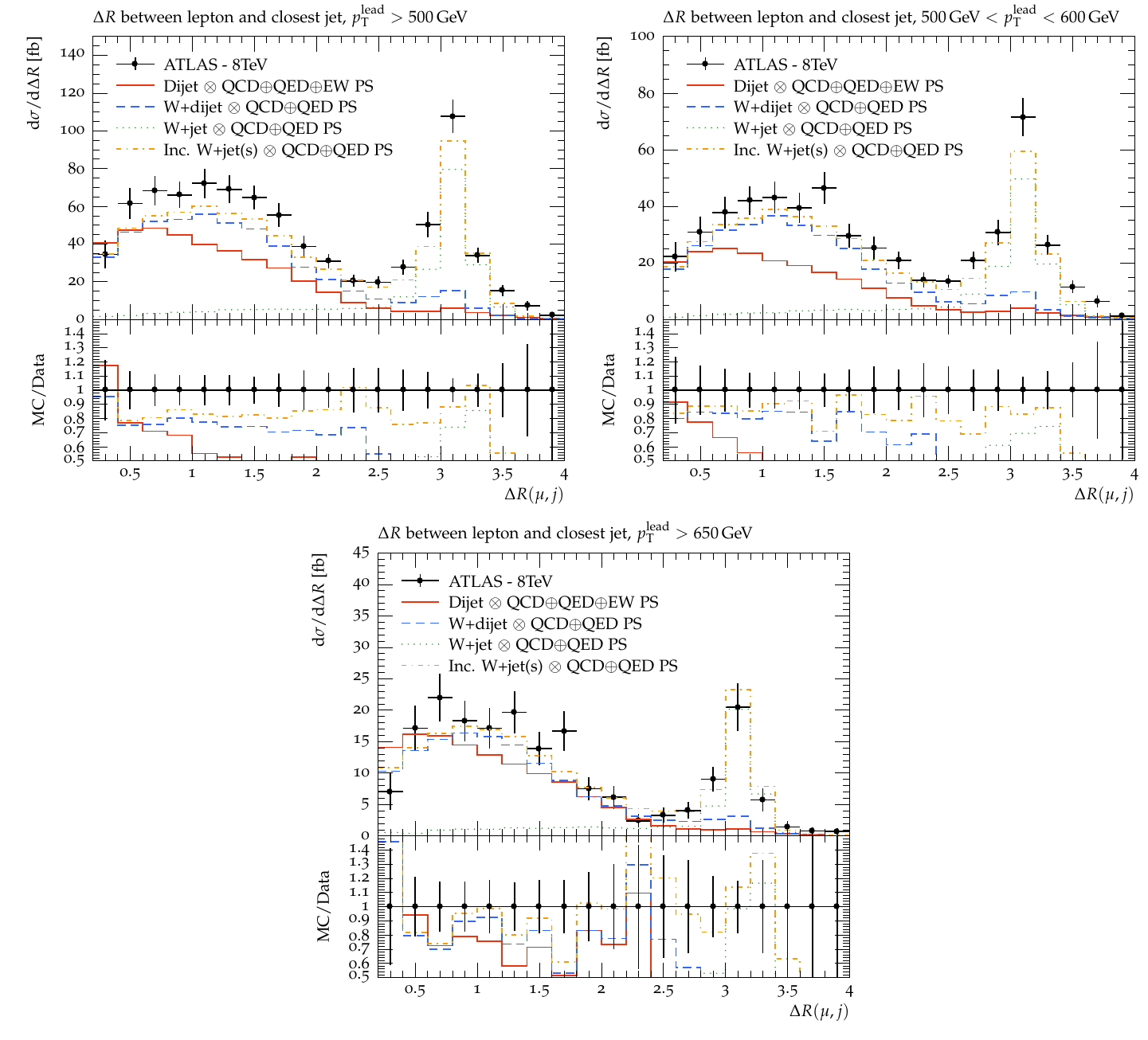}
\caption{Angular distribution of $W^{\pm}$ bosons with high transverse momentum jets at $\sqrt{s}=8$ TeV (ATLAS data \cite{ATLAS:2016jbu}). The dijet sample uses a leading-order hard process, with $W^\pm$ bosons generated by the EW shower without matching to fixed-order $W^\pm+$jets matrix elements.}
\label{\detokenize{review/showers/qtilde:id67}}\label{\detokenize{review/showers/qtilde:fig-ewphysiccheck}}\end{figure}
In Fig.~\ref{\detokenize{review/showers/qtilde:fig-ewphysiccheck}}, the
ME($pp\to jj$) $\oplus$ QCD+QED+EW-shower prediction contains no explicit
$W^\pm$ production in the hard process: the $W^\pm$ bosons are generated
entirely through EW shower emissions. It therefore begins from a less complete
description than the ME($pp\to W^\pm jj$) $\oplus$ QCD+QED-shower calculation.
Nevertheless, the shower-generated $W^\pm$ bosons produce a signature that
closely follows that of explicit $W^\pm$ production across the full kinematic
range, including configurations in which the $W^\pm$ boson and the leading
high-$p_\perp$ jet are nearly back-to-back. This agreement demonstrates that
the full interleaved EW shower reproduces the characteristic kinematic
behaviour of explicit weak-boson production.

\subsection{Dipole shower}
\label{\detokenize{review/showers/dipole:dipole-shower}}\label{\detokenize{review/showers/dipole:sect-dipole-shower}}\label{\detokenize{review/showers/dipole::doc}}

The dipole parton shower algorithm used in Herwig 7 \cite{Platzer:2009jq, Schumann:2007mg}:
\begin{itemize}
\item {} 

maintains exact energy-momentum conservation at each step of the evolution;

\item {} 

ensures colour coherence by careful treatment of the recoils;

\item {} 

uses the dipole splitting functions and kinematics of
\cite{Catani:1996vz, Catani:2002hc} to make matching to next-to-leading
order emissions easier;

\item {} 

has a sophisticated treatment of quark mass effects \cite{Cormier:2018tog};

\item {} 

includes spin correlation effects \cite{Richardson:2018pvo}.

\end{itemize}

In this section, we give a full description of the dipole shower algorithm.

\subsubsection{Structure of the evolution}
\label{\detokenize{review/showers/dipole:structure-of-the-evolution}}\label{\detokenize{review/showers/dipole:dipole-shower-evolution}}

The hard process is showered first, followed by the handling of any unstable
particles. Prior to showering the hard process, the initial conditions required
by the shower must be constructed as described in
\hyperref[\detokenize{review/showers/dipole:dipole-shower-initial-conditions}]{Section \ref{\detokenize{review/showers/dipole:dipole-shower-initial-conditions}}}.
This procedure arranges the external incoming and outgoing coloured particles
into a set of dipole chains, which are colour-singlet groups of particles arranged
into sequentially linked colour dipoles. Dipole chains are the independently evolving
structures in the dipole shower. Partons are emitted from a single parton in
a dipole, the emitter, and another parton, the spectator, is required to
absorb recoil to ensure energy-momentum conservation in each splitting.
Therefore we refer to ‘splittings from dipoles’ in our discussion of the dipole
shower.

It is implicit from the description above that each dipole has two possible
configurations. Either the coloured parton (a parton carrying
colour line, and being one member in the dipole) is the emitter and the anti-coloured parton (a parton carrying an anti-colour
line, being the other member in the dipole) 
is the spectator, or the emitter/spectator identifications are reversed.
For clarity, when we refer to a dipole we always refer to a dipole in a
specific configuration of emitter and spectator, unless we specify otherwise.
In the construction of the initial conditions, the emitter in each dipole
is assigned a starting scale from which to begin the shower evolution.
This scale, referred to as the hard scale of the emitter, is simply the
maximum allowed scale for an emission from the given dipole.

With the initial conditions set, the showering of the process can proceed.
The generation of splittings proceeds according to the veto algorithm,
as described in \hyperref[\detokenize{review/appendix/sudakov:sect-sudakov-solution}]{\ref{\detokenize{review/appendix/sudakov:sect-sudakov-solution}}}.

A splitting from a dipole is described by three variables. The first quantity
is the scale of the splitting which, following the arguments in
\cite{Platzer:2009jq}, is chosen to be the transverse momentum, $q_\perp$,
of the emission. The transverse momentum is the ordering variable in the dipole
shower, \textit{i.e.} subsequent emissions in each dipole chain decrease in this scale,
and it is generated according to the distribution defined by the Sudakov form
factor as defined in \hyperref[\detokenize{review/showers/general:sect-shower-basics}]{Section \ref{\detokenize{review/showers/general:sect-shower-basics}}}.
The form of the branching probabilities, $\mathrm{d}\mathcal{P}_{\mathrm{branching}}$,
for splittings in the dipole shower are given, for each type of dipole,
in \hyperref[\detokenize{review/showers/dipole:dipole-shower-kernels}]{Section \ref{\detokenize{review/showers/dipole:dipole-shower-kernels}}}.
The second variable required is a splitting variable, which we label
$z$, to parameterise the splitting. This variable is identified
in \hyperref[\detokenize{review/showers/dipole:dipole-shower-generalities}]{\ref{\detokenize{review/showers/dipole:dipole-shower-generalities}}}
and is generated according to the $z$ dependence of the corresponding
differential branching probability,
$\mathrm{d}\mathcal{P}_{\mathrm{branching}}$, at the generated value of $q_\perp$.
Finally, we also require the azimuthal angle, $\phi$, of the plane of the
splitting. The azimuthal angle is generated subject to the spin correlation algorithm.

As long as the set of dipole chains to be evolved is not empty the shower
proceeds by selecting a chain to evolve, referred to as the ‘current chain’.
All possible splittings from all of the dipoles in the current chain are
considered to be in competition, therefore all of these possible splittings
are trialled. Accordingly a value of the scale $q_\perp$ is generated
according to the Sudakov form factor for each possible splitting.

If no trial splittings are generated with $q_\perp$ above the infrared
cutoff, $\mu_\mathrm{IR}$, the evolution of the current chain is
terminated. In this case the current chain is removed from the set of chains
to be evolved and the shower algorithm continues with the next chain.

With $q_\perp$ and $z$ determined, a trial value of $\phi$ is sampled
uniformly in the interval $[0,2\pi)$. When spin correlations are enabled, this
trial value is accepted or rejected according to the azimuthal weight
calculated by the spin-correlation algorithm; otherwise, the uniformly sampled
value is retained. The momenta of the splitting products and spectator are then
calculated following the relevant formulation in
\hyperref[\detokenize{review/showers/dipole:dipole-shower-kinematics}]{Section \ref{\detokenize{review/showers/dipole:dipole-shower-kinematics}}}.

A splitting removes dipoles that contained the emitter and produces new
dipoles which contain the splitting products. The hard scale of the splitting
products in the dipoles in which they are emitters is set to $q_\perp$.
As described in detail in \hyperref[\detokenize{review/showers/dipole:dipole-shower-initial-conditions}]{Section \ref{\detokenize{review/showers/dipole:dipole-shower-initial-conditions}}}, if the
selected splitting was a $g \to q\bar{q}$ splitting the structure of
the current dipole chain changes due to the colour structure of this splitting.
In the case that the current dipole chain breaks up into two independent chains,
the newly produced chain is added to the list of chains to be evolved.

Finally the hard scale of the emitter in every dipole in the current chain, and
if applicable the newly created chain, is updated. Simply, for each dipole,
if $q_\perp$ is less than the current hard scale of the emitter in that
dipole the hard scale of the emitter is set to $q_\perp$.

The evolution of the current chain now continues by returning to the first step
in the showering algorithm for the current chain.
The showering of the hard process terminates once all of the dipole chains
have been evolved. The showered partons are now
‘reshuffled’ onto their respective constituent mass-shells,
as defined in \hyperref[\detokenize{review/index:sec-hadronization}]{Section \ref{\detokenize{review/index:sec-hadronization}}}, following the procedure
described in \hyperref[\detokenize{review/showers/dipole:dipole-shower-constituent-reshuffling}]{Section \ref{\detokenize{review/showers/dipole:dipole-shower-constituent-reshuffling}}}.

Following the showering of the hard
process, the dipole shower handles any particles recorded as being
unstable. The treatment of a two-body decay in the dipole shower is
straightforward.
The correct treatment of higher multiplicity decays in the dipole shower
currently requires that such processes take place via a decay tree,
\textit{i.e.} a sequence of two-body decays. For example it is common to perform
top quark decays as a $1\to 3$ process with an internal W-boson,
in order to correctly include off-shell effects for the W-boson.
In this case each $1\to 2$ decay is showered separately,
starting from the particle incoming to the decay tree and working
towards the end of the decay tree. To ensure the correct treatment in this
situation, for any given decay tree, we include only one of the $1\to 2$
decay processes in the record of unstable particles at any point in the shower.
The record is updated following the showering of each decay as described
later in this section.

As long as the record of unstable particles is not empty an
unstable particle is selected. If the selected particle does not have
any associated decay products it will be decayed.
We refer to the incoming decayed particle and the outgoing decay products as the ‘current decay’.

If the current decay contains coloured decay products and the decay
was performed at LO then the user has the option to include an NLO-corrected
first emission from the current decay prior to showering. This is done
using the built-in POWHEG decay corrections available in Herwig
\cite{Richardson:2013nfo}.
As for the hard process, the initial conditions must be constructed for the
current decay prior to showering.
This procedure is described in \hyperref[\detokenize{review/showers/dipole:dipole-shower-initial-conditions}]{Section \ref{\detokenize{review/showers/dipole:dipole-shower-initial-conditions}}}
and again produces a set of dipole chains and sets the starting hard scale
for the emitter in each dipole.
The dipole chains are evolved following the same procedure described above
for the hard process.
Once there are no dipole chains remaining for the current decay, the
partons outgoing from the decay are reshuffled as described in
\hyperref[\detokenize{review/showers/dipole:dipole-shower-constituent-reshuffling}]{Section \ref{\detokenize{review/showers/dipole:dipole-shower-constituent-reshuffling}}}.

Alternatively, if there are no coloured decay products, the shower simply
attempts to produce photon radiation from the current decay according
to the YFS approach \cite{Hamilton:2006xz}, see \hyperref[\detokenize{review/showers/qed:sect-yfs}]{Section \ref{\detokenize{review/showers/qed:sect-yfs}}}.

In contrast to the treatment of the hard process, following the treatment of
a decay process there are additional considerations that must be made to
ensure that any unstable decay products are treated correctly.
There are three possible cases that must be considered.
\begin{itemize}
\item {} 

The simplest case is that the decay products in the current decay are stable,
for example, the decay of a W-boson into two stable quarks. In this case no
further considerations are necessary.

\item {} 

The second case is that one or more of the decay products in the current decay
are unstable but do not have any decay products, for example, the W-boson
produced in a two-body top quark decay. In this case the unstable outgoing
particles are simply added to the record of unstable particles.

\item {} 

The final case is that one or more of the decay products in the current decay
are unstable and already have decay products. For instance, the W-boson produced
in a three-body top quark decay serves as our example.
The momenta of the W-boson and its decay products were originally set in the
three-body decay of the top quark, however the momentum of the W-boson can
change during the treatment of the current decay, \textit{i.e.} the top-bottom-W system.
This change can occur through both the treatment of recoil in shower splittings,
see \hyperref[\detokenize{review/showers/dipole:dipole-shower-kin-decay}]{Section \ref{\detokenize{review/showers/dipole:dipole-shower-kin-decay}}}, and in the reshuffling procedure.
Therefore the momenta of the decay products of the W-boson are updated, through
a straightforward boost, to ensure energy-momentum conservation in the W-boson
decay.
This procedure is implemented for decay trees of any length, such
that, in general, the momenta of all outgoing particles further down the decay
tree are updated, as required, following the treatment of the current decay.
Any unstable decay products in the current decay process, e.g.\ the W-boson
in the example above, are then added to the record of unstable particles.

\end{itemize}

The current decay is now removed from the record of unstable particles and the
algorithm continues by selecting the next particle in the record.

\subsubsection{Initial conditions}
\label{\detokenize{review/showers/dipole:initial-conditions}}\label{\detokenize{review/showers/dipole:dipole-shower-initial-conditions}}

As described in \hyperref[\detokenize{review/showers/dipole:dipole-shower-evolution}]{Section \ref{\detokenize{review/showers/dipole:dipole-shower-evolution}}}, prior to performing
any emissions in the dipole shower, we must compute the initial conditions
required by the shower for the given process. These initial conditions are the
colour structure of the process and the initial starting scales for the shower
evolution.

The procedure for the construction of the required colour structure is
identical for hard processes, secondary processes and decay processes.
In the first step the process is assigned colour-flow information in the
large-$N_C$ limit, which we use to sort the \textit{external} coloured partons
incoming to and outgoing from the process into colour singlets. To achieve this,
we note that a colour singlet is ‘simply connected’,  that is any parton in a
colour singlet can be reached  from another parton in the same singlet by following
colour lines and changing  from a colour line to an anti-colour line when
an external gluon is encountered.

Finally, the partons in each colour singlet are sorted into a sequence in which
colour-connected partons are located in neighbouring positions.
In other words each of the colour singlets is sorted into a series of colour
dipoles. We refer to these sorted colour singlet sequences as ‘dipole chains’. Dipole chains can be ‘circular’ or ‘non-circular’. A dipole chain is called
non-circular if there exists a circular permutation of the partons in it
such that the partons in the first and last positions are not colour-connected.
A dipole chain is called circular if no such permutation exists.
We note that the structure of a given dipole chain is changed in the case of a
$g \to q\bar{q}$ splitting in that chain during showering.
If the dipole chain was circular prior to the splitting, it becomes non-circular
following the spitting. If the current dipole chain was non-circular prior to
the splitting then it breaks up into two independent chains.

As discussed in \hyperref[\detokenize{review/showers/dipole:dipole-shower-evolution}]{Section \ref{\detokenize{review/showers/dipole:dipole-shower-evolution}}}, a dipole consists
of two partons in a given configuration in which one parton is assigned the
role of ‘emitter’ and the other is assigned the role of ‘spectator’. It is
the emitter in a given dipole that can split to produce a shower emission.
Prior to showering, each emitter needs to be assigned an initial scale which
defines the maximum allowed scale for an emission from the given dipole.

The calculation of the initial shower scale for each emitter differs for hard
processes, secondary processes and decay processes. The treatment of hard
processes and secondary processes roughly follow the same procedure.
The maximum possible allowed scale for an emission from a given hard or secondary
process is the partonic centre-of-collision energy of the process, however
there are several cases when a more restrictive maximum emission scale is required.
One option, which is only applicable to hard processes, is to set this maximum
scale to some fixed scale or dynamical scale associated with the hard process,
for example the factorisation scale calculated for the hard process.
Alternatively, if this is not the required behaviour for a given hard process
or if a secondary process is being considered,
a more restrictive scale can be constructed by considering the particles
outgoing from the process. In this case there are two possible cases.
If there are outgoing coloured particles the maximum scale is set to the minimum
transverse mass of the outgoing coloured particles. If there are no coloured
outgoing particles the scale is set to the invariant mass of the system of
outgoing particles.

There are additional scales associated with each emitter that must also be taken
into account. For a given dipole we can calculate the maximum physically
allowed splitting scale for each possible splitting from the dipole.
By comparing the maximum scale for each of the possible splittings we determine
the maximum physically possible scale for an emission from the emitter in the
given dipole.

In addition a ‘veto scale’ can be set for any given particle. If the veto
scale is set for at least one parton in a given dipole chain, the smallest
veto scale of all of the particles in that dipole chain is identified.

For every emitter there are now three relevant scales which we compare:
the maximum allowed scale for emissions from the process, the maximum
physically allowed splitting scale for the given emitter and the
minimum veto scale of all the partons in the dipole chain to which the emitter
belongs.
For each emitter we compare these three scales and set the starting scale
to the smallest of the three.
This is done for the emitter in every dipole in every dipole chain,
\textit{i.e.} every parton that can produce an emission is assigned a scale.

The computation of the initial scales in a decay process is more straightforward.
A decay process can be presented to the shower with or without an NLO-corrected
first emission. In a LO decay process, \textit{i.e.} without an NLO-corrected first emission,
the starting scale for each emitter in the set of dipole chains is set to the
mass of the decayed particle incoming to the process.
Otherwise, if an NLO emission is included, the starting scale for each emitter is
set to the scale of this emission.
In both cases if the maximum physically allowed emission scale for a given emitter
is smaller than the scale stated above, the starting scale for this emitter is
instead set to this maximum scale.

\subsubsection{Shower kinematics and splitting kernels}
\label{\detokenize{review/showers/dipole:shower-kinematics-and-splitting-kernels}}\label{\detokenize{review/showers/dipole:dipole-shower-kinematics-kernels}}

In this section, we present the splitting kernels and the kinematics formulation
used to describe splittings in the dipole shower. The dipole shower for massless
partons has previously been documented in Ref. \cite{Platzer:2009jq}.
We provide the full kinematics formulae here for completeness, incorporating changes
made since the original implementation. The original treatment of massive
quarks in the dipole shower described in Ref. \cite{Ellis:1991qj}, based on
Ref. \cite{Schumann:2007mg}, has been replaced in
Herwig 7.1 by an improved description \cite{Cormier:2018tog} which we present here.
In addition, spin correlations are now included so that the
azimuthal angles are calculated as described in Ref. \cite{Richardson:2018pvo}.
We note that the dipole shower currently does not support dipoles with massive incoming partons.

\paragraph{Notation}
\label{\detokenize{review/showers/dipole:notation}}\label{\detokenize{review/showers/dipole:dipole-shower-notation}}

Throughout the following sections we use a consistent notation to denote
partons and their properties according to their identification as
an emitter, an emission or a spectator and whether they are incoming to, or
outgoing from a dipole. We use the letters $\{a,b\}$ and ${i,j,k}$ to
denote partons incoming to and outgoing from a dipole respectively.
The partons in a dipole prior to a splitting are indicated by a ‘tilde’ and
their momenta are denoted by the letter $p$. The momenta of partons
following a splitting are denoted by the letter $q$. Parton masses
are denoted by the letter $m$.
For clarity we make this notation explicit below:
\begin{itemize}
\item {} 

Prior to splitting:
\begin{itemize}
\item {} 

Outgoing emitter - $\widetilde{ij} \text{, } \tilde{p}_{ij} \text{, } m_{ij}$;
Incoming emitter - $\widetilde{aj} \text{, } \tilde{p}_{aj} \text{, } m_{aj}$;

\item {} 

Outgoing spectator - $\tilde{k} \text{, } \tilde{p}_{k} \text{, } m_{k}$;
Incoming spectator - $\tilde{b} \text{, } \tilde{p}_{b} \text{, } m_{b}$;

\end{itemize}

\item {} 

Following a splitting:
\begin{itemize}
\item {} 

Outgoing emitter - $i \text{, } q_i \text{, } m_{i}$;
Incoming emitter - $a \text{, } q_a \text{, } m_{a}$;

\item {} 

Outgoing spectator - $k \text{, } q_k \text{, } m_{k}$;
Incoming spectator - $b \text{, } q_b \text{, } m_{b}$;

\item {} 

Outgoing emission - $j \text{, } q_j \text{, } m_{j}$;

\end{itemize}

\end{itemize}

where the incoming parton following an emission refers to that incoming
from the beam particle.

The reader should note that, in order to be as general as possible, we have
included the masses of incoming partons in the above list. In all cases where
such a parton is incoming to the hard process these masses are necessarily zero
and are not included in the following. It is also important to note that the
mass of the spectator is necessarily conserved in any given splitting.

Dipole configurations are written in the format
$\text{state}_\text{emitter} - \text{state}_\text{spectator}$ where
$\text{state}_\text{parton}$ is the state, initial or final, of that
parton. There are five dipole configurations included in the dipole shower,
final-final dipoles, final-initial dipoles, initial-final dipoles,
initial-initial dipoles and final-initial decay-dipoles which include
an incoming decayed coloured particle.

\paragraph{Generalities}
\label{\detokenize{review/showers/dipole:generalities}}\label{\detokenize{review/showers/dipole:dipole-shower-generalities}}

The splitting kernels \cite{Catani:1996vz, Catani:2002hc} for each dipole are
parameterized in terms of two splitting variables, or just one in the case
of the initial-initial dipole,
and the definition of these variables depends upon the dipole.
In Herwig, we generate the transverse momentum, $p_\perp$, and the
light-cone momentum fraction, $z$. These variables follow the quasi-collinear
Sudakov parameterization of splitting momenta, as described in \hyperref[\detokenize{review/showers/qtilde:sub-shower-kinematics}]{Section \ref{\detokenize{review/showers/qtilde:sub-shower-kinematics}}},
which we summarize below.
In order to compute the splitting kernels for each dipole, we must write the
splitting variables in terms of the generated variables. We provide
the expressions required to do this in each of the following sections.

In the Sudakov parameterization we choose a light-like vector $n$ to
define the collinear direction of the splitting. In a splitting from a
final-state emitter we write the momentum of the emitted parton,
the ‘emission’, as
\begin{equation}\label{equation:review/showers/dipole:eqn:DS:qjforward}
\begin{split}q_j = (1-z)\tilde{p}_{ij} + \frac{m_j^2 - (1-z)^2 m_{ij}^2 + p_\perp^2}
{2 \tilde{p}_{ij}\cdot n (1-z)} n - k_\perp \ ,\end{split}
\end{equation}

where $k_\perp$ is a space-like vector which satisfies
$k_\perp^2 = -p_\perp^2$ and
$k_\perp \cdot \tilde{p}_{ij} = k_\perp \cdot n = 0$.
Similarly, for a splitting from a massless incoming parton, we write:

\begin{equation}\label{equation:review/showers/dipole:eqn:DS:qjbackward}
\begin{split}q_j = (1-z)q_a + \frac{p_\perp^2}
{2 q_a\cdot n (1-z)} n - k_\perp \ .\end{split}
\end{equation}

Note that our assumption that $m_{aj}=m_{a}=0$ implies also that
$m_{j}=0$ in all relevant cases.

\paragraph{Dipole splitting kinematics}
\label{\detokenize{review/showers/dipole:dipole-splitting-kinematics}}\label{\detokenize{review/showers/dipole:dipole-shower-kinematics}}

In each of the following sections, we first state the definition of the splitting
variables in terms of the physical parton momenta following the splitting. We
also state the momentum-quantity that is conserved in a splitting from the given
dipole. In the original specification of the kernels, an expression of
the parton momenta prior to the splitting is given, for each dipole,
in terms of the momenta following the splitting. Our formulation of the
parton momenta following a splitting must satisfy these expressions.
We present a set of expressions for the parton momenta following a splitting
in terms of either the splitting variables or the generated variables, as
appropriate.

We provide expressions for the splitting variables in terms of the
generated variables and expressions for the limits on both the splitting
variables and the generated variables. Unless otherwise stated, the limits
on the splitting variables are the integration limits given in
Refs. \cite{Catani:1996vz, Catani:2002hc}.
Finally, we describe the Jacobian factors required to express the one-particle
phase-space, given in terms of the splitting variables in
Refs. \cite{Catani:1996vz, Catani:2002hc}, in terms of the generated variables
in Herwig. The one-particle phase-space expressions, and hence the Jacobian terms,
are required to compute the branching probabilities, given in their general form
in \hyperref[\detokenize{review/showers/dipole:dipole-shower-kernels}]{Section \ref{\detokenize{review/showers/dipole:dipole-shower-kernels}}}.

We refer to dipoles which contain no massive partons as
‘massless dipoles’ and dipoles which contain at least one massive
parton as ‘massive dipoles’.

We will present the results for the massive dipoles here, except
for the case of initial-initial dipoles, where there is only a massless dipole which we therefore include,
and present the result for the massless dipoles in \hyperref[\detokenize{review/appendix/masslessDipole:sect-massless-dipoles}]{\ref{\detokenize{review/appendix/masslessDipole:sect-massless-dipoles}}}.
In all cases, the massless dipole results can be recovered as the massless limit
of the massive case. However, in Herwig, we implement them separately to avoid
unnecessary computations.

\subparagraph{Final-final dipole kinematics}
\label{\detokenize{review/showers/dipole:final-final-dipole-kinematics}}\label{\detokenize{review/showers/dipole:dipole-shower-kin-ff}}

The final-final dipole splitting kernels,
given in \hyperref[\detokenize{review/showers/dipole:dipole-shower-kernels-ff}]{Section \ref{\detokenize{review/showers/dipole:dipole-shower-kernels-ff}}},
are written in terms of the splitting variables $z_i$ and
$y_{ij,k}$, defined in terms of the physical momenta as
\begin{equation*}
\begin{split}z_i =\frac{q_i \cdot q_k}{q_i \cdot q_k + q_j \cdot q_k} \ ,\end{split}
\end{equation*}\begin{equation}\label{equation:review/showers/dipole:eqn:DS:FFKin:y}
\begin{split}y_{ij,k} = \frac{q_i \cdot q_j}{q_i \cdot q_j + q_i \cdot q_k + q_j \cdot q_k} \ .\end{split}
\end{equation}
We define the conserved dipole momentum, $Q$, and its
self-product, $s$, as
\begin{equation}\label{equation:review/showers/dipole:eqn:DS:FFKin:Q}
\begin{split}Q = \tilde{p}_{ij} + \tilde{p}_{k} = q_i + q_j + q_k \ ,\end{split}
\end{equation}\begin{equation*}
\begin{split}s = Q^2 \ .\end{split}
\end{equation*}

The final-final dipole with a massive spectator presents a particular challenge.
While it is possible to construct expressions for the physical momenta following
the splitting in terms of the splitting variables and the momenta prior to the
splitting, the expressions become convoluted and a simpler
formulation is preferable.

We present a more convenient formulation of the splitting momenta
based on the standard quasi-collinear formulation, however we replace
$\tilde{p}_{ij}$ in Eq. \eqref{equation:review/showers/dipole:eqn:DS:qjforward} with an alternative choice.
We continue to use $z$ to denote the generated variable
although strictly speaking its definition in terms of invariant
products of momenta changes according to the replacement of
$\tilde{p}_{ij}$.

We first introduce the light-like vectors $n_{ij} \text{ and } n_k$,
\begin{equation*}
\begin{split}2 n_{ij}\cdot n_k \equiv s_{ij,k},\end{split}
\end{equation*}

where $s_{ij,k}$ is an invariant quantity dependent upon the
choice of $n_{ij}$ and $n_k$.
We define these vectors in terms of the emitter and spectator momenta as
\begin{equation*}
\begin{split}n_{ij} &= \frac{s_{ij,k}^2}{s_{ij,k}^2- m_{ij}^2 m_k^2}
\left( p_{ij} - \frac{m_{ij}^2}{s_{ij,k}}p_k \right) \ ,
\\
n_k &= \frac{s_{ij,k}^2}{s_{ij,k}^2- m_{ij}^2 m_k^2}
\left( p_{k} - \frac{m_{k}^2}{s_{ij,k}}p_{ij} \right) \ ,\end{split}
\end{equation*}

and solve to find an explicit expression
\begin{equation*}
\begin{split}s_{ij,k} = \frac{1}{2} \left(Q^2-m_{ij}^2-m_k^2
+ \sqrt{(Q^2-m_{ij}^2-m_k^2)^2-4 m_{ij}^2 m_k^2}\right) \ .\end{split}
\end{equation*}

In addition, we introduce two scaling variables, $x_{ij}$ and $x_k$,
and introduce the scaled four-momenta
\begin{equation}\label{equation:review/showers/dipole:eqn:DS:FFKin:Massqk}
\begin{split}q_{ij} &= x_{ij} n_{ij} + \frac{m_{ij}^2}{x_{ij}s_{ij,k}} n_k \ , \\
q_k &= x_k n_k + \frac{m_k^2}{x_k s_{ij,k}} n_{ij} \ ,\end{split}
\end{equation}

where, following our notation, $q_k$ is the momentum of the spectator
following the splitting. The momenta $q_{ij}$ and $n_k$ are now
inserted into
the quasi-collinear Sudakov parameterization to obtain expressions for the
parton momenta following the splitting
\begin{equation}\label{equation:review/showers/dipole:eqn:DS:FFKin:MassNewMom}
\begin{split}q_i & = z \ q_{ij} + \frac{m_i^2-{z}^2 m_{ij}^2 - k_\perp^2 }
{2q_{ij}\cdot n_k\ z} n_k + k_\perp,
\\
q_j & = (1-z)\ q_{ij} + \frac{m_j^2-(1-z)^2 m_{ij}^2 - k_\perp^2 }
{2q_{ij}\cdot n_k\ (1-z)} n_k - k_\perp .\end{split}
\end{equation}

In order to implement this formulation we require explicit expressions for the
scaling variables $x_{ij}$ and $x_k$. For convenience we first
introduce the scaled propagator invariant
\begin{equation*}
\begin{split}w = \frac{1}{x_{ij}\ z(1-z) s_{ij,k} }
\left[ p_\perp^2 + z m_j^2 + (1-z) m_i^2 - z(1-z) m_{ij}^2  \right]
= \frac{(q_i+q_j)^2 -m_{ij}^2}{2q_{ij}\cdot n_k} \ ,\end{split}
\end{equation*}

and define the variables
\begin{equation*}
\begin{split}\lambda_k = 1 + \frac{m_k^2}{s_{ij,k}} \ ,
\qquad \lambda_{ij} = 1 + \frac{m_{ij}^2}{s_{ij,k}} \ .\end{split}
\end{equation*}

From the momentum-conservation requirement in Eq. \eqref{equation:review/showers/dipole:eqn:DS:FFKin:Q}
\begin{align*}
   x_{ij} &= 1-\frac{m_k^2}{s_{ij,k}}\frac{(1-x_k)}{x_k} \ ,
   \\
   x_k &= \frac{1}{2\lambda_k}
   \left[
   \left( \lambda_k + \frac{m_k^2}{s_{ij,k}}\lambda_{ij} - x_{ij} w \right)
   \pm \sqrt{ \left(\lambda_k + \frac{m_k^2}{s_{ij,k}} \lambda_{ij}
   - x_{ij} w \right)^2
   - 4 \lambda_k \lambda_{ij} \frac{m_k^2}{s_{ij,k}} } \right] \ .
\end{align*}

With these, the outgoing momenta following a splitting can be computed,
given the two generated variables $z$ and $p_\perp^2$.

Following directly from its definition in Eq. \eqref{equation:review/showers/dipole:eqn:DS:FFKin:y},
$y_{ij,k}$ can be written in terms of $z$ and $p_\perp$
\begin{eqnarray}
   y_{ij,k} &=& \frac{1}{\bar{s}}
   \left[ x_{ij} s_{ij,k} w + m_{ij}^2 - m_i^2 - m_j^2 \right] \ ,\nonumber
   \\
   &=& \frac{1}{\bar{s} z (1-z )}
   \left[ p_\perp^2 + (1-z )^2 m_i^2 + {z}^2 m_j^2 \right] \ ,
\end{eqnarray}

where $\bar{s} = s - m_i^2 - m_j^2 - m_k^2$. Given this
expression for $y_{ij,k}$ we obtain an expression for $z_i$ by solving
\begin{equation*}
\begin{split}z_i = \frac{1}{(1-y_{ij,k})\bar{s}} 2 q_i \cdot q_k \ ,\end{split}
\end{equation*}

to give
\begin{equation*}
\begin{split}z_i = \frac{1}{(1-y_{ij,k})\bar{s}}
\left[ x_{ij} x_k s_{ij,k} z +
\frac{m_k^2}{x_{ij} x_k s_{ij,k} z}
\left ( p_\perp^2 + m_i^2 \right) \right] \ ,\end{split}
\end{equation*}

The phase-space limits on the splitting variables are
\begin{align*}
   y_- &= \frac{2 m_i m_j}{s - m_i^2 - m_j^2 - m_k^2} \ ,
   \\
   y_+ &= 1 - \frac{2 m_k(\sqrt{s}-m_k)}{s - m_i^2 - m_j^2 - m_k^2} \ ,
   \\
   z_{i,\pm}(y_{ij,k}) &=
   \frac{ 2 m_i^2 + \left(s - m_i^2 - m_j^2 - m_k^2 \right) y_{ij,k}}
   {2 \left[ m_i^2 + m_j^2
   + \left(s - m_i^2 - m_j^2 - m_k^2 \right) y_{ij,k} \right] }
   \times ( 1 \pm v_{ij,i}v_{ij,k}),
\end{align*}

where the relative velocities $v_{ij,k}$ and $v_{ij,i}$ are
\begin{align*}
   v_{ij,k} &= \frac{ \sqrt{ \left[ 2m_k^2 +
   \left(s - m_i^2 - m_j^2 - m_k^2\right)(1-y_{ij,k}) \right]^2 - 4m_k^2 }}
   {\left(s - m_i^2 - m_j^2 - m_k^2\right)(1-y_{ij,k})} \ ,
   \\
   v_{ij,i} &= \frac{\sqrt{ \left(s - m_i^2 - m_j^2 - m_k^2\right)^2 y_{ij,k}^2
   - 4m_i^2 m_j^2 }}{\left(s - m_i^2 - m_j^2 - m_k^2\right) y_{ij,k} + 2m_i^2} \ .
\end{align*}
Working in a frame where $\vec{Q} = 0$, the kinematic upper
limit on the transverse momentum, $p_{\perp,\mathrm{max}}$, is simply the
magnitude of the emission 3-momentum in the limit that the spectator following
the splitting has zero momentum. The momentum conservation requirement in
Eq. \eqref{equation:review/showers/dipole:eqn:DS:FFKin:Q} can then be rearranged to give
\begin{equation*}
\begin{split}p_{\perp,\mathrm{max}}^2 = \frac{\lambda \left(m_i^2, m_j^2, (\sqrt{s}-m_k)^2 \right)}
{4(\sqrt{s}-m_k)^2}  \ ,\end{split}
\end{equation*}

and the limits on $z$ follow from $y_{ij,k} < y_+$, giving
\begin{equation*}
\begin{split}z_\pm = \frac{1}{2 \left( \sqrt{s}-m_k \right)^2}
\left[ m_i^2 - m_j^2 + (\sqrt{s}-m_k)^2
\vphantom{\sqrt{ 1 - \frac{p_\perp^2}{\left(p_{\perp,\max} \right)^2}}}
\pm \sqrt{  \lambda \left(m_i^2, m_j^2, (\sqrt{s}-m_k)^2 \right) }
\sqrt{ 1 - \frac{p_\perp^2}{p_{\perp,\max}^2} } \right] \ .\end{split}
\end{equation*}

In the case of a massive dipole we require
\begin{equation}\label{equation:review/showers/dipole:eqn:DS:FFKin:MassTildeKin}
\begin{split}\tilde{p}_k &= \frac{ \sqrt{ \lambda \left(s, m_{ij}^2, m_k^2 \right) } }
{ \sqrt{ \lambda \left(s, (q_i+q_j)^2, m_k^2 \right) }}
\left( q_k - \frac{Q\cdot q_k}{s}Q \right)
+ \frac{s + m_k^2 - m_{ij}^2}{2s}Q \ ,
\\
\tilde{p}_{ij} &= Q - \tilde{p}_k \ .\end{split}
\end{equation}

In the massless case it is simple to show explicitly that the  corresponding requirements in
Eq. \eqref{equation:review/appendix/masslessDipole:eqn:DS:FFKin:LightTildeKin} are satisfied by the expressions in
Eq. \eqref{equation:review/appendix/masslessDipole:eqn:DS:FFKin:LightNewMom}.
It would be an involved and complex procedure to show explicitly that the
expressions in Eq. \eqref{equation:review/showers/dipole:eqn:DS:FFKin:MassNewMom} and
Eq. \eqref{equation:review/showers/dipole:eqn:DS:FFKin:Massqk} satisfy this requirement.
It is, however, straightforward to show that momentum conservation
requires that the four-momenta of the partons following a splitting are
parameterized by three variables. One of these variables is the azimuthal angle
of the splitting which is generated as described in Ref. \cite{Richardson:2018pvo}.
Therefore for a given $z_i$ and $y_{ij,k}$ the momenta are fully
constrained. As a cross-check, we have also verified numerically that the
requirement in Eq. \eqref{equation:review/showers/dipole:eqn:DS:FFKin:MassTildeKin} is satisfied by the expressions
in Eq. \eqref{equation:review/showers/dipole:eqn:DS:FFKin:MassNewMom} and Eq. \eqref{equation:review/showers/dipole:eqn:DS:FFKin:Massqk}.

\subparagraph{Single-particle phase-space}
\label{\detokenize{review/showers/dipole:single-particle-phase-space}}

The single-particle phase-space can be expressed in terms of $z_i$ and
$y_{ij,k}$
\begin{equation*}
\begin{split}\mathrm{d}q_j = \frac{1}{16\pi^2}
\frac{ \bar{s}^2}{\sqrt{\lambda\left(s, m_{ij}^2, m_{k}^2 \right)}}
\left(1-y_{ij,k} \right) \mathrm{d}y_{ij,k} \mathrm{d}z_i \frac{\mathrm{d}\phi}{2\pi} \ .\end{split}
\end{equation*}

In order to write this in terms of our generated variables, we need the
Jacobian $J_{p_\perp^2, z \rightarrow z_i,y_{ij,k}}$, such that,
\begin{equation*}
\begin{split}\mathrm{d}y_{ij,k} \mathrm{d}z_i
= \left( J_{p_\perp^2, z \rightarrow z_i,y_{ij,k}} \right)
p_\perp^2 \frac{\mathrm{d} p_\perp^2}{p_\perp^2}
\mathrm{d}z \ .\end{split}
\end{equation*}

This is in general given by
\begin{equation*}
\begin{split}J_{p_\perp^2, z \rightarrow z_i,y_{ij,k}} =
\left\lvert
\frac{\partial z_i}{\partial {p_\perp^2}}
\frac{\partial{y_{ij,k}}}{\partial z}
-
\frac{\partial z_i}{\partial z}
\frac{\partial y_{ij,k}}{\partial {p_\perp^2}}
\right\rvert \ ,\end{split}
\end{equation*}

which can be written as
\begin{equation*}
\begin{split}\frac{\mathrm{d}y_{ij,k}}{y_{ij,k}} \mathrm{d}z_i =
\left[ \frac{p_\perp^2}{p_\perp^2 + (1-z)^2 m_i^2 + z^2 m_j^2 } \right]
\left[ 1  -  2\frac{1}{\bar{s}(1-y_{ij,k})}\frac{m_k^2 Q^2}{x_{ij} x_k s_{ij,k}} \right]
\frac{\mathrm{d} p_\perp^2}{p_\perp^2} \mathrm{d}z \ .\end{split}
\end{equation*}

\subparagraph{Final-initial dipole kinematics}
\label{\detokenize{review/showers/dipole:final-initial-dipole-kinematics}}\label{\detokenize{review/showers/dipole:dipole-shower-kinematics-fi}}

The final-initial dipole splitting kernels,
given in \hyperref[\detokenize{review/showers/dipole:dipole-shower-kernels-fi}]{Section \ref{\detokenize{review/showers/dipole:dipole-shower-kernels-fi}}},
are written in terms of the splitting variables $z_i$ and $x_{ij,b}$,
defined in terms of the physical momenta as
\begin{equation*}
\begin{split}z_i = \frac{q_i \cdot q_b}{q_i \cdot q_b + q_j \cdot q_b} \ ,\end{split}
\end{equation*}
\begin{equation}\label{equation:review/showers/dipole:eqn:DS:FIKin:x}
\begin{split}x_{ij,b} = \frac{q_i \cdot q_b + q_j \cdot q_b - q_i \cdot q_j
+ \frac{1}{2} \left( m_{ij}^2 - m_i^2 - m_j^2 \right)}
{q_i \cdot q_b + q_j \cdot q_b } \ .\end{split}
\end{equation}

In addition, we define the conserved momentum transfer
\begin{equation*}
\begin{split}Q = \tilde{p}_{ij} - \tilde{p}_{b} = q_i + q_j - q_b \ ,\end{split}
\end{equation*}

and the invariant
\begin{equation*}
\begin{split}s_{ij,b} = 2 \tilde{p}_{ij} \cdot \tilde{p}_b \ .\end{split}
\end{equation*}

In both the massless and massive case,
we require that the dipole momenta can be written in terms of the momenta
following the splitting
\begin{align*}
   \tilde{p}_b &= x_{ij,b} q_b \ ,
   \\
   \tilde{p}_{ij} &= q_i + q_j - (1-x_{ij,b}) q_b \ .
\end{align*}

The incoming spectator is necessarily massless, therefore
it is most straightforward to write the new particle momenta in the
quasi-collinear Sudakov parameterization, using the generated variables.
It is straightforward in both cases to rewrite the results in terms of the
splitting variables, as we have done for the other dipole configurations.

Using the quasi-collinear Sudakov
parameterization, the physical momenta following the
splitting are
\begin{equation*}
\begin{split}q_i & =  z\tilde{p}_{ij}
+ \frac{m_i^2 - z^2 m_{ij}^2 + p_\perp^2}{ s_{ij,b} z} \tilde{p}_b
+ k_\perp \ ,
\\
q_j & =  (1-z)\tilde{p}_{ij} +
\frac{m_j^2-(1-z)^2 m_{ij}^2 + p_\perp^2}{ s_{ij,b} (1-z)} \tilde{p}_b
- k_\perp \ ,
\\
q_b & =  \frac{1}{x_{ij,b}}\tilde{p}_b \ .\end{split}
\end{equation*}

As the spectator is massless we have $z_i = z$, and we can write
$x_{ij,b}$ in terms of the generated variables as,
\begin{equation*}
\begin{split}x_{ij,b} = \left[ 1 + \frac{p_\perp^2 + (1-z) m_i^2 + z m_j^2
- z(1-z) m_{ij}^2}{ s_{ij,b} z(1-z)}
\right]^{-1} \ .\end{split}
\end{equation*}

We can derive a lower limit on the spectator momentum fraction $x_{ij,b}$.
We first denote the
momentum of the incoming proton as $P$ and the proton momentum-fraction
carried by the spectator prior to the splitting as $x_s$. We can write
\begin{equation*}
\begin{split}q_b = \frac{1}{x_{ij,b}} \tilde{p}_{b}
= \frac{1}{x_{ij,b}} \left( x_s P \right) < P \ ,\end{split}
\end{equation*}

we therefore have the requirement
\begin{equation}\label{equation:review/showers/dipole:eqn:DS:FI:xlim}
\begin{split}x_{ij,b} > x_s \ .\end{split}
\end{equation}

The phase-space limits on the splitting variables $z_i$ and
$x_{ij,b}$ are
\begin{equation*}
\begin{split}z_{i,\pm}(x_{ij,b}) &= \frac{ s_{ij,b}/x_{ij,b} - s_{ij,b}
+ m_{ij}^2 + m_i^2 - m_j^2
\pm \sqrt{
\left( s_{ij,b}/x_{ij,b} - s_{ij,b} + m_{ij}^2 - m_i^2 - m_j^2 \right)^2
- 4 m_i^2 m_j^2 } }
{ 2\left( s_{ij,b}/x_{ij,b} - s_{ij,b} + m_{ij}^2 \right) } \ ,
\\
x_- &= x_s \ ,
\\
x_+ &= \frac{ s_{ij,b} }{ s_{ij,b} - m_{ij}^2 + (m_i + m_j)^2 } \ ,\end{split}
\end{equation*}

where $x_-$ is simply the limit in \eqref{equation:review/showers/dipole:eqn:DS:FI:xlim}.
Following from the inequality in Eq. \eqref{equation:review/showers/dipole:eqn:DS:FI:xlim}
we obtain the following limits on $z$ and $p_\perp$
\begin{equation*}
\begin{split}p_{\perp,\mathrm{max}}^2 =
\frac{s_{ij,b}^\prime}{4} \lambda
\left(1, \frac{m_i^2}{s_{ij,b}^\prime}, \frac{m_j^2}{s_{ij,b}^\prime} \right) \ ,\end{split}
\end{equation*}\begin{equation*}
\begin{split}z_\pm = \frac{1}{2} \left[ 1 + \frac{m_i^2 - m_j^2}{s_{ij,b}^\prime}
\pm \sqrt{  \lambda
\left(1, \frac{m_i^2}{s_{ij,b}^\prime}, \frac{m_j^2}{s_{ij,b}^\prime} \right) }
\sqrt{1-\frac{p_\perp^2} {p_{\perp,\max}^2} } \right] \ ,\end{split}
\end{equation*}

where $\lambda$ is the standard Källén function and for convenience we
have defined the modified invariant
\begin{equation*}
\begin{split}s_{ij,b}^\prime = s_{ij,b} \left(\frac{1-x_s}{x_s}\right) + m_{ij}^2 \ .\end{split}
\end{equation*}

\subparagraph{Single-particle phase-space}
\label{\detokenize{review/showers/dipole:id27}}

The single-particle phase-space can be expressed in terms of $z_i$ and
$x_{ij,b}$
\begin{equation*}
\begin{split}\mathrm{d}q_j = \frac{1}{16\pi^2} 2 \tilde{p}_\mathrm{ij}\cdot q_\mathrm{b}
\mathrm{d}z_i \mathrm{d}x_\mathrm{ij,b} \frac{\mathrm{d}\phi}{2\pi}.\end{split}
\end{equation*}

Recalling that $z_i = z$, the Jacobian for expressing this phase-space in
terms of the generated variables, $J_{p_\perp^2 \rightarrow x_{ij,b}}$,
is simply
\begin{equation*}
\begin{split}J_{p_\perp^2 \rightarrow x_{ij,b}} =
\left \lvert \frac{\partial x_{ij,b}}{\partial p_\perp^2} \right \rvert \ ,\end{split}
\end{equation*}

such that
\begin{equation*}
\begin{split}dx_{ij,b} \mathrm{d}z = \left( J_{p_\perp^2 \rightarrow x_{ij,b}} \right)
p_\perp^2 \frac{\mathrm{d} p_\perp^2}{p_\perp^2} \mathrm{d}z \ .\end{split}
\end{equation*}

Noting that $z_i = z$, we can express the phase-space integration in terms of the generated
variables using the replacement
\begin{equation*}
\begin{split}\frac{1}{x_{ij,b}(1-x_{ij,b})} \mathrm{d}z_i \mathrm{d}x_{ij,b} &=
 \left[ \frac{p_\perp^2}{ p_\perp^2 + (1-z) m_i^2 + z m_j^2 - z(1-z) m_{ij}^2} \right]\frac{ \mathrm{d}p_\perp^2 }{p_\perp^2} \mathrm{d}z \ .\end{split}
\end{equation*}

\subparagraph{Initial-final dipole kinematics}
\label{\detokenize{review/showers/dipole:initial-final-dipole-kinematics}}\label{\detokenize{review/showers/dipole:dipole-shower-kinematics-if}}

The initial-final dipole splitting kernels,
given in \hyperref[\detokenize{review/showers/dipole:dipole-shower-kernels-if}]{Section \ref{\detokenize{review/showers/dipole:dipole-shower-kernels-if}}},
are written in terms of the splitting variables $u_j$ and $x_{jk,a}$,
defined in terms of the physical momenta
\begin{equation*}
\begin{split}x_{jk,a} = \frac{q_a \cdot q_j + q_a \cdot q_k - q_j \cdot q_k}
{(q_j + q_k) \cdot q_a} \ ,\end{split}
\end{equation*}\begin{equation*}
\begin{split}u_j = \frac{q_a \cdot q_j} {(q_j + q_k) \cdot q_a} \ .\end{split}
\end{equation*}

We also define the conserved momentum transfer
\begin{equation*}
\begin{split}Q = \tilde{p}_k -\tilde{p}_{aj} = q_j + q_k - q_a \ ,\end{split}
\end{equation*}

and the invariant
\begin{equation*}
\begin{split}s_{aj,k} = 2 \tilde{p}_{aj} \cdot \tilde{p}_k \ .\end{split}
\end{equation*}

In both cases, \textit{i.e.} either a massless or massive spectator,
our expressions for the momenta following the splitting must satisfy
\begin{equation}\label{equation:review/showers/dipole:eqn:IFKin:TildeKin}
\begin{split}\tilde{p}_{aj} &= x_{jk,a} q_a \ ,
\\
\tilde{p}_k & = q_j + q_k - (1-x_{jk,a})q_a \ .\end{split}
\end{equation}

It follows from this requirement that, in contrast to the case of a final-state
splitting, the longitudinal recoil is absorbed by the incoming
emitter and the outgoing spectator acquires a transverse momentum.
Following the condition in \eqref{equation:review/showers/dipole:eqn:IFKin:TildeKin}, the physical momenta after
the splitting can be expressed in terms of the splitting variables as
\begin{equation*}
\begin{split}q_a &= \frac{1}{x_{jk,a}} \tilde{p}_{aj} \ ,
\\
q_j &= \left[ \left( \frac{1-x_{jk,a}}{x_{jk,a}} \right) (1-u_j)
- u_j \frac{2 m_k^2}{s_{aj,k}} \right] \tilde{p}_{aj}
     + u_j \tilde{p}_k - k_\perp \ ,
\\
q_k &= \left[ \left( \frac{1-x_{jk,a}}{x_{jk,a}} \right) u_j
+ u_j \frac{2 m_k^2}{s_{aj,k}} \right] \tilde{p}_{aj}
+ (1-u_j) \tilde{p}_k + k_\perp \ .\end{split}
\end{equation*}

We set $n = \tilde{p}_k - (m_k^2/s_{aj,k}) \tilde{p}_{aj}$ in
Eq. \eqref{equation:review/showers/dipole:eqn:DS:qjbackward} and equate this to $q_j$ above to obtain
expressions for the splitting variables in terms of $z$ and
$p_\perp^2$,
\begin{align*}
   x_{jk,a} &= \frac{s_{aj,k}}{2r(s_{aj,k}-m_k^2)}(1-z+r)
   \times \left[ 1 - \sqrt{1 -
   \frac{4r(s_{aj,k}-m_k^2)}{s_{aj,k}}\frac{z(1-z)}{(1-z+r)^2}} \right] \ ,
   \\
   u_j &= x_{jk,a} \left( \frac{r}{1-z} \right)  \ .
\end{align*}

Following an analogous argument to that in \hyperref[\detokenize{review/showers/dipole:dipole-shower-kinematics-fi}]{Section \ref{\detokenize{review/showers/dipole:dipole-shower-kinematics-fi}}},
we obtain the limit on $x_{jk,a}$,
\begin{equation}\label{equation:review/showers/dipole:eqn:IFKin:xLim}
\begin{split}x_{jk,a} > x_e \ ,\end{split}
\end{equation}

where $x_e$ is the proton momentum-fraction carried by the emitter prior
to the splitting.

The phase-space limits on the splitting variables $u_j$ and
$x_{jk,a}$ are
\begin{equation*}
\begin{split}u_- = 0 \ , \qquad u_+ = \frac{1 - x_{jk,a}}{1 - x_{jk,a}(1 - m_k^2/s_{aj,k})} \ ,
\\
x_- = x_e \ ,\qquad x_+ = 1 \ .\end{split}
\end{equation*}

The limits on the generated variables $z$ and $p_\perp$ follow
from the inequality in \eqref{equation:review/showers/dipole:eqn:IFKin:xLim} and are
\begin{equation*}
\begin{split}p_{\perp,\mathrm{max}}^2 &= \frac{{s_{aj,k}^\prime}^2}{4} \left[ \frac{1}{m_k^2 + {s_{aj,k}^\prime}} \right] \ ,
     \\
     z_{\pm} &= \frac{1}{2} \left[ (1+x_e) \pm (1-x_e) \sqrt{1-\frac{p_\perp^2}{ p_{\perp,\max}^2}} \right] \ ,\end{split}
\end{equation*}

where for convenience we have defined the rescaled invariant
\begin{equation*}
\begin{split}{s_{aj,k}^\prime} = s_{aj,k} \left( \frac{1-x_e}{x_e} \right) \ .\end{split}
\end{equation*}

\subparagraph{Single-particle phase-space}
\label{\detokenize{review/showers/dipole:id28}}

The single-particle phase-space can be expressed in terms of $u_j$ and
$x_{jk,a}$ as
\begin{equation*}
\begin{split}\mathrm{d}q_j = \frac{1}{16\pi^2} 2 \tilde{p}_k \cdot q_a
\mathrm{d}x_{jk,a} \mathrm{d}u_j \frac{\mathrm{d}\phi}{2\pi} \ .\end{split}
\end{equation*}

In order to write this in terms of our generated variables, we need the
Jacobian $J_{p_\perp^2, z \rightarrow u_j,x_{jk,a}}$, such that
\begin{equation*}
\begin{split}\mathrm{d}x_{jk,a} \mathrm{d}u_j =
\left(J_{p_\perp^2, z \rightarrow u_j,x_{jk,a}} \right)
p_\perp^2 \frac{\mathrm{d} p_\perp^2}{p_\perp^2} \mathrm{d}z
\ .\end{split}
\end{equation*}

It is easier to rearrange the explicit equations
for $x_{jk,a}(z,p_\perp)$ and $u_j(z,p_\perp)$ to obtain equations
$z(x_{jk,a}, u_j)$ and $p_\perp(x_{jk,a}, u_j)$ and instead compute
$J_{ u_j,x_{jk,a}\rightarrow p_\perp^2, z}$. This is simply the
reciprocal of the required Jacobian
$J_{p_\perp^2, z \rightarrow u_j,x_{jk,a}}$ and is given by
\begin{equation*}
\begin{split}J_{ u_j,x_{jk,a}\rightarrow p_\perp^2, z} =
\left\lvert
\frac{\partial z}{\partial u_j}
\frac{\partial p_\perp^2}{\partial x_{jk,a}}
-
\frac{\partial z}{\partial x_{jk,a}}
\frac{\partial p_\perp^2}{\partial u_j}
\right\rvert \ .\end{split}
\end{equation*}

We can therefore express the phase-space integration in terms of the generated variables using the replacement
\begin{equation*}
\begin{split}\frac1{u_j}\frac1{x_{jk,a}}\mathrm{d}u_j \mathrm{d}x_{jk,a} =
\left[ u_j + x_{jk,a} - 2u_jx_{jk,a}\left(1-\frac{m_k^2}{s_{aj,k}}\right) \right]^{-1}
\frac{\mathrm{d} p_\perp^2}{p_\perp^2} \mathrm{d}z \ .\end{split}
\end{equation*}

\subparagraph{Initial-initial dipole kinematics}
\label{\detokenize{review/showers/dipole:initial-initial-dipoles}}\label{\detokenize{review/showers/dipole:dipole-shower-kin-ii}}

In the case of initial-initial dipoles we only consider the case where all
partons in the dipole, before and after the splitting, are massless.

The initial-initial dipole splitting kernels,
given in \hyperref[\detokenize{review/showers/dipole:dipole-shower-kernels-ii}]{Section \ref{\detokenize{review/showers/dipole:dipole-shower-kernels-ii}}},
are written in terms of the splitting variable $x_{j,ab}$,
defined in terms of the physical momenta
\begin{equation*}
\begin{split}x_{j,ab} = \frac{q_a \cdot q_b - q_j \cdot q_a - q_j \cdot q_b}
{ q_a \cdot q_b } \ .\end{split}
\end{equation*}

In order to express the kinematics and the single-particle phase-space
we require an additional splitting variable
\begin{equation*}
\begin{split}v_j = \frac{q_a \cdot q_j}{q_a \cdot q_b}.\end{split}
\end{equation*}

It is convenient in the case of the initial-initial dipole to leave the momentum
of the incoming spectator unchanged following the splitting,
\textit{i.e.} $\tilde{p}_b = q_b$. Furthermore, as in the case of the initial-final
dipole, the momentum of the incoming emitter changes only by a simple rescaling,
that is it absorbs only longitudinal recoil. The remaining recoil from the
splitting is absorbed by all of the final-state particles, including non-coloured
particles, in the process prior to the splitting.

As this splitting involves all of the particles in the event we do not explicitly
consider a conserved dipole momentum or dipole momentum transfer $Q$ as we
have for the other dipoles. Instead we consider the momentum conservation
requirements of the process before and after the splitting, these are
\begin{equation}\label{equation:review/showers/dipole:eqn:IIKin:MomCon}
\begin{split}\tilde{p}_{aj} + q_b = \sum_n {\tilde{k}_n} \ ,
\\
q_a + q_b = \sum_n {k_n} + q_j \ ,\end{split}
\end{equation}

respectively. We use $\{\tilde{k}_n\} \text{ and } \{k_m\}$ to denote the
momenta of all of the final-state particles, except the emission from the
splitting, before and after the splitting respectively.
For convenience we also define the invariant
\begin{equation*}
\begin{split}s_{aj,b} = 2 \tilde{p}_{aj} \cdot q_b \ .\end{split}
\end{equation*}

The momenta prior to the splitting must satisfy
\begin{equation*}
\begin{split}\tilde{p}_{aj} & =   x_{j,ab} q_a,
\\
\tilde{k}_m & =   \Lambda
\left(q_a + q_b - q_j, \tilde{p}_{aj} + q_b \right) k_m,\end{split}
\end{equation*}

where the Lorentz transformation $\Lambda$ satisfying the momentum
conservation requirements in \eqref{equation:review/showers/dipole:eqn:IIKin:MomCon} is
\begin{equation*}
\begin{split}{\Lambda^\mu}_\nu \left(K,\tilde{K}\right) =
g^\mu{}_\nu - \frac{2 \left(K+\tilde{K}\right)^\mu
\left(K+\tilde{K} \right)_\nu}{\left(K+\tilde{K}\right)^2}
+ \frac{2 \tilde{K}^\mu K_\nu }{K^2} \ .\end{split}
\end{equation*}

With these requirements the physical momenta following the splitting are
\begin{equation*}
\begin{split}q_a & =  \frac{1}{x_{j,ab}} \tilde{p}_{aj},
\\
q_j & =  \frac{1-x_{j,ab}-v_j}{x_{j,ab}} \tilde{p}_{aj} + v_j q_b + k_\perp,
\\
k_m & =  \Lambda
\left( \tilde{p}_{aj} + q_b , q_a + q_b - q_j \right) \tilde{k}_m,\end{split}
\end{equation*}

and the splitting variables $x_{j,ab}$ and $v_j$ are written in
terms of the generated variables
\begin{equation*}
\begin{split}x_{j,ab} = \frac{z(1-z)}{(1-z+r)} \ ,
\\
v_j = x_{j,ab} \frac{r}{1-z} \ ,\end{split}
\end{equation*}

where we have defined
$r = \frac{p_\perp^2}{ s_{aj,b} }$.

Following an analogous argument to that in \hyperref[\detokenize{review/showers/dipole:dipole-shower-kinematics-fi}]{Section \ref{\detokenize{review/showers/dipole:dipole-shower-kinematics-fi}}},
we obtain the limit on $x_{j,ab}$,
\begin{equation*}
\begin{split}x_{j,ab} > x_e \ ,\end{split}
\end{equation*}
where $x_e$ is the proton momentum-fraction carried by the emitter prior
to the splitting. The limits on the splitting variable $x_{j,ab}$ are
\begin{equation*}
\begin{split}x_- = x_e \ , \qquad x_+ = 1 \ ,
\\
v_- = 0 \ , \qquad v_+ = 1 - x \ .\end{split}
\end{equation*}

The limits on the generated variables are
\begin{align*}
   p_{\perp,\mathrm{max}}^2 &= s_{aj,b}
   \frac{(1-x_e)^2}{4 x_e} \ ,
   \\
   z_{\pm} &= \frac{1}{2} \left[ (1+x_e) \pm (1-x_e) \sqrt{1-\frac{p_\perp^2}
   { p_{\perp,\max}^2}} \right] \ .
\end{align*}

\subparagraph{Single-particle phase-space}
\label{\detokenize{review/showers/dipole:id29}}

The single-particle phase-space can be expressed in terms of $v_j$ and
$x_{j,ab}$
\begin{equation*}
\begin{split}\mathrm{d}q_j = \frac{1}{16\pi^2} 2 q_a \cdot q_b
\mathrm{d}x_{j,ab} \mathrm{d}v_j \frac{\mathrm{d}\phi}{2\pi} \ .\end{split}
\end{equation*}

In order to write this in terms of our generated variables, we need the
Jacobian $J_{p_\perp^2, z \rightarrow v_j,x_{j,ab}}$, such that
\begin{equation*}
\begin{split}\mathrm{d}x_{j,ab} \mathrm{d}v_j =
\left(J_{p_\perp^2, z \rightarrow v_j,x_{j,ab}} \right)
p_\perp^2 \frac{\mathrm{d} p_\perp^2}{p_\perp^2} \mathrm{d}z
\ .\end{split}
\end{equation*}

This is given by
\begin{equation*}
\begin{split}J_{p_\perp^2, z \rightarrow v_j,x_{j,ab}} =
\left\lvert
\frac{\partial v_j}{\partial {p_\perp^2}}
\frac{\partial{x_{j,ab}}}{\partial z}
-
\frac{\partial v_j}{\partial z}
\frac{\partial x_{j,ab}}{\partial {p_\perp^2}}
\right\rvert \ ,\end{split}
\end{equation*}

so that
\begin{equation*}
\begin{split}\frac1{x_{j,ab}}\frac1{v_j}\mathrm{d}x_{j,ab} \mathrm{d}v_j =  \frac1z
\frac{\mathrm{d} p_\perp^2}{p_\perp^2} \mathrm{d}z.\end{split}
\end{equation*}

\subparagraph{Decay dipoles}
\label{\detokenize{review/showers/dipole:decay-dipoles}}\label{\detokenize{review/showers/dipole:dipole-shower-kin-decay}}

The final distinct type of dipole which must be considered is that consisting
of an incoming decayed particle and an outgoing colour-connected
partner. The outgoing partner is a parton produced either in the decay of
the incoming particle or the subsequent
showering of the decay system prior to the current splitting.
In principle, such a dipole can be considered with the emitter identified as
either the incoming or outgoing parton, however, we currently only include the
latter case in Herwig 7. Following our terminology for the other dipoles,
we only include final-initial decay dipoles and do not consider
initial-final decay dipoles. This is discussed in \hyperref[\detokenize{review/showers/dipole:dipole-shower-kernels-decay}]{Section \ref{\detokenize{review/showers/dipole:dipole-shower-kernels-decay}}}.

As it is the colour partner of the emitter, we refer to the incoming particle
as the spectator, however we wish to preserve the 4-momentum of the incoming
particle
as its momentum has been set, before its decay, in the showering of the
production process. This is in line with the principle of the narrow-width
approximation. Therefore, in contrast with the majority of the other dipoles
discussed in the previous sections,
the spectator does not absorb any recoil from the splitting.
Instead we choose to absorb the recoil by sharing it amongst all of the
particles outgoing from the decay and any previous emissions from the showering
of the decay system,
excluding the emitter and the emitted parton from the current splitting.
We refer to this set of particles as the ‘recoil system’. This is similar to the
approach used in the case of the initial-initial dipole in
\hyperref[\detokenize{review/showers/dipole:dipole-shower-kin-ii}]{Section \ref{\detokenize{review/showers/dipole:dipole-shower-kin-ii}}}.

We will now show that the kinematics for a splitting from a decay dipole are
identical to those for a splitting from a massive final-final dipole.
Following the notation outlined in \hyperref[\detokenize{review/showers/dipole:dipole-shower-notation}]{Section \ref{\detokenize{review/showers/dipole:dipole-shower-notation}}}, we denote the
momenta
of the incoming decayed particle and the outgoing emitter prior to the splitting
as $\tilde{p}_b$ and $\tilde{p}_{ij}$ respectively. The momentum
of the recoil system is denoted by $\tilde{p}_k$.
Following the splitting the momentum
of the incoming particle is unchanged, $q_b = \tilde{p}_b$, the momenta of
the new outgoing emitter and emission are denoted by $q_i$ and $q_j$
respectively and the momentum of the recoil system is denoted by $q_k$.
It follows from our definition of the recoil system that the incoming particle
momentum $q_b$ is in fact the conserved dipole momentum
\begin{equation}\label{equation:review/showers/dipole:eqn:DS:DecayKin:Q}
\begin{split}Q = q_b & =  \ \tilde{p}_{ij} + \tilde{p}_k\\
& =  \ q_i + q_j + q_k \ .\end{split}
\end{equation}

Comparing Eqs.~\eqref{equation:review/showers/dipole:eqn:DS:DecayKin:Q} and \eqref{equation:review/showers/dipole:eqn:DS:FFKin:Q} it is clear that
the required kinematics are simply those for a splitting from a massive
final-final dipole in \hyperref[\detokenize{review/showers/dipole:dipole-shower-kin-ff}]{Section \ref{\detokenize{review/showers/dipole:dipole-shower-kin-ff}}}.

\paragraph{Splitting kernels}
\label{\detokenize{review/showers/dipole:splitting-kernels}}\label{\detokenize{review/showers/dipole:dipole-shower-kernels}}

The splitting kernels for light and massive dipoles were originally given in
Ref. \cite{Catani:1996vz} and Ref. \cite{Catani:2002hc} respectively,
we give them in the following sections in order to
provide a complete reference. For further details we refer the reader to the
original publications.
For simplicity we here only report on the spin averaged splitting kernels. Furthermore we choose to neglect those multiplicative
factors that are not relevant to the implementation of these kernels
in a parton shower.

For each dipole we present the kernels for both the massless and
massive cases. While the massless case can always be obtained from the massive
result, as with the splitting kinematics we choose to implement the two cases
separately in Herwig in order to avoid unnecessary computations in the case of
splittings from massless dipoles.
Unless otherwise stated, all of the notation used in the expressions of the
splitting kernels is defined in \hyperref[\detokenize{review/showers/dipole:dipole-shower-notation}]{\ref{\detokenize{review/showers/dipole:dipole-shower-notation}}} and the relevant
kinematics sections.

The colour factors, $C_\mathrm{F}$ and $C_\mathrm{A}$ follow their
standard definitions in terms of the number of colours, $N_\mathrm{C}$,
\begin{equation*}
\begin{split}C_\mathrm{F} = \frac{N_\mathrm{C}^2 - 1}{2 N_\mathrm{C}} \ ,
\\
C_\mathrm{A} = N_\mathrm{C} \ .\end{split}
\end{equation*}

We highlight to the reader that the dipole shower currently obeys the strict
large-$N_\mathrm{C}$ limit. We also use the factor
$T_\mathrm{R} = 1/2$. The strong coupling, denoted by
$\alpha_\text{S}$, is described in detail in
\hyperref[\detokenize{review/appendix/alphaS:dipole-shower-strong-coupling}]{\ref{\detokenize{review/appendix/alphaS:dipole-shower-strong-coupling}}}.

Splittings from decay dipoles were not included in the original publication of the
splitting kernels. We present them in \hyperref[\detokenize{review/showers/dipole:dipole-shower-kernels-decay}]{Section \ref{\detokenize{review/showers/dipole:dipole-shower-kernels-decay}}} for the
first time and provide a detailed discussion.

\subparagraph{Final-final dipole kernels}
\label{\detokenize{review/showers/dipole:final-final-dipole-kernels}}\label{\detokenize{review/showers/dipole:dipole-shower-kernels-ff}}

We denote the kernel for the splitting
$\{\tilde{ij}, \tilde{k}\} \to \{ i,j,k \}$ as
$\langle V_{X_i Y_j, k} \left( z_i, y_{ij,k} \right) \rangle$
where $X$ and $Y$ denote the type of parton
\begin{equation*}
\begin{split}\langle V_{q_i g_j,k} \left( z_i, y_{ij,k} \right) \rangle
= 8\pi \alpha_\text{S} C_\text{F}
\left\{ \frac{2}{1-z_i(1-y_{ij,k})} - \frac{\tilde{v}_{ij,k}}{v_{ij,k}}
\left[ 1 + z_i + \frac{2\mu_i^2}{\bar{s}y_{ij,k}} \right] \right\},\end{split}
\end{equation*}\begin{equation*}
\begin{split}\langle V_{q_i \bar{q}_j,k} \left( z_i, y_{ij,k} \right) \rangle
= 8\pi \alpha_\text{S} T_\text{R} \frac{1}{v_{ij,k}}
\left\{ 1 - 2\left[ z_i(1-z_i) - (1-\kappa)z_{i,+}z_{i,-}
- \frac{\kappa \mu_i^2}{ 2 \mu_i^2 + ( 1 - 2 \mu_i^2 - \mu_k^2)y_{ij,k}}
\right] \right\},\end{split}
\end{equation*}\begin{equation*}
\begin{split}\langle V_{g_i g_j,k} \left( z_i, y_{ij,k} \right) \rangle
= 16\pi \alpha_\text{S} C_\text{A}
\left[ \frac{1}{1-z_i(1-y_{ij,k})} + \frac{1}{1-(1-z_i)(1-y_{ij,k})}
- \frac{z_i(1-z_i) - (1-\kappa)z_{i,+}z_{i,-} - 2}{v_{ij,k}} \right],\end{split}
\end{equation*}

where the relative velocity term $\tilde{v}_{ij,k}$, \textit{i.e.} the velocity
between $\tilde{p}_{ij}$ and $\tilde{p}_k$, is written as
\begin{equation*}
\begin{split}\tilde{v}_{ij,k} =
\frac{ \sqrt{ \lambda \left( s, m_{ij}^2, m_k^2 \right) }}
{s - m_{ij}^2 - m_k^2} \ .\end{split}
\end{equation*}

The corresponding massless results are given in \hyperref[\detokenize{review/appendix/masslessDipole:dipole-shower-kernels-ff-masses}]{\ref{\detokenize{review/appendix/masslessDipole:dipole-shower-kernels-ff-masses}}}.
The branching probability for a splitting from a final-final dipole is written
in its general form
\begin{equation*}
\begin{split}\mathrm{d}\mathcal{P}_{\mathrm{branching}} =
\frac{1}{(q_i + q_j)^2 - m_{ij}^2}
\langle V_{X_i Y_j, k} \left( z_i, y_{ij,k} \right) \rangle
\mathrm{d} q_j \ .\end{split}
\end{equation*}

\subparagraph{Final-initial dipole kernels}
\label{\detokenize{review/showers/dipole:final-initial-dipole-kernels}}\label{\detokenize{review/showers/dipole:dipole-shower-kernels-fi}}

We denote the kernel for the splitting
$\{\widetilde{ij}, \tilde{b}\} \to \{ i,j,b \}$ as
$\langle V_{X_i Y_j}^b \left( z_i, x_{ij,b} \right) \rangle$
where $X$ and $Y$ denote the parton types
\begin{equation*}
\begin{split}\langle V_{q_i g_j}^b \left( z_i, x_{ij,b} \right) \rangle
= 8 \pi \alpha_\text{S} C_\text{F}
\left\{ \frac{2}{1-z_i + (1-x_{ij,b})} - (1 + z_i)
- \mu_i^2 \frac{2x_{ij,b}}{\bar{s}(1-x_{ij,b})} \right\},\end{split}
\end{equation*}\begin{equation*}
\begin{split}\langle V_{q_i \bar{q}_j}^b \left( z_i, x_{ij,b} \right) \rangle
= 8 \pi \alpha_\text{S} T_\text{R}
\left[ 1 - 2(z_{i,+} - z_i)(z_i - z_{i,-}) \right].\end{split}
\end{equation*}

The corresponding massless results are given in \hyperref[\detokenize{review/appendix/masslessDipole:dipole-shower-kernels-fi-masses}]{\ref{\detokenize{review/appendix/masslessDipole:dipole-shower-kernels-fi-masses}}}.
The branching probability for a splitting from a final-initial dipole is
written in its general form
\begin{equation*}
\begin{split}\mathrm{d}\mathcal{P}_{\mathrm{branching}} =
\frac{1}{(q_i + q_j)^2 - m_{ij}^2}
\frac{1}{x_{ij,b}}
\frac{f_b(x_s/x_{ij,b})}{f_b(x_s)}
\langle V_{X_i Y_j}^b \left( z_i, x_{ij,b} \right) \rangle
\mathrm{d} q_j \ ,\end{split}
\end{equation*}

here, in general, $f_l(x)$ is the parton density function
of parton $l$.

\subparagraph{Initial-final dipole kernels}
\label{\detokenize{review/showers/dipole:initial-final-dipole-kernels}}\label{\detokenize{review/showers/dipole:dipole-shower-kernels-if}}

We denote the kernel for the splitting
$\{\widetilde{aj}, \tilde{k}\} \to \{ a,j,k \}$ as
$\langle V^{X_a Y_j}_k \left( u_j, x_{jk,a} \right) \rangle$
where $X$ and $Y$ denote the type of parton
\begin{equation*}
\begin{split}\langle V^{q_a g_j}_k \left( u_j, x_{jk,a} \right) \rangle
= 8 \pi \alpha_\text{S} C_\text{F}
\left[ \frac{2}{1-x_{jk,a}+u_j} - (1+x_{jk,a}) \right],\end{split}
\end{equation*}\begin{equation*}
\begin{split}\langle V^{g_a q_j}_k \left( u_j, x_{jk,a} \right) \rangle
= 8 \pi \alpha_\text{S} T_\text{R}
\left[ 1 - 2x_{jk,a}(1-x_{jk,a}) \right],\end{split}
\end{equation*}\begin{equation*}
\begin{split}\langle V^{q_a q_j}_k \left( u_j, x_{jk,a} \right) \rangle
= 8 \pi \alpha_\text{S} C_\text{F}
\left[ x_{jk,a} + \frac{2(1-x_{jk,a})}{x_{jk,a}}
- 2\mu_k^2\frac{u_j}{1-u_j}  \right],\end{split}
\end{equation*}\begin{equation*}
\begin{split}\langle V^{g_a g_j}_k \left( u_j, x_{jk,a} \right) \rangle
= 16 \pi \alpha_\text{S} C_\text{A}
\left[ \frac{1}{1-x_{jk,a}+u_j} + \frac{1-x_{jk,a}}{x_{jk,a}}
- 1 + x_{jk,a}(1-x_{jk,a}) - \mu_k^2\frac{u_j}{1-u_j} \right].\end{split}
\end{equation*}

The corresponding massless results are given in \hyperref[\detokenize{review/appendix/masslessDipole:dipole-shower-kernels-if-massless}]{\ref{\detokenize{review/appendix/masslessDipole:dipole-shower-kernels-if-massless}}}.
The branching probability for a splitting from an initial-final dipole is
written in its general form
\begin{equation*}
\begin{split}\mathrm{d}\mathcal{P}_{\mathrm{branching}} =
\frac{1}{2 q_j \cdot q_a}
\frac{1}{x_{jk,a}}
\frac{f_a(x_e/x_{jk,a})}{f_{\widetilde{aj}}(x_e)}
\langle V^{X_a Y_j}_k \left( u_j, x_{jk,a} \right) \rangle
\mathrm{d} q_j \ .\end{split}
\end{equation*}

\subparagraph{Initial-initial dipole kernels}
\label{\detokenize{review/showers/dipole:dipole-shower-kernels-ii}}\label{\detokenize{review/showers/dipole:id34}}

We denote the kernel for the splitting
$\{\widetilde{aj}, \tilde{b}\} \to \{ a,j,b \}$ as
$\langle V^{X_a Y_j,b} \left( x_{j,ab} \right) \rangle$
where $X$ and $Y$ denote the type of parton
\begin{equation*}
\begin{split}\langle V^{q_a g_j,b} \left( x_{j,ab} \right) \rangle
= 8 \pi \alpha_\text{S} C_\text{F}
\left[ \frac{1+x_{j,ab}^2}{1-x_{j,ab}} \right ],\end{split}
\end{equation*}\begin{equation*}
\begin{split}\langle V^{g_a q_j,b} \left( x_{j,ab} \right) \rangle
= 8 \pi \alpha_\text{S} T_\text{R}
\left[ 1 - 2x_{j,ab}(1-x_{j,ab}) \right],\end{split}
\end{equation*}\begin{equation*}
\begin{split}\langle V^{q_a q_j,b} \left( x_{j,ab} \right) \rangle
= 8 \pi \alpha_\text{S} C_\text{F}
\left[ \frac{1}{x_{j,ab}} + \frac{(1-x_{j,ab})^2}{x_{j,ab}} \right],\end{split}
\end{equation*}\begin{equation*}
\begin{split}\langle V^{g_a g_j,b} \left( x_{j,ab} \right) \rangle
= 16 \pi \alpha_\text{S} C_\text{A}
\left[ \frac{x_{j,ab}}{1-x_{j,ab}} + \frac{1-x_{j,ab}}{x_{j,ab}}
+ x_{j,ab}(1-x_{j,ab}) \right].\end{split}
\end{equation*}

The branching probability for a splitting from an initial-initial dipole is
written in its general form
\begin{equation*}
\begin{split}\mathrm{d}\mathcal{P}_{\mathrm{branching}} =
\frac{1}{2 q_j \cdot q_a}
\frac{1}{x_{j,ab}}
\frac{f_a(x_e/x_{j,ab})}{f_{\widetilde{aj}}(x_e)}
\langle V^{X_a Y_j, b} \left( x_{j,ab} \right) \rangle
\mathrm{d} q_j \ ,\end{split}
\end{equation*}

where there is no parton distribution function term for the spectator
because the momentum of the spectator parton is unchanged by the splitting.

\subparagraph{Decay dipoles}
\label{\detokenize{review/showers/dipole:decay-dipole-splitting-kernels}}\label{\detokenize{review/showers/dipole:dipole-shower-kernels-decay}}

We stated in \hyperref[\detokenize{review/showers/dipole:dipole-shower-kin-decay}]{Section \ref{\detokenize{review/showers/dipole:dipole-shower-kin-decay}}} that we do not include splittings
from initial-final decay dipoles. This is because the kernel for this splitting
contains a large negative term proportional to the mass-squared of the
decaying particle. The kernel is therefore negative in a very large proportion of
trialled splittings. In the case of a negative kernel that trial splitting is
entirely discarded and does not contribute to the overall probability
distribution implicitly used to generate the emission. The splitting kernel
for the final-initial decay dipole does not suffer from this issue. We therefore
only explicitly consider final-initial decay dipoles, however in order to correctly
reproduce the probability distribution for each splitting we must include the
splitting kernel expression for the initial-final decay dipole in that used to
generate splittings from final-initial decay dipoles. The sum of the
two contributions is positive in most trial splittings and therefore correctly
produces the required distribution of emissions.

We refer the reader to \cite{Cormier:2018tog} for a
more detailed discussion of the treatment of emissions from decay dipoles and
the choices made. We also highlight that, as always with splitting
kernels, we can within reason keep, discard and add finite terms to the kernels. 
In the default implementation, we use the kernels given below, obtained by
combining the final-initial and initial-final decay-dipole contributions,
including all finite terms displayed explicitly in these expressions. For the
$g\to q\bar q$ and $g\to gg$ kernels, the free parameter is set to
$\kappa=0$. The default $g\to gg$ implementation uses the asymmetric
partition corresponding to \texttt{AsymmetryOption=0}.

As discussed in
\hyperref[\detokenize{review/showers/dipole:dipole-shower-kin-decay}]{Section \ref{\detokenize{review/showers/dipole:dipole-shower-kin-decay}}} the kinematics in the case of decay dipoles are
identical to the final-final dipole splitting kinematics. The notation
therefore follows that in \hyperref[\detokenize{review/showers/dipole:dipole-shower-kin-ff}]{Section \ref{\detokenize{review/showers/dipole:dipole-shower-kin-ff}}}, with the addition of the
mass of the incoming decayed particle, denoted as $m_b$.
Similarly the Jacobian factors and branching probabilities follow the same
formulation as for the final-final dipole we therefore  do not repeat them
for the decay dipole. The kernels are
\begin{align*}\!\begin{aligned}
\lefteqn{\langle V_{q_i g_j,k}^b \left( z_i, y_{ij,k} \right) \rangle = } &\\& 8 \pi \alpha_\text{S} \mathrm{C}_\mathrm{F}
     \left \{\frac{2 \left( 2 m_i^2 + 2y_{ij,k}\bar{s} + \bar{s} \right)}{(1+y_{ij,k})\bar{s} - z_i(1-y_{ij,k})\bar{s}}
     - \frac{\tilde{v}_{ij,k}}{v_{ij,k}} \left( (1+z_i) + \frac{2 m_i^2}{y \bar{s}} \right) \right.\\
& \left.+ \frac{y_{ij,k}}{1-z_i(1-y_{ij,k})}
    \left[ \frac{2 \left( 2 m_i^2 + 2y_{ij,k}\bar{s} + \bar{s} \right)}
      {(1+y_{ij,k})\bar{s} - z_i(1-y_{ij,k})\bar{s}}
      - \frac{\tilde{v}_{ij,k}}{v_{ij,k}}
      \left( 2 + \frac{2 m_b^2}{\left(1-z_i(1-y_{ij,k})\right) \bar{s}} \right)
      \right]
    \right \},\\\\
\lefteqn{\langle V_{g_i g_j,k}^b \left( z_i, y_{ij,k} \right) \rangle =} &\\& 16 \pi \alpha_\text{S} \mathrm{C}_\mathrm{A}
          \left \{ \vphantom{\frac{1+2y_{ij,k}}{(1+y_{ij,k})-z_i(1-y_{ij,k})}}
      \frac{1+2y_{ij,k}}{(1+y_{ij,k})-z_i(1-y_{ij,k})}   + \frac{1+2y_{ij,k}}{(1+y_{ij,k})-(1-z_i)(1-y_{ij,k})} \right.\\
     &\ \ \ \ \  \ \ \ \ \  \ \ \ \  \ \ \ \ \  \ \ \ \ \ \ \ \  \ \ \ \ +\left.  \frac{1}{v_{ij,k}} \left[ z_i(1-z_i) - (1-\kappa)z_{i,+}z_{i,-} -2 \right]
       \right \}\\
 &\phantom{=}           +8 \pi \alpha_\text{S} \mathrm{C}_\mathrm{F} \left \{
        \frac{y_{ij,k}}{1-z_i(1-y_{ij,k})}
        \left[ \frac{2(1+2y_{ij,k})}{(1+y_{ij,k})-z_i(1-y_{ij,k})}
        - \frac{\tilde{v}_{ij,k}}{v_{ij,k}} \left( 2 + \frac{ 2 m_b^2}{(1-z_i(1-y_{ij,k}))\bar{s}} \right) \right]
          \right.\\
  &\left. +
      \frac{y_{ij,k}}{1-(1-z_i)(1-y_{ij,k})}
      \left[ \frac{2(1+2y_{ij,k})}{(1+y_{ij,k})-(1-z_i)(1-y_{ij,k})} - \frac{\tilde{v}_{ij,k}}{v_{ij,k}}
        \left( 2 + \frac{ 2 m_b^2}{(1-(1-z_i)(1-y_{ij,k}))\bar{s}} \right) \right]\right \},\\\\
\lefteqn{\langle V_{q_i \bar{q}_j,k}^b \left( z_i, y_{ij,k} \right) \rangle =} & \\
 & 8 \pi \alpha_\text{S} \mathrm{T}_\mathrm{R} \left \{
   1 - 2\left( z_i(1-z_i) - (1-\kappa)z_{i,+}z_{i,-} -
   \frac{\kappa m_i^2}{2m_i^2 + \bar{s}y_{ij,k}} \right) \right \}.\\
\end{aligned}\end{align*}

\subsubsection{Colour Matrix Element Corrections}
\label{\detokenize{review/showers/dipole:colour-matrix-element-corrections}}\label{\detokenize{review/showers/dipole:dipole-shower-colour-mec}}

Within the dipole shower algorithm the method of improving the radiation
pattern by subleading-$N_\mathrm{C}$ corrections is available following the work outlined
in \cite{Platzer:2012np, Platzer:2018pmd}.  The colour matrix element
corrections successively adjust the distribution of gluon emissions by a
correction factor
\begin{equation*}
\begin{split}w^{\text{CMEC}}_{ij} = -\frac{{\rm Tr}[{\mathbf T}_i\cdot {\mathbf T}_j
{\mathbf M}_n]}{{\mathbf T}_i^2\ {\rm Tr}[{\mathbf M}_n]}\end{split}
\end{equation*}

to each dipole splitting kernel with emitter $i$ and spectator
$j$, in terms of the colour space density operator for $n$
additional partons, $\mathbf{M}_n$. The initial condition for this
object is $\mathbf{M}_0=|{\cal M}\rangle\langle{\cal M}|$ in terms of
the hard scattering amplitude $|{\cal M}\rangle$, and it is successively
updated using
\begin{equation*}
\begin{split}{\mathbf M}_{n+1} = \sum_{i,j} W_{ij} {\mathbf T}_i {\mathbf M}_n {\mathbf
T}_j^\dagger.\end{split}
\end{equation*}

The colour matrix element corrections heavily rely on the Matchbox
infrastructure and use the interface to the ColorFull package
\cite{Sjodahl:2014opa}. They can be enabled using the \sphinxtitleref{Matchbox/CMEC.in}
snippet, and employ the reweighting facilities \hyperref[\detokenize{review/showers/variations:shower-reweight}]{Section \ref{\detokenize{review/showers/variations:shower-reweight}}} of the
dipole shower to apply the correction factors.

\subsubsection{Constituent reshuffling}
\label{\detokenize{review/showers/dipole:constituent-reshuffling}}\label{\detokenize{review/showers/dipole:dipole-shower-constituent-reshuffling}}

The Herwig cluster hadronization model expects partons with momenta on
their constituent mass shell at the end of the perturbative shower
evolution. Unlike the angular ordered shower which ensures
energy-momentum conservation only at the end of the showering process,
the dipole shower
preserves four momentum at each shower emission. It also strictly
keeps partons on their perturbative mass-shell. A transition to their
constituent mass shell is therefore required and applied at the end of the dipole
shower evolution. As with similar problems elsewhere in the simulation the
method of choice was outlined in the context of the RAMBO phase
space generator \cite{Kleiss:1985gy}. This method, transforming pole to
constituent mass shells, boosts the system of final partons into their
centre-of-mass system, and then scales the three momenta by a factor
$\xi$ solving
\begin{equation*}
\begin{split}\sum_i \sqrt{{\mathbf p}_i^2+m_i^2} = \sum_i \sqrt{\xi^2 {\mathbf p}_i^2+ M_i^2} \ ,\end{split}
\end{equation*}

where $m_i$ are the pole masses, and $M_i$ are the constituent
masses. In the highly unlikely case of an event extremely close to threshold, where
there is no solution of this equation, the showering
is restarted from the hard process. The total momentum of partons originating
from a decay system is preserved independently. The \href{https://herwig.hepforge.org/doxygen/classHerwig\_1\_1ConstituentReshuffler.html}{ConstituentReshuffler}
class of the dipole shower performs the transformation above and allows for a different approach if desired. If the treatment of the constituent reshuffling within the hadronization framework, possibly including a dynamic gluon mass, no reshuffling at this level is needed.

\subsection{Perturbative decays and spin correlations}
\label{\detokenize{review/showers/perturbative_decays:perturbative-decays-and-spin-correlations}}\label{\detokenize{review/showers/perturbative_decays:sec-perturbative-decays}}\label{\detokenize{review/showers/perturbative_decays::doc}}

When calculating the matrix element for a given hard process or decay
one must take into account the effect of spin correlations, as they will
affect the distributions of particles in the final state. In particular,
these correlations are important in the production and decay of the top
quark, for the production of tau leptons in Higgs decays and in models
of BSM physics.
They can also affect the azimuthal angles of emissions in the parton
shower. Other implementations of spin-correlation effects in parton
showers are presented in Refs.~\cite{Karlberg:2021kwr,
Hamilton:2021dyz,Hoche:2025anb}.
In Herwig the decays of the fundamental particles and the unstable
hadrons are calculated in the same way in order that correlation effects
for particles such as the tau lepton, which is produced perturbatively
but decays non-perturbatively, are correctly treated.
The angular-ordered and dipole parton showers now include spin correlations,
requiring that these particles are decayed during the parton-shower stage
of the event.

The spin correlation algorithm of
Refs. \cite{Richardson:2001df, Knowles:1988vs, Collins:1987cp, Richardson:2018pvo}
is use to implement the correct spin correlations between the production
and decay of particles, and any emissions
generated by the parton shower. This algorithm is explained in general in
Refs.~\cite{Richardson:2001df, Knowles:1988vs, Collins:1987cp, Richardson:2018pvo},
here we will explain it using the example of the process
$e^+ e^- {\rightarrow}t \bar{t}$ where the top quark subsequently
decays, via a $W^+$ boson, to a $b$ quark and a pair of
light fermions.

Initially, the outgoing momenta of the $t\,\bar{t}$ pair are
generated according to the usual cross-section integral
\begin{equation*}
\begin{split}\frac{(2\pi)^4}{2s}\int {\frac{d^{3}p_{t}}{(2\pi)^3 2E_{t}}} {\frac{d^{3}p_{\bar{t}}}{(2\pi)^3 2E_{\bar{t}}}} {{\mathcal{M}}^{e^+ e^-{\rightarrow}t \bar{t}}_{{\lambda_{t}}{\lambda_{\bar{t}}}}}
{{\mathcal{M}}^{* e^+ e^-{\rightarrow}t \bar{t}}_{{\lambda_{t}}{\lambda_{\bar{t}}}}},\end{split}
\end{equation*}
where
${{\mathcal{M}}^{e^{+} e^{-}{\rightarrow}t \bar{t}}_{{\lambda_{t}}{\lambda_{\bar{t}}}}}$
is the matrix element for the initial hard process and
$\lambda_{t,\bar{t}}$ are the helicities of the $t$ and
$\bar{t}$ respectively. One of the outgoing particles is then
picked at random, say the top quark, and a spin density matrix calculated
\begin{equation*}
\begin{split}\rho^{t}_{{\lambda_{t}}{\lambda_{t}}^{'}}=\frac{1}{N}
{\mathcal{M}}^{e^{+} e^{-} {\rightarrow}t \bar{t}}_{{\lambda_{t}}{\lambda_{\bar{t}}}}
{\mathcal{M}}^{* e^{+} e^{-} {\rightarrow}t \bar{t}}_{{\lambda_{t}}{\lambda_{\bar{t}}}},\end{split}
\end{equation*}
with $N$ defined such that $\rm{Tr}\,\rho = 1$.

The top quark is decayed and the momenta of the decay products distributed
according to
\begin{equation*}
\begin{split}\frac{(2\pi)^4}{2m_t}\int {\frac{d^{3}p_{b}}{(2\pi)^3 2E_{b}}} {\frac{d^{3}p_{W^+}}{(2\pi)^3 2E_{W^+}}} \rho^{t}_{{\lambda_{t}}{\lambda_{t}}^{'}}
{\mathcal{M}}^{t{\rightarrow}bW^+}_{{\lambda_{t}}{\lambda_{W^+}}} {\mathcal{M}}^{*t{\rightarrow}bW^+}_{{\lambda_{t}}^{'}{\lambda_{W^+}}},\end{split}
\end{equation*}

where the inclusion of the spin density matrix ensures the correct correlation
between the top quark decay products and the beam.

A spin density matrix is calculated for the $W^+$ only, because the $b$ is stable
\begin{equation*}
\begin{split}\rho^{W^+}_{{\lambda_{W^+}}{\lambda_{W^+}}^{'}} = \frac{1}{N}
\rho^{t}_{{\lambda_{t}}{\lambda_{t}}^{'}}
{\mathcal{M}}^{t {\rightarrow}bW^+}_{{\lambda_{t}}{\lambda_{W^+}}}
{\mathcal{M}}^{*t {\rightarrow}bW^+}_{{\lambda_{t}}{\lambda_{W^+}}},\end{split}
\end{equation*}

and the $W^+$ decayed in the same manner as the top quark. Here the
inclusion of the spin density matrix ensures the correct correlations
between the $W^+$ decay products, the beam and the bottom quark.

The decay products of the $W^+$ are stable fermions so the decay
chain terminates here and a decay matrix for the $W^+$
\begin{equation*}
\begin{split}D^{W^+}_{{\lambda_{W^+}}{\lambda_{W^+}}^{'}} = \frac{1}{N}
{\mathcal{M}}^{t{\rightarrow}bW^+}_{{\lambda_{t}}{\lambda_{W^+}}}
{\mathcal{M}}^{*t{\rightarrow}bW^+}_{{\lambda_{t}}{\lambda_{W^+}}},\end{split}
\end{equation*}

is calculated. Moving back up the chain a decay matrix for the top quark
is calculated using the decay matrix of the $W^+$
\begin{equation*}
\begin{split}D^{t}_{{\lambda_{t}}{\lambda_{t}}^{'}} = \frac{1}{N}
{\mathcal{M}}^{t{\rightarrow}bW^+}_{{\lambda_{t}}{\lambda_{W^+}}}
{\mathcal{M}}^{*t{\rightarrow}bW^+}_{{\lambda_{t}}^{'}{\lambda_{W^+}}^{'}}
D^{W^+}_{{\lambda_{W^+}}{\lambda_{W^+}}^{'}}.\end{split}
\end{equation*}

Since the top quark came from the hard scattering process we must now deal
with the $\bar{t}$ in a similar manner but instead of using
$\delta_{\lambda_t \lambda'_t}$ when calculating the initial spin
density matrix, the decay matrix of the top quark is used and the
$\bar{t}$ decay is generated accordingly. The density matrices
pass information from one decay chain to the associated chain thereby
preserving the correct correlations.

\begin{figure}[tp]
\centering
\capstart
\noindent\includegraphics{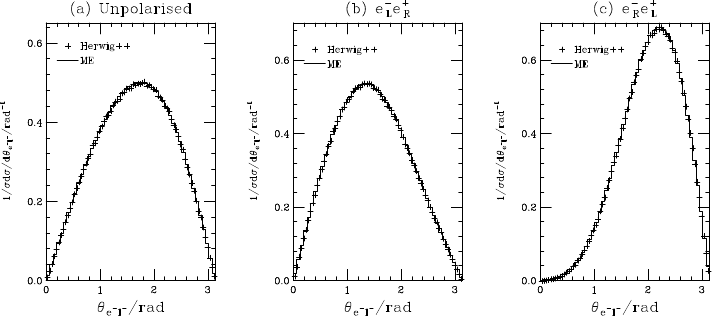}
\caption{Angle between the beam and the outgoing lepton in $e^+e^-\rightarrow
t\bar{t}\rightarrow b\bar{b} l^{+}\nu_l l^{-}\bar{\nu_l}$ in the lab frame for
a centre-of-mass energy of 500 GeV with (a) unpolarized incoming beams,
(b) negatively polarized electrons and positively polarized positrons and
(c) positively polarized electrons and negatively polarized electrons.
The data points show the results of the simulation as production and
decay including spin correlations, while the histograms use the
full matrix element for $e^+e^-\rightarrow t\bar{t}\rightarrow b\bar{b} l^{+}\nu_l l^{-}\bar{\nu_l}$.
Reproduced from \cite{Gigg:2007cr}.}\label{\detokenize{review/showers/perturbative_decays:id4}}\label{\detokenize{review/showers/perturbative_decays:fig-bsm-sub-e-beam}}\end{figure}

The production and decay of the top quark, using the spin correlation
algorithm, is demonstrated in \hyperref[\detokenize{review/showers/perturbative_decays:fig-bsm-sub-e-beam}]{Fig.\@ \ref{\detokenize{review/showers/perturbative_decays:fig-bsm-sub-e-beam}}}, \hyperref[\detokenize{review/showers/perturbative_decays:fig-bsm-sub-e-top}]{Fig.\@ \ref{\detokenize{review/showers/perturbative_decays:fig-bsm-sub-e-top}}} and \hyperref[\detokenize{review/showers/perturbative_decays:fig-bsm-sub-ee}]{Fig.\@ \ref{\detokenize{review/showers/perturbative_decays:fig-bsm-sub-ee}}}.
The hard scattering process and subsequent
decays were generated using the general matrix elements described in
\hyperref[\detokenize{review/index:sect-bsm}]{Section \ref{\detokenize{review/index:sect-bsm}}} rather than the default ones. The results from
the full matrix element calculation are also included to show that the
algorithm has been correctly implemented. The separate plots illustrate
the different stages of the algorithm at work.
\hyperref[\detokenize{review/showers/perturbative_decays:fig-bsm-sub-e-beam}]{Fig.\@ \ref{\detokenize{review/showers/perturbative_decays:fig-bsm-sub-e-beam}}} gives the angle between the beam and the
outgoing lepton. The results from the simulation agree well with the
full matrix element calculation, which demonstrates the consistency of
the algorithm for the decay of the $\bar{t}$.

\hyperref[\detokenize{review/showers/perturbative_decays:fig-bsm-sub-e-top}]{Fig.\@ \ref{\detokenize{review/showers/perturbative_decays:fig-bsm-sub-e-top}}} gives the angle between the top quark
and the produced lepton. This shows the same agreement as the previous
figure and demonstrates the correct implementation of the spin density
matrix for the $\bar{t}$ decay. Finally,
\hyperref[\detokenize{review/showers/perturbative_decays:fig-bsm-sub-ee}]{Fig.\@ \ref{\detokenize{review/showers/perturbative_decays:fig-bsm-sub-ee}}} gives the results for the angle between the
final-state lepton/anti-lepton pair showing the correct implementation
of the decay matrix that encodes the information about the
$\bar{t}$ decay. Distributions for various processes within the
Minimal Supersymmetric Standard Model and for tau production in Higgs boson
decay are shown in Refs. \cite{Grellscheid:2007tt, Gigg:2007cr}.

\begin{figure}[tp]
\centering
\capstart
\noindent\includegraphics{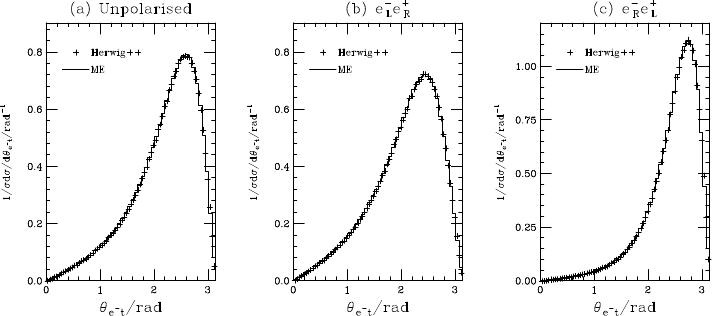}
\caption{Angle between the lepton and the top quark in
$e^+e^-\rightarrow t\bar{t}\rightarrow b\bar{b} l^{+}\nu_l l^{-}\bar{\nu_l}$ in the lab frame for
a centre-of-mass energy of 500 GeV with (a) unpolarized incoming beams,
(b) negatively polarized electrons and positively polarized positrons and
(c) positively polarized electrons and negatively polarized electrons.
The data points show the results of the simulation as production and
decay including spin correlations, while the histograms use the
full matrix element for $e^+e^-\rightarrow t\bar{t}\rightarrow b\bar{b} l^{+}\nu_l l^{-}\bar{\nu_l}$.
Reproduced from \cite{Gigg:2007cr}.}\label{\detokenize{review/showers/perturbative_decays:id5}}\label{\detokenize{review/showers/perturbative_decays:fig-bsm-sub-e-top}}\end{figure}

\begin{figure}[tp]
\centering
\capstart
\noindent\includegraphics{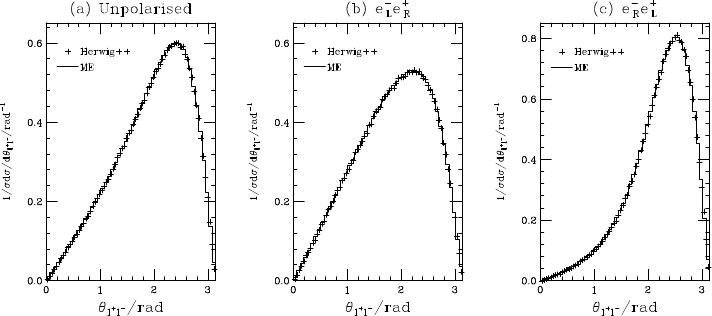}
\caption{Angle between the outgoing lepton and anti-lepton in
$e^+e^-\rightarrow t\bar{t}\rightarrow b\bar{b} l^{+}\nu_l l^{-}\bar{\nu_l}$ in the lab frame for
a centre-of-mass energy of 500 GeV with (a) unpolarized incoming beams,
(b) negatively polarized electrons and positively polarized positrons and
(c) positively polarized electrons and negatively polarized electrons.
The data points show the results of the simulation as production and
decay including spin correlations, while the histograms use the
full matrix element for $e^+e^-\rightarrow t\bar{t}\rightarrow b\bar{b} l^{+}\nu_l l^{-}\bar{\nu_l}$.
Reproduced from \cite{Gigg:2007cr}.}\label{\detokenize{review/showers/perturbative_decays:id6}}\label{\detokenize{review/showers/perturbative_decays:fig-bsm-sub-ee}}\end{figure}

The same algorithm is used regardless of how the particles are produced,
in order to consistently implement the spin correlations in all stages
of the event generation process.

The important perturbative decays in the Standard Model are implemented
in dedicated \href{https://thepeg.hepforge.org/doxygen/classThePEG\_1\_1Decayer.html}{Decayer}
classes used in Herwig to perform the decays of the fundamental
Standard Model particles. The decays of fundamental
particles in new physics models are performed using
\texttt{Decayer}
classes based on the spin structures, as described in \hyperref[\detokenize{review/BSM:sec-bsmdecays}]{Section \ref{\detokenize{review/BSM:sec-bsmdecays}}}.

\subsection{Intrinsic transverse momentum}
\label{\detokenize{review/showers/intrinsic:intrinsic-transverse-momentum}}\label{\detokenize{review/showers/intrinsic:shower-intrinsic}}\label{\detokenize{review/showers/intrinsic::doc}}

The formalism of collinear factorisation assumes that
incoming partons within colliding hadrons are exactly
collinear to their parent beam.
As a high-energy approximation, this is extremely successful, but enforcing exact collinearity
is incompatible with the uncertainty principle.
Simple arguments predict intrinsic transverse momentum of a few hundred MeV.
Within Herwig this is modelled as a tunable distribution
with simple ansätze,
whose precise form depends on the shower used.

\subsubsection{Angular-Ordered shower}
\label{\detokenize{review/showers/intrinsic:angular-ordered-shower}}\label{\detokenize{review/showers/intrinsic:ao-shower-int-trans-mom}}

In the angular-ordered parton shower an intrinsic transverse momentum is added to
the initial-state parton that results from the backward evolution in the
initial-state parton shower, \textit{i.e.} the parton extracted directly from the incoming hadron.
This is then included in the standard kinematic reconstruction described in
\hyperref[\detokenize{review/showers/qtilde:sect-initialrecon}]{Section \ref{\detokenize{review/showers/qtilde:sect-initialrecon}}}.

We model this using a non-perturbative distribution.  Either a Gaussian
\begin{equation*}
\begin{split}\mathrm{d}^2p_\perp \; \frac{(1-\beta)}{\pi{p^G_\perp}^2}
  \exp\left[-\frac{p_\perp^2}{{p^G_\perp}^2}\right],\end{split}
\end{equation*}
an inverse quadratic
\begin{equation*}
\begin{split}\mathrm{d}^2p_\perp\;\beta
 \frac1{\pi\ln\left(1+\frac{p_{T_{\rm{max}}}^2}{\Gamma^2}\right)}\frac{1}{\Gamma^2+p_\perp^2} \,,\end{split}
\end{equation*}
or a combination of both distributions can be used.
The fraction of the inverse quadratic distribution is $\beta$ (\href{https://herwig.hepforge.org/doxygen/QTildeShowerHandlerInterfaces.html\#IntrinsicPtBeta}{IntrinsicPtBeta}), $p_{\perp}$ is the generated
intrinsic transverse momentum, $p^G_\perp$ is the root-mean squared of the Gaussian distribution (\href{https://herwig.hepforge.org/doxygen/QTildeShowerHandlerInterfaces.html\#IntrinsicPtGaussian}{IntrinsicPtGaussian}),
and $\Gamma$ controls the shape of the inverse quadratic distribution (\href{https://herwig.hepforge.org/doxygen/QTildeShowerHandlerInterfaces.html\#IntrinsicPtGamma}{IntrinsicPtGamma}).

By default $\beta=0$, such that a Gaussian distribution is used.

\subsubsection{Dipole shower}
\label{\detokenize{review/showers/intrinsic:dipole-shower}}\label{\detokenize{review/showers/intrinsic:dipole-shower-int-trans-mom}}

Just as for the reshuffling to constituent masses, within the dipole shower
there is no means of adding intrinsic transverse momentum simply on top of the
evolution, and the respective transformation is generated for the initial-state partons at the end of the evolution.

For each of the incoming partons a $p_\perp$ ‘kick’ with magnitude
generated according to the distribution
\begin{equation*}
\begin{split}\frac{{\rm d}P}{{\rm d}p_{\perp}^2} = \frac{1}{2 p_{\perp,S}^2}
\exp\left(-\frac{p_\perp^2}{2 p_{\perp,S}^2}\right)\end{split}
\end{equation*}

and a flat distribution in azimuth is selected. The labels $S=F,A$ here
refer to whether the incoming parton is valence or sea, respectively. The
motivation behind this distinction originates in the phenomenological
assumption that gluons experience a larger transverse spread than (valence)
quarks essentially by a factor of $\sqrt{C_A/C_F}$ though detailed
studies have never been undertaken within this implementation and the values
are set to be equal by default.

The DipoleShowerHandler class delegates this task to an \href{https://herwig.hepforge.org/doxygen/classHerwig\_1\_1IntrinsicPtGenerator.html}{IntrinsicPtGenerator}
class, which implements the procedure outlined above with two \href{https://herwig.hepforge.org/doxygen/IntrinsicPtGeneratorInterfaces.html}{interfaces}
for the mean values $p_{\perp,F}$ (\href{https://herwig.hepforge.org/doxygen/IntrinsicPtGeneratorInterfaces.html\#ValenceIntrinsicPtScale}{ValenceIntrinsicPtScale}) and $p_{\perp,A}$ (\href{https://herwig.hepforge.org/doxygen/IntrinsicPtGeneratorInterfaces.html\#SeaIntrinsicPtScale}{SeaIntrinsicPtScale}), respectively.

\subsection{Forced splitting}
\label{\detokenize{review/showers/forced:forced-splitting}}\label{\detokenize{review/showers/forced::doc}}

After the perturbative shower evolution has terminated, the cluster
hadronization model may necessitate some additional \textit{forced splitting}
of the initial-state parton that results, \textit{i.e.} the parton that extracted from the hadron which
is produced by the backward evolution in the
initial-state parton shower. In hadronic collisions we
require the external initial-state partons, which give rise to the first
hard interaction, to be valence quarks (antiquarks), \textit{i.e.} colour triplet (antitriplet)
states. This allows us to treat each proton (antiproton) remnant as a
diquark (antidiquark) which will be in a colour antitriplet/triplet
state, in order to keep the incoming hadron colour neutral. Modelling
the dissociation in this way allows for a simple, minimal, hadronization
of the remnant in the cluster hadronization model.

Usually, the perturbative evolution, which is guided by the PDFs, will
terminate on a valence quark, since the evolution works towards large
$x$ and small $Q^2$. In the cases where this has not
happened, we force the resulting initial-state parton to undergo one or
two additional splittings. The generation of these additional forced
splittings is largely based on the same principles as that of the
perturbative splittings.

In the perturbative evolution the scale of the PDFs is frozen at a value
$Q_s$ for values $Q<Q_s$. The default value of $Q_s$
is chosen to be small, close to the non--perturbative region but still
above typical values for the parton shower cutoff (\href{https://herwig.hepforge.org/doxygen/ShowerHandlerInterfaces.html\#PDFFreezingScale}{PDFFreezingScale=2.5 GeV}). This freezing
scale leaves a little phase-space for the (non--perturbative) forced
splittings. The forced splittings are generated in much the same vein as
the perturbative splittings. The evolution starts at $Q_s$ and the
next branching scale is distributed according to $\mathrm{d}Q/Q$,
with a lower limit determined by the available phase-space. The
$z$ values are determined from the splitting functions in the same
way as in the perturbative evolution. The splittings are reweighted by
ratios of PDFs as in the perturbative evolution. There is only one
slight difference, the evolution of the PDFs themselves with $Q$
is frozen below $Q_s$. Nevertheless, this reweighting gives the
right flavour content of the initial hadron. For example in the case of
a proton we produce twice as many $u$ quarks as $d$ quarks.
To force the evolution to end up on a valence quark, we only allow one
or two flavours in the evolution:
\begin{enumerate}
\sphinxsetlistlabels{\arabic}{enumi}{enumii}{}{.}%
\item {} 

if the initial parton is a sea quark ($q$) or antiquark
($\bar q$), it is forced to evolve into a gluon, emitting a $\bar q$
or $q$, respectively.

\item {} 

if the initial parton is a gluon, from either the perturbative
evolution or the forced splitting of a seaquark, it is forced to
evolve into a valence quark, emitting the matching antiquark.

\end{enumerate}

In the initial-state showering of additional hard scatters we force the
backward evolution of the colliding partons to terminate on a gluon. We
therefore only need the first kind of forced splitting in this case.
This gluon is assumed to be relatively soft and branches from the
remnant diquark. Again, this allows us to uniquely match up the
final-state partons to the cluster hadronization model.
The emitted partons from these forced splittings, as well as the
remnant diquarks, will show up in the event record as decay products of
a fictitious remnant particle, in order to distinguish them from those
which originate from the perturbative evolution. Additional details
about the colour structure and the event record can be found in
\cite{Bahr:2008dy}.

\subsection{YFS-based QED radiation}
\label{\detokenize{review/showers/qed:yfs-based-qed-radiation}}\label{\detokenize{review/showers/qed:sect-yfs}}\label{\detokenize{review/showers/qed::doc}}

While the angular-ordered parton shower includes QED radiation
we also have the option of simulating QED radiation using the approach of Ref.
\cite{Hamilton:2006ms}. This is included for both particle decays and
unstable $s$-channel resonances produced in the hard process. This
approach is based on the Yennie--Frautschi--Suura (YFS) formalism \cite{Yennie:1961ad}, which
takes into account large double- and single- soft photon logarithms to
all orders. In addition, the leading collinear logarithms are included
to $\mathcal{O}\left(\alpha\right)$ by using the dipole splitting
functions. By default the production of QED radiation is switched off
for both decays and hard processes. It may be included by using the
\href{https://herwig.hepforge.org/doxygen/classHerwig\_1\_1QEDRadiationHandler.html}{QEDRadiationHandler}
in the
\texttt{EventHandler}
as one of the
\href{https://thepeg.hepforge.org/doxygen/EventHandlerInterfaces.html\#PostSubProcessHandlers}{PostSubProcessHandlers}
for the hard process or using the
\href{https://herwig.hepforge.org/doxygen/DecayIntegratorInterfaces.html\#PhotonGenerator}{PhotonGenerator}
interface of the relevant Decayer inheriting from the
\href{https://herwig.hepforge.org/doxygen/classHerwig\_1\_1DecayIntegrator.html}{DecayIntegrator}
class for the decays.

\subsection{Shower variations and reweighting}
\label{\detokenize{review/showers/variations:shower-variations-and-reweighting}}\label{\detokenize{review/showers/variations:sect-shvar}}\label{\detokenize{review/showers/variations::doc}}

For a general parton shower algorithm there are a number of components that
dictate its behaviour. The primary components of a particular algorithm are its
splitting kernels, evolution variable and phase-space constraints. Beyond this is
the dependence on the unphysical scales introduced by the truncation of the
expansion series. The variation of these scales in the parton shower serves to
quantify missing higher-logarithmic contributions. The typical recipe to assess
these perturbative uncertainties is to vary the chosen scales by factors of
$1/2$ and $2$ around the central value as an estimate of these
missing contributions.

Typically, these variations are performed via independent runs of the Monte Carlo
after varying the scales detailed in \hyperref[\detokenize{review/showers/variations:shower-scale-variations}]{Section \ref{\detokenize{review/showers/variations:shower-scale-variations}}}. However, it is
also possible to use shower reweighting, as described in \hyperref[\detokenize{review/showers/variations:shower-reweight}]{Section \ref{\detokenize{review/showers/variations:shower-reweight}}},
to perform these variations during a single run. The shower reweighting
framework can also be employed for other applications, which are detailed in
\hyperref[\detokenize{review/showers/variations:shower-reweight-examples}]{Section \ref{\detokenize{review/showers/variations:shower-reweight-examples}}}.

\subsubsection{Shower scale variations}
\label{\detokenize{review/showers/variations:shower-scale-variations}}\label{\detokenize{review/showers/variations:id1}}

There are three main scales that can be varied within a parton shower algorithm.
The first two are the familiar renormalization and factorization scales, which,
respectively, set the argument $\alpha_\text{S}$ and the PDFs in the
shower. The other scale is the hard-veto scale, which is the characteristic scale
that bounds the hardness of emissions. Each of these scales can be varied
independently of the others.

For both parton shower algorithms the values of the rescaling factors
are controlled by the \href{https://herwig.hepforge.org/doxygen/ShowerHandlerInterfaces.html\#RenormalizationScaleFactor}{RenormalizationScaleFactor},  \href{https://herwig.hepforge.org/doxygen/ShowerHandlerInterfaces.html\#FactorizationScaleFactor}{FactorizationScaleFactor} and \href{https://herwig.hepforge.org/doxygen/ShowerHandlerInterfaces.html\#HardScaleFactor}{HardScaleFactor} interfaces of the \href{https://herwig.hepforge.org/doxygen/classHerwig\_1\_1ShowerHandler.html}{ShowerHandler} class. In addition for processes simulated using the POWHEG scheme using Matchbox with the angular-ordered parton shower the \texttt{HardScaleFactor}  also needs to be set in the matching.

Additionally, one can also vary the infrared cutoff of the parton shower as an
indication of regions expected to be sensitive to the interplay between the
parton shower and hadronization.

\subsubsection{Profile scale choices}
\label{\detokenize{review/showers/variations:profile-scale-choices}}
\begin{figure}[tp]
\centering
\capstart
\noindent\includegraphics[width=0.600\linewidth]{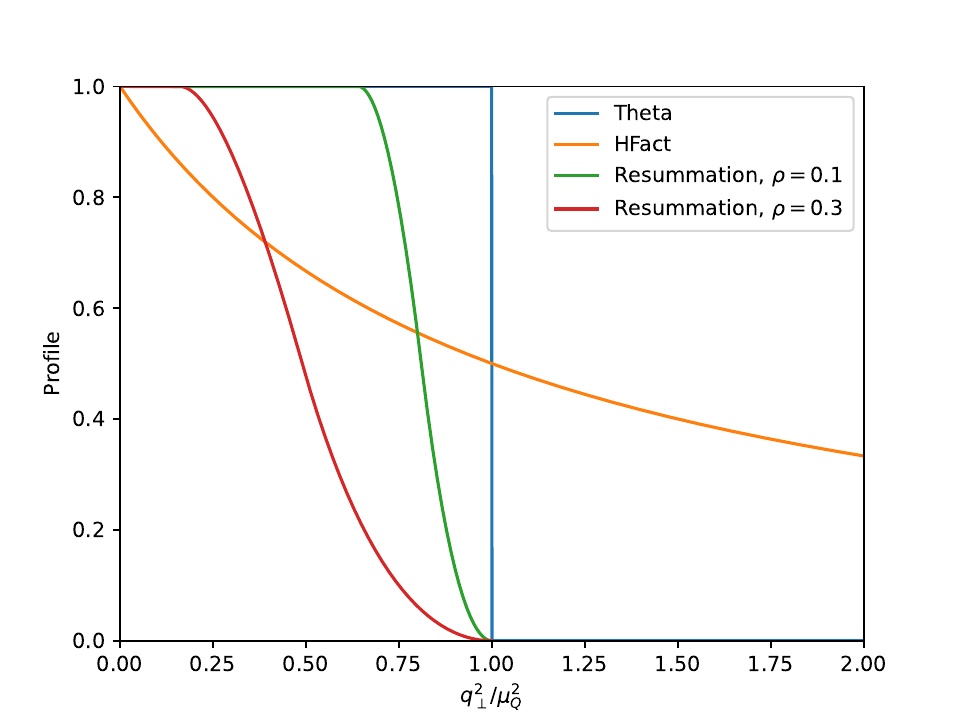}
\caption{Comparison of the different profile scales.}\label{\detokenize{review/showers/variations:id9}}\label{\detokenize{review/showers/variations:fig-profile}}\end{figure}

Within the parton shower, we include the freedom to weight emissions with an
arbitrary function $\kappa$, the \textit{profile scale choice}. This
function modifies the Sudakov form factor such that
\begin{equation*}
\begin{split}-\ln \Delta_{K_\perp^2}(p_\perp^2|\mu_Q^2) =
\int_{p_\perp^2}^{R_\perp^2} \frac{{\rm d}q_\perp^2 }{q_\perp^2}
\kappa(\mu_Q^2,q_\perp^2)\int {\rm d}z \, P_{K_\perp^2}(q^2_\perp,z),\end{split}
\end{equation*}

where $R_\perp$ is the scale at which all of the phase-space is available
to emissions, and $K_\perp$ gives the boundary condition for $z$.

In order for the shower to retain the desired resummation properties the profile scales must
satisfy:
\begin{itemize}
\item {} 

$\kappa(\mu_Q^2,p_\perp^2) \to 1$ as $p_\perp^2 \ll \mu_Q^2$;

\item {} 

$\kappa(\mu_Q^2,p_\perp^2) \to 0$ as $p_\perp^2 \sim R_\perp^2 \gg \mu_Q^2$.

\end{itemize}

These conditions ensure that the small and large $p_\perp$ limits are
retained.

Beyond these constraints we also require that $K_\perp^2 \sim \mu_Q^2$,
such that the correct logarithmic resummation is obtained, and that
$\kappa(\mu_Q^2,p_\perp^2)$ is slowing varying when $p_\perp$ is far
from $\mu_Q$ such that terms involving the derivative of the profile scale
are subleading.

In Herwig a number of choices are implemented:
\begin{itemize}
\item {} 

the \href{https://herwig.hepforge.org/doxygen/HardScaleProfileInterfaces.html\#ProfileType}{Theta} profile scale
implements a hard cutoff at the veto scale, and has the functional form
\begin{equation*}
\begin{split}\kappa(\mu_Q^2,q_\perp^2) = \theta(\mu_Q^2 - q_\perp^2);\end{split}
\end{equation*}
\item {} 

the \href{https://herwig.hepforge.org/doxygen/HardScaleProfileInterfaces.html\#ProfileType}{Resummation} profile
is similar to the \texttt{Theta}  profile,
but smoothly approaches the cutoff via quadratic interpolation
\begin{equation*}
\begin{split}\kappa(\mu_Q^2,q_\perp^2) =
\left\{
\begin{array}{ll}
1 & q_\perp/\mu_Q \le 1-2\rho, \\
1 - \frac{(1-2\rho-q_\perp/\mu_Q)^2}{2\rho^2} & q_\perp/\mu_Q \in
(1-2\rho,1-\rho], \\
\frac{(1-q_\perp/\mu_Q)^2}{2\rho^2}& q_\perp/\mu_Q \in
(1-\rho,1], \\
0 & q_\perp/\mu_Q > 1,
\end{array}
\right.\end{split}
\end{equation*}

instead of a hard cutoff, where the $\rho$ \href{https://herwig.hepforge.org/doxygen/HardScaleProfileInterfaces.html\#ProfileRho}{parameter}
controls the region over which the profile changes. For $\rho=0$ the Resummation and Theta profiles
are equivalent;

\item {} 

rather than the cutoff type behaviour of the previous two profiles, the
\href{https://herwig.hepforge.org/doxygen/HardScaleProfileInterfaces.html\#ProfileType}{HFact}
profile dampens emissions around the veto scale, and allows emissions above the
veto scale. In this profile, the veto scale is the scale at which
$\kappa=1/2$. The \texttt{HFact}
profile is described by the equation
\begin{equation*}
\begin{split}\kappa(\mu_Q^2,q_\perp^2) = \left(1+q_\perp^2/\mu_Q^2\right)^{-1}.\end{split}
\end{equation*}
\item {} 

the final profile scale choice corresponds to the Power shower, \textit{i.e.} it
imposes no constraints on emissions aside from those inherent to the shower algorithm
\begin{equation*}
\begin{split}\kappa(\mu_Q^2,q_\perp^2) = 1,\end{split}
\end{equation*}

it is equivalent to having no profile scale and is therefore used
by setting the \href{https://herwig.hepforge.org/doxygen/ShowerHandlerInterfaces.html\#HardScaleProfile}{HardScaleProfile}
to NULL.

\end{itemize}

A comparison of the different profile scale choices is shown in \hyperref[\detokenize{review/showers/variations:fig-profile}]{Fig.\@ \ref{\detokenize{review/showers/variations:fig-profile}}} and
their impact on some observables can be found in a phenomenological study \cite{Bellm:2016rhh}.

In addition both the HFact and Power choices require that restrictions on the phase-space
for emission in the parton shower are switched off, \textit{i.e.}
\href{https://herwig.hepforge.org/doxygen/ShowerHandlerInterfaces.html\#RestrictPhasespace}{RestrictPhasespace=Off}.

\subsubsection{Parton shower variations reweighting}
\label{\detokenize{review/showers/variations:parton-shower-variations-reweighting}}\label{\detokenize{review/showers/variations:shower-reweight}}

The method typically used to evaluate uncertainties arising from scale choices in the parton shower
is to perform a full Monte Carlo simulation of the process in question for each set of
scales of interest. Clearly, this method becomes very computationally intensive if there are multiple
scale variations to be evaluated, as is often the case in a phenomenological study.
Alternatively, the same scale variations can be evaluated by performing a single run
using a central set of scales and applying a reweighting factor, calculated for each set of
scale variations, to the central result. Using this method, the hard process evaluation
and hadronization steps are performed only once, which can lead to a significant
reduction in runtime compared to performing many separate runs. Automated implementations
of this approach are also available in \textsc{Pythia~8}~\cite{Mrenna:2016sih} and \textsc{Sherpa}~\cite{Bothmann:2016nao}.

In Herwig 7, we have implemented reweighting for variations of the renormalisation and
factorisation scales used in the shower. 
The automated scale reweighting varies these scales only by multiplicative
factors relative to their central values; it does not change their functional
dependence on the shower kinematics. Variations of the functional form require
a separate shower configuration.
We evolve a reweighting factor for each set
of scales on a splitting-by-splitting basis such that when it is applied to the result
from the central run, we produce the distribution expected from showering using the
scales of interest. In \hyperref[\detokenize{review/showers/variations:sect-weighted-sudakov}]{Section \ref{\detokenize{review/showers/variations:sect-weighted-sudakov}}}
we first review the standard Sudakov veto algorithm
used in parton showers and then present the modified veto algorithm for performing shower
reweighting. Additionally, in \hyperref[\detokenize{review/showers/variations:sect-veto-application}]{\ref{\detokenize{review/showers/variations:sect-veto-application}}}, we discuss the key points
relevant to the implementation of this algorithm in Herwig 7. More details of
our approach can be found in Ref. \cite{Bellm:2016voq}.

\paragraph{The weighted Sudakov veto algorithm}
\label{\detokenize{review/showers/variations:the-weighted-sudakov-veto-algorithm}}\label{\detokenize{review/showers/variations:sect-weighted-sudakov}}

Parton shower evolution is performed using the Sudakov veto algorithm,
see \hyperref[\detokenize{review/appendix/sudakov:sect-sudakov-solution}]{ \ref{\detokenize{review/appendix/sudakov:sect-sudakov-solution}}} for more details and references.
We start showering from a given starting scale $Q$ and need to generate the scale
of the next emission $q$ and the $d$ additional associated splitting
variables $x$. These are generated according to the distribution
\begin{equation}\label{equation:review/showers/variations:eq:ShVar:probDist}
\begin{split}{\rm d}S_P (\mu, x_\mu|q,x|Q) &=  {\rm d}q\ {\rm d}^dx  \biggl[ \Delta_P(\mu|Q) \, \delta(q-\mu) \,\delta(x-x_\mu)\\
&  \phantom{ {\rm d}q\ {\rm d}^dx \biggl[} + P(q,x) \, \theta(Q-q) \, \theta(q-\mu) \, \Delta_P(q|Q) \biggr],\end{split}
\end{equation}

where $x_\mu$ is a parameter point associated to the infra-red cutoff scale $\mu$,
the splitting kernel is $P(q,x)$ and the Sudakov form factor is
\begin{equation*}
\begin{split}\Delta_P(q|Q) = \exp\left(-\int_q^Q {\rm d}k \int {\rm d}^dz\ P(k,z)\right) \ .\end{split}
\end{equation*}
The distribution $S_P$ is normalized to unity.

We consider an overestimate of the kernel, $R(q,x)$, such that
\begin{equation}\label{equation:review/showers/variations:eq:ShVar:overest}
\begin{split}R(q,x) \geq P(q,x) \ \ \ \ \forall\ \ \  q,x \ .\end{split}
\end{equation}

The overestimate is chosen to be integrable and invertible so that
$q,x$ can easily be generated according to the overestimated distribution
\begin{equation*}
\begin{split}{\rm d}S_R (\mu,x_\mu|q,x|Q) &= {\rm d}q\ {\rm d}^dx \biggl[ \Delta_R(\mu|Q) \, \delta(q-\mu) \, \delta(x-x_\mu) \\
                             & \phantom{{\rm d}q\ {\rm d}^dx \biggl[} + R(q,x) \, \theta(Q-q) \, \theta(q-\mu) \,\Delta_R(q|Q)\biggr],\end{split}
\end{equation*}
with a Sudakov form factor
\begin{equation*}
\begin{split}\Delta_R(q|Q) = \exp\left(-\int_q^Q {\rm d}k \int {\rm d}^dz\ R(k,z)\right) \ .\end{split}
\end{equation*}

Starting at a scale $k=Q$, the standard Sudakov veto algorithm proceeds as follows:
\begin{enumerate}
\sphinxsetlistlabels{\arabic}{enumi}{enumii}{}{.}%
\item {} 

a trial splitting scale and variables, $q,x$, are generated according to
$S_R (\mu,x_\mu|q,x|k)$;

\item {} 

if the scale $q=\mu$ then there is no emission and the cut-off scale,
$\mu$, and associated parameter point $x_\mu$ are returned;

\item {} 

The trial scale and splitting variables are accepted with probability
\begin{equation*}
\begin{split}\frac{P(q,x)}{R(q,x)} \ ,\end{split}
\end{equation*}
otherwise the process is repeated with $k=q$.

\end{enumerate}

This algorithm reproduces the distribution in Eq. \eqref{equation:review/showers/variations:eq:ShVar:probDist}, see Refs. \cite{Buckley:2011ms, Platzer:2011dq}.

The standard veto algorithm presented above can be generalised, see
Refs. \cite{Bellm:2016voq, Platzer:2011dq, Hoeche:2009xc, Lonnblad:2012hz, Mrenna:2016sih} to include weights
and to relax the requirements so that $P$ is not required to be positive and
removing the restriction on $R$, Eq. \eqref{equation:review/showers/variations:eq:ShVar:overest}. We note that $S_P$
is still normalized to unity. To achieve this generalization we
introduce an acceptance probability $\epsilon(q,x|k,y)$ such that
\begin{equation*}
\begin{split}0 \le \epsilon(q,x|k,y) < 1 \ .\end{split}
\end{equation*}

We start with a weight $w=1$, which will be evolved on a splitting-by-splitting basis. Again we start at a scale $k=Q$ and the modified veto algorithm proceeds as follows:
\begin{enumerate}
\sphinxsetlistlabels{\arabic}{enumi}{enumii}{}{.}%
\item {} 

A trial splitting scale and variables, $q,x$, are generated according to
$S_R (\mu,x_\mu|q,x|k)$;

\item {} 

If the scale $q=\mu$ then there is no emission and the cut-off scale,
$\mu$, and associated parameter point $x_\mu$ are returned with weight $w$;

\item {} 

The trial splitting variables $q,x$ are accepted with probability
$\epsilon(q,x|k,y)$ and the returned weight is
\begin{equation}\label{equation:review/showers/variations:eq:ShVar:weightAccept}
\begin{split}w \times \frac{1}{\epsilon(q,x|k,y)} \times \frac{P(q,x)}{R(q,x)} \ ;\end{split}
\end{equation}
\item {} 

otherwise the weight becomes
\begin{equation}\label{equation:review/showers/variations:eq:ShVar:weightReject}
\begin{split}w\times \frac{1}{1-\epsilon(q,x|k,y)}\times \left(1-\frac{P(q,x)}{R(q,x)}\right) \ ,\end{split}
\end{equation}

and the algorithm continues with $k=q$.

\end{enumerate}

The proof that this algorithm produces the correct probability distribution is given in Ref. \cite{Bellm:2016voq}.

\paragraph{Application of the modified veto algorithm}
\label{\detokenize{review/showers/variations:application-of-the-modified-veto-algorithm}}\label{\detokenize{review/showers/variations:sect-veto-application}}

The primary motivation for the development of the weighted veto algorithm presented
above is to enable more efficient evaluation of shower uncertainties
due to scale choices. Using the modified veto algorithm we can perform a shower
splitting using a ‘default’ splitting kernel and at the same time calculate
the weights for one or several different splitting kernels. In order to evaluate scale
variations, we choose our ‘default’ kernel, $P_0 (q,x)$, to be a given splitting
kernel evaluated at some chosen ‘central’ renormalization and factorisation scales,
$\mu_{\rm{R},0} \text{ and } \mu_{\rm{F},0}$. We generate splitting variables
for the shower with this central scale choice, therefore the overestimate, $R (q,x)$,
is necessarily calculated with these scales. We then choose the acceptance probability to be
\begin{equation}\label{equation:review/showers/variations:eq:acceptance:prob}
\begin{split}\epsilon(q,x|k,y) = \frac{P_0(q,x)}{R(q,x)}.\end{split}
\end{equation}

With this choice the central, unweighted, result is identical to the result obtained using the standard
veto algorithm for the default kernel. If we denote the kernel using a different set of scales,
$\mu_{\rm{R},i} \text{ and } \mu_{\rm{F},i}$, as $P_i (q,x)$, we can then evaluate the weight
for this variation by substituting $P(q,x) \to P_i (q,x)$ in Eqs. \eqref{equation:review/showers/variations:eq:ShVar:weightAccept} and \eqref{equation:review/showers/variations:eq:ShVar:weightReject}.

There are several additional potential uses of the weighted Sudakov veto algorithm which
we will not discuss in detail here. One example is its application to the generation of
shower emissions using more complicated, potentially negative, splitting kernels.

There are two considerations that a user should be aware of when
using the reweighting approach to evaluate shower uncertainties.

The first consideration is that it should take less time to calculate the results of
a given set of scale variations using shower reweighting than by performing a separate
event generator run for each scale choice. In cases where the other stages of event generation take
significantly more time to run than the shower generation, reweighting should give a sizeable
time saving. However, if the process of interest is simple and the hard process simulation is
quick, the two approaches can be comparable.

When evaluating uncertainties in the dipole shower, it is virtually always quicker to use
reweighting than to perform separate runs, regardless of the complexity of the process.
When using the angular-ordered shower, however, it can be the case that for very simple hard
processes the time consumed using the reweighting method is comparable to that used
by generating separate runs. This is due to differences in the sampling methods used in the two showers.

The dipole shower uses an adaptive-sampling method in which only one acceptance probability is calculated
for each trial splitting. The angular-ordered parton shower uses a method in which the computation
of the acceptance probability, Eq. \eqref{equation:review/showers/variations:eq:acceptance:prob} is split into several parts,
as described in \hyperref[\detokenize{review/appendix/sudakov:sect-sudakov-solution}]{\ref{\detokenize{review/appendix/sudakov:sect-sudakov-solution}}}.
As an example, for space-like evolution
the splitting kernel is
\begin{equation*}
\begin{split}P(q,z) = \frac1q\frac{\alpha_{s}(z(1-z)\tilde{q})}{2\pi}P_{\widetilde{ij}\to ij}(z,\tilde{q})
\frac{\frac{x}zf(x/z,\tilde{q})}{xf(x,\tilde{q})} \ ,\end{split}
\end{equation*}

where $\tilde{q}$ is the angular-ordered evolution variable, $x$ is the momentum
fraction of the branching parton and $P_{\widetilde{ij}\to ij}(z,\tilde{Q})$ is the
splitting function given in Eq. \eqref{equation:review/showers/qtilde:eq:AP}. A simple overestimate can be written as
\begin{equation*}
\begin{split}R(q,z) = \frac1q\frac{\alpha^{\rm over}_S}{2\pi}P^{\rm over}_{\widetilde{ij}\to ij}(z)
{\rm PDF}^{\rm over}(z) \ ,\end{split}
\end{equation*}

where $\alpha^{\rm over}_S$, $P^{\rm over}_{\widetilde{ij}\to ij}(z)$ and
${\rm PDF}^{\rm  over}(z)$ are the overestimates of
$\alpha_{s}(z(1-z)q)$, $P_{\widetilde{ij}\to ij}(z,q)$ and $\frac{\frac{x}zf(x/z,q)}{xf(x,q)}$,
respectively.  The veto is applied separately for the three weights
\begin{equation*}
\begin{split}w_1 = \frac{\alpha_{s}(z(1-z)q)}{\alpha^{\rm over}_S} \ ,
\qquad w_2 = \frac{P_{\widetilde{ij}\to ij}(z,q)}{P^{\rm over}_{\widetilde{ij}\to ij}(z)} \ ,
 \qquad w_3 = \frac{\frac{\frac{x}zf(x/z,q)}{xf(x,q)}}{{\rm PDF}^{\rm over}(z)} \ .\end{split}
\end{equation*}

The evaluation of $w_3$ is the most time-consuming piece of the calculation. In the standard algorithm,
the calculation is organised such that $w_3$ is
only evaluated if the emission is accepted following the tests on $w_1 \text{ and } w_2$.
The reweighting method however requires that the complete weights are calculated for the central
run and all of the variation runs for each trial emission. This is because the weights are evolved according to
Eqs. \eqref{equation:review/showers/variations:eq:ShVar:weightAccept} and \eqref{equation:review/showers/variations:eq:ShVar:weightReject} for accepted and vetoed emissions respectively. If the time taken
to evaluate these weights is significant compared to the rest of the evolution it is possible that,
for the given process, the reweighting method could be slower than performing separate runs.

In practice when all parts of the simulation, e.g.\ hadronization and decays,
are included, reweighting is faster than performing separate runs even for the
majority of simple processes. Furthermore, performance will improve rapidly with
increasing complexity of the hard process.

The second point to consider is that if the variation of the weights about 1 is large for one or
more of the scale variations, then a large number of reweighted events will be required in order
for the reweighted result to converge, within an acceptable error, on that obtained by directly
simulating the events. Furthermore, there may be regions of phase-space which would be populated
with the varied scales but which are not filled for the central value, and hence have infinite
weight.

A very efficient sequence of the veto algorithm at the central scale is one where the overestimate
is very similar to the kernel, i.e.~$\epsilon(q,x|k,y) \sim 1$. It follows from
Eq. \eqref{equation:review/showers/variations:eq:ShVar:weightReject} that a very efficient central evolution is likely to significantly broaden
the weight distributions of the reweighted results. There is therefore a trade-off between achieving
an efficient algorithm for the central scale choice, which reduces run time, and producing a narrow
weight distribution for the reweighted runs, which reduces the number of events required.

We can make the veto algorithm for the central run less efficient by introducing a ‘detuning’
parameter $\lambda > 1$ which multiplies the overestimate in the denominator of the
acceptance probability, Eq. \eqref{equation:review/showers/variations:eq:acceptance:prob}.
The only requirement on the overestimate used to compute the acceptance probability is that it is
greater than the splitting kernel in the numerator at all points in phase-space. Therefore we can
increase the denominator in Eq. \eqref{equation:review/showers/variations:eq:acceptance:prob}, to reduce the efficiency of the algorithm
without affecting its validity.

Using the reweighting method, multiple scale variations can be specified and evaluated in a single
run. The \href{https://herwig.hepforge.org/doxygen/ShowerHandlerInterfaces.html\#AddVariation}{AddVariation}
interface of the \texttt{ShowerHandler} class
can be used to add variations. Each variation requires a name, NameOfVariation,
and a specification of the scale factors to
apply to the central renormalization and factorisation scales, $\mu_R$ and $\mu_F$ respectively. Finally we
can choose to apply the scale variation in the showering of the hard process only, the showering of
secondary interactions only or in both by specifying Hard, Secondary or All respectively.
The detuning parameter discussed above can be set using the
\href{https://herwig.hepforge.org/doxygen/SplittingGeneratorInterfaces.html}{Detuning} interface of
the \href{https://herwig.hepforge.org/doxygen/classHerwig\_1\_1SplittingGenerator.html}{SplittingGenerator} class
for the angular-ordered parton shower and the \texttt{Detuning}
interface of the \href{https://herwig.hepforge.org/doxygen/classHerwig\_1\_1DipoleShowerHandler.html}{DipoleShowerHandler} class
for the dipole shower.

\subsubsection{Parton shower reweighting}
\label{\detokenize{review/showers/variations:parton-shower-reweighting}}\label{\detokenize{review/showers/variations:shower-reweight-examples}}

Beyond uncertainty variations, shower vetoing and reweighting provide the
mechanisms needed to modify the shower evolution for matching and
matrix-element-correction procedures. In the angular-ordered shower, vetoes are
used, for example, to consistently shower Les Houches events for
$t\bar{t}$ production that already contain radiative corrections in the top
decay, as generated by the POWHEG BOX RES framework
\cite{Ravasio:2018lzi,Jezo:2015aia}. In the dipole shower, splitting and
cascade-reweighting factors are used by the KrkNLO matching and colour
matrix-element-correction methods.

\paragraph{Angular-Ordered shower}
\label{\detokenize{review/showers/variations:angular-ordered-shower}}

There are two mechanisms provided in the angular-ordered parton shower to allow either individual emissions, or the whole
parton shower to be vetoed so that different matching techniques can be implemented, or the event generation optimised
by requiring the shower to have specific features, for example the presence of a $b\bar{b}$ pair.
Both types of veto can be implemented by inheriting from the relevant base class and then implementing
the veto requirement.
\begin{itemize}
\item {}

\sphinxstylestrong{Shower Veto} Classes inheriting from the \href{https://herwig.hepforge.org/doxygen/classHerwig\_1\_1ShowerVeto.html}{ShowerVeto} class
only have access to the parton initiating the parton shower, the parton currently undergoing showering and the kinematics
of the trial emission. The inheriting classes have the option of vetoing the individual emission, the full parton shower, or the
whole event.

\item {}

\sphinxstylestrong{Full Shower Veto} Classes inheriting from the
\href{https://herwig.hepforge.org/doxygen/classHerwig\_1\_1FullShowerVeto.html}{FullShowerVeto} class have access to the full
event and therefore can apply a veto to the whole showering process. An example can be found in \texttt{Contrib/ShowerVeto},
which vetoes parton showers that have not produced a $b\bar{b}$ pair, so that $b\bar{b}$ final states can be efficiently
simulated.

\end{itemize}

\paragraph{Dipole shower}
\label{\detokenize{review/showers/variations:dipole-shower}}

There are two mechanisms provided in the Dipole Shower that allow for the reweighting of either a single splitting or the
whole parton shower cascade. Both mechanisms are available to classes inheriting from
the \href{https://herwig.hepforge.org/doxygen/classHerwig\_1\_1DipoleEventReweight.html}{DipoleEventReweight}.
\begin{itemize}
\item {}

\sphinxstylestrong{Splitting Reweighting} is accomplished by the \href{https://herwig.hepforge.org/doxygen/classHerwig\_1\_1DipoleEventReweight.html\#a25e857470814fa7415f05ce6b98a4591}{weight}
member function of the inheriting classes. This has access to the implementation of the strong couplings
used in the parton shower and the kinematics of both the hard process and the splitting (\textit{i.e.}, the weight can be a function of the splitting kinematics).
The weight can be applied to all or only selected branchings of the parton shower.

\item {}

\sphinxstylestrong{Cascade Reweighting} is accomplished by the \href{https://herwig.hepforge.org/doxygen/classHerwig\_1\_1DipoleEventReweight.html\#ab5158887873376038d34f26ed1276d90}{weightCascade}
member function. As in the case of Splitting Reweighting, this weight can be a function of the splitting kinematics. However, this weight is evaluated once after the complete dipole-shower cascade, including cascades with no emissions, and contributes a single multiplicative weight to the event. It is not applied separately to the initial- and final-state parts of the cascade, although the reweighting class may be configured to act on the primary hard interaction or on secondary interactions.

\end{itemize}

\subsection{Code structure}
\label{\detokenize{review/showers/code:code-structure}}\label{\detokenize{review/showers/code:sect-showercode}}\label{\detokenize{review/showers/code::doc}}

The Herwig shower module has one main
\texttt{ShowerHandler} class
which inherits from the
\href{https://thepeg.hepforge.org/doxygen/classThePEG\_1\_1CascadeHandler.html}{CascadeHandler} class
of ThePEG and implements the core functionality which is
independent of the nature of the shower.  It has responsibility for the overall
administration of showering and the multiple interactions, including:
\begin{itemize}
\item {} 

Identifying the partons which should be showered;

\item {} 

Using the inheriting classes to generate the radiation from the primary hard process;

\item {} 

Performing the decays of any particles whose lifetime is less than the hadronization
time-scale, or identify any cases where external programs have supplied partonic configurations
which include particle decays, and then using the inheriting classes to perform the parton shower;

\item {} 

Generating any additional scattering processes using a class inheriting from the
\href{https://herwig.hepforge.org/doxygen/classHerwig\_1\_1UEBase.html}{UEBase}
class  as described in \hyperref[\detokenize{review/index:sect-ue}]{Section \ref{\detokenize{review/index:sect-ue}}}, and using the inheriting classes to perform the
parton shower;

\item {} 

Performing the forced splitting and handling of the beam remnants using the
\href{https://herwig.hepforge.org/doxygen/classHerwig\_1\_1HwRemDecayer.html}{HwRemDecayer}
class.

\end{itemize}

The \texttt{ShowerHandler}
class also provides \href{https://herwig.hepforge.org/doxygen/ShowerHandlerInterfaces.html}{switches}
to either enable or disable radiation, in the initial or final
state, for different interactions:
\begin{itemize}
\item {} 

\href{https://herwig.hepforge.org/doxygen/ShowerHandlerInterfaces.html\#DoISR}{DoISR} switch
on (\texttt{Yes}) and
off (\texttt{No})
initial-state radiation.

\item {} 

\href{https://herwig.hepforge.org/doxygen/ShowerHandlerInterfaces.html\#DoFSR}{DoFSR} switch
on (\texttt{Yes}) and
off (\texttt{No})
final-state radiation.

\item {} 

\href{https://herwig.hepforge.org/doxygen/ShowerHandlerInterfaces.html\#SplitHardProcess}{SplitHardProcess}
whether or not to try and split the hard process into production and decay processes.

\item {} 

\href{https://herwig.hepforge.org/doxygen/ShowerHandlerInterfaces.html\#RestrictPhasespace}{RestrictPhasespace}
switches on (\texttt{Yes}) or off (\texttt{No}) restrictions on the maximum
hardness of shower emissions. When enabled, the corresponding upper scales are
determined from the hard-process kinematics before the shower evolution begins.
In the angular-ordered shower, trial emissions above these limits are vetoed,
while in the dipole shower the limits enter the maximum evolution scales of the
dipoles. This option is used in matched calculations, including POWHEG
showering, but is not specific to such applications.

\item {} 

\href{https://herwig.hepforge.org/doxygen/ShowerHandlerInterfaces.html\#MaxPtIsMuF}{MaxPtIsMuF}
Whether or not to use the factorization scale for the hard process as the maximum transverse momentum for
emission in the parton shower.

\end{itemize}

In addition the scales used in the parton shower can be varied using the interfaces described in \hyperref[\detokenize{review/showers/variations:sect-shvar}]{Section \ref{\detokenize{review/showers/variations:sect-shvar}}}.

There are two inheriting classes which implement the details of the angular-ordered and dipole shower
using a number of helper classes. Both use the
\href{https://herwig.hepforge.org/doxygen/structHerwig_1_1ShowerHandler_1_1ShowerTriesVeto.html}{ShowerTriesVeto}
to signal that complete showering of a given event failed a predefined number of times. This is handled together with
the generation of multiple interactions.

We will only describe the structure of the code, \textit{i.e.} how the various
classes work together to generate the parton shower evolution. Detailed
documentation of all the classes can be found in Doxygen.

\subsubsection{Angular-Ordered shower}
\label{\detokenize{review/showers/code:angular-ordered-shower}}

The angular-ordered shower consists of a large number of classes and is
designed to implement the improved
angular-ordered shower based on \cite{Gieseke:2003rz} and described
above.
The main class implementing the angular-ordered parton shower is the
\href{https://herwig.hepforge.org/doxygen/classHerwig\_1\_1QTildeShowerHandler.html}{QTildeShowerHandler}, which
inherits from the \texttt{ShowerHandler}.
This class performs the showering of primary and secondary hard scattering
processes, and the generation of any radiation produced in the decay of unstable fundamental particles.
The \texttt{QTildeShowerHandler}
uses a number of helper classes to implement various parts of the
algorithm together with some data storage classes, which hold
information needed to generate the parton shower.
This proceeds as follows:
\begin{itemize}
\item {} 

The particles supplied by the ShowerHandler are converted
from \href{https://thepeg.hepforge.org/doxygen/classThePEG\_1\_1Particle.html}{Particle}
objects, which store particle information in ThePEG, to
\href{https://herwig.hepforge.org/doxygen/classHerwig\_1\_1ShowerParticle.html}{ShowerParticle} objects,
which inherit from Particle and include the storage of the additional
information, such as the evolution scales and colour partners, needed
to generate the parton shower. Each particle in a hard process, be
that the primary scattering process or the subsequent decay of a
fundamental particle, is assigned to a
\href{https://herwig.hepforge.org/doxygen/classHerwig\_1\_1ShowerProgenitor.html}{ShowerProgenitor}
object containing references
to the particle together with additional information required for
particles that initiate a parton shower. For each hard process a
\href{https://herwig.hepforge.org/doxygen/classHerwig\_1\_1ShowerTree.html}{ShowerTree}
object is created containing the objects for all the particles in the
hard process and the information required to shower that process.

\item {} 

The \href{https://herwig.hepforge.org/doxygen/classHerwig\_1\_1PartnerFinder.html}{PartnerFinder}
is then used to find the evolution partners and initial scale
for the parton shower from each particle, as described in
\hyperref[\detokenize{review/showers/qtilde:sect-showerinitial}]{Section \ref{\detokenize{review/showers/qtilde:sect-showerinitial}}}.

\item {}

For hard processes, if the matrix element used to generate the process
inherits from the \texttt{HwMEBase} class and implements either a hard
matrix-element correction or real-emission generation in the POWHEG scheme,
the corresponding corrected or real-emission configuration is generated at
this stage. Similarly, for decay processes, the corresponding correction is
generated if the \texttt{Decayer} class inherits from
\texttt{DecayIntegrator} and implements either of these procedures.
This is described in more detail in
\hyperref[\detokenize{review/matching/matrix-element-corrections:matrix-element-corrections}]{Section \ref{\detokenize{review/matching/matrix-element-corrections:matrix-element-corrections}}}.

\item {} 

The intrinsic $p_{\perp}$ of
incoming partons in hadronic collisions is also generated at this stage.

\item {} 

Given the initial scale, the evolution of the particles proceeds as
described in \hyperref[\detokenize{review/showers/qtilde:sub-final-state-radiation}]{Section \ref{\detokenize{review/showers/qtilde:sub-final-state-radiation}}}-
\hyperref[\detokenize{review/showers/qtilde:sub-radiation-in-particle}]{Section \ref{\detokenize{review/showers/qtilde:sub-radiation-in-particle}}}, using the
\texttt{SplittingGenerator}
class to generate the types
and scales of the branchings. In turn the
\texttt{SplittingGenerator} uses the
\href{https://herwig.hepforge.org/doxygen/classHerwig\_1\_1SudakovFormFactor.html}{SudakovFormFactor} to
generate the possible evolution scales for each allowed type of
branching and then selects the branching with the highest scale, as
described in \hyperref[\detokenize{review/showers/qtilde:sub-final-state-radiation}]{Section \ref{\detokenize{review/showers/qtilde:sub-final-state-radiation}}}.

\item {} 

The new \href{https://herwig.hepforge.org/doxygen/classHerwig\_1\_1ShowerParticle.html}{ShowerParticle} objects produced
in the branching are then evolved until no further branching is
possible. When all the particles have been evolved the
\href{https://herwig.hepforge.org/doxygen/classHerwig\_1\_1KinematicsReconstructor.html}{KinematicsReconstructor}
reconstructs the momentum of all the particles in the shower (
\hyperref[\detokenize{review/showers/qtilde:sub-final-state-radiation}]{Section \ref{\detokenize{review/showers/qtilde:sub-final-state-radiation}}}- \hyperref[\detokenize{review/showers/qtilde:sub-radiation-in-particle}]{Section \ref{\detokenize{review/showers/qtilde:sub-radiation-in-particle}}}).

\item {} 

Finally, once the parton shower has been
generated the
\texttt{QTildeShowerHandler}
inserts them into the \href{https://thepeg.hepforge.org/doxygen/classThePEG\_1\_1Event.html}{Event} object.

\end{itemize}

The \texttt{QTildeShowerHandler}
class is primarily administrative, the actual physics is
implemented in the various helper classes.
In turn many of the helper classes used by the main classes implementing
the shower have their own helper classes for various parts of the
simulation.

The \texttt{SplittingGenerator}
class holds lists of available branchings. They are used to generate the shower
variables associated with each branching using objects. The
SplittingGenerator and \texttt{SudakovFormFactor} classes use the following
helper classes:
\begin{itemize}
\item {} 

The \href{https://herwig.hepforge.org/doxygen/classHerwig\_1\_1SplittingFunction.html}{SplittingFunction}
class is the base class for defining splitting functions used in the
shower evolution. This includes the calculation of the splitting
function together with the overestimate, integral and inverse integral
of it required to implement the veto algorithm as described in
\hyperref[\detokenize{review/showers/qtilde:sub-final-state-radiation}]{Section \ref{\detokenize{review/showers/qtilde:sub-final-state-radiation}}} and
\hyperref[\detokenize{review/showers/qtilde:sub-initial-state-radiation}]{Section \ref{\detokenize{review/showers/qtilde:sub-initial-state-radiation}}}. The splitting functions implemented
in Herwig are listed in \hyperref[\detokenize{review/showers/qtilde:sub-shower-dynamics}]{Section \ref{\detokenize{review/showers/qtilde:sub-shower-dynamics}}}.

\item {} 

The \href{https://herwig.hepforge.org/doxygen/classHerwig\_1\_1ShowerAlpha.html}{ShowerAlpha}
class is the base class implementing the running couplings used in the
shower evolution.

\end{itemize}

The \texttt{QTildeShowerHandler}
uses the \texttt{ShowerVeto}
class to provide a general interface to veto emission
attempts by the shower. The veto may be applied to either a single
emission (resetting the evolution scale for the particle to the
attempted branching scale), an attempt to shower a given event, or the
overall event generation.

Two exception classes are used inside the angular-ordered parton shower
module, mainly to communicate exceptional events or configurations,
rather than signalling a serious error during event generation. In
particular we use
\href{https://herwig.hepforge.org/doxygen/structHerwig_1_1VetoShower.html}{VetoShower}
to communicate vetoing of a complete shower attempt and
\href{https://herwig.hepforge.org/doxygen/structHerwig_1_1KinematicsReconstructionVeto.html}{KinematicsReconstructionVeto}
is used to signal an exceptional configuration that cannot be handled by
the  \href{https://herwig.hepforge.org/doxygen/classHerwig\_1\_1KinematicsReconstructor.html}{KinematicsReconstructor},
resulting in the shower being restarted from the original event.

\paragraph{Shower options and interface}
\label{\detokenize{review/showers/code:shower-options-and-interface}}\label{\detokenize{review/showers/code:sect-ewinterface}}

After Herwig-7.3.0 the default choice for parton shower for all input files is the AO \textit{QCD+QED+EW} shower with the \textit{dot-product} preserving recoil scheme. If not enabled by default or if a change in the shower setting is needed, the \texttt{{ShowerHandler}} interface can be explicitly configured with:

\begin{sphinxVerbatim}[commandchars=\\\{\}]
\PYG{n}{cd} \PYG{o}{/}\PYG{n}{Herwig}\PYG{o}{/}\PYG{n}{Shower}
\PYG{n+nb}{set} \PYG{n}{ShowerHandler}\PYG{p}{:}\PYG{n}{Interactions} \PYG{n}{ALL}
\end{sphinxVerbatim}

The \texttt{{ALL}} option corresponds to the \textit{QCD+QED+EW} shower scheme. Other available options include \texttt{{QEDQCD}}, \texttt{{QCD}}, \texttt{{QED}}, and \texttt{{EWOnly}}. Note that the $\gamma \to W^+ W^-$ EW branching is part of the QED parton shower and can be accessed using either the \texttt{{ALL}} or \texttt{{QED}} options.

Also, the recoil schemes for both final-state and initial-state radiation are introduced: transverse-momentum preserving, virtuality-preserving, and dot-product preserving scheme. The evolution scheme can be altered using the commands:

\begin{sphinxVerbatim}[commandchars=\\\{\}]
\PYG{n}{cd} \PYG{o}{/}\PYG{n}{Herwig}\PYG{o}{/}\PYG{n}{Shower}
\PYG{n+nb}{set} \PYG{n}{ShowerHandler}\PYG{p}{:}\PYG{n}{EvolutionScheme} \PYG{n}{DotProduct}
\end{sphinxVerbatim}

which is the default, while the other schemes can be chosen via the \texttt{{Q2}} or \texttt{{pT}} switches.

\subsubsection{Dipole shower}
\label{\detokenize{review/showers/code:dipole-shower}}

The primary class which steers the dipole shower is the
\texttt{DipoleShowerHandler},
which inherits from the
\texttt{ShowerHandler}
class.
The \texttt{DipoleShowerHandler}
controls the showering of primary and secondary hard scatterings,
and the showering of the decays of any unstable fundamental particles.
The decay processes are showered in the
\texttt{DipoleShowerHandler},
directly after the showering of a primary hard scattering.
The \texttt{DipoleShowerHandler}
uses several helper classes and data storage classes to perform
and record the showering procedure.

The dipole shower proceeds as follows:
\begin{itemize}
\item {} 

The \href{http://herwig.hepforge.org/doxygen/classHerwig\_1\_1DipoleEventRecord.html}{DipoleEventRecord}
contains the data storage objects for the showering of an event and is
responsible for updating these objects throughout the showering procedure.
In the first step of showering a hard scattering, the
\texttt{DipoleEventRecord}
takes the particles provided by the
\texttt{ShowerHandler}
and stores them as required by the dipole shower.
The \texttt{DipoleEventRecord}
sorts the incoming and outgoing particles for the given
process into \href{http://herwig.hepforge.org/doxygen/classHerwig\_1\_1DipoleChain.html}{DipoleChains},
in which neighbouring pairs of partons form \href{http://herwig.hepforge.org/doxygen/classHerwig\_1\_1Dipole.html}{Dipole}
objects as described in \hyperref[\detokenize{review/showers/dipole:dipole-shower-initial-conditions}]{Section \ref{\detokenize{review/showers/dipole:dipole-shower-initial-conditions}}}.
In the case of a decay process, if the
\texttt{Decayer}
class inherits from the
\texttt{DecayIntegrator}
and implements the QCD real emission in the POWHEG scheme, this first emission
is generated and the
\texttt{DipoleEventRecord}
is updated accordingly.

\item {} 

The \texttt{DipoleShowerHandler}
assigns an initial scale to each parton in each dipole, as described in detail
in \hyperref[\detokenize{review/showers/dipole:dipole-shower-initial-conditions}]{Section \ref{\detokenize{review/showers/dipole:dipole-shower-initial-conditions}}}.
The \href{http://herwig.hepforge.org/doxygen/classHerwig\_1\_1DipoleEvolutionOrdering.html}{DipoleEvolutionOrdering}
class is used to return the maximum physically allowed splitting scale for each dipole.

\item {} 

The \href{http://herwig.hepforge.org/doxygen/classHerwig\_1\_1DipoleSplittingGenerator.html}{DipoleSplittingGenerator}
class interfaces to the ExSample library \cite{Platzer:2011dr}
and directs the sampling of the branching probability and Sudakov form
factor for a given splitting.
The \href{http://herwig.hepforge.org/doxygen/classHerwig\_1\_1DipoleIndex.html}{DipoleIndex}
class identifies a \textit{type} of dipole based on the properties of the emitter
and spectator partons.
The dipole shower algorithm proceeds as described in \hyperref[\detokenize{review/showers/dipole:dipole-shower-evolution}]{Section \ref{\detokenize{review/showers/dipole:dipole-shower-evolution}}},
whereby all possible splittings from each dipole under consideration are tried.
For each dipole the
\texttt{DipoleIndex}
is constructed and, if the
\texttt{DipoleIndex}
has not previously been encountered, a
\texttt{DipoleSplittingGenerator}
object is constructed, and stored, for each possible splitting.

\item {} 

The \href{http://herwig.hepforge.org/doxygen/classHerwig\_1\_1DipoleSplittingInfo.html}{DipoleSplittingInfo}
class is a data-storage class used to store the information required to
generate a splitting from a particular dipole.
\texttt{DipoleSplittingInfo}
objects are created and used as required throughout the showering procedure.

\item {} 

For each trial splitting the
\texttt{DipoleEvolutionOrdering}
class uses output from the
\texttt{DipoleSplittingGenerator}
object to return the splitting scale for that splitting. The
\texttt{DipoleShowerHandler}
selects the trial splitting with the highest splitting scale.
The chosen splitting is performed by the
\texttt{Dipole}
object which uses the
\href{http://herwig.hepforge.org/doxygen/structHerwig\_1\_1DipolePartonSplitter.html}{DipolePartonSplitter}
class to update the parent-child relationships of the partons involved.
The showering of a process finishes once all of the
\texttt{DipoleChains}
have been evolved.

\item {} 

The generation of the intrinsic $p_\perp$ of incoming partons in
hadronic collisions is performed by the
\texttt{IntrinsicPtGenerator},
according to the procedure described in \hyperref[\detokenize{review/showers/intrinsic:dipole-shower-int-trans-mom}]{Section \ref{\detokenize{review/showers/intrinsic:dipole-shower-int-trans-mom}}}.

\item {} 

Following the showering of a hard scattering or decay process, the
\texttt{ConstituentReshuffler}
class is used to put the outgoing partons on their constituent mass shell according
to the procedure described in \hyperref[\detokenize{review/showers/dipole:dipole-shower-constituent-reshuffling}]{Section \ref{\detokenize{review/showers/dipole:dipole-shower-constituent-reshuffling}}}.

\item {} 

Finally the
\texttt{DipoleEventRecord}
updates the
\texttt{Event}
object to include the completed shower.

\end{itemize}

In the above we have stated that
\texttt{DipoleSplittingGenerator}
objects are stored.
The numerical sampling of a multi-dimensional distribution is computationally
expensive, whereas the memory requirements of storing these distributions are,
in the case at hand, very small.
\texttt{DipoleSplittingGenerator}
objects are therefore created and stored once and are used throughout each generator run.
Furthermore the ExSample library implements an adaptive sampling method, which
necessarily requires the storage and reuse of the distributions to provide a
benefit over non-adaptive sampling.

The
\texttt{ConstituentReshuffler},
\texttt{IntrinsicPtGenerator}
and \texttt{DipoleEvolutionOrdering}
objects used in the shower are all set via interfaces to the
\texttt{DipoleShowerHandler}.
Currently only a single default implementation is available for each of these
classes, however this structure enables future development and investigation.
The
\texttt{DipoleEvolutionOrdering}
class is a pure abstract class and in practice the inheriting
\href{http://herwig.hepforge.org/doxygen/classHerwig\_1\_1DipoleChainOrdering.html}{DipoleChainOrdering}
class, which implements the default $p_\perp$-ordered scheme, is used.

There are two other important classes that are necessary for the implementation of
the dipole shower which, for clarity, we have avoided referring to explicitly in the description above:
\begin{itemize}
\item {} 

The \href{http://herwig.hepforge.org/doxygen/classHerwig\_1\_1DipoleSplittingKernel.html}{DipoleSplittingKernel}
class is the base class for defining splitting kernels used within the dipole
shower. There is an inheriting class corresponding to each possible splitting
in the dipole shower, each of which includes the calculation of the appropriate
splitting function, see \hyperref[\detokenize{review/showers/dipole:dipole-shower-kernels}]{Section \ref{\detokenize{review/showers/dipole:dipole-shower-kernels}}}.
Each \texttt{DipoleSplittingGenerator}
object contains a
\texttt{DipoleSplittingKernel}
object, which it uses to construct the branching probability for the given splitting.

\item {} 

The \href{http://herwig.hepforge.org/doxygen/classHerwig\_1\_1DipoleSplittingKinematics.html}{DipoleSplittingKinematics}
class is the base class for implementing the kinematical prescription for a
splitting from a dipole. There is an inheriting class for each type of dipole discussed
in \hyperref[\detokenize{review/showers/dipole:dipole-shower-kinematics}]{Section \ref{\detokenize{review/showers/dipole:dipole-shower-kinematics}}}, each of which includes the computation of
the momenta of the new partons following a given splitting and the calculation
of several important scales.
Additionally, the
\texttt{DipoleSplittingKinematics}
class computes the Jacobians required by the
\texttt{DipoleSplittingGenerator}
to construct the branching probabilities.

\end{itemize}

\subsubsection{Decays}
\label{\detokenize{review/showers/code:decays}}

The code structure for particle decays in Herwig is described in more
detail in \hyperref[\detokenize{review/index:sect-hadron-sub-decay}]{Section \ref{\detokenize{review/index:sect-hadron-sub-decay}}} for the hadronic
decays. All of the
\texttt{Decayer}
classes for fundamental particles inherit from the
\texttt{DecayIntegrator}
class in order to use the multi-channel phase-space integration it
provides.

There are a small number of decays of fundamental Standard Model
particles currently implemented. These are implemented as
\texttt{Decayer}
classes for top quark, $W^\pm$ and $Z^0$, and Higgs boson
decays. The following classes are available:
\begin{itemize}
\item {} 

the
\href{https://herwig.hepforge.org/doxygen/classHerwig\_1\_1SMTopDecayer.html}{SMTopDecayer}
implements the three-body decay of the top quark to the bottom quark
and a Standard Model fermion-antifermion pair, via an intermediate
$W^+$ boson. This class implements the soft and hard matrix-element corrections
to top decay for use with the angular-order parton shower \cite{Hamilton:2006ms}.
In addition the
\texttt{SMTopDecayer} class
also implements the hardest emission in the POWHEG scheme for use with either
parton shower module \cite{Richardson:2013nfo}.

\item {} 

the
\href{https://herwig.hepforge.org/doxygen/classHerwig\_1\_1SMWDecayer.html}{SMWDecayer}
class implements the decay of the $W^\pm$
boson to a Standard Model fermion-antifermion pair including the soft and hard matrix-element corrections
for use with the angular-ordered parton shower. The \texttt{SMWDecayer}
class also implements the  hardest emission in the POWHEG scheme for use with either
parton shower module.

\item {} 

the
\href{https://herwig.hepforge.org/doxygen/classHerwig\_1\_1SMZDecayer.html}{SMZDecayer}
class implements the decay of the $Z^0$
bosons to a Standard Model fermion-antifermion pair including the soft and hard matrix-element corrections
for use with the angular-ordered parton shower.
The \texttt{SMZDecayer}
class also implements the  hardest emission in the POWHEG scheme for use with either
parton shower module.

\item {} 

the
\href{https://herwig.hepforge.org/doxygen/classHerwig\_1\_1SMHiggsFermionsDecayer.html}{SMHiggsFermionsDecayer}
class implements the decay of the Higgs boson to a Standard Model
fermion-antifermion pair, \textit{i.e.} $h^0\to f \bar{f}$. The
\texttt{SMHiggsFermionsDecayer}
class implements the hardest emission and next-to-leading order rate in the POWHEG scheme for use with either
parton shower module \cite{Richardson:2012bn}.

\item {} 

the
\href{https://herwig.hepforge.org/doxygen/classHerwig\_1\_1SMHiggsWWDecayer.html}{SMHiggsWWDecayer}
implements the decay of the Higgs boson to $W^\pm$ or
$Z^0$ bosons, \textit{i.e.} $h^0\to W^+W^-,Z^0Z^0$, including
the decay of the gauge bosons.

\item {} 

the
\href{https://herwig.hepforge.org/doxygen/classHerwig\_1\_1SMHiggsGGHiggsPPDecayer.html}{SMHiggsGGHiggsPPDecayer}
implements the decay of the Higgs boson to a pair of either gluons or
photons.

\end{itemize}

In many cases off-shell effects for the EW gauge bosons are
included by generating the gauge bosons as intermediate particles, for
example in top quark and Higgs boson decays. In general, external top
quarks and $W^\pm$ and $Z^0$ bosons are produced off
mass-shell using the approach described in Ref. \cite{Gigg:2008yc}.
Given the observed mass of the Standard Model Higgs boson, the mass
of the Higgs boson is now generated with a simple Breit-Wigner lineshape,
rather than the more sophisticated approach described
in Ref. \cite{Seymour:1995qg}, which was used in versions
of Herwig before the discovery of the Higgs boson.

The
\href{https://herwig.hepforge.org/doxygen/classHerwig\_1\_1SMHiggsMassGenerator.html}{SMHiggsMassGenerator}
implements the generation of the mass of off-shell Higgs bosons using
the running width implemented in the
\href{https://herwig.hepforge.org/doxygen/classHerwig\_1\_1SMHiggsWidthGenerator.html}{SMHiggsWidthGenerator}
class. These classes inherit from the
\href{https://herwig.hepforge.org/doxygen/classHerwig\_1\_1GenericMassGenerator.html}{GenericMassGenerator}
and
\href{https://herwig.hepforge.org/doxygen/classHerwig\_1\_1GenericWidthGenerator.html}{GenericWidthGenerator}
classes of Herwig in order to have access to the full infrastructure
for the simulation of off-shell particles described in
\hyperref[\detokenize{review/index:sect-hadron-sub-decay}]{Section \ref{\detokenize{review/index:sect-hadron-sub-decay}}}.

\subsubsection{YFS-based QED radiation}
\label{\detokenize{review/showers/code:yfs-based-qed-radiation}}

The structure of the code for the simulation of QED radiation in
particle decays is designed to be general, so that other approaches can
be implemented. The generation of the radiation is handled by a class
inheriting from the abstract
\href{https://herwig.hepforge.org/doxygen/classHerwig\_1\_1DecayRadiationGenerator.html}{DecayRadiationGenerator}
class. Currently only the YFS approach, as described in Ref.
\cite{Hamilton:2006ms}, is implemented in the
\href{https://herwig.hepforge.org/doxygen/classHerwig\_1\_1SOPHTY.html}{SOPHTY}
class, which uses the helper
\href{https://herwig.hepforge.org/doxygen/classHerwig\_1\_1FFDipole.html}{FFDipole}
and
\href{https://herwig.hepforge.org/doxygen/classHerwig\_1\_1IFDipole.html}{IFDipole}
classes for radiation from final-final and initial-final dipoles,
respectively. In addition the
\texttt{QEDRadiationHandler}
is included to allow the
\texttt{DecayRadiationGenerator}
to be used to generate radiation in the decay of particles generated as
$s$-channel resonances in the hard process.

\clearpage

\section{Matching and merging}
\label{\detokenize{review/index:matching-and-merging}}

\subsection{Overview}
\label{\detokenize{review/matching/general:matching-and-merging-overview}}\label{\detokenize{review/matching/general:matching-merging}}\label{\detokenize{review/matching/general::doc}}

The Matchbox framework described in
\hyperref[\detokenize{review/index:hard-processes}]{Section \ref{\detokenize{review/index:hard-processes}}}
is able to construct and calculate the cross section for a given process with leading or next-to-leading order accuracy. NLO calculations include, as well as virtual corrections,  contributions given by the matrix elements for the process with one additional particle in the final state, i.e.\ real corrections. The parton shower algorithms described in 
\hyperref[\detokenize{review/index:parton-showers}]{Section \ref{\detokenize{review/index:parton-showers}}}
also generate real corrections to arbitrary processes, but based on approximations to the exact matrix elements. Thus, parton showers cannot be trivially combined with NLO-accurate hard processes, without risking double-counting of this real radiation. The overlap of the two calculations must be carefully understood and the double-counting corrected for, before an NLO calculation can be showered, and we call this process \emph{matching}. The result is a parton-showered description of the final state of a process, with formal NLO accuracy of its normalizing cross section.

Since parton showers are applied iteratively, they generate multi-emission processes. Because the matrix elements they use are based on the soft and collinear approximations, they describe multiple emissions well when there is a hierarchy of scales between them, as in the bulk of the distribution, but if one is interested in final states with multiple hard well-separated emissions, a parton shower algorithm cannot be expected to do well. We would thus like to supplement the parton shower with information from exact higher order matrix elements, where available. Again, to do this, we have to ensure that the parton shower and matrix elements are combined in a way that avoids double-counting between them. We call this the \emph{merging} of a sample of matrix-element events of varying emission multiplicities with the parton shower description of their final states.

The state of the art, as implemented in Herwig~7 through the Matchbox module, is to \emph{merge} samples of a range of multiplicities, each with its NLO-accurate cross section \emph{matched} with the subsequent parton shower.

The biggest challenge in the matching algorithm is to set up an expression for the real-emission distribution produced by the parton shower, truncated at NLO, which we automate using Matchbox, and describe in \hyperref[\detokenize{review/matching/matching-subtractions:matching-subtractions}]{Section \ref{\detokenize{review/matching/matching-subtractions:matching-subtractions}}}.
The biggest challenge in the merging algorithm is to ensure that the cutoff that defines the multiplicities of the multi-emission samples is properly accounted for by the parton shower algorithm, so that if this cutoff is taken small, it is reflected in the Sudakov suppression of additional emission.
This is essential to ensuring that the merging does not compromise the formal logarithmic accuracy of the parton shower itself, and described in
\hyperref[\detokenize{review/matching/merging:multijet-merging}]{Section \ref{\detokenize{review/matching/merging:multijet-merging}}}.
The success of matching and merging algorithms hinges on many details matching correctly between the matrix element and parton shower descriptions, and parts of this section are therefore, necessarily, rather technical.

To be concrete, we use the language of NLO QCD, although much of our approach is general to other shower sectors.
The NLO QCD matching construction, covered in
\hyperref[\detokenize{review/matching/matching-subtractions:matching-subtractions}]{Section \ref{\detokenize{review/matching/matching-subtractions:matching-subtractions}}}, is available for both the
angular-ordered shower and the dipole shower, while multijet merging,
using both LO and NLO QCD multijet cross sections, can currently only
be combined with the dipole shower and is described in
\hyperref[\detokenize{review/matching/merging:multijet-merging}]{Section \ref{\detokenize{review/matching/merging:multijet-merging}}}.
We also describe an alternative approach to matching, KrkNLO, in
\hyperref[\detokenize{review/matching/krk-matching:krknlo}]{Section \ref{\detokenize{review/matching/krk-matching:krknlo}}}
and the simpler matrix-element corrections approach in
\hyperref[\detokenize{review/matching/matrix-element-corrections:matrix-element-corrections}]{Section \ref{\detokenize{review/matching/matrix-element-corrections:matrix-element-corrections}}.}
The matching and merging modules are closely
integrated with the Matchbox module and to some extent with the dipole
shower as far as the merging is concerned, and are described in \hyperref[\detokenize{review/matching/code-structure:matching-code-structure}]{Section \ref{\detokenize{review/matching/code-structure:matching-code-structure}}}.

\subsection{Handling of matching subtractions}
\label{\detokenize{review/matching/matching-subtractions:handling-of-matching-subtractions}}\label{\detokenize{review/matching/matching-subtractions:matching-subtractions}}\label{\detokenize{review/matching/matching-subtractions::doc}}

\subsubsection{General structure}
\label{\detokenize{review/matching/matching-subtractions:general-structure}}\label{\detokenize{review/matching/matching-subtractions:matching-subtractions-general}}

The NLO matching paradigm in Matchbox \cite{Platzer:2011bc} is rather general, incorporating both the well-known subtractive (MC@NLO-like) and multiplicative (POWHEG-like) methods as particular cases.
It is driven
by solving a matching condition where the combination of NLO cross
section and parton-shower evolution reproduces the NLO cross section
exactly, up to terms that are either higher order, or, unavoidably, suppressed by
the parton shower infrared cutoff.  In the following sections, we will
elaborate on the available matching algorithms and their
implementation, and on uncertainties from scale variations in matched
predictions. We will also address details connected to shower and hard
process scale choices, and how they impact the matching
uncertainties.

Before we go on to explain how the shower subtractions are exactly
combined with the NLO real-emission subtraction terms in the subtraction paradigm, we quickly repeat the
basic ingredients of the latter itself.  We write the partonic cross
section for the hard process at leading order as
\begin{equation*}
\begin{split}\sigma^{\text{LO}}[u]
= \int {\rm d}\sigma^B(\phi_n) \; {\rm d}f \; u(\phi_n)\;,\end{split}
\end{equation*}
where ${\rm d}f$ denotes the parton distribution functions,
and ${\rm d}\sigma^B(\phi_n)$ and $u(\phi_n)$ represent the Born
cross section and a generic observable as a function of the Born phase-space
$\phi_n=\{p_a,p_b\to p_1,...,p_n\}$.
For more details see \hyperref[\detokenize{review/hardprocess/tree-level:cross-section}]{Section \ref{\detokenize{review/hardprocess/tree-level:cross-section}}}.

For the NLO calculation, carried out in the subtraction method as detailed in
\hyperref[\detokenize{review/hardprocess/nlo:nlo-cross-section}]{Section \ref{\detokenize{review/hardprocess/nlo:nlo-cross-section}}}, the cross section reads
\begin{equation*}
\begin{split}\sigma^{\text{NLO}}[u]
= \sigma^{\text{LO}}[u]
+ \sigma^{V+A+C}[u]
+ \sigma^{R-A}[u]\;,\end{split}
\end{equation*}
containing the leading-order cross section
$\sigma_{\text{LO}}$ and the finite combination
$\sigma^{V+A+C}$ of virtual corrections, analytically integrated
subtraction terms and collinear counterterms, all defined over the Born
phase-space point $\phi_n$ and handled accordingly. For further details
see \hyperref[\detokenize{review/hardprocess/nlo:nlo-virtual-cross-section}]{Section \ref{\detokenize{review/hardprocess/nlo:nlo-virtual-cross-section}}}.
Instead, the finite combination
$\sigma^{R-A}$ of real and subtraction terms is defined over the real-emission phase-space point $\phi_{n+1}$. It is given by
\begin{equation*}
\begin{split}\sigma^{R-A}[u]
= \int
\Big[
  {\rm d}\sigma^R(\phi_{n+1}) \; u(\phi_{n+1})
- \tsum_i {\rm d}\sigma^A_{(i)}(\phi_{n+1} ) \; u\left(\rmPhiTilde_{n|(i)}(\phi_{n+1})\right)
\Big]{\rm d}f\;,\end{split}
\end{equation*}
where the subtraction terms ${\rm d}\sigma^A_{(i)}(\phi_{n+1})$
and the real-emission contributions ${\rm
d}\sigma^R(\phi_{n+1})$ are functions of the real-emission phase space and each of the dipole configurations
$(i)$ is associated with a particular kinematic mapping
$\rmPhiTilde_{n|(i)}(\phi_{n+1})$ onto the corresponding
underlying (called tilde) Born phase-space point.

It is important for the construction of matching subtractions that the kinematic mappings are invertible. Together with the tilde Born phase-space point, a given real phase-space point also defines a set of emission variables, $R_{(i)}(\phi_{n+1})$, typically chosen to be the scale of
the emission, a momentum fraction and an azimuthal variable. Thus the inverse transformation is
\begin{equation*}
\begin{split}
  \rmPhiTilde_{n+1|(i)}(\phi_n,r)=\phi_{n+1}
  \quad\Leftrightarrow\quad
  \phi_n=\rmPhiTilde_{n|(i)}(\phi_{n+1});\;
  r=R_{(i)}(\phi_{n+1}).
\end{split}
\end{equation*}
These mappings admit
phase-space convolutions which can be cast into phase-space
factorizations upon introducing suitably adapted parton distribution
functions,
\begin{equation*}
\begin{split}\left. {\rm d}\phi_{n+1} \; {\rm d}f \right|_{\phi_{n+1}=\rmPhiTilde_{n+1|(i)}(\phi_n,r)}
= {\cal J}_{(i)}(\phi_n,r) \; {\rm d}\phi_{n} \; {\rm d}f_{(i)} \; {\rm d}r\;,\end{split}
\end{equation*}
where ${\cal J}_{(i)}(\phi_n,r)$ is a Jacobian factor.

For more details see \hyperref[\detokenize{review/hardprocess/nlo:nlo-real-cross-section}]{Section \ref{\detokenize{review/hardprocess/nlo:nlo-real-cross-section}}}.
As further described in \hyperref[\detokenize{review/hardprocess/nlo:nlo-subtraction}]{Section \ref{\detokenize{review/hardprocess/nlo:nlo-subtraction}}}, Matchbox uses diagrammatic
information to deduce which subtraction terms need to be included, and
automatically sets up the NLO calculation in the form described above.

The matching of parton showers to NLO calculations generically proceeds through subtracting a fixed-order expansion of the parton shower from the hard process calculation. As our parton shower algorithms, including those supplemented by a matrix element correction, are based on unitarity, virtual contributions from the parton shower appear in a similar form as the subtraction terms in the NLO calculation. Owing to this, we focus in particular on the NLO subtracted real-emission cross section, $\sigma^{R-A}$, and construct a similar cross section from the parton shower distribution.

\subsubsection{Matching subtractions}
\label{\detokenize{review/matching/matching-subtractions:id2}}

The parton-shower action can formally be described in terms of a parton-shower
operator as
\begin{equation*}
\begin{split}\sigma[u]\to \sigma[{\rm PS}_{\mu_{\text{IR}}}[u]]\;,\end{split}
\end{equation*}
with the parton-shower operator itself being defined by
\begin{equation}\label{equation:review/matching/matching-subtractions:eq:psopfull}
\begin{split}{\rm PS}_{\mu_{\text{IR}}}[u](\phi_n)
&= \Delta(\phi_n,\mu_{\text{IR}}) \; u(\phi_n) \\
&+ \sum_i {\rm d}P_{(i)}(\phi_n,r) \, \kappa(Q_{(i)}(\phi_n),p_\perp(r)) \;
\theta(q(r)-\mu_{\text{IR}}) \\
&\quad\;\;\times \Delta(\phi_n,q(r)) \; {\rm PS}_{\mu_{\text{IR}}}[u](\rmPhiTilde_{n+1|(i)}(\phi_n,r))\;,\end{split}
\end{equation}
where ${\rm d}P_{(i)}(\phi_n,r)$ is the differential parton-shower probability for a splitting configuration $(i)$ from phase-space point $\phi_n$ with emission variables $r$. We describe the emission in terms of an evolution variable $q(r)$, which is limited from above by a hard scale $Q_{(i)}(\phi_{n})$ and from below by the infrared cutoff $\mu_{\text{IR}}$.
The constraint on the hard scale is in general not a sharp cutoff, but might be imposed in different ways, by what is called the profile scale choice $\kappa(Q_{(i)}(\phi_{n}),p_\perp(r))$, which is assumed to be theta-function-like, but may potentially be smooth. It is further discussed below and in \hyperref[\detokenize{review/showers/variations:shower-scale-variations}]{Section \ref{\detokenize{review/showers/variations:shower-scale-variations}}}.

The Sudakov form factor for the phase-space point $\phi_n$ is given by the product of Sudakov form factors for each splitting configuration,
\begin{equation*}
\begin{split}\Delta(\phi_n,q) &= \prod_j\Delta_{(j)}(\phi_n,q) \;, \end{split}
\end{equation*}
where the no-emission probability for a single splitting configuration is given by
\begin{equation*}
\begin{split}-\ln \Delta_{(i)}(\phi_n,q) =
\int{\rm d}P_{(i)}(\phi_n,r) \, \kappa(Q_{(i)}(\phi_{n}),p_\perp(r)) \; \theta(q(r)-q)\;.\end{split}
\end{equation*}
A parton shower that produces exactly zero or one emissions is given by Eq.~\eqref{equation:review/matching/matching-subtractions:eq:psopfull} but with the subsequent parton shower replaced by unity and the lower limit of the subsequent Sudakov form factors replaced by the infrared cutoff,
\begin{equation*}
\begin{split}{\rm PS}_{\mu_{\text{IR}}}[u](\phi_n)
&= \prod_i\Delta_{(i)}(\phi_n,\mu_{\text{IR}}) \; u(\phi_n) \\
&+ \sum_i {\rm d}P_{(i)}(\phi_n,r) \, \kappa(Q_{(i)}(\phi_{n}),p_\perp(r)) \;
\theta(q(r)-\mu_{\text{IR}}) \\
&\quad\;\;\times \prod_j\Delta_{(j)}(\phi_n,\mu_{\text{IR}}) \; u(\rmPhiTilde_{n+1|(i)}(\phi_n,r))\;.\end{split}
\end{equation*}
Note that we choose the same kinematic mapping as for the
dipole subtraction terms. In fact, the kinematic reconstruction, as
well as the kinematics used in the dipole shower and the POWHEG
correction discussed further below, are for one emission equal to the
dipole subtraction kinematics, and we do not consider any additional
Jacobian factors because of it. We stress, however, that the Matchbox
module does allow the inclusion of such additional Jacobian factors, and
that this is not a conceptual limitation. Notice that we have not
specified what shower we consider here -- the structure outlined can
cover both the angular ordered, as well as the dipole shower; they
might also include matrix element corrections, which we view in the
same way.

At this point we can expand the shower action to first order in
$\alpha_{s}$.
We recast both the integrand of
the Sudakov exponent and the emission rate times Born cross
section into the form of a cross section using the inverse
kinematic mapping,
\begin{equation}\label{equation:review/matching/matching-subtractions:eq:dsigps}
\begin{split}{\rm d}\sigma_{(i)}^{\text{PS}}(\phi_{n+1}) \; {\rm d}f
\equiv\Big[\;
{\rm d}\sigma^B \; {\rm d}f_{(i)} \; {\rm d}P_{(i)}(\phi_n,r) \, \kappa(Q_{(i)}(\phi_{n}),p_\perp(r))
\;\Big]_{\phi_{n}=\rmPhiTilde_{n|(i)}(\phi_{n+1}),r = R_{(i)}(\phi_{n+1})}\;,\end{split}
\end{equation}
where we have not yet imposed the infrared cutoff for reasons which
will become clear below. The ${\cal O}(\alpha_{s})$ parton-shower cross section is then
\begin{equation*}
\begin{split}\sigma^{R-A,\text{PS}}[u]
=\sum_i  \int {\rm d}\sigma_{(i)}^{\text{PS}}(\phi_{n+1}) \; {\rm d}f \;
\theta(q_{(i)}(\phi_{n+1})-\mu_{\text{IR}})\left(u(\phi_{n+1})-u(\rmPhiTilde_{n|(i)}(\phi_{n+1}))\right)\;,\end{split}
\end{equation*}
with the shorthand $q_{(i)}(\phi_{n+1})=q(R_{(i)}(\phi_{n+1}))$.
From this, we define the matching-subtracted NLO cross section as
\begin{equation*}
\begin{split}\sigma^{\text{NLO},\text{matched}}[u]
=\sigma^{\text{NLO}}[u]-\sigma^{R-A,\text{PS}}[u]\;.\end{split}
\end{equation*}
We can now check that this satisfies the crucial property
\begin{equation*}
\begin{split}\sigma^{\text{NLO},\text{matched}}[{\rm
PS}_{\mu_{\text{IR}}}[u]] =\sigma^{\text{NLO}}[u]+\text{h.o.}\end{split}
\end{equation*}
That is, the total parton-showered matched cross section is identical to the total unshowered NLO cross section, up to higher-order corrections.

Combining with the dipole subtraction terms, which enumerate the
possible shower branchings assuming for simplicity that the shower configurations match the dipole configurations $(i)$, we can write
\begin{equation*}
\begin{split}\sigma^{\text{NLO},\text{matched}}[u]
= \sigma^{S}[u]+\sigma^{H}[u]\;,\end{split}
\end{equation*}
with
\begin{equation*}
\begin{split}\sigma^S[u]
&= \sigma^{\text{LO}}[u] + \sigma^{V+A+C}[u] \\
&+ \sum_i \int \left(
{\rm d}\sigma_{(i)}^{\text{PS}}(\phi_{n+1}) \; \theta(q_{(i)}(\phi_{n+1})-\mu_{\text{IR}})
- {\rm d}\sigma_{(i)}^{A}(\phi_{n+1})
\right){\rm d}f \; u(\rmPhiTilde_{n|(i)}(\phi_{n+1}))\;,\end{split}
\end{equation*}
which constitutes Born-type configurations, referred to as $S$ events, and
\begin{equation*}
\begin{split}\sigma^H[u]
= \int \Big(
{\rm d}\sigma^R(\phi_{n+1})
- \sum_i {\rm d}\sigma_{(i)}^{\text{PS}}(\phi_{n+1}) \; \theta(q_{(i)}(\phi_{n+1})-\mu_{\text{IR}})
\Big)\ {\rm d}f\ u(\phi_{n+1})\;,\end{split}
\end{equation*}
to provide real-emission type configurations, referred to as $H$
events.

Note that these contributions are not yet finite owing to the
presence of the parton shower infrared cutoff.  Even if the
parton-shower approximated cross section reproduces
the full singularity structure of the real emission, which is not guaranteed to be the case, uncancelled
divergences still remain unless the parton shower cutoff is removed
from the matching subtraction. In order to achieve an infrared-finite result, we introduce
an additional auxiliary cross section,
\begin{equation*}
\begin{split}\sigma^{X}[u]
= \sum_i  \int {\rm d}\sigma_{(i)}^{X}(\phi_{n+1}) \; {\rm d}f \, \theta(\mu_{\text{IR}}-q_{(i)}(\phi_{n+1}))
\left(u(\rmPhiTilde_{n|(i)}(\phi_{n+1}))-u(\phi_{n+1})\right)\;,\end{split}
\end{equation*}
which is added to the matched cross section and yields modified
versions of $\sigma^S$ and $\sigma^H$, which can be
employed to generate events. In practice, we use the dipole
subtraction terms themselves to facilitate this, i.e.~${\rm
d}\sigma^X={\rm d}\sigma^A$. Note that, for infrared-safe observables
$u$, $\sigma^{X}$ only adds power corrections below the
infrared cutoff.
Technically, the combination of the auxiliary cross
section, the shower kernels and the NLO subtraction terms are combined
into one modified subtraction. The matching modules, in particular
the ShowerApproximation class instances, hold pointers to the
respective shower objects such that they will consistently adjust
changes to the showers in evaluating the matching subtractions.

\subsubsection{Showers without matrix element correction -- subtractive matching}
\label{\detokenize{review/matching/matching-subtractions:showers-without-matrix-element-correction-subtractive-matching}}\label{\detokenize{review/matching/matching-subtractions:matching-subtractions-subtractive}}

Both the angular-ordered and the dipole showers fit into the framework
outlined above, which constitutes the subtractive, or MC@NLO-type,
matching in Herwig 7, and the sole task is to determine the shower
matching subtraction ${\rm d}\sigma^{R-A,\text{PS}}$, which we
have implemented in a process-independent way in Matchbox. These
subtractions are very similar to the dipole subtraction terms, but
averaged over azimuthal angle and for colour correlators evaluated in
the large-$N_c$ limit. With the recent development of
spin-correlation algorithms in both shower modules
\cite{Richardson:2018pvo}, spin correlations can be restored in these
subtractions, and full colour correlations can be justified when using
colour matrix-element corrections
\cite{Platzer:2012np, Platzer:2018pmd}, at least for the dipole shower
algorithm. For more details on the two showers see
\hyperref[\detokenize{review/showers/qtilde:sect-angular-shower}]{Section \ref{\detokenize{review/showers/qtilde:sect-angular-shower}}} and \hyperref[\detokenize{review/showers/dipole:sect-dipole-shower}]{Section \ref{\detokenize{review/showers/dipole:sect-dipole-shower}}}. It is
important to stress that at this point no additional Jacobian factors
accounting for different kinematic mappings than the dipole mapping are
needed, since there are choices to distribute recoil in the angular
ordered shower in a way fully compatible with the Catani--Seymour
mapping for a single emission. In fact, we find that for
initial-final colour connections there is no alternative due to the
constraints imposed, while for final state connections the reshuffling
procedure turns out to reproduce the subtraction term mapping even in
the fully massive case, and an additional choice for the
initial-initial reconstruction has been introduced to reproduce the
dipole mapping. Note, however, that this is not a general limitation
and all of the structures are in place to account for a broader
variety of kinematic mappings and the associated Jacobian factors. Extensions of the matching objects to accommodate different mappings are possible.

\subsubsection{Matching with matrix element corrections -- multiplicative matching}
\label{\detokenize{review/matching/matching-subtractions:matching-with-matrix-element-corrections-multiplicative-matching}}\label{\detokenize{review/matching/matching-subtractions:matching-subtractions-multiplicative}}

In the previous section we have discussed the subtractive, or MC@NLO-type
matching in Herwig 7. Another choice is the multiplicative, or POWHEG-type
matching, for which we employ a hardest emission generator, which performs a
shower emission using a modified splitting function, or matrix-element
correction. This is determined from the real-emission and Born matrix elements, by partitioning them using functions $w_{(i)}(\phi_{n+1})$, as
\begin{equation*}
\begin{split}P_{(i)}(\phi_n,r) \rightarrow
\frac{w_{(i)}(\rmPhiTilde_{n+1|(i)}(\phi_n,r))}{\sum_j w_{(j)}(\rmPhiTilde_{n+1|(j)}(\phi_n,r))}
\frac{| \mathcal{R}(\rmPhiTilde_{n+1|(i)}(\phi_n,r))|^2}{| \mathcal{B}(\phi_n)|^2}\ ,\end{split}
\end{equation*}
for which the full divergent behaviour is reproduced by the sum over $(i)$, by construction.
For
the $w_{(i)}$ in practice we use dipole-type partitioned Eikonal
factors, and the ExSample library \cite{Platzer:2011dr} to generate
emissions according to the Sudakov form factor obtained from the above
matrix-element correction. Notice that the matrix element correction
simply constitutes another kind of shower, which we handle on equal
footing to the other matching subtractions. In particular the
remarks regarding scale choices and phase-space restrictions apply
equally well to all cases. Matrix element corrections using built-in
matrix elements are discussed in detail in \hyperref[\detokenize{review/matching/matrix-element-corrections:matrix-element-corrections}]{Section \ref{\detokenize{review/matching/matrix-element-corrections:matrix-element-corrections}}}.

If the parton shower is ordered in transverse momentum, as Herwig's dipole shower is, showering these events is straightforward: the matrix element correction supplies the hardest emission and the shower continues from there. But if, like the angular-ordered shower, the ordering condition is not transverse momentum, a more detailed procedure is needed. Firstly, a veto is needed to ensure that no emission is generated that is harder than the matrix-element emission. And, secondly, the parton that was assigned the matrix-element emission has to be allowed to emit at wider angles (but smaller transverse momenta) than the matrix-element emission, i.e.\ before it in the shower ordering. Such earlier emissions are generated by what is known as a truncated shower. In Herwig 7, this truncated shower is implemented by constructing the shower variables that would correspond to the matrix-element emission, evolving the shower subject to the shower veto until the evolution scale of the matrix-element emission is reached, and then inserting the prescribed emission with those shower variables. At the matching level, whether a truncated shower is requested is controlled by the \texttt{TruncatedShower} switch of \texttt{MEMatching}; it is enabled in the default angular-ordered POWHEG setup and disabled for the dipole-shower setup.

\subsubsection{Advanced choices}
\label{\detokenize{review/matching/matching-subtractions:advanced-choices}}\label{\detokenize{review/matching/matching-subtractions:matching-advanced}}

The infrastructure employed by Matchbox to facilitate the matching
subtractions and the generation of the hard emission is also
available to other combinations such as pure matrix element
corrections for which real emission corrections to the shower are
included; virtual corrections are then typically deduced only by
employing unitarity. This provides an alternative to study matrix
element corrections in a different context and to investigate options
such as the scale choices and profiles discussed below in the context
of a leading-order merging type procedure. It can simply be achieved
by turning off the virtual corrections when using the multiplicative
matching paradigm; one then obtains a simulation with leading-order
cross section in which the parton shower has been supplemented by a
matrix element correction according to the POWHEG method.

\subsubsection{Scales in matching}
\label{\detokenize{review/matching/matching-subtractions:scales-in-matching}}\label{\detokenize{review/matching/matching-subtractions:matching-subtractions-scale-choices}}

The matching algorithm depends critically on several scales, related to the hard process, the shower, and the matching between the two. In this section, we review in more detail the roles of these different scales, and uncertainties due to their variation.

\paragraph{Hard process and veto scales}
\label{\detokenize{review/matching/matching-subtractions:hard-process-and-veto-scales}}

The two shower modules in Herwig imply an upper limit, the hard veto
scale $\vetoScale$, which limits the transverse momentum that is
available to an emission by the shower.  For the $p_\perp$-ordered
dipole shower this hard veto scale is identical to the starting scale of
the dipole shower, while for the angular-ordered shower it is
implemented as a veto on the reconstructed transverse momentum . For
more details on the two shower modules see \hyperref[\detokenize{review/showers/dipole:sect-dipole-shower}]{Section \ref{\detokenize{review/showers/dipole:sect-dipole-shower}}} and
\hyperref[\detokenize{review/showers/qtilde:sect-angular-shower}]{Section \ref{\detokenize{review/showers/qtilde:sect-angular-shower}}}.

The hard veto scale is not generally fixed but should be of the order of
the hard scale that is set by the hard process.  The default choice for
the hard veto scale is to be equal to the hard process scale,
$\vetoScale=\hardProcScale$, which in turn is typically set to the
factorization and renormalization scale,
$\hardProcScale=\mu_\mathrm{R}=\mu_\mathrm{F}$. However, all these
scales may also be chosen independently in Herwig 7, and in general,
depending on the choice for $\hardProcScale$, different choices
for $\vetoScale$ can have differing and significant effects on
predicted observables.

For NLO matched predictions, the $S$ and $H$ events (see previous section)
separately undergo showering. The $S$ events constitute Born-type events
and are treated as such.  For $H$ events, however, several complications may arise.
The subtracted real-emission cross section could possibly still
contain power corrections in the regions where the real emission is soft
or collinear.  However, there is no strict requirement of exact
cancellation between the real-emission matrix element and the
subtraction term in any region of phase-space away from these
limits. Correspondingly, we can expect to obtain a number of $H$ events
with a soft or collinear emission.  For such $H$ events it does not, however,
seem reasonable to choose the hard veto scale to be of the order of
the small transverse momentum of the real emission.

We want to set the hard veto scale $\vetoScale$ to a fairly
general choice of scale for the final states that may arise from the
hard process.  For relatively hard real emissions this scale will
probably be of the order of the transverse momentum of that emission.
On the other hand this is to be avoided for very soft emissions where we
would probably associate the general hard scale of the hard process, as
e.g.~given by an invariant of the hard final scale.  There are choices
for $\hardProcScale$ that, despite being reminiscent of the scale
of the real emission, are relatively large over a wide range of real
emission scales. If $\hardProcScale$ is large enough, the maximum
scale for the first emission will be the maximum scale that is allowed
for the given splitting kinematics. In this case, while
$\hardProcScale$ may be directly affected by the scale of the real
emission, the scale of the real emission will have only a small impact
on the subsequent showering.  In such cases one should consider using an
alternative choice for $\vetoScale$.  There are various scale
choices implemented in Herwig 7, that can be used directly, further
details on these choices can be found in \hyperref[\detokenize{review/hardprocess/scales:scale-choices}]{Section \ref{\detokenize{review/hardprocess/scales:scale-choices}}}.

\paragraph{Profile scale functions}
\label{\detokenize{review/matching/matching-subtractions:profile-scale-functions}}

The hard veto scale from the parton shower, $\vetoScale$ should be
large enough to account for the hardness of the process and should not
be too small for real emissions in order to remain in the perturbative
domain.  The balance of these was discussed above, but if
$\vetoScale$ is too large, we find another effect, the summation of
an unphysical tower of logarithms in the Sudakov exponent.  Here, we
think that as this stems entirely from hard emissions, these logarithms
should be controlled by fixed-order computations and not the parton
shower.

In order to avoid these contributions we introduce a profile scale
function $\kappa(Q_{(i)}(\phi_n),p_\perp(r))$ (cf
Eqs. \eqref{equation:review/matching/matching-subtractions:eq:psopfull} to \eqref{equation:review/matching/matching-subtractions:eq:dsigps}). This function, also
described in \hyperref[\detokenize{review/showers/variations:sect-shvar}]{Section \ref{\detokenize{review/showers/variations:sect-shvar}}}, facilitates an upper limit for the scale
of hard parton shower emissions.  This upper limit is controlled by the
profile scale function in that it interpolates between regions of phase
space where too hard emissions are completely vetoed and where they are
taken into account completely with a smooth transition between the
regions, i.e.~for any emission it is some function of the phase-space
$Q_{(i)}(\phi_n)=\vetoScale$.

The same function will be needed to modify the kernels in the matching
subtraction, and in turn regulates how much of the real emission
contributions we include as a new, hard process, and how much we let the
shower populate regions of hard emission. Note that to first order in
the strong coupling, by definition, different profile scales do not
generate a difference.  However beyond this there will be visible
differences, and in particular the transition region between the shower
and hard jet phase-space will be affected.  In \cite{Bellm:2016rhh}
several possible parametrizations of the profile scale choices were
investigated for leading-order plus parton-shower predictions.  There it
was pointed out that the choice of the profile scale, i.e.\ how to
approach the boundary of hard emissions, is non-trivial and highly
relevant in the context of NLO plus parton-shower matching.

The choice function that determines the profile scale for any emission
is constrained by consistency conditions on central predictions and
uncertainties.  It should not modify the input
distributions of the hard process. Uncertainties should only become large in unreliable regions or regions where
hadronization effects are dominant.  And finally, stable results should be obtained in the Sudakov region.  It was found in \cite{Bellm:2016rhh}
that the hfact profile does not admit results compatible with these
criteria. Instead, using the resummation profile it was found that the
angular-ordered and dipole showers are compatible with each other, both
in central predictions and uncertainties (despite their very different
nature).  Further studies, comparing the resummation and hfact profiles
in next-to-leading-order plus parton-shower predictions were performed
in \cite{Cormier:2018tog}.

In Herwig 7, we set the upper limit on the transverse momentum available
to the shower to the hard process scale,
$\vetoScale=\hardProcScale$, by default.  This in turn is normally
set to the factorization and renormalization scale,
$\hardProcScale=\mu_\mathrm{R}=\mu_\mathrm{F}$, however, the hard
process, factorization and renormalization scales may also be chosen
independently in Herwig 7 as discussed above.

\paragraph{Scale uncertainties}
\label{\detokenize{review/matching/matching-subtractions:scale-uncertainties}}

It is sometimes useful to be able to quantify uncertainties for event
generator predictions.  To this end variations of the scales involved as
well as different central choices and distributions of the scales may be
pursued.  In order to understand matched predictions it is useful to
discriminate the effects of scale uncertainties in different parts of
the simulation.  This will give a complete picture for understanding
systematic uncertainty estimates in more sophisticated setups,
like multi-jet merging.

Variations of the scales involved in the hard
production process as well as the scales involved in the subsequent parton
showering may be pursued. Following the approach in \cite{Bellm:2016rhh}
and \cite{Cormier:2018tog}, variations of three scales may be considered:
\begin{itemize}
\item {} 

the hard process scale $\hardProcScale$, which we typically set to the
factorization and renormalization scale in the hard process,
i.e.~$\hardProcScale=\mu_\mathrm{R}=\mu_\mathrm{F}$;

\item {} 

the parton shower hard scale, or hard veto scale, $\vetoScale$;

\item {} 

the arguments of $\alpha_{s}$ and the PDFs in the parton shower,
which we collectively denote as shower scale $\showerScale$%
\begin{footnote}\sphinxAtStartFootnote
We are concerned only with variations of the arguments of
$\alpha_{s}$ and the PDFs in the parton showers. Therefore, even
though they can differ, we use the common terminology ‘shower scale’
for these scales.  In the angular-ordered shower the argument of the
strong coupling is related to the transverse momentum of the emitted
parton and differs for initial- and final-state evolution, while the
argument of the PDFs is simply the ordering variable for
initial-state evolution \cite{Gieseke:2003rz}.  In the dipole shower
the transverse momentum of the emitted parton is used for both
scales.
\end{footnote}.

\end{itemize}

For a full set of scale variations, consisting of 27 different
combinations for three scales, one central and one up and down for each,
we typically apply multiplicative factors of 0.5, 1 and 2 to each of the
corresponding central scales. For more details see also
\hyperref[\detokenize{review/showers/variations:sect-shvar}]{Section \ref{\detokenize{review/showers/variations:sect-shvar}}}.

In addition to scale variations around these central values, one should further consider the impact
of different choices for the dependence of the central value of the hard veto scale on the hard kinematics, the choice
of the hard process scale, and the impact of choosing different profile
scale functions, as also described in
\cite{Bellm:2016rhh} and \cite{Cormier:2018tog}.

\subsection{KrkNLO}
\label{\detokenize{review/matching/krk-matching:krknlo}}\label{\detokenize{review/matching/krk-matching:krk-matching}}\label{\detokenize{review/matching/krk-matching::doc}}

The KrkNLO method for parton shower matching \cite{Jadach:2016qti, Jadach:2015mza, Jadach:2016acv, Sarmah:2024hdk,Sarmah:2025vnb}
is able to provide NLO matching for colour-singlet processes.
In contrast with MC@NLO and POWHEG, the KrkNLO method achieves NLO accuracy using a
special-purpose factorisation scheme for the parton distribution functions
(PDFs), the ‘Krk’ scheme \cite{Jadach:2016acv} (formerly called the ‘MC’ scheme)
instead of the $\overline{\mathrm{MS}}$ scheme.
This is defined to absorb into the PDFs collinear convolution terms, similar in form to
$\sigma^{C}[u]$, which cancel those arising from the NLO expansion of the dipole shower
Sudakov factor,
allowing NLO accuracy to be achieved through a simple multiplicative reweight alone.

Concretely, the method can be expressed as
\begin{equation*}
\begin{split}\sigma^{\text{KrkNLO}}[u]
&= \int {\rm d} \sigma^B(\phi_n)    \left( 1
+ \frac{\alpha_{s}(\mu_{\text{R}})}{2\pi}
  {\mathcal{B}_{n}(\phi_n)}^{-1}
  \left( \mathcal{V}_{n}(\phi_n)
  + \mathrm{I}(\phi_n)
    + \Delta_{0}^{\text{Krk}}
  \right)
  \right)
  \Biggl[ \Delta(\phi_n,\mu_{\text{IR}})
  \,
u(\phi_n) \\
& \qquad {} + \sum_i {\rm d}P_{(i)}(\phi_n,r)
\; \theta(Q_{\max{}}^{(i)}(\phi_{n}) - p_\perp(r))
\, \theta(p_{\perp}(r) - \mu_{\text{IR}})
\; \Delta(\phi_n, p_{\perp}(r)) \\
& \qquad \quad \times \frac{{\mathcal{R}_{n+1}}(\phi_{n+1|(i)}(\phi_n,r))}{\sum_{j} \mathcal{B}_n(\rmPhiTilde_{n|(j)}(\rmPhiTilde_{n+1|(i)})) \, P_j(\rmPhiTilde_{n+1|(i)}) }
\, {\rm{PS}}_{\mu_{\text{IR}}} [ u ] ( \rmPhiTilde_{n+1|(i)}(\phi_n,r) )
\Biggr],\end{split}
\end{equation*}

where
$\mathrm{I}$ denotes the integrated dipole contributions detailed in \hyperref[\detokenize{review/hardprocess/nlo:vi}]{Section \ref{\detokenize{review/hardprocess/nlo:vi}}},
$\Delta_{0}^{\text{Krk}}$ denotes a virtual-like correction arising from the factorisation scheme transition,
and
$Q_{\max{}}^{(i)}$ is the maximum kinematically-permitted scale for an emission from
splitting \sphinxtitleref{(i)}.
For more details please consult \cite{Sarmah:2024hdk}.

This gives NLO accuracy only when convolved with parton distributions
in the Krk factorisation scheme, due to the otherwise-uncancelled $\mathcal{O}(\alpha_{s})$
collinear contributions from the phase-space integral within the Sudakov factor in the first line.

Two versions of the Krk factorisation scheme exist: KrkDY, in which only the quark and antiquark PDFs
are transformed (the gluon PDF remains identical to the input $\overline{\mathrm{MS}}$ scheme
gluon PDF), and the full Krk scheme which additionally transforms the gluon PDF.
Further details about the properties of the Krk factorisation scheme may be found in \cite{Delorme:2025teo}.

In Herwig, the KrkNLO method is implemented on top of the Dipole Shower using the \texttt{DipoleEventReweight}
\href{https://herwig.hepforge.org/doxygen/classHerwig\_1\_1DipoleEventReweight.html\#a25e857470814fa7415f05ce6b98a4591}{Splitting-} and \href{https://herwig.hepforge.org/doxygen/classHerwig\_1\_1DipoleEventReweight.html\#ab5158887873376038d34f26ed1276d90}{Cascade-} reweight functions.
The former is used to apply a matrix-element-correction after the first shower-generated emission,
which corrects the weight implied by the Born matrix-element and dipole shower splitting-function to the
weight implied by the real-emission matrix element (the third line above).
The latter is used to apply a virtual-type correction to the entire shower cascade, which (in particular)
corrects the weight of the no-emission events from the weight implied by the Born matrix element
to that implied by the Born and virtual matrix elements and the compensatory terms required by the
factorisation scheme transformation (the first line above).

In Herwig 7.3.0, only the Drell-Yan and Higgs production (via gluon-fusion) processes are currently supported.%
\begin{footnote}\sphinxAtStartFootnote
Additional processes will be supported from Herwig 7.4;
see \cite{Sarmah:2024hdk,Sarmah:2025vnb}.
\end{footnote}
Process selection within KrkNLO is controlled by the \href{https://herwig.hepforge.org/doxygen/KrknloEventReweightInterfaces.html\#Mode}{Mode} switch of the \href{https://herwig.hepforge.org/doxygen/classHerwig\_1\_1KrknloEventReweight.html}{KrknloEventReweight} class:
\begin{itemize}
\item {} 

\sphinxstylestrong{Drell--Yan} \cite{Jadach:2015mza} The process $p p \to Z \to e^+ e^-$ is supported, and can be used with both versions of the PDF factorisation scheme: KrkDY and Krk. The real weight is calculated in an approximation which factorises production from decay, neglecting spin-correlations between the initial- and final-states.
This mode is selected by choosing \texttt{{p p -\textgreater{} e+ e-}} as your process in Herwig, and setting \href{https://herwig.hepforge.org/doxygen/KrknloEventReweightInterfaces.html\#Mode}{Mode=Z}.

\item {} 

\sphinxstylestrong{Higgs production via Gluon Fusion} \cite{Jadach:2016qti, Jadach:2016acv} The process $g g \to H$ is supported and can be used only with the Krk scheme PDFs. The calculation is performed in the large top-mass limit.
This mode is selected by choosing \texttt{{p p -\textgreater{} H}} as your process in Herwig, and setting \href{https://herwig.hepforge.org/doxygen/KrknloEventReweightInterfaces.html\#Mode}{Mode=H},
as well as \texttt{{read Matchbox/HiggsEffective.in}}. Note that the $qq$-initiated tree-level diagram must be generated separately and added manually.

\end{itemize}

\begin{figure}[tp]
\centering
\capstart
\begin{subfigure}{0.999\textwidth}
\centering
\noindent\includegraphics[width=0.325\linewidth]{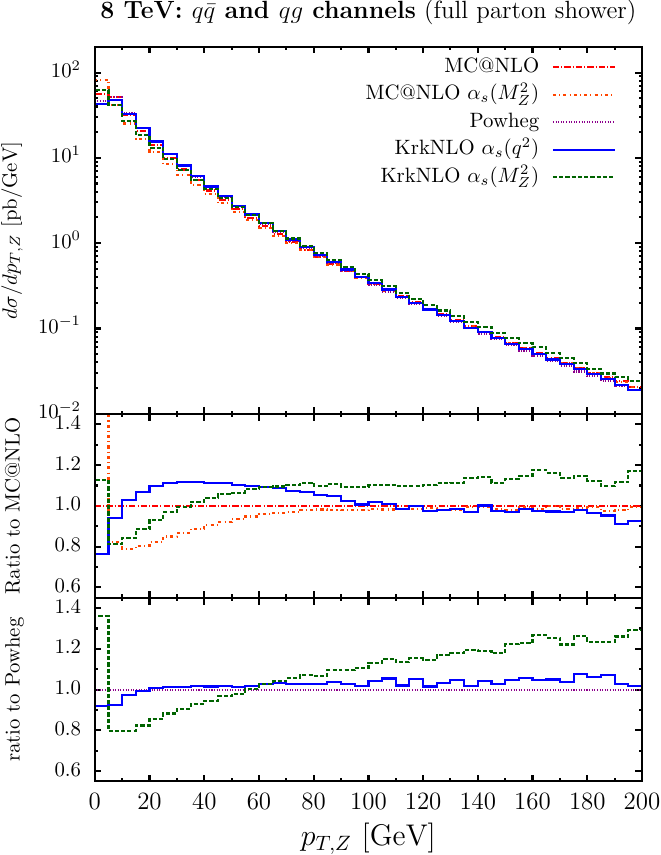}
\noindent\includegraphics[width=0.325\linewidth]{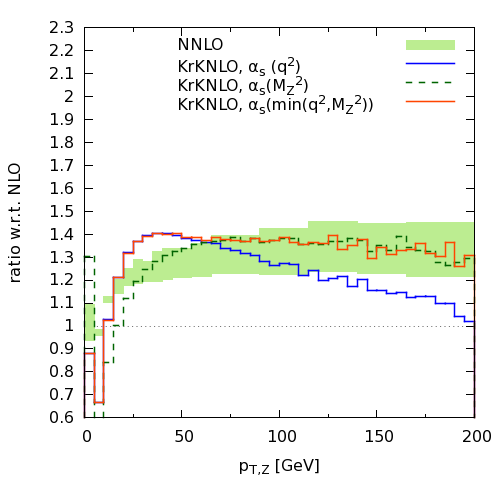}
\noindent\includegraphics[width=0.325\linewidth]{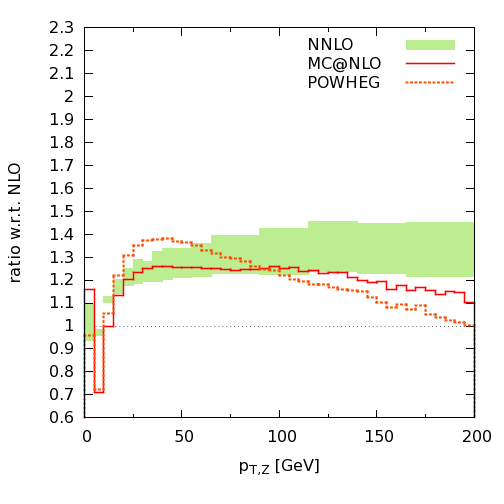}
\caption{Drell--Yan predictions generated by the KrkNLO method,
         compared with alternative matching methods MC@NLO and Powheg.
         Here $q^2$ is the shower ordering scale, and
         the higher-order scale-dependence is compared with both the matching-scheme dependence, and the 
         higher-order corrections from the corresponding NNLO Drell--Yan calculation
         (i.e., an NLO $p_{\mathrm{T}}^{Z}$ distribution).
         From \cite{Jadach:2015mza}.}
\end{subfigure}

\begin{subfigure}{0.999\textwidth}
\centering
\noindent\includegraphics[width=0.325\linewidth]{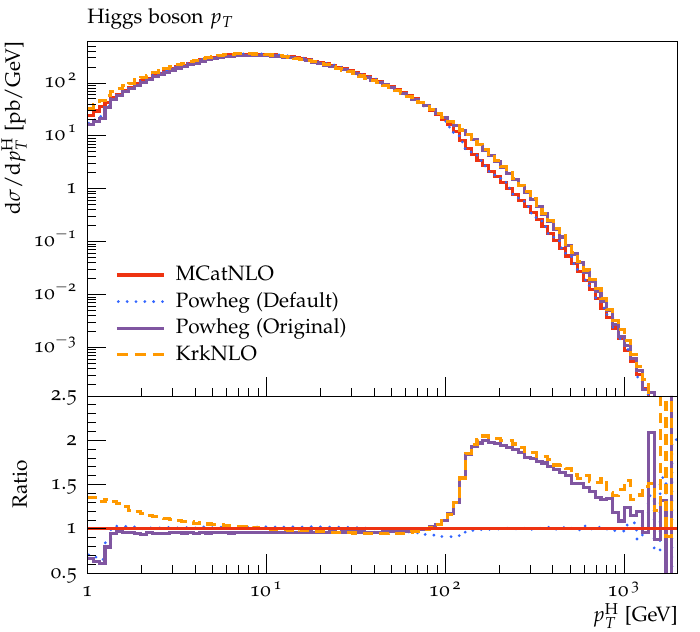}
\noindent\includegraphics[width=0.325\linewidth]{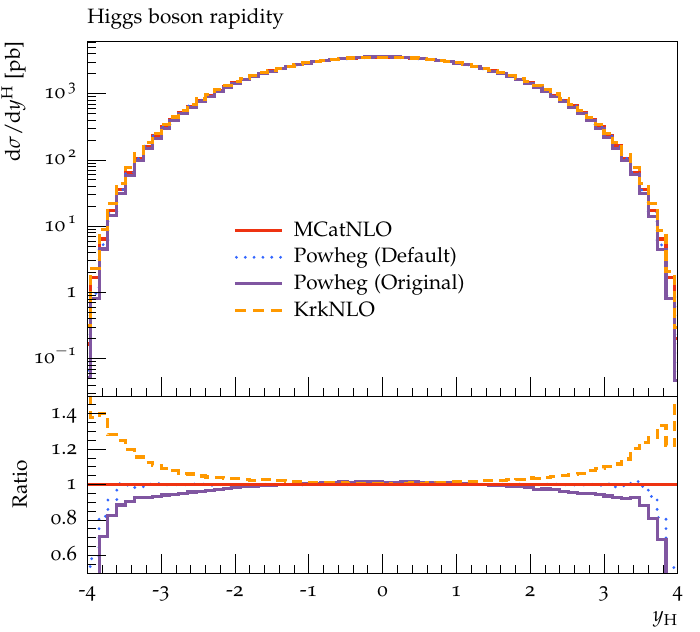}
\noindent\includegraphics[width=0.325\linewidth]{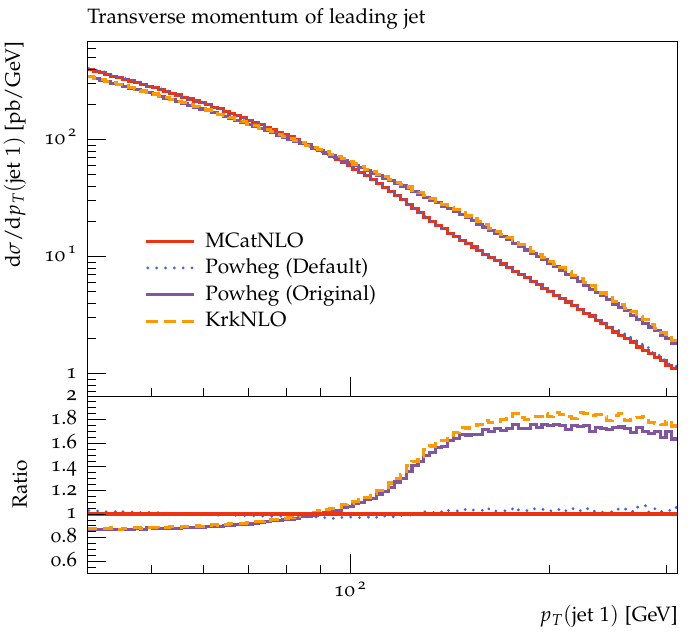}

\caption{Higgs-production predictions generated by the KrkNLO method,
         compared with alternative matching methods MC@NLO and Powheg computed within Herwig 7.
         Here `Powheg (Default)' restricts shower emissions to have transverse momentum smaller than $M_H$, whereas
         `Original' lifts the restriction.
         From \cite{Jadach:2016qti}.}
\end{subfigure}

\caption{Examples of the KrkNLO method for the processes available in Herwig 7.3.}
\label{\detokenize{review/matching/krknlo}}
\end{figure}

Example results from both processes are shown in \hyperref[\detokenize{review/matching/krknlo}]{Fig.\@ \ref{\detokenize{review/matching/krknlo}}}.
It can be seen that within the family of NLO-accurate schemes, there remains
a `matching uncertainty' that can lead to observable-dependent differences in
both shape and normalisation.

Note that the KrkNLO method requires the use of the factorisation scale $\mu_\mathrm{F} = \sqrt{\hat{s}_{12}}$.
The renormalisation scale choice used for the real emission is by default the emission scale
of the dipole shower, but can be changed to the virtuality of the vector boson
or the physical mass of the boson. The renormalisation scale choice used for the virtual (and factorisation scheme) corrections is by default the virtuality of the boson, but can be changed
to its physical mass.

A selection of PDF sets, transformed from the $\overline{\mathrm{MS}}$ to the Krk scheme,
are available in LHAPDF6 format from \href{https://krknlo.hepforge.org/}{krknlo.hepforge.org}.
For others, or to request a specific set, please contact the authors.

\subsection{Multijet merging}
\label{\detokenize{review/matching/merging:multijet-merging}}\label{\detokenize{review/matching/merging:id1}}\label{\detokenize{review/matching/merging::doc}}

The (N)LO merging in Herwig is thoroughly described in
\cite{Bellm:2017ktr, Platzer:2012bs}.  Here we highlight the main
features and some aspects of the algorithms.  While in the case of
next-to-leading order matching there is essentially no ambiguity once
a parton shower to match to has been identified, the multi-jet merging
requires an additional resolution criterion to separate
regions of hard jet production from regions of jet evolution performed
by the parton shower.
For one additional emission, one can view this
procedure as a different choice of hard profile scale to be used
within the parton shower, but for higher multiplicities this
intuitive picture is no longer applicable.

To ensure consistency between the
hard multi-jet (matrix-element) regime
and the
parton-shower-emission regime,
the higher jet multiplicities must explicitly be weighted
by Sudakov factors, which account for the exclusiveness of the jet
configuration in the hard, matrix-element, region.
As described in detail in
\cite{Bellm:2017ktr, Platzer:2012bs}, 
such a combination does not
leave a spurious logarithmic dependence on the resolution criterion,
provided that several key criteria are met.
At next-to-leading order, however, a naive generalization leads to
sub-leading logarithmic contributions which eventually spoil the NLO
accuracy of the inclusive cross sections with fewer jets,
due to the parton shower itself not containing NLO corrections
in a flexible and differential way \cite{Platzer:2012bs}
(some NLO corrections are accounted for inclusively through the CMW
scheme).

These spurious contributions can be removed by explicitly subtracting
them. The subtraction can be calculated by enforcing that
the inclusive cross section is reproduced exactly at its input value.
In \cite{Bellm:2017ktr} we have proposed an extension of this scheme,
which classifies configurations with additional jets depending on
whether we assume that they will lead to a logarithmic enhancement
(if they could have been generated by a strongly-ordered
parton shower evolution), or should rather be considered an
additional, finite correction to the process of interest.
In the former case, they will be removed by subtraction; in the latter case
they will be explicitly presented to the shower as an additional hard
process.
This criterion, in particular, allows us to also consider
processes with jets at the level of the hard process, and so does not
limit the applicability of the merging
for hadron colliders to the case of colour-singlet production.

The hard, matrix element (ME) region is defined by requiring all
clusterings that correspond to QCD vertices and a branching in the
dipole shower algorithm to yield transverse momenta larger than the
merging scale $\rho$, such that in the ordering inherent to the
shower evolution they will be produced at scales larger than
$\rho$. The parton shower (PS) region is complementary to the ME
region, and will be populated by parton shower emissions subject to a
veto in the ME region. Since we are interested in combining different
jet multiplicities, each hard matrix element configuration needs to be
interpreted as if it had been produced by the parton shower, and a
history is extracted up to the point where we could not have obtained
the configuration by an ordered parton shower evolution. This
configuration is classified as a core hard process to start the
evolution from. The entire process of clustering into a parton
shower history is described in \hyperref[\detokenize{review/matching/merging:historyextraction}]{Section \ref{\detokenize{review/matching/merging:historyextraction}}}.

A complication arises at NLO: while the subtracted virtual corrections
are straightforwardly interpreted as a Born-type configuration at a
fixed jet multiplicity, the real configurations could now be
attributed to either an additional jet, or an unresolved correction to
the jet multiplicity associated with the Born process. In the ME region,
the real emission is indeed accounted for as an additional jet, while in the PS region,
we perform a ‘standard’ NLO matching, by
subtracting the parton shower action expanded to the relevant order in
the strong coupling. Attaching vetoed parton shower evolution to the
configurations thus obtained completes our multijet merging
algorithm. We note that momentum conservation with a recoil scheme on
an emission-by-emission basis is needed to properly define the
phase-space boundaries between the ME and PS regions, and to decide on
what should be interpreted as the hard process, and how possible jet
selections would act on the hard process definition. Thus,
unfortunately, multijet merging is only possible with the dipole
shower, which conserves momentum on a per-emission basis.

\subsubsection{General formalism}
\label{\detokenize{review/matching/merging:general-formalism}}

In order to decide whether an emission from the parton shower
should be kept as a parton shower emission or would fall into the ME
phase-space region we use what we call a transparent (vetoed) parton shower, as
outlined in \cite{Bellm:2017ktr}. It is defined by the symbolic notation
\begin{equation*}
\begin{split}\PS_\mu [u(\phi_n,Q)]= \tPSV_\mu [\tPS_\rho[u(\phi_n,Q)]]\ ,\end{split}
\end{equation*}
where $\tPS_\rho$ only emits in the ME region while $\tPSV_\mu$
only emits into the PS region.
This shower is aware of the possibility that the emission scale
$q_i$ for a given emission might be different if that
emission would have been made from another leg and hence may still fall
into the PS region even though it would not for this particular
splitting.  This allows us to ensure that parton shower emissions from $\tPS_\rho$ truly
populate the matrix element phase-space above the merging scale.
With this definition we can replace
emissions in the ME region from $\tPS_\rho$ by emissions from the
full matrix elements in order to achieve a merging algorithm.  We
thereby ensure that the emission phase-space is completely covered by
all emissions and all regions are only accessible by a single type of
emission.

In this equation, we think of $u(\phi_n,q)$ as a functional representation of an event record: operators that operate on it can obtain the momenta of the $n$ final-state particles, and the upper limit for their subsequent evolution,~$q$. The function we considered in the earlier matching section is $u(\phi_n)=u(\phi_n,0)$ or equivalently $u(\phi,\mu)$, since there is no evolution below~$\mu$.

Once we have constructed states according to $\tPS_\rho$, showering them with $\tPSV_\mu$ is relatively straightforward: this is the standard shower, augmented with a special veto on emission above $\rho$. Emission is only vetoed if it falls into the matrix element region, i.e.~if it is above $\rho$ according to \emph{all} emission configurations. Emission from a given leg can still be produced above~$\rho$, if it happens to be in a phase-space region that is below $\rho$ with respect to some other leg.

Using (N)LO matrix elements to construct $\tPS_\rho$ states is, however, more involved. We use an algorithm that preserves, as far as possible, the unitarity that would be present if the states were generated by the parton shower $\tPS_\rho$ itself. By unitarity, we mean that the sum and integral over states that could be produced from the core hard process by a parton shower with at least a given number of emissions, does reproduce the inclusive cross section given by the fixed-order matrix element for the core hard process plus that number of additional partons. However, as discussed in detail in \cite{Bellm:2017ktr}, strictly enforcing unitarity can lead to a loss of logarithmic accuracy, because of matrix element configurations that cannot be produced by the parton shower algorithm. As a concrete example, consider in a sample of Z~+~jets events, an event in which two jets have high $p_t>M_Z$, roughly balanced, and the Z has low $p_t$. In principle, such an event could be considered a Z emission from a QCD $2\to2$ scattering, but in the absence of a multi-sector merging algorithm, we have to acknowledge that this configuration cannot come about as a QCD shower from a simple Z event. We therefore treat it as an additional hard process~-- it should not contribute to the unitarization of the lower jet cross sections, but higher jet multiplicities do contribute to its unitarization. A similar argument applies in cases where despite a strong ordering of scales, the flavour structure is such that the configuration could not have come from a shower splitting.

Our merging algorithm unitarizes with respect to shower-accessible configurations and takes into account these additional hard processes, and can merge NLO or LO matrix elements for each jet multiplicity.

In order to explain how this works, we first describe the algorithm as if all configurations can be produced by showering a single core hard process. Then, afterwards, we describe where this should be modified to account for additional hard process contributions.

The starting point is the merged $n$-jet cross section,
\begin{equation*}
\begin{split}
{\rm d}\sigma_n^{\text{merged}} \; u(\phi_n,q_n) 
&=
{\rm d}\sigma_n \; u(\phi_n,q_n) 
\left(1-
\left.\frac{\partial w^n_H}{\partial \alpha_{s}}\right|  
\right)
\\&\quad
- \int_\rho^{q_n} \hspace{-1mm} {\rm d}q
\left.
    \sum_{\alpha} \frac{w_{C,\alpha}}{\sum_\beta w_{C,\beta}}
    \; u(\phi^{\alpha}_n,q^{\alpha}_n) \; {\rm d}\sigma_{n+1} \; 
    \Delta_n(q_n|q_{n+1},\phi_n) \;
    \delta(q-q_{n+1})
    \right|_{\phi_n^\alpha=\phi_n;\,q_n^\alpha=q_n}
    \; ,
\end{split}
\end{equation*}
where ${\rm d}\sigma_n$ and ${\rm d}\sigma_{n+1}$ are the LO or NLO cross sections for $n$ and $n+1$ additional partons respectively.
The sum over $\alpha$ is over emission configurations, in practice dipoles, so $\alpha=(i)$ from the matching section. The $w_{C,\alpha}$ are cluster weights, in practice dipole emission factors, $\phi^{\alpha}_n$ is a projected phase space point, equivalent to $\rmPhiTilde_{n|(i)}(\phi_{n+1})$ and $q^{\alpha}_n$ is the corresponding emission scale, equivalent to $q_{(i)}(\phi_{n+1})$. $\Delta_n(q_n|q_{n+1},\phi_n)$ is the Sudakov form factor for non-emission at scales between $q_n$ and $q_{n+1}$, including a PDF ratio for incoming legs. The integral is over all configurations that have a scale $q_{n+1}>\rho$ that project onto the configuration $\phi_n$. If the parton shower were a perfect approximation to the higher order matrix element, the result of the integral would be exactly ${\rm d}\sigma_n \; u(\phi_n,q_n) \; (1-\Delta_n(q_n|\rho,\phi_n))$.

$w^n_H$ is called the history weight of this configuration, and will be defined and discussed in more detail shortly. The $\left.\frac{\partial}{\partial \alpha_{s}}\right|$ notation implies the first term of a Taylor expansion in $\alpha_s=\alpha_s(\mu_R)$, which can be obtained by writing all occurrences of scale-dependent functions (running couplings and PDFs) as expansions in $\alpha_s$ of their evolution equations. This is only needed when ${\rm d}\sigma_n$ is NLO, and is discussed in more detail in \hyperref[\detokenize{review/matching/merging:history-expansion}]{Section \ref{\detokenize{review/matching/merging:history-expansion}}}.

${\rm d}\sigma_n$ and ${\rm d}\sigma_{n+1}$ are obtained from fixed-order matrix-element calculations. In order to convert ${\rm d}\sigma_n^{\text{merged}}$ into the cross section for events in which the $n$ additional partons are the hardest $n$ partons, which can then be showered further, it has to be multiplied by the Sudakov form factor for no emission at higher scales. And, in order to match the logarithmic accuracy of the shower, we have to evaluate the running coupling in the CMW scheme, as well as the parton distribution functions, at the dynamic scale of emission. These extra factors are incorporated by multiplying by the  history weight~$w^n_H$. To calculate this, we have to construct the tree by which, if the parton shower had generated the same configuration, it would have done so. In the (usual) case that more than one tree can lead to the same configuration, their relative probabilities are calculated from dipole splitting functions as specified by the chooseHistory switch described in \hyperref[\detokenize{review/matching/merging:merging-user-choices}]{Section \ref{\detokenize{review/matching/merging:merging-user-choices}}}. With this tree, the history weight is calculated according to Eq. \eqref{equation:review/matching/merging:historyextractionweights} in the next section. If the configuration is one that could be generated by the parton shower from the core hard process, then $k$ in Eq. \eqref{equation:review/matching/merging:historyextractionweights} is equal to $n$ and the algorithm is complete.

As mentioned earlier, the unitarization inherent in this algorithm is not consistent with the logarithmic accuracy of the shower if the matrix element sample contains configurations that cannot be generated from the core hard process by parton showering. If the history construction finds a step that could not have been generated by a shower, it terminates and the clustered configuration it found is treated as a new hard process. It may have its own choice of renormalization and factorization scales (for example, in the Z~+~2 jet configuration mentioned earlier, one might choose the jet $H_T$, rather than $M_Z$ as a suitable scale). The rest of the algorithm operates in the same way, with $k$ in Eq. \eqref{equation:review/matching/merging:historyextractionweights} equal to the number of steps taken from $n$ to the new hard process. This ensures that additional emissions from this new hard process are unitarized.

\subsubsection{History extraction and weights}
\label{\detokenize{review/matching/merging:history-extraction-and-weights}}\label{\detokenize{review/matching/merging:historyextraction}}

To construct the shower history of the process with $n$ additional
legs, a tree of cluster nodes is constructed.
The nodes are determined by the possible dipoles in the CS subtraction
procedure. Each node is able to find the next steps in the reduced
multiplicities recursively.
With this tree at hand, each node can be asked to find the next set of ordered histories.

Once all the possible next-ordered nodes are extracted, the algorithm decides probabilistically which node to choose.
The probability can be chosen by the user as described in \hyperref[\detokenize{review/matching/merging:merginguserchoice}]{Section \ref{\detokenize{review/matching/merging:merginguserchoice}}}.

With the full history at hand, the weights are calculated for a shower history with $k$ steps, as
\begin{equation}\label{equation:review/matching/merging:historyextractionweights}
\begin{split}w_H^k =   \frac{f_k(\eta_{k},q_k)}{f_{k}(\eta_{k},\mu_F)} \prod_{i=0}^{k-1}\frac{\alpha_{s}(q_i)}{\alpha_{s}(\mu_R)}\frac{f_i(\eta_{i},q_i)}{f_i(\eta_{i},q_{i+1})}  w_N(q_i|q_{i+1},\phi_i)\;,\end{split}
\end{equation}
where $w_N$ is the non-emission weight (called $\Delta_i(q_a|q_b,\phi_i)$ above), given by the product of Sudakov form factors for each leg,
\begin{equation*}
\begin{split}w_N(q_a|q_b,\phi_i)=
\Pi^{(1)}(q_a|q_b,\phi_i) \; 
\Pi^{(2)}(q_a|q_b,\phi_i)
\prod_f \Delta^{(f)}(q_a|q_b,\phi_i)\;,
\end{split}
\end{equation*}
and $\Pi$ is the backward-evolution Sudakov, equal to the standard Sudakov with a PDF ratio in the exponent.

In constructing the tree, if any ordering scales are below the merging scale, this is not a ME event and should be given weight zero.
If any ordering scales are disordered, the step before the step that has the disordered scale is called the hard process.
If the final step before the hard process has an ordering variable greater than the scale of the hard process~-- calculated from the ScaleChoice object~-- this is again considered a new hard process.

\subsubsection{History expansion}
\label{\detokenize{review/matching/merging:history-expansion}}

For the fixed order accuracy it is crucial to expand the impact of the shower/resummation
to the same accuracy as it is added with the improved full calculation.
Compared to the matching, where the expansion of the shower leads to
the shower approximation only, the handling in the NLO merging includes more
parts. To claim NLO fixed order accuracy one has to expand not only the same
multiplicity and the following as in NLO matching but also the full history of
states that contributed to the process in question.
In \hyperref[\detokenize{review/matching/merging:historyextraction}]{Section \ref{\detokenize{review/matching/merging:historyextraction}}} the algorithm to extract the history is described.
For each step in this evolution history the weights in Eq. \eqref{equation:review/matching/merging:historyextractionweights}
need to be expanded to their respective $\alpha_{s}$-expansion.
Code-wise, similar functions are used to calculate the expansions and the actual
weights and only minor modifications are needed.
To preserve the accuracy of the shower but also the fixed order expansion
care needs to be taken that the virtual and real corrections are reweighted similar
as the expansion weights, however, there are still ambiguities of how to
define the $\alpha_{s}$-expansion (e.g.\ multiplicative or
additive).  As every weight is a product of a PDF-weight ($f$), an
$\alpha_{s}$-weight ($\alpha$) and a Sudakov-weight
($\Delta$) we can formally subtract the NLO contributions \textit{e.g.} at the level of each factor or globally.
Variations should only affect terms of order $\alpha^2_S$. To be able
to perform such a variation several schemes ($s$) are defined:
\begin{itemize}
\item {} 

No subtraction of $\alpha_{s}$ terms ($s=0$).  This is
formally incorrect and is kept for comparison only.

\item {} 

The $\alpha_{s}$-terms from each weight are subtracted straight
away ($s=1$).  This choice seems least ambiguous.

\item {} 

The $\Delta$-weight is kept fixed in the subtraction
($s=2$).

\item {} 

All weights are kept fixed in the subtraction ($s=3$).

\item {} 

The $\alpha$-weight is kept fixed in the subtraction
($s=4$).

\end{itemize}

Further details and a comparison of these schemes can be found in
\cite{Bellm:2017ktr}.  While formally all schemes (except $s=0$!)
are at the same level of accuracy there are notably numerical
differences which should be considered an estimate for this source of
uncertainty.

The parameter that controls those expansions is
\href{https://herwig.hepforge.org/doxygen/classHerwig\_1\_1MergerBase.html}{ShowerExpansionWeights}
that belongs to the \href{https://herwig.hepforge.org/doxygen/classHerwig\_1\_1MergerBase.html}{Merger} and can be set to the schemes $s$ as
described above.

\subsubsection{MC-scheme treatment}
\label{\detokenize{review/matching/merging:mc-scheme-treatment}}\label{\detokenize{review/matching/merging:mcscheme-merging}}

The so-called MC-scheme \cite{Catani:1990rr} that is commonly used in parton
showers is used to capture effects of higher-order corrections for
final state gluon emissions.
Those scheme-dependent corrections, however, are also included if NLO
corrections are calculated for the additional emissions. In particular,
the modification that is proposed in
\cite{Catani:1990rr} to make the transition from $\overline{\mathrm{MS}}$ to MC will have an impact.
Including the actual NLO correction and in addition taking into account the
main effects of the approximated correction can, therefore, lead to double counting.
To circumvent the issue the NLO-merging treats the transition from $\overline{\mathrm{MS}}$ to MC
as an $\alpha_{s}$ effect that is taken care of in the
$\alpha_{s}$-expansion of the history weight.

\subsubsection{Merging user choices}
\label{\detokenize{review/matching/merging:merging-user-choices}}\label{\detokenize{review/matching/merging:merginguserchoice}}

As described in \hyperref[\detokenize{review/matching/merging:historyextraction}]{Section \ref{\detokenize{review/matching/merging:historyextraction}}} a choice has to be made in the extraction of the
shower history. To allow variations the weights to choose the history can be modified with the switch,
\begin{itemize}
\item {} 

chooseHistory:
The weight to choose the history are calculated from the
0: dipole XS (default):
1: dipole/born:
2: flat:
3: $1/p_{t, \rm dip}$

\item {} 

MergingScale:
The merging scale is the key scale that is used to define the matrix element region.
The scale can be modified with the interface, e.g.\ ‘set Merger:MergingScale 20*GeV’.

\item {} 

MergingScaleSmearing:
In general the merging scale is an unphysical parameter and the effects of the scale are mild.
To minimize further dependencies it is possible to smear the scale. For this, the parameter
belonging to the interface Merger: MergingScaleSmearing can be used. e.g.\ a value of 0.1 would mean a
10\% variation.

\item {} 

CMWScheme:
As described in \hyperref[\detokenize{review/matching/merging:mcscheme-merging}]{Section \ref{\detokenize{review/matching/merging:mcscheme-merging}}} the scheme we use to describe the strong coupling is
of importance to be consistent with the NLO merging. The merging and the dipoles can be used with three
schemes. For comparison, the MC-scheme can be switched off, but the Linear,
\begin{equation*}
\begin{split}\alpha^{\mathrm{MC}}_s(q) = \alpha^{\overline{\mathrm{MS}}}_s(q) \; \left( 1+K_g \,
\frac{\alpha^{\overline{\mathrm{MS}}}_s(q)}{2\pi} \right) \mathrm{\;\;with\;\;}
K_g=C_A\left(\frac{67}{18}-\frac{\pi^2}{6}\right)-\frac{5}{9}N_f\ .\end{split}
\end{equation*}

or Factor,
\begin{equation*}
\begin{split}\alpha^{\mathrm{MC}}_s(q)  = \alpha^{\overline{\mathrm{MS}}}_s(k q) \mathrm{\;\;with\;\;}
k=\exp\left(-\frac{67-3\pi^2-\frac{10}{3} N_f}{33-2N_f}\right)\end{split}
\end{equation*}

schemes are preferred.
To steer the behaviour the input files Merging/FactorCMWScheme.in or Merging/LinearCMWScheme.in should be read.

\item {} 

ShowerExpansionWeights: see \hyperref[\detokenize{review/matching/merging:history-expansion}]{Section \ref{\detokenize{review/matching/merging:history-expansion}}}.

\end{itemize}

\subsubsection{MergingFactory}
\label{\detokenize{review/matching/merging:mergingfactory}}

As in the case of the matching of NLO calculations to the showers in Herwig,
the merging procedure is set up in a SubprocessHandler, here called MergingFactory.
The syntax to initiate a process is kept close to that used for matching.
Compared to the matching
an additional square-bracket is included in the process definition, to indicate
additional jet multiplicities, e.g.\ “p p -\textgreater{} l+ l- {[}j j{]}” would generate the Drell--Yan process
with two additional, merged, Born-like processes ($Z + j$, $Z + 2j$) at LO.

These additional merged processes can also be calculated at NLO, using the parameter NLOProcesses
within the MergingFactory.
NLOProcesses is an integer corresponding to the number of merged processes to be calculated at NLO.  
The counting starts with the underlying process (i.e.~excluding the square-brackets),
such that a NLOProcesses value of 1 includes only
NLO corrections to the underlying Born-like process (“p p -\textgreater{} l+ l-” in the example above).
Example results from just such a calculation, with NLOProcesses set to 2, are shown in 
\hyperref[\detokenize{review/matching/examples:DYCMS}]{Fig.\@ \ref{\detokenize{review/matching/examples:DYCMS}}}
and
\hyperref[\detokenize{review/matching/examples:ZjCMS}]{Fig.\@ \ref{\detokenize{review/matching/examples:ZjCMS}}},
compared to NLO matched results from Matchbox, and CMS data from
\cite{CMS:2019raw,CMS:2022ubq}.

As the number of generated subprocesses can be large, it is possible to split up the integration- and run-steps into
chunks of subsets.  If the MergingFactory:Chunk parameter is set to an integer larger than zero,
the integration- and run-steps
are split into smaller pieces. Each input-file with an integer parameter MergingFactory:ChunkPart with values
in {[}0,Chunk{]} contains a subset of processes. The result of each of the subsets needs to be merged by
the user at a later stage. Further, the uniqueness and completeness of the sets must be ensured by the user.

\begin{figure}[tp]
\centering
\capstart
\begin{subfigure}{0.49\textwidth}
\centering

\noindent\includegraphics[width=0.999\linewidth]{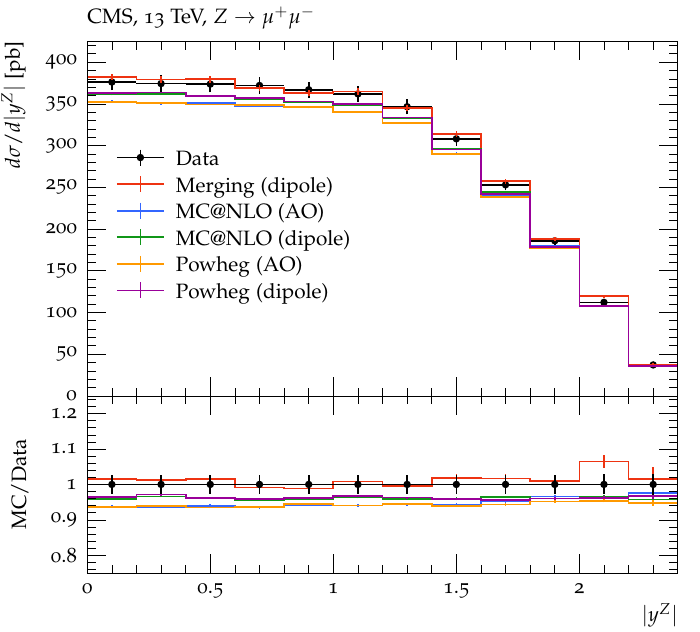}
\caption{Rapidity distribution of the reconstructed $Z$-boson.}
\end{subfigure}
\begin{subfigure}{0.49\textwidth}
\centering

\noindent\includegraphics[width=0.999\linewidth]{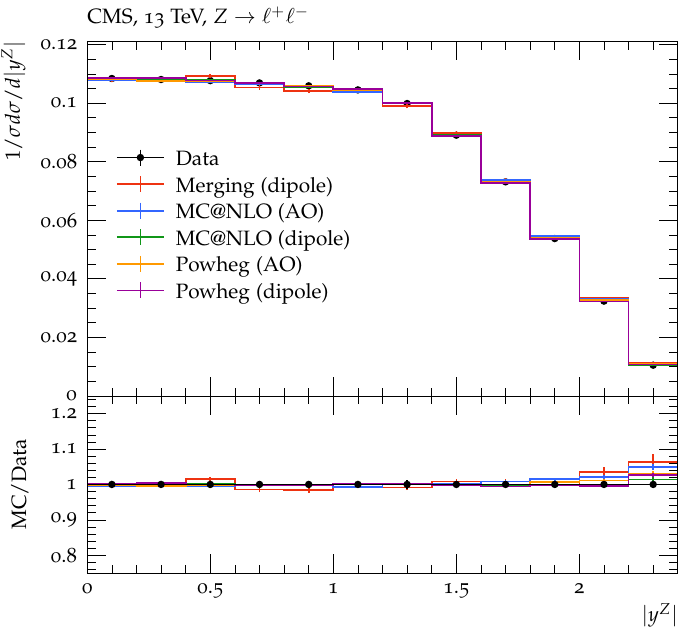}
\caption{Shape of the rapidity distribution.}
\end{subfigure}
\caption{Matching and merging with Matchbox for the Drell--Yan process at the LHC (\cite{CMS:2019raw};
\texttt{{CMS\_2019\_I1753680}}).}\label{\detokenize{review/matching/examples:DYCMS}}

\capstart
\begin{subfigure}[t]{0.49\textwidth}
\centering
\noindent\includegraphics[width=0.999\linewidth]{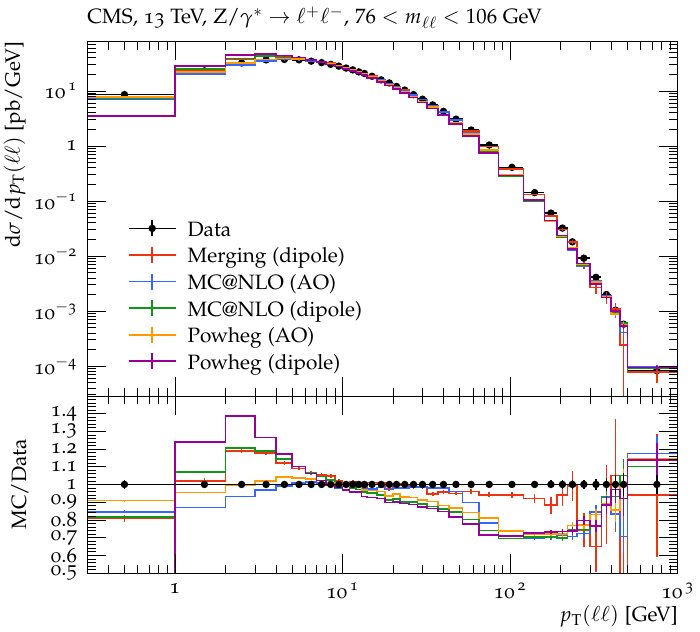}
\caption{Transverse momentum of the reconstructed $Z$-boson, double-differential in its invariant mass.}
\end{subfigure}
\begin{subfigure}[t]{0.49\textwidth}
\centering
\noindent\includegraphics[width=0.999\linewidth]{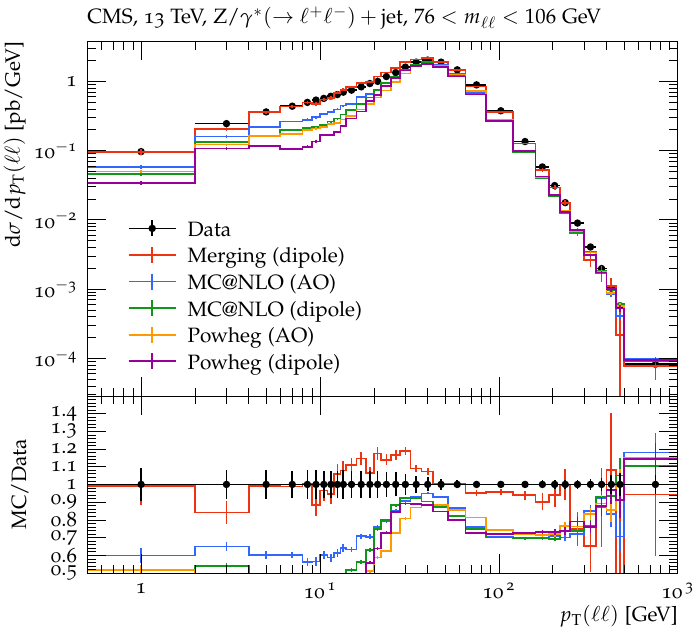}
\caption{The same distribution, subject to an additional jet cut.}
\end{subfigure}
\caption{Matching and merging with Matchbox,
applied to the Drell--Yan process at the LHC (\cite{CMS:2022ubq}; \texttt{CMS\_2022\_I2079374}).
Even with the additional jet cut, the merged prediction retains
NLO accuracy for the transverse-momentum distribution.
This could also be obtained from an NLO-matched $Z+j$ prediction.}\label{\detokenize{review/matching/examples:ZjCMS}}
\end{figure}

\subsection{Matrix element corrections}
\label{\detokenize{review/matching/matrix-element-corrections:matrix-element-corrections}}\label{\detokenize{review/matching/matrix-element-corrections:id1}}\label{\detokenize{review/matching/matrix-element-corrections::doc}}

For a few processes, we provide a simpler matrix element improvement
to the angular-ordered parton shower, which we call matrix element
corrections. These can be thought of as a simple form of matching, in
that next-order real emission matrix elements are used in phase-space
regions that cannot be generated by the parton shower (`hard'
corrections) and are used to reweight the parton shower distributions
via the Sudakov veto algorithm in the region that they do cover (`soft
corrections'). However, no virtual corrections are included, so the
normalization is still to the leading order cross section.

These corrections are available for all processes involving a pair of
colour-connected partons, i.e.\ $e^{+}e^{-}\rightarrow q\bar{q}$
and related processes like EW vector boson decay; deep
inelastic scattering; Drell--Yan processes; and top quark decay.

In this section, we briefly recap the main features of these hard and
soft matrix element corrections. More details can be obtained from
Ref. \cite{Seymour:1994df}.

\subsubsection{Hard matrix element corrections}
\label{\detokenize{review/matching/matrix-element-corrections:hard-matrix-element-corrections}}\label{\detokenize{review/matching/matrix-element-corrections:sub-hard-matrix-corrections}}

Since angular-ordered parton showers only generate radiation within
some angularly-defined subset of phase-space, it is common that the
first emission cannot cover the whole of phase-space, as is the case
for the $\tilde{q}$-shower implemented in Herwig 7. Soft emission
from the individual emitters is possible at all angles, and their
emission regions just ‘touch’ each other, meaning that soft emission
is counted once and only once. Collinear emission from each parton
populates the full available hard collinear phase-space. However, hard
emission cannot be emitted at all angles, and there is a region of
phase-space that the shower misses, which we call the dead region.

Since the dead region does not include any soft or collinear emission,
the integral of the real-emission matrix element over it is
finite. Moreover, it is typically small. Thus it is consistent to
ignore any resummation / form factor effects and to simply generate
phase-space points across this region weighted by the tree-level
matrix element. These are then used as the starting point for a
subsequent parton shower. Since the partons in these configurations
are always hard and well-separated there is no merging ambiguity or
double-counting with this subsequent emission.

\subsubsection{Soft matrix element corrections}
\label{\detokenize{review/matching/matrix-element-corrections:soft-matrix-element-corrections}}\label{\detokenize{review/matching/matrix-element-corrections:sub-soft-matrix-element-corrections}}

The parton shower approximation is, by construction, accurate in the
soft and collinear limits, and is not guaranteed to remain accurate
away from those limits. The idea of the soft matrix element correction
is to use the next-order real emission matrix element as a
multiplicative reweighting of the emission distribution, similar to
elements of the POWHEG and KrkNLO methods. The additional element is
that it is applied to \textit{every emission that is the hardest} (highest
transverse momentum) \textit{so far}. Thus, not only is the hardest emission
in the event corrected by the hard matrix element as in those
approaches, but when there are multiple widely-separated emissions,
they are all emitted independently according to the real emission
matrix element.

The veto algorithm used by the shower is then simply augmented by an
additional veto with probability given by the ratio of the matrix
element distribution over the analytically-calculated parton shower
distribution. In cases that this ratio can be greater than unity, a
simple extension of the algorithm multiplies the shower emission
probability by the smallest integer that bounds the ratio (in
practice, $2$ is always sufficient) and then the veto brings
this down to the correct value.

\subsection{Matching and merging with \texttt{MadGraph5\_aMC@NLO} LHE files}
\label{\detokenize{review/matching/mg5merging:mg5merging}}\label{\detokenize{review/matching/mg5merging::doc}}

The \texttt{FxFx} merging module provides support for the NLO multi-jet merging method of~\cite{Frederix:2012ps}, via Les Houches-accord event files generated by \texttt{MadGraph5\_aMC@NLO}~\cite{Alwall:2011uj}. The framework also
provides an interface for merging of tree-level events generated
either by \texttt{MadGraph5\_aMC@NLO} or \texttt{AlpGen}~\cite{Mangano:2002ea} via the MLM
technique. The relevant input files for the \texttt{FxFx} merging and tree-level merging are now \texttt{LHE-FxFx.in} and \texttt{LHE-MGMerging.in}
respectively. We emphasize that it is essential to include the MC@NLO matching settings for \texttt{MadGraph5\_aMC@NLO} when performing the FxFx merging, as given in \texttt{LHE-MCatNLO.in}. These settings
should not be included when merging tree-level events. The tree-level merging functionality via \texttt{MadGraph5\_aMC@NLO} events uses the event tags in the appropriately-generated LHE files and requires the option \texttt{MergeMode} to be set to \texttt{TreeMG5}, as is done by default in \texttt{LHE-MGMerging.in}. To enable merging with events generated via \texttt{AlpGen},  \texttt{MergeMode} should be switched to \texttt{Tree}.

\begin{figure}[tp]
\centering
\capstart
\begin{subfigure}{0.49\textwidth}
\centering

\noindent\includegraphics[width=0.999\linewidth]{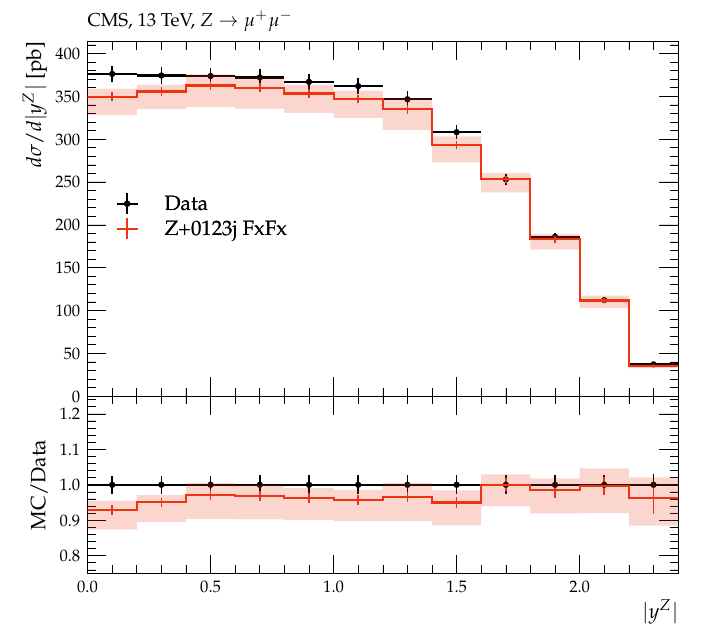}
\caption{Rapidity distribution of the reconstructed $Z$-boson.}
\end{subfigure}
\begin{subfigure}{0.49\textwidth}
\centering

\noindent\includegraphics[width=0.999\linewidth]{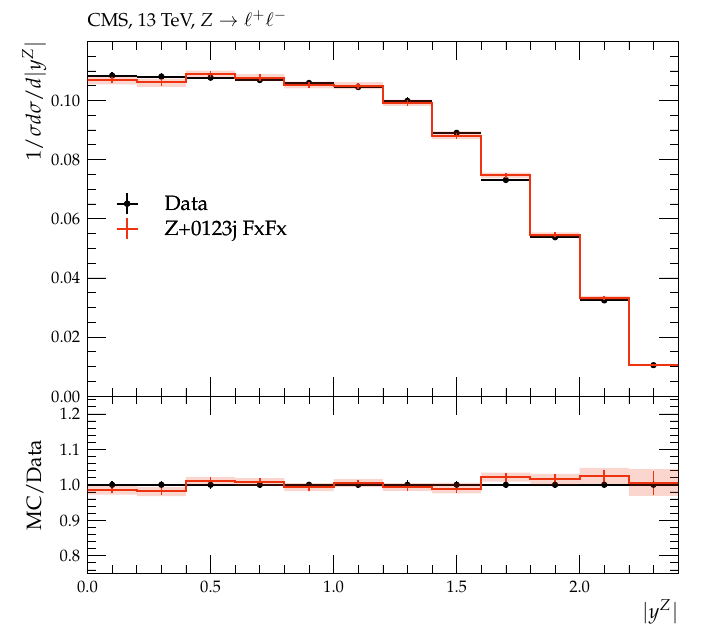}
\caption{Shape of the rapidity distribution.}
\end{subfigure}
\caption{Matching and merging of \texttt{MadGraph5\_aMC@NLO} LHE files, for the Drell--Yan process at the LHC (\cite{CMS:2019raw}; \texttt{{CMS\_2019\_I1753680}}). A merging scale of 30~GeV was employed, and the bands represent hard process scale variations by a factor of 2 up or down.}\label{\detokenize{review/matching/examples:fxfx1}}

\capstart
\begin{subfigure}[t]{0.49\textwidth}
\centering
\noindent\includegraphics[width=0.999\linewidth]{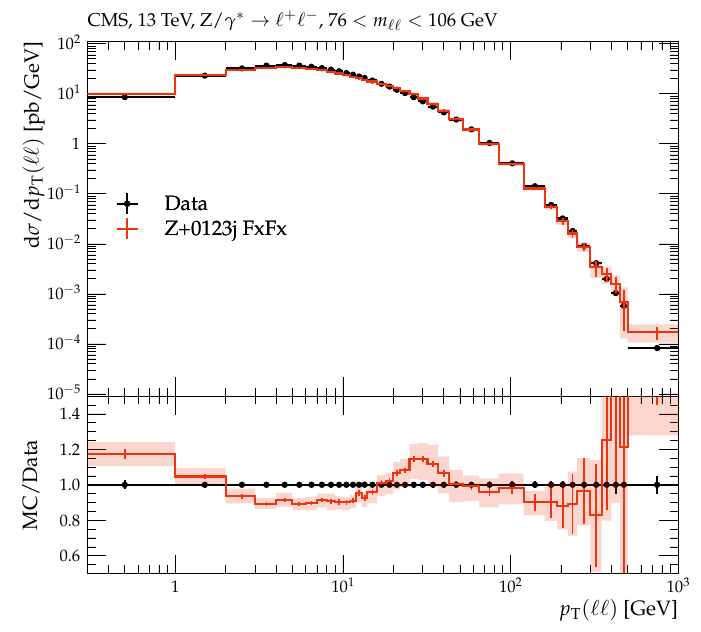}
\caption{Transverse momentum of the reconstructed $Z$-boson, double-differential in its invariant mass.}
\end{subfigure}
\begin{subfigure}[t]{0.49\textwidth}
\centering
\noindent\includegraphics[width=0.999\linewidth]{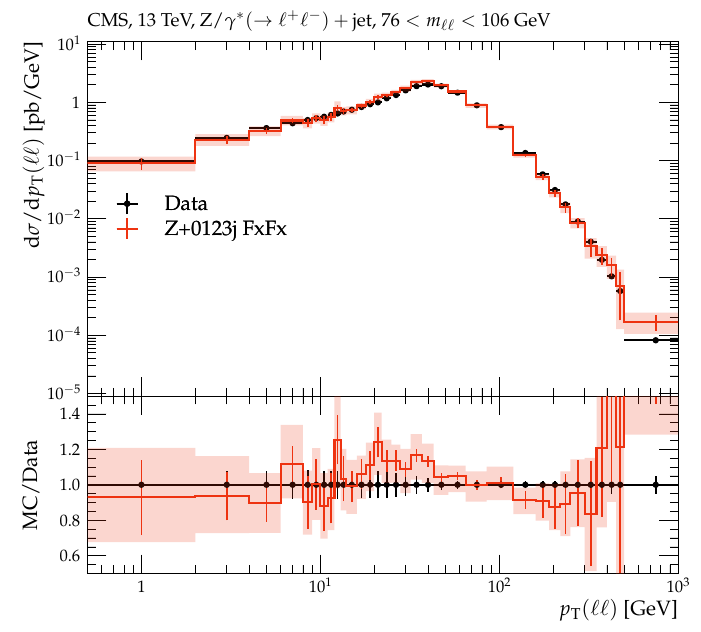}
\caption{The same distribution, subject to an additional jet cut.}
\end{subfigure}
\caption{Matching and merging of \texttt{MadGraph5\_aMC@NLO} LHE files, applied to the Drell--Yan process at the LHC~(\cite{CMS:2022ubq}; \texttt{CMS\_2022\_I2079374}). A merging scale of 30~GeV was employed, and the bands represent hard process scale variations by a factor of 2 up or down.}\label{\detokenize{review/matching/examples:fxfx2}}
\end{figure}

The FxFx functionality was tested thoroughly for $\mathrm{W+jets}$ and $\mathrm{Z+jets}$ events in~\cite{Frederix:2015eii}, where it was compared against LHC data at 7 and 8 TeV. Here, we present results at 13~TeV, matched and merged at NLO including 0, 1, 2, 3 jet multiplicities at matrix-element level, with the rapidity distribution of the reconstructed $Z$-boson shown in 
\hyperref[\detokenize{review/matching/examples:fxfx1}]{Fig.\@ \ref{\detokenize{review/matching/examples:fxfx1}}}, compared to CMS data. The transverse momentum of the reconstructed $Z$-boson is shown in 
\hyperref[\detokenize{review/matching/examples:fxfx1}]{Fig.\@ \ref{\detokenize{review/matching/examples:fxfx2}}}. A merging scale of 30~GeV was employed, along with hard process scale variations by a factor of 2 up or down. 

We note that no tuning was performed using events generated via this interface.

\subsection{Code structure}
\label{\detokenize{review/matching/code-structure:code-structure}}\label{\detokenize{review/matching/code-structure:matching-code-structure}}\label{\detokenize{review/matching/code-structure::doc}}

The key task of the matching-specific code within Matchbox is to construct explicitly the
parton-shower approximation to the real-emission matrix-element in a manner consistent
with that constructed implicitly by the active
\texttt{ShowerHandler}
object, which generates the first-emission of the parton-shower algorithm
(and then the others) as described in \hyperref[\detokenize{review/showers/code:sect-showercode}]{Section \ref{\detokenize{review/showers/code:sect-showercode}}}.

This shower-approximation contribution is then combined consistently with the other
contributions to the matched cross section in the SubtractionDipole object as
described in \hyperref[\detokenize{review/hardprocess/code-structure:matchbox-code-structure}]{Section \ref{\detokenize{review/hardprocess/code-structure:matchbox-code-structure}}}.

\subsubsection{Shower approximation}
\label{\detokenize{review/matching/code-structure:shower-approximation}}

The
\href{https://herwig.hepforge.org/doxygen/classHerwig\_1\_1ShowerApproximation.html}{ShowerApproximation}
class is a base class storing information required by both the angular-ordered and the dipole-shower
matching, including pointers to the real and Born XComb objects, the SubtractionDipole object
responsible for the same emission as is under consideration, and the TildeKinematics and
InvertedTildeKinematics objects governing the phase-space mappings between the
Born and real phase-space configurations.

It also contains methods which may be overridden by shower-specific implementations in the
DipoleMatching
and
QTildeMatching
subclasses, as described below.

Methods which are common to all concrete implementations (aside from ‘getters’ and ‘setters’)
are primarily concerned with identifying the scales and PDF weights associated with
the respective Born and real phase-space configurations.

\subsubsection{Subtractive matching}
\label{\detokenize{review/matching/code-structure:subtractive-matching}}

As described in \hyperref[\detokenize{review/matching/matching-subtractions:matching-subtractions-subtractive}]{Section \ref{\detokenize{review/matching/matching-subtractions:matching-subtractions-subtractive}}}, Matchbox allows ‘subtractive’,
MC@NLO-type matching to both the angular-ordered and dipole showers.
These are handled by the  QTildeMatching and DipoleMatching subclasses of the ShowerApproximation
base class respectively.

In both cases, for each relevant splitting identified by the MatchboxFactory,
QTildeMatching or DipoleMatching checks if the specified real phase-space point is
within the phase-space region that could be generated via that splitting from the
underlying Born configuration, including whether it is above the cut-off scale
and below the starting ‘hard’ scale.

If so, it calculates the product of the relevant splitting function with the
underlying Born matrix-element, dresses it with the appropriate Born PDF and coupling
factors, and prepares it as a contribution to the matched cross section to be included
alongside the others in SubtractionDipole.

\paragraph{Matching to the angular-ordered shower}
\label{\detokenize{review/matching/code-structure:matching-to-the-angular-ordered-shower}}

The
\href{https://herwig.hepforge.org/doxygen/classHerwig\_1\_1QTildeMatching.html}{QTildeMatching}
class governs subtractive matching to the angular-ordered shower.
Concretely, it contains an implementation of the splitting functions associated with
each splitting and uses these to construct the shower-approximation.

The shower IR-cutoff is propagated from the QTildeShower and applied consistently
to the shower approximation, subject to a technical cut-off specified by SafeCut.

\paragraph{Matching to the dipole shower}
\label{\detokenize{review/matching/code-structure:matching-to-the-dipole-shower}}

The
\href{https://herwig.hepforge.org/doxygen/classHerwig\_1\_1DipoleMatching.html}{DipoleMatching}
class governs subtractive matching to the Herwig dipole shower.
In this case, the implementation uses the SubtractionDipole objects to calculate the dipole functions.

The shower-approximation IR cutoff is set separately from that used within the dipole shower
itself, again subject to the SafeCut technical cut-off.

\subsubsection{Multiplicative matching}
\label{\detokenize{review/matching/code-structure:multiplicative-matching}}

As described in \hyperref[\detokenize{review/matching/matching-subtractions:matching-subtractions-multiplicative}]{Section \ref{\detokenize{review/matching/matching-subtractions:matching-subtractions-multiplicative}}}, Matchbox allows multiplicative matching
in which the emission is sampled according to the ratio of real-emission to Born matrix-elements.
In practice, this is implemented as a special case of the ShowerApproximation class in
MEMatching, which uses the information provided by the MatchboxFactory to assemble the possible
splittings together with the corresponding real and Born matrix-elements.

\paragraph{Shower approximation kernels}
\label{\detokenize{review/matching/code-structure:shower-approximation-kernels}}

The
\href{https://herwig.hepforge.org/doxygen/classHerwig\_1\_1ShowerApproximationKernel.html}{ShowerApproximationKernel}
class implements a splitting generator for sampling from a splitting
kernel.
For the sampling itself, it uses an instance of the ExSample adaptive sampler \cite{Platzer:2011dr}
to construct an adaptive overestimate function for each possible splitting.
Since the effective ‘splitting functions’ in the matrix-element corrections case also depend
on the Born phase-space,
through the Born matrix-element in the denominator of the ratio,
the sampler also adapts to the random numbers used to generate the Born phase-space,
to help improve the sampling efficiency.

\paragraph{Shower approximation generator}
\label{\detokenize{review/matching/code-structure:shower-approximation-generator}}

The
\href{https://herwig.hepforge.org/doxygen/classHerwig\_1\_1ShowerApproximationGenerator.html}{ShowerApproximationGenerator} class implements a single step of the parton shower,
by performing a competition between the ShowerApproximationKernel splittings
identified by the MatchboxFactory as being possible from the underlying Born phase-space configuration.
These are stored in theKernelMap for each Born partonic configuration.

The real-emission matrix-element is partitioned among splitting channels according to
dipole-type eikonal factors computed in the channelWeight method of the ShowerApproximation
base class.

Once the competition has been performed and the winning splitting with the highest-scale
emission identified, the event record is updated to correspond to the chosen real-emission
subprocess.

\subsubsection{Merging}
\label{\detokenize{review/matching/code-structure:merging}}

The \href{https://herwig.hepforge.org/doxygen/classHerwig\_1\_1MergingFactory.html}{MergingFactory}
organizes the matrix elements for all involved subprocesses at different
orders and provides events with the individual weights.  All
calculations are carried out by the \texttt{Merger}.
Here the merging histories are determined and assigned weights as
described above.  Both classes inherit from \href{https://herwig.hepforge.org/doxygen/classHerwig\_1\_1MergerBase.html}{MergerBase}
that provides some common interfaces.

\clearpage

\section{Physics Beyond the Standard Model}
\label{\detokenize{review/index:physics-beyond-the-standard-model}}\label{\detokenize{review/index:sect-bsm}}

No one knows what kind of new physics, if any, will be encountered in the LHC era, and a wide variety of new physics models must be studied. This motivated the design of the Herwig program, with the inclusion of a general framework for the implementation of new physics models. Using this framework, new models can be realized quickly and efficiently. This method is described in full in Refs.~\cite{Gigg:2007cr, Gigg:2008yc} and will be reviewed and updated here.

In describing the features needed to simulate Beyond the Standard Model (BSM) processes, we need only to concern ourselves with the hard collisions, either producing known particles through modified couplings or the exchange of new particles, or producing new particles in the final state, and with decays of the new particles. All other steps of event generation are handled in the same way as for Standard Model processes%
\begin{footnote}\sphinxAtStartFootnote
Other features do emerge in certain models, for example the hadronization of new long-lived coloured particles, which is not yet fully implemented in Herwig, but for the majority of new physics models under active study this is the case.
\end{footnote}.
Both of these steps involve calculating an amplitude, which in turn relies on knowledge of the Feynman rules within the model being used. In Herwig, the Feynman rules are implemented as a series of Vertex classes that inherit from the generic classes of ThePEG. These Vertex classes are based on the HELAS formalism \cite{Murayama:1992gi}, with each class able to evaluate the vertex as a complex number or, given different information, an off-shell wavefunction that can be used as input for another calculation. Each Feynman diagram contributing to a given process is evaluated in terms of these vertex building blocks, and the sum of the resulting contributions is squared to give the matrix element.

While a number of models are available internally in Herwig, we also provide an interface that can convert models in the UFO format \cite{Degrande:2011ua} into a library usable by Herwig. We therefore do not expect any further models to be directly implemented in Herwig.

In this section, we start by briefly describing the generation of the hard processes and decays in models of new physics. This is followed by a description of the models currently implemented in Herwig, including the Standard Model, the interface to UFO models, and finally the structure of the code.

\subsection{Hard process}
\label{\detokenize{review/BSM:hard-process}}\label{\detokenize{review/BSM:sec-bsmhardprocess}}\label{\detokenize{review/BSM::doc}}

\hyperref[\detokenize{review/hardprocess/me:builtin-matrix-elements}]{Section \ref{\detokenize{review/hardprocess/me:builtin-matrix-elements}}} gave details on the default matrix elements
available for generating Standard Model processes in Herwig. These
classes are based on specific particle interactions whereas the classes
used for BSM models are determined by the external spin structure of a
$2\to2$ scattering process. To generate a specific process the
user specifies the desired states that are to participate in the hard
interaction, using configuration files, and the code then generates
the relevant diagrams and a MatrixElement object for each process%
\begin{footnote}\sphinxAtStartFootnote
It is only necessary to specify a single outgoing particle as the
code will produce all processes with this particle in the final
state.
\end{footnote}.
In addition to the general $2\to2$ scattering processes a small
number of higher-multiplicity processes are available for neutral scalar
bosons, including associated production with either a $W^\pm$ or
$Z^0$ boson or heavy quark-antiquark pair, and for the vector-boson
fusion process.

\subsection{Decays}
\label{\detokenize{review/BSM:decays}}\label{\detokenize{review/BSM:sec-bsmdecays}}

In order to decay the BSM states, the possible decay modes must first
be known. If a supersymmetric model is required a spectrum
generator can be used to produce not only the required spectrum, in accordance with
the SUSY Les Houches Accord \cite{Skands:2003cj}, but also a decay
table. Herwig is designed to be able to read this information and set
up the appropriate decay modes for later use. Other models do not have
such programs and therefore the list of possible two- and three-body
decays is generated automatically, together with
the option of using hadronic-currents for decays where the
mass splitting is small and four-body decays of scalars to fermions.

A multi-particle final state may be generated either as a sequence of two-body decays through an unstable intermediate particle or directly by a dedicated multi-body \texttt{Decayer}. In the sequential treatment, the production and subsequent decay of the intermediate are represented by separate \texttt{DecayMode}s. In the direct treatment, the complete multi-body matrix element is evaluated, including the relevant intermediate propagators and interference terms where implemented.

For decay modes constructed automatically for BSM models, the treatment of potentially resonant contributions is controlled by the \texttt{RemoveOnShell} option of the corresponding \texttt{NBodyDecayConstructorBase}. By default, diagrams containing an intermediate particle that can be produced and decay on shell are removed from the direct multi-body mode, since they are instead treated as a sequence of two-body decays. The user may retain these diagrams by setting \texttt{RemoveOnShell=No}; the \texttt{Production} option removes them whenever on-shell production of the intermediate is kinematically possible. Separately, the \texttt{GenerateIntermediates} option controls whether intermediate particles occurring in a direct multi-body decay are included explicitly in the event record, without changing the underlying matrix element.

When generating the possible decays automatically we also need to be
able to calculate the partial width of a given mode so that the
branching fraction and total width can be calculated. For a general
two-body decay, the matrix element only depends on the mass-square
values of each particle so the phase-space factor can be integrated
separately and the partial width is given by
\begin{equation*}
\begin{split}\Gamma(a\to b,c) = \frac{|\overline{\mathcal{M}}|^2 p_{cm}}
        {8\pi m_a^2},\end{split}
\end{equation*}

where $|\overline{\mathcal{M}}|^2$ is the matrix element squared
summed over final-state colours and spins and averaged over
initial-state colours and spins and $p_{cm}$ is the centre-of-mass
momentum
\begin{equation*}
\begin{split}p_{cm} = \frac{1}{2 m_a}\left[ \left( m_a^2 - (m_b + m_c)^2 \right)
    \left( m_a^2 - (m_b - m_c)^2 \right) \right]^{1/2}\!.\end{split}
\end{equation*}

A three-body decay has a partial width given by
\begin{equation}\label{equation:review/BSM:eqn:threebodypartial}
\begin{split}  \Gamma(a\to b,c,d) =
  \frac{1}{(2\pi)^3}\frac{1}{32m_a^3}
  \int^{(m_a - m_d)^2}_{(m_b + m_c)^2} \mathrm{d}m_{bc}^2
  \int^{(m_{cd}^2)_{\mathrm{max}}}_{(m_{cd}^2)_{\mathrm{min}}} \mathrm{d}m_{cd}^2 \;
  |\overline{\mathcal{M}}|^2,\end{split}
\end{equation}

with
  \begin{eqnarray}
    (m_{cd}^2)_{\mathrm{max}} & = &(E_c^\ast + E_d^\ast)^2 -
    \left(\sqrt{E_c^{*2} - m_c^2} - \sqrt{E_d^{*2} - m_d^2}\right)^2, \\
    (m_{cd}^2)_{\mathrm{min}} & = &(E_c^\ast + E_d^\ast)^2 -
    \left(\sqrt{E_c^{*2} - m_c^2} + \sqrt{E_d^{*2} - m_d^2}\right)^2,
  \end{eqnarray}

where $E^\ast_c = (m_{bc}^2 - m_b^2 + m_c^2)/2m_{bc}$ and
$E^\ast_d = (m_a^2 - m_{bc}^2 - m_d^2)/2m_{bc}$ are the energies
of $c$ and $d$ respectively, in the $(bc)$ rest frame.
In general, the
phase-space integration can no longer be performed analytically, because
the matrix element is a complicated function of the invariant mass
combinations $m_{bc}$ and $m_{cd}$, therefore it must be
performed numerically. Given the low number of dimensions of the
phase-space integrals in Eq. \eqref{equation:review/BSM:eqn:threebodypartial}, they are
performed using standard techniques rather than by the Monte Carlo
method. The total width of the parent is simply the sum of the partial
widths.

To compute the momenta of the decay products, we need to be able to
calculate the matrix element for a selected decay mode. When each mode
is created, it is assigned a
\texttt{Decayer}
object that is capable of calculating the value of
$|{\mathcal{M}}|^2$ for that process. This is done in a similar
manner to the hard matrix element calculations, \textit{i.e.} using the
helicity libraries of ThePEG.

In decays involving coloured particles that have more than one possible
colour flow, the colour is treated in exactly the same way as described
in \hyperref[\detokenize{review/BSM:sec-bsmhardprocess}]{Section \ref{\detokenize{review/BSM:sec-bsmhardprocess}}} for hard processes.

In addition for two-body decays we include the real emission correction in the
POWHEG matching scheme \cite{Richardson:2013nfo}.

\subsection{Off-shell effects}
\label{\detokenize{review/BSM:off-shell-effects}}

The production and decay processes described above have their external
particles on mass shell throughout. This assumes that the narrow width
approximation, defined by the following assumptions:
\begin{enumerate}
\sphinxsetlistlabels{\arabic}{enumi}{enumii}{}{.}%
\item {} 

the resonance has a small width $\Gamma$ compared with its pole
mass $M$, $\Gamma \ll M$;

\item {} 

we are far from threshold, $\sqrt{s} - M \gg \Gamma$, where
$\sqrt{s}$ denotes the centre-of-mass energy;

\item {} 

the propagator is separable;

\item {} 

the mass of the parent is much greater than the mass of the decay
products;

\item {} 

there are no significant non-resonant contributions;

\end{enumerate}

is a valid approximation. In general, given that we do not have a
specific mass spectrum, this is not a good enough approximation. In
particular, if processes occur at or close to threshold, there can be
large corrections that we need to take into account.

To improve our simulation, we provide an option to include the weight factor
\begin{equation}\label{equation:review/BSM:eqn:ofswgt}
\begin{split}
\frac{1}{\pi}\int_{m^2_{\mathrm{min}}}^{m^2_{\mathrm{max}}}\mathrm{d}m^2
\frac{m\Gamma(m)}{(m^2-M^2)^2 + m^2\Gamma^2(m)}.
\end{split}
\end{equation}
This factor is incorporated through the generation of the off-shell mass rather than being retained as an additional event weight. A trial mass is generated from a fixed-width Breit-Wigner distribution and corrected to the desired running-width distribution by an acceptance-rejection step. The corresponding weight therefore enters the unweighting of the generated mass, and accepted events carry no additional weight from Eq.~\eqref{equation:review/BSM:eqn:ofswgt}. Here, $\Gamma(m)$ is the running width of the particle to be considered off shell, $M$ is the pole mass, and $m_{\mathrm{min},\mathrm{max}}$ are defined such that the maximum deviation from the pole mass is a constant times the on-shell width. A derivation of this factor can be found in the appendix of Ref.~\cite{Gigg:2008yc}.

This procedure could, if naively applied, lead to violations of gauge invariance in production processes. Instead, the distribution \eqref{equation:review/BSM:eqn:ofswgt} is used for each particle in the kinematics, but the masses $m_3$ and $m_4$ of particle-(anti)particle pairs are projected onto their average value, $(m_3+m_4)/2$, for use in the matrix-element evaluation, as discussed and validated in Ref.~\cite{Gigg:2008yc}.

\subsection{Model descriptions}
\label{\detokenize{review/BSM:model-descriptions}}

This section gives a description of the models that are included in
Herwig. In general within Herwig, the implementation of a physics model
consists of a main class, which inherits from the
\href{https://herwig.hepforge.org/doxygen/classHerwig\_1\_1StandardModel.html}{StandardModel}
class and implements the calculation of any parameters required by the
model or, for a SUSY model, reads them from an input SUSY Les Houches
Accord (SLHA) \cite{Skands:2003cj, Allanach:2008qq}
file. In addition, there are various classes that inherit from the
general Vertex classes of ThePEG, which implement the Feynman rules of
the model. There may also be some classes implementing other features of
the model, for example the running couplings in the specific model.

\subsubsection{Standard Model}
\label{\detokenize{review/BSM:standard-model}}

The implementation of the Standard Model in Herwig inherits from the
\href{https://thepeg.hepforge.org/doxygen/classThePEG\_1\_1StandardModelBase.html}{StandardModelBase}
class of ThePEG. ThePEG includes classes to implement the running strong
and electromagnetic couplings, together with the CKM matrix.

In Herwig we include our own implementations of the running
electromagnetic coupling, in the
\href{https://herwig.hepforge.org/doxygen/classHerwig\_1\_1AlphaEM.html}{AlphaEM}
class, and the running strong coupling in the
\href{https://herwig.hepforge.org/doxygen/classHerwig\_1\_1O2AlphaS.html}{O2AlphaS}
class. However, by default we use the implementations of the running couplings
from ThePEG and the Herwig implementations are only provided to allow
us to make exact comparisons with the FORTRAN HERWIG program.

In order to perform helicity amplitude calculations we need access to
the full CKM matrix. However, the
\href{https://thepeg.hepforge.org/doxygen/classThePEG\_1\_1CKMBase.html}{CKMBase}
class of ThePEG only provides the squares of its components. The
\href{https://herwig.hepforge.org/doxygen/classHerwig\_1\_1StandardCKM.html}{StandardCKM}
class therefore provides access to the matrix elements as well and is
used in all our helicity amplitude calculations.

Also included is a structure for the implementation of running mass
calculations. The
\href{https://herwig.hepforge.org/doxygen/classHerwig\_1\_1RunningMassBase.html}{RunningMassBase}
class provides a base class and the two-loop QCD running mass is
implemented in the
\href{https://herwig.hepforge.org/doxygen/classHerwig\_1\_1RunningMass.html}{RunningMass}
class.

The Standard Model input parameters in Herwig do not form a minimal
set, because it is possible to independently set the value of the weak
mixing angle, the $W^\pm$ masses and the $Z^0$ boson mass,
without satisfying the tree-level relationship between them.
A number of formally consistent schemes are however also supported
using the \href{https://thepeg.hepforge.org/doxygen/StandardModelBaseInterfaces.html\#EW/Scheme}{EW/Scheme}
switch.
The
EW parameters we use are:
\begin{itemize}
\item {} 

the value of the electromagnetic coupling at zero momentum transfer,
\href{https://thepeg.hepforge.org/doxygen/StandardModelBaseInterfaces.html\#EW/AlphaEM}{EW/AlphaEM};

\item {} 

the value of $\sin^2\theta_W$,
\href{https://thepeg.hepforge.org/doxygen/StandardModelBaseInterfaces.html\#EW/Sin2ThetaW}{EW/Sin2Theta};

\item {} 

the masses of the $W^\pm$, $M_W=80.403\,\rm{GeV}$, and
$Z^0$, $M_Z=91.1876\,\rm{GeV}$, bosons, which are taken
from their
\href{https://thepeg.hepforge.org/doxygen/classThePEG\_1\_1ParticleData.html}{ParticleData}
objects;

\item {} 

the mixing angles, 
\href{https://herwig.hepforge.org/doxygen/classHerwig_1_1StandardCKM.html#theta_12}{$\theta_{12}$},
\href{https://herwig.hepforge.org/doxygen/classHerwig_1_1StandardCKM.html#theta_13}{$\theta_{13}$},
and 
\href{https://herwig.hepforge.org/doxygen/classHerwig_1_1StandardCKM.html#theta_22}{$\theta_{23}$},
and phase, 
\href{https://herwig.hepforge.org/doxygen/classHerwig_1_1StandardCKM.html#delta}{$\delta$},
of the CKM matrix.

\end{itemize}

In addition, many of the Standard Model couplings to the $Z^0$
boson can be changed to simulate non-Standard Model effects if desired.

\subsubsection{Supersymmetric Models}
\label{\detokenize{review/BSM:supersymmetric-models}}\label{\detokenize{review/BSM:sect-susy-models}}

\paragraph{Minimal Supersymmetric Standard Model}
\label{\detokenize{review/BSM:minimal-supersymmetric-standard-model}}

The Minimal Supersymmetric Standard Model (MSSM) is the most studied
supersymmetric model and as such it should be included in any generator
attempting to simulate BSM physics. As its name suggests, it contains the
smallest number of additional fields required for the theory to be
consistent. The additional particle content over that of the Standard
Model is listed in \hyperref[\detokenize{review/BSM:tab-bsm-mssmspectrum}]{Table \ref{\detokenize{review/BSM:tab-bsm-mssmspectrum}}}.

\begin{savenotes}\sphinxattablestart
\sphinxthistablewithglobalstyle
\centering
\sphinxcapstartof{table}
\sphinxthecaptionisattop
\sphinxcaption{The additional particle content of the MSSM contained in Herwig. The particle’s PDG codes are the standard ones given by the Particle Data Group \cite{Tanabashi:2018oca}.}\label{\detokenize{review/BSM:id48}}\label{\detokenize{review/BSM:tab-bsm-mssmspectrum}}
\sphinxaftertopcaption
\begin{tabulary}{\linewidth}[t]{TT}
\sphinxtoprule
\sphinxstyletheadfamily 

Spin
&\sphinxstyletheadfamily 

Particles
\\
\sphinxmidrule
\sphinxtableatstartofbodyhook

$0$
&

$\tilde{d}_L,\tilde{u}_L,\tilde{s}_L,\tilde{c}_L,\tilde{b}_1,\tilde{t}_1$
\\
\hline&

$\tilde{e}_L,\tilde{\nu}_{eL},\tilde{\mu}_{L},\tilde{\nu}_{\mu L},\tilde{\tau}_1,\tilde{\nu}_{\tau L}$
\\
\hline&

$\tilde{d}_R,\tilde{u}_R,\tilde{s}_R,\tilde{c}_R,\tilde{b}_2,\tilde{t}_2$
\\
\hline&

$\tilde{e}_R,\tilde{\mu}_{R},\tilde{\tau}_2$
\\
\hline&

$H^0,\,A^0,\,H^{+}$
\\
\hline

$1/2$
&

$\tilde{g}, \,\tilde{\chi}^0_1,$
$\tilde{\chi}^0_2,\,\tilde{\chi}^0_3,\,\tilde{\chi}^0_4,\,\tilde{\chi}^{+}_1,\,\tilde{\chi}^{+}_2$
\\
\sphinxbottomrule
\end{tabulary}
\sphinxtableafterendhook\par
\sphinxattableend\end{savenotes}

The additional particles must have masses and couplings to be of any use
in an event simulation. For supersymmetric models various programs are
available that, given some set of input parameters, produce a spectrum
containing all of the other parameters necessary to be able to calculate
physical quantities within the model. As stated in the previous section
the output from such a generator must comply with the SLHA
\cite{Skands:2003cj, Allanach:2008qq} for it to be used with Herwig.

While reading the information from an SLHA file is straightforward,
there is a minor complication when dealing with particle masses that
have a mixing matrix associated with them. For example, consider the
neutralinos, which are an admixture of the bino $\tilde{b}$, third
wino $\tilde{w}_3$ and 2 higgsinos $\tilde{h}_1$ and
$\tilde{h}_2$. The physical eigenstates $\tilde{\chi}^0_i$
are given by
\begin{equation*}
\begin{split}\tilde{\chi}^0_i = N_{ij}\tilde{\psi}^0_j,\end{split}
\end{equation*}
where $N_{ij}$ is the neutralino mixing matrix in the
$\tilde{\psi}^0=(-i\tilde{b},-i\tilde{w},\tilde{h}_1,\tilde{h}_2)^T$
basis. The diagonalized mass term for the gauginos is then
$N^{*}\mathcal{M}_{\tilde{\psi}^0}N^{\dagger}$, which in general
can produce complex mass values. To keep the mass values real the phase
is instead absorbed into the definition of the corresponding field
thereby yielding a strictly real mass and mixing matrix. There is
however a price to be paid for this — while the masses are kept real
they can become negative. For an event generator, a negative mass for a
physical particle does not make sense so we instead choose a
complex-valued mixing matrix along with real and non-negative masses. If
a negative mass is encountered while reading an SLHA file, the
physical mass is taken as the absolute value and the appropriate row of
the mixing matrix is multiplied by a factor of $i$. This approach
is used in order to facilitate the implementation of extended
supersymmetric models in the future.

\paragraph{Next-to-Minimal Supersymmetric Standard Model}
\label{\detokenize{review/BSM:next-to-minimal-supersymmetric-standard-model}}

The next-to-minimal Supersymmetric Standard Model (NMSSM) includes an additional Higgs singlet
and therefore leads to an additional scalar Higgs boson, an additional pseudoscalar
Higgs boson and an additional neutralino state. Many of the Feynman rules are the same as
those of the MSSM and hence Vertex classes can be adapted by just adding
the additional states in the mixing matrices. As with the MSSM the parameters
of the model are usually generated using external programs according to the second version of the
SLHA \cite{Allanach:2008qq}.

\paragraph{R-parity Violating Supersymmetric Models}
\label{\detokenize{review/BSM:r-parity-violating-supersymmetric-models}}

The R-parity violating supersymmetric model includes additional terms in the superpotential
\begin{equation}\label{equation:review/BSM:eqn:RPV_super}
\begin{split}{\bf W_{\not R_p}} &=  \kappa_i\varepsilon^{ab}L_a^{i}H_b^2+\frac12\lambda_{ijk}\varepsilon^{ab}L_{a}^{i}L_{b}^{j}\overline{E}^{k}
     \\
  & \phantom{=} + \lambda_{ijk}'\varepsilon^{ab}L_{a}^{i}Q_{b}^{j}\overline{D}^{k} +\frac{1}{2}\lambda_{ijk}''\varepsilon^{\alpha\beta\gamma}\overline{U}_{\alpha}^{i}
   \overline{D}_{\beta}^{j}\overline{D}_{\gamma}^{k},\end{split}
\end{equation}

where $L_i$, $E_i$, $Q_i$, $U_i$, $D_i$ are respectively
the lepton doublet, lepton singlet, quark doublet, up and down quark singlet $SU(2)$ superfields
for the $i$ th generation. The couplings $\lambda_{i,j,k}$, $\lambda_{i,j,k}'$ and
$\lambda_{i,j,k}''$ couple the different generations and $H^2$ is the hypercharge $Y=1$
Higgs boson supermultiplet. The indices $a,b$ are $SU(2)$ indices and $\alpha,\beta,\gamma$
are $SU(3)$ indices.

The first three terms in Eq. \eqref{equation:review/BSM:eqn:RPV_super} violate the conservation of lepton number. The first bilinear term
is the most complicated of these to handle as it involves mixing between the neutralinos and Standard Model neutrinos,
and charginos and Standard Model charged leptons. The final term in the superpotential also proves complicated as it involves
the violation of baryon number and the antisymmetric tensor in the colours of the interacting quarks.
The simulation of these models is described in Ref. \cite{Dreiner:1999qz}. The model allows the
inclusion of either all the terms, or just the trilinear couplings.

\subsubsection{Extra-Dimensional Models}
\label{\detokenize{review/BSM:extra-dimensional-models}}

\paragraph{Randall-Sundrum Model}
\label{\detokenize{review/BSM:randall-sundrum-model}}

The first models proposed with extra dimensions were of the
Randall-Sundrum (RS) \cite{Randall:1999ee} type where a tensor
particle, namely the graviton, is included and is allowed to propagate
in the extra dimensions. All other matter, however, is restricted to our
usual 4D brane and as a result all of the SM couplings are left
unchanged. The only extra couplings required are those of the graviton
to ordinary matter, which depend on a single parameter
$\Lambda_\pi$.

Two parameters can be controlled in the Randall-Sundrum model; the
cutoff $\Lambda_\pi$ and the mass of the graviton. The default
mass of the graviton is $500\,\textrm{GeV}$, which can be
changed via the
\href{https://thepeg.hepforge.org/doxygen/ParticleDataInterfaces.html\#NominalMass}{NominalMass}
interface of its
\texttt{ParticleData}
object. The cutoff is set via the
\href{https://herwig.hepforge.org/doxygen/classHerwig_1_1RSModel.html#Lambda_pi}{Lambdapi} interface
of the
\href{https://herwig.hepforge.org/doxygen/classHerwig\_1\_1RSModel.html}{RSModel}
object and has a default value of $10\,\textrm{TeV}$.

\paragraph{Minimal Universal Extra Dimensions Model}
\label{\detokenize{review/BSM:minimal-universal-extra-dimensions-model}}

We also include a model based on the idea of universal extra dimensions
where all fields are allowed to propagate in the bulk. Following similar
lines to supersymmetry, the model included in Herwig is of a minimal
type and has a single compact extra dimension of radius $R$
\cite{Hooper:2007qk}.

Compactifying the extra dimension and allowing all fields to propagate
in it leads to a rich new structure within the theory. Analogous to the
particle-in-a-box scenario, one obtains an infinite number of
excitations of the fields all characterized by a quantity called the
Kaluza-Klein (KK)
number. This is most easily demonstrated by showing how a scalar
field $\Phi$ would be decomposed after compactification
\begin{equation*}
\begin{split}\Phi(x^\mu,y) = \frac{1}{\sqrt{\pi R}}\left[\Phi_0(x^\mu) +
    \sqrt{2}\sum_{n=1}^{\infty} \Phi_n(x^\mu)\cos\left(\frac{ny}{R}\right) \right],\end{split}
\end{equation*}

where $x^\mu$ are the 4D coordinates, $y$ is the position in
the 5th dimension and $n$ is the KK-number of the mode with
$n=0$ identified as the SM mode. In general, in some
compactification schemes, it is possible to have KK-number-violating
interactions but in the Minimal Universal Extra Dimensions (MUED)
framework in Herwig we include only those interactions that conserve
KK-parity $P=(-1)^n$ and also limit ourselves to $n=1$.

\hyperref[\detokenize{review/BSM:tab-bsm-mued-spectrum}]{Table \ref{\detokenize{review/BSM:tab-bsm-mued-spectrum}}} shows the MUED
particle content contained in Herwig along with the particle ID
codes used, which have not been standardized by the Particle Data Group
\cite{Tanabashi:2018oca}. Unlike the MSSM, there are no external programs
available that calculate the mass spectrum, so this is performed
internally by the
\href{https://herwig.hepforge.org/doxygen/classHerwig\_1\_1UEDBase.html}{UEDBase}
class, which implements the UED model. At tree level, the mass of any
level-$n$ particle is simply given by
$(m^2_0 + (n/R)^2)^{1/2}$, where $m_0$ is the mass of the
SM particle, and $1/R$ is generally much larger than the SM mass
so the spectrum is highly degenerate and no decays can occur. This
situation changes once radiative corrections are taken into account and
a spectrum that can be phenomenologically similar to the MSSM arises.
The full set of radiative corrections, as derived in Ref.
\cite{Cheng:2002iz}, is incorporated in the
\texttt{UEDBase}
class to give a realistic spectrum.

\begin{savenotes}\sphinxattablestart
\sphinxthistablewithglobalstyle
\centering
\sphinxcapstartof{table}
\sphinxthecaptionisattop
\sphinxcaption{The MUED particle spectrum contained in Herwig along with their ID codes. \protect$^\bullet\protect$ denotes a doublet under SU(2) and \protect$^\circ\protect$ a singlet. As with the standard PDG codes an antiparticle is given by the negative of the number in the table.}\label{\detokenize{review/BSM:id49}}\label{\detokenize{review/BSM:tab-bsm-mued-spectrum}}
\sphinxaftertopcaption
\begin{tabulary}{\linewidth}[t]{TTTTTT}
\sphinxtoprule
\sphinxstyletheadfamily 

Spin
&\sphinxstyletheadfamily 

Particle
&\sphinxstyletheadfamily 

ID code
&\sphinxstyletheadfamily 

Spin
&\sphinxstyletheadfamily 

Particle
&\sphinxstyletheadfamily 

ID code
\\
\sphinxmidrule
\sphinxtableatstartofbodyhook

$0$
&

$h^0_1$
&

5100025
&

$1$
&

$g_1^*$
&

5100021
\\
\hline&

$A^0_1$
&

5100036
&&

$\gamma_1^*$
&

5100022
\\
\hline&

$H_1^{+}$
&

5100037
&&

$Z_1^{0\,*}$
&

5100023
\\
\hline&&&&

$W_1^{+\,*}$
&

5100024
\\
\hline

$1/2$
&

$d^{\bullet}_1$
&

5100001
&

$1/2$
&

$d^{\circ}_1$
&

6100001
\\
\hline&

$u^{\bullet}_1$
&

5100002
&&

$u^{\circ}_1$
&

6100002
\\
\hline&

$s^{\bullet}_1$
&

5100003
&&

$s^{\circ}_1$
&

6100003
\\
\hline&

$c^{\bullet}_1$
&

5100004
&&

$c^{\circ}_1$
&

6100004
\\
\hline&

$b^{\bullet}_1$
&

5100005
&&

$b^{\circ}_1$
&

6100005
\\
\hline&

$t^{\bullet}_1$
&

5100006
&&

$t^{\circ}_1$
&

6100006
\\
\hline&

$e^{-\,\bullet}_1$
&

5100011
&&

$e^{-\,\circ}_1$
&

6100011
\\
\hline&

$\nu^{\bullet}_{e1}$
&

5100012
&&&\\
\hline&

$\mu^{-\,\bullet}_1$
&

5100013
&&

$\mu^{-\,\circ}_1$
&

6100013
\\
\hline&

$\nu^{\bullet}_{\mu 1}$
&

5100014
&&&\\
\hline&

$\tau^{-\,\bullet}_1$
&

5100015
&&

$\tau^{-\,\circ}_1$
&

6100015
\\
\hline&

$\nu^{\bullet}_{\tau 1}$
&

5100016
&&&\\
\sphinxbottomrule
\end{tabulary}
\sphinxtableafterendhook\par
\sphinxattableend\end{savenotes}

There are three parameters that can be set to control the UED model: the
inverse of the radius of compactification $R^{-1}$; the cutoff
scale $\Lambda$; and the mass of the Higgs boson at the boundary
of the compactified dimension $\overline{m}_h$. These are
controlled through the interfaces:
\begin{itemize}
\item {} 

\href{https://herwig.hepforge.org/doxygen/UEDBaseInterfaces.html\#InverseRadius}{InverseRadius}
the value of $R^{-1}$, the default value is $500\,\textrm{GeV}$;

\item {} 

\href{https://herwig.hepforge.org/doxygen/UEDBaseInterfaces.html\#LambdaR}{LambdaR}
the dimensionless number $\Lambda R$, the default value is $20$;

\item {} 

\href{https://herwig.hepforge.org/doxygen/UEDBaseInterfaces.html\#HiggsBoundaryMass}{HiggsBoundaryMass}
the value of the Higgs mass at the boundary, the default value is
$0\,\textrm{GeV}$.

\end{itemize}

\paragraph{ADD Model}
\label{\detokenize{review/BSM:add-model}}

In the ADD Model
\cite{ArkaniHamed:1998rs,ArkaniHamed:1998nn,Lykken:1996fj,Witten:1996mz,Horava:1995qa,Horava:1996ma,Antoniadis:1990ew},
gravity propagates in $\delta$ extra spatial dimensions, which have a flat
metric. The large size of these extra dimensions leads to a tower of
Kaluza-Klein excitations of the graviton, which can either contribute as
virtual particles to Standard Model processes, or be produced leading to
missing energy signatures. This model is implemented using the
conventions of Ref. \cite{Giudice:1998ck}.

Three parameters can be controlled in the ADD model:
\begin{itemize}
\item {} 

the number of extra dimensions $\delta$, which is controlled by
the
\href{https://herwig.hepforge.org/doxygen/ADDModelInterfaces.html\#Delta}{Delta}
interface of the
\href{https://herwig.hepforge.org/doxygen/classHerwig\_1\_1ADDModel.html}{ADDModel};

\item {} 

the $d$-dimensional Planck mass $M_D$, which can be set
using the
\href{https://herwig.hepforge.org/doxygen/ADDModelInterfaces.html\#DdPlanckMass}{DdPlanckMass}
interface of the
\texttt{ADDModel};

\item {} 

the scale for virtual processes $\Lambda_T$, which is
controlled by the
\href{https://herwig.hepforge.org/doxygen/ADDModelInterfaces.html\#LambdaT}{LambdaT}
interface of the
\texttt{ADDModel}.

\end{itemize}

\subsubsection{Little Higgs Models}
\label{\detokenize{review/BSM:little-higgs-models}}

The main ideas behind Little Higgs models are \cite{Han:2003wu}:
\begin{itemize}
\item {} 

the Higgs fields are the Goldstone bosons associated with global symmetry breaking at a higher scale;

\item {} 

the Higgs bosons become pseudo-Goldstone bosons and acquire a mass via symmetry breaking
at the EW scale;

\item {} 

the masses of the Higgs bosons are protected by the approximate global symmetry, which ensures
the Higgs bosons remain light.

\end{itemize}

We currently include two models, the Littlest Higgs Model, which contains only new gauge bosons and
a top quark partner, and the Little Higgs Model with T-parity, which includes partners for
all the Standard Model particles.

\paragraph{Littlest Higgs Model}
\label{\detokenize{review/BSM:littlest-higgs-model}}

The Littlest Higgs Model is implemented using the Feynman rules described in Ref. \cite{Han:2003wu}.
The additional particles are the additional heavy gauge boson states $A_H$, $Z_H$,
$W_H^\pm$, for each of the Standard Model EW bosons, the additional partner
of the top quark $T$, and additional scalars $\phi^0$, $\phi^P$, $\phi^\pm$
and $\phi^{\pm\pm}$.

The main parameters of the model are:
\begin{itemize}
\item {} 

$\cot\theta$ the value of the mixing between the $W$ bosons,
specified using the \href{https://herwig.hepforge.org/doxygen/LHModelInterfaces.html\#CotTheta}{CotTheta} interface;

\item {} 

$\tan\theta'$ the value of the mixing between the $B$ bosons, set using
the \href{https://herwig.hepforge.org/doxygen/LHModelInterfaces.html\#TanThetaPrime}{TanThetaPrime} interface;

\item {} 

$f$ the symmetry breaking scale is set using \href{https://herwig.hepforge.org/doxygen/LHModelInterfaces.html\#f}{f};

\item {} 

$\frac{\lambda_1}{\lambda_2}$ the ratio of the two quark couplings is set
using \href{https://herwig.hepforge.org/doxygen/LHModelInterfaces.html\#LambdaRatio}{LambdaRatio};

\item {} 

$v'/v$ the ratio of the vacuum expectation values is set
using \href{https://herwig.hepforge.org/doxygen/LHModelInterfaces.html\#VEVRatio}{VEVRatio};

\item {} 

$m_H$ the mass of the Higgs boson is set using \href{https://herwig.hepforge.org/doxygen/LHModelInterfaces.html\#mH}{mH}.

\end{itemize}

\paragraph{Little Higgs Model with T-parity}
\label{\detokenize{review/BSM:little-higgs-model-with-t-parity}}

In order to evade the constraints from EW precision data, the original Little Higgs models required fine-tuning.
One solution to this problem is to add a discrete symmetry (analogous to R-parity of SUSY) called T-parity.
This removes most of the constraints from EW precision data, gives a dark matter candidate, and leads to
a spectrum of mirror fermion partners of the Standard Model particles. This model is implemented based on Ref.
\cite{Hubisz:2004ft} using the Feynman rules from \cite{Hubisz:2004ft, Belyaev:2006jh, Hubisz:2005bd}.
In addition to the particle content of the Little Higgs model, there are T-parity odd partners of
the Standard Model fermions and of the additional top quark partner.

The main parameters of the model are:
\begin{itemize}
\item {} 

$f$ the symmetry breaking scale for the non-linear $\sigma$-model is set using
\texttt{f};

\item {} 

$\sin\alpha$ the mixing in the top quark sector is set using
\href{https://herwig.hepforge.org/doxygen/LHTPModelInterfaces.html\#SinAlpha}{SinAlpha};

\item {} 

$\kappa_q$ the parameter controlling the masses of the T-odd quarks is set using
\href{https://herwig.hepforge.org/doxygen/LHTPModelInterfaces.html\#KappaQuark}{KappaQuark};

\item {} 

$\kappa_\ell$ the parameter controlling the masses of the T-odd leptons is set using
\href{https://herwig.hepforge.org/doxygen/LHTPModelInterfaces.html\#KappaLepton}{KappaLepton};

\item {} 

$m_H$ the mass of the Higgs boson is set using \href{https://herwig.hepforge.org/doxygen/LHTPModelInterfaces.html\#HiggsMass}{HiggsMass};

\end{itemize}

\subsubsection{Other Models}
\label{\detokenize{review/BSM:other-models}}

\paragraph{Leptoquarks}
\label{\detokenize{review/BSM:leptoquarks}}

Fermion masses may arise from the mixing of elementary fermions with
composite, fermionic resonances of a strong
sector \cite{Kaplan:1991dc} responsible for the breaking of the
$SU(2)_L \times U(1)_Y$ EW symmetry of the SM. It follows that
this strongly-coupled sector must also be charged under colour $SU(3)$
and must contain, at the very least, colour-triplet fermionic
resonances that can mix with the elementary colour triplets to make
the observed quarks. It is reasonable to expect that such a
strongly-coupled sector will contain other coloured resonances. These
may be bosonic and, depending on their charges, may couple to a quark
and a lepton. These leptoquark resonances may be light if they arise
as pseudo-Nambu Goldstone bosons and make an ideal target for LHC
searches. They will decay exclusively to third-generation fermions due
to suppression of the couplings to light fermions.

The present implementation includes non-derivatively coupled leptoquarks, which couple to Standard Model
fermions as in Eq. (2.4) of Ref. \cite{Gripaios:2010hv} and single-derivatively coupled
leptoquarks such as those in Eq. (2.5) of Ref. \cite{Gripaios:2010hv}. In the case of the
derivatively coupled leptoquarks, the simplification that the
primed lepton and primed quark couplings are equal has been made.

\paragraph{Sextet}
\label{\detokenize{review/BSM:sextet}}

The colour sextet diquark is, in group theory language, a rank 2 symmetric tensor formed from the direct
product of two fundamental representations $3\otimes3 = 6 \oplus \bar{3}$. As such it is the
lowest colour representation that has not been observed and is therefore worthy of
study at the LHC. The Lagrangian, from Refs. \cite{Zhang:2010kr,Arik:2001bc,Cakir:2005iw,Atag:1998xq,Ma:1998pi},
is
\begin{equation*}
\begin{split}\mathcal{L}= &
\left(g_{1L}\overline{q_{L}^{c}}i\tau_{2}q_{L}+g_{1R}\overline{u_{R}^{c}}d_{R}\right)\Phi_{1,1/3}
\:+\: g_{1R}^{\prime}\overline{d_{R}^{c}}d_{R}\Phi_{1,-2/3}
\:+\: g_{1R}^{\prime\prime}\overline{u_{R}^{c}}u_{R}\Phi_{1,4/3} \:+ \\
& g_{3L}\overline{q_{L}^{c}}i\tau_{2}\tau q_{L}\cdot\Phi_{3,1/3}
\:+\: g_{2}\overline{q_{L}^{c}}\gamma_{\mu}d_{R}V_{2,-1/6}^{\mu}
\:+\: g_{2}^{\prime}\overline{q_{L}^{c}}\gamma_{\mu}u_{R}V_{2,5/6}^{\mu}
\:+\: h.c. \, ,\end{split}
\end{equation*}

where $q_L$ is the left-handed quark doublet, $u_{R}$ and $d_{R}$ are the
right-handed quark singlet fields, and $q^{c}\equiv C\bar{q}^{T}$ is
the charge conjugate quark field. The colour and generation indices
are omitted to give a more compact notation and the subscripts on the
scalar, $\Phi$, and vector, $V^{\mu}$, fields denote the SM EW
gauge quantum numbers: ($SU(2)_L$, $U(1)_Y$). The
Lagrangian is
assumed to be flavour diagonal to avoid any flavour-changing
currents arising from the new interactions. The main parameters are the couplings
of the model, which can be set using the interfaces of the
\href{https://herwig.hepforge.org/doxygen/classHerwig\_1\_1SextetModel.html}{SextetModel}
class, and the masses of the sextet diquarks, which can be set via their
\texttt{ParticleData} objects.

The simulation of the sextet colour structure in both hard production processes,
decays and the parton shower are described in Ref. \cite{Richardson:2011df} together
with the phenomenology of the model.

\paragraph{Transplanckian}
\label{\detokenize{review/BSM:transplanckian}}

This model describes the $2\rightarrow 2$ scattering of partons at
high energy,
in the transplanckian regime, using the eikonal approximation, as investigated in
\cite{Giudice:2001ce}. The approximation is valid in the high-energy, low-angle
scattering regime, where the centre-of-mass scattering angle of the incoming
parton $\hat{\theta} \rightarrow 0$ or, in terms of Mandelstam variables,
$-\hat{t} / \hat{s} \rightarrow 0$ . The matrix elements for the scattering
are proportional to the functions $F_n$ that appear in \cite{Giudice:2001ce}, which we have
calculated numerically and tabulated. The implementation allows variation of
the Planck scale (\href{https://herwig.hepforge.org/doxygen/METRP2to2Interfaces.html\#PlanckMass}{PlanckMass}) as well as the number of extra dimensions (\href{https://herwig.hepforge.org/doxygen/METRP2to2Interfaces.html\#NumberExtraDimensions}{NumberExtraDimensions}), up to a maximum of 6.

\paragraph{TTbAsymm}
\label{\detokenize{review/BSM:ttbasymm}}

The addition of this model was motivated by the initially anomalously large, mass-dependent forward-backward asymmetry in $t\bar{t}$ production, observed at the Tevatron CDF experiment, \textit{e.g.} \cite{Aaltonen:2011kc}. The discrepancy has since been reduced through the inclusion of higher-order corrections in the SM calculation \cite{Czakon:2014xsa,Czakon:2016ckf,Aaltonen:2017efp}. Such asymmetries may invoke new interactions in the top sector. In this implementation, we have included four types of new interaction that have been shown to reproduce the measured asymmetry (see, \textit{e.g.} \cite{Gripaios:2013rda}):
\begin{itemize}
\item {} 

a flavour-changing $W'$ (prime) vector boson that couples top quarks to down quarks
\begin{equation*}
\begin{split}\mathcal{L} \supset \bar{t} \gamma^\mu ( g_L P_L + g_R P_R ) d~ W'_\mu + \mathrm{h.c.} \;,\end{split}
\end{equation*}

where $g_{L,R}$ are the left- and right-handed couplings, \textit{i.e.} those corresponding to the left- and right-handed projection operators $P_{L,R}$, respectively;

\item {} 

an Abelian $Z$-prime vector boson that couples top quarks
to up quarks
\begin{equation*}
\begin{split}\mathcal{L} \supset g_{Z'}^{(R,L)} \bar{u} \gamma^\mu P_{R,L} t Z'_\mu + h^{(R,L)}_{Z',i} \bar{u}_i \gamma^\mu P_{R,L} u_i Z'_\mu + \mathrm{h.c.}\;,\end{split}
\end{equation*}

where $g_{Z'}^{(R,L)}$ are the right- and left-handed flavour-changing couplings respectively, and $h^{(R,L)}_i$ are flavour-conserving couplings for the $i^{\mathrm{th}}$ generation;

\item {} 

an axial heavy gluon that couples to $\bar{q}q$ and $\bar{t}t$
\begin{equation*}
\begin{split}\mathcal{L} \supset g_s \left[ \bar{q} T^A \gamma^\mu ( g^q_L P_L + g^q_R P_R ) q + \bar{t} T^A \gamma^\mu (g^t_L P_L + g^t_R P_R) t \right] G_\mu^{'A}\;,\end{split}
\end{equation*}

where $g_s$ is the QCD strong coupling, $T^A$ ($A \in \{1,8\}$) are
the $SU(3)$ generator matrices in the adjoint representation, $g^q_{L,R}$ are the left- and right-handed couplings
to $q\bar{q}$ (excluding the top quark), and $g^t_{L,R}$ is the left- and right-handed coupling to $t\bar{t}$;

\item {} 

a model that includes an additional, non-Abelian, $SU(2)_X$ gauge interaction, for further details see Ref. \cite{Jung:2011zv}.

\end{itemize}

The “active” model can be chosen through the
\href{https://herwig.hepforge.org/doxygen/TTbAModelInterfaces.html\#modelselect}{modelselect} interface of the \href{https://herwig.hepforge.org/doxygen/classHerwig\_1\_1TTbAModel.html}{TTbAModel} %
\begin{footnote}\sphinxAtStartFootnote
Note that if the axial gluon model is selected, the line in the corresponding input file that excludes the “Ag” particle should be commented out.
\end{footnote}.

\paragraph{Zprime}
\label{\detokenize{review/BSM:zprime}}

This simple model describes a heavy vector boson ($Z$-prime),
which is neutral under $U(1)_\mathrm{e.m.}$. The interactions are
flavour-conserving and the corresponding Lagrangian has the form
\begin{equation*}
\begin{split}\mathcal{L} \supset g_{q_i}^{(R,L)} \bar{q}_i \gamma^\mu P_{R,L} q_i Z'_\mu + g^{(R,L)}_{\ell_i} \bar{\ell}_i \gamma^\mu P_{R,L} \ell_i Z'_\mu + \mathrm{h.c.}\;,\end{split}
\end{equation*}

where $g_{q_i}^{(R,L)}$ and $g_{\ell_i}^{(R,L)}$ are the right- and left-handed couplings to the quarks and leptons of the $i^{\mathrm{th}}$ generation respectively. The couplings can be
modified, see \href{https://herwig.hepforge.org/doxygen/ZprimeModelInterfaces.html}{ZprimeModel} for more details.

\subsubsection{UFO}
\label{\detokenize{review/BSM:ufo}}

There are a number of programs available (\textit{e.g.} LanHEP \cite{Semenov:1996es, Semenov:2010qt}, FeynRules \cite{Christensen:2008py, Alloul:2013bka} and Sarah \cite{Staub:2008uz, Staub:2012pb}) that can output the Feynman rules for a new physics model in the Universal FeynRules Output (UFO) \cite{Degrande:2011ua} format. To enable these Feynman rules to be used with Herwig 7 we provide a program, \texttt{{ufo2herwig}}, that can convert the UFO format into a set of \texttt{{C++}} classes that can be compiled into a model library. As UFO is a \texttt{{python}} format, it was convenient to write the conversion program in \texttt{{python}} as well. The model library produced by this conversion can be used in the same way as the internal models described above. It is compiled as a separate shared library and loaded dynamically at run time, so Herwig itself does not need to be recompiled. The library can therefore be incorporated into an experimental software framework, provided that it is compiled against a compatible Herwig and ThePEG installation.

If possible we map the Lorentz structures of these vertices to the generic perturbative forms implemented in ThePEG, so that we only have to implement the couplings in the specific model. However, there are a number of models where the Lorentz structure does not have the perturbative form. Starting from version 7.2 the \texttt{{ufo2herwig}} converter will now generate all the code required to implement arbitrary Lorentz structures, including for spin-$\frac32$ and spin-$2$ particles, making use of the \texttt{{sympy}} package \cite{sympy}. Most colour structures are also supported, at least for phenomenologically relevant interactions. Some four-point vertices still cannot be handled, however these are rarely phenomenologically relevant, i.e.\ they involve at least three new particles.

\subsection{Code structure}
\label{\detokenize{review/BSM:code-structure}}

The
\href{https://herwig.hepforge.org/doxygen/classHerwig\_1\_1ModelGenerator.html}{ModelGenerator}
class is responsible for setting up the new MatrixElement objects, which
inherit from the
\href{https://herwig.hepforge.org/doxygen/classHerwig\_1\_1GeneralHardME.html}{GeneralHardME}
class, and
\href{https://thepeg.hepforge.org/doxygen/classThePEG\_1\_1DecayMode.html}{DecayMode}
objects for a new physics model. Helper classes aid in the creation of
these objects, they are:

\subsubsection{HardProcessConstructor}
\label{\detokenize{review/BSM:hardprocessconstructor}}

The \texttt{ModelGenerator}
uses one or more classes inheriting from the
\href{https://herwig.hepforge.org/doxygen/classHerwig\_1\_1HardProcessConstructor.html}{HardProcessConstructor}
base class to create the diagrams for the requested processes and construct the appropriate matrix elements. The inheriting classes are.
\begin{description}
\sphinxlineitem{TwoToTwoProcessConstructor}

The \href{https://herwig.hepforge.org/doxygen/classHerwig\_1\_1TwoToTwoProcessConstructor.html}{TwoToTwoProcessConstructor} is
responsible for constructing the diagrams for $2\to2$ scattering processes and creating an object
inheriting from the \texttt{GeneralHardME} class
based on the spin structure of the process. The full range of supported spin structures is given in \hyperref[\detokenize{review/BSM:tab-general-me}]{Table \ref{\detokenize{review/BSM:tab-general-me}}}.

\sphinxlineitem{ResonantProcessConstructor}

The \href{https://herwig.hepforge.org/doxygen/classHerwig\_1\_1ResonantProcessConstructor.html}{ResonantProcessConstructor}
is of a similar design to the
\texttt{HardProcessConstructor}
but it only constructs the diagrams for a process with a given $s$-channel resonance.

\sphinxlineitem{HiggsVBFProcessConstructor}

The \href{https://herwig.hepforge.org/doxygen/classHerwig\_1\_1HiggsVBFProcessConstructor.html}{HiggsVBFProcessConstructor} class constructs the diagrams
for the production of a neutral (colour and electrically) scalar boson via the vector boson fusion (VBF) process and creates an object of the
\href{https://herwig.hepforge.org/doxygen/classHerwig\_1\_1GeneralfftoffH.html}{GeneralfftoffH} class to calculate matrix elements for the process.

\sphinxlineitem{HiggsVectorBosonProcessConstructor}

The \href{https://herwig.hepforge.org/doxygen/classHerwig\_1\_1HiggsVectorBosonProcessConstructor.html}{HiggsVectorBosonProcessConstructor} class constructs the diagrams
for the production of a neutral (colour and electrically) scalar boson in association with a Standard Model EW vector boson ($W^\pm,Z^0$) including
the decay products of the vector boson. An object of the
\href{https://herwig.hepforge.org/doxygen/classHerwig\_1\_1GeneralfftoVH.html}{GeneralfftoVH} class is created to calculate matrix elements for the process.

\sphinxlineitem{QQHiggsProcessConstructor}

The \href{https://herwig.hepforge.org/doxygen/classHerwig\_1\_1QQHiggsProcessConstructor.html}{QQHiggsProcessConstructor} class constructs the diagrams for the production of
a neutral (colour and electrically) scalar boson in association with a heavy quark-antiquark pair and constructs an object of the
\href{https://herwig.hepforge.org/doxygen/classHerwig\_1\_1GeneralQQHiggs.html}{GeneralQQHiggs} class to calculate matrix elements for  the process.

\end{description}

\subsubsection{DecayConstructor}
\label{\detokenize{review/BSM:decayconstructor}}

The
\href{https://herwig.hepforge.org/doxygen/classHerwig\_1\_1DecayConstructor.html}{DecayConstructor}
stores a collection of objects that inherit from the
\href{https://herwig.hepforge.org/doxygen/classHerwig_1_1NBodyDecayConstructorBase.html}{NBodyDecayConstructorBase}
class. Each of these is responsible for constructing the decay modes for
the $n$-body decays. Currently the following classes are implemented:
\begin{description}
\sphinxlineitem{\href{https://herwig.hepforge.org/doxygen/classHerwig\_1\_1TwoBodyDecayConstructor.html}{TwoBodyDecayConstructor},}

for two-body decays;

\sphinxlineitem{\href{https://herwig.hepforge.org/doxygen/classHerwig\_1\_1ThreeBodyDecayConstructor.html}{ThreeBodyDecayConstructor},}

for three-body decays;

\sphinxlineitem{\href{https://herwig.hepforge.org/doxygen/classHerwig\_1\_1FourBodyDecayConstructor.html}{FourBodyDecayConstructor},}

for four-body decays;

\sphinxlineitem{\href{https://herwig.hepforge.org/doxygen/classHerwig\_1\_1WeakCurrentDecayConstructor.html}{WeakCurrentDecayConstructor},}

for weak decays using the weak currents from \hyperref[\detokenize{review/decays:sect-weakcurrents}]{Section \ref{\detokenize{review/decays:sect-weakcurrents}}} for decays where two particles are almost mass
degenerate.

\end{description}

While all the important type of two- and three-body decays are currently implemented only the
most phenomenologically relevant four body decay, of a scalar boson to four fermions, is implemented. This decay mode can be
important for Higgs bosons decays via off-shell gauge bosons and the lightest stau in R-parity violating SUSY models.

In addition, the
\texttt{ModelGenerator}
class is responsible for setting up objects of
\href{https://herwig.hepforge.org/doxygen/classHerwig\_1\_1BSMWidthGenerator.html}{BSMWidthGenerator}
and
\texttt{GenericMassGenerator}
type so that off-shell effects can be simulated. To achieve this
either, \texttt{ParticleData}
objects are added to the
\href{https://herwig.hepforge.org/doxygen/ModelGeneratorInterfaces.html\#Offshell}{Offshell}
interface so that the selected particles are treated as off shell, or
the
\href{https://herwig.hepforge.org/doxygen/ModelGeneratorInterfaces.html\#WhichOffshell}{WhichOffshell}
interface is set to \sphinxstylestrong{All} so that all BSM particles are treated as off
shell.

The matrix element classes all inherit from the
\texttt{GeneralHardME}
class and implement the matrix element for a particular spin
configuration. The classes inheriting from the
\texttt{GeneralHardME}
class and the spin structures they implement are given in
\hyperref[\detokenize{review/BSM:tab-general-me}]{Table \ref{\detokenize{review/BSM:tab-general-me}}}.

The on-shell decayer classes inherit from the
\href{https://herwig.hepforge.org/doxygen/classHerwig\_1\_1GeneralTwoBodyDecayer.html}{GeneralTwoBodyDecayer},
\href{https://herwig.hepforge.org/doxygen/classHerwig\_1\_1GeneralThreeBodyDecayer.html}{GeneralThreeBodyDecayer}
or \href{https://herwig.hepforge.org/doxygen/classHerwig\_1\_1GeneralFourBodyDecayer.html}{GeneralFourBodyDecayer}
class and each is responsible for calculating the value of the matrix
element for that particular set of spins. A
\href{https://herwig.hepforge.org/doxygen/classHerwig_1_1VectorCurrentDecayer.html}{VectorCurrentDecayer}
class also exists for decay modes created with the
\texttt{WeakCurrentDecayConstructor}
class. The
\texttt{Decayer}
classes implemented in Herwig and the types of decay they implement
are given in \hyperref[\detokenize{review/BSM:tab-general-decay}]{Table \ref{\detokenize{review/BSM:tab-general-decay}}}.

\begin{savenotes}\sphinxattablestart
\sphinxthistablewithglobalstyle
\centering
\sphinxcapstartof{table}
\sphinxthecaptionisattop
\sphinxcaption{The general hard process matrix elements, based on spin structures, implemented in Herwig.}\label{\detokenize{review/BSM:id50}}\label{\detokenize{review/BSM:tab-general-me}}
\sphinxaftertopcaption
\begin{tabulary}{\linewidth}[t]{TT}
\sphinxtoprule
\sphinxstyletheadfamily 

Class Name
&\sphinxstyletheadfamily 

Hard Process
\\
\sphinxmidrule
\sphinxtableatstartofbodyhook

\href{https://herwig.hepforge.org/doxygen/classHerwig\_1\_1MEff2ff.html}{MEff2ff}
&

Fermion fermion to fermion fermion.
\\
\hline

\href{https://herwig.hepforge.org/doxygen/classHerwig\_1\_1MEff2ss.html}{MEff2ss}
&

Fermion fermion to scalar scalar.
\\
\hline

\href{https://herwig.hepforge.org/doxygen/classHerwig\_1\_1MEff2sv.html}{MEff2sv}
&

Fermion fermion to scalar vector.
\\
\hline

\href{https://herwig.hepforge.org/doxygen/classHerwig\_1\_1MEff2tv.html}{MEff2tv}
&

Fermion fermion to tensor vector.
\\
\hline

\href{https://herwig.hepforge.org/doxygen/classHerwig\_1\_1MEff2vs.html}{MEff2vs}
&

Fermion fermion to vector scalar.
\\
\hline

\href{https://herwig.hepforge.org/doxygen/classHerwig\_1\_1MEff2vv.html}{MEff2vv}
&

Fermion fermion to vector vector.
\\
\hline

\href{https://herwig.hepforge.org/doxygen/classHerwig\_1\_1MEfv2fs.html}{MEfv2fs}
&

Fermion vector to fermion scalar.
\\
\hline

\href{https://herwig.hepforge.org/doxygen/classHerwig\_1\_1MEfv2tf.html}{MEfv2tf}
&

Fermion vector to tensor fermion.
\\
\hline

\href{https://herwig.hepforge.org/doxygen/classHerwig\_1\_1MEfv2vf.html}{MEfv2vf}
&

Fermion vector to vector fermion.
\\
\hline

\href{https://herwig.hepforge.org/doxygen/classHerwig\_1\_1MEvv2ff.html}{MEvv2ff}
&

Vector vector to fermion fermion.
\\
\hline

\href{https://herwig.hepforge.org/doxygen/classHerwig\_1\_1MEvv2ss.html}{MEvv2ss}
&

Vector vector to scalar scalar.
\\
\hline

\href{https://herwig.hepforge.org/doxygen/classHerwig\_1\_1MEvv2tv.html}{MEvv2tv}
&

Vector vector to tensor vector.
\\
\hline

\href{https://herwig.hepforge.org/doxygen/classHerwig\_1\_1MEvv2vv.html}{MEvv2vv}
&

Vector vector to vector vector.
\\
\hline

\href{https://herwig.hepforge.org/doxygen/classHerwig\_1\_1MEvv2vs.html}{MEvv2vs}
&

Vector vector to vector scalar.
\\
\sphinxbottomrule
\end{tabulary}
\sphinxtableafterendhook\par
\sphinxattableend\end{savenotes}

\begin{savenotes}\sphinxattablestart
\sphinxthistablewithglobalstyle
\centering
\sphinxcapstartof{table}
\sphinxthecaptionisattop
\sphinxcaption{The general decays based on spin structures implemented in Herwig.}\label{\detokenize{review/BSM:id51}}\label{\detokenize{review/BSM:tab-general-decay}}
\sphinxaftertopcaption
\begin{tabulary}{\linewidth}[t]{TT}
\sphinxtoprule
\sphinxstyletheadfamily 

Class Name
&\sphinxstyletheadfamily 

Decay
\\
\sphinxmidrule
\sphinxtableatstartofbodyhook

\href{https://herwig.hepforge.org/doxygen/classHerwig\_1\_1FFSDecayer.html}{FFSDecayer}
&

Fermion to fermion scalar decay.
\\
\hline

\href{https://herwig.hepforge.org/doxygen/classHerwig\_1\_1FFVDecayer.html}{FFVDecayer}
&

Fermion to fermion vector decay.
\\
\hline

\href{https://herwig.hepforge.org/doxygen/classHerwig\_1\_1FRVDecayer.html}{FRVDecayer}
&

Fermion to spin-$\frac32$ fermion vector decay.
\\
\hline

\href{https://herwig.hepforge.org/doxygen/classHerwig\_1\_1FRSDecayer.html}{FRSDecayer}
&

Fermion to spin-$\frac32$ fermion scalar decay.
\\
\hline

\href{https://herwig.hepforge.org/doxygen/classHerwig\_1\_1SFFDecayer.html}{SFFDecayer}
&

Scalar to fermion fermion decay.
\\
\hline

\href{https://herwig.hepforge.org/doxygen/classHerwig\_1\_1SRFDecayer.html}{SRFDecayer}
&

Scalar to spin-$\frac32$ fermion fermion decay.
\\
\hline

\href{https://herwig.hepforge.org/doxygen/classHerwig\_1\_1SSSDecayer.html}{SSSDecayer}
&

Scalar to two scalar decay.
\\
\hline

\href{https://herwig.hepforge.org/doxygen/classHerwig\_1\_1SSVDecayer.html}{SSVDecayer}
&

Scalar to scalar vector decay.
\\
\hline

\href{https://herwig.hepforge.org/doxygen/classHerwig\_1\_1SVVDecayer.html}{SVVDecayer}
&

Scalar to two vector decay.
\\
\hline

\href{https://herwig.hepforge.org/doxygen/classHerwig\_1\_1StoFFFFDecayer.html}{StoFFFFDecayer}
&

Scalar to four fermion decay.
\\
\hline

\href{https://herwig.hepforge.org/doxygen/classHerwig\_1\_1VFFDecayer.html}{VFFDecayer}
&

Vector to two fermion decay.
\\
\hline

\href{https://herwig.hepforge.org/doxygen/classHerwig\_1\_1VSSDecayer.html}{VSSDecayer}
&

Vector to two scalar decay.
\\
\hline

\href{https://herwig.hepforge.org/doxygen/classHerwig\_1\_1VVVDecayer.html}{VVSDecayer}
&

Vector to vector scalar decay.
\\
\hline

\href{https://herwig.hepforge.org/doxygen/classHerwig\_1\_1VVVDecayer.html}{VVVDecayer}
&

Vector to two vector decay.
\\
\hline

\href{https://herwig.hepforge.org/doxygen/classHerwig\_1\_1TFFDecayer.html}{TFFDecayer}
&

Tensor to two fermion decay.
\\
\hline

\href{https://herwig.hepforge.org/doxygen/classHerwig\_1\_1TSSDecayer.html}{TSSDecayer}
&

Tensor to two scalar decay.
\\
\hline

\href{https://herwig.hepforge.org/doxygen/classHerwig\_1\_1TVVDecayer.html}{TVVDecayer}
&

Tensor to two vector decay.
\\
\hline

\href{https://herwig.hepforge.org/doxygen/classHerwig\_1\_1FtoFFFDecayer.html}{FtoFFFDecayer}
&

Fermion to three fermion decay.
\\
\hline

\href{https://herwig.hepforge.org/doxygen/classHerwig\_1\_1FtoFVVDecayer.html}{FtoFVVDecayer}
&

Fermion to fermion and two vector decay.
\\
\hline

\href{https://herwig.hepforge.org/doxygen/classHerwig\_1\_1StoSFFDecayer.html}{StoSFFDecayer}
&

Scalar to scalar and two fermion decay.
\\
\hline

\href{https://herwig.hepforge.org/doxygen/classHerwig\_1\_1StoFFVDecayer.html}{StoFFVDecayer}
&

Scalar to two fermion and vector decay.
\\
\hline

\href{https://herwig.hepforge.org/doxygen/classHerwig\_1\_1VtoFFVDecayer.html}{VtoFFVDecayer}
&

Vector to two fermion and vector decay.
\\
\hline

\href{https://herwig.hepforge.org/doxygen/classHerwig\_1\_1FFVCurrentDecayer.html}{FFVCurrentDecayer}
&

Fermion to fermion vector decay with the vector off-shell
\\
\hline&

and decaying via a weak current from \hyperref[\detokenize{review/decays:sect-weakcurrents}]{Section \ref{\detokenize{review/decays:sect-weakcurrents}}}.
\\
\sphinxbottomrule
\end{tabulary}
\sphinxtableafterendhook\par
\sphinxattableend\end{savenotes}

The specification of the particles involved in the hard process is
achieved through
the \href{https://herwig.hepforge.org/doxygen/HardProcessConstructorInterfaces.html\#Incoming}{Incoming}
and
\href{https://herwig.hepforge.org/doxygen/HardProcessConstructorInterfaces.html\#Outgoing}{Outgoing}
interfaces of the
\texttt{HardProcessConstructor}.
Both interfaces are lists of
\texttt{ParticleData}
objects. The switch
\href{https://herwig.hepforge.org/doxygen/HardProcessConstructorInterfaces.html\#IncludeEW}{IncludeEW}
can be set to \sphinxstylestrong{No} to include only the strong coupling diagrams.

In order to pass spin correlations through the decay stage,
\texttt{DecayIntegrator}
objects must be created. This is achieved by populating a list held in
the
\texttt{ModelGenerator}
class, which can be accessed through the
\href{https://herwig.hepforge.org/doxygen/ModelGeneratorInterfaces.html\#DecayParticles}{DecayParticles}
interface. The particles in this list will have spin correlation
information passed along when their decays are generated. If a decay
table is read in for a SUSY model then the
\href{https://herwig.hepforge.org/doxygen/TwoBodyDecayConstructorInterfaces.html\#CreateDecayModes}{CreateDecayModes}
interface should be set to \sphinxstylestrong{No} so that only the decay modes listed in
the externally generated decay table are created %
\begin{footnote}\sphinxAtStartFootnote
If a decay table is being used with a SUSY model then the
DecayParticles list must still be populated so that the decays will
have spin correlation information included.
\end{footnote}. For all other
models the possible decay modes are also created from the particles in
the
\texttt{DecayParticles}
list.

In addition, we provide a number of classes to allow the loop-mediated but phenomenologically important
couplings of the Higgs bosons in a model to photon or gluon pairs to be automatically generated from
the vertices of the model. The \href{https://herwig.hepforge.org/doxygen/classHerwig\_1\_1VVSLoopVertex.html}{VVSLoopVertex}
class implements the loop calculations while the \href{https://herwig.hepforge.org/doxygen/classHerwig\_1\_1GenericHPPVertex.html}{GenericHPPVertex}
and \href{https://herwig.hepforge.org/doxygen/classHerwig\_1\_1GenericHGGVertex.html}{GenericHGGVertex} classes implement the couplings to
photon and gluon pairs, respectively.

In addition to the code that handles the calculation of the matrix
elements for the decays and scattering cross sections each model
requires a number of classes to implement the model.

The Standard Model is implemented in the
\texttt{StandardModel}
class, which inherits from the
\texttt{StandardModelBase}
class of ThePEG and implements access to the helicity Vertex classes and
some additional couplings, such as the running mass, used by Herwig.
The Vertex classes that implement the Standard Model interactions are
given in \hyperref[\detokenize{review/BSM:tab-sm-vertices}]{Table \ref{\detokenize{review/BSM:tab-sm-vertices}}}.

\begin{savenotes}\sphinxattablestart
\sphinxthistablewithglobalstyle
\centering
\sphinxcapstartof{table}
\sphinxthecaptionisattop
\sphinxcaption{Herwig Vertex classes for the Standard Model.}\label{\detokenize{review/BSM:id52}}\label{\detokenize{review/BSM:tab-sm-vertices}}
\sphinxaftertopcaption
\begin{tabulary}{\linewidth}[t]{TT}
\sphinxtoprule
\sphinxstyletheadfamily 

Class
&\sphinxstyletheadfamily 

Interaction
\\
\sphinxmidrule
\sphinxtableatstartofbodyhook

\href{https://herwig.hepforge.org/doxygen/classHerwig\_1\_1SMFFGVertex.html}{SMFFGVertex}
&

Interaction of the gluon with the SM fermions
\\
\hline

\href{https://herwig.hepforge.org/doxygen/classHerwig\_1\_1SMFFPVertex.html}{SMFFPVertex}
&

Interaction of the photon with the SM fermions
\\
\hline

\href{https://herwig.hepforge.org/doxygen/classHerwig\_1\_1SMFFWVertex.html}{SMFFWVertex}
&

Interaction of the $W^\pm$ boson with the SM fermions
\\
\hline

\href{https://herwig.hepforge.org/doxygen/classHerwig\_1\_1SMFFZVertex.html}{SMFFZVertex}
&

Interaction of the $Z^0$ boson with the SM fermions
\\
\hline

\href{https://herwig.hepforge.org/doxygen/classHerwig\_1\_1SMFFHVertex.html}{SMFFHVertex}
&

Interaction of the Higgs boson with the SM fermions
\\
\hline

\href{https://herwig.hepforge.org/doxygen/classHerwig\_1\_1SMGGGVertex.html}{SMGGGVertex}
&

Triple gluon vertex
\\
\hline

\href{https://herwig.hepforge.org/doxygen/classHerwig\_1\_1SMGGGGVertex.html}{SMGGGGVertex}
&

Four gluon vertex
\\
\hline

\href{https://herwig.hepforge.org/doxygen/classHerwig\_1\_1SMWWWVertex.html}{SMWWWVertex}
&

Triple EW gauge boson vertex
\\
\hline

\href{https://herwig.hepforge.org/doxygen/classHerwig\_1\_1SMWWWWVertex.html}{SMWWWWVertex}
&

Four EW gauge boson vertex
\\
\hline

\href{https://herwig.hepforge.org/doxygen/classHerwig\_1\_1SMWWHVertex.html}{SMWWHVertex}
&

Interaction of the Higgs boson with the EW gauge bosons
\\
\hline

\href{https://herwig.hepforge.org/doxygen/classHerwig\_1\_1SMWWHHVertex.html}{SMWWHHVertex}
&

Two Higgs bosons, two EW gauge boson vertex
\\
\hline

\href{https://herwig.hepforge.org/doxygen/classHerwig\_1\_1SMHHHVertex.html}{SMHHHVertex}
&

Triple Higgs boson couplings
\\
\hline

\href{https://herwig.hepforge.org/doxygen/classHerwig\_1\_1SMHGGVertex.html}{SMHGGVertex}
&

Higgs boson coupling to two gluons via quark loops
\\
\hline

\href{https://herwig.hepforge.org/doxygen/classHerwig\_1\_1SMHPPVertex.html}{SMHPPVertex}
&

Higgs boson coupling to two photons via fermion and boson loops
\\
\sphinxbottomrule
\end{tabulary}
\sphinxtableafterendhook\par
\sphinxattableend\end{savenotes}

Most of the BSM models in Herwig implement a class describing the model, which inherits from the
\href{https://herwig.hepforge.org/doxygen/classHerwig\_1\_1BSMModel.html}{BSMModel} class, which in turn
inherits from the
\texttt{StandardModel}
class. This allows an instance of this class, which implements any additional particles used in the
model, to be used instead of an instance of the \texttt{StandardModel}
when simulating this model. The \texttt{BSMModel} class also
implements the reading of particle masses and decay modes in the SLHA format so that this can be used in all BSM models.

The structure of the implementation of the SUSY model is designed to allow the implementation of
extended SUSY models. The
\href{https://herwig.hepforge.org/doxygen/classHerwig\_1\_1SusyBase.html}{SusyBase}
class, which inherits from the
\texttt{BSMModel}
class, is designed to read in the SLHA files specifying the SUSY
spectrum.

\begin{savenotes}\sphinxattablestart
\sphinxthistablewithglobalstyle
\centering
\sphinxcapstartof{table}
\sphinxthecaptionisattop
\sphinxcaption{Herwig Vertex classes for the MSSM.}\label{\detokenize{review/BSM:id53}}\label{\detokenize{review/BSM:tab-susy-vertices}}
\sphinxaftertopcaption
\begin{tabulary}{\linewidth}[t]{TT}
\sphinxtoprule
\sphinxstyletheadfamily 

Class
&\sphinxstyletheadfamily 

Interaction
\\
\sphinxmidrule
\sphinxtableatstartofbodyhook

\href{https://herwig.hepforge.org/doxygen/classHerwig\_1\_1SSNFSVertex.html}{SSNFSVertex}
&

Neutralino with a SM fermion and a sfermion
\\
\hline

\href{https://herwig.hepforge.org/doxygen/classHerwig\_1\_1SSCFSVertex.html}{SSCFSVertex}
&

Chargino with a SM fermion and a sfermion
\\
\hline

\href{https://herwig.hepforge.org/doxygen/classHerwig\_1\_1SSGFSVertex.html}{SSGFSVertex}
&

Gluino with a quark and squark
\\
\hline

\href{https://herwig.hepforge.org/doxygen/classHerwig\_1\_1SSNNZVertex.html}{SSNNZVertex}
&

A pair of neutralinos with a $Z^0$ boson
\\
\hline

\href{https://herwig.hepforge.org/doxygen/classHerwig\_1\_1SSCCZVertex.html}{SSCCZVertex}
&

A pair of charginos with a $Z^0$ boson
\\
\hline

\href{https://herwig.hepforge.org/doxygen/classHerwig\_1\_1SSCNWVertex.html}{SSCNWVertex}
&

Chargino with a neutralino and a $W^\pm$ boson
\\
\hline

\href{https://herwig.hepforge.org/doxygen/classHerwig\_1\_1SSGSGSGVertex.html}{SSGSGSGVertex}
&

SM gluon with a pair of gluinos
\\
\hline

\href{https://herwig.hepforge.org/doxygen/classHerwig\_1\_1SSGSSVertex.html}{SSGSSVertex}
&

SM gluon with a pair of squarks
\\
\hline

\href{https://herwig.hepforge.org/doxygen/classHerwig\_1\_1SSWSSVertex.html}{SSWSSVertex}
&

SM gauge boson with a pair of sfermions
\\
\hline

\href{https://herwig.hepforge.org/doxygen/classHerwig\_1\_1SSFFHVertex.html}{SSFFHVertex}
&

A pair of SM fermions with a Higgs boson
\\
\hline

\href{https://herwig.hepforge.org/doxygen/classHerwig\_1\_1SSWHHVertex.html}{SSWHHVertex}
&

SM EW gauge boson with a pair of Higgs bosons
\\
\hline

\href{https://herwig.hepforge.org/doxygen/classHerwig\_1\_1SSWWHVertex.html}{SSWWHVertex}
&

A pair of gauge bosons with a Higgs boson
\\
\hline

\href{https://herwig.hepforge.org/doxygen/classHerwig\_1\_1SSWWHHVertex.html}{SSWWHHVertex}
&

A pair of gauge bosons with a pair Higgs bosons
\\
\hline

\href{https://herwig.hepforge.org/doxygen/classHerwig\_1\_1SSGOGOHVertex.html}{SSGOGOHVertex}
&

A pair of gauginos with a Higgs boson
\\
\hline

\href{https://herwig.hepforge.org/doxygen/classHerwig\_1\_1SSHSFSFVertex.html}{SSHSFSFVertex}
&

A Higgs boson with a pair of sfermions
\\
\hline

\href{https://herwig.hepforge.org/doxygen/classHerwig\_1\_1SSHHHVertex.html}{SSHHHVertex}
&

Triple Higgs boson self coupling
\\
\hline

\href{https://herwig.hepforge.org/doxygen/classHerwig\_1\_1SSHGGVertex.html}{SSHGGVertex}
&

A Higgs boson with a pair of gluons via quark and squark loops
\\
\hline

\href{https://herwig.hepforge.org/doxygen/classHerwig\_1\_1SSHPPVertex.html}{SSHPPVertex}
&

A Higgs boson with a pair of photons loops diagrams
\\
\hline

\href{https://herwig.hepforge.org/doxygen/classHerwig\_1\_1SSGGSQSQVertex.html}{SSGGSQSQVertex}
&

A pair of gluons with a pair of squarks
\\
\hline

\href{https://herwig.hepforge.org/doxygen/classHerwig\_1\_1SSGVFSVertex.html}{SSGVFSVertex}
&

Gravitino with a fermion and a scalar boson
\\
\hline

\href{https://herwig.hepforge.org/doxygen/classHerwig\_1\_1SSGVNVVertex.html}{SSGVNVVertex}
&

Gravitino with a neutralino and vector boson
\\
\hline

\href{https://herwig.hepforge.org/doxygen/classHerwig\_1\_1SSGVNHVertex.html}{SSGVNHVertex}
&

Gravitino with a neutralino and Higgs boson
\\
\hline

\href{https://herwig.hepforge.org/doxygen/classHerwig\_1\_1SSGNGVertex.html}{SSGNGVertex}
&

Gluino with a neutralino and gluon via loop diagrams
\\
\hline

\href{https://herwig.hepforge.org/doxygen/classHerwig\_1\_1SSNNPVertex.html}{SSNNPVertex}
&

Neutralino with a neutralino and a photon via loop diagrams
\\
\hline

\href{https://herwig.hepforge.org/doxygen/classHerwig\_1\_1SSNCTVertex.html}{SSNCTVertex}
&

Flavour changing neutralino, charm stop coupling
\\
\sphinxbottomrule
\end{tabulary}
\sphinxtableafterendhook\par
\sphinxattableend\end{savenotes}

The details of the MSSM are implemented in the
\href{https://herwig.hepforge.org/doxygen/classHerwig\_1\_1MSSM.html}{MSSM}
class, which inherits from the
\texttt{SusyBase}
class. The Vertex classes for the MSSM are given in
\hyperref[\detokenize{review/BSM:tab-susy-vertices}]{Table \ref{\detokenize{review/BSM:tab-susy-vertices}}}. A spectrum file in SLHA format must be
supplied or the MSSM model cannot be used. MSSM in Herwig is designed
to allow extended SUSY models to be added  and for example supports additional
neutralino states to make the implementation of the NMSSM model easier.

The details of the NMSSM model are implemented in the
\href{https://herwig.hepforge.org/doxygen/classHerwig\_1\_1NMSSM.html}{NMSSM} class,
which inherits from the
\texttt{MSSM}
class. Some vertices are the same as those in the MSSM, or the MSSM vertices can be easily
modified, for example by including the additional neutralino state, so we make use of the
MSSM Vertex classes. The additional Vertex classes for the NMSSM are given in \hyperref[\detokenize{review/BSM:tab-nmssm-vertices}]{Table \ref{\detokenize{review/BSM:tab-nmssm-vertices}}}.

\begin{savenotes}\sphinxattablestart
\sphinxthistablewithglobalstyle
\centering
\sphinxcapstartof{table}
\sphinxthecaptionisattop
\sphinxcaption{Herwig Vertex classes for the NMSSM.}\label{\detokenize{review/BSM:id54}}\label{\detokenize{review/BSM:tab-nmssm-vertices}}
\sphinxaftertopcaption
\begin{tabulary}{\linewidth}[t]{TT}
\sphinxtoprule
\sphinxstyletheadfamily 

Class
&\sphinxstyletheadfamily 

Interaction
\\
\sphinxmidrule
\sphinxtableatstartofbodyhook

\href{https://herwig.hepforge.org/doxygen/classHerwig\_1\_1NMSSMFFHVertex.html}{NMSSMFFHVertex}
&

A pair of SM fermions with a Higgs boson
\\
\hline

\href{https://herwig.hepforge.org/doxygen/classHerwig\_1\_1NMSSMWHHVertex.html}{NMSSMWHHVertex}
&

SM EW gauge boson with a pair of Higgs bosons
\\
\hline

\href{https://herwig.hepforge.org/doxygen/classHerwig\_1\_1NMSSMWWHVertex.html}{NMSSMWWHVertex}
&

A pair of gauge bosons with a Higgs boson
\\
\hline

\href{https://herwig.hepforge.org/doxygen/classHerwig\_1\_1NMSSMWWHHVertex.html}{NMSSMWWHHVertex}
&

A pair of gauge bosons with a pair Higgs bosons
\\
\hline

\href{https://herwig.hepforge.org/doxygen/classHerwig\_1\_1NMSSMGOGOHVertex.html}{NMSSMGOGOHVertex}
&

A pair of gauginos with a Higgs boson
\\
\hline

\href{https://herwig.hepforge.org/doxygen/classHerwig\_1\_1NMSSMHSFSFVertex.html}{NMSSMHSFSFVertex}
&

A Higgs boson with a pair of sfermions
\\
\hline

\href{https://herwig.hepforge.org/doxygen/classHerwig\_1\_1NMSSMHHHVertex.html}{NMSSMHHHVertex}
&

Triple Higgs boson self coupling
\\
\sphinxbottomrule
\end{tabulary}
\sphinxtableafterendhook\par
\sphinxattableend\end{savenotes}

The details of the R-parity violating SUSY model are implemented in the
\href{https://herwig.hepforge.org/doxygen/classHerwig\_1\_1RPV.html}{RPV} class,
which inherits from the
\texttt{MSSM}
class. Many vertices are the same as those in the MSSM and therefore we make use of the
MSSM Vertex classes. The additional Vertex classes for the R-parity violating model are given in \hyperref[\detokenize{review/BSM:tab-rpv-vertices}]{Table \ref{\detokenize{review/BSM:tab-rpv-vertices}}}.

\begin{savenotes}\sphinxattablestart
\sphinxthistablewithglobalstyle
\centering
\sphinxcapstartof{table}
\sphinxthecaptionisattop
\sphinxcaption{Herwig Vertex classes for the R-parity violating SUSY model.}\label{\detokenize{review/BSM:id55}}\label{\detokenize{review/BSM:tab-rpv-vertices}}
\sphinxaftertopcaption
\begin{tabulary}{\linewidth}[t]{TT}
\sphinxtoprule
\sphinxstyletheadfamily 

Class
&\sphinxstyletheadfamily 

Interaction
\\
\sphinxmidrule
\sphinxtableatstartofbodyhook

\href{https://herwig.hepforge.org/doxygen/classHerwig\_1\_1RPVFFZVertex.html}{RPVFFZVertex}
&

A pair of fermions with the $Z^0$ boson
\\
\hline

\href{https://herwig.hepforge.org/doxygen/classHerwig\_1\_1RPVFFWVertex.html}{RPVFFWVertex}
&

A pair of fermions with the $W^\pm$ boson
\\
\hline

\href{https://herwig.hepforge.org/doxygen/classHerwig\_1\_1RPVFFSVertex.html}{RPVFFSVertex}
&

A pair of fermions with a scalar fermion
\\
\hline

\href{https://herwig.hepforge.org/doxygen/classHerwig\_1\_1RPVWSSVertex.html}{RPVWSSVertex}
&

A pair of scalar fermions with the $W^\pm$ boson
\\
\hline

\href{https://herwig.hepforge.org/doxygen/classHerwig\_1\_1RPVWWHVertex.html}{RPVWWHVertex}
&

A pair of $W^\pm$ bosons with the Higgs bosons
\\
\hline

\href{https://herwig.hepforge.org/doxygen/classHerwig\_1\_1RPVSSSVertex.html}{RPVSSSVertex}
&

Three scalar bosons
\\
\hline

\href{https://herwig.hepforge.org/doxygen/classHerwig\_1\_1RPVLLEVertex.html}{RPVLLEVertex}
&

Three leptons via the R-parity violating LLE term
\\
\hline

\href{https://herwig.hepforge.org/doxygen/classHerwig\_1\_1RPVLQDVertex.html}{RPVLQDVertex}
&

A lepton and two quarks via the R-parity violating LQD term
\\
\hline

\href{https://herwig.hepforge.org/doxygen/classHerwig\_1\_1RPVUDDVertex.html}{RPVUDDVertex}
&

Three quarks via the R-parity violating UDD term
\\
\sphinxbottomrule
\end{tabulary}
\sphinxtableafterendhook\par
\sphinxattableend\end{savenotes}

\begin{savenotes}\sphinxattablestart
\sphinxthistablewithglobalstyle
\centering
\sphinxcapstartof{table}
\sphinxthecaptionisattop
\sphinxcaption{Herwig Vertex classes for the Randall-Sundrum model.}\label{\detokenize{review/BSM:id56}}\label{\detokenize{review/BSM:tab-rs-vertices}}
\sphinxaftertopcaption
\begin{tabulary}{\linewidth}[t]{TT}
\sphinxtoprule
\sphinxstyletheadfamily 

Class
&\sphinxstyletheadfamily 

Interaction
\\
\sphinxmidrule
\sphinxtableatstartofbodyhook

\href{https://herwig.hepforge.org/doxygen/classHerwig\_1\_1RSModelFFGRVertex.html}{RSModelFFGRVertex}
&

Coupling of the graviton to SM fermions
\\
\hline

\href{https://herwig.hepforge.org/doxygen/classHerwig\_1\_1RSModelSSGRVertex.html}{RSModelSSGRVertex}
&

Coupling of the graviton to a Higgs boson pair
\\
\hline

\href{https://herwig.hepforge.org/doxygen/classHerwig\_1\_1RSModelFFGGRVertex.html}{RSModelFFGGRVertex}
&

Coupling of the graviton to two SM fermions and a gluon
\\
\hline

\href{https://herwig.hepforge.org/doxygen/classHerwig\_1\_1RSModelFFWGRVertex.html}{RSModelFFWGRVertex}
&

Coupling of the graviton to two SM
\\
\hline&

fermions and an EW gauge boson
\\
\hline

\href{https://herwig.hepforge.org/doxygen/classHerwig\_1\_1RSModelVVGRVertex.html}{RSModelVVGRVertex}
&

Coupling of the graviton to two gauge bosons
\\
\hline

\href{https://herwig.hepforge.org/doxygen/classHerwig\_1\_1RSModelGGGGRVertex.html}{RSModelGGGGRVertex}
&

Coupling of the graviton to three gluons
\\
\hline

\href{https://herwig.hepforge.org/doxygen/classHerwig\_1\_1RSModelWWWGRVertex.html}{RSModelWWWGRVertex}
&

Coupling of the graviton to three EW gauge bosons
\\
\sphinxbottomrule
\end{tabulary}
\sphinxtableafterendhook\par
\sphinxattableend\end{savenotes}

The
\texttt{RSModel}
class inherits from the
\texttt{BSMModel}
class and implements the calculations needed for the Randall-Sundrum
model. We have only implemented the vertices that are phenomenologically
relevant and therefore some four-point vertices that are not important
for resonance graviton production are not included. The Vertex classes
implemented for the Randall-Sundrum model are given in \hyperref[\detokenize{review/BSM:tab-rs-vertices}]{Table \ref{\detokenize{review/BSM:tab-rs-vertices}}}.

\begin{savenotes}\sphinxattablestart
\sphinxthistablewithglobalstyle
\centering
\sphinxcapstartof{table}
\sphinxthecaptionisattop
\sphinxcaption{Herwig Vertex classes for the UED model.}\label{\detokenize{review/BSM:id57}}\label{\detokenize{review/BSM:tab-ued-vertices}}
\sphinxaftertopcaption
\begin{tabulary}{\linewidth}[t]{TT}
\sphinxtoprule
\sphinxstyletheadfamily 

Class
&\sphinxstyletheadfamily 

Interaction
\\
\sphinxmidrule
\sphinxtableatstartofbodyhook

\href{https://herwig.hepforge.org/doxygen/classHerwig\_1\_1UEDF1F1P0Vertex.html}{UEDF1F1P0Vertex}
&

SM photon with a pair of KK-1 fermions
\\
\hline

\href{https://herwig.hepforge.org/doxygen/classHerwig\_1\_1UEDF1F1W0Vertex.html}{UEDF1F1W0Vertex}
&

SM $W^\pm$ boson with a pair of KK-1 fermions
\\
\hline

\href{https://herwig.hepforge.org/doxygen/classHerwig\_1\_1UEDF1F1Z0Vertex.html}{UEDF1F1Z0Vertex}
&

SM $Z^0$ boson with a pair of KK-1 fermions
\\
\hline

\href{https://herwig.hepforge.org/doxygen/classHerwig\_1\_1UEDF1F1G0Vertex.html}{UEDF1F1G0Vertex}
&

SM gluon with a pair of KK-1 fermions
\\
\hline

\href{https://herwig.hepforge.org/doxygen/classHerwig\_1\_1UEDF1F0W1Vertex.html}{UEDF1F0W1Vertex}
&

KK-1 fermion with an EW KK-1 boson and a SM fermion
\\
\hline

\href{https://herwig.hepforge.org/doxygen/classHerwig\_1\_1UEDF1F0G1Vertex.html}{UEDF1F0G1Vertex}
&

KK-1 fermion with a KK-1 gluon and a SM fermion
\\
\hline

\href{https://herwig.hepforge.org/doxygen/classHerwig\_1\_1UEDF1F0H1Vertex.html}{UEDF1F0H1Vertex}
&

KK-1 fermion with a KK-1 Higgs boson and a SM fermion
\\
\hline

\href{https://herwig.hepforge.org/doxygen/classHerwig\_1\_1UEDP0H1H1Vertex.html}{UEDP0H1H1Vertex}
&

SM photon with a pair of KK-1 charged Higgs bosons
\\
\hline

\href{https://herwig.hepforge.org/doxygen/classHerwig\_1\_1UEDW0W1W1Vertex.html}{UEDW0W1W1Vertex}
&

A pair of KK-1 gauge bosons with a SM $W^\pm$ or $Z^0$ boson
\\
\hline

\href{https://herwig.hepforge.org/doxygen/classHerwig\_1\_1UEDG1G1G0Vertex.html}{UEDG1G1G0Vertex}
&

A pair of KK-1 gluons with a SM gluon
\\
\hline

\href{https://herwig.hepforge.org/doxygen/classHerwig\_1\_1UEDG0G0G1G1Vertex.html}{UEDG0G0G1G1Vertex}
&

A pair of SM gluons with a pair of KK-1 gluons
\\
\hline

\href{https://herwig.hepforge.org/doxygen/classHerwig\_1\_1UEDW0A1H1Vertex.html}{UEDW0A1H1Vertex}
&

SM $W^\pm$ boson with a KK-1 charged Higgs boson and a
\\
\hline&

KK-1 pseudoscalar Higgs boson
\\
\hline

\href{https://herwig.hepforge.org/doxygen/classHerwig\_1\_1UEDZ0H1H1Vertex.html}{UEDZ0H1H1Vertex}
&

SM $Z^0$ boson with a pair of KK-1 charged Higgs bosons
\\
\hline

\href{https://herwig.hepforge.org/doxygen/classHerwig\_1\_1UEDZ0A1h1Vertex.html}{UEDZ0A1h1Vertex}
&

SM $Z^0$ boson with a KK-1 pseudoscalar Higgs boson and
\\
\hline&

a KK-1 scalar Higgs boson
\\
\sphinxbottomrule
\end{tabulary}
\sphinxtableafterendhook\par
\sphinxattableend\end{savenotes}

The UED model is implemented in the
\texttt{UEDBase}
class, which inherits from the
\texttt{BSMModel}
class and implements the calculation of the parameters of the model. The
Vertex classes for the UED model are given in \hyperref[\detokenize{review/BSM:tab-ued-vertices}]{Table \ref{\detokenize{review/BSM:tab-ued-vertices}}}.

The
\texttt{ADDModel}
class inherits from the
\texttt{BSMModel}
class and implements the calculations needed in this model. We have only
implemented the vertices that are phenomenologically relevant and
therefore the five--point vertices that are not important for $2\to2$
scattering processes are not included. The Vertex classes implemented
for the ADD model are given in \hyperref[\detokenize{review/BSM:tab-add-vertices}]{Table \ref{\detokenize{review/BSM:tab-add-vertices}}}.
In addition the
\href{https://herwig.hepforge.org/doxygen/classHerwig\_1\_1GravitonMassGenerator.html}{GravitonMassGenerator}
class is used to generate the mass of external gravitons, due to the sum
over the tower of Kaluza-Klein states.

\begin{savenotes}\sphinxattablestart
\sphinxthistablewithglobalstyle
\centering
\sphinxcapstartof{table}
\sphinxthecaptionisattop
\sphinxcaption{Herwig Vertex classes for the ADD model.}\label{\detokenize{review/BSM:id58}}\label{\detokenize{review/BSM:tab-add-vertices}}
\sphinxaftertopcaption
\begin{tabulary}{\linewidth}[t]{TT}
\sphinxtoprule
\sphinxstyletheadfamily 

Class
&\sphinxstyletheadfamily 

Interaction
\\
\sphinxmidrule
\sphinxtableatstartofbodyhook

\href{https://herwig.hepforge.org/doxygen/classHerwig\_1\_1ADDModelFFGRVertex.html}{ADDModelFFGRVertex}
&

Coupling of the graviton to SM fermions
\\
\hline

\href{https://herwig.hepforge.org/doxygen/classHerwig\_1\_1ADDModelSSGRVertex.html}{ADDModelSSGRVertex}
&

Coupling of the graviton to a Higgs boson pair
\\
\hline

\href{https://herwig.hepforge.org/doxygen/classHerwig\_1\_1ADDModelFFGGRVertex.html}{ADDModelFFGGRVertex}
&

Coupling of the graviton to two SM
\\
\hline&

quarks and a gluon
\\
\hline

\href{https://herwig.hepforge.org/doxygen/classHerwig\_1\_1ADDModelFFWGRVertex.html}{ADDModelFFWGRVertex}
&

Coupling of the graviton to two SM
\\
\hline&

fermions and an EW gauge boson
\\
\hline

\href{https://herwig.hepforge.org/doxygen/classHerwig\_1\_1ADDModelVVGRVertex.html}{ADDModelVVGRVertex}
&

Coupling of the graviton to two gauge bosons
\\
\hline

\href{https://herwig.hepforge.org/doxygen/classHerwig\_1\_1ADDModelGGGGRVertex.html}{ADDModelGGGGRVertex}
&

Coupling of the graviton to three gluons
\\
\hline

\href{https://herwig.hepforge.org/doxygen/classHerwig\_1\_1ADDModelWWWGRVertex.html}{ADDModelWWWGRVertex}
&

Coupling of the graviton to three EW gauge bosons
\\
\sphinxbottomrule
\end{tabulary}
\sphinxtableafterendhook\par
\sphinxattableend\end{savenotes}

The
\href{https://herwig.hepforge.org/doxygen/classHerwig\_1\_1LHModel.html}{LHModel}
class inherits from the
\texttt{BSMModel}
class and implements the calculations needed in the Little Higgs model. We have only
implemented the vertices that are phenomenologically relevant, which are described in
\hyperref[\detokenize{review/BSM:tab-lh-vertices}]{Table \ref{\detokenize{review/BSM:tab-lh-vertices}}}.

\begin{savenotes}\sphinxattablestart
\sphinxthistablewithglobalstyle
\centering
\sphinxcapstartof{table}
\sphinxthecaptionisattop
\sphinxcaption{Herwig Vertex classes for the Little Higgs model.}\label{\detokenize{review/BSM:id59}}\label{\detokenize{review/BSM:tab-lh-vertices}}
\sphinxaftertopcaption
\begin{tabulary}{\linewidth}[t]{TT}
\sphinxtoprule
\sphinxstyletheadfamily 

Class
&\sphinxstyletheadfamily 

Interaction
\\
\sphinxmidrule
\sphinxtableatstartofbodyhook

\href{https://herwig.hepforge.org/doxygen/classHerwig\_1\_1LHFFGVertex.html}{LHFFGVertex}
&

Interaction of the gluon with the fermions
\\
\hline

\href{https://herwig.hepforge.org/doxygen/classHerwig\_1\_1LHFFPVertex.html}{LHFFPVertex}
&

Interaction of the photon with the fermions
\\
\hline

\href{https://herwig.hepforge.org/doxygen/classHerwig\_1\_1LHFFWVertex.html}{LHFFWVertex}
&

Interaction of the $W^\pm$ boson with the fermions
\\
\hline

\href{https://herwig.hepforge.org/doxygen/classHerwig\_1\_1LHFFZVertex.html}{LHFFZVertex}
&

Interaction of the $Z^0$ boson with the fermions
\\
\hline

\href{https://herwig.hepforge.org/doxygen/classHerwig\_1\_1LHFFHVertex.html}{LHFFHVertex}
&

Interaction of the Higgs boson with the fermions
\\
\hline

\href{https://herwig.hepforge.org/doxygen/classHerwig\_1\_1LHWHHVertex.html}{LHWHHVertex}
&

Two Higgs bosons, EW gauge boson vertex
\\
\hline

\href{https://herwig.hepforge.org/doxygen/classHerwig\_1\_1LHWWWVertex.html}{LHWWWVertex}
&

Triple EW gauge boson vertex
\\
\hline

\href{https://herwig.hepforge.org/doxygen/classHerwig\_1\_1LHWWWWVertex.html}{LHWWWWVertex}
&

Four EW gauge boson vertex
\\
\hline

\href{https://herwig.hepforge.org/doxygen/classHerwig\_1\_1LHWWHVertex.html}{LHWWHVertex}
&

Interaction of the Higgs boson with the EW gauge bosons
\\
\hline

\href{https://herwig.hepforge.org/doxygen/classHerwig\_1\_1LHWWHHVertex.html}{LHWWHHVertex}
&

Two Higgs bosons, two EW gauge boson vertex
\\
\sphinxbottomrule
\end{tabulary}
\sphinxtableafterendhook\par
\sphinxattableend\end{savenotes}

The
\href{https://herwig.hepforge.org/doxygen/classHerwig\_1\_1LHTPModel.html}{LHTPModel}
class inherits from the
\texttt{BSMModel}
class and implements the parameters calculations needed in the Little Higgs model with T-parity conservation. We have only
implemented the vertices that are phenomenologically relevant, which are described in
\hyperref[\detokenize{review/BSM:tab-lhtp-vertices}]{Table \ref{\detokenize{review/BSM:tab-lhtp-vertices}}}.

\begin{savenotes}\sphinxattablestart
\sphinxthistablewithglobalstyle
\centering
\sphinxcapstartof{table}
\sphinxthecaptionisattop
\sphinxcaption{Herwig Vertex classes for the Little Higgs model with T-parity conservation.}\label{\detokenize{review/BSM:id60}}\label{\detokenize{review/BSM:tab-lhtp-vertices}}
\sphinxaftertopcaption
\begin{tabulary}{\linewidth}[t]{TT}
\sphinxtoprule
\sphinxstyletheadfamily 

Class
&\sphinxstyletheadfamily 

Interaction
\\
\sphinxmidrule
\sphinxtableatstartofbodyhook

\href{https://herwig.hepforge.org/doxygen/classHerwig\_1\_1LHTPFFGVertex.html}{LHTPFFGVertex}
&

Interaction of the gluon with the fermions
\\
\hline

\href{https://herwig.hepforge.org/doxygen/classHerwig\_1\_1LHTPFFPVertex.html}{LHTPFFPVertex}
&

Interaction of the photon with the fermions
\\
\hline

\href{https://herwig.hepforge.org/doxygen/classHerwig\_1\_1LHTPFFWVertex.html}{LHTPFFWVertex}
&

Interaction of the $W^\pm$ boson with the fermions
\\
\hline

\href{https://herwig.hepforge.org/doxygen/classHerwig\_1\_1LHTPFFZVertex.html}{LHTPFFZVertex}
&

Interaction of the $Z^0$ boson with the fermions
\\
\hline

\href{https://herwig.hepforge.org/doxygen/classHerwig\_1\_1LHTPFFHVertex.html}{LHTPFFHVertex}
&

Interaction of the Higgs boson with the fermions
\\
\hline

\href{https://herwig.hepforge.org/doxygen/classHerwig\_1\_1LHTPWHHVertex.html}{LHTPWHHVertex}
&

Two Higgs bosons, EW gauge boson vertex
\\
\hline

\href{https://herwig.hepforge.org/doxygen/classHerwig\_1\_1LHTPWWHVertex.html}{LHTPWWHVertex}
&

Interaction of the Higgs boson with the EW gauge bosons
\\
\hline

\href{https://herwig.hepforge.org/doxygen/classHerwig\_1\_1LHTPWWWVertex.html}{LHTPWWWVertex}
&

Triple EW gauge boson vertex
\\
\hline

\href{https://herwig.hepforge.org/doxygen/classHerwig\_1\_1LHTPHHHVertex.html}{LHTPHHHVertex}
&

Triple Higgs boson couplings
\\
\sphinxbottomrule
\end{tabulary}
\sphinxtableafterendhook\par
\sphinxattableend\end{savenotes}

The
\texttt{SextetModel}
class inherits from the
\texttt{BSMModel}. The vertices
for the interactions of the sextet particles are given in \hyperref[\detokenize{review/BSM:tab-sextet-vertices}]{Table \ref{\detokenize{review/BSM:tab-sextet-vertices}}}.

\begin{savenotes}\sphinxattablestart
\sphinxthistablewithglobalstyle
\centering
\sphinxcapstartof{table}
\sphinxthecaptionisattop
\sphinxcaption{Herwig Vertex classes for the Sextet model}\label{\detokenize{review/BSM:id61}}\label{\detokenize{review/BSM:tab-sextet-vertices}}
\sphinxaftertopcaption
\begin{tabulary}{\linewidth}[t]{TT}
\sphinxtoprule
\sphinxstyletheadfamily 

Class
&\sphinxstyletheadfamily 

Interaction
\\
\sphinxmidrule
\sphinxtableatstartofbodyhook

\href{https://herwig.hepforge.org/doxygen/classHerwig\_1\_1SextetFFSVertex.html}{SextetFFSVertex}
&

Interaction of two fermions with a scalar Sextet
\\
\hline

\href{https://herwig.hepforge.org/doxygen/classHerwig\_1\_1SextetFFVVertex.html}{SextetFFVVertex}
&

Interaction of two fermions with a vector Sextet
\\
\hline

\href{https://herwig.hepforge.org/doxygen/classHerwig\_1\_1SextetGSSVertex.html}{SextetGSSVertex}
&

Interaction of the gluon with two scalar Sextets
\\
\hline

\href{https://herwig.hepforge.org/doxygen/classHerwig\_1\_1SextetGVVVertex.html}{SextetGVVVertex}
&

Interaction of the gluon with two vector Sextets
\\
\hline

\href{https://herwig.hepforge.org/doxygen/classHerwig\_1\_1SextetGGSSVertex.html}{SextetGGSSVertex}
&

Interaction of two gluons with two scalar Sextets
\\
\hline

\href{https://herwig.hepforge.org/doxygen/classHerwig\_1\_1SextetGGVVVertex.html}{SextetGGVVVertex}
&

Interaction of two gluons with two vector Sextets
\\
\sphinxbottomrule
\end{tabulary}
\sphinxtableafterendhook\par
\sphinxattableend\end{savenotes}

The Transplanckian scattering matrix elements are implemented in the \href{https://herwig.hepforge.org/doxygen/classHerwig\_1\_1METRP2to2.html}{METRP2to2}
class for $2\to2$ transplanckian scattering.

The
\texttt{TTbAModel}
class inherits from the
\texttt{BSMModel}. The vertices
for the interactions of the particles are given in \hyperref[\detokenize{review/BSM:tab-ttba}]{Table \ref{\detokenize{review/BSM:tab-ttba}}}.

\begin{savenotes}\sphinxattablestart
\sphinxthistablewithglobalstyle
\centering
\sphinxcapstartof{table}
\sphinxthecaptionisattop
\sphinxcaption{Herwig Vertex classes for the TTbAModel.}\label{\detokenize{review/BSM:id62}}\label{\detokenize{review/BSM:tab-ttba}}
\sphinxaftertopcaption
\begin{tabulary}{\linewidth}[t]{TT}
\sphinxtoprule
\sphinxstyletheadfamily 

Class
&\sphinxstyletheadfamily 

Interaction
\\
\sphinxmidrule
\sphinxtableatstartofbodyhook

\href{https://herwig.hepforge.org/doxygen/classHerwig\_1\_1TTbAModelSU2XVertex.html}{TTbAModelSU2XVertex}
&

Interaction of two fermions in the SU2X model
\\
\hline

\href{https://herwig.hepforge.org/doxygen/classHerwig\_1\_1TTbAModelZPQQVertex.html}{TTbAModelZPQQVertex}
&

Interaction of two fermions with a Z prime
\\
\hline

\href{https://herwig.hepforge.org/doxygen/classHerwig\_1\_1TTbAModelAGQQVertex.html}{TTbAModelAGQQVertex}
&

Interaction of the axial gluon with two quarks
\\
\hline

\href{https://herwig.hepforge.org/doxygen/classHerwig\_1\_1SextetGVVVertex.html}{TTbAModelWPTDVertex}
&

Interaction of two fermions with a W prime
\\
\sphinxbottomrule
\end{tabulary}
\sphinxtableafterendhook\par
\sphinxattableend\end{savenotes}

The
\href{https://herwig.hepforge.org/doxygen/classHerwig\_1\_1LeptoquarkModel.html}{LeptoquarkModel} class inherits from the
\texttt{BSMModel}. The vertices
for the interactions of the leptoquarks are given in \hyperref[\detokenize{review/BSM:tab-lepto}]{Table \ref{\detokenize{review/BSM:tab-lepto}}}.

\begin{savenotes}\sphinxattablestart
\sphinxthistablewithglobalstyle
\centering
\sphinxcapstartof{table}
\sphinxthecaptionisattop
\sphinxcaption{Herwig Vertex classes for the LeptoQuark model.}\label{\detokenize{review/BSM:id63}}\label{\detokenize{review/BSM:tab-lepto}}
\sphinxaftertopcaption
\begin{tabulary}{\linewidth}[t]{TT}
\sphinxtoprule
\sphinxstyletheadfamily 

Class
&\sphinxstyletheadfamily 

Interaction
\\
\sphinxmidrule
\sphinxtableatstartofbodyhook

\href{https://herwig.hepforge.org/doxygen/classHerwig\_1\_1LeptoquarkModelSLQSLQGGVertex}{LeptoquarkModelSLQSLQGGVertex}
&

Interaction of two leptoquarks and two gluons
\\
\hline

\href{https://herwig.hepforge.org/doxygen/classHerwig\_1\_1LeptoquarkModelSLQSLQGVertex}{LeptoquarkModelSLQSLQGVertex}
&

Interaction of two leptoquarks and one gluon
\\
\hline

\href{https://herwig.hepforge.org/doxygen/classHerwig\_1\_1LeptoquarkModelSLQFFVertex}{LeptoquarkModelSLQFFVertex}
&

Interaction of the leptoquark, a lepton and a quark
\\
\sphinxbottomrule
\end{tabulary}
\sphinxtableafterendhook\par
\sphinxattableend\end{savenotes}

The
\texttt{ZprimeModel} class inherits from the
\texttt{BSMModel}. The vertex
is implemented in the \href{https://herwig.hepforge.org/doxygen/classHerwig\_1\_1ZprimeModelZPQQVertex.html}{ZprimeModelZPQQVertex} class.

The full list of interfaces for all the classes is provided in the
\href{https://herwig.hepforge.org/doxygen/index.html}{Doxygen}
documentation.

\clearpage

\section{Hadronization}
\label{\detokenize{review/index:hadronization}}\label{\detokenize{review/index:sec-hadronization}}

Hadronization occurs in the step after the conclusion of the parton shower:
the resulting quarks and gluons are assembled into the observed hadrons.
The cluster model \cite{Webber:1983if} used in Herwig 7
is based on the principle of colour preconfinement \cite{Amati:1979fg},
satisfied by coherent parton showers;
the quarks and gluons emerging from the shower are organised
in clusters, whose mass distribution is independent
of both the hard process considered, and the energy scale of the hard scattering.

The cluster model is local in the colour of the partons,
and asymptotically independent of both the hard
process and the centre-of-mass energy of the collision
\cite{Webber:1983if, Marchesini:1987cf}.
It is motivated by consideration of the mass spectrum of
hadronic resonances of a given (quark-antiquark) flavour. The
lowest-lying states are very narrow, but higher resonances become
increasingly broad until one can imagine, rather than individual
resonances, a continuum of completely overlapping hadronic states.

The preconfined colour singlets which emerge from the parton shower
are projected onto this continuum, and called clusters. In the simplest
model, they decay directly to narrower-resonance lighter mesons. Because the
clusters are considered to be the overlap of multiple states of
different spins, the simplest model assumes that this decay is
isotropic.

This simple model has been refined and extended, so that for a single event
hadronization proceeds as:
\begin{enumerate}
\sphinxsetlistlabels{\arabic}{enumi}{enumii}{}{.}%
\item {} 

cluster formation, from final-state quarks and gluons (\hyperref[\detokenize{review/hadronization:sect-clusterformation}]{Section \ref{\detokenize{review/hadronization:sect-clusterformation}}});

\item {} 

colour reconnection, to rearrange the constituents of the primordial clusters (\hyperref[\detokenize{review/hadronization:sect-colourreconnection}]{Section \ref{\detokenize{review/hadronization:sect-colourreconnection}}});

\item {} 

cluster fission, to split heavy clusters into lighter clusters (\hyperref[\detokenize{review/hadronization:sect-clusterfission}]{Section \ref{\detokenize{review/hadronization:sect-clusterfission}}});

\item {} 

cluster decay, into observed hadrons (\hyperref[\detokenize{review/hadronization:sect-clusterdecay}]{Section \ref{\detokenize{review/hadronization:sect-clusterdecay}}}).

\end{enumerate}

Each step is described in further detail in the indicated section below and represented pictorially in \hyperref[\detokenize{review/hadronization:fig-hadronizationherwig-steps}]{Fig.\@ \ref{\detokenize{review/hadronization:fig-hadronizationherwig-steps}}}.

\begin{figure}[bp]
\centering
\capstart
\noindent\includegraphics[width=0.600\linewidth]{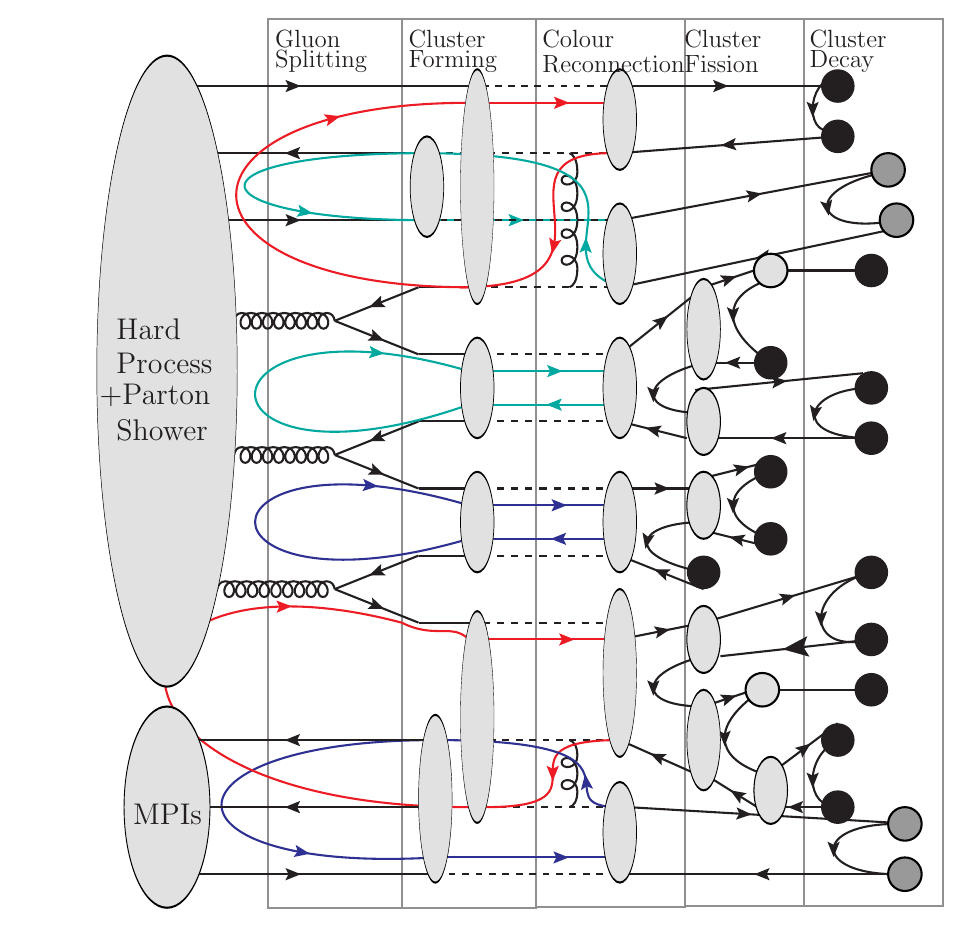}
\caption{Pictorial representation of the Herwig cluster hadronization model.}\label{\detokenize{review/hadronization:id74}}\label{\detokenize{review/hadronization:fig-hadronizationherwig-steps}}\end{figure}

The optional inclusion of spin effects for heavy mesons and baryons is described in \hyperref[\detokenize{review/hadronization:sect-spinhadronization}]{Section \ref{\detokenize{review/hadronization:sect-spinhadronization}}};
the application of the hadronization model to models containing BSM physics is described in \hyperref[\detokenize{review/hadronization:sect-bsmhadronization}]{Section \ref{\detokenize{review/hadronization:sect-bsmhadronization}}}.
Finally, Herwig 7.3 also supports the Lund string model via an interface
to Pythia. This is described in \hyperref[\detokenize{review/hadronization:sect-stringhad}]{Section \ref{\detokenize{review/hadronization:sect-stringhad}}}.

\subsection{Cluster formation}
\label{\detokenize{review/hadronization:cluster-formation}}\label{\detokenize{review/hadronization:sect-clusterformation}}\label{\detokenize{review/hadronization::doc}}

Clusters are formed by non-perturbatively splitting
the gluons in the final-state of the parton shower into quark-antiquark pairs.
For this to be possible, a constituent mass is assigned to the gluons at the end of the parton shower or as a first step of the hadronization handler, which must be heavier than twice the constituent mass of the lightest quark, so that the quarks after the splitting are on the constituent
mass shell.\begin{footnote}\sphinxAtStartFootnote
We typically set the constituent masses of the up and down quarks to
be equal, although they can in principle differ.
\end{footnote}
In the angular-ordered parton shower the gluons are given the constituent mass as part
of the kinematic reconstruction process (see \hyperref[\detokenize{review/showers/qtilde:sect-finalrecon}]{Section \ref{\detokenize{review/showers/qtilde:sect-finalrecon}}}). In the dipole shower
a momentum reshuffling is performed after the parton shower (see
\hyperref[\detokenize{review/showers/dipole:dipole-shower-constituent-reshuffling}]{Section \ref{\detokenize{review/showers/dipole:dipole-shower-constituent-reshuffling}}}),  This option is also available as an alternative within the angular ordered shower, such that the result of the angular ordered shower is accessible without being perturbed by the constituent mass, which is central to event-by-event hadronization corrections to be discussed in \hyperref[\detokenize{review/hadronization:event-by-event-hadronization-corrections}]{Section \ref{\detokenize{review/hadronization:event-by-event-hadronization-corrections}}}.
As part of newly developed models \cite{Hoang:2024zwl}, a dynamic gluon mass generation
can now be implemented using the \texttt{{GluonMassGenerator}} class.
The gluon is allowed to decay into any of the accessible light quark flavours,
with a probability given by the available phase-space for the decay.
\begin{footnote}\sphinxAtStartFootnote
The option of gluon decay into diquarks, which was available in
FORTRAN HERWIG, is no longer supported. Diquarks are therefore
present only as remnants of incoming baryons, or from baryon number
violating processes (see \hyperref[\detokenize{review/hadronization:sec-bnv}]{Section \ref{\detokenize{review/hadronization:sec-bnv}}}).
\end{footnote}

The gluon is allowed to decay into any of the accessible light quark flavours, with a probability given by the available phase-space for the decay%
\begin{footnote}\sphinxAtStartFootnote
The option of gluon decay into diquarks, which was available in FORTRAN HERWIG, is no longer supported which is a specific choice of the current implementation, though this might change in future versions. Diquarks are therefore present only as remnants of incoming baryons, or from baryon number violating processes (see \hyperref[\detokenize{review/hadronization:sec-bnv}]{Section \ref{\detokenize{review/hadronization:sec-bnv}}}).
\end{footnote}.

The gluon decays isotropically and, following this, the event contains solely colour-connected%
\begin{footnote}\sphinxAtStartFootnote
Note that quarks and antiquarks represent a start or endpoint, respectively, of the colour-connected pair.
\end{footnote}
quark and anti-quark pairs or (anti-)diquark and (anti-)quark pairs. The colour singlets formed by these colour-connected parton pairs are formed into clusters, which are each assigned a momentum corresponding to the total momentum of the constituent partons.

The mass distribution of primary clusters is shown in \hyperref[\detokenize{review/hadronization:fig-clustermasses}]{Fig.\@ \ref{\detokenize{review/hadronization:fig-clustermasses}}}, and confirms that the shower algorithm in Herwig 7 satisfies the preconfinement property to a good approximation at a centre-of-mass energy of 100 GeV, and is clearly invariant beyond that.

\subsection{Colour reconnection}
\label{\detokenize{review/hadronization:colour-reconnection}}\label{\detokenize{review/hadronization:sect-colourreconnection}}

The cluster formation process outlined above does not allow clusters to contain
partons from colour-disconnected regions, or from different stages of the event.
Colour reconnection reshuffles the cluster constituents to allow partons to
form a cluster regardless of their origin within the event.
This is of particular relevance for the simulation of hadronic collisions with
multiple partonic scatters.

The properties of a cluster depend on the
invariant cluster mass $M$, which is given by
\begin{equation*}
\begin{split}M^2 = (p_1 + p_2)^2,\end{split}
\end{equation*}

where $p_1$ and $p_2$ are the four momenta of the cluster
constituents, which are either $q\bar{q}'$ mesonic clusters or
$q(q',q'')_s/\bar{q}(\bar{q}',\bar{q}'')_s$ (anti-)diquark containing clusters.
The subsequent fission and decay of the clusters depend on
this invariant cluster mass $M$,
which directly influences the multiplicity of the final state particles.

However, these primordial clusters have been formed using the leading colour approximation
and would therefore hadronize completely independently in our cluster model. In particular,
two clusters with very close-by constituents in phase space would hadronize independently, even though a more
reasonable colour connection could be found by rearrangement of these clusters.
Restoring more meaningful colour connections to the clusters is the job of colour reconnection,
for which \texttt{{Herwig}} provides several different algorithms.

Each algorithm, which can be chosen using the interface \texttt{{ColourReconnector:Algorithm}},
attempts to minimize a closeness measure between the cluster constituents
over the permutation space of allowed rearrangements of these constituents into colour-singlet clusters.

Four colour-reconnection algorithms are currently implemented in Herwig:
\begin{enumerate}
\sphinxsetlistlabels{\roman}{enumi}{enumii}{(}{)}%
\item {} 

‘plain’ colour reconnection (\hyperref[\detokenize{review/hadronization:sect-plaincr}]{Section \ref{\detokenize{review/hadronization:sect-plaincr}}});

\item {} 

‘statistical’ colour reconnection \cite{Gieseke:2012ft} (\hyperref[\detokenize{review/hadronization:sect-statcr}]{Section \ref{\detokenize{review/hadronization:sect-statcr}}});

\item {} 

‘baryonic’ colour reconnection \cite{Gieseke:2017clv} (\hyperref[\detokenize{review/hadronization:sect-barycr}]{Section \ref{\detokenize{review/hadronization:sect-barycr}}});

\item {} 

‘baryonic--mesonic’ colour reconnection \cite{Zimmermann} (\hyperref[\detokenize{review/hadronization:sect-barymescr}]{Section \ref{\detokenize{review/hadronization:sect-barymescr}}}).

\end{enumerate}

A full description of each implemented colour-reconnection model
can be found in Refs. \cite{Gieseke:2012ft, Gieseke:2017clv, Zimmermann} respectively.

The first two algorithms try to identify exchanges of cluster constituents that
reduce the sum of the invariant cluster masses,
while baryonic colour reconnection uses
the relative rapidities of the constituents to quantify their
closeness and identify alternative clusterings as either baryonic or mesonic.

The default algorithm is (iii), baryonic colour reconnection.
It is possible to switch between the algorithms using the
\href{https://herwig.hepforge.org/doxygen/ColourReconnectorInterfaces.html\#Algorithm}{Algorithm}
switch with the values \texttt{{Plain}}, \texttt{{Statistical}}, \texttt{{Baryonic}}, and \texttt{{BaryonicMesonic}},
respectively.

In all cases, we do not allow reconnections that would connect a quark-antiquark pair
that was originally in a colour-octet state, \textit{i.e.} that came from the same gluon.
This is controlled by the \href{https://herwig.hepforge.org/doxygen/ColourReconnectorInterfaces.html\#OctetTreatment}{OctetTreatment}
switch:
either only quark-antiquark pairs from a non-perturbative gluon splitting are prevented from being reconnected
(\texttt{{OctetTreatment=Final}}),
or all quark--antiquarks in an octet state are prevented from being reconnected
(\texttt{{OctetTreatment=All}}).
The latter is set as the default in Herwig 7.3 to improve the simulation of gluon jets \cite{Reichelt:2017hts}.

\subsubsection{Plain colour reconnection}
\label{\detokenize{review/hadronization:plain-colour-reconnection}}\label{\detokenize{review/hadronization:sect-plaincr}}

The closeness measure for the \texttt{{Plain}} colour-reconnection algorithm is defined as
\begin{equation*}
\begin{split}\lambda = \sum_{i=1}^{N_{\mathrm{cl}}} M_i^2.\end{split}
\end{equation*}

This involves reclustering the pairs of quarks/antiquarks to reduce their invariant mass,
and hence their contribution to $\lambda$.

‘Plain’ colour reconnection proceeds by choosing a reference quark at random
from all cluster constituents;
the cluster containing this quark is compared to all other clusters,
each of which is considered for reclustering.

For each cluster,
the sum of the invariant cluster masses of the original cluster configuration, $M_A + M_B$,
and of the masses of the possible new clusters, $M_C + M_D$, are calculated.
The alternative cluster configuration that results in the lowest sum of cluster masses
is accepted for reconnection, with probability $p_{\mathrm{Reco}}$.
If the reconnection is accepted, the original clusters are replaced by the new clusters.
This is iterated for each cluster in the list such that each cluster on the list has had the chance to be reconnected.

This algorithm results in modified clusters with smaller invariant masses than the original configuration and systematically replaces heavier clusters with lighter ones,
resulting in a shift of the invariant cluster mass.

\subsubsection{Statistical colour reconnection}
\label{\detokenize{review/hadronization:statistical-colour-reconnection}}\label{\detokenize{review/hadronization:sect-statcr}}

In principle, in order to minimise colour length
among all possible clusters of the final-state partons in an event,
all possible permutations of cluster configurations have to be considered.

In hadronic $pp$ collisions, such a ‘brute force’ approach would be
computationally prohibitive i.e.\ for $N$ clusters the time complexity
would be $\mathcal{O}(N!)$, due to the large number of clusters
originating from the hard process and multiparton interactions. To tackle this
problem, a statistical model for colour reconnection was implemented in Ref.
\cite{Gieseke:2012ft}.

The statistical colour-reconnection model samples cluster configurations to
find configurations of clusters with successively lower values of $\lambda$.
The approach is based on a simulated annealing algorithm,
which selects random pairs of clusters and accepts the
reconnection if the possible new cluster configuration lowers $\lambda$.

To allow the configuration to escape local minima of colour length,
if the reconnection would increase $\lambda$ it is accepted with probability
\begin{equation*}
\begin{split}P = \exp{\left(-\frac{\lambda_2 - \lambda_1}{T}\right)},\end{split}
\end{equation*}

where $T$ is the control parameter of the simulated annealing algorithm,
analogous to temperature.

The parameter $T$ is progressively reduced from a starting value that is
chosen from the median of 10 randomly chosen values for $|\Delta
\lambda_{ij}| = |\lambda_i - \lambda_j|$ of the particular event, multiplied by
a tuneable input parameter (\texttt{{InitialTemperature}}). It is stepped down by
a factor (\texttt{{AnnealingFactor}}) in each step; the number of steps
(\texttt{{AnnealingSteps}}), and the number of colour swaps tried in each step
(\texttt{{TriesPerStepFactor}}), can be set by the user.

\subsubsection{Baryonic colour reconnection}
\label{\detokenize{review/hadronization:baryonic-colour-reconnection}}\label{\detokenize{review/hadronization:sect-barycr}}

This algorithm is the default within Herwig 7.3. In this case, we additionally
allow baryonic-type cluster configurations, containing three quarks or three
antiquarks.

\begin{figure}
\centering
\capstart
\noindent\includegraphics[width=0.600\linewidth]{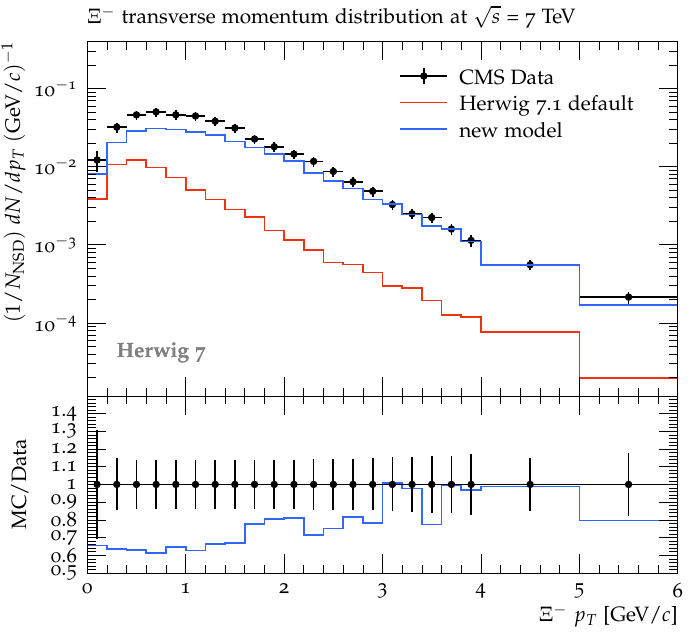}
\caption{$p_T$ spectrum of $\Xi^-$ baryons for minimum bias events measured by CMS
\cite{CMS:2011jlm} at $\sqrt{s}=7$ TeV from \cite{Gieseke:2018gff} for the old colour-reconnection model and the tuned baryonic model, labelled as `new model'.}\label{\detokenize{review/hadronization:id75}}\label{\detokenize{review/hadronization:fig-bcr-plotxi}}\end{figure}

This allows for a different baryon production mechanism especially for
multiply strange/heavy baryons as can be seen e.g.~in \hyperref[\detokenize{review/hadronization:fig-bcr-plotxi}]{Fig.\@ \ref{\detokenize{review/hadronization:fig-bcr-plotxi}}}. Furthermore,
for high-multiplicity events it gives a further mechanism to reduce multiplicity
even stronger than regular mesonic colour reconnection as can be seen in \hyperref[\detokenize{review/hadronization:fig-bcr-plotmulti}]{Fig.\@ \ref{\detokenize{review/hadronization:fig-bcr-plotmulti}}}.

\begin{figure}[htp]
\centering
\capstart

\noindent\includegraphics[width=0.600\linewidth]{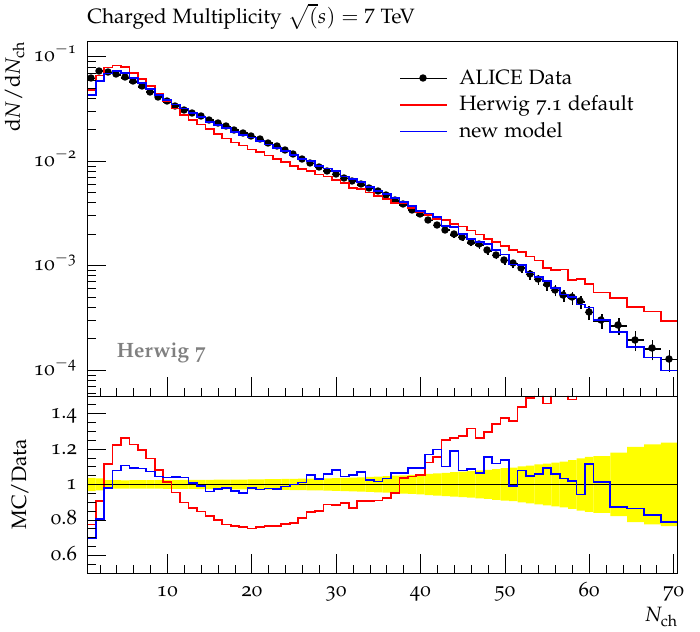}
\caption{Multiplicity distribution for minimum bias events measured by ALICE
\cite{ALICE:2010mty} at $\sqrt{s}=7$ TeV from \cite{Gieseke:2018gff} for the old colour
reconnection model and the tuned baryonic model, labelled as `new model'.}\label{\detokenize{review/hadronization:id76}}\label{\detokenize{review/hadronization:fig-bcr-plotmulti}}\end{figure}

Instead of finding quark combinations that directly lower the invariant
cluster mass, we consider a simple geometric picture of nearest
neighbours that populate approximately the same phase-space region based
on their relative rapidities motivated by \cite{Gieseke:2018gff}.

As with the other colour-reconnection models, the
starting point of the algorithm is the predefined colour configuration
that emerges once the parton shower evolution has terminated, and the
remaining gluons are split non-perturbatively into quark-antiquark
pairs.

A cluster $C_A$ is then picked randomly from the list of clusters and we
boost into the rest frame of $C_A$ where the direction of the
antiquark in the rest frame of the cluster is defined as the
positive axis.

In the next step we loop over all remaining clusters $C_B$ and calculate
the rapidities of the cluster constituents with respect to this
axis in the rest frame of the cluster $C_A$.
Based on the calculated rapidities, the clusters fall into one of three categories,
according to the alignment of the two quark-antiquark pairs:
\begin{itemize}
\item {} 

Mesonic: $y(q) > 0 > y(\bar{q})$.

\item {} 

Baryonic: $y(\bar{q}) > 0 > y(q)$.

\item {} 

None of the above.

\end{itemize}

If the cluster falls into the last category it is not considered for
reconnection. In the next step the label of the category and the sum
of the absolute values $|y(q)|+|y(\bar{q})|$ is saved for
the clusters with the two largest sums.
If the cluster with the largest sum is labeled \textit{mesonic}, the reconnection
is accepted with a probability given by the parameter $p_{\mathrm{Reco}}$ (\texttt{{ReconnectionProbability}}) as shown in \hyperref[\detokenize{review/hadronization:fig-mesonicreco}]{Fig.\@ \ref{\detokenize{review/hadronization:fig-mesonicreco}}}.
If the clusters with the two largest sums are labeled \textit{baryonic} they are
considered for baryonic reconnection with the probability
$p_{\mathrm{RecoBaryonic}}$ (\texttt{{ReconnectionProbabilityBaryonic}})
and the three mesonic clusters are rearranged
to form two baryonic-type clusters as shown in \hyperref[\detokenize{review/hadronization:fig-baryonicreco}]{Fig.\@ \ref{\detokenize{review/hadronization:fig-baryonicreco}}}.

\begin{figure}[tp]
\centering
\capstart
\noindent\includegraphics[width=0.500\linewidth]{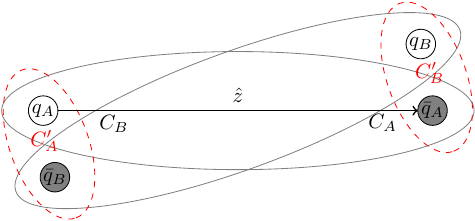}
\caption{Pictorial representation of a cluster configuration that would lead to a
mesonic reconnection from $C_A,C_B$ to $C_A',C_B'$.}\label{\detokenize{review/hadronization:id77}}\label{\detokenize{review/hadronization:fig-mesonicreco}}\end{figure}

\begin{figure}[tp]
\centering
\capstart
\noindent\includegraphics[width=0.500\linewidth]{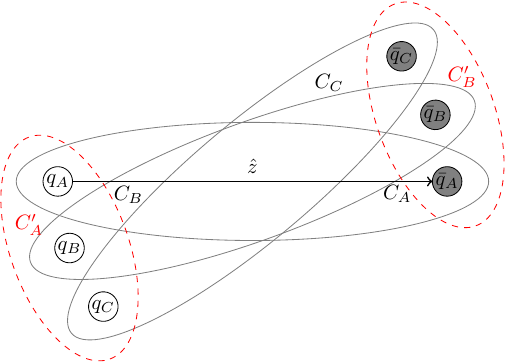}
\caption{Pictorial representation of a cluster configuration that would lead to a
baryonic reconnection from $C_A,C_B,C_C$ to $C_A',C_B'$.}\label{\detokenize{review/hadronization:id78}}\label{\detokenize{review/hadronization:fig-baryonicreco}}\end{figure}

Once a baryonic-type cluster is formed, all participating clusters are removed
from the list and not considered for further reconnections. We run this algorithm
until we have every cluster $C_A$ from the list of clusters has had the
chance to be reconnected. Further details of the algorithm are described in
\cite{Gieseke:2018gff}.

Note that all baryonic clusters are reduced to diquark-quark at the end of the
algorithm resulting in two component clusters. This reduction is done by
grouping the smallest-invariant-mass quark pair to a diquark and rescaling the
momentum to their invariant mass. Optionally, one can force the mass of the
reduced diquarks to be the same as the diquarks constituent mass via the
interface \texttt{{ClusterFinder:DiQuarkOnShell=Yes}}.

\subsubsection{Baryonic-Mesonic colour reconnection \sphinxfootnotemark[5]}
\label{\detokenize{review/hadronization:baryonic-mesonic-colour-reconnection}}\label{\detokenize{review/hadronization:sect-barymescr}}%
\begin{footnotetext}[5]\sphinxAtStartFootnote
Note that in Herwig 7.3 the baryonic-mesonic colour-reconnection model is not tuned.
This colour-reconnection model depends strongly on the lab frame,
which means for observables which are not tied to the lab frame,
results obtained using this model might not be well under control.
\end{footnotetext}\ignorespaces 

In this algorithm based on \cite{Zimmermann} we consider many different colour-reconnection (CR) topologies of the types summarized in \hyperref[\detokenize{review/hadronization:table-cr-topology-baryonic-mesonic}]{Table \ref{\detokenize{review/hadronization:table-cr-topology-baryonic-mesonic}}} alongside with the name of the static reconnection probabilities. Note that here we have denoted mesonic clusters consisting of a quark and an antiquark with ${M},{M}'$ and (anti)baryonic clusters which consist of three (anti)quarks as $B,B'$ ($\bar{B},\bar{B}'$). Similarly to the baryonic colour reconnection this algorithm allows for the coalescence of three mesonic cluster to a baryonic and antibaryonic cluster. However here we allow also all possible back-reactions as displayed in table \hyperref[\detokenize{review/hadronization:table-cr-topology-baryonic-mesonic}]{Table \ref{\detokenize{review/hadronization:table-cr-topology-baryonic-mesonic}}}.

In order to include all possible colour-reconnection topologies we need to impose a selection algorithm, which is described in the flowchart shown in \hyperref[\detokenize{review/hadronization:fig-baryonic-mesonic-algorithm1}]{Fig.\@ \ref{\detokenize{review/hadronization:fig-baryonic-mesonic-algorithm1}}}. This selection algorithm draws the following possible sets of clusters from all clusters available. The possible subsets of clusters fall into one of the following categories:
\begin{itemize}
\item {} 

Three mesonic clusters (3M), which can reconnect to three different mesonic clusters (3M’) or to a baryonic-antibaryonic cluster pair (BbarB)

\item {} 

One mesonic and one (anti)-baryonic cluster (MB), which can reconnect only to a different set of one mesonic and one (anti)-baryonic cluster (M’B’)

\item {} 

Two (anti)-baryonic clusters (2B), which can only reconnect to two different (anti)-baryonic clusters (2B’)

\item {} 

One anti-baryonic and one baryonic cluster (BbarB), which can only reconnect to three mesonic clusters (3M)

\end{itemize}

With the selection algorithm described in the flowchart in \hyperref[\detokenize{review/hadronization:fig-baryonic-mesonic-algorithm1}]{Fig.\@ \ref{\detokenize{review/hadronization:fig-baryonic-mesonic-algorithm1}}} we can draw randomly possible subsets of clusters of the above categories. The baryonic-mesonic colour-reconnection algorithm then proceeds as follows in order to accept a new colour configuration subject to a distance measure $\Delta R_\text{tot}$ defined later on:
\begin{enumerate}
\sphinxsetlistlabels{\arabic}{enumi}{enumii}{}{.}%
\item {} 

Select a subset of clusters $\mathcal{C}_\text{ini}$ randomly (where only the types of subsets in \hyperref[\detokenize{review/hadronization:table-cr-topology-baryonic-mesonic}]{Table \ref{\detokenize{review/hadronization:table-cr-topology-baryonic-mesonic}}} are allowed) according to the selection algorithm.

\item {} 

Minimize a measure of distance in phase-space $\Delta R_\text{tot}$ in the lab frame on parton level with the constraint of having only mesonic and baryonic clusters at the end and no remaining partons. Call the configuration that minimizes this $\mathcal{C}_\text{proposal}$

\item {} 

Allow the initial configuration to reconnect to the minimal CR configuration with a static probability of $P_{\mathcal{C}_\text{ini}\rightarrow \mathcal{C}_\text{proposal}}$, which is a tunable parameter of the model

\item {} 

Go back to 1. and repeat the process for $f_\text{step}N_\text{Clusters}$ times, where $N_\text{Clusters}$ are the initial number of clusters before CR and $f_\text{step}$ is a tunable parameter.

\end{enumerate}

The proposal configuration $\mathcal{C}_\text{proposal}$ is determined by minimizing the distance measure $\Delta R_\text{tot}$,
\begin{equation*}
\begin{split}\Delta R_\text{tot}(C)
&=
\omega_\text{MTBF} \sum_{\text{mesons }m} \Delta R_{q_m\bar{q}_m} +\sum_{\text{baryons }b} \sum_{i\in b} \Delta R_{i\langle b\rangle}\\\end{split}
\end{equation*}

which can be seen to comprise a contribution from the mesonic clusters $m$ in
$C$, and a contribution inspired by baryon-junctions (i.e.\ the reconnection of strings into vertices connecting three quarks) from the baryonic clusters $b$.
The parameter $\omega_{\text{MTBF}}$ is tunable (\texttt{{MesonToBaryonFactor}}),
and controls the relative distance scale between the two types of cluster.
The mean rapidity and azimuth of a baryonic cluster
as indicated by $\langle b\rangle$ are defined as
the arithmetic mean and circular mean (better-defined for periodic variables \cite{PatternRecognition}) respectively,
\begin{equation*}
\begin{split}\bar{y}&= \frac{1}{3} \sum_i y_{q_i} \\
\bar{\phi}&= \atantwo\left(\frac{1}{3}\sum_i \sin (\phi_{q_i}),\frac{1}{3}\sum_i \cos (\phi_{q_i})\right)\end{split}
\end{equation*}

and the rapidity-azimuth distance measure is defined as
\begin{equation*}
\begin{split}\Delta R = \sqrt{(\Delta y)^2 + (\Delta \phi)^2}.\end{split}
\end{equation*}

It is also possible to use a different definition of the baryonic distance measure,
which instead of the above baryon junction-like measure of distance
uses all permutation distances.
This changes the selection criterion for baryons such that $\Delta R_\text{baryonic}(C_b)$ becomes:
\begin{equation*}
\begin{split}\Delta R_\text{baryonic}(C_b)&=\Delta R_{q_1,q_2}+\Delta R_{q_2,q_3}+\Delta R_{q_1,q_3}\end{split}
\end{equation*}

This option can be turned on by \href{https://herwig.hepforge.org/doxygen/ColourReconnectorInterfaces.html\#Junction}{Junction=Off}.
Note that the same vetoing applies if we would propose an octet mesonic cluster as previously discussed depending on the option \texttt{OctetTreatment}.

\begin{figure}[tp]
\centering
\capstart
\noindent\includegraphics[width=0.800\linewidth]{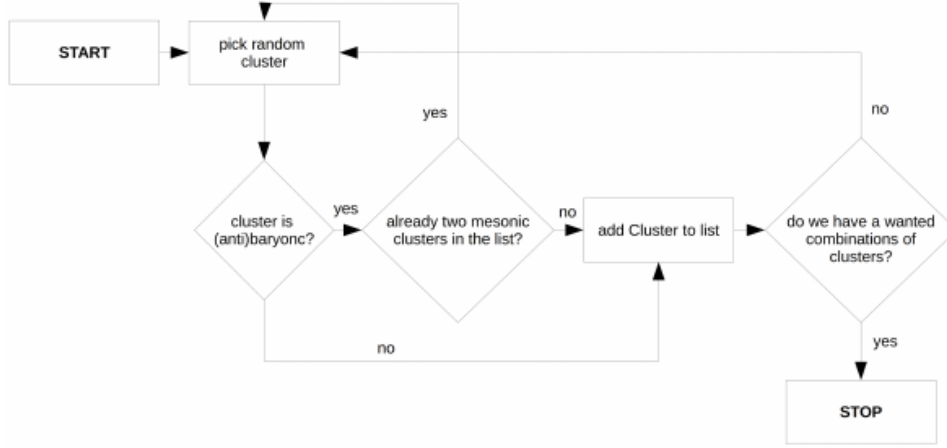}
\caption{Flowchart of selection algorithm for cluster sets of the algorithm (from \cite{Zimmermann}).}\label{\detokenize{review/hadronization:id79}}\label{\detokenize{review/hadronization:fig-baryonic-mesonic-algorithm1}}\end{figure}

\begin{savenotes}\sphinxattablestart
\sphinxthistablewithglobalstyle
\centering
\sphinxcapstartof{table}
\sphinxthecaptionisattop
\sphinxcaption{In this table we summarize the different colour reconnection topologies and their corresponding static probabilities. Here we denote mesonic clusters as \protect$M\protect$ and (anti)-baryonic clusters as \protect$B(\bar{B})\protect$.}\label{\detokenize{review/hadronization:id80}}\label{\detokenize{review/hadronization:table-cr-topology-baryonic-mesonic}}
\sphinxaftertopcaption
\begin{tabulary}{\linewidth}[t]{TT}
\sphinxtoprule
\sphinxtableatstartofbodyhook

Colour-reconnection topology
&

Reconnection probability
\\
\hline

$3 M\rightarrow 3M'$
&

\href{https://herwig.hepforge.org/doxygen/ColourReconnectorInterfaces.html}{ReconnectionProbability3Mto3M}
\\
\hline

$3 M\rightarrow B\bar{B}$
&

\href{https://herwig.hepforge.org/doxygen/ColourReconnectorInterfaces.html}{ReconnectionProbability3MtoBBbar}
\\
\hline

$M,B\rightarrow M',B'$
&

\href{https://herwig.hepforge.org/doxygen/ColourReconnectorInterfaces.html}{ReconnectionProbabilityMBtoMB}
\\
\hline

$M,\bar{B}\rightarrow M',\bar{B}'$
&

\texttt{ReconnectionProbabilityMBtoMB}
\\
\hline

$\bar{B},B \rightarrow 3 M$
&

\href{https://herwig.hepforge.org/doxygen/ColourReconnectorInterfaces.html}{ReconnectionProbabilityBbarBto3M}
\\
\hline

$2 B\rightarrow 2 B'$
&

\href{https://herwig.hepforge.org/doxygen/ColourReconnectorInterfaces.html}{ReconnectionProbability2Bto2B}
\\
\hline

$2 \bar{B}\rightarrow 2 \bar{B}'$
&

\texttt{ReconnectionProbability2Bto2B}
\\
\sphinxbottomrule
\end{tabulary}
\sphinxtableafterendhook\par
\sphinxattableend\end{savenotes}

\subsection{Cluster fission}
\label{\detokenize{review/hadronization:cluster-fission}}\label{\detokenize{review/hadronization:sect-clusterfission}}

The cluster model is based on the
observation that because the cluster mass spectrum is both universal and
peaked at low masses, as shown in \hyperref[\detokenize{review/hadronization:fig-clustermasses}]{Fig.\@ \ref{\detokenize{review/hadronization:fig-clustermasses}}}, the clusters
can be regarded as highly excited hadron resonances and decayed,
according to phase-space, into the observed hadrons.
However, a small fraction of clusters are too heavy for this to be a
reasonable approach.
These heavy clusters are therefore first split into
lighter clusters before they decay.

A cluster is split into two clusters if its mass, $M$, is such
that
\begin{equation}\label{equation:review/hadronization:eqn:clustersplit}
\begin{split}M^{\bf Cl_{pow}} \geq {\bf Cl_{max}}^{\bf Cl_{pow}}+(m_1+m_2)^{\bf Cl_{pow}},\end{split}
\end{equation}

where ${\bf Cl_{max}}$ and ${\bf Cl_{pow}}$ are parameters
of the model, and $m_{1,2}$ are the masses of the constituent
partons of the cluster. In practice,
in order to improve the description of the production of bottom
and charm hadrons, we include separate values of both
${\bf Cl_{max}}$ (\href{https://herwig.hepforge.org/doxygen/ClusterFissionerInterfaces.html\#ClMaxLight}{ClMaxLight},
\href{https://herwig.hepforge.org/doxygen/ClusterFissionerInterfaces.html\#ClMaxCharm}{ClMaxCharm} and \href{https://herwig.hepforge.org/doxygen/ClusterFissionerInterfaces.html\#ClMaxBottom\#ClMaxBottom}{ClMaxBottom}) and ${\bf Cl_{pow}}$ (\href{https://herwig.hepforge.org/doxygen/ClusterFissionerInterfaces.html\#ClPowLight}{ClPowLight},
\href{https://herwig.hepforge.org/doxygen/ClusterFissionerInterfaces.html\#ClPowCharm}{ClPowCharm} and \href{https://herwig.hepforge.org/doxygen/ClusterFissionerInterfaces.html\#ClPowBottom\#ClPowBottom}{ClPowBottom}) for
clusters containing light, charm and bottom quarks, respectively. The
default values of these and other important hadronization parameters are
given in \hyperref[\detokenize{review/hadronization:table-clusterparam}]{Table \ref{\detokenize{review/hadronization:table-clusterparam}}}, at the end of this section.

\begin{figure}[htp]
\centering
\capstart
\begin{subfigure}{0.49\textwidth}
\centering

\noindent\includegraphics[width=0.999\linewidth]{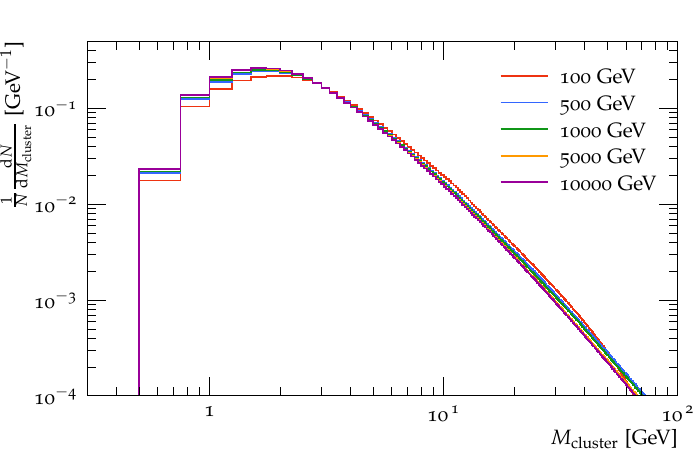}
\caption{The mass spectrum of primary clusters.}
\end{subfigure}
\begin{subfigure}{0.49\textwidth}
\centering

\noindent\includegraphics[width=0.999\linewidth]{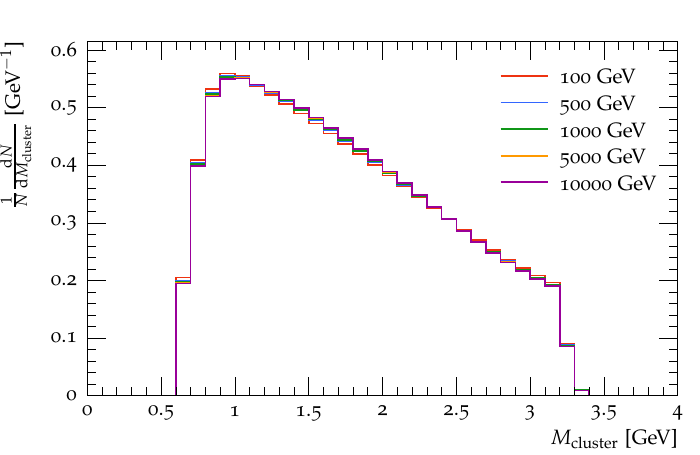}
\caption{The mass spectrum of clusters after cluster fission.}
\end{subfigure}
\caption{The mass spectrum of clusters, generated using
$e^+e^-\to d\bar{d}$ collisions at the displayed centre-of-mass energies.
Only clusters containing light quarks are shown.}\label{\detokenize{review/hadronization:fig-clustermasses}}\end{figure}

For clusters that need to be split, a $q\bar{q}$ pair is selected
to be popped from the vacuum. Only up, down and strange quarks are
chosen with probabilities given by the parameters
, ${\bf Pwt}_q$ %
\begin{footnote}\sphinxAtStartFootnote
We use ${\bf Pwt}_q$ to denote the probability of selecting a
given quark or diquark. The same parameters are used both for cluster
fission and cluster decay, with only \href{https://herwig.hepforge.org/doxygen/HadronSelectorInterfaces.html\#PwtDquark}{PwtDquark}, \href{https://herwig.hepforge.org/doxygen/HadronSelectorInterfaces.html\#PwtUquark}{PwtUquark} and \href{https://herwig.hepforge.org/doxygen/HadronSelectorInterfaces.html\#PwtSquark}{PwtSquark} used for fission as well as \href{https://herwig.hepforge.org/doxygen/HadronSelectorInterfaces.html\#PwtCquark}{PwtCquark} and \href{https://herwig.hepforge.org/doxygen/HadronSelectorInterfaces.html\#PwtBquark}{PwtBquark} for decay. Much like the gluon branching at the beginning of the hadronization, di-quarks are currently not available in the cluster fission, though this might be change in future versions.
For diquarks in cluster decay, the product of the diquark probability
\href{https://herwig.hepforge.org/doxygen/HadronSelectorInterfaces.html\#PwtDIquark}{PwtDIquark}, the
probabilities of the quarks forming the diquark, and a symmetry
factor, for diquarks.
\end{footnote}, where $q$ is the flavour of the quark.
Once a pair is selected, the cluster is decayed into two new clusters
with one of the original partons in each cluster. Unless one of the
partons is a remnant of the incoming beam particle the mass distribution
of the new clusters is given by
\begin{equation}\label{equation:review/hadronization:eqn:clusterspect}
\begin{split}M_1 &= m_1+(M-m_1-m_q)\,\mathcal{R}_1^{1/P},\\
M_2 &= m_2+(M-m_2-m_q)\,\mathcal{R}_2^{1/P},\end{split}
\end{equation}

where $m_q$ is the mass of the parton popped from the vacuum,
$M_{1,2}$ are the masses of the clusters formed by the splitting
and $\mathcal{R}_{1,2}$ are uniformly distributed random numbers
on the range $(0,1)$.
The distribution of the masses of the clusters is controlled by the
parameter $P$, which is \href{https://herwig.hepforge.org/doxygen/ClusterFissionerInterfaces.html\#PSplitLight}{PSplitLight}, \href{https://herwig.hepforge.org/doxygen/ClusterFissionerInterfaces.html\#PSplitCharm}{PSplitCharm} or
\href{https://herwig.hepforge.org/doxygen/ClusterFissionerInterfaces.html\#PSplitBottom}{PSplitBottom} for clusters containing light, charm
or bottom quarks.

In addition to selecting the mass according to Eq. \eqref{equation:review/hadronization:eqn:clusterspect}, the masses of the daughter clusters are required to satisfy either a \texttt{{Static}}
\begin{equation*}
\begin{split}M > M_1 + M_2, \quad M_1 > m_q + m_1, \quad M_2 > m_q + m_2,\end{split}
\end{equation*}

or a \texttt{{Dynamic}} mass hierarchy criterion \cite{Masouminia:2023zhb}:
\begin{equation*}
\begin{split}M^2 > M_1^2 + M_2^2, \quad M_1^2 > m_q^2 + m_1^2 + \delta_{\rm th}, \quad M_2^2 > m_q^2 + m_2^2 + \delta_{\rm th}.\end{split}
\end{equation*}

$\delta_{\rm th}$ is the kinematic threshold shift parameter. The \texttt{{Dynamic}} kinematic threshold scheme also incorporates a probability distribution function, $P_{\rm cluster}$, as a decision-making device:

\begin{equation*}
\begin{split}P_{\rm cluster} = \left( 1 + \left| \frac{M - \delta}{M_{\rm th}} \right|^{r} \right)^{-1}, \qquad P_{\rm cluster} > \mathcal{R},\end{split}
\end{equation*}

to either accept or reject the cluster splitting through comparison against a randomly generated number $\mathcal{R}$ distributed uniformly on (0,1). This function also has two new tunable parameters: $r$ as the probability power factor and $\delta$ as the probability shift. Here, $M_{th}$ is the mass threshold for the cluster,
defined as the sum of the masses of the constituent quarks plus the mass of the spawned di-quark. This scheme provides a smooth distribution for dynamic threshold cuts based on the cluster mass and its kinematic properties \cite{Masouminia:2023zhb}. The \texttt{{Dynamic}} threshold option is the default choice, which can be changed via \href{https://herwig.hepforge.org/doxygen/classHerwig_1_1ClusterFissioner.html#kinematicThresholdChoice}{KinematicThreshold} switch:

\begin{sphinxVerbatim}[commandchars=\\\{\}]
\PYG{n+nb}{cd}\PYG{+w}{ }/Herwig/Hadronization
\PYG{n+nb}{set}\PYG{+w}{ }ClusterFissioner:KinematicThreshold\PYG{+w}{ }\PYGZlt{}Static/Dynamic\PYGZgt{}
\end{sphinxVerbatim}

The spectrum of the cluster masses after the cluster splitting is shown in \hyperref[\detokenize{review/hadronization:fig-clustermasses}]{Fig.\@ \ref{\detokenize{review/hadronization:fig-clustermasses}}}.

For clusters that contain a remnant of the beam particle in hadronic
collisions a soft distribution is used for the masses of the clusters
produced in the splitting.
The \href{https://herwig.hepforge.org/doxygen/ClusterFissionerInterfaces.html\#RemnantOption}{RemnantOption} switch controls whether the soft distribution is used for both daughter clusters (\href{https://herwig.hepforge.org/doxygen/ClusterFissionerInterfaces.html\#RemnantOption}{RemnantOption=0}) or only the daughter
cluster containing the remnant (\href{https://herwig.hepforge.org/doxygen/ClusterFissionerInterfaces.html\#RemnantOption}{RemnantOption=1}), the default. The mass of the soft
clusters is given by
\begin{equation*}
\begin{split}M_i = m_i+m_q+x,\end{split}
\end{equation*}

where $x$ is distributed between $0$ and
$M-m_1-m_2-2m_q$ according to
\begin{equation*}
\begin{split}\frac{{\rm d}P}{{\rm d} x^2} \propto \exp \left( -bx\right),\end{split}
\end{equation*}

where
$b=2/{\mbox{\href{https://herwig.hepforge.org/doxygen/ClusterFissionerInterfaces.html\#SoftClusterFactor}{{\bf SoftClusterFactor}}}}$.

\subsection{Cluster decays}
\label{\detokenize{review/hadronization:cluster-decays}}\label{\detokenize{review/hadronization:sect-clusterdecay}}

The final step of the cluster hadronization model is the decay of the
cluster into a pair of hadrons. For a cluster of a given flavour
$(q_1,\bar{q}_2)$ a quark-antiquark or antidiquark-diquark pair
$(q,\bar{q})$ is extracted from the vacuum and a pair of hadrons
with flavours $(q_1,\bar{q})$ and $(q,\bar{q}_2)$ formed.
The hadrons are selected from all the possible hadrons with the
appropriate flavour based on the available phase-space, spin and flavour
of the hadrons. While the general approach is the same in all cluster
models, several variations are implemented in Herwig 7:
the original model of Ref.
\cite{Webber:1983if} used in FORTRAN
HERWIG \cite{Corcella:2000bw, Corcella:2002jc}; the approach of Ref.
\cite{Kupco:1998fx}, which was designed to solve the problem of isospin
violation in the original model if incomplete $\mathrm{SU}(2)$
multiplets of hadrons are included; and a new variant that addresses the
issue of the low rate of baryon production in the approach of Ref.
\cite{Kupco:1998fx}.

In all these approaches the weight for the production of the hadrons
$a_{(q_1,\bar{q})}$ and $b_{(q,\bar{q}_2)}$ is
\begin{equation*}
\begin{split}W(a_{(q_1,\bar{q})},b_{(q,\bar{q}_2)}) = P_qw_as_aw_bs_bp^*_{a,b},\end{split}
\end{equation*}

where $P_q$ is the weight for the production of the given
quark-antiquark or diquark-antidiquark pair, $w_{a,b}$ are the
weights for the production of individual hadrons and $s_{a,b}$ are
the suppression factors for the hadrons, which allow the production
rates of individual meson multiplets, and singlet and decuplet baryons
to be adjusted. The momentum of the hadrons in the rest frame of the
decaying cluster,
\begin{equation*}
\begin{split}p^*_{a,b} = \frac1{2M}\left[\left(M^2-(m_a+m_b)^2\right)
                            \left(M^2-(m_a-m_b)^2\right)\right]^{\frac12},\end{split}
\end{equation*}

measures the phase-space available for two-body decay. If the masses of
the decay products are greater than the mass of the cluster then the
momentum is set to zero. The weight for the individual hadron is
\begin{equation*}
\begin{split}w_h = w_{\rm mix}(2J_h+1),\end{split}
\end{equation*}

where $w_{\rm mix}$ is the weight for the mixing of the neutral
light mesons %
\begin{footnote}\sphinxAtStartFootnote
$w_{\rm mix}=1$ for all other particles.
\end{footnote} and $J_h$ is the spin of the hadron.

The different approaches vary in how they implement the selection of the
cluster decay products based on this probability.

In the approach of Ref. \cite{Webber:1983if} the probability is
generated in a number of pieces. First the flavour of the
quark-antiquark, or diquark-antidiquark, pair popped from the vacuum is
selected with probability
\begin{equation*}
\begin{split}P_q = \frac{{\bf Pwt}_q}{\sum_{q'}{\bf Pwt}_{q'}}.\end{split}
\end{equation*}

Each of the hadrons produced in the cluster decay is then selected from
the available hadrons of the appropriate flavours using the weight
\begin{equation*}
\begin{split}P_h = \frac{s_hw_h}{sw_{\max(q,\bar{q'})}},\end{split}
\end{equation*}

where $sw_{\max(q,\bar{q'})}$ is the maximum value of the
suppression factor times the weight for a given flavour combination.

A weight is calculated for this pair of hadrons
\begin{equation*}
\begin{split}W = \frac{p^*_{a,b}}{p^*_{\rm max}},\end{split}
\end{equation*}

where $p^*_{a,b}$ is the momentum of the hadrons in the cluster
rest frame and $p^*_{\rm max}$ is the maximum momenta of the decay
products for hadrons with the relevant flavour%
\begin{footnote}\sphinxAtStartFootnote
That is, the momentum with the lightest possible choices for
$a$ and $b$.
\end{footnote}.
The hadrons
produced are then retained according to this weight. If they are not
retained, the algorithm restarts from the choice of the flavour that is
popped out of the vacuum.

This procedure gives a probability of approximately
\begin{equation*}
\begin{split}P(a_{(q_1,\bar{q})},b_{(q,\bar{q}_2)}|q_1,\bar{q}_2) \propto P_q\frac1{N_{(q_1,\bar{q})}}\frac1{N_{(q,\bar{q}_2)}}
  \frac{s_aw_a}{sw_{\max(q_1,\bar{q})}}\frac{s_bw_b}{sw_{\max(q,\bar{q}_2)}}\frac{p^*_{a,b}}{p^*_{\rm max}}\end{split}
\end{equation*}

of choosing hadrons $a_{(q_1,\bar{q})}$ and
$b_{(q,\bar{q}_2)}$. The number of hadrons with flavour
$(q_1,\bar{q}_2)$ is $N_{(q_1,\bar{q}_2)}$.

Kup\v{c}o \cite{Kupco:1998fx} pointed out one problem with this approach:
as new hadrons with a given flavour are added, the production of the
existing hadrons with the same flavour is suppressed. In order to
rectify this problem he proposed a new approach for choosing the decay
products of the cluster. Instead of splitting the probability into
separate parts, as in Ref. \cite{Webber:1983if}, a single weight was
calculated for each combination of decay products
\begin{equation*}
\begin{split}W(a_{(q_1,\bar{q})},b_{(q,\bar{q}_2)}|q_1,\bar{q}_2) = P_qw_aw_bs_as_bp^*_{a,b},\end{split}
\end{equation*}

which gives the probability of selecting the combination
\begin{equation*}
\begin{split}P(a_{(q_1,\bar{q})},b_{(q,\bar{q}_2)}|q_1,\bar{q}_2) =
\frac{W(a_{(q_1,\bar{q})},b_{(q,\bar{q}_2)}|q_1,\bar{q}_2)}{\sum_{c,d,q'}W(c_{(q_1,\bar{q}')},d_{(q',\bar{q}_2)}|q_1,\bar{q}_2)}.\end{split}
\end{equation*}

The addition of new hadrons now increases the probability of choosing a
particular flavour, however because these new hadrons are usually heavy
they will not contribute for the majority of light clusters.

The main problem with this approach is that because many more mesons are
included in the simulation than baryons not enough baryons are produced.
In order to address this problem in Herwig 7, if a cluster mass is
sufficiently large that it can decay into the lightest baryon-antibaryon
pair the parameter ${\bf Pwt}_{qq}$ is used to decide whether to
select a mesonic or baryonic decay of the cluster. The probabilities of
selecting a mesonic decay or baryonic decay are
$\frac1{1+{\bf Pwt}_{qq}}$ and
$\frac{{\bf Pwt}_{qq}}{1+{\bf Pwt}_{qq}}$. This modification not
only increases the number of baryons produced but gives direct control
over the rate of baryon production.
Tetraquark states are not implemented in Herwig 7, so for the baryonic
cluster that results from a hadron remnant, only ‘mesonic’ decays
$q(qq)\to q\bar{q}+qqq$ are allowed.

Once the decay products of the cluster are selected, the cluster is
decayed. In general the cluster decay products are isotropically
distributed in the cluster rest frame. However, hadrons that contain a
parton produced in the perturbative stage of the event retain the
direction of the parton in the cluster rest frame, apart from a possible
Gaussian smearing of the direction. This is controlled by the \sphinxstylestrong{ClDir}
parameter, which by default {[}\sphinxstylestrong{ClDir=true{]}} retains the parton
direction, and the \sphinxstylestrong{ClSmr} parameter, which controls the Gaussian
smearing through an angle $\theta_{\rm smear}$ where
\begin{equation}\label{equation:review/hadronization:eqn:hadronsmear}
\begin{split}\cos\theta_{\rm smear} = 1+ {\bf ClSmr}\log\mathcal{R},\end{split}
\end{equation}

where $\mathcal{R}$ is another uniformly distributed random number
on the range $(0,1)$.
The azimuthal angle relative to the parton direction is distributed
uniformly. To provide greater control the parameters \sphinxstylestrong{ClDir} (\href{https://herwig.hepforge.org/doxygen/ClusterDecayerInterfaces.html\#ClDirLight}{ClDirLight}, \href{https://herwig.hepforge.org/doxygen/ClusterDecayerInterfaces.html\#ClDirCharm}{ClDirCharm} and \href{https://herwig.hepforge.org/doxygen/ClusterDecayerInterfaces.html\#ClDirBottom}{ClDirBottom})
and \sphinxstylestrong{ClSmr} (\href{https://herwig.hepforge.org/doxygen/ClusterDecayerInterfaces.html\#ClSmrLight}{ClSmrLight}, \href{https://herwig.hepforge.org/doxygen/ClusterDecayerInterfaces.html\#ClSmrCharm}{ClSmrCharm} and \href{https://herwig.hepforge.org/doxygen/ClusterDecayerInterfaces.html\#ClSmrBottom}{ClSmrBottom}) can be set independently for clusters containing
light, charm and bottom quarks.

In practice there is always a small fraction of clusters that are too
light to decay into two hadrons. These clusters are therefore decayed to
a single hadron, with the appropriate flavours, together with a
small reshuffling of energy and momentum with the neighbouring clusters
to allow the hadron to be given the correct physical mass. Available undecayed clusters, excluding those containing beam remnants, are considered as recoil partners in increasing order of $d_{ij}=\sqrt{|(x_i-x_j)^2|}$, where $x_i^\mu$ and $x_j^\mu$ are their assigned space-time vertices. For a cluster $i$ converted to a hadron of mass $m_h$, a candidate partner $j$ of mass $M_j$ is accepted if the combined invariant mass $M_{ij}=\sqrt{(p_i+p_j)^2}$ satisfies $M_{ij}>m_h+M_j$ and $M_j$ exceeds the lightest allowed two-hadron threshold for that partner. Otherwise, the partner can also be converted to its lightest flavour-compatible hadron, of mass $m_{h_j}$, provided $M_{ij}>m_h+m_{h_j}$. The first candidate satisfying either condition is used. The momenta are reconstructed using two-body kinematics, conserving the pair's total four-momentum and preserving the original directions in its centre-of-mass frame. In addition, for clusters containing a bottom or charm quark, in
order to improve the behaviour at the threshold, the option exists of
allowing clusters above the threshold mass for the production of two
hadrons, $M_{\rm threshold}$, to decay into a single hadron.
A single hadron can be formed for masses
\begin{equation}\label{equation:review/hadronization:eqn:singlehadron}
\begin{split}M<M_{\rm limit} = (1+{\bf SingleHadronLimit})M_{\rm threshold}.\end{split}
\end{equation}

The probability of such a single-meson cluster decay is assumed to
decrease linearly for $M_{\rm threshold}<M<M_{\rm limit}$. The
parameters \href{https://herwig.hepforge.org/doxygen/HadronSelectorInterfaces.html\#SingleHadronLimitCharm}{SingleHadronLimitCharm} and \href{https://herwig.hepforge.org/doxygen/HadronSelectorInterfaces.html\#SingleHadronLimitBottom}{SingleHadronLimitBottom}
control the limit on the production of single clusters
for charm and bottom clusters respectively. Increasing the limit has the
effect of hardening the momentum spectrum of the heavy mesons.
Until Herwig 7.1 the lightest hadron with the correct flavours was always chosen. However
there are cases, for example:
\begin{enumerate}
\sphinxsetlistlabels{\arabic}{enumi}{enumii}{}{.}%
\item {} 

$(u\bar{b})$ clusters where the $B^+$ and $B^{*+}$ mesons both have masses which are
below the threshold for the decay of the cluster to $B\pi$;

\item {} 

$(c\bar{c})$ clusters where there are many charmonium states with masses below the threshold for
the cluster decay to open charm;

\end{enumerate}

where there are many states below the threshold. If we only select the lightest state it can lead
to over-production of the lightest state, and under-production of the other states which lie below the threshold.
In the most extreme case for $(c\bar{c})$ clusters it means that the $J/\psi$ meson is never produced during
hadronization. In Herwig 7.1 we therefore introduced a new switch
\href{https://herwig.hepforge.org/doxygen/HadronSelectorInterfaces.html\#BelowThreshold}{BelowThreshold}
to control the treatment of clusters with masses below the two hadron threshold and by default
select from all the hadrons below the threshold according to their spin weights (\href{https://herwig.hepforge.org/doxygen/HadronSelectorInterfaces.html\#BelowThreshold}{BelowThreshold=All}) to pick the hadron used for single hadron decay.

\subsubsection{Kinematic strangeness production}
\label{\detokenize{review/hadronization:kinematic-strangeness-production}}

In Herwig strangeness can be produced at each non-perturbative step of
the hadronization stage (gluon splitting, cluster fissioning, and
cluster decay). Instead of having a flat number as a weight for the
probability to produce a strange and an antistrange quark, the option of
kinematic strangeness production uses the available kinematic
information of the relevant surrounding colour-singlet system in order
to calculate the probability to produce a strange quark-antiquark pair.
This procedure makes the production of strangeness more dynamic and
dependent on the event topology.

At each of the hadronization stages, the flat weights are replaced by the following functional form:
\begin{equation*}
\begin{split}w_s(M)^2 = \exp\left(-\frac{M_0^2}{M^2}\right),\end{split}
\end{equation*}
where $M_0^2$ is the characteristic mass scale for each stage and $M^2$ is the invariant mass
of the relevant colour singlet system on which the production of strangeness depends. An additional option
for the mass measure is the threshold production measure $\lambda$ which is defined by
\begin{equation*}
\begin{split}\lambda = M_{cs}^2 - \left( \sum_i M_i \right)^2.\end{split}
\end{equation*}
Here, $M_{cs}^2$ is the total invariant mass of the the colour singlet system and
$M_i$ are the invariant masses of the endpoints for pre-clusters or the constituent
partons in a normal cluster.

While at the cluster fissioning stage and at the cluster decay stage the relevant colour singlet system
is a normal cluster, we introduce the concept of a pre-cluster at the stage of non-perturbative gluon splitting.
Instead of immediately splitting the gluons into $q\bar{q}$ pairs we collect the various colour singlet systems
in the event and form so called \textit{pre-clusters}. A pictorial representation of such a pre-cluster is shown in \hyperref[\detokenize{review/hadronization:fig-coloursinglet}]{Fig.\@ \ref{\detokenize{review/hadronization:fig-coloursinglet}}}.

\begin{figure}[tp]
\centering
\capstart
\noindent\includegraphics[width=0.500\linewidth]{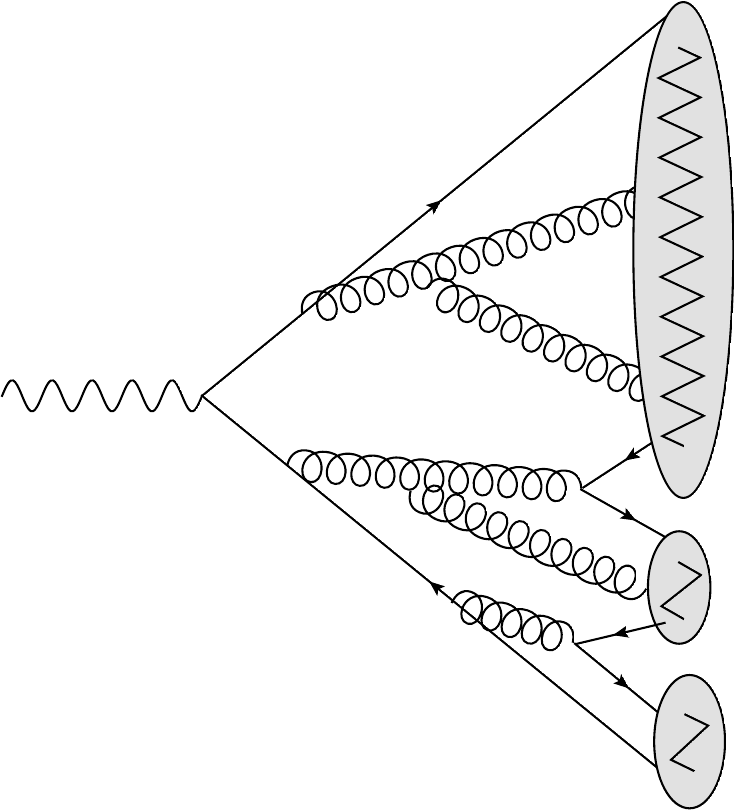}
\caption{Schematic figure of a colour singlet which can occur before the perturbative gluons
from the parton shower are split non-perturbatively into $q\bar{q}$ pairs.}\label{\detokenize{review/hadronization:id81}}\label{\detokenize{review/hadronization:fig-coloursinglet}}\end{figure}

In order to use the kinematic strangeness production one simply has to set the interface \texttt{{EnhanceSProb=Exponential}}
for the gluon splitting stage (\href{https://herwig.hepforge.org/doxygen/PartonSplitterInterfaces.html\#EnhanceSProb}{PartonSplitter}), the cluster fissioner stage (\href{https://herwig.hepforge.org/doxygen/ClusterFissionerInterfaces.html\#EnhanceSProb}{ClusterFissioner})
and finally the cluster decay stage which is done via the hadron selector (\href{https://herwig.hepforge.org/doxygen/HadronSelectorInterfaces.html\#EnhanceSProb}{HadronSelector}). Additionally one can choose
between two mass measures used for the exponential scaling (\texttt{{MassMeasure=Mass}} and \texttt{{MassMeasure=Lambda}}).

The option for kinematic strangeness production is explicitly recommended when
trying to simulate flavour sensitive observables. A detailed description of the
model can be found in \cite{Duncan:2018gfk}.

\subsubsection{Mixing weights}
\label{\detokenize{review/hadronization:mixing-weights}}

For neutral mesons that only contain the light (up, down and strange)
quarks there is mixing. If we consider the wavefunctions of the neutral
mesons, which we write for the $\phantom{.}^1\rm{S}_0$ meson
multiplet but the treatment applies to an arbitrary
$\mathrm{SU}(3)$ flavour multiplet, then
\begin{equation*}
\begin{split}\pi^0 &= \frac1{\sqrt{2}}\left(d\bar{d}-u\bar{u}\right), \\
\eta  &= \psi_8\cos\theta-\psi_1\sin\theta,\\
\eta' &= \psi_8\sin\theta+\psi_1\cos\theta,\end{split}
\end{equation*}

where $\theta$ is the nonet mixing angle and the wavefunctions for
the octet and singlet components are
\begin{equation*}
\begin{split}\psi_8 &= \frac1{\sqrt{6}}\left(u\bar{u}+d\bar{d}-2s\bar{s}\right),\\
\psi_1 &= \frac1{\sqrt{3}}\left(u\bar{u}+d\bar{d}+ s\bar{s}\right).\end{split}
\end{equation*}

The probabilities of finding a given quark-antiquark inside a particular
neutral meson can be calculated, which gives the mixing weights for the
neutral light mesons
\begin{equation*}
\begin{split}w^{\pi^0}_{u\bar{u}}  = w^{\pi^0}_{d\bar{d}} = \frac12,   w^{\pi^0}_{s\bar{s}}=0,\\
w^{\eta }_{u\bar{u}}  = w^{\eta }_{d\bar{d}} = \frac12\cos^2(\theta+\phi),
w^{\eta }_{s\bar{s}} = \sin^2(\theta+\phi),\\
w^{\eta'}_{u\bar{u}}  = w^{\eta'}_{d\bar{d}} = \frac12\sin^2(\theta+\phi),
w^{\eta'}_{s\bar{s}} = \cos^2(\theta+\phi),\end{split}
\end{equation*}
where $\phi=\tan^{-1}\sqrt{2}$ is the ideal mixing angle.

In the approach of Ref. \cite{Webber:1983if} the factor of
$\frac12$ in the weights for the $u\bar{u}$ and
$d\bar{d}$ components was omitted as this is approximately given
by the ratio of the number of charged mesons containing up and down
quarks to neutral ones, which is exactly two for ideal mixing where the
$s\bar{s}$ mesons do not mix with those containing up and down
quarks.

In practice the mixing angles can be adjusted for each meson multiplet
that is included in the simulation although with the exception of the
lightest pseudoscalar, vector, tensor and spin-3 multiplets the
assumption of ideal mixing is used.

\subsection{HQET and spin-hadronization}
\label{\detokenize{review/hadronization:hqet-and-spin-hadronization}}\label{\detokenize{review/hadronization:sect-spinhadronization}}

Heavy quark effective theory (HQET) is a framework that systematically describes the behaviour and interactions of heavy quarks within hadrons, i.e.\ when $m_Q \gg \Lambda_{\rm QCD}$, while simultaneously assuming that the heavy quark moves at a sub-relativistic velocity. HQET leverages this mass hierarchy and velocity separation to simplify the mathematical description of heavy quark interactions within hadrons. It assumes the velocity and spin of heavy quarks can be treated as independent variables, resulting in a decoupling of the heavy quark’s dynamics from the lighter degrees of freedom. This allows for a separation of scales, where interactions involving the heavy quark are analysed using a perturbative expansion, while interactions involving lighter degrees of freedom are treated non-perturbatively. This \textit{heavy quark spin-flavour symmetry} allows for model-independent predictions of certain properties of heavy hadrons, such as their spectra and decay rates \cite{Falk:1993rf, Falk:1992cx, Masouminia:2023zhb}.

A hadronic state can be considered a non-recoiling source of colour when it consists of a heavy quark ($Q$) and light degrees of freedom ($q$) \cite{Shuryak:1980pg, Caswell:1985ui, Eichten:1987xu, Lepage:1987gg, Isgur:1989vq}. In this regime, since $m_Q \gg m_q$ the heavy quark’s colour magnetic moment decouples from hadronic properties, and a heavy quark \textit{spin-flavour symmetry} emerges \cite{Falk:1993rf}. This symmetry can be used to produce model-independent predictions of heavy hadron spectra, weak matrix elements, and strong decay rates. An interesting consequence of this symmetry is the ability to distinguish between short- and long-distance interactions in a hadronization sequence, such as during a procedurally generated parton shower.

Generally, the strongly interacting particles produced during hard processes and subsequent parton showers in a simulated event have relatively large momenta. Therefore, the perturbative interactions during these stages occur over short time scales. Non-perturbative fragmentation processes that finalise these branches occur over longer time scales. Hadronization processes, regardless of model dependencies, are expected to occur at length scales of the order of $\mathcal{O}(\Lambda_{\rm QCD}^{-1})$. This means the resulting energy redistribution is much smaller than $m_Q$, making it plausible to assume that the velocity, mass, and spin of the heavy quark state (determined by short-distance physics) remain unchanged, indicating a decoupling between the dynamics of the heavy and light quark states \cite{Falk:1990cz, Mannel:1990up, Cohen:1992hu}.

This argument extends to the production of excited heavy mesons and heavy
baryons in a hadronization event, where it is possible to tag the produced
hadrons with the spin polarisation information of their heavy quark components \cite{Falk:1993rf}.
This comes with a couple of caveats:
\begin{enumerate}
\sphinxsetlistlabels{\roman}{enumi}{enumii}{(}{)}%
\item {} 

If the light-quark state’s phase-space allows its angular momentum to become of the order of $\mathcal{O}(m_Q^{-1})$, the process can redistribute this angular momentum, meaning the outgoing heavy quark polarisation will depend on the polarisation of the light degree of freedom created in the fragmentation process.

\item {} 

While conserving parity, fragmentation can produce anisotropic light degrees of freedom along its axis. The alignment of a light degree of freedom with spin $j$ can be characterized by a model-dependent dimensionless parameter, $\omega_j$ \cite{Falk:1993rf}.

\end{enumerate}

\subsubsection{Kinematics of fragmentation}
\label{\detokenize{review/hadronization:kinematics-of-fragmentation}}\label{\detokenize{review/hadronization:sect-fragmentationkin}}

To properly translate the implications of HQET and the spin-flavour symmetry, one must start by projecting the kinematics and time scales of heavy quark fragmentation. Here, we closely follow \cite{Falk:1993rf, Masouminia:2023zhb} by working in the rest frame of the heavy quark and identifying the preferred direction as the momentum of the parton shower progenitor. In this setup, the only interaction that can manipulate the spin of the heavy quark, i.e.~the colour magnetic moment of the light degrees of freedom, is sufficiently suppressed by a factor of $m_Q^{-1}$. Collectively, the hadron can possess a total spin of $j_{\pm} = j_q \pm 1/2$, with $j_q$ being the spin of the light state. We can therefore identify the $j_{\pm}$ configurations with a spin multiplet ($H$, $H^\star$), and assume a small mass splitting $\Delta m = m_{H^\star} - m_{H}$.

Due to the heavy quark constituent, the $H$ and $H^\star$ states are considered unstable. They can decay weakly to lighter hadrons, where it is safe to assume an identical decay rate:
\begin{equation*}
\begin{split}\Gamma(H \to X) = \Gamma(H^\star \to X).\end{split}
\end{equation*}

Alternatively, $H^\star$ can strongly or radiatively decay to $H$, with rate:
\begin{equation*}
\begin{split}\gamma(H^\star \to H X) \propto \Phi_{\rm phase-space}^{H^\star \to H X} \times |\mathcal{M}(H^\star \to H X)|^2 \sim \mathcal{O}(m_Q^{-(2+n)}),\end{split}
\end{equation*}

where $n \geq 1$, suggesting $\Delta m \gg \gamma$. Meanwhile, the relation between $\Gamma$, $\Delta m$, and $\gamma$ determines the nature of the fragmentation. Specifically, the case of $\Gamma \gg \Delta m \gg \gamma$ allows for rapid decay of heavy hadrons while decoupling the colour magnetic moments of the heavy and light degrees of freedom. This holds true for both strong or weak decays, even for $\Gamma \sim \Lambda_{\rm QCD}$, where the heavy quark can undergo partial hadronization before it decays.

The other possible cases, namely $\Delta m \gg \Gamma \gg \gamma$ and $\Delta m \gg \gamma \gg \Gamma$, result in depolarisation of the heavy quark from its initial orientation. The above argument suggests that under most conditions, the angular distribution of decay products provides no information on the polarisation of the heavy quark, unless the condition $\Gamma \gg \Delta m \gg \gamma$ is met. This is the so-called Falk--Peskin “no-win” theorem, which suggests that the polarisation of heavy quarks can remain intact only through strong, weak, or radiative decays of heavy excited mesons and heavy baryons \cite{Falk:1993rf}.

\subsubsection{Polarisation of excited heavy mesons}
\label{\detokenize{review/hadronization:polarisation-of-excited-heavy-mesons}}\label{\detokenize{review/hadronization:sect-mesonpol}}

To explore the implications of HQET and spin-flavour symmetry regarding the polarisations of heavy excited mesons, we use the charm sector as an example, particularly the observed excited charmed mesons $D_0^{\star}$, $D_1$, $D_1'$, and $D_2^{\star}$. Assuming the initial $c$ quark to be left-handed, a heavy charmed meson can be formed by combining with light degrees of freedom with $j_q = {1/2}$. The colour magnetic interaction is decoupled, leaving the spin orientation of the light degrees independent of the charm quark and distributed uniformly, i.e.~$j_q^{(3)} = \pm 1 / 2$ with equal probabilities; ${\left|\downarrow\,\rangle \right.}_c \; {\left|\downarrow\,\rangle \right.}_q$ and ${\left|\downarrow\,\rangle \right.}_c \; {\left| \uparrow\,\rangle \right.}_q$. This scenario becomes more intriguing when considering the $D_1$ and $D_2^{\star}$ meson states. With the assumed left-handed helicity of the charm quark, we now have light degrees of freedom with $j_q = 3/2$, which can manifest in any of the four possible helicity states. Parity invariance dictates that the probability of forming a specific helicity state cannot depend on the sign of its helicity, $j_3$, although probabilities may differ for states with distinct helicity magnitudes, $|j_q^{(3)}|$. Defining the parameter $\omega_j$ ($0 \leq \omega_j \leq 1$), which is the likelihood of fragmentation leading to a state with the maximum value of $|j_q^{(3)}|$ in a system with light degrees of freedom with spin $j_q$, one can derive the probabilities for the emergence of different helicity states of the light degrees as

\begin{savenotes}\sphinxattablestart
\sphinxthistablewithglobalstyle
\centering
\sphinxcapstartof{table}
\sphinxthecaptionisattop
\sphinxcaption{Probability distribution of the helicity of the light degrees of freedom in excited heavy mesons.}\label{\detokenize{review/hadronization:id82}}
\sphinxaftertopcaption
\begin{tabulary}{\linewidth}[t]{TTTTT}
\sphinxtoprule
\sphinxstyletheadfamily 

$j_q^{(3)}$
&\sphinxstyletheadfamily 

-3/2
&\sphinxstyletheadfamily 

-1/2
&\sphinxstyletheadfamily 

1/2
&\sphinxstyletheadfamily 

3/2
\\
\sphinxmidrule
\sphinxtableatstartofbodyhook

Probability
&

${1 \over 2} \omega_{3 \over 2}$
&

${1 \over 2} (1-\omega_{3 \over 2})$
&

${1 \over 2} (1-\omega_{3 \over 2})$
&

${1 \over 2} \omega_{3 \over 2}$
\\
\sphinxbottomrule
\end{tabulary}
\sphinxtableafterendhook\par
\sphinxattableend\end{savenotes}

The combination of the left-handed $c$ spin state with a specific light degrees of freedom helicity $j_q^{(3)}$ results in a coherent linear superposition of the charmed states with helicity $j = j_q + j_Q$. With this rationale one can calculate the probabilities of each helicity states getting populated as was outlined in Ref. \cite{Masouminia:2023zhb}.

\begin{savenotes}\sphinxattablestart
\sphinxthistablewithglobalstyle
\centering
\sphinxcapstartof{table}
\sphinxthecaptionisattop
\sphinxcaption{Probabilities for the population of the helicity states of \protect$D\protect$, \protect$D^{\star}\protect$, \protect$D_1\protect$, and \protect$D_2^{\star}\protect$. The probabilities sum to unity over both mesons and all helicities within each doublet, \protect$(D,D^{\star})\protect$ and \protect$(D_1,D_2^{\star})\protect$.}\label{\detokenize{review/hadronization:id83}}\label{\detokenize{review/hadronization:tab-helicity-probabilities-6-3}}
\sphinxaftertopcaption
\begin{tabulary}{\linewidth}[t]{TTTTTT}
\sphinxtoprule
\sphinxstyletheadfamily $j^{(3)}$ &\sphinxstyletheadfamily -2 &\sphinxstyletheadfamily -1 &\sphinxstyletheadfamily 0 &\sphinxstyletheadfamily +1 &\sphinxstyletheadfamily +2 \\
\sphinxmidrule
\sphinxtableatstartofbodyhook
$D$ & -- & -- & ${1 \over 4}$ & -- & -- \\
\hline
$D^{\star}$ & -- & ${1 \over 2}$ & ${1 \over 4}$ & 0 & -- \\
\hline
$D_1$ & -- & ${1 \over 8} (1-\omega_{3 \over 2})$ & ${1 \over 4} (1-\omega_{3 \over 2})$ & ${3 \over 8} \omega_{3 \over 2}$ & -- \\
\hline
$D_2^{\star}$ & ${1 \over 2} \omega_{3 \over 2}$ & ${3 \over 8} (1-\omega_{3 \over 2})$ & ${1 \over 4} (1-\omega_{3 \over 2})$ & ${1 \over 8} \omega_{3 \over 2}$ & 0 \\
\sphinxbottomrule
\end{tabulary}
\sphinxtableafterendhook\par
\sphinxattableend\end{savenotes}

To evaluate the parameter $\omega_j$, we consider the amplitude for the production of a pion at $\theta,\phi$ from a $H^{\star} \to H\pi$ type meson decay, which is proportional to the spherical harmonics $Y_{j}^{\ell}(\theta, \phi)$ ($\ell$ being the angular momentum quantum number of $H^{\star}$):
\begin{equation}\label{equation:review/hadronization:eq:Weight}
\begin{split}{d\Gamma(H^{\star} \to H\pi) \over d\cos\theta} \propto \int d\phi \sum_{j}  P_{H^{\star}}(j)  \bigl| Y_{j}^{\ell}(\theta,\phi) \bigr|^2,\end{split}
\end{equation}

where $P_{H^{\star}}(j)$ are the probabilities given in Table~\ref{\detokenize{review/hadronization:id83}}. Additionally, $\theta$ and $\phi$ are the angles of emission for the produced pion. Considering the case of $D_2^{\star} \to D\pi$, Eq. \eqref{equation:review/hadronization:eq:Weight} can be rewritten as \cite{Masouminia:2023zhb}:
\begin{equation*}
\begin{split}{1 \over \Gamma}{d\Gamma(D_2^{\star} \to D\pi) \over d\cos\theta} = {1 \over 4}
\left[ 1 + 3 \cos^2 \theta - 6 \; \omega_{3 \over 2} \left( \cos^2 \theta - {1 \over 3} \right)\right].\end{split}
\end{equation*}
To implement the above arguments in the context of HQET and spin-flavour symmetry principles in Herwig 7, we first generalise our calculations to an arbitrary heavy quark helicity, $\rho_Q$:
\begin{equation*}
\begin{split}\hat{\rho}_{D} &= 1, \\
\hat{\rho}_{D^{\star}} &= \mathrm{diag}\Big[\frac{1}{2}(1-\rho_Q),\; \frac{1}{2},\; \frac{1}{2}(1+\rho_Q)\Big], \\
\hat{\rho}_{D_1} &= \mathrm{diag}\Big[\frac{1}{16}[1-\rho_Q + \omega_{3 \over 2}(3-5\rho_Q)],\; \frac{1}{4}(1-\omega_{3 \over 2}),\; \frac{1}{16}[1+\rho_Q + \omega_{3 \over 2}(3+5\rho_Q)]\Big], \\
\hat{\rho}_{D_2^{\star}} &= \mathrm{diag}\Big[\frac{1}{4} \omega_{3 \over 2} (1-\rho_Q),\; \frac{3}{16}(1-\rho_Q) - \frac{1}{8}\omega_{3 \over 2}(1-\rho_Q),\; \frac{1}{4}(1-\omega_{3 \over 2}), \\ &
\frac{3}{16}(1+\rho_Q) - \frac{1}{8}\omega_{3 \over 2}(1+\rho_Q),\; \frac{1}{4} \omega_{3 \over 2} (1+\rho_Q)\Big].\end{split}
\end{equation*}

Here, $\hat{\rho}_H$ are the diagonal spin density matrices of the produced mesons $H$, with their components $\rho_{i,i}$, where $i=0,1,2,3,4$ runs from the most negative to the most positive valid helicity states. The diagonal structure of $\hat{\rho}$ arises due to the statistical nature of hadronization, where quantum coherence between helicity states is lost. During hadronization, the strong interaction operates at non-perturbative scales, leading to the randomisation of relative phases between spin states. As a result, only diagonal elements remain, representing independent populations of helicity states. Within HQET and spin-flavour symmetry, the diagonal form of $\hat{\rho}$ reflects parity conservation and the approximate decoupling of the heavy quark’s spin from the light degrees of freedom. Off-diagonal terms, which would correspond to spin interference effects, vanish due to averaging over unobserved spin correlations in the fragmentation process. Consequently, the second index in $\rho_{i,i}$ does not denote an intrinsic matrix structure but rather labels the helicity components, ordered from the most negative to the most positive eigenstates. To ensure numerical consistency in Herwig 7, the default parameter $\omega_{3 \over 2} = 0.20$ is introduced, governing the relative population of helicity states for mesons with light degrees of freedom in a $j_q = 3/2$ configuration.

In Herwig 7, the above implementation is achieved through the introduction of the \href{https://herwig.hepforge.org/doxygen/classHerwig_1_1SpinHadronizer.html}{SpinHadronizer} class. In particular, its \href{https://herwig.hepforge.org/doxygen/classHerwig_1_1SpinHadronizer.html#mesonSpin}{mesonSpin} function systematically handles the assignment of spin information and polarisation to generated mesons based on their spin characteristics and heavy constituent quark flavours. Starting with checks on the meson’s parentage and its constituents, \texttt{{mesonSpin}} employs spin information from the heavy quark to construct the spin properties of the meson, pertinent to the excited heavy mesons (in addition to their complex conjugates). The \texttt{{mesonSpin}} function also calculates the spin polarisation of the quark, updates the average polarisation for the specific flavour, and then assigns spin density matrix elements to the meson according to its spin value. It accommodates different spin types and meson categories, tailoring the spin behaviour and polarisation calculations to match the distinct characteristics of various excited heavy mesons.

\subsubsection{Polarisation of heavy baryons}
\label{\detokenize{review/hadronization:polarisation-of-heavy-baryons}}\label{\detokenize{review/hadronization:sect-baryonpol}}

The ground state of a heavy baryon is formed by combining a heavy quark with a diquark system having a helicity arrangement of $j_{qq} = 0$. In this setup, there is no angular momentum transferred to the heavy quark, preserving the initial polarisation without dilution. This means that the polarisation of the initial heavy quarks can affect the ground state of heavy baryons. The relative probabilities of finding these states during heavy sector fragmentation are governed by two parameters, $\omega_a$ and $\omega_j$. Here, $\omega_a$ represents the relative probability of producing a $j_{qq} = 1$ diquark compared to the ground state $j_{qq} = 0$ configuration. In the cluster hadronization model of Herwig 7, we can set $\omega_a = 1$, since there is no preference between spin-0 and spin-1 diquarks \cite{Masouminia:2023zhb}.

Similar to excited heavy mesons, we consider the fragmentation of a left-handed polarised $c$ quark. Then, the combination of the left-handed $c$ spin state with a specific light degrees of freedom helicity $j_q^{(3)}$ results in a coherent linear superposition of the charmed states with helicity $j = j_q + j_Q$. The probabilities of each helicity state being populated are shown in the table below \cite{Falk:1993rf}:

\begin{savenotes}\sphinxattablestart
\sphinxthistablewithglobalstyle
\centering
\sphinxcapstartof{table}
\sphinxthecaptionisattop
\sphinxcaption{Probabilities for the population of the possible helicity states of \protect$\Lambda_c\protect$, \protect$\Sigma_c\protect$, and \protect$\Sigma_c^\star\protect$.}\label{\detokenize{review/hadronization:id84}}
\sphinxaftertopcaption
\begin{tabulary}{\linewidth}[t]{TTTTT}
\sphinxtoprule
\sphinxstyletheadfamily 

$j^{(3)}$
&\sphinxstyletheadfamily 

-3/2
&\sphinxstyletheadfamily 

-1/2
&\sphinxstyletheadfamily 

1/2
&\sphinxstyletheadfamily 

3/2
\\
\sphinxmidrule
\sphinxtableatstartofbodyhook

$\Lambda_c$
&

$--$
&

${1 \over 1+\omega_a}$
&

0
&

--
\\
\hline

$\Sigma_c$
&

$--$
&

${(1-\omega_1) \omega_a \over 3 (1+\omega_a)}$
&

${ \omega_1 \omega_a \over 3 (1+\omega_a)}$
&

--
\\
\hline

$\Sigma_c^\star$
&

${\omega_1 \omega_a \over 2 (1+\omega_a)}$
&

${2 (1-\omega_1) \omega_a \over 3 (1+\omega_a)}$
&

${\omega_1 \omega_a \over 6 (1+\omega_a)}$
&

0
\\
\sphinxbottomrule
\end{tabulary}
\sphinxtableafterendhook\par
\sphinxattableend\end{savenotes}

The parameter $\omega_1$ can be estimated similarly to $\omega_{3 \over 2}$ by expanding and normalising the differential decay width equations for the observed decay modes $\Sigma_c \to \Lambda_c \pi$ and $\Sigma_c^{\star} \to \Lambda_c \pi$:
\begin{equation*}
\begin{split}{1 \over \Gamma}{d\Gamma(\Sigma_c \to \Lambda_c \pi) \over d\cos\theta} = {1 \over 2} \ ,\end{split}
\end{equation*}\begin{equation*}
\begin{split}{1 \over \Gamma}{d\Gamma(\Sigma_c^{\star} \to \Lambda_c \pi) \over d\cos\theta} = {1 \over 4}
\left[ 1 + 3 \cos^2 \theta - {9 \over 2} \omega_{1} \left( \cos^2 \theta - {1 \over 3} \right)\right] \ ,\end{split}
\end{equation*}

This results in $\omega_{1} = {2/3}$. The above arguments can be generalised for arbitrary heavy quark helicity:
\begin{equation*}
\begin{split}\rho_{\Lambda_c} &= \mathrm{diag}\Big[\frac{1}{2}(1-\rho_Q), \frac{1}{2}(1+\rho_Q)\Big], \\
\rho_{\Sigma_c} &= \mathrm{diag}\Big[\frac{1}{2}(1-\rho_Q) + \omega_{1} \rho_Q, \frac{1}{2}(1+\rho_Q) - \omega_{1} \rho_Q\Big], \\
\rho_{\Sigma_c^\star} &= \mathrm{diag}\Big[\frac{3}{8}\omega_{1}(1-\rho_Q), \frac{1}{2}(1-\rho_Q) - \frac{1}{8}\omega_{1}(3-5\rho_Q), \frac{1}{2}(1+\rho_Q) - \frac{1}{8}\omega_{1}(3+5\rho_Q),
\nonumber \\ &
\frac{3}{8}\omega_{1}(1+\rho_Q)\Big]\end{split}
\end{equation*}

The polarisation information of the heavy baryons is passed to the \texttt{{SpinHadronizer}} transitional class through the \href{https://herwig.hepforge.org/doxygen/classHerwig_1_1SpinHadronizer.html#baryonSpin}{baryonSpin} function. This function begins by checking the baryon’s parentage and its constituents. It then uses the spin information from the heavy quark to determine the spin characteristics of the baryon, particularly for excited heavy baryons.

\subsubsection{Spin-hadronization in Herwig 7}
\label{\detokenize{review/hadronization:spin-hadronization-in-herwig-7}}

Spin handling in Herwig 7 is managed by the \texttt{{SpinHadronizer}} class, which assigns spin information to hadronized particles. The \texttt{{mesonSpin}} and \texttt{{baryonSpin}} functions extract spin information from the heavy quark, update the average polarisation, and assign a diagonal spin density matrix to the produced heavy hadrons.

The \texttt{{SpinHadronizer}} class is inserted into the hadronization sequence via:

\begin{sphinxVerbatim}[commandchars=\\\{\}]
cd /Herwig/Hadronization
insert EventHandler:PostHadronizationHandlers 0 SpinHadronizer
\end{sphinxVerbatim}

ensuring that spin assignments occur after hadronization. The \textit{src/defaults/HerwigDefaults.in} file integrates spin handling into the event generation framework by defining:

\begin{sphinxVerbatim}[commandchars=\\\{\}]
cd /Herwig/Hadronization
newdef EventHandler:HadronizationHandler ClusterHadHandler\\
insert EventHandler:PostHadronizationHandlers 0 SpinHadronizer
\end{sphinxVerbatim}

To illustrate the impact of the spin-hadronization implementation, we present efficiency-corrected decay rates of the $D_{s1}^+$ meson, comparing different configurations of Herwig.

\begin{figure}[tp]
\centering
\capstart
\noindent\includegraphics[width=1.000\linewidth]{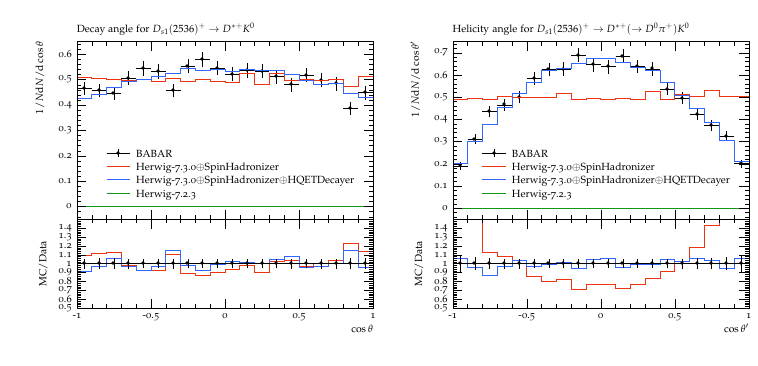}
\caption{Efficiency-corrected decay rates of the $D_{s1}^+$ meson. The left panel shows the $D_{s1} \to D^{\star +} K^0$ decay rate as a function of the lab-frame angle $\theta$, while the right panel presents the same distribution in the $D_{s1}^+$ centre-of-mass frame, denoted by $\theta'$. Data is sourced from \cite{BaBar:2011vbs}. The red histograms represent predictions from Herwig-7.3.0 with only \texttt{{SpinHadronizer}} enabled. The blue histograms include both \texttt{{SpinHadronizer}} and \texttt{{HQETDecayer}}, while the green histograms correspond to Herwig-7.2.3, which lacks HQET enhancements.}\label{\detokenize{review/hadronization:id85}}\end{figure}

\begin{figure}[tp]
\centering
\capstart
\noindent\includegraphics[width=1.000\linewidth]{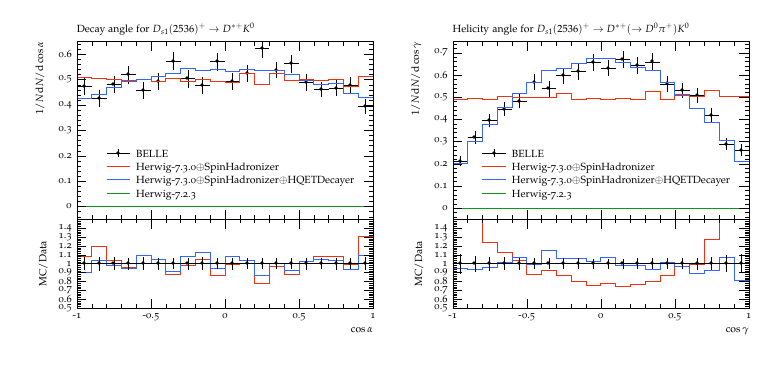}
\caption{Further efficiency-corrected decay rate analysis for the $D_{s1}^+$ meson. The left panel shows the decay rate as a function of the lab-frame angle $\alpha$ and the right panel displays the angular dependence of $\gamma$, the angle between $\pi^+$ and $K^0$ in the $D^{\star +}$ rest frame. Data is taken from \cite{Belle:2007kff}, and histogram annotations follow those in Figure~\ref{\detokenize{review/hadronization:id85}}.}\label{\detokenize{review/hadronization:id86}}\end{figure}

In the above figures, we examine how HQET and spin-flavour symmetry improve predictions of $e^+e^-$ data for polarisation-sensitive $D_{s1}$ meson decay measurements. The figures present normalised, efficiency-corrected rates for the $D_{s1} \to D^{\star +} K^0$ decay mode across different angular variables. The effect of \texttt{{SpinHadronizer}} and \texttt{{HQETDecayer}} in Herwig-7.3.0 is studied by enabling them separately to isolate their contributions. For comparison, we include predictions from Herwig-7.2.3, which does not incorporate HQET-based spin handling. As expected, Herwig-7.2.3 fails to predict either the existence or the angular dependence of the decay rates. In contrast, Herwig-7.3.0 correctly reproduces the mean values of the $s$-wave contributions when only \texttt{{SpinHadronizer}} is active, but does not fully capture the angle-dependent structure. Notably, with both \texttt{{SpinHadronizer}} and \texttt{{HQETDecayer}} enabled, Herwig-7.3.0 achieves excellent agreement with experimental data.

\subsection{Hadronization in BSM models}
\label{\detokenize{review/hadronization:hadronization-in-bsm-models}}\label{\detokenize{review/hadronization:sect-bsmhadronization}}

In most cases the hadronization of events involving new physics, using
the cluster model, proceeds in the same way as for Standard Model
events. There are however some classes of new physics model that require
special treatment, in particular:

\sphinxstylestrong{Stable strongly interacting particles},
if there are strongly interacting particles that are stable on the
hadronization timescale, these particles will hadronize before they
decay. If the new particles are in the fundamental representation of
colour $\mathrm{SU}(3)$ then their hadronization proceeds in the
same way as for quarks, however if they are in the octet representation
the situation is more complicated \cite{Kilian:2004uj} and is not yet
implemented.

\sphinxstylestrong{Baryon number violation (BNV)},
there are models of new physics in which the conservation of baryon
number is violated. This typically occurs at a vertex that has the
colour tensor $\epsilon^{ijk}$ leading to three quarks, or
antiquarks, that are colour-connected to each other after the parton
shower and gluon splitting.

The Herwig 7 hadronization module is designed so that both stable
coloured particles and baryon number violation are correctly treated as
described below.

\subsubsection{Stable strongly interacting particles}
\label{\detokenize{review/hadronization:stable-strongly-interacting-particles}}

Currently only the hadronization of objects in the fundamental
representation of the $\mathrm{SU}(3)$ group of the strong force
is supported. Provided that the relevant hadrons exist the hadronization
of these particles is handled in the same way as for quarks. In an upcoming development, hadron spectra can be specified for hypothetical new interactions or for theoretical investigations. This functionality is developed in the context of studying strongly interacting dark sectors \cite{Kulkarni:2024okx}.

\subsubsection{Baryon number violation}
\label{\detokenize{review/hadronization:baryon-number-violation}}\label{\detokenize{review/hadronization:sec-bnv}}

The treatment of QCD radiation and hadronization in models that violate
baryon number conservation is described in Refs. \cite{Dreiner:1999qz}
and \cite{Gibbs:1995bt} and was implemented in the
FORTRAN HERWIG program. In events where baryon number is violated there
are typically two situations that can arise.
\begin{enumerate}
\sphinxsetlistlabels{\arabic}{enumi}{enumii}{}{.}%
\item {} 

The baryon-number-violating vertices are unconnected, leading to
three quarks, or antiquarks, connected to each BNV vertex after the
gluon splitting. These (anti)quarks must be formed into a cluster,
which decays to give a (anti)baryon and a meson, giving the expected
baryon number violation. In the approach of Refs.
\cite{Dreiner:1999qz, Gibbs:1995bt} this is modelled by first
combining two of the (anti)quarks into a (anti)diquark, which is in
the (anti)-triplet representation of colour $\mathrm{SU}(3)$.
The (anti)quark and (anti)diquark can then be formed into a colour
singlet cluster, which can be handled by the hadronization module in
the normal way.

\item {} 

Two baryon number-violating vertices are colour-connected to each
other, leading to two quarks connected to one vertex and two
antiquarks connected to the second, after gluon splitting. In this
case two clusters must be formed by pairing one of the quarks with
one of the antiquarks at random and then pairing up the remaining
pair.

\end{enumerate}

The handling of these colour flows in both the shower and hadronization
is fully supported.

\subsection{Interface to string hadronization model}
\label{\detokenize{review/hadronization:interface-to-string-hadronization-model}}\label{\detokenize{review/hadronization:sect-stringhad}}

\begin{figure}[t]
\centering
\capstart
\noindent
\includegraphics[width=0.45\linewidth]{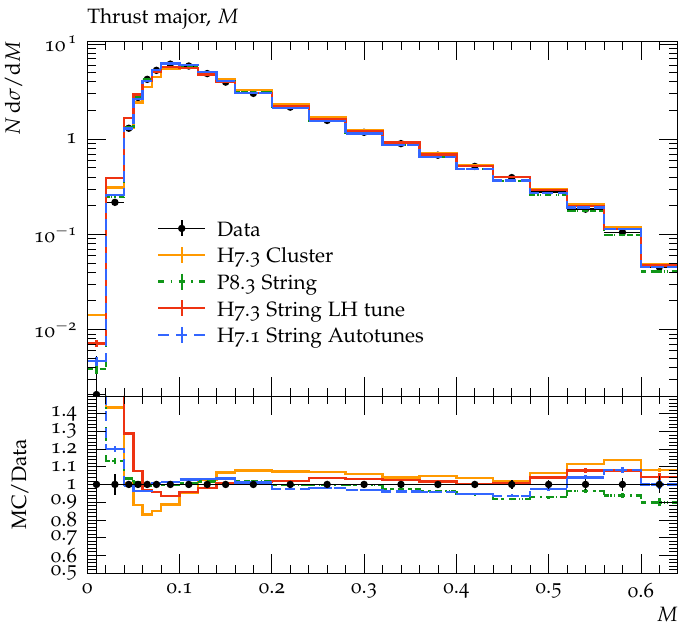}
\includegraphics[width=0.45\linewidth]{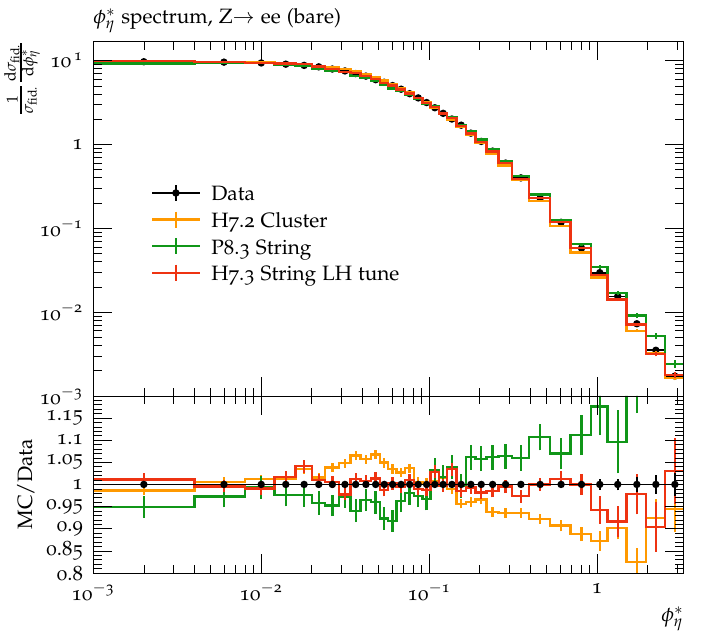}
\caption{Left panel: Thrust major observables measured by the DELPHI experiment at LEP~\cite{DELPHI:1996sen}.
Right panel: Normalized differential cross section of $Z$ boson production as a function of the $\phi^*_\eta$ parameter
for $Z \to ee$ decays~\cite{ATLAS:2012ewf} measured by ATLAS experiment.}\label{fig:LHTuneComparison}
\end{figure}

With version~7.3, Herwig includes the option to use the Lund string model
for hadronization in both $e^+e^-$ and $pp$ collisions.
The interface to the Pythia~8 string model~\cite{Bierlich:2022pfr}
is provided through the TheP8I C++ package~\cite{PIeight}.
While the default implementation of TheP8I is sufficient for hadronization in electron–positron collisions, it has been extended to include colour reconnection, which is required for a realistic description of hadron–hadron collisions.
The Lund string model, together with the angular-ordered parton shower and
colour reconnection model introduced in~\cite{Christiansen:2015yqa}
(the `QCD-based scheme', the only model currently validated through the interface),
has been tuned to LEP, Tevatron, and LHC data~\cite{Divisova:2025xqp}.
The obtained tune, referred to as the \textit{Les Houches} tune (LH Tune),
provides competitive predictions for both lepton--lepton and hadron--hadron collisions. 
It enables the study of uncertainties associated with non-perturbative hadronization effects
by fixing all other components of the simulation in Herwig and varying only the
hadronization model; see, for example, Fig.~\ref{fig:LHTuneComparison}, which shows the distribution of the Thrust
Major in $e^+e^-$ collisions measured by DELPHI experiment and the $\phi^*_\eta$ observable in
$pp$ collisions measured by ATLAS experiment, together with predictions obtained using different
hadronization models (the orange line corresponds to Herwig with the cluster model and the
red line to Herwig with the string model).

\subsection{Event-by-event hadronization corrections}
\label{\detokenize{review/hadronization:event-by-event-hadronization-corrections}}

A new strategy has been developed to transfer the assignment of
constituent masses entirely into the hadronization, which is available
as an option to the hadronization. Not only does this allow to have a
consistent physics interface to the string model, but it also provides
the possibility to extract event-by-event hadronization corrections in
a clean way, i.e.\ a parton level which is not ‘contaminated’ by the
non-perturbative constituent mass parameters. This is achieved by
reshuffling the partonic ensemble to different mass shells at the
beginning of the hadronization. While this can, in principle, be done
across the entire event, the more physical choice is to perform the
reshuffling within colour singlet subsystems which will branch into
clusters. The algorithm used is the same reshuffling algorithm as we
use it elsewhere, in particular it is the one previously used in the
dipole shower to assign new mass shells after showering (see \hyperref[\detokenize{review/showers/dipole:dipole-shower-constituent-reshuffling}]{Section \ref{\detokenize{review/showers/dipole:dipole-shower-constituent-reshuffling}}}). It can be
enabled using the \href{https://herwig.hepforge.org/doxygen/classHerwig_1_1ClusterHadronizationHandler.html#Reshuffle}{Reshuffle}
switch of the \href{https://herwig.hepforge.org/doxygen/classHerwig_1_1ClusterHadronizationHandler.html}{ClusterHadronizationHandler}
class, together with choosing the \href{https://herwig.hepforge.org/doxygen/classHerwig_1_1ClusterHadronizationHandler.html#ReshuffleMode}{ReshuffleMode}
option to equal \sphinxtitleref{ColourConnected}. At the same time, the
\texttt{ShowerHandler} needs to
be instructed to not put partons on their constituent mass shells
using the \href{https://herwig.hepforge.org/doxygen/classHerwig_1_1ShowerHandler.html\#UseConstituentMasses}{UseConstituentMasses}
switch similarly to when the string model is in use.  The intermediate
partonic state is then tagged by a status code which can be chosen
through the \href{https://herwig.hepforge.org/doxygen/classHerwig_1_1ShowerHandler.html\#TagIntermediates}{TagIntermediates}
interface of the \texttt{ShowerHandler} class.

\subsection{Code structure}
\label{\detokenize{review/hadronization:code-structure}}

The \texttt{ClusterHadronizationHandler}
inherits from the \href{https://thepeg.hepforge.org/doxygen/classThePEG\_1\_1HadronizationHandler.html}{HadronizationHandler} of
ThePEG and implements the cluster hadronization
model. The \texttt{ClusterHadronizationHandler}
makes use of a number of helper classes to implement different parts of the model. The helper classes,
in the order they are called, are:
\begin{itemize}
\item {} 

The \texttt{PartonSplitter}
performs the non-perturbative splitting of the gluons to
quark-antiquark pairs.

\item {} 

The \href{https://herwig.hepforge.org/doxygen/classHerwig\_1\_1ClusterFinder.html}{ClusterFinder}
is responsible for taking the partons after the gluon splitting and
forming them into colour singlet clusters as particles.

\item {} 

It is possible that rather than using the leading $N_c$ colour
structure of the event there is some rearrangement of the colour
connections and this is performed by \href{https://herwig.hepforge.org/doxygen/classHerwig\_1\_1ColourReconnector.html}{ColourReconnector}.

\item {} 

The \texttt{ClusterFissioner}
class is responsible for splitting large mass clusters into lighter
ones as described in \hyperref[\detokenize{review/hadronization:sect-clusterfission}]{Section \ref{\detokenize{review/hadronization:sect-clusterfission}}}.

\item {} 

The \href{https://herwig.hepforge.org/doxygen/classHerwig\_1\_1LightClusterDecayer.html}{LightClusterDecayer}
decays any clusters for which the decay to two hadrons is
kinematically impossible into a single hadron with the correct
flavour together with the reshuffling of momentum with neighbouring
clusters, which is required to conserve energy and momentum, as
described at the end of \hyperref[\detokenize{review/hadronization:sect-clusterdecay}]{Section \ref{\detokenize{review/hadronization:sect-clusterdecay}}}.

\item {} 

The \href{https://herwig.hepforge.org/doxygen/classHerwig\_1\_1ClusterDecayer.html}{ClusterDecayer}
decays the remaining clusters into pairs of hadrons as described in
\hyperref[\detokenize{review/hadronization:sect-clusterdecay}]{Section \ref{\detokenize{review/hadronization:sect-clusterdecay}}}.

\end{itemize}

In addition to these classes the \texttt{ClusterDecayer}
makes use of a \texttt{HadronSelector} to select
the hadrons produced in the cluster decay%
\begin{footnote}\sphinxAtStartFootnote
The \texttt{LightClusterDecayer}
also makes use of this class to select a single hadron, with a given flavour below the appropriate threshold, for
cluster decay to a single hadron.
\end{footnote}. In order to support the
different options described in \hyperref[\detokenize{review/hadronization:sect-clusterdecay}]{Section \ref{\detokenize{review/hadronization:sect-clusterdecay}}} the base
HadronSelector implements much of the functionality needed to select the
hadrons in the cluster model but the \texttt{{chooseHadronPair()}} method, which is
used to select the hadrons, is virtual and must be implemented in
inheriting classes that implement specific variants of the cluster
model. The FORTRAN HERWIG algorithm is implemented in the \href{https://herwig.hepforge.org/doxygen/classHerwig\_1\_1Hw64Selector.html}{Hw64Selector} class and the
Kup\v{c}o and Herwig++ methods in the \href{https://herwig.hepforge.org/doxygen/classHerwig\_1\_1HwppSelector.html}{HwppSelector} class. Newer versions will start to refer to \texttt{HadronSpectrum} objects, to be described in an update of this manual.

\begin{table}
\caption{Important hadronization parameters for the Herwig~7.3.0 setup described in this review. For all parameters other
than the light parton constituent masses, the limits given
are enforced by the interface. For the light partons, the
limits are not enforced but give a sensible range over which
the parameters can be varied. For the gluon, the upper limit
we give is about the largest value we would consider
reasonable, although it is not a hard limit. The up and down
masses must be less than half the gluon mass, otherwise the
non-perturbative gluon decays are impossible.  The same holds
for the strange quark mass as the gluon decay into strange
quarks has been found important to describe recent data on
strange particle production.\strut}\label{review/hadronization:table-clusterparam}
\centering
\begin{tabular}{llll}
\hline
Parameter & Default & Allowed & Description \\
\hline
& Value & Range & \\
\hline
\multicolumn{4}{l}{\texttt{ColourReconnector}} \\
\hline
\href{https://herwig.hepforge.org/doxygen/ColourReconnectorInterfaces.html\#ReconnectionProbability}{ReconnectionProbability} & 0.95 & 0-1 & Probability for colour reconnection \\
\hline
\href{https://herwig.hepforge.org/doxygen/ColourReconnectorInterfaces.html\#ReconnectionProbabilityBaryonic}{ReconnectionProbabilityBaryonic} & 0.70 & 0-1 & Probability for baryonic reconnection \\
\hline
\href{https://herwig.hepforge.org/doxygen/ColourReconnectorInterfaces.html\#AnnealingFactor}{AnnealingFactor} & 0.90 & 0-1 & Parameter for statistical reconnection \\
\hline
\href{https://herwig.hepforge.org/doxygen/ColourReconnectorInterfaces.html\#AnnealingSteps}{AnnealingSteps} & 50 & 1-10000 & Parameter for statistical reconnection \\
\hline
\href{https://herwig.hepforge.org/doxygen/ColourReconnectorInterfaces.html\#InitialTemperature}{InitialTemperature} & 0.10 & $10^{-5}$-100 & Parameter for statistical reconnection \\
\hline
\href{https://herwig.hepforge.org/doxygen/ColourReconnectorInterfaces.html\#TriesPerStepFactor}{TriesPerStepFactor} & 5.0 & 0-100 & Parameter for statistical reconnection \\
\hline
\multicolumn{4}{l}{\texttt{HadronSelector}} \\
\hline
\texttt{PwtDquark} & 1.0 & 0-10 & Weight for choosing a down quark \\
\hline
\texttt{PwtUquark} & 1.0 & 0-10 & Weight for choosing an up quark \\
\hline
\texttt{PwtSquark} & 0.37 & 0-10 & Weight for choosing a strange quark \\
\hline
\texttt{PwtDIquark} & 0.33 & 0-10 & Weight for choosing a diquark \\
\hline
\href{https://herwig.hepforge.org/doxygen/HadronSelectorInterfaces.html\#SngWt}{SngWt} & 0.89 & 0-10 & Weight for singlet baryons \\
\hline
\href{https://herwig.hepforge.org/doxygen/HadronSelectorInterfaces.html\#DecWt}{DecWt} & 0.42 & 0-10 & Weight for decuplet baryons \\
\hline
\texttt{SingleHadronLimitBottom} & 0.0 & 0-10 & Bottom cluster to 1 hadron param. \\
\hline
\texttt{SingleHadronLimitCharm} & 0.0 & 0-10 & Charm cluster to 1 hadron param. \\
\hline
\href{https://herwig.hepforge.org/doxygen/HadronSelectorInterfaces.html\#SingleHadronLimitCharm}{DecayMassScale} & 0.0 & 0.1-50 & Mass scale for kinematic strangeness \\
\hline
\multicolumn{4}{l}{\texttt{ClusterDecayer}} \\
\hline
\texttt{ClDirLight} & 1 & 0/1 & Orientation of light cluster decays \\
\hline
\texttt{ClDirBottom} & 1 & 0/1 & Orientation of bottom cluster decays \\
\hline
\texttt{ClDirCharm} & 1 & 0/1 & Orientation of charm clusters \\
\hline
\texttt{ClSmrLight} & 0.78 & 0--2 & Smearing of light cluster decays \\
\hline
\texttt{ClSmrBottom} & 0.08 & 0--2 & Smearing of bottom cluster decays \\
\hline
\texttt{ClSmrCharm} & 0.16 & 0--2 & Smearing of charm cluster decays \\
\hline
\href{https://herwig.hepforge.org/doxygen/ClusterDecayerInterfaces.html\#OnShell}{OnShell} & 0 & 0/1 & Masses of produced hadrons \\
\hline
\multicolumn{4}{l}{\texttt{ClusterFissioner}} \\
\hline
\texttt{ClMaxLight} & 3.53 & 0--10 & Max. mass for light clusters (GeV) \\
\hline
\texttt{ClMaxBottom} & 3.76 & 0--10 & Max. mass for bottom clusters (GeV) \\
\hline
\texttt{ClMaxCharm} & 3.95 & 0--10 & Max. mass for charm clusters (GeV) \\
\hline
\texttt{ClPowLight} & 1.85 & 0--10 & Mass exponent for light clusters \\
\hline
\texttt{ClPowBottom} & 0.55 & 0--10 & Mass exponent for bottom clusters \\
\hline
\texttt{ClPowCharm} & 2.56 & 0--10 & Mass exponent for charm clusters \\
\hline
\texttt{PSplitLight} & 0.91 & 0--10 & Splitting param. for light clusters \\
\hline
\texttt{PSplitBottom} & 0.63 & 0--10 & Splitting param. for bottom clusters \\
\hline
\texttt{PSplitCharm} & 0.99 & 0--10 & splitting param. for charm clusters \\
\hline
\texttt{RemnantOption} & 1 & 0/1 & Treatment of remnant clusters \\
\hline
\href{https://herwig.hepforge.org/doxygen/ClusterFissionerInterfaces.html\#SoftClusterFactor}{SoftClusterFactor} & 1 & 0.1--10 & Remnant mass param. (GeV) \\
\hline
\href{https://herwig.hepforge.org/doxygen/ClusterFissionerInterfaces.html\#ProbablityPowerFactor}{ProbablityPowerFactor} & 6.48 & 1--10 & power factor in cluster splitting \\
\hline
\href{https://herwig.hepforge.org/doxygen/ClusterFissionerInterfaces.html\#ProbablityShift}{ProbablityShift} & -0.88 & -10--10 & shift in cluster splitting threshold \\
\hline
\href{https://herwig.hepforge.org/doxygen/ClusterFissionerInterfaces.html\#KineticThresholdShift}{KineticThresholdShift} & 0.09 & -10--10 & shift in cluster splitting probability \\
\hline
\multicolumn{4}{l}{\href{https://thepeg.hepforge.org/doxygen/ConstituentParticleDataInterfaces.html}{Constituent Masses}} \\
\hline
gluon & 0.95 & 0--1 & Gluon constituent mass (GeV) \\
\hline
up & 0.325 & 0--$m_g/2$ & Up quark constituent mass (GeV) \\
\hline
down & 0.325 & 0--$m_g/2$ & Down quark constituent mass (GeV) \\
\hline
strange & 0.45 & 0--$m_g/2$ & Strange quark constituent mass (GeV) \\
\hline
\end{tabular}
\end{table}

There are a number of switches and parameters that control the
hadronization. Here we merely give a summary of the most important ones.
All the parameters are described in full in the Doxygen documentation of
the relevant classes.

The main choice is which variant of the cluster model to use. This can
be controlled by using either the \texttt{Hw64Selector}
for the original model of Ref. \cite{Webber:1983if} or the
\texttt{HwppSelector}
class for the Kup\v{c}o and Herwig++ variants.
The choice of whether to use the \texttt{Hw64Selector} or
\texttt{HwppSelector} is
controlled by setting the \texttt{HadronSelector}
interface of the \texttt{ClusterDecayer} and
\texttt{LightClusterDecayer} classes. In addition, for
the \texttt{HwppSelector} the
\texttt{Mode}
switch controls whether the Kup\v{c}o (\href{https://herwig.hepforge.org/doxygen/HwppSelectorInterfaces.html\#Mode}{Mode=Kupco}) or
Herwig++ (\href{https://herwig.hepforge.org/doxygen/HwppSelectorInterfaces.html\#Mode}{Mode=Hwpp}), the default,
variant is used. The production of specific hadrons by the cluster model
can be forbidden via the \href{https://herwig.hepforge.org/doxygen/HadronSelectorInterfaces.html\#Forbidden}{Forbidden}
interface of the \texttt{HadronSelector}: this option is currently
only used to forbid the production of the $\sigma$ and
$\kappa$ resonances, which are only included in the simulation to
model low-mass $s$-wave $\pi\pi$ and $K\pi$ systems in
certain particle decays.

In addition the mixing angles for the various multiplets can be changed
in the \texttt{HadronSelector} as can the suppression weights for different
$\mathrm{SU}(3)$ meson flavour multiplets.

If the option of using the soft underlying event model
\cite{Alner:1986is} is used, as described in \hyperref[\detokenize{review/ue:sect-ua5}]{Section \ref{\detokenize{review/ue:sect-ua5}}},
then the \href{https://herwig.hepforge.org/doxygen/ClusterHadronizationHandlerInterfaces.html\#UnderlyingEventHandler}{UnderlyingEventHandler}
interface needs to be set to the
\href{https://herwig.hepforge.org/doxygen/classHerwig\_1\_1UA5Handler.html}{UA5Handler}, by default this is set to the NULL
pointer and the multiple scattering model of the underlying event
described in \hyperref[\detokenize{review/index:sect-ue}]{Section \ref{\detokenize{review/index:sect-ue}}} used.

The other main parameters of the cluster model, and their default
values, are given in \hyperref[\detokenize{review/hadronization:table-clusterparam}]{Table \ref{\detokenize{review/hadronization:table-clusterparam}}}.

Finally the \href{https://thepeg.hepforge.org/doxygen/ConstituentParticleDataInterfaces.html\#ConstituentMass}{ConstituentMass}
of the gluon and, to a lesser extent the light quarks, which
can be set in their \texttt{ParticleData}
objects, have a major effect on the hadronization
since they set the scale for the cluster mass distribution.

\clearpage

\section{Underlying Event and beam remnants}
\label{\detokenize{review/index:underlying-event-and-beam-remnants}}\label{\detokenize{review/index:sect-ue}}

The underlying event model of Herwig is based on the eikonal model
discussed in \cite{Butterworth:1996zw} and described in detail in
\cite{Bahr:2008dy}. It models the underlying event activity as
additional semi-hard and soft partonic scatters. In doing so, it allows
the description of minimum bias events as well as the underlying event
in hard scattering processes.  In case of minimum bias interactions a
substantial fraction of events will be diffractive, these events are
modeled differently.  

\begin{figure}[htp]
\centering
\capstart
\noindent\includegraphics[width=0.900\linewidth]{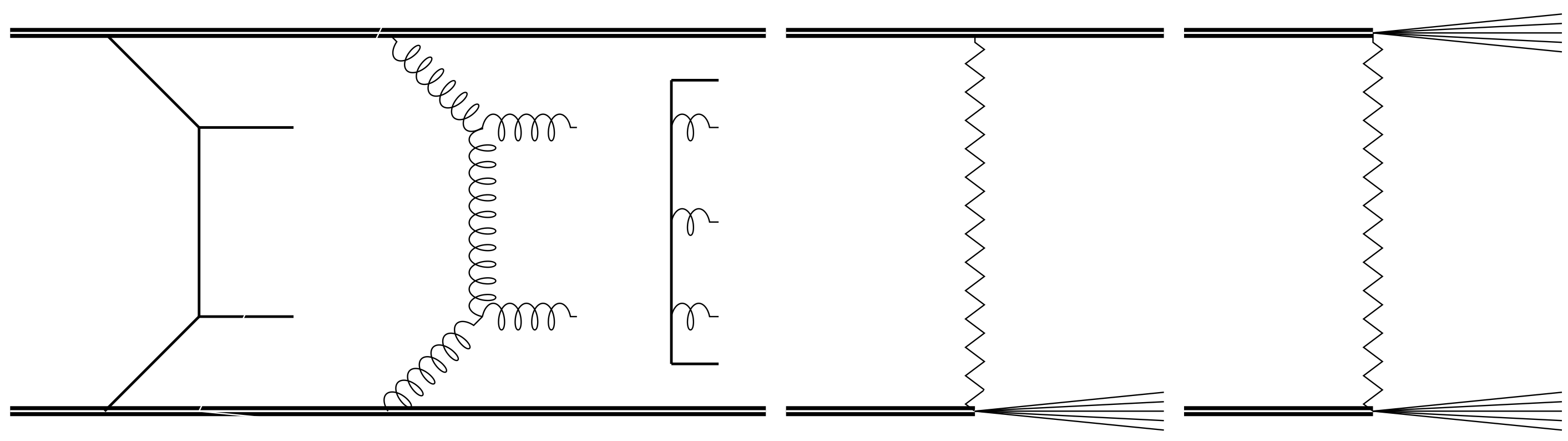}
\caption{Contributions to particle production in Herwig: dummy process to initiate a minimum bias event, semi-hard scattering, soft particle production from gluon ladder exchange (left part).  Single and double diffraction (right)}\label{\detokenize{review/ue:id26}}\label{\detokenize{review/ue:fig-softprocesses}}
\end{figure}

In Fig.~\ref{\detokenize{review/ue:fig-softprocesses}} we show the different contributions to particle production that will be described below.  In Minimum-bias events the event generation is initiated with a dummy process that merely breaks up the proton by extracting two partons with zero transverse momentum (leftmost).  This process will not contribute to particle production which will be performed by multiple semi-hard or soft scatters.  Semi-hard scatters are described by perturbative partonic two-to-two subprocesses down to a minimum transverse momentum.  Processes with very small transverse momentum are modeled as exchange of a gluon ladder.  All of these processes may contribute to minimum bias events or are used to model the underlying event in a hard scattering.  For minimum bias events, a fraction of events will be diffractive.  These are modeled separately but still contribute to particle production in events with low multiplicities.  

The perturbative part of the models provides very
similar functionality to FORTRAN HERWIG + JIMMY with some minor
improvements. The non-perturbative part has never been part of the
JIMMY implementation.  Our first implementation of soft interactions
\cite{Bahr:2009ek, Yennie:1961ad},
where these were simple models of two soft gluon production has been
superseded with a model where a chain of soft gluons is produced with
multiperipheral kinematics.

In this section, we briefly discuss the basics of how to calculate the
multiplicities of the semi-hard scatterings, before mentioning the
details of the soft additional scatters and explaining the integration
into the full Monte Carlo simulation. Finally we will describe the code
structure, which implements these ideas. A more detailed explanation of
the semi-hard model can be found in Ref. \cite{Bahr:2008dy}.  The soft
and diffractive model have been introduced in \cite{Gieseke:2016fpz},
where additional details can be found.

\subsection{Semi-hard partonic scatters}
\label{\detokenize{review/ue:semi-hard-partonic-scatters}}\label{\detokenize{review/ue::doc}}

The starting point is the observation that the cross section for QCD jet
production may exceed the total $pp$ or $p\bar p$ cross
section already at an intermediate energy range and eventually violates
unitarity. For example, for QCD jet production with a minimum of 2 GeV
this already happens at $\sqrt{s} \sim 1$ TeV. This cutoff should
however be large enough to ensure that we can calculate the cross
section using pQCD. The reason for the rapid increase of the cross
section turns out to be the strong rise of the proton structure function
at small $x$, since the $x$ values probed decrease with
increasing centre of mass energy. This proliferation of low $x$
partons may lead to a non-negligible probability of having more than one
partonic scattering in the same hadronic collision. This is not in
contradiction with the definition of the standard parton distribution
function as the \textit{inclusive} distribution of a parton in a hadron, with
all other partonic interactions summed and integrated out. It does,
however, signal the onset of a regime in which the simple interpretation
of the pQCD calculation as describing the only partonic scattering must
be unitarized by additional scatters.

In   principle, predicting the rate of multi-parton scattering processes
requires multi-parton distribution functions, about which we have almost
no experimental information. However, the fact that the standard parton
distribution functions describe the inclusive distribution gives a
powerful constraint, which we can use to construct a simple model. The
eikonal model used in Refs.  \cite{Durand:1987yv, Durand:1988ax, Durand:1988cr}
derives from the assumption that at fixed impact
parameter, $b$, individual scatterings are independent and that
the distribution of partons in hadrons factorizes with respect to the
$b$ and $x$ dependence. This implies that the average number
of partonic collisions at a given $b$ value is
\begin{equation}\label{equation:review/ue:eqn:avgN}
\begin{split}  \avgN = A(b) \ \sigmahard(s;\ptmin)\, ,\end{split}
\end{equation}

where $A(b)$ is the partonic overlap function of the colliding
hadrons, with
\begin{equation*}
\begin{split}\int \db A(b) = 1 \, ,\end{split}
\end{equation*}

and $\sigmahard(s;\ptmin)$ is the inclusive cross section to
produce a pair of partons with $\pt > \pt^{\rm min}$. We model the
impact parameter dependence of partons in a hadron by the
electromagnetic form factor, resulting in an overlap function for
$pp$ and $p\bar{p}$ collisions of
\begin{equation}\label{equation:review/ue:eqn:overlap}
\begin{split}    A(b;\mu) = \frac{\mu^2}{96 \pi} (\mu b)^3 K_3(\mu b) \, ,\end{split}
\end{equation}

where $\mu$ is the inverse proton radius and $K_3(x)$ is the
modified Bessel function of the third kind. We do not fix $\mu$ at
the value determined from elastic $ep$ scattering, but rather
treat it as a free parameter, because the spatial parton distribution is
assumed to be similar to the distribution of charge, but not necessarily
identical.

The assumption that different scatters are uncorrelated leads to the
Poissonian distribution for the number of scatters, $n$, at fixed
impact parameter,
\begin{equation}\label{equation:review/ue:eqn:probN}
\begin{split}  \mathcal{P}_n(b,s) = \frac{\avgN^n}{n!} \ \myexp{ -\avgN } \, .\end{split}
\end{equation}

Using \eqref{equation:review/ue:eqn:probN} the unitarized cross section can now be
written as
\begin{equation}\label{equation:review/ue:eq:sigma_inel}
\begin{split}  \sigma_{\rm inel}(s) = \int \db \sum_{k=1}^{\infty} \mathcal{P}_k(b,s)
  = \int \db \left[ 1 - \myexp{- \avgN} \right] \, ,\end{split}
\end{equation}

which properly takes multiple scatterings into account. The key
ingredient for the Monte Carlo implementation is then the probability of
having $n$ scatterings given there is at least one, integrated
over impact parameter space. This expression reads
\begin{equation}\label{equation:review/ue:eq:prob}
\begin{split}  \mathrm{P}_{n}(s) = \frac{\int \db \mathcal{P}_n(b,s)}{\int \db \sum_{k=1}^{\infty}
  \mathcal{P}_k(b,s)} \, .\end{split}
\end{equation}

It is worth noting that this distribution, after integration over
$b$, is much broader than Poissonian and has a long tail to high
multiplicities.

Equation  \eqref{equation:review/ue:eq:prob} is used as the basis of the multi-parton
scattering generator for events in which the hard process is identical
to the one used in the underlying event, \textit{i.e.}  QCD $2\to2$
scattering. For scatterings of more than one type of hard process, the
formulae can be easily generalised, but in fact for the realistic case
in which all other cross sections are small compared to the jet cross
section, they saturate at a simple form,
\begin{equation}\label{equation:review/ue:eq:prob1}
\begin{split}  \mathrm{P}_{n}(s) = \frac{n}{\sigmahard} \int \db \mathcal{P}_n(b,s) \, ,\end{split}
\end{equation}

which allows for a more efficient generation of additional scatterings.
It is worth noting that the fact that we have ‘triggered on’ a process
with a small cross section leads to a bias in the $b$ distribution
and hence a higher multiplicity of additional scatters than in the pure
QCD $2\to2$ scattering case. A slight further modification to the
distribution is needed when the small cross section process is a subset
of the large one, for example QCD $2\to2$ scattering restricted to
the high $\pt$ region.

As described so far, the $n$ scatters are completely independent,
which is expected to be a good approximation in the region in which
multiple scattering dominates, \textit{i.e.} small momentum fractions.
However, some fraction of events come from higher $x$ values and
must lead to correlations between the scatters at some level. At the
very least, the total momentum and flavour must be conserved: the total
$x$ value of all partons extracted from a hadron cannot exceed
unity and each valence parton can only be extracted once. In Herwig
these correlations are included in the simplest possible way, by vetoing
any scatters that would take the total extracted energy above unity and
by only evolving the first scatter back to a valence parton and all the
others back to a gluon.

It has to be noted that in addition to these correlations of events from
momentum constraints there will be correlations in colour space as well.
At this point the assignment of colour lines to each scatter is
completely ad-hoc.  Cross-talk between different jets will be introduced
by applying colour reconnections that colour connect nearby partons from
different scatters and thereby lower the overall multiplicity and harden
the transverse momenta of individual partons.

\subsection{Soft partonic scatters}
\label{\detokenize{review/ue:soft-partonic-scatters}}\label{\detokenize{review/ue:sec-softpartonic}}

The elastic scattering amplitude, $a(\vect{b},s)$, in impact
parameter space can be expressed in terms of a real eikonal function
$\eik{}$, as
\begin{equation*}
\begin{split}  a(\vect{b},s) = \frac{1}{2 i} \left[ \myexp{-\eik{}} - 1 \right] \, .\end{split}
\end{equation*}
The elastic scattering amplitude, $\mathcal{A}(s,t)$, is the
Fourier transform of $a(\vect b, s)$ and therefore the total
$pp$ ($p\bar p$) cross section as well as the elastic cross
section can be obtained from that parameterisation as,
\begin{equation*}
\begin{split}    \stot(s) = 2 \int \db\ \left[ 1 - \myexp{-\eik{}} \right]  \, , \quad\quad
    \sela(s) = \int \db\ \abs{ 1 - \myexp{-\eik{}} }^2 \, .\end{split}
\end{equation*}
The inelastic cross section is obtained as the difference between the
two cross sections,
\begin{equation*}
\begin{split}  \begin{split}
    \sigma_{\rm inel} = \ & \stot - \sigma_{\rm el} \\
    = \ & \int \db \ \left[ 1 - \myexp{-2\eik{}} \right] \, .
  \end{split}\end{split}
\end{equation*}
The elastic $t$-slope parameter at zero momentum transfer is also
calculable within this framework and yields \cite{Block:1984ru}
\begin{equation}\label{equation:review/ue:eq:slope_eik}
\begin{split}  \slope = \frac{1}{\stot} \int \db b^2 \ \left[ 1 - \myexp{-\eik{}}
  \right] \, .\end{split}
\end{equation}

To reproduce the results from \eqref{equation:review/ue:eq:sigma_inel}, we choose
\begin{equation*}
\begin{split}  \eik{} = \br{2} \avgN \, .\end{split}
\end{equation*}
However we want to introduce additional scatters below the transverse
momentum cut-off. Therefore, we identify this as the \textit{hard} part of a
universal eikonal function, which then has the form,
\begin{equation}\label{equation:review/ue:eq:fulleik}
\begin{split}  \chi_{\rm tot}(\vect b,s) = \chi_{\rm QCD}(\vect b,s) + \chi_{\rm soft}(\vect b,s) \, ,\end{split}
\end{equation}

with the perturbative part
\begin{equation*}
\begin{split}  \chi_{\rm QCD}(\vect{b},s) = \frac{1}{2} A(\vect{b}; \mu) \
  \sigmahard(s;\ptmin) \, ,\end{split}
\end{equation*}
as in \eqref{equation:review/ue:eqn:avgN}.

In the models of Refs. \cite{Durand:1988ax, Durand:1988cr, Borozan:2002fk},
the soft eikonal function has the form
\begin{equation*}
\begin{split}  \chi_{\rm soft}(\vect{b},s) = \frac{1}{2} A_{\rm soft}(\vect{b};
  \mu_{\rm soft}) \ \sigmasoft \, ,\end{split}
\end{equation*}
where $\sigmasoft$ is the purely non-perturbative cross section
below $\ptmin$, which is a free parameter of the model. That is,
we assume that soft scatters are the result of partonic interactions
that are local in impact parameter. Previous Monte Carlo implementations
used the simplest assumption about the partonic overlap function probed
by the soft scatters, $A_{\rm soft}(\vect b) \equiv A(\vect b)$,
i.e.\ an identical distribution to the one probed by semi-hard
scatters. In Ref. \cite{Bahr:2008wk} it was shown that measurements on
the elastic $t$-slope confine the allowed parameter space of such
models vastly. The remaining parameter space seems to be in
contradiction with constraints obtained from measurements of the
effective cross section in double parton scattering events
\cite{Abe:1997bp, Abe:1997xk}. We therefore introduced the option of
relaxing the condition of identical overlap distributions in
Herwig. This enables the dynamical determination of the soft overlap
distribution, $A_{\rm soft}(\vect b)$. In this case, which is the
default setting, we use \eqref{equation:review/ue:eqn:overlap} but allow an independent
radius parameter for the soft overlap function. The parameter
$\mu_{\rm soft}$ is then dynamically fixed by the requirement of a
correct description of the elastic $t$-slope from
\eqref{equation:review/ue:eq:slope_eik} at the current centre-of-mass energy. At the same
time we fix the second free parameter in the soft sector,
$\sigmasoft$, by choosing it such that the total cross section,
evaluated with the parametrisation from Ref.  \cite{Donnachie:1992ny}
is correctly described.  At this point we also have to take into account
the possibility that the inelastic cross section may contain diffractive
scatters.
Therefore, we fit the parameters  $\sigmasoft$ and  $\mu_{\rm soft}$
such that the inelastic cross section will give a fraction  $1 - x_{\rm NSD}$
of the experimental value, see Section 7.4.

We will discuss this point below.  Measurements of the total
cross section may deviate from the prediction in Ref.
\cite{Donnachie:1992ny} and therefore the parameter can be used to set
the total cross section at the current centre-of-mass energy explicitly.

With the full eikonal from \eqref{equation:review/ue:eq:fulleik}, we can construct our model for
additional semi-hard and soft scatters, by imposing the additional
assumptions,
\begin{itemize}
\item {} 

The probability distributions of semi-hard and soft scatters are
independent

\item {} 

Soft scatters are uncorrelated and therefore obey Poissonian
statistics like the semi-hard scatters

\end{itemize}

The probability $\Pcal_{h,n}(\vect b,s)$, for having exactly
$h$ semi-hard and $n$ soft scatters at impact parameter
$\vect b$ and centre-of-mass energy $s$ is then given by,
\begin{equation}\label{equation:review/ue:eq:prob_b_final}
\begin{split}  \Pcal_{h,n}(\vect b,s) = \frac{(2\eik{QCD})^h}{h!} \
  \frac{(2\eik{soft})^n}{n!} \myexp{-2\eik{total}} \, .\end{split}
\end{equation}

From \eqref{equation:review/ue:eq:prob_b_final} we can now deduce the cross section for
having exactly $h$ semi-hard and $n$ soft scatters as,
\begin{equation}\label{equation:review/ue:eq:sigma_final}
\begin{split}  \sigma_{h,n}(s) = \int \db \Pcal_{h,n}(\vect b, s) \, .\end{split}
\end{equation}

The cross section for an inelastic collision (either semi-hard or soft),
is obtained by summing over the appropriate multiplicities and yields
\begin{equation}\label{equation:review/ue:eq:inel_final}
\begin{split}\begin{aligned}
  \sinel(s) &= \int \db \sum_{h+n \geq 1} \Pcal_{h,n}(\vect b, s) \nonumber\\
  &= \int \db \left[ 1 - \myexp{-\eik{total}} \right] \, .\end{aligned}\end{split}
\end{equation}

The inelastic cross section for at least one semi-hard scattering is
\begin{equation*}
\begin{split}\begin{aligned}
  \sigma_{\rm inel}^{\rm semi-hard}(s) &= \int \db \sum_{h\geq1, n\geq
  0}\Pcal_{h,n}(\vect b, s) \nonumber\\
  &= \int \db \left[ 1 - \myexp{-\eik{QCD}}
  \right] \, .\end{aligned}\end{split}
\end{equation*}

With the cross sections from \eqref{equation:review/ue:eq:sigma_final} and
\eqref{equation:review/ue:eq:inel_final} we can construct the basis of our multiple soft and
semi-hard scattering model, the probability, $P_{h,n}$, of having
exactly $h$ semi-hard and $n$ soft scatters in an inelastic
event ($h+n\geq1$). It is given by
\begin{equation}\label{equation:review/ue:eq:prob_final}
\begin{split}  \mathrm{P}_{h,n}(s) = \frac{\sigma_{h,n}(s)}{\sinel(s)} = \frac{\int \db
  \Pcal_{h,n}(\vect b, s)}{\int \db \left[ 1 - \myexp{-\eik{total}}
  \right]} \, , \quad h+n\geq1 \, ,\end{split}
\end{equation}

which is analogous to \eqref{equation:review/ue:eq:prob} for the case of solely semi-hard
additional scatterings. Equation  defines a matrix of probabilities for
individual multiplicities. This matrix is evaluated at the beginning of
each run and the corresponding multiplicities are drawn for each event
from this matrix according to their probability.

Equation \eqref{equation:review/ue:eq:prob_final} leads to very inefficient generation of
additional scatters in cases where a rare hard scattering, with cross
section $\sigma_{\rm rare}$, takes place. Equation \eqref{equation:review/ue:eq:prob1}
has been deduced for this case, by exploiting the independence of
different scatters. The presence of soft scatters does not alter that
result as our assumption is that the soft scatters are independent from
each other and from the other scatterings. Hence, the probability for
$h$ hard scatters (from which one is distinct, i.e.~$h=m+1$)
and $n$ soft scatters is given by
\begin{equation}\label{equation:review/ue:eq:prob1_final}
\begin{split}\begin{aligned}
  \mathrm{P}_{h=m+1,n}(s) &= \frac{\int \db \Pcal_{m,n}(\vect b, s) \
  \frac{(A(b)\sigma_{\rm rare})^1}{1!}  \myexp{-A(b)\sigma_{\rm
  rare}}}{\int \db A(b)\sigma_{\rm rare}}
   \\ &\approx \int \db \Pcal_{m,n}(\vect b, s) A(b)
   \\ &= \frac{h}{\sigmahard} \ \int \db
  \Pcal_{h,n}(\vect b, s) \, .  \end{aligned}\end{split}
\end{equation}

The probability for $m$ semi-hard ($\pt \geq \ptmin$) and
$n$ soft additional scatters is multiplied with the probability
for exactly one scattering with an inclusive cross section of
$\sigma_{\rm rare}$. The denominator is the inclusive cross
section for this distinct scattering, i.e.\ summed over all
multiplicities for additional semi-hard and soft scatters. By
approximating the exponential with unity and exploiting the
normalisation of $A(b)$ ($\int \db A(b) = 1$), we finally
deduce \eqref{equation:review/ue:eq:prob1_final}.

Once the number of semi-hard scatters is known, they are simulated like
“ordinary” hard scatters, the only exception is that we use a fast
implementation of the QCD $2 \to 2$ matrix elements as they are
used in every hadronic collision.  After the semi-hard scatters are
sampled, the parton showering and hadronization are applied as described
above.  More details can be found in \cite{Bahr:2008dy}.

\subsubsection{Transverse momentum of soft scatters}
\label{\detokenize{review/ue:transverse-momentum-of-soft-scatters}}

The transverse momentum of soft scatters is connected to the soft
inelastic cross section described above.  We take this connection quite
literally for historical reasons of our implementation.  We do not
necessarily have to make this connection now as in the end we do not
directly model the transverse momenta of soft hadrons.  They only come
as a result of the full soft model described below and are secondary
particles only after cluster decay.  We begin the discussion with the
construction that the transverse momentum distribution must not exceed
the lower bound of hard transverse momenta, $\ptmin$.  The
functional form of it is not predetermined but a Gaussian distribution
seems well-motivated by the fact that hadrons in soft hadronic
interactions roughly follow a Gaussian in transverse momentum as,
\begin{equation*}
\begin{split}\mydiff{\sigmasoft}{\pt^2} = A \myexp{-\beta \, \pt^2} \, ,\end{split}
\end{equation*}

to parameterise it. To fix the free parameters $A$ and
$\beta$, we impose the following constraints:
\begin{itemize}
\item {} 

The resulting soft cross section has to match the total soft cross
section, which has been fixed to describe the total inelastic cross
section and the slope parameter,
\begin{equation*}
\begin{split}\int \dpt \mydiff{\sigmasoft}{\pt^2} \ \stackrel{!}{=} \
\sigmasoft.\end{split}
\end{equation*}
\item {} 

The transverse momentum distribution of semi-hard and soft
scatterings should be continuous at the matching scale,
\begin{equation*}
\begin{split}\begin{aligned}
      H(s;\ptmin) := \left. \mydiff{\sigmahard}{\pt^2} \right|_{\pt = \ptmin} \
      \stackrel{!}{=} \ & \left. \mydiff{\sigmasoft}{\pt^2} \right|_{\pt =
        \ptmin} \, ,
    \end{aligned}\end{split}
\end{equation*}

where we introduced $H$ as shorthand for the hard inclusive jet
cross section at $\pt = \ptmin$.

\end{itemize}

These conditions are fulfilled by the parameterisation,
\begin{equation}\label{equation:review/ue:eq:transverse}
\begin{split}  \mydiff{\sigmasoft}{\pt^2} = H(s;\ptmin) \myexp{-\beta (\pt^2 - {\ptmin}^2)}
  \, ,\end{split}
\end{equation}

where the slope, $\beta$, must satisfy,
\begin{equation*}
\begin{split}  \frac{\myexp{\,\beta {\ptmin}^2} -1}{\beta} = \frac{\sigmasoft}{H(s;\ptmin)}
  \, .\end{split}
\end{equation*}

\hyperref[\detokenize{review/ue:fig-pt-spectrum}]{Fig.\@ \ref{\detokenize{review/ue:fig-pt-spectrum}}} shows the transverse momentum
spectrum for two different cut-off values. The slope, $\beta$, is
chosen such that both curves correspond to the same integrated cross
section.

\begin{figure}[htp]
\centering
\capstart

\noindent\includegraphics[width=0.600\linewidth]{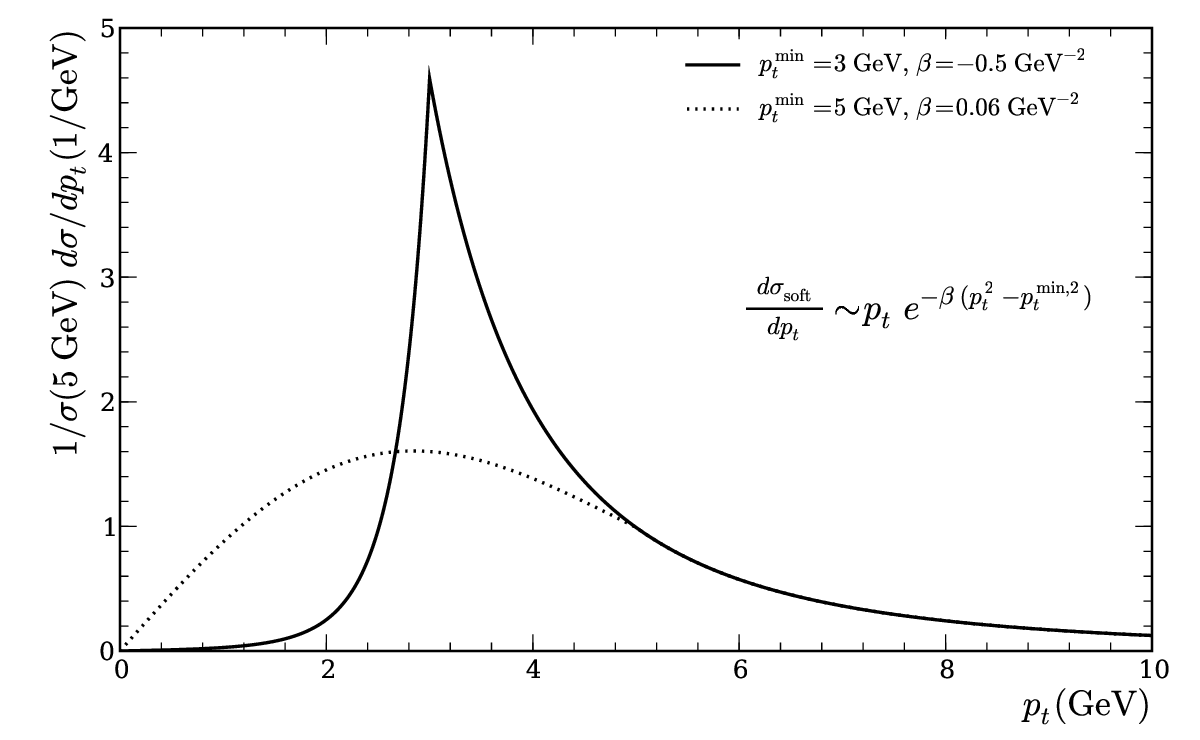}
\caption{Transverse momentum distribution of additional scatters}\label{\detokenize{review/ue:id27}}\label{\detokenize{review/ue:fig-pt-spectrum}}\end{figure}

One peculiarity is that also negative values of $\beta$ are
possible when the soft cross section becomes small.  This results in a
spike-like shape in the differential partonic cross section.   We make
use of this distribution of transverse momenta for all soft particles
generated as outlined in the next section. It was found that the parameter
$\ptmin$ should not remain fixed for simulations at different centre-of-mass energies
but rather evolve softly with energy,
\begin{equation}\label{equation:review/ue:eq:ptmin}
\begin{split}  \ptmin = \ptmino \Big(\frac{b+\sqrt{s}}{E_0^2}\Big)^p\end{split}
\end{equation}

\begin{figure}[tp]
\centering
\capstart
\noindent\includegraphics[width=0.600\linewidth]{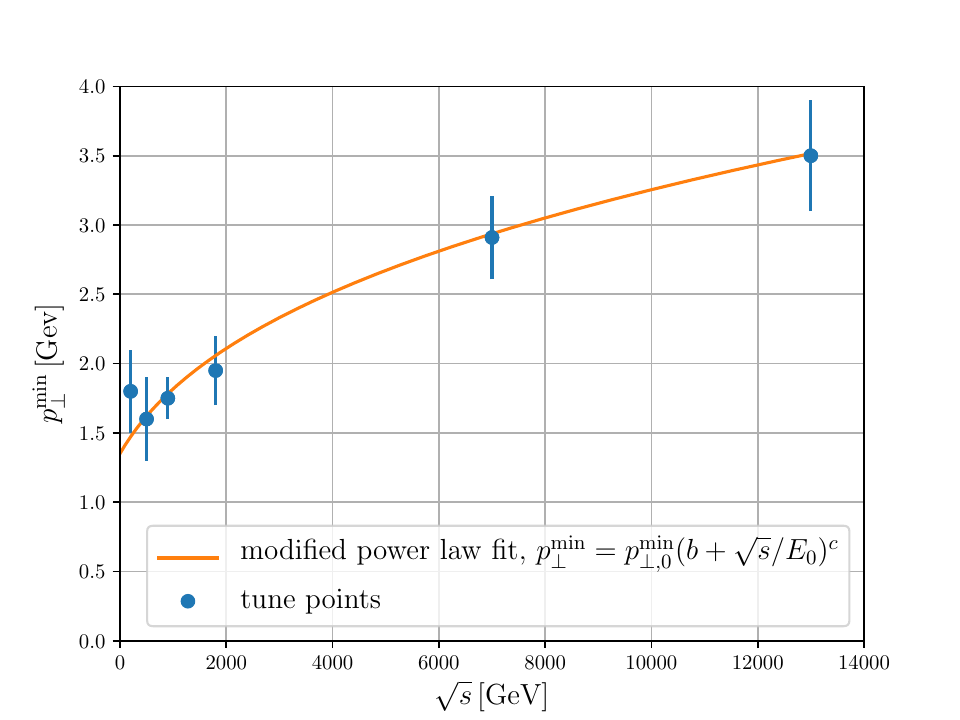}
\caption{Determination of the parameters in \eqref{equation:review/ue:eq:ptmin} from runs at various energies, cf.~\cite{Bellm:2019icn}.}\label{\detokenize{review/ue:id28}}\label{\detokenize{review/ue:fig-ptminfit}}
\end{figure}

The parameters $\ptmino$, $b$ and $p$ have been fixed from tuning the Underlying Event at
different CM energy, see Fig.~\ref{\detokenize{review/ue:fig-ptminfit}}.  Note that $E_0$ should be kept fixed for such a set of tunes and only acts
as a reference scale.

\subsection{Soft particle production model}
\label{\detokenize{review/ue:soft-particle-production-model}}

In this section we describe the implementation of the model for soft
interactions in Herwig, \textit{i.e.} how we set up the $s$ soft
scatters.  The current model was developed in order to tackle some of
the shortcomings of the old model and is described in more detail in
Reference \cite{Gieseke:2016fpz}.

The model for additional soft interactions is based on the assumption
that we expect soft particle production to be flat in rapidity and
narrow in transverse momenta.  It is inspired by the model for
multiperipheral particle production of
\cite{Baker:1976cv}. Multiperipheral particle production is a $2
\rightarrow N$ process where the $N$ resulting particles are
ordered in rapidity. A process with multiperipheral particle production
is shown in \hyperref[\detokenize{review/ue:fig-multi-ladder}]{Fig.\@ \ref{\detokenize{review/ue:fig-multi-ladder}}} where $p_A$ and
$p_B$ are the incoming particles that interact with each other and
the $p_i$ are the outgoing particles.

\begin{figure}[tp]
\centering
\capstart

\noindent\includegraphics[width=0.300\linewidth]{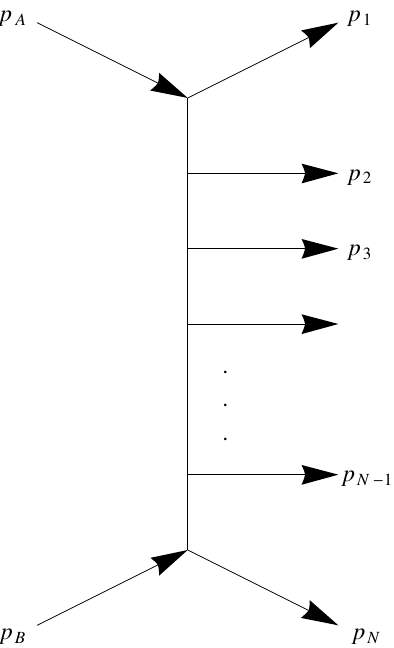}
\caption{Multiperipheral particle production.}\label{\detokenize{review/ue:id29}}\label{\detokenize{review/ue:fig-multi-ladder}}\end{figure}

The internal virtual particles are denoted by $q_i$ and in the
following we will refer to a $2 \rightarrow N$ process with
multiperipheral particle production as a particle ladder.  In Herwig the
final particles whose kinematics is constructed using this model will be
partons. More precisely proton remnants, sea quarks and gluons.  As
outlined in the previous section, the soft process follows after a
number of quasi hard processes where a valence quark or gluons with only
longitudinal momentum are selected from the proton.  The remnant takes
the rest of the momentum.  The total energy available to perform the
multiperipheral particle production is given by the remaining energy of
the proton remnants. The incoming momenta of the remnants are denoted by
$p_{r1}$ and $p_{r2}$.  The number of the final partons in
the ladder is then sampled from a Poissonian distribution with mean at
\begin{equation*}
\begin{split}\langle N \rangle \approx N_{\mathrm{ladder}} \times \ln
\frac{(p_{r1} + p_{r2})^2}{m_{\mathrm{rem}}^2} + \mathrm{b},\end{split}
\end{equation*}

where $N_{\rm{ladder}}$, $b$ are parameters of the model that roughly
determine the number density of gluons per unit rapidity and $m_{\mathrm{rem}}$ is
of the order of the constituent mass of the proton remnant and it is currently set
to the constituent mass of the gluon.

\begin{savenotes}\sphinxattablestart
\sphinxthistablewithglobalstyle
\centering
\sphinxcapstartof{table}
\sphinxthecaptionisattop
\sphinxcaption{Important diffraction and underlying event model parameters and their allowed ranges in Herwig~7.3.0.}\label{\detokenize{review/ue:id30}}
\sphinxaftertopcaption
\begin{tabulary}{\linewidth}[t]{TTT}
\sphinxtoprule
\sphinxtableatstartofbodyhook

Parameter
&

Allowed range
&

Description
\\
\hline

\href{https://herwig.hepforge.org/doxygen/HwRemDecayerInterfaces.html\#ladderMult}{ladderMult}
&

0-10
&

The ladder multiplicity factor.
\\
\hline

\href{https://herwig.hepforge.org/doxygen/HwRemDecayerInterfaces.html\#DiffractiveRatio}{DiffractiveRatio}
&

0-1
&

Fraction of diffractive cross section in inelastic cross section.
\\
\hline\sphinxmultirow{2}{10}{%
\begin{varwidth}[t]{\sphinxcolwidth{1}{3}}

\href{https://herwig.hepforge.org/doxygen/MPIHandlerInterfaces.html\#InvRadius}{InvRadius}
\par
\vskip-\baselineskip\vbox{\hbox{\strut}}\end{varwidth}%
}%
&\sphinxmultirow{2}{11}{%
\begin{varwidth}[t]{\sphinxcolwidth{1}{3}}

0.2-4.0
\par
\vskip-\baselineskip\vbox{\hbox{\strut}}\end{varwidth}%
}%
&\sphinxmultirow{2}{12}{%
\begin{varwidth}[t]{\sphinxcolwidth{1}{3}}

The inverse hadron radius squared used in the overlap function.
\par
\vskip-\baselineskip\vbox{\hbox{\strut}}\end{varwidth}%
}%
\\
\sphinxvlinecrossing{1}\sphinxvlinecrossing{2}\sphinxfixclines{3}\sphinxtablestrut{10}&\sphinxtablestrut{11}&\sphinxtablestrut{12}\\
\hline\sphinxmultirow{2}{13}{%
\begin{varwidth}[t]{\sphinxcolwidth{1}{3}}

\href{https://herwig.hepforge.org/doxygen/MPIHandlerInterfaces.html\#pTmin0}{pTmin0}
\par
\vskip-\baselineskip\vbox{\hbox{\strut}}\end{varwidth}%
}%
&\sphinxmultirow{2}{14}{%
\begin{varwidth}[t]{\sphinxcolwidth{1}{3}}

0-10
\par
\vskip-\baselineskip\vbox{\hbox{\strut}}\end{varwidth}%
}%
&\sphinxmultirow{2}{15}{%
\begin{varwidth}[t]{\sphinxcolwidth{1}{3}}

The pTmin at the reference scale for power law extrapolation of pTmin.
\par
\vskip-\baselineskip\vbox{\hbox{\strut}}\end{varwidth}%
}%
\\
\sphinxvlinecrossing{1}\sphinxvlinecrossing{2}\sphinxfixclines{3}\sphinxtablestrut{13}&\sphinxtablestrut{14}&\sphinxtablestrut{15}\\
\hline\sphinxmultirow{2}{16}{%
\begin{varwidth}[t]{\sphinxcolwidth{1}{3}}

\href{https://herwig.hepforge.org/doxygen/MPIHandlerInterfaces.html\#Power}{Power}
\par
\vskip-\baselineskip\vbox{\hbox{\strut}}\end{varwidth}%
}%
&\sphinxmultirow{2}{17}{%
\begin{varwidth}[t]{\sphinxcolwidth{1}{3}}

0-10
\par
\vskip-\baselineskip\vbox{\hbox{\strut}}\end{varwidth}%
}%
&\sphinxmultirow{2}{18}{%
\begin{varwidth}[t]{\sphinxcolwidth{1}{3}}

The power for power law extrapolation of the pTmin cut-off.
\par
\vskip-\baselineskip\vbox{\hbox{\strut}}\end{varwidth}%
}%
\\
\sphinxvlinecrossing{1}\sphinxvlinecrossing{2}\sphinxfixclines{3}\sphinxtablestrut{16}&\sphinxtablestrut{17}&\sphinxtablestrut{18}\\
\hline\sphinxmultirow{2}{19}{%
\begin{varwidth}[t]{\sphinxcolwidth{1}{3}}

\href{https://herwig.hepforge.org/doxygen/MPIHandlerInterfaces.html\#Offset}{Offset}
\par
\vskip-\baselineskip\vbox{\hbox{\strut}}\end{varwidth}%
}%
&\sphinxmultirow{2}{20}{%
\begin{varwidth}[t]{\sphinxcolwidth{1}{3}}

500-1000
\par
\vskip-\baselineskip\vbox{\hbox{\strut}}\end{varwidth}%
}%
&\sphinxmultirow{2}{21}{%
\begin{varwidth}[t]{\sphinxcolwidth{1}{3}}

The offset used in the  power law extrapolation of the pTmin cut-off.
\par
\vskip-\baselineskip\vbox{\hbox{\strut}}\end{varwidth}%
}%
\\
\sphinxvlinecrossing{1}\sphinxvlinecrossing{2}\sphinxfixclines{3}\sphinxtablestrut{19}&\sphinxtablestrut{20}&\sphinxtablestrut{21}\\
\hline\sphinxmultirow{2}{22}{%
\begin{varwidth}[t]{\sphinxcolwidth{1}{3}}

\href{https://herwig.hepforge.org/doxygen/MPIHandlerInterfaces.html\#ReferenceScale}{ReferenceScale}
\par
\vskip-\baselineskip\vbox{\hbox{\strut}}\end{varwidth}%
}%
&\sphinxmultirow{2}{23}{%
\begin{varwidth}[t]{\sphinxcolwidth{1}{3}}

0-2000
\par
\vskip-\baselineskip\vbox{\hbox{\strut}}\end{varwidth}%
}%
&\sphinxmultirow{2}{24}{%
\begin{varwidth}[t]{\sphinxcolwidth{1}{3}}

The reference energy for power law energy extrapolation of pTmin.
\par
\vskip-\baselineskip\vbox{\hbox{\strut}}\end{varwidth}%
}%
\\
\sphinxvlinecrossing{1}\sphinxvlinecrossing{2}\sphinxfixclines{3}\sphinxtablestrut{22}&\sphinxtablestrut{23}&\sphinxtablestrut{24}\\
\sphinxbottomrule
\end{tabulary}
\sphinxtableafterendhook\par
\sphinxattableend\end{savenotes}

In \hyperref[\detokenize{review/ue:fig-soft-ladder}]{Fig.\@ \ref{\detokenize{review/ue:fig-soft-ladder}}} we
illustrate a case with $N=6$ where we have the two remnants a sea
quark and an antiquark and the gluons.  The two sea quarks, which are
generated by the soft event, are needed in order to create the correct
colour connections between the partons in the ladder. The flavour and
the orientation of the sea quarks is randomly chosen from up and down
quarks with respective unit weights.

\begin{figure}[htp]
\centering
\capstart

\noindent\includegraphics[width=0.300\linewidth]{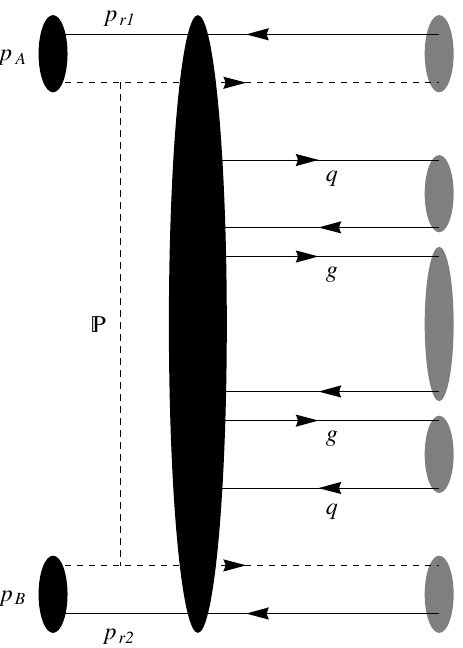}
\caption{Cluster formation in the multiperipheral final state.}\label{\detokenize{review/ue:id31}}\label{\detokenize{review/ue:fig-soft-ladder}}\end{figure}

The kinematics of the partons in the ladder are generated with a
modified Jadach algorithm \cite{Jadach:2001mp} the partons’
rapidities are distributed evenly between the rapidities of the
remnants.  The transverse momentum of one randomly selected parton is
chosen according to Equation \eqref{equation:review/ue:eq:transverse}. The transverse momenta
of the remaining partons in the ladder is then sampled from a flat
probability distribution below the $p_T$ of the first
parton. Additionally all partons are flat in azimuthal angle.

Since the first quark is extracted from the proton by using a PDF we
have to ensure that its rapidity is close to the second parton in the
ladder, which is in our case a sea quark. This is done by choosing the
proper value for $x_{\mathrm{min}}$ of the PDF.  This algorithm
guarantees the exponential fall off of the differential cross section
for large values of rapidity separation $\Delta \eta$. It also
produces a roughly flat distribution of the clusters’ rapidities and the
subsequently produced final state particles.

For all of the soft scatters, the number $s$ of which was obtained
as described above from \eqref{equation:review/ue:eq:prob_final}, one of these ladders is
generated.  As the rapidity spans are reduced ladder-by-ladder the phase
space shrinks and the partons successively populate more central
rapidity regions.

\subsection{Diffraction}
\label{\detokenize{review/ue:diffraction}}

A completely new feature in the current version of Herwig is the model
for the simulation of diffractive events.  The diffractive events are
generated according to the differential cross sections for single and
double diffraction which can be described by Regge theory and the
generalised optical theorem \cite{Mueller:1970fa,Barone:2002cv}.  The
process for single diffraction can be depicted by $A+B \rightarrow
X+B$ where $A$ and $B$ are the incoming hadrons and
$X$ is the hadronic final state in the limit $s\gg M^2\gg
|t|$. $s$ is the total centre of mass energy of the incoming
particles, $M$ is the invariant mass of the dissociated hadron and
$-t$ is the momentum transfer.  The same holds true for the double
diffractive process $A+B \rightarrow X_A + X_B$.  So far, we have only
considered diffraction among protons.  The ratio between single and double
diffraction is not fixed. It is chosen roughly according to the
measurements of the cross section in \cite{Abelev:2012sea}.

In order to connect the diffraction model to the other parts of the MPI
model in Herwig we have to ensure that the cross sections for semi-hard
and soft interactions only sum up to a certain fraction $x_{\rm
NSD}$ of the total cross section when we fix the parameters of the soft
model. The diffractive events are then generated as an independent
process.

The kinematics are generated by first sampling the momentum transfer
$t$, and the diffractive masses $M_A$, $M_B$ (for
single diffraction one of them is the proton mass $m_p$). The
scattering angle between the incoming protons and the diffractive
systems is then calculated according to
\begin{equation*}
\begin{split}\cos(\theta) = \frac{s(s+2t-2m_p^2-M_A^2-M_B^2)}{\lambda(s,M_A^2,M_B^2)\lambda(s,m_p^2,m_p^2)}.\end{split}
\end{equation*}

where $\lambda(x,y,z)$ is the Källen function.

With the invariant masses and the scattering angle the outgoing momenta
can be constructed. The dissociated proton is then decayed into a
quark-diquark pair which moves collinear to the direction of the
dissociated proton.  This final diffractive state is treated fully
non-perturbatively and then handled by the hadronization model where the
cluster will decay into the final state hadrons.

We point out that a fraction of the diffractive events for very low
diffractive masses are modelled directly as the excitation of the proton
to a $\Delta$ baryon as a final state instead of quark-diquark
pair.  $p p \rightarrow \Delta p$ for single and $p p
\rightarrow \Delta \Delta$ for double diffractive events. The
$\Delta$ is then handled by the hadronic decay handler.

\subsection{Connection to different simulation phases}
\label{\detokenize{review/ue:connection-to-different-simulation-phases}}

The model introduced so far is entirely formulated at the parton level.
However, an event generator aims for a full description of the event at
the level of hadrons. This implies that the implementation of
multi-parton scattering must be properly connected to the parton shower
and hadronization models, a few details of which we discuss in the
following.

\subsubsection{Parton showers and hard matrix elements}
\label{\detokenize{review/ue:parton-showers-and-hard-matrix-elements}}

After generating the hard process and invoking parton showers on its
coloured particles, the number of additional scatters is calculated
according to Eq. \eqref{equation:review/ue:eq:prob_final} or Eq. \eqref{equation:review/ue:eq:prob1_final} respectively. After the initial-state
shower has terminated, the incoming partons are extracted out of the
beam particles in the usual way.

The requested additional scatters are then generated using a similar but
completely independent infrastructure from the one of the hard process.
Dedicated hard matrix elements with hand-coded formulae summed over
parton spins are used for greater speed, as mentioned in
\hyperref[\detokenize{review/hardprocess/me:sec-specificmes}]{Section \ref{\detokenize{review/hardprocess/me:sec-specificmes}}}. This also has the advantage that specific
cuts, different to those used for the main hard process, can be
specified.  e.g.\ the lower bound on the transverse momenta should be the
very same $\ptmin$ as used in the computations for the number of
scatters.

For each additional scattering, parton showers evolve the produced
particles down to the hadronic scale. The backward evolution of
additional scatters is forced to terminate on a gluon. After
termination, these gluons are extracted out of the beam particles. If
this process leads to a violation of four-momentum conservation, the
scattering cannot be established. It is therefore regenerated until the
desired multiplicity has been reached. If a requested scattering can
never be generated without leading to violation of momentum
conservation, the program eventually gives up, reducing the multiplicity
of scatters.

\subsubsection{Minimum bias process}
\label{\detokenize{review/ue:minimum-bias-process}}

Herwig simulates minimum bias collisions as events in which there is
effectively no hard process. However, to maintain a uniform structure
with the simulation of standard hard process events, we have implemented
a matrix element class, that generates a ‘hard’ process with as minimal
an effect as possible. It extracts only light (anti)quarks ($d$,
$u$, $\bar d$ or $\bar u$) from the hadrons and allows
them to ‘scatter’ through colourless exchange at zero transverse
momentum, with a matrix element set to unity, so that their longitudinal
momentum is determined only by their parton distribution functions. To
give a predominantly valence-like distribution, a cut on their
longitudinal momentum fraction $x>10^{-2}$ is recommended. Note
that because the matrix element is set to unity, the cross section that
is printed to the output file at the end of the run is meaningless.

\subsubsection{Hadronization}
\label{\detokenize{review/ue:hadronization}}

The underlying event and beam remnant treatment are closely connected
because the generation of additional scatters requires the extraction of
several partons out of the proton. As described before, all additional
partons are extracted from the incoming beam particles. This is
different from the procedure that was used in FORTRAN JIMMY, where the
successive partons were always extracted from the previous beam remnant,
a difference in the structure of the event record that should not lead
to significant differences in physical distributions.

The cluster hadronization described in the previous section can only act
on (anti)quarks or (anti)diquarks. However, naively extracting several
partons from a hadron will not leave a flavour configuration that is
amenable to such a description in general. Therefore, the strategy we
use, as already mentioned, is to terminate the backward evolution of the
hard process on a valence parton of the beam hadron and additional
scatterings on gluons, giving a structure that can be easily iterated
for an arbitrary number of scatters. This structure is essentially the
same as in FORTRAN JIMMY.

\subsection{Older soft models}
\label{\detokenize{review/ue:older-soft-models}}\label{\detokenize{review/ue:sect-ua5}}

While the new multiple interaction model provides a better description
of the underlying event and minimum bias data, and is recommended for
all realistic physics studies, Herwig still contains the original soft
model of the underlying event used before version 2.1, and also the
soft model that was used with the early implementations of the MPI model
in Herwig.  Please refer to \cite{Bahr:2008pv} for details.

\subsection{Code structure}
\label{\detokenize{review/ue:code-structure}}

In addition to being the main class responsible for the administration
of the shower, the \sphinxstylestrong{ShowerHandler}, described in
\hyperref[\detokenize{review/showers/code:sect-showercode}]{Section \ref{\detokenize{review/showers/code:sect-showercode}}}, is also responsible for the generation of the
additional semi-hard scattering processes. It has a reference to the
\sphinxstylestrong{MPIHandler} in the input files, which is used to actually
create the additional scattering processes. It invokes the parton shower
on all the available scatters and connects them properly to the incoming
beam particles. This includes potential re-extraction of the incoming
parton if it is changed as a result of initial-state radiation. A number
of classes are used by the \sphinxstylestrong{MPIHandler} to generate the
additional scattering processes. Soft additional scatters are generated
in the \sphinxstylestrong{HwRemDecayer} class.

The \sphinxstylestrong{MPIHandler} administers the calculation of the underlying
event activity. It uses to sample the phase-space of the processes that
are connected to it. Using that cross section the probabilities for the
individual multiplicities of additional scatters are calculated during
initialisation. The method MPIHandler::multiplicity() samples a number
of extra scatters from that pretabulated probability distribution. The
method MPIHandler::generate() creates one subprocess according to the
phase-space and returns it.

The \sphinxstylestrong{MPISampler} performs the phase-space sampling for the
additional scatterings. It inherits from \sphinxstylestrong{SamplerBase} and
implements the Auto Compensating Divide-and-Conquer phase-space
generator, \sphinxstylestrong{ACDCGen}.

The \sphinxstylestrong{HwRemDecayer} is responsible for decaying the
\sphinxstylestrong{RemnantParticle} to something that can be processed by the
cluster hadronization, \textit{i.e.} (anti)quarks or (anti)diquarks.  This
includes the forced splittings to valence quarks and gluons
respectively. Also the colour connections between the additional
scatters and the remnants are set here. If additional soft partonic
interactions, i.e.\ the non-perturbative part of the underlying event,
are enabled, they are generated inside this class after the remnants
have been decayed to the (anti)diquarks.

The \sphinxstylestrong{MPIPDF} class is used to modify the PDF’s used for the
initial state shower of additional scatters. All sorts of rescaling are
possible but currently the mode that is used is the one where the
valence part of the PDF is removed. The objects are instantiated inside
and set to the default PDF’s using \texttt{{void
ThePEG::CascadeHandler::resetPDFs(...)}}

The most important interfaces to set parameters for the above mentioned
classes are described here. An exhaustive description of all interfaces
is provided by our Doxygen documentation.
\sphinxstylestrong{MPIHandler} has the following important interfaces:

\sphinxstylestrong{SubProcessHandlers} is a vector of references to
\sphinxstylestrong{SubProcessHandler} objects. The first element is reserved for
the underlying event process. Additional references can be set to
simulate additional hard processes in a single collision. See
\hyperref[\detokenize{review/hardprocess/general:sect-hard-process}]{Section \ref{\detokenize{review/hardprocess/general:sect-hard-process}}} for details of how to use this functionality.

\sphinxstylestrong{Cuts} Vector of references to \sphinxstylestrong{Cuts} objects. The first
element is used to impose the minimal transverse momentum of the
additional scatters, $\ptmin$.
This is one of the two main
parameters of the model. The current default, obtained from a fit to
Tevatron data is $4.0\,\mathrm{GeV}$. See Ref. \cite{Bahr:2008wk}
for details.
Additional cuts object may be defined for additional hard
processes that should be simulated in the same event.

\sphinxstylestrong{additionalMultiplicities} Vector of integer values to specify
the multiplicity of additional hard scattering processes in a single
collision. See \hyperref[\detokenize{review/hardprocess/general:sect-hard-process}]{Section \ref{\detokenize{review/hardprocess/general:sect-hard-process}}} for an example.

\sphinxstylestrong{InvRadius} sets the value of the inverse beam particle radius
squared, $\mu^2$. The current default is
$1.5\,\mathrm{GeV}^{2}$, obtained from the above mentioned fit.
\begin{description}
\sphinxlineitem{\sphinxstylestrong{pTmin0}, \sphinxstylestrong{Offset}, \sphinxstylestrong{Power}}

describe the energy evolution of the transverse momentum cutoff
$\ptmino$, as described in  Eq. \eqref{equation:review/ue:eq:ptmin}.

\sphinxlineitem{\sphinxstylestrong{DiffractiveRatio},}

gives the fraction of diffractive events for a given CM energy as described above as $1 - x_{\rm NSD}$.
Note that this parameter is generally energy dependent as the diffractive cross section will evolve with energy
differently from the total cross section.

\end{description}

\sphinxstylestrong{IdenticalToUE} An integer parameter specifying which element of
the list of \sphinxstylestrong{SubProcessHandler}’s
is identical to the underlying event process. Zero means the the
conventional hard subprocess is QCD jet production. -1 means that no
process is identical. Any number $>0$ means that one of the
additional hard scatterings is QCD jet production, where the exact
number specifies the position in the vector. The default is -1, which is
appropriate as long as no QCD jet production is simulated.

\sphinxstylestrong{softIn} Switch to turn the inclusion of non-perturbative
scatters to the underlying event model on (\texttt{{Yes}}) or off (\texttt{{No}}). The
current default is \texttt{{Yes}}.

\sphinxstylestrong{twoComp} Switch to toggle between an independent overlap
function for soft additional scatters (\texttt{{Yes}}) and identical ones
$A_{\rm soft}(\vect b) \equiv A(\vect b)$ (\texttt{{No}}). If the
two-component model is used, the parameters of the soft sector are
automatically chosen to describe the total cross section as well as the
elastic $t$-slope correctly.

\sphinxstylestrong{DLMode} Integer number $\in \{1,2,3\}$ to choose between
three different parametrizations of the total cross section as a
function of the centre-of-mass energy:
\begin{enumerate}
\sphinxsetlistlabels{\arabic}{enumi}{enumii}{}{.}%
\item {} 

Parametrisation of Ref. \cite{Donnachie:1992ny}.

\item {} 

Parametrisation of Ref. \cite{Donnachie:1992ny} but with rescaled
normalisation to match the central value of the measurement
\cite{sigma_tot_CDF} by CDF. \texttt{{Default}}

\item {} 

Parametrization of Ref. \cite{Donnachie:2004pi}.

\end{enumerate}

\sphinxstylestrong{MeasuredTotalXSec} Parameter to set the total cross section (in
mb) explicitly. If this parameter is used, it will overwrite the
parametrisation selected with the previous switch. This is intended for
first data on the total cross section and should be used instead of the
parametrisation, which may deviate substantially.

In the \sphinxstylestrong{ShowerHandler} there is a reference to the
\sphinxstylestrong{MPIHandler} . To switch multiple parton interactions off, this
reference has to be set to \texttt{{NULL}}.

\sphinxstylestrong{HwRemDecayer} has just one important interface:

\sphinxstylestrong{ladderMult}, the number of gluons per rapidity in the soft particle production model.

\clearpage

\section{Hadronic decays}
\label{\detokenize{review/index:hadronic-decays}}\label{\detokenize{review/index:sect-hadron-sub-decay}}

Herwig uses a sophisticated model of hadronic decays, as described in
Ref. \cite{Grellscheid:2007tt}. The simulation of decays
in Herwig is designed to have the following properties:
\begin{itemize}
\item {} 

a unified treatment of the decays of both fundamental particles
and the unstable hadrons, this is of particular importance for
particles like the $\tau$ lepton, which, while a fundamental
particle, is more akin to the unstable hadrons in the way it decays;

\item {} 

up-to-date particle properties, \textit{i.e.} masses, widths, lifetimes,
decay modes and branching ratios together with a new database to
store these properties to make updating the properties easier and the
choices made in deriving them clearer;

\item {} 

full treatment of spin correlation effects using the algorithm of
Refs.
\cite{Richardson:2001df, Knowles:1988vs, Collins:1987cp}
for the decay of all unstable particles, it is important that the
same algorithm is used consistently in all stages of the program so
that correlations between the different stages can be correctly
included;

\item {} 

a sophisticated treatment of off-shell effects for both unstable
hadrons and fundamental particles;

\item {} 

a large range of matrix elements for hadron and tau decays including
both general matrix elements based on the spin structures of the
decays and specific matrix elements for important decay modes;

\item {} 

the accurate simulation of QED radiation in the particle decays using
the Yennie--Frautschi--Suura (YFS) formalism \cite{Hamilton:2006xz}.

\item {} 

an interface to EvtGen \cite{Lange:2001uf} for the decay of
bottom and charm mesons.

\end{itemize}

In this section we describe the simulation of hadron and tau lepton decays in
Herwig. We start by discussing the physical properties of the hadrons
used in the simulation and how they are determined. In ThePEG framework
these physical properties are stored using the
\texttt{ParticleData} class,
which has one
instance for each particle used in the simulation. In turn the
properties of a given decay mode are stored using the
\texttt{DecayMode} class, which
contains both the particles involved in the decay and a reference to a \texttt{Decayer}
object that can be used to generate the kinematics of the decay
products. The \href{https://thepeg.hepforge.org/doxygen/classThePEG\_1\_1DecayHandler.html}{DecayHandler} class uses these \texttt{DecayMode} objects to select a decay
of a given particle, according to the probabilities given by the
branching ratios for the different decay modes, and then generates the
kinematics using the relevant
\texttt{Decayer} specified by the \texttt{DecayMode}.

Following a brief discussion of the treatment of off-shell effects we
therefore discuss the different \texttt{Decayer} classes available in Herwig for the
decay of tau leptons, strong and electromagnetic hadron decays and then weak
hadron decays. This is followed by a discussion of the code structure.

\subsection{Particle properties}
\label{\detokenize{review/decays:particle-properties}}\label{\detokenize{review/decays::doc}}

The information in the Particle Data Group’s (PDG) compilation
\cite{ParticleDataGroup:2022pth} of experimental data is sufficient in many cases to
determine the properties of the hadrons used in Herwig. However, there
are some particles for which the data are incomplete or too inaccurate
to be used. Equally, there are a number of particles that are necessary
for the simulation but have not been observed, particularly excited
bottom and charm hadrons, which should perhaps be regarded as part of
the hadronization model affecting the momentum spectrum of lighter
states, rather than as physical states. A large number of choices
therefore have to be made in constructing the particle data tables used
in the event generator, based on the data in Ref. \cite{ParticleDataGroup:2022pth}.

In the past the data were stored as either a text file or the contents of a FORTRAN COMMON block. However, due to the relatively large amount of data that needs to be stored we decided to adopt a database approach based on the MySQL database system. The event generation still uses text files to read in the particle properties but these files are now automatically generated from the database. This provides us with a range of benefits: the data can be edited using a web interface; additional information describing how the particle properties were determined is stored in the database, both improving the long-term maintenance and allowing the user to understand the uncertainties and assumptions involved. The database and its web interface are maintained internally by the Herwig collaboration and are not publicly accessible. The principal experimental input is the 2022 Particle Data Group compilation, Ref.~\cite{ParticleDataGroup:2022pth}, supplemented where necessary by the additional experimental and theoretical sources listed below.

\begin{figure}[htp]
\centering
\capstart
\fbox{\noindent\includegraphics[trim=0 40mm 0 0,clip]{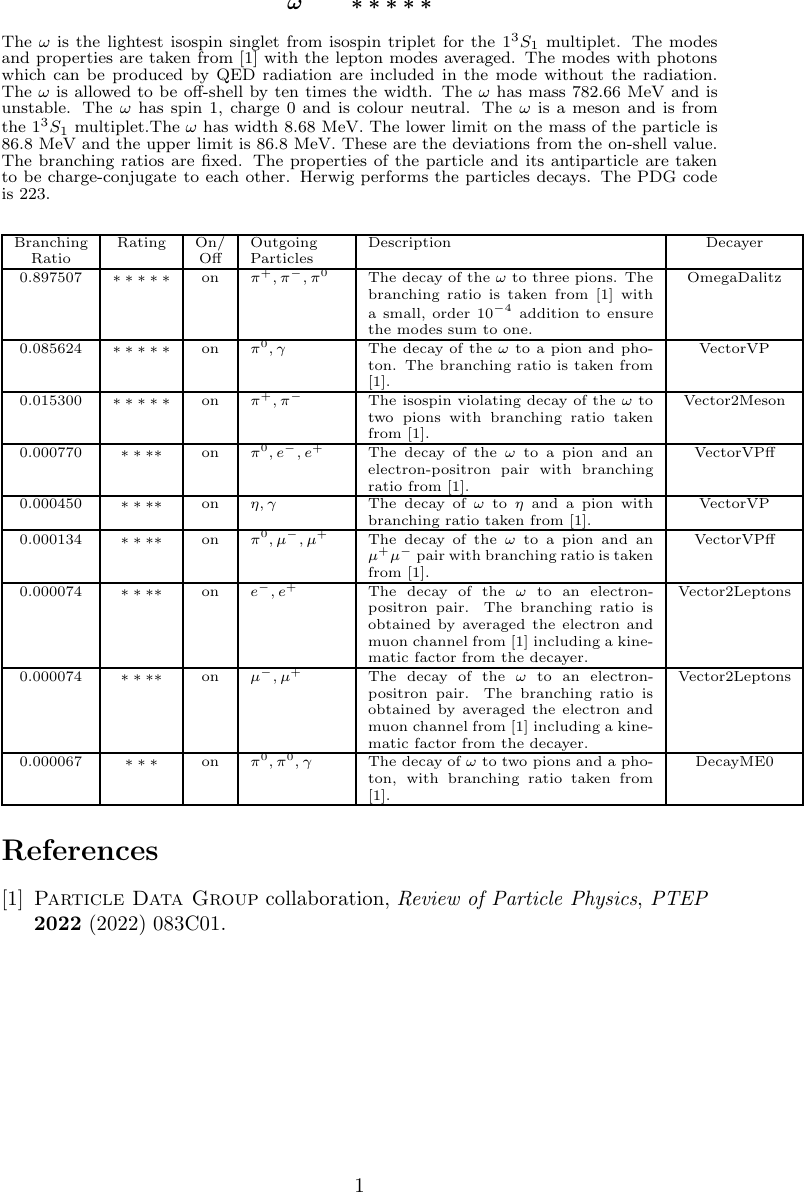}}
\caption{An example of the particle properties in Herwig, in this case for the $\omega$ meson. The properties of the particle, including the mass, width, decay modes and branching ratios, are given together with comments on how properties were determined. In the full web version, links are included to the descriptions of the objects responsible for generating the kinematics for the various decay modes.}
\label{\detokenize{review/decays:id320}}\label{\detokenize{review/decays:fig-omegaeg}}
\end{figure}

An example of the output from the database for the properties of the $\omega$ meson is shown in \hyperref[\detokenize{review/decays:fig-omegaeg}]{Fig.\@ \ref{\detokenize{review/decays:fig-omegaeg}}}. This includes the basic properties of the particle together with an explanation of how they were derived. In addition there is a star rating between one and five, which gives an indication of how reliable the properties of the particle and the modelling of individual decay modes are.

In general we used the following philosophy to determine the particle
properties used in Herwig:
\begin{itemize}
\item {} 

The properties of the light mesons in the lowest lying multiplets
were taken from Ref. \cite{ParticleDataGroup:2022pth}. In some cases we used either
lepton universality or the phase-space factors from our \texttt{Decayer}s to average
the branching ratios for poorly measured modes.

\item {} 

Where possible, the properties of the excited light mesons were taken
from Ref. \cite{ParticleDataGroup:2022pth} together with some additional
interpretation of the data. The mesons
needed to fill the $1^1{\rm S}_0$, $1^3{\rm S}_1$,
$1^1{\rm P}_1$, $1^3{\rm P}_0$, $1^3{\rm P}_1$,
$1^3{\rm P}_2$, $1^1{\rm D}_2$, $1^3{\rm D}_1$,
$1^3{\rm D}_3$, $2^1{\rm S}_0$, $1^1{\rm S}_0$ and
$2^3{\rm S}_1$ $\mathrm{SU}(3)$ multiplets are included
together with the $K$ mesons from the $1^3{\rm D}_2$
multiplet.

\item {} 

We use the EvtGen package \cite{Lange:2001uf} for the decays
of the $D_{u,d,s}$ mesons and $B_{u,d,s}$ mesons.

\item {} 

The mass and lifetime of the $B_c$ meson is taken from Ref.
\cite{ParticleDataGroup:2022pth}. The branching ratios were taken from the
theoretical calculations of Ref. \cite{Kiselev:2003mp} together with
some partonic modes to ensure that the branching ratios sum to one.

\item {} 

The properties and decay modes of the charmonium resonances were
taken from Ref. \cite{ParticleDataGroup:2022pth} where possible, together with the
use of partonic decays, to $ggg$, $gg$ or
$q\bar{q}$, to model the unobserved inclusive modes. For some
of the particles, in particular the $h_c$ and
$\eta_c(2S)$, the results of Ref. \cite{Skwarnicki:2003wn}
were used and the $\eta_c(2S)$ branching ratios were taken from
the theoretical calculation of Ref. \cite{Eichten:2002qv}.

\item {} 

The properties and decay modes of the bottomonium resonances were
taken from Ref. \cite{ParticleDataGroup:2022pth} where possible. In addition we
have added a large number of states that are expected to have small
widths, \textit{i.e.} the mass is expected to be below the $B\bar{B}$
threshold, using the theoretical calculations of Refs.
\cite{Kwong:1988ae, Godfrey:2002rp, Eichten:1994gt, Ebert:2002pp, Kwong:1987ak}
for many of the properties.

\item {} 

The properties of the excited $D$ and $D_s$ mesons were
taken from Ref. \cite{ParticleDataGroup:2022pth}. For many of the mesons we were
forced to assume that the observed modes saturated the total width in
order to obtain the branching ratios using the results in Ref.
\cite{ParticleDataGroup:2022pth}.

\item {} 

The properties of the excited $B_{u,d,s}$ mesons are uncertain.
The $B^*_{u,d,s}$, together with one ${B_1}_{uds}$ state and the ${B_2}_{uds}$,
have been observed and there is evidence in
Ref. \cite{ParticleDataGroup:2022pth} for further excited states, however
it was unclear which states have been observed. We therefore take the
properties of the remaining unobserved states from Ref.
\cite{DiPierro:2001uu}.

\item {} 

The mass of the $B_c(2S)^+$ is taken from \cite{ParticleDataGroup:2022pth},
while the masses of the unobserved excited $B_c$ mesons
are taken from the lattice results in Ref.
\cite{Brambilla:2004wf}, which agree with potential model
calculations. The widths and branching ratios were taken from the
theoretical calculation of Ref. \cite{Godfrey:2004ya}.

\item {} 

The properties of the light baryons were taken from Ref.
\cite{ParticleDataGroup:2022pth} where possible. In general, we have included all
the light baryons from the first $(56,0^+_0)$ octet and
decuplet multiplets. We include the light baryons from the next
$\frac12^+$ $(56,0^+_2)$, $\frac12^-$
$(70,1^-_1)$, and $\frac32^-$ $(70,1^-_1)$
multiplets, although in some cases we have used higher $\Xi$
resonances whose properties are better determined, rather than those
given in Ref. \cite{Yao:2006px}. In addition, the singlet
$\Lambda(1405)$ and $\Lambda(1520)$ are also included. By
default the $\frac32^-$ $(70,1^-_1)$ multiplet and
$\Lambda(1520)$ are not produced in the hadronization stage in
order to improve the agreement with LEP data.

\item {} 

The properties of the weakly decaying charm baryons were taken from
\cite{Yao:2006px} together with a number of decay modes with
theoretically-calculated branching ratios from \cite{Korner:1992wi}
and partonic decay modes in order to saturate the total width.

\item {} 

The experimental data on the weakly decaying bottom baryons is
limited. Where possible, this data, taken from Ref.
\cite{ParticleDataGroup:2022pth}, was used together with a number of theoretical
calculations
\cite{Ivanov:1996fj,Datta:2003yk,Leibovich:2003tw,Ivanov:1997ra,Huang:2000xw,Cheng:1996cs}
for the branching ratios to exclusive modes. The masses were
calculated using the equivalent splitting in the charm system and the
$\Lambda_b$ mass where they have not been measured. In addition
to the exclusive modes a number of partonic modes are included to
model the unobserved exclusive decays.

\item {} 

The properties of the strongly and radiatively decaying charm
baryons, \textit{i.e} $\Sigma_c$, $\Xi'_c$, and excited
$\Lambda_c$ and $\Xi_c$, are taken from Ref.
\cite{Yao:2006px} together with some results from Ref.
\cite{Ivanov:1999bk} for branching ratios and widths where the
experimental data is insufficient.

\item {} 

The baryons containing a single charm or bottom quark from the multiplets
containing the $\Lambda(1405)$ and $\Lambda(1520)$ have
been observed and are included with the properties taken from Ref.
\cite{ParticleDataGroup:2022pth} where possible and Ref. \cite{Ivanov:1999bk} for
some widths.

\item {} 

We include the baryons containing two heavy quarks, either $cc$, $bb$ or $bc$,
required to complete the lightest two hadronic multiplets.
We possible the properties are taken from Ref.
\cite{ParticleDataGroup:2022pth}, otherwise we take the masses from Ref. \cite{Ebert:2004ck}
and lifetimes from \cite{Berezhnoy:2018bde}. We use partonic decays to model
the decay of the ground states and given the small mass splittings radiative decays
for the excited states. Baryons containing more than one heavy quark cannot be
produced in the cluster hadronization model, however these states are included to
support future studies of these states.

\item {} 

No tetraquark or pentaquarks, are currently included in the particle properties.

\end{itemize}

\subsection{Line shapes}
\label{\detokenize{review/decays:line-shapes}}

In general, if we wish to include the off-shell effects for an outgoing
external particle in a hard production or decay process we need to
include the following factor in the calculation of the matrix element
\begin{equation}\label{equation:review/decays:eqn:offshell}
\begin{split}W_{\rm off} = \frac1\pi\int {\rm d}m^2 \frac{m\Gamma(m)}{(M^2-m^2)^2+m^2\Gamma^2(m)},\end{split}
\end{equation}

where $M$ is the physical mass of the particle, $m$ is the
off-shell mass and $\Gamma(m)$ is the running width evaluated at
scale $m$. In practice other effects can be included to improve
this simple formula, for example we include the Flatté lineshape
\cite{Flatte:1976xu} for the $a_0(980)$ and $f_0(980)$
mesons. In Herwig, we calculate the running width of the particle based
on its decay modes. The \texttt{Decayer} responsible for each decay mode specifies the
form of the running partial width, $\Gamma_i(m)$, for the decay
mode either in a closed analytic form for two-body decays or as a \href{https://herwig.hepforge.org/doxygen/classHerwig\_1\_1WidthCalculatorBase.html}{WidthCalculatorBase}
object, which is capable of calculating the partial width numerically
and is used to construct an interpolation table. The running width for a
given particle is then the sum of the partial widths
\begin{equation*}
\begin{split}\Gamma(m) = \sum_i \Gamma_i(m).\end{split}
\end{equation*}

This both gives a sophisticated model of the running width based on the
decay modes and allows us to use the partial widths to normalize the
weights for the phase-space integration of the decays to improve
efficiency close to the kinematic threshold for the decay.

In some cases, where the partial width varies significantly over the
mass range allowed in the simulation, we can choose to use a variable
branching ratio
\begin{equation*}
\begin{split}{\rm BR}_i(m) = \frac{\Gamma_i(m)}{\Gamma(m)},\end{split}
\end{equation*}

both to prevent the production of kinematically unavailable modes and to
improve the physics of the simulation. The classic examples are the
decays of the $f_0$ and $a_0$ scalar mesons, which lie close
to the $K\bar{K}$ threshold. This means that, depending on their
mass, they decay to either $\pi\pi$ or $\eta\pi$
respectively below the threshold or with a significant $K\bar{K}$
branching fraction above the $K\bar{K}$ threshold.

The weight in Eq. \eqref{equation:review/decays:eqn:offshell} is automatically included for
all the \texttt{Decayer}s inheriting from the
\texttt{DecayIntegrator} class, which is the case for vast majority
of the Herwig \texttt{Decayer}s. The \texttt{GenericWidthGenerator} calculates the running widths using
information from the Herwig \texttt{Decayer}s inheriting from the
\texttt{DecayIntegrator} class. For decayers inheriting from the \href{https://herwig.hepforge.org/doxygen/classHerwig\_1\_1Baryon1MesonDecayerBase.html}{Baryon1MesonDecayerBase} class the running
width is calculated using the \href{https://herwig.hepforge.org/doxygen/classHerwig\_1\_1BaryonWidthGenerator.html}{BaryonWidthGenerator} class.
The \texttt{GenericMassGenerator} class is responsible for calculating the
weight in Eq. \eqref{equation:review/decays:eqn:offshell} or generating a mass according to
this distribution.

\subsection{Tau decays}
\label{\detokenize{review/decays:tau-decays}}

The simulation of $\tau$ lepton decays in Herwig is described in
detail in Ref. \cite{Grellscheid:2007tt}, together with a detailed
comparison between the results of Herwig and
TAUOLA \cite{Jadach:1993hs,Golonka:2003xt}. Here we simply describe the
basic formalism for the decays of the tau and the different models
available for the different decays, together with the structure of the
code.

The matrix element for the decay of the $\tau$ lepton can be
written as
\begin{equation*}
\begin{split}\mathcal{M} = \frac{G_F}{\sqrt{2}}\,L_\mu\,J^\mu,\qquad
L_\mu       = \bar{u}(p_{\nu_\tau})\,\gamma_\mu(1-\gamma_5)\,
        u(p_{\tau}),\end{split}
\end{equation*}
where $p_\tau$ is the momentum of the $\tau$ and
$p_{\nu_\tau}$ is the momentum of the neutrino produced in the
decay. The information on the decay products of the virtual $W$
boson is contained in the hadronic current, $J^\mu$. This
factorization allows us to implement the leptonic current $L_\mu$
for the decaying tau and the hadronic current separately and then
combine them to calculate the $\tau$ decay matrix element.

In Herwig, this factorization is used to define a \href{https://herwig.hepforge.org/doxygen/classHerwig\_1\_1TauDecayer.html}{TauDecayer} class, which implements
the calculation of the leptonic current for the $\tau$ decay and
uses a class inheriting from the \href{https://herwig.hepforge.org/doxygen/classHerwig\_1\_1WeakCurrent.html}{WeakCurrent} class to calculate the hadronic current for a
given decay mode. This factorization allows us to reuse the hadronic
currents in other applications, for example in weak meson decay using
the naïve factorization approximation or in the decay of the lightest
chargino to the lightest neutralino in Anomaly Mediated SUSY
Breaking (AMSB) models where there is a small mass difference between
the neutralino and chargino. We also make use of these currents, with
an appropriate isospin rotation, to allow the calculation of the cross
sections for exclusive hadronic final states in low energy $e^+e^-$
collisions \cite{Plehn:2019jeo}.

\subsubsection{Hadronic currents}
\label{\detokenize{review/decays:hadronic-currents}}\label{\detokenize{review/decays:sect-weakcurrents}}

We have implemented a number of hadronic currents, which are mainly used
for the simulation of $\tau$ decays
or the exclusive cross sections for the production of hadrons in low-energy $e^+e^-$ collisions.
These are all based on the \texttt{WeakCurrent}
class.
In many cases the same currents can be used to describe both $e^+e^-\to\mathrm{hadrons}$ and
hadronic decays. In $e^+e^-\to\mathrm{hadrons}$ however there can be additional isospin zero
contributions. Also due to the requirement that the final state has the quantum numbers of the photon the
three pion final-state is described by different currents in $\tau$ decays and  $e^+e^-$
annihilation.
In this section we list the available hadronic currents together
with a brief description. A more detailed description can be found in
either Refs. \cite{Grellscheid:2007tt, Plehn:2019jeo} or the Doxygen documentation.

There is one current available to describe the leptonic decays of the $\tau$ lepton.
\begin{itemize}
\item {} 

\href{https://herwig.hepforge.org/doxygen/classHerwig\_1\_1LeptonNeutrinoCurrent.html}{LeptonNeutrinoCurrent}
The current for weak decay to a lepton and the associated anti-neutrino
is given by
\begin{equation*}
\begin{split}J^\mu = \bar{u}(p_\ell)\gamma^\mu(1-\gamma_5)v(p_{\bar{\nu}}),\end{split}
\end{equation*}

where $p_{\bar{\nu}}$ is the momentum of the anti-neutrino and
$p_\ell$ is the momentum of the charged lepton.

\end{itemize}

We provide two currents to describe the production of any spin-zero or -one mesons.
\begin{itemize}
\item {} 

\href{https://herwig.hepforge.org/doxygen/classHerwig\_1\_1ScalarMesonCurrent.html}{ScalarMesonCurrent} The simplest hadronic current is that for the production of a
pseudoscalar meson, \textit{e.g.} the current for the production of
$\pi^\pm$ in the decay of the tau. The hadronic current can be
written as
\begin{equation*}
\begin{split}J^\mu = f^{}_P\, p^\mu_P,\end{split}
\end{equation*}

where $p^\mu_P$ is the momentum of the pseudoscalar meson and
$f_P$ is the pseudoscalar meson decay constant.

\item {} 

\href{https://herwig.hepforge.org/doxygen/classHerwig\_1\_1VectorMesonCurrent.html}{VectorMesonCurrent} The current for the production of a vector meson is given by
\begin{equation*}
\begin{split}J^\mu = \sqrt{2}g_{V}\epsilon^{*\mu}_V,\end{split}
\end{equation*}

where $\epsilon^{*\mu}_V$ is the polarisation vector for the
outgoing meson and $g_V$ is the decay constant of the vector
meson.

\end{itemize}

We also have one base class designed to allow the easy implementation of three meson currents.
\begin{itemize}
\item {} 

\href{https://herwig.hepforge.org/doxygen/classHerwig\_1\_1ThreeMesonCurrentBase.html}{ThreeMesonCurrentBase}
In order to simplify the implementation of a number of standard currents
for the production of three pseudoscalar mesons we define the current in
terms of several form factors. The current is defined to be
\cite{Jadach:1993hs}
\begin{equation*}
\begin{split}J^\mu &= \left(g^{\mu\nu}-\frac{q^\mu q^\nu}{q^2}\right)
   \left[F_1(p_2-p_3)^\mu +F_2(p_3-p_1)^\mu+F_3(p_1-p_2)^\mu\right]  +q^\mu F_4
   +iF_5\epsilon^{\mu\alpha\beta\gamma}p_1^\alpha p_2^\beta p_3^\gamma,\end{split}
\end{equation*}

where $p_{1,2,3}$ are the momenta of the mesons in the order given
below and $F_{1\to5}$ are the form factors. We use this approach
for a number of three-meson modes that occur in $\tau$ decays:
$\pi^-  \pi^-    \pi^+$; $\pi^0  \pi^0    \pi^-$;
$K^-   \pi^-    K^+$; $K^0   \pi^-    \bar{K}^0$;
$K^-   \pi^0    K^0$; $\pi^0  \pi^0    K^-$;
$K^-   \pi^-    \pi^+$; $\pi^-  \bar{K}^0  \pi^0$;
$\pi^-  \pi^0    \eta$; $K^0_S\pi^-K^0_S$;
$K^0_L\pi^-K^0_L$; $K^0_S\pi^-K^0_L$. The current is
implemented in terms of these form factors in a base class so that any
model for these currents can be implemented by inheriting from this
class and specifying the form factors.

\end{itemize}

There are a number of currents available which can be used to describe either $\tau$ decays or
the isospin rotated state in $e^+e^-$ collisions.
\begin{itemize}
\item {} 

\href{https://herwig.hepforge.org/doxygen/classHerwig\_1\_1TwoPionRhoCurrent.html}{TwoPionRhoCurrent}
The weak current for production of either two pions or two kaons
via the $\rho$ resonances has the form
\begin{equation*}
\begin{split}J^\mu =(p_1-p_2)_\nu\left(g^{\mu\nu}-\frac{q^\mu q^\nu}{q^2}\right)
\frac1{\sum_k\alpha_k}\sum_k \alpha_k \mathrm{BW}_k(q^2),\end{split}
\end{equation*}

where $p_{1,2}$ are the momenta of the outgoing mesons,
$q=p_1+p_2$, $\mathrm{BW}_k(q^2)$ is the Breit-Wigner
distribution for the intermediate vector meson $k$ and
$\alpha_k$ is the weight for the resonance, which can be complex.
The Breit-Wigner terms are summed over the $\rho$
resonances that can contribute to a given decay mode.

The models of either Kühn and Santamaria \cite{Kuhn:1990ad}, which uses
a Breit-Wigner distribution with a $p$-wave running width, or
Gounaris and Sakurai \cite{Gounaris:1968mw} are supported for the shape
of the Breit-Wigner distribution.

\item {} 

\href{https://herwig.hepforge.org/doxygen/classHerwig\_1\_1TwoPionCzyzCurrent.html}{TwoPionCzyzCurrent}
We also supply the alternative model of Ref. \cite{Czyz:2010hj} to describe the production of two pions via
the $\rho$ resonance, and its excited states. The parameters of this model are fitted to
$e^+e^-$ annihilation data.

\item {} 

\href{https://herwig.hepforge.org/doxygen/classHerwig\_1\_1TwoKaonCzyzCurrent.html}{TwoKaonCzyzCurrent}
We also supply the alternative model of Ref. \cite{Czyz:2010hj},
with the parameters from the fit of Ref. \cite{Plehn:2019jeo} to
$e^+e^-$ annihilation data,
for the current for the production of two kaons.

\item {} 

\href{https://herwig.hepforge.org/doxygen/classHerwig\_1\_1PionPhotonCurrent.html}{PionPhotonCurrent}
We use the model of Ref. \cite{SND:2016drm}, via intermediate $\rho, \omega, \phi$ and
$\omega^\prime$ final states, to describe the current for the production of $\pi\gamma$
final states.

\item {} 

\href{https://herwig.hepforge.org/doxygen/classHerwig\_1\_1OmegaPionSNDCurrent.html}{OmegaPionSNDCurrent}
The implementation of the model of Ref. \cite{Achasov:2016zvn} to describe the production of
$\omega\pi$ final states. This model is provided to make comparisons and we use models
including the decay of the $\omega$ meson by default.

\item {} 

\href{https://herwig.hepforge.org/doxygen/classHerwig\_1\_1PhiPiCurrent.html}{PhiPiCurrent}
We use the simple model of  Refs. \cite{Plehn:2019jeo,BaBar:2007ceh}
to describe the production of $\phi\pi$ final states.

\item {} 

\href{https://herwig.hepforge.org/doxygen/classHerwig\_1\_1TwoPionPhotonCurrent.html}{TwoPionPhotonCurrent}
The branching ratio for the decay $\tau^-\to\omega\pi^-\nu_\tau$
is 1.95\% \cite{ParticleDataGroup:2022pth}. The majority of this decay is modelled as
an intermediate state in the four-pion current described below. However
there is an 8.35\% \cite{ParticleDataGroup:2022pth} branching ratio of the
$\omega$ into $\pi^0\gamma$, which must also be modelled. We
do this using a current for $\pi^-\pi^0 \gamma$ via an
intermediate $\omega$. The hadronic current for this mode,
together with the masses, widths and other parameters, are taken from
Ref. \cite{Jadach:1993hs}.

\item {} 

\href{https://herwig.hepforge.org/doxygen/classHerwig\_1\_1TwoPionPhotonCurrent.html}{TwoPionPhotonSNDCurrent}
The implementation of the model of Ref. \cite{Achasov:2016zvn} to describe the production of
two pions and a photon via the decay of the $\omega$ meson.

\item {} 

\href{https://herwig.hepforge.org/doxygen/classHerwig\_1\_1FourPionCzyzCurrent.html}{FourPionCzyzCurrent}
We use the model of \cite{Czyz:2008kw} to describe the cross section for four pion production in
electron-positron annihilation.

\item {} 

\href{https://herwig.hepforge.org/doxygen/classHerwig\_1\_1FourPionNovosibirskCurrent.html}{FourPionNovosibirskCurrent}
We use the model of Ref. \cite{Bondar:2002mw} %
\begin{footnote}\sphinxAtStartFootnote
It should be noted that there were a number of mistakes in this
paper, which were corrected in Ref. \cite{Golonka:2003xt}.
\end{footnote} to model the decay
of the $\tau$ to four pions. The model is based on a fit to
$e^+e^-$ data from Novosibirsk.

\item {} 

\href{https://herwig.hepforge.org/doxygen/classHerwig\_1\_1EtaPiPiCurrent.html}{EtaPiPiCurrent}
We use the fit of Ref. \cite{Plehn:2019jeo} using the model of Ref. \cite{Czyz:2013xga}
which uses the $\rho$ and its excited states to describe the production
of $\eta\pi\pi$ final states.

\item {} 

\href{https://herwig.hepforge.org/doxygen/classHerwig\_1\_1EtaPiPiDefaultCurrent.html}{EtaPiPiDefaultCurrent}
Before version 7.3 of Herwig our default model of the current for $\eta\pi\pi$ final state was that
of Ref. \cite{Decker:1992kj}, however from version 7.3 we use the more recent fit described above.

\item {} 

\href{https://herwig.hepforge.org/doxygen/classHerwig\_1\_1EtaPrimePiPiCurrent.html}{EtaPrimePiPiCurrent}
We use the fit of Ref. \cite{Plehn:2019jeo} using the model of Ref. \cite{Czyz:2013xga}
which uses the $\rho$ and its excited states to describe the production
of $\eta^\prime\pi\pi$ final states.

\end{itemize}

We also have a number of currents which can only be used to describe hadronic $\tau$ decays.
These currents usually either include a large pseudovector component which is not present in
$e^+e^-$ annihilations, or have final states which are forbidden in $e^+e^-$ annihilations.
\begin{itemize}
\item {} 

\href{https://herwig.hepforge.org/doxygen/classHerwig\_1\_1KPiKStarCurrent.html}{KPiKStarCurrent}
The weak current for production of a kaon and a pion uses the same form as the
\texttt{TwoPionRhoCurrent}
described above.

\item {} 

\href{https://herwig.hepforge.org/doxygen/classHerwig\_1\_1KPiCurrent.html}{KPiCurrent}
Unlike the $\pi^+\pi^0$ decay of the tau the $K\pi$ decay
mode can occur via either intermediate scalar or vector mesons. We
therefore include a model for the current for the $K\pi$ decay
mode including the contribution of both vector and scalar resonances
based on the model of Ref. \cite{Finkemeier:1996dh}. The current is
given by
\begin{equation*}
\begin{split}J^\mu = c_V(p_1-p_2)_\nu\frac1{\sum_k\alpha_k}\sum_k\alpha_k{\rm BW}_k(q^2)
        \left(g^{\nu\mu}-\frac{q^\nu q^\mu}{M^2_k}\right)+c_Sq^\mu\frac1{\sum_k\beta_k}\sum_k\beta_k{\rm BW}_k(q^2),\end{split}
\end{equation*}

where $p_{1,2}$ are the momenta of the outgoing mesons,
$q=p_1+p_2$, ${\rm BW}_k(q^2)$ is the Breit-Wigner
distribution for the intermediate resonance $k$ and $\alpha_k$ is its
weight. The sum over the resonances is over the vector
$K^*$ states in the first, vector, part of the current and the
excited scalar $K^*$ resonances in the second, scalar, part of the
current. By default the vector part of the current includes the
$K^*(892)$ and $K^*(1410)$ states and the scalar part of the
current includes the $K^*_0(1430)$ together with the option of
including the $\kappa(800)$ to model any low-mass enhancement in
the mass of the $K\pi$ system, although additional resonances can
be included if necessary.

\item {} 

\href{https://herwig.hepforge.org/doxygen/classHerwig\_1\_1ThreePionDefaultCurrent.html}{ThreePionDefaultCurrent}
This is the implementation of the model of Refs.
\cite{Jadach:1993hs, Kuhn:1990ad, Decker:1992kj}, which uses the form of
Ref. \cite{Kuhn:1990ad} for the $a_1$ width, for the current for the production of three pions in $\tau$
lepton decays. The form factors for
the different modes are given in Refs. \cite{Jadach:1993hs, Decker:1992kj}.

\item {} 

\href{https://herwig.hepforge.org/doxygen/classHerwig\_1\_1ThreePionCLEOCurrent.html}{ThreePionCLEOCurrent}
This is the implementation of the model of Ref. \cite{Asner:1999kj} for
the weak current for three pions. This model includes $\rho$
mesons in both the $s$- and $p$-wave, the scalar
$\sigma$ resonance, the tensor $f_2$ resonance and scalar
$f_0(1370)$. The form factors for the $\pi^0\pi^0\pi^-$ mode
are given in Ref. \cite{Asner:1999kj} and the others can be obtained by
isospin rotation.

\item {} 

\href{https://herwig.hepforge.org/doxygen/classHerwig_1_1OneKaonTwoPionDefaultCurrent.html}{OneKaonTwoPionMesonDefaultCurrent}
This current implements the $K\pi\pi$ currents using the model of Refs.
\cite{Jadach:1993hs, Kuhn:1990ad, Decker:1992kj}, which uses the form of
Ref. \cite{Kuhn:1990ad} for the $a_1$ width. The form factors for
the different modes are given in Refs.
\cite{Jadach:1993hs, Decker:1992kj}.

\item {} 

\href{https://herwig.hepforge.org/doxygen/classHerwig\_1\_1OneKaonTwoPionCurrent.html}{OneKaonTwoPionMesonCurrent}
This current implements the $K\pi\pi$ currents using the model of Ref.
\cite{Finkemeier:1995sr}. Like the model of Ref. \cite{Decker:1992kj} this model
is designed to reproduce the correct chiral
limit for tau decays to three mesons. However, this model makes a
different choice of the resonances to use away from this limit for the
decays involving at least one kaon and in the treatment of the
$K_1$ resonances. The form factors for the different modes are
given in Ref. \cite{Finkemeier:1995sr}.

\item {} 

\href{https://herwig.hepforge.org/doxygen/classHerwig_1_1OneKaonTwoPionDefaultCurrent.html}{TwoKaonOnePionMesonDefaultCurrent}
This current implements the $KK\pi$ currents using the model of Refs.
\cite{Jadach:1993hs, Kuhn:1990ad, Decker:1992kj} described above.

\item {} 

\href{https://herwig.hepforge.org/doxygen/classHerwig\_1\_1OneKaonTwoPionCurrent.html}{TwoKaonOnePionMesonCurrent}
This current implements the $KK\pi$ currents using the model of Ref. \cite{Finkemeier:1995sr}
described above.

\item {} 

\href{https://herwig.hepforge.org/doxygen/classHerwig\_1\_1FivePionCurrent.html}{FivePionCurrent}
We use the model of Ref. \cite{Kuhn:2006nw}, which includes
$\rho\omega$ and $\rho\sigma$ intermediate states, via the
$a_1$ meson, to model the five-pion decay modes of the
$\tau$.

\end{itemize}

In addition we have a number of currents which are primarily used
to describe low energy $e^+e^-$ annihilation into hadrons.
These primarily have isospin zero, or contain isospin zero and one components but not
the pseudovector contributions required to describe $\tau$ lepton decay.
\begin{itemize}
\item {} 

\href{https://herwig.hepforge.org/doxygen/classHerwig\_1\_1EtaOmegaCurrent.html}{EtaOmegaCurrent}
We use the model of Ref. \cite{Achasov:2016qvd} which proceeds via excited $\omega$ states to
describe the production of $\eta\omega$ final states.

\item {} 

\href{https://herwig.hepforge.org/doxygen/classHerwig\_1\_1EtaPhiCurrent.html}{EtaPhiCurrent}
We use the model of Ref. \cite{Achasov:2018ygm} which uses the $\phi(1680)$ intermediate state to
describe the production of $\eta\phi$ final states.

\item {} 

\href{https://herwig.hepforge.org/doxygen/classHerwig\_1\_1EtaPhotonCurrent.html}{EtaPhotonCurrent}
This is the implementation of the model of \cite{Achasov:2006dv} for the production of
$\eta\gamma$ final state via intermediate $\rho, \omega, \phi$ and $\omega^\prime$ states.

\item {} 

\href{https://herwig.hepforge.org/doxygen/classHerwig_1_1ThreePionCzyzCurrent.html}{ThreePionCzyzCurrent}
We use the model of Ref. \cite{Czyz:2005as} to describe the isospin zero current for the
production of three pions, via $\omega$ and $\phi$ resonances, and their excited states.

\item {} 

\href{https://herwig.hepforge.org/doxygen/classHerwig_1_1KKPiCurrent.html}{KKPiCurrent}
This is a simple model for the production of $KK\pi$ final state in $e^+e^-$ collisions
via $\rho$ and $\phi$ resonances (and the relevant excited states) to
and intermediate $K^*K$ intermediate state, see  Ref. \cite{Plehn:2019jeo} for more details.

\item {} 

\href{https://herwig.hepforge.org/doxygen/classHerwig_1_1OmegaPiPiCurrent.html}{OmegaPiPiCurrent}
This is the implementation of the simple model described in  Ref. \cite{Plehn:2019jeo} for the production
of the $\omega\pi\pi$ final states.

\end{itemize}

Finally we also provide a wrapper class
\href{https://herwig.hepforge.org/doxygen/classHerwig\_1\_1WeakBaryonCurrent.html}{WeakBaryonCurrent}
which allows the baryon form factors described in \hyperref[\detokenize{review/decays:sect-weakexclusive}]{Section \ref{\detokenize{review/decays:sect-weakexclusive}}} to
be used as a weak current in order to describe the low energy production of baryon-antibaryon
pairs in $e^+e^-$ collisions.

\subsection{Strong and electromagnetic hadron decays}
\label{\detokenize{review/decays:strong-and-electromagnetic-hadron-decays}}

The vast majority of the strong and electromagnetic decays in Herwig
are simulated using a few simple models based on the spin structure of
the decay. These simple models are supplemented with a small number of
specialized models, usually from experimental fits, for specific decay
modes. In this section we describe the different models we use for these
decays for the scalar, vector and tensor mesons. All of these are
implemented in \texttt{Decayer}
classes that inherit from the \texttt{DecayIntegrator}
class of Herwig.

For a number of the decays of bottomonium and charmonium resonances we
use inclusive electromagnetic and strong decays to $q\bar{q}$,
$gg$, $ggg$ and $gg\gamma$, which are described in a
separate section.

A number of decays are still performed using a phase-space distribution
generated using the \href{https://herwig.hepforge.org/doxygen/classHerwig\_1\_1Hw64Decayer.html}{Hw64Decayer},
which implements the same models as were available
in the FORTRAN HERWIG program. In addition we use the MAMBO algorithm,
\cite{Kleiss:1991rn}, implemented in the \href{https://herwig.hepforge.org/doxygen/classHerwig\_1\_1MamboDecayer.html}{MamboDecayer}
class, to generate the momenta
of the decay products according to a phase-space distribution for a
number of high-multiplicity modes.

In the matrix elements for all the decays given below $0$ refers to the decaying particle, and then
$1\ldots$ to the outgoing particles in the order specified by the name of the \texttt{Decayer}. The 4-momentum of any particle in given by $p^\mu_i$, its mass by $m_i$, the polarization vector of spin-1 particles
by $\epsilon^\mu_i$, the polarization tensor of spin-2 particles by $\epsilon^{\mu\nu}_i$ and the
rank-3 polarization tensor for spin-3 particles by $\epsilon^{\mu\nu\rho}_i$, where the index $i$ gives the particle.
The 4-dimension Levi-Civita tensor is given by $\epsilon^{\alpha\beta\gamma\delta}$ and $g$ is the coupling for
a specific decay channel.

\subsubsection{Scalar mesons}
\label{\detokenize{review/decays:scalar-mesons}}

While the majority of the scalar meson decays are performed using
general \texttt{Decayer}s
based on the spin structures, there are a number of models
implemented for the rare radiative decays of the light pseudoscalar
mesons, three-body decays of the $\eta$ and $\eta'$, and the
decay $\pi^0\to e^+e^-e^+e^-$.
\begin{itemize}
\item {} 

\href{https://herwig.hepforge.org/doxygen/classHerwig\_1\_1EtaPiGammaGammaDecayer.html}{EtaPiGammaGammaDecayer}
We use the Vector-Meson Dominance (VMD)-based model of Ref.
\cite{Holstein:2001bt} for the decays
$\eta,\eta'\to \pi^0 \gamma \gamma$. In practice we use a running
width for the $\rho$ to include the $\eta'$ decay as well as
the $\eta$ decay and take the parameters from Ref.
\cite{Holstein:2001bt}.

\item {} 

\href{https://herwig.hepforge.org/doxygen/classHerwig\_1\_1EtaPiPiGammaDecayer.html}{EtaPiPiGammaDecayer}
We use either a VMD type model or a model using either the theoretical
or experimental form of the Omnes function %
\begin{footnote}\sphinxAtStartFootnote
Our default choice is to use the experimental form of the Omnes
function.
\end{footnote} taken from Refs.
\cite{Holstein:2001bt, Venugopal:1998fq} for the decay of the
$\eta$ or $\eta'$ to $\pi^+\pi^-\gamma$.

\item {} 

\href{https://herwig.hepforge.org/doxygen/classHerwig\_1\_1EtaPiPiPiDecayer.html}{EtaPiPiPiDecayer}
The decay of a pseudoscalar meson, for example the $\eta$ or
$\eta'$, to two charged and one neutral or three neutral pions, or
of the $\eta'$ to two pions and the $\eta$, is performed
using a parameterization of the matrix element squared taken from Ref.
\cite{Beisert:2003zs}. The experimental results of Refs.
\cite{Gormley:1970qz} and \cite{Tippens:2001fm} are used for the
$\eta\to\pi^+\pi^-\pi^0$ and $\eta\to\pi^0\pi^0\pi^0$ decays
respectively. The theoretical values from Ref. \cite{Beisert:2003zs}
are used for the other decays.

\item {} 

\href{https://herwig.hepforge.org/doxygen/classHerwig\_1\_1EtaPiPiFermionsDecayer.html}{EtaPiPiFermionsDecayer}
We use the model described above in the
\texttt{EtaPiPiGammaDecayer}
together with the electromagnetic branching of the photon to describe the decay of the
$\eta$ or $\eta'$ to $\pi^+\pi^-\ell^+\ell^-$.

\item {} 

\href{https://herwig.hepforge.org/doxygen/classHerwig\_1\_1PScalar4FermionsDecayer.html}{PScalar4FermionsDecayer}
As the $\pi^0$ is so copiously produced it is one of the small
number of particles for which we include branching ratios below the
level of $10^{-4}$. The matrix element for the sub-leading decay
$\pi^0\to e^+e^-e^+e^-$ is taken to be the combination of the
standard matrix element for $\pi^0\to\gamma\gamma$ and the
branching of the photons into $e^+e^-$.

\item {} 

\href{https://herwig.hepforge.org/doxygen/classHerwig\_1\_1PScalarPScalarVectorDecayer.html}{PScalarPScalarVectorDecayer}
This matrix element is used to simulate the decay of the 2S pseudoscalar
mesons to a vector meson and a 1S pseudoscalar meson. It is also used
for the decay of some scalar mesons to vector mesons and another scalar
meson, which has the same spin structure. The matrix element has the
form
\begin{equation*}
\begin{split}\mathcal{M} = g\epsilon_2^\mu(p_0+p_1)_\mu.\end{split}
\end{equation*}
\item {} 

\href{https://herwig.hepforge.org/doxygen/classHerwig\_1\_1PScalarVectorFermionsDecayer.html}{PScalarVectorFermionsDecayer}
There are a number of decays of a pseudoscalar meson to either a vector
meson or the photon and a lepton-antilepton pair. The classic example is
the Dalitz decay of the neutral pion, $\pi^0\to\gamma e^+e^-$. We
take the propagator of the off-shell photon to be
$\frac1{m^2_{f\bar{f}}}$, where $m_{f\bar{f}}$ is the mass
of the fermion-antifermion pair. The option of including a vector meson
dominance form factor is included.

\item {} 

\href{https://herwig.hepforge.org/doxygen/classHerwig\_1\_1PScalarVectorVectorDecayer.html}{PScalarVectorVectorDecayer}
In practice the vast majority of the decays of pseudoscalar mesons to
two spin-1 particles are of the form $P\to\gamma\gamma$ for which,
because the photon is stable, it is not as important to have a good
description of the matrix element. There are however some decays, \textit{e.g.}
$\eta'\to\omega\gamma$, for which this matrix element is needed.

The matrix element is taken to be
\begin{equation*}
\begin{split}\mathcal{M} = g\epsilon^{\mu\nu\alpha\beta}
                   p_{1\mu}   \epsilon_{1\nu}
                  p_{2\alpha}\epsilon_{2\beta}.\end{split}
\end{equation*}
\item {} 

\href{https://herwig.hepforge.org/doxygen/classHerwig\_1\_1PseudoScalar2FermionsDecayer.html}{PseudoScalar2FermionsDecayer}
This is a simple model for the decay of a pseudoscalar meson to a fermion-antifermion pair. It is used to simulate the strong
decays of the $\eta_c$ meson into baryon-antibaryon pairs.

\item {} 

\href{https://herwig.hepforge.org/doxygen/classHerwig\_1\_1ScalarMesonTensorScalarDecayer.html}{ScalarMesonTensorScalarDecayer}
There are a limited number of decays of a (pseudo)scalar meson to a
tensor meson and another (pseudo)scalar meson. The matrix element takes
the form
\begin{equation*}
\begin{split}\mathcal{M} = g\epsilon^{\alpha\beta} p_{0\alpha} p_{2\beta}.\end{split}
\end{equation*}
\item {} 

\href{https://herwig.hepforge.org/doxygen/classHerwig\_1\_1ScalarScalarScalarDecayer.html}{ScalarScalarScalarDecayer}
The decay of a scalar meson to two scalar mesons has no spin structure
and we assume that the matrix element is simply constant, \textit{i.e.}
\begin{equation*}
\begin{split}\mathcal{M} = g.\end{split}
\end{equation*}

We still include a matrix element for this decay in order to simulate
both the off-shell effects in the decay and to give the correct partial
width to be used in the running width calculation for the incoming
particle.

\item {} 

\href{https://herwig.hepforge.org/doxygen/classHerwig\_1\_1ScalarVectorVectorDecayer.html}{ScalarVectorVectorDecayer}
A number of the scalar mesons decay to two vector mesons. The matrix
element is taken to have the form
\begin{equation*}
\begin{split}\mathcal{M} =g\left[ p_1 \cdot p_2 \epsilon_1 \cdot \epsilon_2
                        -p_1 \cdot \epsilon_2 p_2 \cdot\epsilon_1\right].\end{split}
\end{equation*}
\item {} 

\href{https://herwig.hepforge.org/doxygen/classHerwig\_1\_1Scalar2FermionsDecayer.html}{Scalar2FermionsDecayer}
We implement the decay of a scalar meson to a fermion-antifermion pair in order to simulate
the decay of scalar quarkonium state to baryon-antibaryon pairs.

\end{itemize}

We also make use of the  \href{https://herwig.hepforge.org/doxygen/classHerwig\_1\_1ScalarTo3ScalarDalitz.html}{ScalarTo3ScalarDalitz}
class which is primarily intended to implement three-body weak decays, see \hyperref[\detokenize{review/decays:weak-dalitz}]{Section \ref{\detokenize{review/decays:weak-dalitz}}} for more details, to simulate a number of
three-body decays of the $\eta_c$ meson. Models for the following decays are currently implemented:
\begin{itemize}
\item {} 

$\eta_c\to K^+K^-\eta$ using the model of Ref. \cite{BaBar:2014asx}, available as \texttt{{EtacKpKmEtaBABAR}};

\item {} 

$\eta_c\to K^+K^-\eta^\prime$ using the model of Ref. \cite{BaBar:2021fkz}, available as \texttt{{EtacKpKmEtaPBABAR}};

\item {} 

$\eta_c\to K^0_{S,K}K^0_{S,K}\eta$ using the model of Ref. \cite{BaBar:2014asx}, available as \texttt{{EtacKS0KS0EtaBABAR}} and \texttt{{EtacKL0KL0EtaBABAR}};

\item {} 

$\eta_c\to K^0_{S,K}K^0_{S,K}\eta^\prime$ using the model of Ref.  \cite{BaBar:2021fkz}, available as \texttt{{EtacKL0KL0EtaPBABAR}} and \texttt{{EtacKS0KS0EtaPBABAR}};

\item {} 

$\eta_c\to \pi^+\pi^-\eta$ using the model of Ref.  \cite{BaBar:2021fkz} , available as \texttt{{EtacPipPimEtaBABAR}};

\item {} 

$\eta_c\to \pi^+\pi^-\eta^\prime$ using the model of Ref.  \cite{BaBar:2021fkz} , available as \texttt{{EtacPipPimEtaPBABAR}};

\item {} 

$\eta_c\to \pi^0\pi^0\eta$ using the model of Ref.  \cite{BaBar:2021fkz} , available as \texttt{{EtacPi0Pi0EtaBABAR}};

\item {} 

$\eta_c\to \pi^0\pi^0\eta^\prime$ using the model of Ref.  \cite{BaBar:2021fkz}, available as \texttt{{EtacPi0Pi0EtaPBABAR}};

\item {} 

$\eta_c\to K^0_{S,K}K^0_{S,K}\pi^0$ using the model of Ref.   \cite{BaBar:2015kii}, available as \texttt{{EtacKL0KL0Pi0BABAR}} and \texttt{{EtacKS0KS0Pi0BABAR}};

\item {} 

$\eta_c\to K^+K^-\pi^0$ using the model of Ref.  \cite{BaBar:2015kii}, available as \texttt{{EtacKpKmPi0BABAR}};

\item {} 

$\eta_c\to K^0_{S,L}K^\pm\pi^\mp$ using the model of Ref.   \cite{BaBar:2015kii}, available as \texttt{{EtacKS0KpPimBABAR}} and \texttt{{EtacKL0KpPimBABAR}}.

\end{itemize}

\subsubsection{Vector mesons}
\label{\detokenize{review/decays:vector-mesons}}

With the exception of the three-pion decay modes of the $\omega$,
$\phi$ and $a_1$ mesons, some four-pion decays, and the decays of onium
resonances to two-pion and a lighter onium state, we use general \texttt{Decayer}s
based on the spin structure for all the
strong and electromagnetic vector and pseudovector meson decays.
\begin{itemize}
\item {} 

\href{https://herwig.hepforge.org/doxygen/classHerwig\_1\_1a1SimpleDecayer.html}{a1SimpleDecayer}
This class implements the model of Kühn and Santamaria
\cite{Kuhn:1990ad} for the decay of the $a_1$ meson to three
pions and only includes the lightest two $\rho$ meson multiplets
in the modelling of the decay.

\item {} 

\href{https://herwig.hepforge.org/doxygen/classHerwig\_1\_1a1ThreePionCLEODecayer.html}{a1ThreePionCLEODecayer}
This class implements the model of CLEO \cite{Asner:1999kj} for
$a_1$ decay to three pions, which is a fit to CLEO data on
$\tau^-\to\pi^0\pi^0\pi^-\nu_\tau$. The model includes the
coupling of the $a_1$ to the $\rho$, $\rho(1450)$,
$f_0(1370)$, $f_2(1270)$ and $\sigma$ mesons.

\item {} 

\href{https://herwig.hepforge.org/doxygen/classHerwig\_1\_1a1ThreePionDecayer.html}{a1ThreePionDecayer}
This class implements a model of $a_1$ decay to three pions based
on the modelling of the $a_1$ used in the $4\pi$ currents
for tau decays presented in Ref. \cite{Bondar:2002mw} and includes the
$\rho$ and $\sigma$ resonances.

\item {} 

\href{https://herwig.hepforge.org/doxygen/classHerwig\_1\_1f1FourPiDecayer.html}{f1FourPiDecayer}
The decay of the $f_1$ meson to four pions is complicated as many of the intermediate
states are either close to or below the production threshold. We include this model of the
decay via the $a_1$ and $\rho$ mesons to investigate these effects.

\item {} 

\href{https://herwig.hepforge.org/doxygen/classHerwig\_1\_1f1RhoPiPiDecayer.html}{f1RhoPiPiDecayer}
This is an alternative model for the simulation of the decay of the $f_1$ meson to four pions
via the $a_1$ meson  where the decay of the $\rho$ meson is not included.

\item {} 

\href{https://herwig.hepforge.org/doxygen/classHerwig\_1\_1OniumToOniumPiPiDecayer.html}{OniumToOniumPiPiDecayer}
The decay of onium resonances to lighter states and a pion pair, $\mathcal{O}'\to\mathcal{O}\pi\pi$, uses
the matrix element \cite{Brown:1975dz}
\begin{equation*}
\begin{split}\mathcal{M} = \epsilon_1\cdot\epsilon_0\left[
  \mathcal{A}\left(q^2-2m^2_\pi\right)+\mathcal{B}E_1E_2\right]
 +\mathcal{C}\left((\epsilon_1\cdot q_1)(\epsilon_0\cdot q_2)+
                      (\epsilon_1\cdot q_2)(\epsilon_0\cdot q_1)\right),\end{split}
\end{equation*}

where $\mathcal{A}$, $\mathcal{B}$ and
$\mathcal{C}$ are complex couplings, $m_\pi$ is the pion
mass, $E_{1,2}$ are the pion energies, $q_{1,2}$ are the
pion momenta and $q$ is the momentum of the $\pi\pi$ system.

The results of BES \cite{Bai:1999mj} are used for
$\psi'\to J/\psi$ and CLEO \cite{Cronin-Hennessy:2007sj} for
$\Upsilon(3S)$ and $\Upsilon(2S)$ decays. The remaining
parameters are chosen to approximately reproduce the distributions from
BaBar \cite{Aubert:2006bm} and CLEO \cite{Adam:2005mr} for
$\Upsilon(4S)$ and $\psi(3770)$ decays, respectively.

\item {} 

\href{https://herwig.hepforge.org/doxygen/classHerwig_1_1VectorMeson2FermionDecayer.html}{VectorMeson2FermionDecayer}
Most of the decays of the vector mesons to a fermion-antifermion pair
are the decays of the light vector mesons to electron and muon pairs,
and of the bottomonium and charmonium resonances to all the charged
leptons.

The matrix element is taken to have the form
\begin{equation*}
\begin{split}\mathcal{M} = g\epsilon_{0\mu} \bar{u}(p_f)\gamma^\mu v(p_{\bar{f}}),\end{split}
\end{equation*}

where $p_f$ is the
four-momentum of the outgoing fermion and $p_{\bar{f}}$ is the
four-momentum of the outgoing antifermion.

\item {} 

\href{https://herwig.hepforge.org/doxygen/classHerwig\_1\_1VectorMeson2MesonDecayer.html}{VectorMeson2MesonDecayer}
The matrix element for the decay of a vector meson to two scalar mesons\_(or other decays with the same parity) is given by
\begin{equation*}
\begin{split}\mathcal{M} = g\epsilon_0 \cdot (p_1-p_2).\end{split}
\end{equation*}
\item {} 

\href{https://herwig.hepforge.org/doxygen/classHerwig\_1\_1VectorMesonPScalarFermionsDecayer.html}{VectorMesonPScalarFermionsDecayer}
The decay of a vector meson to a pseudoscalar meson and a
fermion-antifermion pair is simulated using a matrix element based on
that for the $V\to VP$ vertex combined with the branching of the
vector, which is in reality always a photon, into a fermion-antifermion
pair.

\item {} 

\href{https://herwig.hepforge.org/doxygen/classHerwig\_1\_1VectorMesonTensorVectorDecayer.html}{VectorMesonTensorVectorDecayer}
The matrix element for the decay of a vector meson to a tensor and vector meson is given by
\begin{equation*}
\begin{split}\mathcal{M}= g\epsilon_0^{\beta_1\beta_2} \epsilon_1^{\alpha} \epsilon_2^{\gamma}
\left(g_{\beta_1\alpha}+\frac{p_{0\beta_1}p_{1\alpha}}{p_0\cdot p_1-m_0m_1}\right)
\left(g_{\beta_2\gamma}+\frac{p_{2\beta_2}p_{1\gamma}}{p_0\cdot p_1-m_1m_2}\right).\end{split}
\end{equation*}
\item {} 

\href{https://herwig.hepforge.org/doxygen/classHerwig\_1\_1VectorMesonVectorPScalarDecayer.html}{VectorMesonVectorPScalarDecayer}
The decay of a vector meson to another spin-1 particle and a
pseudoscalar meson is common in both the radiative decay of the 1S
vector mesons and the decay of higher vector multiplets to the 1S vector
mesons. The matrix element for the decay is
\begin{equation*}
\begin{split}\mathcal{M} = g\epsilon^{\mu\nu\alpha\beta}
     \epsilon_{0\mu} p_{0\nu}  p_{1\alpha} \epsilon_{1\beta}.\end{split}
\end{equation*}
\item {} 

\href{https://herwig.hepforge.org/doxygen/classHerwig\_1\_1VectorMesonVectorScalarDecayer.html}{VectorMesonVectorScalarDecayer}
We include a number of decays of the vector mesons to a scalar meson and
either the photon or another vector meson. In practice the vast majority
of these decays are radiative decays involving scalar mesons. The
remaining decays use the $\sigma$ meson as a model for four-pion
decays of the excited $\rho$ multiplets.

The matrix element for the decay is
\begin{equation*}
\begin{split}\mathcal{M}=g\epsilon_{0}\mu\left[ p_1 \cdot p_0 \epsilon_1^\mu
                    -p_1^\mu \epsilon_1 \cdot p_0\right].\end{split}
\end{equation*}
\item {} 

\href{https://herwig.hepforge.org/doxygen/classHerwig\_1\_1VectorMesonVectorVectorDecayer.html}{VectorMesonVectorVectorDecayer}
There are a small number of decays of excited $\rho$ multiplets to
$\rho$ mesons included in the simulation. We model these decays
using the matrix element
\begin{equation*}
\begin{split}\mathcal{M}= \frac{g}{m_0^2}
              ( p_{0\nu}\epsilon_0^\alpha-p_{0\alpha} \epsilon_0^\nu)\left[
              (p_{1\nu} \epsilon_1^\beta- p_1^\beta \epsilon_{1\nu})
              (p_{2\alpha} \epsilon_{2\beta}- p_{2\beta} \epsilon_{2\alpha})
             -(\nu \leftrightarrow\alpha)\right].\end{split}
\end{equation*}
\item {} 

\href{https://herwig.hepforge.org/doxygen/classHerwig\_1\_1VectorMeson2SpinHalfBaryonsDecayer.html}{VectorMeson2SpinHalfBaryonsDecayer}
In recent years the BES collaboration have measured the properties of a large number spin-$\frac12$ baryon-antibaryon
decay modes of the $J/\psi$ ($p\bar{p}$ \cite{BESIII:2012ion}, $n\bar{n}$ \cite{BESIII:2012ion},
$\Lambda^0\bar\Lambda^0$ \cite{BESIII:2022lsz,BESIII:2018cnd,BESIII:2017kqw,BESIII:2022qax}, $\Sigma^0\bar\Sigma^0$ \cite{BESIII:2017kqw},
$\Xi^0\bar{\Xi}^0$  \cite{BESIII:2016nix,BESIII:2023drj}, $\Xi^-\bar{\Xi}^+$  \cite{BESIII:2016ssr,BESIII:2021ypr}, $\Sigma^+\bar\Sigma^-$ \cite{BESIII:2020fqg})
and $\psi(2S)$ ($p\bar{p}$ \cite{BESIII:2018flj}, $n\bar{n}$ \cite{BESIII:2018flj},
$\Lambda^0\bar\Lambda^0$ \cite{BESIII:2017kqw}, $\Sigma^0\bar\Sigma^0$ \cite{BESIII:2017kqw},
$\Xi^0\bar{\Xi}^0$  \cite{BESIII:2016nix,BESIII:2023lkg}, $\Xi^-\bar{\Xi}^+$  \cite{BESIII:2016ssr,BESIII:2022lsz},
$\Sigma^+\bar\Sigma^+$ \cite{BESIII:2020fqg}, $\Sigma^-\bar\Sigma^+$ \cite{BESIII:2022upt}) states
and also the decay $\psi(3770)\to\Lambda^0\bar\Lambda^0$ \cite{BESIII:2021cvv}. We implement these decays
using the most recent measurements of the asymmetry and phase.

\item {} 

\href{https://herwig.hepforge.org/doxygen/classHerwig\_1\_1VectorMeson2SpinThreeHalfBaryonsDecayer.html}{VectorMeson2SpinThreeHalfBaryonsDecayer}
In recent years the BES collaboration have also measured the properties of a number spin-$\frac32$ baryon-antibaryon
decay modes of the $J/\psi$ ($\Sigma^{*0}\bar\Sigma^{*0}$  \cite{BESIII:2016nix}, $\Sigma^{*\mp}\bar\Sigma^{*\pm}$  \cite{BESIII:2016ssr})
and $\psi(2S)$ ($\Sigma^{*0}\bar\Sigma^{*0}$  \cite{BESIII:2016nix}, $\Sigma^{*\mp}\bar\Sigma^{*\pm}$  \cite{BESIII:2016ssr},
$\Xi^{*0}\bar{\Xi}^{*0}$ \cite{BESIII:2021gca}, $\Xi^{*-}\bar{\Xi}^{*+}$ \cite{BESIII:2019dve}) states. We implement these decays
using the most recent measurements of the asymmetry and phase.

\item {} 

\href{https://herwig.hepforge.org/doxygen/classHerwig\_1\_1VectorTo3PseudoScalarDalitz.html}{VectorTo3PseudoScalarDalitz}
We use the same approach as described in \hyperref[\detokenize{review/decays:weak-dalitz}]{Section \ref{\detokenize{review/decays:weak-dalitz}}} for scalar Dalitz decays to simulate the
decays of vector mesons to three pseudoscalar mesons. The
\texttt{VectorTo3PseudoScalarDalitz}
implements the decays but a range of intermediate states and Lorentz structures for the matrix element can be specified
in order to describe a range of decays. Currently we supply three models:
\begin{enumerate}
\sphinxsetlistlabels{\arabic}{enumi}{enumii}{}{.}%
\item {} 

\texttt{{OmegaDalitz}} for the decay $\omega\to\pi^+\pi^-\pi^0$;

\item {} 

\texttt{{PhiDalitz}}  for the decay  $\phi\to\pi^+\pi^-\pi^0$;

\item {} 

\texttt{{JpsiPipPimPi0BABAR}} for the decay $J/\psi\to\pi^+\pi^-\pi^0$.

\end{enumerate}

For the $\omega$ and $\phi$ we assumed the decay is dominated by the
production of the lowest lying $\rho$ multiplet. Our default model
for the matrix element for this decay is
\begin{equation*}
\begin{split}\mathcal{M} = g\epsilon^{\mu\alpha\beta\nu}\epsilon_\mu p_{1\alpha} p_{2\beta} p_{3\nu}
            \left[d+\sum_if_i\left[\mathrm{BW}_i(s_{12})+\mathrm{BW}_i(s_{13})+\mathrm{BW}_i(s_{23})\right]\right]\!,\end{split}
\end{equation*}

where $s_{ij}=(p_i+p_j)^2$, $d$ is a complex coupling for the direct interaction,
$f_i$ is the coupling of the $i$th $\rho$ multiplet
and $\mathrm{BW}_i(s)$ is a Breit-Wigner distribution with a
$p$-wave running width. This is an extension of the model used by
KLOE \cite{Aloisio:2003ur} to include higher $\rho$ multiplets.

For the decay of the $J/\psi$ we use the model of BABAR from Ref. \cite{BaBar:2017dwm}.

\end{itemize}

\subsubsection{Tensor mesons}
\label{\detokenize{review/decays:tensor-mesons}}

Only a relatively small number of tensor meson states are included in
the simulation, compared to the vector and scalar mesons. All their
decays are simulated using a small number of matrix elements based on
the spin structure of the decays. Many of the multi-body decays of the
tensor mesons are simulated using these two-body matrix elements with
off-shell vector and scalar mesons.
\begin{itemize}
\item {} 

\href{https://herwig.hepforge.org/doxygen/classHerwig\_1\_1PseudoTensorMesonTensorVectorDecayer.html}{PseudoTensorMesonTensorVectorDecayer}
The decay of a pseudotensor meson to a tensor and vector meson is described using the matrix element
\begin{equation*}
\begin{split}\mathcal{M}=g\epsilon_0^{\alpha_1\alpha_2} \epsilon_1^{\beta_1\beta_2}
\epsilon_{\alpha_1\beta_1\gamma_1p_2}\epsilon_{2\gamma_1}
\left(g_{\alpha_2\beta_2}+\frac{p_{1\alpha_2}p_{0\beta_2}}{p_0\cdot p_1-m_0m_1}\right).\end{split}
\end{equation*}
\item {} 

\href{https://herwig.hepforge.org/doxygen/classHerwig\_1\_1PseudoTensorMesonVectorVectorDecayer.html}{PseudoTensorMesonVectorVectorDecayer}
The matrix element for a pseudotensor meson decaying to two vector meson is taken to be
\begin{equation*}
\begin{split}\mathcal{M}=g\epsilon_0^{\alpha_1\alpha_2} \epsilon_1^{\beta} \epsilon_2^{\gamma}  \epsilon^{\mu\alpha_1\beta'\gamma'}(p_1-p_2)_{\alpha_2}
G_{\beta\beta'} G_{\gamma\gamma'},\end{split}
\end{equation*}
where
\begin{equation*}
\begin{split}G_{\mu\nu} = g_{\mu\nu} + \frac1X\left(-p_1\cdot p_2(p_{2\mu}p_{1\nu}p_{1\mu}p_{2\nu}) + m_2^2p_{1\mu}p_{1\nu} + m_1^2p_{2\mu}p_{2\nu}\right),\end{split}
\end{equation*}
where $X=(p_1\cdot p_2)^2-m^2_1m^2_2$.

\item {} 

\href{https://herwig.hepforge.org/doxygen/classHerwig\_1\_1TensorMeson2PScalarDecayer.html}{TensorMeson2PScalarDecayer}
The matrix element for the decay of a tensor meson to two
pseudoscalar (or scalar) mesons is
\begin{equation*}
\begin{split}\mathcal{M} = g\epsilon_0^{\mu\nu}p_{1\mu}p_{2\nu}.\end{split}
\end{equation*}
\item {} 

\href{https://herwig.hepforge.org/doxygen/classHerwig\_1\_1TensorMesonVectorPScalarDecayer.html}{TensorMesonVectorPScalarDecayer}
There are a number of decays of tensor mesons to a spin-1 particle,
either a vector meson or the photon, and a pseudoscalar meson, examples
include $a_2\to\rho\pi$ and $a_2\to\pi\gamma$. The matrix
element is taken to be
\begin{equation*}
\begin{split}\mathcal{M}=g\epsilon_0^{\mu\nu}p_{2\mu} \epsilon_{\nu\alpha\beta\gamma}
                  p_1^\alpha \epsilon_1^\beta p_2^\gamma.\end{split}
\end{equation*}
\item {} 

\href{https://herwig.hepforge.org/doxygen/classHerwig\_1\_1TensorMesonSpin3VectorDecayer.html}{TensorMesonSpin3VectorDecayer}
We take the matrix element for the decay of a spin-2 meson to a spin-3 meson and a vector meson to be
\begin{equation*}
\begin{split}\mathcal{M} &=  g \epsilon_1^{\alpha_1\alpha_2\alpha_3}\epsilon_0^{\beta_1\beta_2} \epsilon_2^\gamma
   \left(g_{\alpha_1\beta_1} - \frac{p_{1\alpha_1}p_{0\beta_1}}{p_0\cdot p_1-m_0m_1}\right)
   \left(g_{\alpha_2\beta_2} - \frac{p_{1\alpha_2}p_{0\beta_2}}{p_0\cdot p_1-m_0m_1}\right) \\
&  \left(g_{\alpha_3\gamma } - \frac{p_{2\alpha_3}p_{1\gamma }}{p_0\cdot p_1-m_0m_1}\right).\end{split}
\end{equation*}
\item {} 

\href{https://herwig.hepforge.org/doxygen/classHerwig\_1\_1TensorMesonTensorPScalarDecayer.html}{TensorMesonTensorPScalarDecayer}
The matrix element for the decay of a spin-2 meson to a spin-2 meson and a pseudoscalar meson is taken to be
\begin{equation*}
\begin{split}\mathcal{M}=g\epsilon_0^{\alpha_1\alpha_2} \epsilon_1^{\beta_1\beta_2}
\left(g_{\alpha_1\beta_1}+\frac{p_{1\alpha_1}p_{0\beta_1}}{p_0\cdot p_1-m_0m_1}\right)
\epsilon_{\mu\nu\alpha_2\beta_2} p_{1\mu}p_{2\nu}.\end{split}
\end{equation*}
\item {} 

\href{https://herwig.hepforge.org/doxygen/classHerwig\_1\_1TensorMesonTensorScalarDecayer.html}{TensorMesonTensorScalarDecayer}
The matrix element for the decay of a spin-2 meson to a spin-2 meson and a scalar meson is taken to be
\begin{equation*}
\begin{split}\mathcal{M}=g\epsilon_0^{\alpha_1\alpha_2} \epsilon_1^{\beta_1\beta_2}
\left(g_{\alpha_1\beta_1}+\frac{p_{1\alpha_1}p_{0\beta_1}}{p_0\cdot p_1-m_0m_1}\right)
\left(g_{\alpha_2\beta_2}+\frac{p_{1\alpha_2}p_{0\beta_2}}{p_0\cdot p_1-m_0m_1}\right).\end{split}
\end{equation*}
\item {} 

\href{https://herwig.hepforge.org/doxygen/classHerwig\_1\_1TensorMesonVectorScalarDecayer.html}{TensorMesonVectorScalarDecayer}
We take the matrix element for the decay of a tensor meson to a vector and pseudoscalar meson to be
\begin{equation*}
\begin{split}\mathcal{M}=g\epsilon_0^{\alpha_1\alpha_2} \epsilon_1^{\beta}
\left(g_{\alpha_1\beta}+\frac{p_{1\alpha_1}p_{0\beta}}{p_0\cdot p_1-m_0m_1}\right)(p_1-p_2)_{\alpha_2}.\end{split}
\end{equation*}
\item {} 

\href{https://herwig.hepforge.org/doxygen/classHerwig\_1\_1TensorMesonVectorVectorDecayer.html}{TensorMesonVectorVectorDecayer}
We have based our matrix element for the decay of a tensor meson to two
vector mesons on the perturbative graviton decay matrix element
\cite{Han:1998sg} in such a way that it vanishes if the polarisations
of the outgoing vectors are replaced with their momenta. The matrix
element is
\begin{equation*}
\begin{split}\mathcal{M} &=  g \Big[
\epsilon_{0\mu\nu}\left\{
\left(\epsilon_{1\alpha} p_1^\mu - \epsilon_1^\mu p_{1\alpha}\right)
\left(\epsilon_2^\alpha  p_2^\nu - \epsilon_2^\nu p_2^\alpha\right)
+\left(\mu\leftrightarrow\nu\right)\right\}
\nonumber \\
&-\frac12\epsilon^\mu_\mu
\left(\epsilon_{1\alpha} p_{1\beta}- \epsilon_{1\beta} p_{1\alpha}\right)
\left(\epsilon_2^\alpha  p_2^\beta - \epsilon_2^\beta p_2^\alpha\right)\Big].\end{split}
\end{equation*}

In practice this matrix element is mainly used with off-shell
vector mesons to model three- and four-body decays of the tensor mesons.

\end{itemize}

\subsubsection{Spin-Three mesons}
\label{\detokenize{review/decays:spin-three-mesons}}

Starting with version 7.3 we also include a small number of matrix elements for the decay of spin-3 mesons
based on the spin structures of the decays.
\begin{itemize}
\item {} 

\href{https://herwig.hepforge.org/doxygen/classHerwig\_1\_1Spin3Meson2PScalarDecayer.html}{Spin3Meson2PScalarDecayer}
The matrix element for the decay of a spin-3 meson to two (pseudo)scalar mesons is taken to be
\begin{equation*}
\begin{split}\mathcal{M} = g\epsilon_0^{\mu\nu\rho}(p_1-p_2)_\mu(p_1-p_2)_\nu(p_1-p_2)_\rho.\end{split}
\end{equation*}
\item {} 

\href{https://herwig.hepforge.org/doxygen/classHerwig\_1\_1Spin3MesonTensorPScalarDecayer.html}{Spin3MesonTensorPScalarDecayer}
The matrix element for the decay of a spin-3 meson to spin-2 meson and a pseudoscalar is taken to be
\begin{equation*}
\begin{split}\mathcal{M} =  g \epsilon_0^{\alpha_1\alpha_2\alpha_3}\epsilon_1^{\beta_1\beta_2} \epsilon_{\mu\nu\alpha_1\beta_1}\left(g_{\alpha_2\beta_2} - \frac{p_{1\alpha_2}p_{0\beta_2}}{p_0\cdot p_1-m_0m_1}\right)p_0^\mu p_1^\nu(p_1-p_2)_{\alpha_3}.\end{split}
\end{equation*}
\item {} 

\href{https://herwig.hepforge.org/doxygen/classHerwig\_1\_1Spin3MesonTensorVectorDecayer.html}{Spin3MesonTensorVectorDecayer}
The matrix element for the decay of a spin-3 meson to spin-2 meson and a vector meson is taken to be
\begin{equation*}
\begin{split}\mathcal{M} &=  g \epsilon_0^{\alpha_1\alpha_2\alpha_3}\epsilon_1^{\beta_1\beta_2} \epsilon_2^\gamma
  \left(g_{\alpha_1\beta_1} - \frac{p_{1\alpha_1}p_{0\beta_1}}{p_0\cdot p_1-m_0m_1}\right)
  \left((g_{\alpha_2\beta_2} - \frac{p_{1\alpha_2}p_{0\beta_2}}{p_0\cdot p_1-m_0m_1}\right) \\
&  \left((g_{\alpha_3\gamma } - \frac{p_{2\alpha_3}p_{0\gamma }}{p_0\cdot p_2-m_0m_2}\right).\end{split}
\end{equation*}
\item {} 

\href{https://herwig.hepforge.org/doxygen/classHerwig\_1\_1Spin3MesonVectorPScalarDecayer.html}{Spin3MesonVectorPScalarDecayer}
The matrix element for the decay of a spin-3 meson to vector meson and a pseudoscalar meson is taken to be
\begin{equation*}
\begin{split}\mathcal{M} =  g \epsilon_0^{\mu\nu\rho}\epsilon_1^\beta \epsilon_{\alpha\gamma\mu\beta}(p_1-p_2)_\nu (p_1-p_2)_\rho p_0^\alpha p_1^\beta.\end{split}
\end{equation*}
\item {} 

\href{https://herwig.hepforge.org/doxygen/classHerwig\_1\_1Spin3MesonVectorScalarDecayer.html}{Spin3MesonVectorScalarDecayer}
The matrix element for the decay of a spin-3 meson to vector meson and a scalar meson is taken to be
\begin{equation*}
\begin{split}\mathcal{M} =  g \epsilon_0^{\mu\nu\rho}\epsilon_1^\beta \left(g_{\mu\beta} - \frac{p_{1\mu}p_{0\beta}}{p_0\cdot p_1-m_0m_1}\right) (p_1-p_2)_\nu (p_1-p_2)_\rho.\end{split}
\end{equation*}
\end{itemize}

\subsubsection{Excited heavy mesons}
\label{\detokenize{review/decays:excited-heavy-mesons}}

In Ref. \cite{Masouminia:2023zhb}, a systematic approach to transmitting the spin information of heavy hadron constituents was detailed, aiming to simulate the polarisation states of excited heavy mesons and heavy baryons within Herwig 7 using the principles of HQET and spin-flavour symmetry. HQET has been already utilised for the strong and radiative decays of heavy baryons, as evidenced by works in references \cite{Ivanov:1999bk, Ivanov:1998wj} and its incorporation into Herwig++ \cite{Bahr:2008pv, Bellm:2015jjp}. However, there remains a pressing need to model the decays of excited heavy mesons to better align predictions with the experimental observations. Here we focus on the interactions specific to charm mesons, though the derived results are similarly applicable to bottom mesons. Initially, we consider the $s$- and $p$-wave mesons and identify three meson multiplets under heavy quark symmetry, as presented in \cite{Falk:1992cx}:
\begin{itemize}
\item {} 

The ground state ($J^P=0^-, 1^-$ doublet) includes $D$ and $D^{\star}$ mesons, characterised by $l=0$ for the light degrees of freedom.

\item {} 

The $J^P=1^+, 2^+$ doublet consists of $D_1$ and $D_2^{\star}$, distinguished by $l=1$ and $j=\frac32$ for the light degrees of freedom.

\item {} 

The $J^P=0^+, 1^+$ doublet includes $D_0^{\star}$ and $D_1^\prime$, defined by $l=1$ and $j=\frac12$ for the light degrees of freedom.

\end{itemize}

Additionally, there is potential for mixing between $D_1$ and $D_1^\prime$, arising from sub-leading corrections within the heavy quark limit.

To determine the relevant matrix elements for these decays, we follow the methodology and notation detailed in Refs. \cite{Falk:1992cx, Falk:1995th}, aiming to balance a coherent representation of experimental data with theoretical accuracy. We prioritise the leading terms within the heavy quark limit for the interactions, while also accommodating the mixing between the $D_1$ and $D_1^\prime$ mesons, $\theta_q$. This yields the following expressions for the decay matrix elements:
\begin{equation*}
\begin{split}\mathcal{M}(D^{\star}\to D\pi) &= -\frac{2g}{f_\pi} \left( m_D m_{D^{\star}} \right)^{1/2} p_0 \cdot \epsilon_0, \\
\mathcal{M}(D_2^{\star}\to D\pi) &= -\frac{2h}{f_\pi \Lambda} \left( m_{D_2} m_D^{\star} \right)^{1/2} \epsilon_0^{\mu\nu} p_{0,\mu} p_{0,\nu}, \\
\mathcal{M}(D_2^{\star}\to D^{\star}\pi) &= -i \frac{2h}{f_\pi \Lambda} \left( \frac{m_{D^{\star}}}{m_{D_2}} \right)^{1/2}  \epsilon^{\alpha\beta\mu\nu} \epsilon^0_{\alpha\gamma} p_0^\gamma p_{0,\mu} p_{1\nu} \epsilon_{1\beta}, \\
\mathcal{M}(D_1\to D^{\star}\pi) &= \frac{h}{f_\pi\Lambda} \left( \frac{2}{3} m_{D_1} m_D \right)^{1/2} \left[ \epsilon_0 \cdot \epsilon_1 \left(p_0^2 - \left(\frac{p_0 \cdot p_1}{m_0}\right)^2\right) - 3 \epsilon_0 \cdot p_0 \epsilon_1 \cdot p_0 \right], \\
\mathcal{M}(D_0^{\star}\to D\pi) &= \frac{f^{\prime\prime}}{f_\pi} \left( m_{D_0^{\star}} m_D \right)^{1/2} p_0 \cdot \left(\frac{p_1}{m_{D^{\star}_0}} + \frac{p_2}{m_D}\right), \\
\mathcal{M}(D_1^\prime \to D^{\star}\pi) &= -\frac{f^{\prime\prime}}{f_\pi} \left( m_{D_1^\prime} m_D \right)^{1/2} \left[ -p_0 \cdot \left(\frac{p_1}{m_{D^{\star}_0}} + \frac{p_2}{m_D}\right) \epsilon_0 \cdot \epsilon_1 \right. \\
& \left. + \frac{1}{m_{D'_1}} \epsilon_1 \cdot p_1 \epsilon_0 \cdot p_0 + \frac{1}{m_D} \epsilon_0 \cdot p_2 \epsilon_1 \cdot p_0 \right].\end{split}
\end{equation*}

Here, $p_i$ and $\epsilon_i$ are the momenta and polarisation vectors of the hadrons, with $i=0$ indicating the parent hadron and $i=1,2$ the heavy and light child hadrons, respectively. $m_H$ is the mass of hadron $H$, and $g$, $h$, $\Lambda$, $f_\pi$, and $f^{\prime\prime}$ are decay parameters. We also introduce $\theta_q$ as the mixing angles between the $(D_1, D'_1)$ and $(D_{s1}, D'_{s1})$ mesons.

Having the matrix elements of the decays, we can calculate the partial widths using the Fermi golden rule for two-body decays:
\begin{equation*}
\begin{split}\Gamma(H^{\star} \to H \pi) = \frac{\Theta(H^{\star} \to H \pi)}{2 m_{H^{\star}}} \left| \mathcal{M}(H^{\star} \to H \pi) \right|^2,\end{split}
\end{equation*}

with $\Theta$ being the two-body phase-space factor. To ensure consistency between the theoretical calculations and our implementation in Herwig 7, we retain some sub-leading terms when considering the heavy quark limit. Explicitly, we use the following prescription for our calculations:
\begin{equation*}
\begin{split}\Gamma(H^{\star} \to H \pi) = \frac{1}{8 \pi m_{H^{\star}}^2} \left| \mathcal{M}(H^{\star} \to H \pi) \right|^2 p_{\rm CM},\end{split}
\end{equation*}

where $p_{\rm CM}$ denotes the momentum of the parent hadron in the kinematic centre-of-mass frame of the two-body decay. The resulting partial widths are outlined below:
\begin{equation*}
\begin{split}\Gamma(D^{\star} \to D\pi) &= \frac{g^2}{6\pi f_\pi^2} \frac{m_D}{m_{D^{\star}}} p_{\text{cm}}^3, \\
\Gamma(D_2^{\star} \to D^{\star}\pi) &= \frac{h^2}{15\pi f_\pi^2 \Lambda^2} \frac{m_{D^{\star}}}{m_{D_2^{\star}}} p_{\text{cm}}^5, \\
\Gamma(D_2^{\star} \to D\pi) &= \frac{h^2}{10 \pi f_\pi^2 \Lambda^2} \frac{m_D}{m_{D_2^{\star}}} p_{\text{cm}}^5, \\
\Gamma(D_1 \to D^{\star}\pi) &= \frac{h^2}{144 \pi f_\pi^2 \Lambda^2} \frac{\left[-2 m_{D_1}^2 (m_{D^{\star}}^2 - 5 m_\pi^2) + (m_\pi^2 - m_{D^{\star}}^2)^2 + 25 m_{D_1}^4 \right]}{m_{D^{\star}} m_{D_1}^3} p_{\text{cm}}^5, \\
\Gamma(D_0^{\star} \to D\pi) &= \frac{f^{\prime\prime 2}}{32 \pi f_\pi^2} \frac{\left(m_{D_0^{\star}} - m_D\right)^2}{m_D m_{D_0^{\star}}^3} \left[ \left(m_{D_0^{\star}} + m_D\right)^2 - m_\pi^2 \right]^2 p_{\text{cm}}, \\
\Gamma(D_1^\prime \to D^{\star}\pi) &= \frac{f^{\prime\prime 2}}{32 \pi f_\pi^2} \frac{\left(m_{D_1^\prime} - m_{D^{\star}}\right)^2}{m_{D^{\star}} m_{D_1^\prime}^3} \left[ \left(m_{D_1^\prime} + m_{D^{\star}}\right)^2 - m_\pi^2 \right]^2 p_{\text{cm}}.\end{split}
\end{equation*}

To extract the numerical values of the decay parameters, we utilise the latest observed masses and decay widths of the charmed mesons from BaBar \cite{BaBar:2010zpy, BaBar:2011vbs} and LHCb \cite{LHCb:2013jjb} collaborations. The complete list of utilised masses and widths for the $(0^-,1^-), (1^+,2^+), (0^+,1^+)$ multiplets is given in Table 6 of Ref. \cite{Masouminia:2023zhb}, resulting in the best fits:

\begin{savenotes}\sphinxattablestart
\sphinxthistablewithglobalstyle
\centering
\sphinxcapstartof{table}
\sphinxthecaptionisattop
\sphinxcaption{Fitted decay parameters from the Herwig~7.3.0 study of Ref.~\cite{Masouminia:2023zhb}, using the charm meson masses and decay widths from \cite{BaBar:2010zpy, BaBar:2011vbs, LHCb:2013jjb}.}\label{\detokenize{review/decays:id321}}
\sphinxaftertopcaption
\begin{tabulary}{\linewidth}[t]{TT}
\sphinxtoprule
\sphinxstyletheadfamily 

Parameter
&\sphinxstyletheadfamily 

Fitted Value
\\
\sphinxmidrule
\sphinxtableatstartofbodyhook

$f^{''}$
&

$-0.465 \pm 0.017$
\\
\hline

$f_\pi$
&

$\phantom{-}0.130 \pm 0.001$ {[}GeV{]}
\\
\hline

$h$
&

$\phantom{-}0.824 \pm 0.007$
\\
\hline

$\Lambda$
&

$\phantom{-}1.000 \pm 0.000$ {[}GeV{]}
\\
\hline

$g$
&

$\phantom{-}0.565 \pm 0.006$
\\
\hline

$\theta_{u,d}$
&

$\phantom{-}0.000 \pm 0.100$
\\
\hline

$\theta_s$
&

$-0.047 \pm 0.002$
\\
\sphinxbottomrule
\end{tabulary}
\sphinxtableafterendhook\par
\sphinxattableend\end{savenotes}

Besides the strong isospin-conserving decays, certain excited mesons also undergo electromagnetic decays, as well as strong isospin-violating ones. These latter decays are particularly prominent in cases where the strong isospin-conserving decays are either kinematically inhibited or strongly suppressed due to lack of sufficient rest energy, threshold effects, angular momentum conservation, particular selection rules in the given event handler, hierarchy of coupling constants, or presence of other dominant channels. These are predominantly observed in the following cases:
\begin{itemize}
\item {} 

$D^{\star}$ mesons: Strong isospin-conserving decays are kinematically limited, making radiative modes crucial,

\item {} 

$B^{\star}$ mesons: Strong isospin-conserving decays cannot occur due to kinematic constraints. Thus, only the radiative mode emerges as a possibility,

\item {} 

$D^{+}_s$, $D^{+}_{s0}$, and $D^{+}_{s1}(2460)$ mesons: Both radiative and isospin-violating decay modes hold significance since the strong isospin-conserving $DK$ modes are kinematically proscribed,

\item {} 

$B_s^{*0}$ meson: Only the radiative mode is feasible from a kinematic perspective.

\end{itemize}

Again, our focus can be on the $D^{\star}$ system, as it represents the most intricate scenario, exhibiting a plethora of excited mesons below the strong decay threshold. However, our arguments can be effortlessly extended to other cases. The amplitude for the radiative decay of the $D^{\star}$ mesons, as detailed in \cite{Cho:1992nt}, is given by:
\begin{equation*}
\begin{split}\mathcal{M}(D^{\star}\to D\gamma) = A \left[ 64\pi \frac{m_D}{m_{D^{\star}}} \right]^{1/2} \epsilon^{\alpha\beta\mu\nu} \epsilon_{\gamma\alpha} p_{\gamma\beta} \epsilon_{D^{\star}\mu} p_{D^{\star}\nu}.\end{split}
\end{equation*}

Here, the coupling $A$ is expressed as:
\begin{equation*}
\begin{split}A = \frac{e_Q}{4m_Q} \alpha(m_Q)^{1/2} + \frac{c_H}{\Lambda} e_q \alpha(\Lambda)^{1/2},\end{split}
\end{equation*}

with $e_Q$ and $e_q$ being the electric charges of the heavy and light quarks, respectively, and $m_Q$ being the mass of the heavy quark. $\alpha$ is the electromagnetic coupling constant, while $c_H = -1.058$ is the electromagnetic coefficient for heavy meson decays. The scale $\Lambda$ is the same as in the case of strong decays. Consequently, the partial width becomes \cite{Bardeen:2003kt, Cho:1994zu}:
\begin{equation*}
\begin{split}\Gamma(D^{\star}\to D\gamma) = \frac{2 m_D |A|^2}{3 m_{D^{\star}}} \left( \frac{m_{D^{\star}}^2 - m_D^2}{m_{D^{\star}}} \right)^3.\end{split}
\end{equation*}

To implement the strong and radiative decays of excited heavy mesons in Herwig 7, we introduced two specialised classes: \href{https://herwig.hepforge.org/doxygen/classHerwig\_1\_1HQETStrongDecayer.html}{HQETStrongDecayer} and \href{https://herwig.hepforge.org/doxygen/classHerwig\_1\_1HQETRadiativeDecayer.html}{HQETRadiativeDecayer}. The decay parameters highlighted in this section are defined as user-adjustable variables. This design provides flexibility, facilitating potential tuning and refinement based on future insights or requirements. In the subsequent section, we will evaluate the robustness of our implementation by comparing it against the available experimental data.

\subsubsection{Baryon Decays}
\label{\detokenize{review/decays:baryon-decays}}

The strong and electromagnetic decays of the baryons are modelled in
Herwig using models based on either heavy quark effective theory, for
the baryons containing a bottom or charm quark, or
$\mathrm{SU}(3)$ symmetry for the light baryons.

All the strong decays, and many of the weak hadronic decays, involve the
decay of a spin-$\frac12$ or $\frac32$ baryon to either a
pseudoscalar meson or a vector particle and another
spin-$\frac12$ or $\frac32$ baryon. Lorentz invariance
restricts the possible form of the matrix elements. We use the following
matrix elements, which are implemented in the \texttt{Baryon1MesonDecayerBase}
class from which the
\texttt{Decayer}s inherit.
\begin{equation*}
\begin{split}\begin{aligned}
\mathcal{M} &= \bar{u}(p_1)(A+B\gamma_5)u(p_0) & \frac12\to\frac12+0; \\
\mathcal{M} &= \bar{u}(p_1)\epsilon^{*\beta}\left[
                          \gamma_\beta(A_1+B_1\gamma_5)
                                      +p_{0\beta}(A_2+B_2\gamma_5)\right]u(p_0)
&\frac12\to\frac12+1; \\
   \mathcal{M} &= \bar{u}^\alpha(p_1) p_{0\alpha}\left[A+B\gamma_5\right]u(p_0)
   &\frac12\to\frac32+0; \\
   \mathcal{M} &= \bar{u}^\alpha(p_1)\epsilon^{*\beta}\left[
         g_{\alpha\beta}(A_1+B_1\gamma_5) +p_{0\alpha}(A_2+B_2\gamma_5)
        +p_{0\alpha}p_{0\beta}(A_3+B_3\gamma_5)
        \right]u(p_0)
   & \frac12\to\frac32+1;
\end{aligned}\end{split}
\end{equation*}

for spin-$\frac12$ decays and:
\begin{equation*}
\begin{split}\begin{aligned}
\mathcal{M} &= \bar{u}(p_1) p_{1\alpha}\left[A+B\gamma_5\right]u^\alpha(p_0)
& \frac32\to\frac12+0;\\
\mathcal{M} &= \bar{u}(p_1)\epsilon^{*\beta}\left[
        g_{\alpha\beta}(A_1+B_1\gamma_5)
       +p_{1\alpha}(A_2+B_2\gamma_5)
       +p_{1\alpha}p_{0\beta}(A_3+B_3\gamma_5)
        \right]u^\alpha(p_0)
& \frac32\to\frac12+1;\\
\mathcal{M} &= \bar{u}^\alpha(p_1)\left[(A_1+B_1\gamma_5)g_{\alpha\beta}
                  +p_{0\alpha}p_{1\beta}(A_2+B_2\gamma_5)\right]u^\beta(p_0)
& \frac32\to\frac32+0;
\end{aligned}\end{split}
\end{equation*}

for spin-$\frac32$ decays. In general $u(p_0)$ is the
spinor of a decaying spin-$\frac12$ baryon, $u^\beta(p_0)$
is the spinor of a decaying spin-$\frac32$ baryon,
$\bar{u}(p_1)$ is the spinor for an outgoing
spin-$\frac12$ baryon and $\bar{u}^\beta(p_1)$ is the
spinor for an outgoing spin-$\frac32$ baryon. The momentum of
the decaying baryon is $p_0$, of the outgoing baryon is
$p_1$ and of the outgoing meson is $p_2$. All the matrix
elements are parameterized in terms of a number of coefficients
$A$ and $B$, which can in principle depend on the momentum
transferred in the decay.
\begin{itemize}
\item {} 

\href{https://herwig.hepforge.org/doxygen/classHerwig\_1\_1RadiativeDoublyHeavyBaryonDecayer.html}{RadiativeDoublyHeavyBaryonDecayer}
Decay of the excited doubly heavy baryons via photon emission to the weakly decaying states.

\item {} 

\href{https://herwig.hepforge.org/doxygen/classHerwig\_1\_1RadiativeHeavyBaryonDecayer.html}{RadiativeHeavyBaryonDecayer}
There are a number of radiative decays of heavy baryons included in the
simulation. Apart from some transitions of charm baryons, \textit{e.g.}
$\Xi'_c\to\Xi_c\gamma$, these transitions have not been observed
and are included based on model calculations based on heavy quark
effective theory \cite{Ivanov:1999bk}.

\item {} 

\href{https://herwig.hepforge.org/doxygen/classHerwig_1_1RadiativeHyperonDecayer.html}{RadiativeHyperonDecayer}
The radiative decays of hyperons are simulated using the model of Ref.
\cite{Borasoy:1999nt}.

\item {} 

\href{https://herwig.hepforge.org/doxygen/classHerwig\_1\_1StrongHeavyBaryonDecayer.html}{StrongHeavyBaryonDecayer}
The \texttt{StrongHeavyBaryonDecayer} class implements the strong decays of bottom and charm baryons using
the results of Ref. \cite{Ivanov:1999bk}.

\item {} 

\href{https://herwig.hepforge.org/doxygen/classHerwig\_1\_1SU3BaryonDecupletOctetPhotonDecayer.html}{SU3BaryonDecupletOctetPhotonDecayer}
The decay of a decuplet baryon to an octet baryon and a photon is
assumed to occur via the $\mathrm{SU}(3)$ conserving term in the
chiral Lagrangian.

\item {} 

\href{https://herwig.hepforge.org/doxygen/classHerwig\_1\_1SU3BaryonDecupletOctetScalarDecayer.html}{SU3BaryonDecupletOctetScalarDecayer}
This Decayer is based on $\mathrm{SU}(3)$ symmetry for the decay
of a decuplet baryon to an octet baryon and a scalar meson.

\item {} 

\href{https://herwig.hepforge.org/doxygen/classHerwig\_1\_1SU3BaryonOctetDecupletScalarDecayer.html}{SU3BaryonOctetDecupletScalarDecayer}
This
performs the decay of excited octet baryons to decuplet baryons and
a scalar meson using a Lagrangian based on $\mathrm{SU}(3)$
symmetry.

\item {} 

\href{https://herwig.hepforge.org/doxygen/classHerwig\_1\_1SU3BaryonOctetOctetPhotonDecayer.html}{SU3BaryonOctetOctetPhotonDecayer}
This
models the radiative decay of excited octet baryons using a
Lagrangian based on $\mathrm{SU}(3)$ symmetry.

\item {} 

\href{https://herwig.hepforge.org/doxygen/classHerwig\_1\_1SU3BaryonOctetOctetScalarDecayer.html}{SU3BaryonOctetOctetScalarDecayer}
This
simulates the strong decay of excited octet baryons using a
Lagrangian based on $\mathrm{SU}(3)$ symmetry.

\item {} 

\href{https://herwig.hepforge.org/doxygen/classHerwig\_1\_1SU3BaryonSingletOctetPhotonDecayer.html}{SU3BaryonSingletOctetPhotonDecayer}
This
models the radiative decay of singlet baryons using a Lagrangian
based on $\mathrm{SU}(3)$ symmetry.

\item {} 

\href{https://herwig.hepforge.org/doxygen/classHerwig\_1\_1SU3BaryonSingletOctetScalarDecayer.html}{SU3BaryonSingletOctetScalarDecayer}
This
simulates the strong decay of singlet baryons using a Lagrangian
based on $\mathrm{SU}(3)$ symmetry.

\end{itemize}

\subsubsection{Inclusive strong and electromagnetic decays}
\label{\detokenize{review/decays:inclusive-strong-and-electromagnetic-decays}}\label{\detokenize{review/decays:sect-stronginclusive}}

For a number of bottomonium and charmonium resonances we make use of
partonic decays of the mesons to model the unobserved inclusive modes
needed to saturate the branching ratios. These decays are modelled using
the \href{https://herwig.hepforge.org/doxygen/classHerwig\_1\_1QuarkoniumDecayer.html}{QuarkoniumDecayer}
class, which implements the decay of the onium resonances to
$q\bar{q}$ and $gg$ according to a phase-space distribution,
and the decay to $ggg$ and $gg\gamma$ according to the
Ore-Powell matrix element \cite{Ore:1949te}. This class inherits from
the \href{https://herwig.hepforge.org/doxygen/classHerwig\_1\_1PartonicDecayerBase.html}{PartonicDecayerBase}, which uses the cluster model to hadronize the resulting partonic
final state with a veto to ensure that there is no double counting with
the exclusive modes.

\subsection{Weak hadronic decays}
\label{\detokenize{review/decays:weak-hadronic-decays}}

There are five classes of weak mesonic decays currently included in the
simulation:
\begin{enumerate}
\sphinxsetlistlabels{\arabic}{enumi}{enumii}{}{.}%
\item {} 

weak exclusive semi-leptonic decays of bottom and charm mesons;

\item {} 

weak exclusive hadronic decays of bottom and charm mesons;

\item {} 

weak inclusive decays;

\item {} 

weak leptonic decay of pseudoscalar mesons;

\item {} 

weak inclusive $b\to s\gamma$ mediated decays.

\end{enumerate}

The different approaches we adopt for each of these decays are described
below.

\subsubsection{Exclusive semi-leptonic decays}
\label{\detokenize{review/decays:exclusive-semi-leptonic-decays}}\label{\detokenize{review/decays:sect-weakexclusive}}

The matrix element for exclusive semi-leptonic decays of heavy mesons,
$X\to Y\ell\nu$, can be written as
\begin{equation*}
\begin{split}\mathcal{M} = \frac{G_F}{\sqrt{2}}  \langle X|(V-A)_\mu|Y\rangle
        \bar{u}(p_\nu)\gamma^\mu(1-\gamma_5)u(p_\ell),\end{split}
\end{equation*}

where $p_\ell$ is the momentum of the outgoing charged lepton,
$p_\nu$ is the momentum of the neutrino and $G_F$ is the
Fermi constant. The hadronic current
$\langle X|(V-A)_\mu|Y\rangle$ can be written as a general Lorentz
structure, for a particular type of decay, with a number of unknown form
factors.

We have implemented a number of form-factor models based on experimental
fits, QCD sum rule calculations and quark models. The form factors for
the weak decay of pseudoscalar mesons are implemented using the general
Lorentz-invariant form. In each case the momentum of the decaying
particle, $X$, is $p_X$ while the momentum of the decay
product, $Y$, is $p_Y$. In general the form factors are
functions of the momentum transfer $q^2$ where $q=p_X-p_Y$.
The masses of the decaying particle and hadronic decay product are
$m_X$ and $m_Y$ respectively.

The scalar-scalar transition matrix element is defined by
\begin{equation*}
\begin{split}\langle Y(p_Y)|(V-A)_\mu|X(p_X)\rangle =
(p_X+p_Y)_\mu f_+(q^2)
     +\left\{\frac{m_X^2-m_Y^2}{q^2}\right\}q_\mu\left[f_0(q^2)-f_+(q^2)\right],\end{split}
\end{equation*}

where $f_+(q^2)$ and $f_0(q^2)$ are the form factors for the
transition. In general the terms proportional to $q_\mu$ give rise
to contributions proportional to the lepton mass for semi-leptonic
decays and therefore only contribute to the production of tau leptons.

The scalar-vector transition matrix element is defined to be
\begin{equation*}
\begin{split}\langle Y(p_Y)|(V-A)_\mu|X(p_X)\rangle & =  -i\epsilon^*_\mu(m_X+m_Y)A_1(q^2)
 +i(p_X+p_Y)_\mu\epsilon^*\cdot q \frac{A_2(q^2)}{m_X+m_Y}\\
&
 +iq_\mu\epsilon^*\cdot q \frac{2m_Y}{q^2}\left(A_3(q^2)-A_0(q^2)\right)
 +\epsilon_{\mu\nu\rho\sigma}\epsilon^{*\nu}p_X^\rho p_Y^\sigma \frac{2V(q^2)}{m_X+m_Y}\nonumber,\end{split}
\end{equation*}

where the form factor $A_3(q^2)$ can be defined in terms of
$A_1$ and $A_2$ using
\begin{equation*}
\begin{split}A_3(q^2) = \frac{m_X+m_Y}{2m_Y}A_1(q^2)-\frac{m_X-m_Y}{2m_Y}A_2(q^2)\end{split}
\end{equation*}

and $A_0(0)=A_3(0)$. The independent form factors are
$A_0(q^2)$, $A_1(q^2)$, $A_2(q^2)$ and $V(q^2)$.

The transition matrix element for the scalar-tensor transition is
\begin{equation*}
\begin{split}\langle Y(p_Y)|(V-A)_\mu|X(p_X)\rangle &=
i h(q^2) \epsilon_{\mu\nu\lambda\rho} \epsilon^{*\nu\alpha} p_{Y\alpha}
   (p_X+p_Y)^\lambda(p_X-p_Y)^\rho
   -k(q^2)\epsilon^*_{\mu\nu}p_Y^\nu \\
&
   -b_+(q^2)\epsilon^*_{\alpha\beta}p_X^\alpha p_X^\beta(p_X+p_Y)_\mu
   -b_-(q^2)\epsilon^*_{\alpha\beta}p_X^\alpha p_X^\beta(p_X-p_Y)_\mu,\end{split}
\end{equation*}

where $h(q^2)$, $k(q^2)$, $b_-(q^2)$ and
$b_+(q^2)$ are the Lorentz invariant form factors.

The combination of the form factors and the leptonic current is handled
by the \href{https://herwig.hepforge.org/doxygen/classHerwig\_1\_1SemiLeptonicScalarDecayer.html}{SemiLeptonicScalarDecayer}
class, which combines the form factor and the current to
calculate the matrix element and uses the methods available in the
\texttt{DecayIntegrator}
class, from which it inherits, to generate the momenta of the decay
products.

In addition to the form factors for the standard weak current we include
the form factors needed for weak radiative decays where available,
although these are not currently used in the simulation.

The various form factors that are implemented in Herwig are described
below. They all inherit from the \href{https://herwig.hepforge.org/doxygen/classHerwig\_1\_1ScalarFormFactor.html}{ScalarFormFactor}
class and implement the relevant
virtual member functions for the calculation of the form factors.
\begin{itemize}
\item {} 

\href{https://herwig.hepforge.org/doxygen/classHerwig\_1\_1BallZwickyScalarFormFactor.html}{BallZwickyScalarFormFactor}
This is the implementation of the QCD sum rule calculation of the form
factors of Ref. \cite{Ball:2004ye} for the decay of a $B$-meson
to a light pseudoscalar meson.

\item {} 

\href{https://herwig.hepforge.org/doxygen/classHerwig\_1\_1BallZwickyVectorFormFactor.html}{BallZwickyVectorFormFactor}
This is the implementation of the QCD sum rule calculation of the form
factors of Ref. \cite{Ball:2004rg} for the decay of a $B$-meson
to a light vector meson.

\item {} 

\href{https://herwig.hepforge.org/doxygen/classHerwig\_1\_1HQETFormFactor.html}{HQETFormFactor}
This implements the form factors for the transitions between mesons
containing bottom and charm quarks in the heavy quark limit. The
parameterization of Ref. \cite{Caprini:1997mu} for the finite-mass
corrections is used together with the experimental results of Refs.
\cite{Aubert:2007rs,Snyder:2007qn}.

\item {} 

\href{https://herwig.hepforge.org/doxygen/classHerwig\_1\_1ISGWFormFactor.html}{ISGWFormFactor}
The ISGW form factor model \cite{Isgur:1988gb} is one of the original
quark models for the form factors and is included in the simulation
mainly for comparison with the later, ISGW2 \cite{Scora:1995ty}, update
of this model. This set of form factors has the advantage that it
includes form factors for most of the transitions required in the
simulation. The form factors are taken from Ref. \cite{Isgur:1988gb}
together with the form factors that are suppressed by the lepton mass
from Refs. \cite{Isgur:1990jf,Scora:1989ys}.

\item {} 

\href{https://herwig.hepforge.org/doxygen/classHerwig\_1\_1ISGW2FormFactor.html}{ISGW2FormFactor}
The ISGW2 form factors \cite{Scora:1995ty} are an update of the
original ISGW form factors \cite{Isgur:1988gb}. As with the original
model they are based on a quark model and supply most of the form
factors we need for the simulation.

\item {} 

\href{https://herwig.hepforge.org/doxygen/classHerwig\_1\_1KiselevBcFormFactor.html}{KiselevBcFormFactor}
This is the implementation of the form factors of Ref.
\cite{Kiselev:2002vz} for the weak decays of $B_c$ mesons. This
model is used as the default model for weak $B_c$ decays as the
branching ratios for the $B_c$ meson used in the simulation are
calculated using the same model.

\item {} 

\href{https://herwig.hepforge.org/doxygen/classHerwig\_1\_1MelikhovFormFactor.html}{MelikhovFormFactor}
This is the implementation of the relativistic quark model form factors
of Ref. \cite{Melikhov:1996ge} for $B\to\pi,\rho$.

\item {} 

\href{https://herwig.hepforge.org/doxygen/classHerwig\_1\_1MelikhovStechFormFactor.html}{MelikhovStechFormFactor}
This is the implementation of the model of Ref. \cite{Melikhov:2000yu},
which is an update of the model of Ref. \cite{Melikhov:1996ge}
including additional form factors.

\item {} 

\href{https://herwig.hepforge.org/doxygen/classHerwig\_1\_1WSBFormFactor.html}{WSBFormFactor}
This is the implementation of the form factor model of Ref.
\cite{Wirbel:1985ji} for the semi-leptonic form factors. It includes
form factors for a number of $D$, $B$ and $D_s$
decays. In practice the parameters of the model were taken from Ref.
\cite{Bauer:1986bm}, which includes a number of transitions that were
not considered in the original paper.

This form factor model is included both to give an alternative for many
modes to the ISGW models and for use in the factorization approximation
for hadronic charm meson decays.

\end{itemize}

We also include exclusive semi-leptonic decays of heavy baryons in the
same way. The transition matrix elements are given below for the decay
$X(p_X)\to Y(p_Y)$ with $q_\mu=(p_X-p_Y)_\mu$, as for the
mesonic case. The transition matrix for the $\frac12\to\frac12$
transition is defined as
\begin{equation*}
\begin{split}\langle Y(p_Y)|(V-A)_\mu|X(p_X)\rangle &=
    \bar{u}(p_Y) \left[  \gamma_\mu \left(F^V_1+F^A_1 \gamma_5\right)
     +\frac{i}{(m_X+m_Y)}\sigma_{\mu\nu}q^\nu\left(F^V_2+F^A_2\gamma_5\right)
\right.\nonumber\\
&\ \ \ \ \ \ \ \ \ \ \ \ \  \ \ \ \ \
\left.     +\frac1{(m_X+m_Y)}q_\mu\left(F^V_3+F^A_3\gamma_5\right)\right] u(p_X),\end{split}
\end{equation*}

where we have suppressed the dependence of the form factors
$F^{V,A}_{1,2,3}$ on the momentum transfer $q^2$.

The transition matrix element for the $\frac12\to\frac32$
transition is
\begin{equation*}
\begin{split}\langle Y(p_Y)|(V-A)_\mu|X(p_X)\rangle &=
 \bar{u}^\alpha(p_Y) \left[ g_{\alpha\mu}\left(G^V_1+G^A_1 \gamma_5\right)
      +\frac1{(m_X+m_Y)}p_{X\alpha}\gamma_\mu\left(G^V_2+G^A_2\gamma_5\right)
      \right.\nonumber\\
 & \left.
      +\frac1{(m_X+m_Y)^2}p_{X\alpha}p_{Y\mu}\left(G^V_3+G^A_3\gamma_5\right)
      +\frac1{(m_X+m_Y)^2}p_{X\alpha}q_\mu\left(G^V_4+G^A_4\gamma_5\right)\right]
    \gamma_5 u(p_X),\end{split}
\end{equation*}

where again the dependence of the form factors $G^{V,A}_{1,2,3,4}$
on the momentum transfer $q^2$ has been suppressed. These
definitions differ from those in the literature because we have divided
some terms by the sum of the baryon masses to ensure that the
form factors are all dimensionless.

We have implemented a number of different models for the baryon form
factors, mainly based on quark model calculations. All the form factor
classes inherit from the
\href{https://herwig.hepforge.org/doxygen/classHerwig\_1\_1BaryonFormFactor.html}{BaryonFormFactor}
class and implement the calculation of the form
factors in the specific model. The
\href{https://herwig.hepforge.org/doxygen/classHerwig\_1\_1SemiLeptonicBaryonDecayer.html}{SemiLeptonicBaryonDecayer}
class handles the combination of the
form factor and the leptonic current to calculate the partial width and
decay kinematics.

The models we have implemented are:
\begin{itemize}
\item {} 

\href{https://herwig.hepforge.org/doxygen/classHerwig\_1\_1BaryonSimpleFormFactor.html}{BaryonSimpleFormFactor}
This is a simple form factor model for the semi-leptonic decay of
the light baryons. The form factors are assumed to be constant and are
taken from the quark model results of \cite{Donoghue:1981uk}.

\item {} 

\href{https://herwig.hepforge.org/doxygen/classHerwig\_1\_1BaryonThreeQuarkModelFormFactor.html}{BaryonThreeQuarkModelFormFactor}
This model is the implementation of the relativistic three-quark model
calculation of \cite{Ivanov:1996fj} for the form factors of baryons
containing a heavy quark.

As the only formulae in the paper that are in a form that can be
implemented in
the simulation are for the heavy-to-heavy, \textit{i.e.} bottom to charm, decays,
these are the only modes included, although the paper also includes
charm decays and bottom decays to light quarks. The form factors are
calculated by numerically computing the integrals from
\cite{Ivanov:1996fj} to obtain the coefficients for an expansion of the
form factors in $\omega$.

\item {} 

\href{https://herwig.hepforge.org/doxygen/classHerwig\_1\_1ChengHeavyBaryonFormFactor.html}{ChengHeavyBaryonFormFactor}
This is a quark model calculation \cite{Cheng:1995fe,Cheng:1996cs} of
form factors for bottom and charm baryons. It is used for some bottom
and charm baryon semi-leptonic decays. However it is mainly intended to
implement the factorization approximation results of
\cite{Cheng:1996cs} for non-leptonic decays.

\item {} 

\href{https://herwig.hepforge.org/doxygen/classHerwig\_1\_1CzyzNucleonFormFactor.html}{CzyzNucleonFormFactor}
This is the implementation of the  model of Ref. \cite{Czyz:2014sha}
and is intended to describe the $p\bar{p}$ and $n\bar{n}$ cross section in low energy $e^+e^-$ collisions.

\item {} 

\href{https://herwig.hepforge.org/doxygen/classHerwig\_1\_1KornerKurodaFormFactor.html}{KornerKurodaFormFactor}
This is a simple model of the baryon form factors based on Ref. \cite{Korner:1976hv} which we use to
simulate the produce of baryon-antibaryon pairs in low-energy $e^+e^-$ collisions.

\item {} 

\href{https://herwig.hepforge.org/doxygen/classHerwig\_1\_1LambdabExcitedLambdacSumRuleFormFactor.html}{LambdabExcitedLambdacSumRuleFormFactor}
This is the QCD sum rule based calculation of \cite{Huang:2000xw} for
the form factors for the decay of the $\Lambda_b^0$ to excited
$\Lambda^+_c$ states. This is used for the semi-leptonic decay of
the $\Lambda_b^0$ to excited $\Lambda^+_c$ states to
model the part of the total semi-leptonic branching ratio of the
$\Lambda_b^0$ not accounted for by the production of the
$\Lambda_c^+$.

\item {} 

\href{https://herwig.hepforge.org/doxygen/classHerwig\_1\_1LightBaryonQuarkModelFormFactor.html}{LightBaryonQuarkModelFormFactor}
This is a relativistic quark model calculation \cite{Schlumpf:1994fb}
of the form factors for the decay of baryons containing light
quarks.

\item {} 

\href{https://herwig.hepforge.org/doxygen/classHerwig\_1\_1SingletonFormFactor.html}{SingletonFormFactor}
This model is a quark model calculation \cite{Singleton:1990ye} of the
form factors of spin-$\frac12$ baryons containing a bottom or
charm quark.

\end{itemize}

\subsubsection{Exclusive hadronic decays}
\label{\detokenize{review/decays:exclusive-hadronic-decays}}

We include three types of simulations for exclusive weak hadronic decays:
\begin{enumerate}
\sphinxsetlistlabels{\arabic}{enumi}{enumii}{}{.}%
\item {} 

using the naïve factorization approximation;

\item {} 

using isobar models for three-body Dalitz decays of bottom and charm mesons;

\item {} 

various theoretical models of weak baryon decay.

\end{enumerate}

\paragraph{Naïve factorization}
\label{\detokenize{review/decays:naive-factorization}}

We include two types of simulation of exclusive weak meson decays. The first is based on the naïve factorization approximation. If we consider, for example, the decay of a charm meson, the effective Lagrangian for the interaction can be written as \cite{Bauer:1986bm}
\begin{equation*}
\begin{split}
\mathcal{L}_{\rm eff} = \frac{G_F}{\sqrt2}V_{ud}V_{sc}
\left[
a_1\left(\bar{u}\gamma_\mu P_Ld\right)\left(\bar{s}\gamma_\mu P_Lc\right)+
a_2\left(\bar{s}\gamma_\mu P_Ld\right)\left(\bar{u}\gamma_\mu P_Lc\right)
\right],
\end{split}
\end{equation*}
where $G_F$ is the Fermi constant, $V_{ud}$ and $V_{sc}$ are the relevant CKM matrix elements and $a_{1,2}$ are scale-dependent coefficients. The remainder of the expression involves the currents for the quark fields. When we consider the transition between mesonic states the matrix element can be written in terms of the form factors for the $c\to s$ or $c\to u$ transitions, and weak currents for $\left(\bar{u}\gamma_\mu P_Ld\right)$ or $\left(\bar{s}\gamma_\mu P_Ld\right)$.

This allows us to simulate weak hadronic decays using the form factors we have already implemented for semi-leptonic meson decays together with the weak currents from tau decays. The kinematic range of the current in such a hadronic decay is determined by the available momentum transfer, $q^2\leq (m_X-m_Y)^2$, and can therefore extend beyond the range accessible in tau decay. No universal restriction to $q^2\leq m_\tau^2$ is imposed. Consequently, where the current is based only on a parametrization validated in tau decay, its use outside that range constitutes a model-dependent extrapolation.

The combination of the form factor classes, which inherit from \texttt{ScalarFormFactor}, and weak currents, which inherit from \texttt{WeakCurrent}, is handled by the \href{https://herwig.hepforge.org/doxygen/classHerwig\_1\_1ScalarMesonFactorizedDecayer.html}{ScalarMesonFactorizedDecayer} class for the simulation of hadronic weak meson decays in the factorization approximation. Similarly the combination of weak form factors inheriting from the \texttt{BaryonFormFactor} class and weak currents is handled by the \href{https://herwig.hepforge.org/doxygen/classHerwig\_1\_1BaryonFactorizedDecayer.html}{BaryonFactorizedDecayer} class for the simulation of hadronic weak baryon decays in the factorization approximation.

\paragraph{Isobar models for Dalitz decays}
\label{\detokenize{review/decays:isobar-models-for-dalitz-decays}}\label{\detokenize{review/decays:weak-dalitz}}

Prior to the release of Herwig 7.3 we included a small number of
classes for the simulation of $D\to K\pi\pi$ Dalitz decays based on various experimental fits.
In Herwig 7.3 these simulation of these decays has been restructured so that
we have a base class \href{https://herwig.hepforge.org/doxygen/classHerwig\_1\_1DalitzBase.html}{DalitzBase}
which handles the specification of the possible intermediate channels via Interfaces and
the \texttt{ScalarTo3ScalarDalitz}
which implements the Lorentz structures required to calculate the matrix element. This structure
is designed so that further three body decays can use the same base classes, such as the
\texttt{VectorTo3PseudoScalarDalitz}
class which implements the Dalitz decays of vector mesons.

There are a large number of decay models implemented for specific decay modes.

For the $D^+$ meson the following decays and models are available:
\begin{itemize}
\item {} 

$D^+\to K^-\pi^+\pi^-$ the models of Ref. \cite{CLEO:2008jus} as \texttt{{DpKmPipPipCLEO}} isobar model (Herwig default)
and the model independent partial-wave analysis (MIPWA) as \texttt{{DpKmPipPipCLEOMIPWA}},
Ref. \cite{E691:1992rwf} as \texttt{{DpKmPipPipE691}}, Ref. \cite{E791:2002xlc} both the MIPWA as \texttt{{DpKmPipPipE791MIPWA}}
and isobar model as \texttt{{DpKmPipPipE791Isobar}}, Ref. \cite{FOCUS:2007mcb} both the isobar model as \texttt{{DpKmPipPipFOCUSIsobar}} and
K-matrix analysis as \texttt{{DpKmPipPipFOCUSKMatrix}},  Ref. \cite{FOCUS:2009bwp} as \texttt{{DpKmPipPipFOCUSMIPWA}} and
Ref. \cite{MARK-III:1987qok} as \texttt{{DpKmPipPipMarkIII}};

\item {} 

$D^+\to \bar{K}^0\pi^+\pi^0$ the models of Ref. \cite{MARK-III:1987qok} as \texttt{{DpKbar0PipPi0MarkIII}}
and  Ref. \cite{BESIII:2014oag} as \texttt{{DpKbar0PipPi0BES}} (Herwig default);

\item {} 

$D^+\to \pi^+\pi^+\pi^-$ the model of Ref. \cite{E791:2000vek} as \texttt{{DpPipPipPimE791}};

\item {} 

$D^+\to K^+\bar{K}^0\pi^0$ the model of Ref. \cite{BESIII:2021dmo} as \texttt{{DpKpKbar0Pi0BES}};

\item {} 

$D^+\to K^+K^-\pi^+$ the model of Ref. \cite{CLEO:2008msk} as \texttt{{DpKpKmPipCLEO}};

\item {} 

$D^+\to K^+K^+K^-$ the model of Ref. \cite{LHCb:2019tdw} as \texttt{{DpKpKpKmLHCB}};

\item {} 

$D^+\to K^+\pi^+\pi^-$ the model of   Ref. \cite{FOCUS:2004muk} as \texttt{{DpKpPipPimFOCUS}}.

\end{itemize}

For the $D^0$ meson the following decays and models are available:
\begin{itemize}
\item {} 

$D^0\to K^-\pi^+\pi^0$ the models of Ref. \cite{CLEO:2000fvk} as  \texttt{{D0KmPipPi0CLEO}} (Herwig default),
Ref. \cite{E691:1992rwf} as \texttt{{D0KmPipPi0E691}} and Ref. \cite{MARK-III:1987qok} as \texttt{{D0KmPipPi0MarkIII}};

\item {} 

$D^0\to \bar{K}^0\pi^+\pi^-$ the models of Ref. \cite{CLEO:2002uvu} as \texttt{{D0KL0PipPimCLEO}} and \texttt{{D0KS0PipPimCLEO}},
Ref. \cite{E691:1992rwf} as \texttt{{D0Kbar0PipPimE691}},
Ref. \cite{MARK-III:1987qok} as \texttt{{D0Kbar0PipPimMarkIII}},
Ref. \cite{CLEO:2003sah} as \texttt{{D0KS0PipPimCLEO2}} and
Ref. \cite{BaBar:2018cka} as \texttt{{D0KS0PipPimBFactory}};

\item {} 

$D^0\to \bar{K}^0\pi^0\pi^0$ the model of Ref. \cite{CLEO:2011cnt} as \texttt{{D0KS0Pi0Pi0CLEO}};

\item {} 

$D^0\to \pi^+\pi^-\pi^0$ the models of
Ref. \cite{CLEO:2005uoz}  as \texttt{{D0PipPimPi0CLEO}},
Ref. \cite{BaBar:2007dro} as \texttt{{D0PipPimPi0BABAR}}  and
Ref. \cite{BaBar:2016kvp} as \texttt{{D0PipPimPi0BABAR2}};

\item {} 

$D^0\to K^0_{S,L}\pi^+K^-$ the models of
Ref. \cite{CLEO:2012obf} as \texttt{{D0KS0PipKmCLEO}}  and \texttt{{D0KS0PimKpCLEO}} and
Ref. \cite{LHCb:2015lnk} as \texttt{{D0KS0PipKmLHCB}} and  \texttt{{D0KS0PimKpLHCB}};

\item {} 

$D^0\to K^+K^-\pi^0$ the model of Ref. \cite{BaBar:2007soq} as \texttt{{D0KpKmPi0BABAR}};

\item {} 

$D^0\to K^0_{S,L}K^+K^-$ the model of Ref. \cite{BaBar:2010nhz} as \texttt{{D0KS0KpKmBABAR}};

\item {} 

$D^0\to K^0_{S,L}\pi^0\eta$ the model of Ref. \cite{CLEO:2004umu} as \texttt{{D0KS0Pi0EtaCLEO}};

\item {} 

$D^0\to K^-\pi^+\eta$ the model of  Ref. \cite{Belle:2020fbd} as \texttt{{D0KmPipEtaBELLE}}.

\end{itemize}

For the $D^+_s$ meson the following decays and models are available:
\begin{itemize}
\item {} 

$D_s^+\to K^+K^-\pi^+$ the models of
Ref. \cite{BaBar:2010wqe} as \texttt{{DsKpKmPipBABAR}},
Ref. \cite{CLEO:2009nuz} as \texttt{{DsKpKmPipCLEO}} and
Ref. \cite{BESIII:2020ctr} as \texttt{{DsKpKmPipBES}};

\item {} 

$D_s^+\to \pi^+\pi^-\pi^+$  the models of
Ref. \cite{E791:2000lzz}   as \texttt{{DsPipPipPimE791}},
Ref. \cite{BESIII:2021jnf} as \texttt{{DsPipPipPimBES}},
Ref. \cite{BaBar:2008nlp}  as \texttt{{DsPipPipPimBABAR}} and
Ref. \cite{FOCUS:2003tdy}  as \texttt{{DsPipPipPimFOCUS}};

\item {} 

$D_s^+\to \pi^+\pi^0\pi^0$  the model of Ref. \cite{BESIII:2021eru} as \texttt{{DsPipPi0Pi0BES}};

\item {} 

$D_s^+\to K^+\pi^+\pi^-$ the models of Ref. \cite{BESIII:2022vaf} as \texttt{{DsKpPipPimBES}} and
Ref. \cite{FOCUS:2004muk} as \texttt{{DsKpPipPimFOCUS}};

\item {} 

$D_s^+\to K^0\pi^+\pi^0$  the model of Ref. \cite{BESIII:2021xox} as \texttt{{DsPipPi0K0BES}};

\item {} 

$D_s^+\to K^0_{S,L}K^0_{S,L}\pi^+$  the model of Ref. \cite{BESIII:2021anf} as \texttt{{DsKL0KL0PipBES}} and \texttt{{DsKS0KS0PipBES}};

\item {} 

$D_s^+\to K^0_{S,L}K^+\pi^0$ the model of  Ref. \cite{BESIII:2022npc} as \texttt{{DsKS0KpPi0BES}} and \texttt{{DsKL0KpPi0BES}};

\item {} 

$D_s^+\to\pi^+\pi^0\eta$ the model of Ref. \cite{BESIII:2019jjr} as \texttt{{DsPipPi0EtaBES}};

\item {} 

$D_s^+\to\pi^+\pi^0\eta^\prime$ the model of Ref. \cite{BESIII:2022ewq} of \texttt{{DsPipPi0EtaPBES}}.

\end{itemize}

\paragraph{Weak baryonic decays}
\label{\detokenize{review/decays:weak-baryonic-decays}}

While some of the exclusive weak baryonic decays are simulated using the
factorization approximation we also use a number of other models that
include non-factorizable contributions. These all inherit from the
\texttt{Baryon1MesonDecayerBase}
class which
performs the calculation of the matrix elements.
\begin{itemize}
\item {} 

\href{https://herwig.hepforge.org/doxygen/classHerwig\_1\_1KornerKramerCharmDecayer.html}{KornerKramerCharmDecayer}
This is the implementation of the results of the spectator quark model
of \cite{Korner:1992wi} for the non-leptonic weak decay of charm
baryons, \textit{i.e.} $\Lambda_c^+$, $\Xi_c^0$, $\Xi^+_c$
and $\Omega_c^0$.

This model provides branching ratios and decay matrix elements for a
large number of modes and is used as the default simulation for many of
the hadronic decay modes of the weakly decaying charm baryons. In
addition, since for many of these baryons not all the decay modes have been
observed, in some cases the branching ratio calculations are used to add
these modes.

\item {} 

\href{https://herwig.hepforge.org/doxygen/classHerwig\_1\_1NonLeptonicHyperonDecayer.html}{NonLeptonicHyperonDecayer}
We use the results of \cite{Borasoy:1999md} for the matrix elements for
the weak, non-leptonic, decay of a number of hyperons, \textit{i.e.}
$\Sigma^{\pm,0}$, $\Xi^{0,-}$.
The matrix element for the decay is given in terms of the invariant
amplitudes
\begin{equation*}
\begin{split}\mathcal{L}=\bar{u}_{B_j} \left\{A+B\gamma_5\right\}u_{B_i},\end{split}
\end{equation*}

where $B_j$ is the outgoing baryon and $B_i$ is the incoming
baryon.
The default amplitudes are taken from the fit in \cite{Borasoy:1999md}.

\item {} 

\href{https://herwig.hepforge.org/doxygen/classHerwig\_1\_1NonLeptonicOmegaDecayer.html}{NonLeptonicOmegaDecayer}
We use the model of \cite{Borasoy:1999md} for the non-leptonic weak
decays of the $\Omega^-$ to a baryon from the lightest
$\mathrm{SU}(3)$ octet and a pseudoscalar meson. Due to problems
with the size of the $d$-wave term in this model, and more recent measurements
having given the opposite sign for the $\alpha$ parameter, we have set
this term to zero in the simulation.

\item {} 

\href{https://herwig.hepforge.org/doxygen/classHerwig\_1\_1OmegaXiStarPionDecayer.html}{OmegaXiStarPionDecayer}
We use the model of \cite{Duplancic:2004dy} for the weak decay of the
$\Omega^-$ to the $\Xi^*$ and a pion. This decay has a very
low branching ratio and the model is mainly included to test the code
involving the decay of a spin-$\frac32$ particle to another
spin-$\frac32$ particle.

\end{itemize}

\subsubsection{Weak inclusive decays}
\label{\detokenize{review/decays:weak-inclusive-decays}}\label{\detokenize{review/decays:sect-weakinc}}

In addition to the exclusive weak decays of the hadrons to specific
final states we include a number of models of the decay of hadrons
containing a heavy, \textit{i.e.} bottom or charm, quark, based on the partonic
decay of the heavy quark. The Herwig cluster hadronization model is
then applied to the resulting partonic final state to produce hadrons.
This approach is primarily used for the bottom hadrons where there are
insufficient exclusive modes to saturate the branching ratios. All of
the classes implementing partonic decay models inherit from the
\texttt{PartonicDecayerBase}
class to use the cluster hadronization model to hadronize the partonic final state.

The
\href{https://herwig.hepforge.org/doxygen/classHerwig\_1\_1HeavyDecayer.html}{HeavyDecayer}
class implements the weak decays of hadrons using either the weak
$V-A$ matrix element or flat phase-space. The
\href{https://herwig.hepforge.org/doxygen/classHerwig\_1\_1WeakPartonicDecayer.html}{WeakPartonicDecayer} class
includes additional
features to simulate decays intended to increase the rate of baryon
production and gluonic penguin decays and handling the polarization of decaying baryons
as described in Ref. \cite{Masouminia:2023zhb}.

In addition, the
\href{https://herwig.hepforge.org/doxygen/classHerwig\_1\_1BtoSGammaDecayer.html}{BtoSGammaDecayer}
class for weak penguin-mediated decays, described in
\hyperref[\detokenize{review/decays:sect-btosgamma}]{Section \ref{\detokenize{review/decays:sect-btosgamma}}}, and the
\texttt{QuarkoniumDecayer}
class for the decay of bottomonium and
charmonium resonances, described in \hyperref[\detokenize{review/decays:sect-stronginclusive}]{Section \ref{\detokenize{review/decays:sect-stronginclusive}}},
also perform partonic decays and inherit from the
\texttt{PartonicDecayerBase}
class.

\subsubsection{Leptonic decays}
\label{\detokenize{review/decays:leptonic-decays}}

There are a small number of decays of pseudoscalar mesons to a charged
lepton and a neutrino, \textit{e.g.} $\pi\to\ell\nu$ and
$D_s\to\ell\nu$. For most of these decays the inclusion of the
matrix element is superfluous as the decay products are stable. However
the $B$ and $D_s$ mesons can decay in this way to the
$\tau$ and therefore we include the
\href{https://herwig.hepforge.org/doxygen/classHerwig\_1\_1PScalarLeptonNeutrinoDecayer.html}{PScalarLeptonNeutrinoDecayer}
class to simulate these decays
using the matrix element
\begin{equation*}
\begin{split}\mathcal{M} = \frac1{\sqrt{2}}f_PG_FV_{CKM}m_l\bar{u}(p_{\ell})(1-\gamma_5)v(p_\nu),\end{split}
\end{equation*}

where $f_P$ is the pseudoscalar decay constant, $G_F$ is the
Fermi constant, $V_{CKM}$ is the relevant CKM matrix element,
$m_\ell$ is the mass of the lepton, $p_\ell$ is the momentum
of the charged lepton and $p_\nu$ is the momentum of the neutrino.

\subsubsection{\texorpdfstring{$b\to s\gamma$}{b → sγ}}
\label{\detokenize{review/decays:b-to-s-gamma}}\label{\detokenize{review/decays:sect-btosgamma}}

There are a range of decays, both inclusive and exclusive, mediated by
the $b\to s\gamma$ transition. We currently only include modelling
of the inclusive decay. These decays are simulated by using a partonic
decay of the $B$ meson to a photon and a hadronic system, composed
of a quark and antiquark, which recoils against the photon. The mass
spectrum of the hadronic system is calculated using a theoretical model.

The calculation of the mass spectrum is handled by a class inheriting
from the
\href{https://herwig.hepforge.org/doxygen/classHerwig\_1\_1BtoSGammaHadronicMass.html}{BtoSGammaHadronicMass}
class. Different models of the mass spectrum can then be
implemented by inheriting from this class. Currently we have only
implemented two such models. The first,
\href{https://herwig.hepforge.org/doxygen/classHerwig\_1\_1BtoSGammaFlatEnergy.html}{BtoSGammaFlatEnergy},
is solely designed for testing
and generates a mass spectrum such that the photon energy distribution
is flat. The second model,
\href{https://herwig.hepforge.org/doxygen/classHerwig\_1\_1BtoSGammaKagan.html}{BtoSGammaKagan},
which is the default, implements the
theoretical calculation of Ref. \cite{Kagan:1998ym}. The
\texttt{BtoSGammaDecayer}
class then uses the
calculation of the hadronic mass spectrum to simulate the partonic decay
as a model of the inclusive mode. As with the
\texttt{Decayer}s
described in \hyperref[\detokenize{review/decays:sect-weakinc}]{Section \ref{\detokenize{review/decays:sect-weakinc}}} this inherits from the
\texttt{PartonicDecayerBase}
class to use the cluster model
to perform the hadronization of the partonic final state.

\begin{figure}[htp]
\centering
\capstart

\noindent\includegraphics[width=0.600\linewidth]{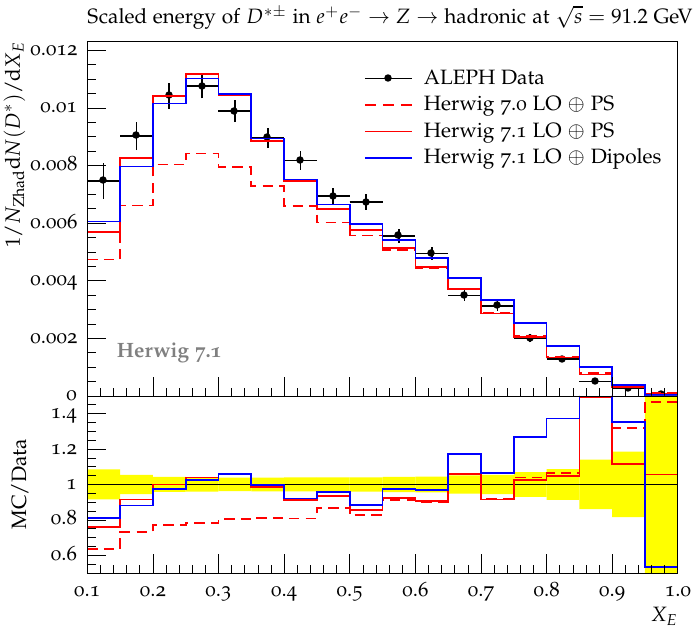}
\caption{The spectrum of $D^*$ mesons measured by the ALEPH experiment
\cite{Barate:1999bg} compared to Herwig. As an example, we show LO plus PS
predictions, however as expected these are not significantly changed in
the presence of higher order corrections. The Herwig 7.0 line
corresponds to the internal decayers, while the Herwig 7.1 lines use
EvtGen for bottom meson decays.}\label{\detokenize{review/decays:id322}}\label{\detokenize{review/decays:fig-charm}}\end{figure}

\subsection{EvtGen interface}
\label{\detokenize{review/decays:evtgen-interface}}

It has proven impossible to provide a good tune to data for the decay of bottom and
charm mesons, largely due to the lack of published distributions. Given that the
EvtGen package \cite{Lange:2001uf} has been tuned to non-public data from the B-factory
experiments and internally uses similar algorithms to include spin correlations
in particle decays, we include an interface to EvtGen. This
communicates the spin information between the two programs, ensuring that the
full correlations are generated. EvtGen is now the default for the decay of
bottom and charm mesons. As there is less data for bottom and charm baryons
and our modelling of baryonic form-factors is more sophisticated, the decays of
heavy baryons continue to be performed by Herwig. This leads to the improvement
of a number of distributions, \textit{e.g.} the momentum distribution of $D^*$
mesons \cite{Barate:1999bg}, \hyperref[\detokenize{review/decays:fig-charm}]{Fig.\@ \ref{\detokenize{review/decays:fig-charm}}}, where there is a
significant contribution from $D^*$ mesons produced in bottom meson decays.

The \href{https://herwig.hepforge.org/doxygen/classHerwig\_1\_1EvtGenInterface.html}{EvtGenInterface}
class handles the conversion between the two programs, while the
\href{https://herwig.hepforge.org/doxygen/classHerwig\_1\_1EvtGenDecayer.html}{EvtGenDecayer}
class is used by Herwig to perform decays using EvtGen. The
\href{https://herwig.hepforge.org/doxygen/classHerwig\_1\_1EvtGenRandom.html}{EvtGenRandom}
class wraps the Herwig random number generator to ensure that the same random number stream is use
by both programs.

As stated above, EvtGen is the default for the decays of bottom- and charm-mesons, while heavy-baryon decays are performed by Herwig; the interface uses \texttt{EvtGenInterface}/\texttt{EvtGenDecayer} and shares the random stream via \texttt{EvtGenRandom}.
A frequent point of confusion concerns stability flags (e.g.\ \verb|set pi0:Stable Stable|). With the default settings, when a $B/D$ meson is decayed by EvtGen, the subsequent decays of its daughters are also performed by EvtGen; therefore, Herwig-level stability settings are not enforced on particles produced inside the EvtGen decay chain (so a $\pi^0$ may still decay).
If users wish EvtGen to decay only the parent heavy meson and to hand all daughters back to Herwig (so that Herwig's stability/decay settings apply), they may set:
\begin{sphinxVerbatim}[commandchars=\\\{\}]
set /Herwig/Decays/EvtGenDecayer:Option Parent
\end{sphinxVerbatim}
As an alternative, one can disable the use of EvtGen for heavy-meson decays entirely and revert to Herwig's internal $B$ decays:
\begin{sphinxVerbatim}[commandchars=\\\{\}]
cd /Herwig/Particles
read defaults/HerwigBDecays.in
\end{sphinxVerbatim}
These options address cases such as $\pi^0$ stability in decay chains while preserving the tuned default for heavy-meson decays when desired.

\subsection{Code structure}
\label{\detokenize{review/decays:code-structure}}\phantomsection\label{\detokenize{review/decays:sect-hadron-sub-structure}}

The \href{https://herwig.hepforge.org/doxygen/classHerwig\_1\_1HwDecayHandler.html}{HwDecayHandler}
class, which inherits from the
\texttt{DecayHandler}
class of ThePEG, is responsible for
handling all particle decays in Herwig. It uses the DecaySelector from the
\texttt{ParticleData}
object of the decaying particle to select a
\texttt{DecayMode}
object corresponding to
a specific decay, according to the probabilities given by the branching
ratios for the different modes. The \texttt{DecayMode} object then specifies a
\texttt{Decayer}
object that is responsible for generating the kinematics of the decay
products for a specific decay.

All of the \texttt{Decayer} classes in Herwig inherit from the
\href{https://herwig.hepforge.org/doxygen/classHerwig\_1\_1HwDecayerBase.html}{HwDecayerBase}
class, which in
turn inherits from the
\texttt{Decayer}
class of ThePEG. In turn, with the exception of
the \texttt{Hw64Decayer}
and \texttt{MamboDecayer}
classes, which implement general phase-space distributions for
the decay products, all the \texttt{Decayer} classes in Herwig inherit from
either the \texttt{DecayIntegrator}
or \texttt{PartonicDecayerBase}
classes.

The \texttt{DecayIntegrator}
class provides a sophisticated multi-channel phase-space integrator
to perform the integration over the phase-space for the decays. This
means that the calculation of the matrix element and specification of
the phase-space channels are all that is required to implement a new
decay model. The majority of the matrix elements are calculated as
helicity amplitudes, which allows the spin-propagation algorithm of
Refs.
\cite{Collins:1987cp,Knowles:1988vs,Knowles:1988hu,Richardson:2001df}
to be implemented. The structure of the Herwig \texttt{Decayer} classes
and the \texttt{HwDecayHandler}
class is designed
so that these correlations are automatically included, provided the
helicity amplitudes for the matrix elements are supplied.

The \texttt{PartonicDecayerBase}
class provides a structure so that the decay products of a partonic
hadron decay can be hadronized using the cluster model, while at the
same time ensuring that there is no overlap with the particle’s
exclusive decay modes. All classes implementing partonic decays in
Herwig inherit from the \texttt{PartonicDecayerBase} class.

The \texttt{DalitzBase} class is
designed to enable the implementation of a range of $1\to3$ decay processes. The base class
makes use of a number of helper classes to specify the different channels and intermediates for the decay, and
how they are to be treated. The inheriting classes then calculate the matrix element for the specific type of
decay. Currently there are two inheriting classes
\texttt{ScalarTo3ScalarDalitz}
for the decay of (pseudo)scalar mesons to three  (pseudo)scalar mesons and
\texttt{VectorTo3PseudoScalarDalitz}
for the decay of vector mesons to three pseudoscalar mesons. The
\href{https://herwig.hepforge.org/doxygen/classHerwig\_1\_1DalitzResonance.html}{DalitzResonance}
class specifies how the resonances in a particular channel should be handled.
The \texttt{DalitzResonance} implements
the standard Breit-Wigner form but can be overridden in inheriting classes. Currently we provide:
\href{https://herwig.hepforge.org/doxygen/classHerwig\_1\_1DalitzGS.html}{DalitzGS} implementing the form of
Gounaris and Sakurai \cite{Gounaris:1968mw};
\href{https://herwig.hepforge.org/doxygen/classHerwig\_1\_1DalitzKMatrix.html}{DalitzKMatrix} allowing
the use of a K-matrix inheriting from the  \href{https://herwig.hepforge.org/doxygen/classHerwig\_1\_1KMatrix.html}{KMatrix} class;
\href{https://herwig.hepforge.org/doxygen/classHerwig\_1\_1DalitzLASS.html}{DalitzLASS} for the LASS parameterization of Ref. \cite{Aston:1987ir}
for s-wave $K\pi$ scattering;
\href{https://herwig.hepforge.org/doxygen/classHerwig\_1\_1DalitzSigma.html}{DalitzSigma} implementing the model of Ref. \cite{Bugg:1996ki} for
scalar resonances;
\href{https://herwig.hepforge.org/doxygen/classHerwig\_1\_1FlatteResonance.html}{FlatteResonance} implementing the Flatté lineshape \cite{Flatte:1976xu};
\href{https://herwig.hepforge.org/doxygen/classHerwig\_1\_1MIPWA.html}{MIPWA} to allow the implementation of MPIWA models;
\href{https://herwig.hepforge.org/doxygen/classHerwig\_1\_1PiPiI2.html}{PiPiI2}  the model of Ref. \cite{Achasov:2003xn} for the $I=2$
component of $\pi\pi$ scattering.

Certain \texttt{Decayer} classes also make use of helper classes to implement the
decays. The main examples are:
\begin{itemize}
\item {} 

the \texttt{WeakCurrent}
class provides a base class for the implementation of weak
hadronic currents and is used by the \texttt{TauDecayer},
\texttt{SemiLeptonicScalarDecayer},
\texttt{SemiLeptonicBaryonDecayer},
\texttt{ScalarMesonFactorizedDecayer} and
\texttt{BaryonFactorizedDecayer}
classes, which implement
tau decays, semi-leptonic meson and baryon decays and hadronic weak
meson and baryon decays using the naïve factorization approximation,
respectively;

\item {} 

the \texttt{ScalarFormFactor}
class provides a base class for the implementation of the
scalar form factors and is used by the
\texttt{SemiLeptonicScalarDecayer} and
\texttt{ScalarMesonFactorizedDecayer}
classes, which implement
semi-leptonic meson decays and hadronic weak meson decays using the
naïve factorization approximation, respectively;

\item {} 

the \texttt{BaryonFormFactor}
class provides a base class for the implementation of the
baryon form factors and is used by the
\texttt{SemiLeptonicBaryonDecayer} and
\texttt{BaryonFactorizedDecayer}
classes, which implement
semi-leptonic baryon decays and hadronic weak baryon decays using the
naïve factorization approximation, respectively;

\item {} 

the \texttt{BtoSGammaHadronicMass}
provides a model of the hadronic mass spectrum in
inclusive $b\to s\gamma$ decays performed by the
\texttt{BtoSGammaDecayer}
class.

\item {} 

the \href{https://herwig.hepforge.org/doxygen/classHerwig\_1\_1OmnesFunction.html}{OmnesFunction} class provides
as base class for the implementation of different models as of the Omnes function which are used by various
decay models. There are currently two inheriting classes one implementing a version based on
experimental measurements \href{https://herwig.hepforge.org/doxygen/classHerwig\_1\_1ExperimentalOmnesFunction.html}{ExperimentalOmnesFunction}
and one using an analytic form
\href{https://herwig.hepforge.org/doxygen/classHerwig\_1\_1AnalyticOmnesFunction.html}{AnalyticOmnesFunction}.

\item {} 

the \texttt{KMatrix} class provides
a base class for the implementation of K-matrix models which are used by various decay models,
primarily of three-body bottom and charm decays. Currently there are three inheriting classes
\href{https://herwig.hepforge.org/doxygen/classHerwig\_1\_1KPiIHalfFOCUSKMatrix.html}{KPiIHalfFOCUSKMatrix}
implementing the $I=\frac12$ and
\href{https://herwig.hepforge.org/doxygen/classHerwig\_1\_1KPiIThreeHalfFOCUSKMatrix.html}{KPiIThreeHalfFOCUSKMatrix}
implementing the $I=\frac32$ model of Ref. \cite{FOCUS:2007mcb} for $K\pi$ scattering and
\href{https://herwig.hepforge.org/doxygen/classHerwig\_1\_1PiPiAnisovichKMatrix.html}{PiPiAnisovichKMatrix}
implementing model of \cite{Anisovich:2002ij} for $pi\pi$ scattering.

\end{itemize}

The vast majority of the decay models have a large number of parameters,
all of which are accessible via the Interfaces of the classes. A more
detailed description of both the physics models used in the code and
their parameters can be found in the Doxygen documentation and Ref.
\cite{Grellscheid:2007tt}.

There are a number of classes that are designed to include the off-shell
weight given in Eq. \eqref{equation:review/decays:eqn:offshell} in the generation of the
particle decays. The \texttt{GenericWidthGenerator}
class is designed to use the information on the partial
widths for the different decay modes supplied by the \texttt{Decayer}
classes that inherit from \texttt{DecayIntegrator},
to calculate the running width for a given particle. The
\texttt{GenericMassGenerator}
class then uses the running width to allow the weight given in Eq.
\eqref{equation:review/decays:eqn:offshell} to be included when generating the particle decays.
The inheriting
\href{https://herwig.hepforge.org/doxygen/classHerwig\_1\_1ScalarMassGenerator.html}{ScalarMassGenerator}
class implements the Flatté lineshape
\cite{Flatte:1976xu} for the $a_0(980)$ and $f_0(980)$
mesons.

For decays where the decay products can be off-shell, and three-body
decays, integrals over either the masses of the decay products or the
three-body phase-space must be performed in order to calculate the
running partial widths. In order to facilitate the calculation of the
partial widths a number of classes inheriting from the
\texttt{WidthCalculatorBase}
class are
implemented to calculate the partial widths for various decays:
\begin{itemize}
\item {} 

the \href{https://herwig.hepforge.org/doxygen/classHerwig\_1\_1TwoBodyAllOnCalculator.html}{TwoBodyAllOnCalculator} class
returns the partial width for a two-body decay where both the
decay products are on mass-shell;

\item {} 

the \href{https://herwig.hepforge.org/doxygen/classHerwig\_1\_1OneOffShellCalculator.html}{OneOffShellCalculator} class
returns the partial width for a decay where one of the outgoing
particles is off mass-shell;

\item {} 

the \href{https://herwig.hepforge.org/doxygen/classHerwig\_1\_1TwoOffShellCalculator.html}{TwoOffShellCalculator} class
returns the partial width for a decay where two of the outgoing
particles are off mass-shell;

\item {} 

the \href{https://herwig.hepforge.org/doxygen/classHerwig\_1\_1ThreeBodyAllOnCalculator.html}{ThreeBodyAllOnCalculator} class
returns the partial width for a three-body decay where all the
decay products are on mass-shell by performing the two non-trivial
integrals over the phase-space variables;

\item {} 

the \href{https://herwig.hepforge.org/doxygen/classHerwig\_1\_1ThreeBodyAllOn1IntegralCalculator.html}{ThreeBodyAllOn1IntegralCalculator} class
returns the partial width for a three-body decay where all the
decay products are on mass-shell, by performing one integral over the
phase-space variables. This requires that the second integral has
already been performed analytically.

\end{itemize}

\clearpage

\section{Summary}
\label{\detokenize{review/index:summary-ms}}

In this paper we have presented a consolidated account of the physics content, software architecture and validated performance of Herwig~7, version~7.3, the C++ successor to the Fortran-based HERWIG and the Herwig++~2.x series. The program provides a coherent, modular framework for precision collider phenomenology in lepton-lepton, lepton-hadron and hadron-hadron environments. Hard-scattering matrix elements are automated within Matchbox, interfacing external amplitude providers at LO, NLO and for loop-induced processes, and furnishing general infrared subtraction, multi-channel phase-space sampling, dynamic scale choices and two complementary matching paradigms (POWHEG- and MC@NLO-type). Consistent multijet merging at LO and NLO is supported within the same framework, enabling inclusive-to-exclusive descriptions with controlled logarithmic structure.

Parton radiation is simulated by two shower algorithms with distinct systematics. The AO shower implements QCD coherence via angular ordering, includes heavy-quark mass effects (dead-cone suppression) and azimuthal spin correlations, and interleaves QCD, QED and EW emissions. The dipole shower is designed for local recoils, colour-coherent emissions and robust matching/merging to higher-order calculations. Perturbative accuracy is augmented by matrix-element corrections to key processes and by coordinated scale-variation and fast reweighting facilities that quantify higher-order and shower uncertainties with negligible additional event-generation cost. Multiple-soft QED effects and mixed QCD-EW radiation are handled via a Yennie--Frautschi--Suura formalism and an EW shower, providing a consistent treatment of photon and weak-boson emissions in production and decay stages.

Non-perturbative dynamics are described primarily by an advanced cluster hadronization model. Gluons are non-perturbatively split into $q\bar{q}$ pairs, colour singlets are formed into clusters and then undergo fission and decay with spin-sensitive treatments and heavy-flavour-aware options. Several colour-reconnection schemes are available, including baryonic reconnection (default in version~7.3), which improves the description of baryon yields and high-multiplicity final states by allowing three-quark (or three-antiquark) cluster topologies. Enhanced strangeness production, consistent handling of the shower infrared cutoff, and options for alternative hadronization are provided; in particular, an interface to the Lund string model enables cross-checks against string-based fragmentation where appropriate.

The underlying event is modelled by an extended eikonal multiple-partonic-scattering framework with semi-hard and soft components and a dedicated diffraction sector, delivering realistic minimum-bias and UE phenomenology and ensuring smooth integration with colour reconnection and hadronization. A unified decay framework covers hadronic and leptonic decays, including detailed $\tau$ decays, off-shell effects, spin correlations and hard QED/EW radiation in both SM and BSM channels. The BSM infrastructure allows import of models specified via Feynman rules (with general many-body decays and colour-sextet diquarks), while retaining consistent shower-hadronization matching and uncertainty evaluation throughout.

Herwig~7.3 thus constitutes a precision-ready event generator whose perturbative and non-perturbative components are systematically improvable and tightly integrated. In comparisons to LHC measurements across benchmark processes (vector-boson, top and multijet final states), matched/merged predictions with either shower provide competitive agreement and complementary uncertainty envelopes, reflecting the differing recoil and logarithmic-accuracy systematics of AO and dipole showers. The unified treatment of QCD, QED and EW radiation, together with modern MPI, colour reconnection and decay models, underpins reliable Standard Model predictions and broad BSM sensitivity at current and future facilities.

Technical documentation, Doxygen class references, validated examples, tutorials and FAQs are available at
\mbox{\sphinxurl{https://herwig.hepforge.org}}.
To report issues, please contact the authors at
\href{mailto:herwig@projects.hepforge.org}{herwig@projects.hepforge.org}.
We rely on user feedback to refine the present release and to steer future development; we therefore encourage users to
report their experience (positive or negative),
provide minimal working examples illustrating issues or unexpected behaviour, and
report bugs and suggest additional features or improvements.
Precise, reproducible reports that isolate problems within Herwig~7 (as opposed to external tools) enable rapid diagnosis and fixes.
Where appropriate, please also consult the designated references for this manual series and the primary physics papers underlying specific modules when citing Herwig in physics studies.

\clearpage

\section*{Acknowledgements}
\label{\detokenize{review/index:acknowledgements}}

This work was supported by the Science and Technology Facilities
Council, formerly the Particle Physics and Astronomy Research Council,
the European Union Marie Curie Training Networks
MCnet (contract MRTN-CT-2006-035606),
MCnetITN (contract PITN-GA-2012-315877) and
MCnetITN3 (contract 722104). 
This work has been supported in part by the BMFTR (formerly BMBF) under
grant agreements 05H21VKCCA and 05H24VKB. 
M.R.~Masouminia is supported by the UK Science and Technology Facilities Council (grant numbers ST/T001011/1 and ST/X000745/1). 
A.~Papaefstathiou acknowledges support by the National Science Foundation under Grant No.\ PHY 2210161 and the U.S. Department of Energy, Office of Science, Office of Nuclear Physics under Award Number DE-SC0025728.
S.~Sule would like to thank the UK Science and Technology Facilities Council (STFC) for the award of a studentship. M.~H.~Seymour also acknowledges STFC's support through grants ST/T001038/1 and ST/X00077X/1.
The work of A.~Siódmok, P.~Sarmah and J.~Whitehead was supported by the National Science Centre Poland grant No 2019/34/E/ST2/00457.
A.~Siódmok is also supported
by the Priority Research Area Digiworld under the program ‘Excellence Initiative – Research University’ at the Jagiellonian University in Krakow and the OpenMAPP project via National Science Centre, Poland under CHIST-ERA programme
(grant No. NCN 2022/04/Y/ST2/00186).
A.~Siódmok and J.~Whitehead gratefully acknowledge the Polish high-performance
computing infrastructure PLGrid (HPC Centre: ACK Cyfronet AGH)
for providing computing facilities and support within computational
grants PLG/2024/017664 and PLG/2024/017366.

Development of Herwig 7 would not have been possible without the early
work of Alberto Ribon and Phil Stephens, as well as continued support 
provided by Leif L\"onnblad and Andy Buckley.
We have received technical advice and support from the HepForge project, which hosts
the Herwig 7 development environment and provides a variety of related services. 
We acknowledge the use of computing resources provided by the Worldwide LHC Computing Grid (WLCG) 
and the Institute for Particle Physics Phenomenology (IPPP), Durham University, as well as at the Particle Physics Group at the University of Vienna and the HPC resources at the University of Graz.
We also acknowledge the support provided by Michael Spannowsky, Jeppe Andersen and the IPPP IT management team, Adam Boutcher and Paul Clark.

\clearpage
\appendix
\settocdepth{section}

\section{Tuning of model parameters}

\label{\detokenize{review/appendix/tuning:tuning}}\label{\detokenize{review/appendix/tuning:sect-tuning}}\label{\detokenize{review/appendix/tuning::doc}}

Herwig 7 has been tuned to a wide range of experimental data. In general
we tune the majority of parameters to data from $e^+e^-$ collisions
and then those parameters which only, or predominately, affect hadronic
collisions to data from hadron-hadron collisions. We have so far considered that parton shower parameters, {\it i.e.} the value of the strong coupling and the infrared cutoff should be considered tuning parameters. Parameters of the hadronization model are then tuned together with the parton shower parameters at LEP data, for which colour reconnection is not considered. Below we describe in detail, how different physics effects are disentangled between light and heavy quarks and other domains of the hadronization. In the future, this workflow might change also as many new observables now become available and tuning tools become more versatile. Parameters of multi-parton interactions and the colour reconnection model are then considered in comparison to hadron collider data. With the tunes we have been considering we do not consider an optimization for a specific class of processes or a specific class of observables, but we try to work in as general a way as possible to strive for the best overall data description of virtually all available data that we can sensibly describe.

\subsection{Tuning philosophy in detail}

The original approach used to tune the angular-ordered parton shower and hadronization parameters is presented in detail in Ref. \cite{Reichelt:2017hts} and adopted in Ref. \cite{Bewick:2019rbu}, where few modifications to the showering algorithm were introduced. The new strategy for tuning angular-ordered parton shower was discussed in Ref. \cite{Masouminia:2023zhb}.

The interested reader should consult Refs. \cite{Reichelt:2017hts}, \cite{Bewick:2019rbu} and \cite{Masouminia:2023zhb} for more results. The Rivet program \cite{Buckley:2010ar,Bierlich:2019rhm} was used to analyse the simulated events and compare the results with the experimental measurements. The Professor program \cite{Buckley:2009bj} was then used to interpolate the shower response and tune the parameters by minimising the $\chi^{2}$%
\begin{footnote}\sphinxAtStartFootnote
While tuning the parameters sensitive to bottom quarks it proved impossible to get a reliable interpolation of
the generator response with Professor and therefore a random scan of the bottom parameters
was performed and the values adjusted by hand about the minimum to minimise the $\chi^{\prime2}$.
\end{footnote}.

In general we use a heuristic chi-squared function
\begin{equation*}
\begin{split}\chi^{\prime2}(p)  = \sum_{\mathcal{O}} w_{\mathcal{O}}  \sum_{b\in\mathcal{O}}
\frac{\left(f_b(p)-\mathcal{R}_b\right)^2}{\Delta_b^2}\end{split}
\end{equation*}
where $p$ is the set of parameters being tuned, $\mathcal{O}$ are the observables used, each with weight $w_{\mathcal{O}}$, and $b$ are the different bins in each observable distribution, with associated experimental measurement $\mathcal{R}_b$, error $\Delta_b$ and Monte Carlo prediction $f_b(p)$. Weighting of those observables for which a good description of the experimental result is important is used in most cases. The parameterisation of the event generator response, $f(p)$, is used to minimise $\chi^{\prime2}$ and find the optimum parameter values. We take $w_{\mathcal{O}}=1$ in most cases except for charged-particle multiplicity distributions, average total charged-particle multiplicities and identified-hadron production rates, defined as the mean numbers produced per event, for which we use $w_{\mathcal{O}}=10$, $w_{\mathcal{O}}=50$ and $w_{\mathcal{O}}=100$, respectively. This ensures that multiplicity and identified-hadron yield measurements influence the result of the fit despite the much larger quantity of event-shape and spectrum data used in the tuning. We also use $w_{\mathcal{O}}=10$ for data on gluon jets in order to avoid the fit being dominated by the large quantity of data sensitive to quark jets. In addition, as we do not expect a Monte Carlo event generator to give a perfect description of all the data, and in order to avoid the fit being dominated by a few observables with very small experimental errors, we use
\begin{equation*}
\begin{split}
\Delta^{\rm eff}_b = \max(0.05\times\mathcal{R}_b,\Delta_b),
\end{split}
\end{equation*}
rather than the true experimental error, $\Delta_b$, in the fit.

The standard procedure which was adopted to tune the shower
and hadronization parameters of the
Herwig 7 parton shower and hadronization model to data is:
\begin{itemize}
\item {} 

first the shower and those
hadronization parameters which are primarily sensitive to light quark-initiated processes
are tuned to measurements of event shapes, the average charged particle multiplicity and
charged particle multiplicity distribution, and identified particle
spectra and rates which only involve light quark mesons and baryons;

\item {} 

the hadronization parameters for bottom quarks are tuned to
the bottom quark fragmentation function measured by LEP1 and SLD together with
LEP1 and SLD measurements of event shapes and identified particle spectra from bottom events and measurements of the charged particle multiplicity in bottom initiated events;

\item {} 

the hadronization parameters involving charm quarks are then tuned
to identified particle spectra, from both the B-factories and LEP1,
and LEP1 and SLD measurements of event shapes and identified particle spectra from charm events and measurements of the charged particle multiplicity in charm initiated events;

\item {} 

the light quark parameters are then retuned using the new values of the
bottom and charm parameters.

\end{itemize}

The tune described in \ref{\detokenize{review/appendix/tuning:herwig-7-3-general-tunes}} is the default angular-ordered shower and hadronization tune distributed with Herwig~7.3.0. The subsequent subsections document earlier or alternative tuning studies, including the comparison of angular-ordered recoil schemes and tunes developed specifically for the dipole shower. These studies were performed independently and are retained as distinct developments in their own right; their quoted parameter values should not be interpreted as the Herwig~7.3.0 default unless explicitly stated.

\subsection{Default Herwig 7.3 general tune}
\label{\detokenize{review/appendix/tuning:herwig-7-3-general-tunes}}

For Herwig 7.3, we utilised $e^-e^+$ data from LEP, PETRA, SLAC, SLC, and TRISTAN, encompassing over 9,200 separate data bins \cite{Masouminia:2023zhb}. The data were weighted around both light and heavy hadron production rates and multiplicities, focusing on the most dominant processes. We attempted a multi-layered, brute-force approach using the \texttt{{prof2-chisq}} module from Professor II \cite{Buckley:2009bj}, minimising the $\chi^2$ as an indicator of the best tune. In this context, we analysed a total of 12 parameters, comprising 10 for cluster hadronization and 2 for the parton shower, specifically \texttt{{AlphaIn}} and \texttt{{pTmin}}. Below is a detailed breakdown of these parameters:
\begin{itemize}
\item {} 

\texttt{{ClMaxLight}}: Maximum allowable cluster mass for light quarks.

\item {} 

\texttt{{ClPowLight}}: Power exponent for the mass of clusters with light quarks.

\item {} 

\texttt{{PSplitLight}}: Parameter affecting the mass splitting for clusters with light quarks.

\item {} 

\texttt{{PwtSquark}}: Probability for a $s\bar{s}$ quark pair to be spawned during cluster splittings.

\item {} 

\texttt{{PwtDIquark}}: Probability for quarks forming a di-quark.

\item {} 

\texttt{{SngWt}}: Weighting factor for singlet baryons in hadronization.

\item {} 

\texttt{{DecWt}}: Weighting factor for decuplet baryons in hadronization.

\item {} 

\href{https://herwig.hepforge.org/doxygen/classHerwig_1_1ClusterFissioner.html#probPowFactor}{ProbabilityPowerFactor}: Exponential factor in the \texttt{{ClusterFissioner}} probability function.

\item {} 

\href{https://herwig.hepforge.org/doxygen/classHerwig_1_1ClusterFissioner.html#probShift}{ProbabilityShift}: Offset in the \texttt{{ClusterFissioner}} probability function.

\item {} 

\href{https://herwig.hepforge.org/doxygen/classHerwig_1_1ClusterFissioner.html#kinThresholdShift}{KinematicThresholdShift}: Adjustment to the kinematic threshold in \texttt{{ClusterFissioner}}.

\item {} 

\texttt{{AlphaIn}}: Initial value for the strong coupling constant at $M_{Z^0}=91.1876$ GeV%
\begin{footnote}\sphinxAtStartFootnote
The reader should note that after the Herwig-7.3 release, the \texttt{{AlphaIn}} has been split into two separate parameters, \texttt{{AlphaQCD:AlphaIn}} and \texttt{{AlphaQCDFSR:AlphaIn}}, that control the strong coupling constants for initial- and final-state radiations respectively. The tuned value we suggest here points to \texttt{{AlphaQCDFSR:AlphaIn}}, as it is the one relevant to our $e^+e^-$ tuning strategy.
\end{footnote}.

\item {} 

\texttt{{pTmin}}: Minimum transverse momentum in the parton shower.

\end{itemize}

Following the above description and methodology, we tuned the hadronization and parton shower parameters in Herwig 7.3. The results are given in the table below \cite{Masouminia:2023zhb}.

\begin{savenotes}\sphinxattablestart
\sphinxthistablewithglobalstyle
\centering
\sphinxcapstartof{table}
\sphinxthecaptionisattop
\sphinxcaption{The values of tuned parameters in Herwig 7.3 compared to Herwig 7.2.}\label{\detokenize{review/appendix/tuning:tuned-parameters}}
\sphinxaftertopcaption
\begin{tabulary}{\linewidth}[t]{TTT}
\sphinxtoprule
\sphinxstyletheadfamily 

Tuned Parameter
&\sphinxstyletheadfamily 

Herwig-7.3.0
&\sphinxstyletheadfamily 

Herwig-7.2.0
\\
\sphinxmidrule
\sphinxtableatstartofbodyhook

\texttt{{ClMaxLight}} {[}GeV{]}
&

3.529
&

3.649
\\
\hline

\texttt{{ClPowLight}}
&

1.849
&

2.780
\\
\hline

\texttt{{PSplitLight}}
&

0.914
&

0.899
\\
\hline

\texttt{{PwtSquark}}
&

0.374
&

0.292
\\
\hline

\texttt{{PwtDIquark}}
&

0.331
&

0.298
\\
\hline

\texttt{{SngWt}}
&

0.891
&

0.740
\\
\hline

\texttt{{DecWt}}
&

0.416
&

0.620
\\
\hline

\texttt{{ProbabilityPowerFactor}}
&

6.486
&

--
\\
\hline

\texttt{{ProbabilityShift}}
&

-0.879
&

--
\\
\hline

\texttt{{KinematicThresholdShift}} {[}GeV{]}
&

0.088
&

--
\\
\hline

\texttt{{AlphaIn}}
&

0.102
&

0.126
\\
\hline

\texttt{{pTmin}} {[}GeV{]}
&

0.655
&

0.958
\\
\sphinxbottomrule
\end{tabulary}
\sphinxtableafterendhook\par
\sphinxattableend\end{savenotes}

Despite its computational intensity, our brute-force approach can be considered a reasonable success, achieving a substantial improvement in the overall $\chi^2$ values. Specifically, the tuned version of Herwig 7.3 shows a $\sim 50.75\%$ improvement in $\chi^2$ compared to its untuned counterpart and a $\sim 12.76\%$ improvement when compared to Herwig-7.2.3. In the following section, we will examine the effects of this new tune on the production rates of heavy mesons and baryons.

Here we discuss the tuning of the parton shower and hadronization parameters,
we then go on and consider the tuning of the intrinsic $p_\perp$ and
parameters of the multiple parton and soft scattering models, together
with the colour reconnection model. In this section we focus on the tuning of the angular ordered shower. While older tunings are available for the dipole shower, including some using the autotunes framework \cite{Bellm:2019owc}, a general tune for the dipole shower will be in preparation once all next-to-leading logarithmic improvements following \cite{Forshaw:2020wrq} are available.

\subsection{AO recoil-scheme tunes}
\label{\detokenize{review/appendix/tuning:angular-ordered-parton-shower-tune-for-different-recoil-schemes}}

In order to tune the shower and light quark hadronization parameters we used data on jet rates and event shapes for centre-of-mass energies between 14 and 44 GeV \cite{Braunschweig:1990yd,MovillaFernandez:1997fr,Pfeifenschneider:1999rz}, at LEP1 and SLD \cite{Abreu:1996na,Barate:1996fi,Pfeifenschneider:1999rz,Abbiendi:2004qz,Heister:2003aj} and LEP2 \cite{Pfeifenschneider:1999rz,Heister:2003aj,Abbiendi:2004qz}, particle multiplicities \cite{Abreu:1996na,Barate:1996fi} and spectra \cite{Akers:1994ez,Alexander:1995gq,Alexander:1995qb,Abreu:1995qx,Alexander:1996qj,Abreu:1996na,Barate:1996fi,Ackerstaff:1997kj,Abreu:1998nn,Ackerstaff:1998ap,Ackerstaff:1998ue,Abbiendi:2000cv,Heister:2001kp} at LEP 1, identified particle spectra below the $\Upsilon(4S)$ from Babar \cite{Lees:2013rqd}, the charged particle multiplicity \cite{Ackerstaff:1998hz,Abe:1996zi} and particle spectra \cite{Ackerstaff:1998hz,Abe:1998zs,Abe:2003iy} in light quark events at LEP1 and SLD, the charged particle multiplicity in light quark events at LEP2 \cite{Abreu:2000nt,Abbiendi:2002vn}, the charged particle multiplicity distribution at LEP 1 \cite{Decamp:1991uz}, and hadron multiplicities at the Z-pole \cite{Amsler:2008zzb}. We also implemented in Rivet and made use of the data on the properties of gluon jets \cite{Abbiendi:2003gh,Abbiendi:2004pr} for the first time.

The hadronization parameters for charm quarks were tuned using the charged particle multiplicity in charm events at SLD \cite{Abe:1996zi} and LEP2 \cite{Abreu:2000nt,Abbiendi:2002vn}, the light hadron spectra in charm events at LEP1 and SLD \cite{Ackerstaff:1998hz,Abe:1998zs,Abe:2003iy}, the multiplicities of charm hadrons at the Z-pole \cite{Abreu:1996na,Amsler:2008zzb}, and charm hadron spectra below the $\Upsilon(4S)$ \cite{Seuster:2005tr,Aubert:2006cp} and at LEP1 \cite{Barate:1999bg}.

The hadronization parameters for bottom quarks were tuned using the charged particle multiplicity in bottom events at SLD \cite{Abe:1996zi} and LEP2 \cite{Abreu:2000nt,Abbiendi:2002vn}, the light hadron spectra in bottom events at LEP1 and SLD \cite{Ackerstaff:1998hz,Abe:1998zs,Abe:2003iy}, the multiplicities of charm and bottom hadrons at the Z-pole \cite{Abreu:1996na,Amsler:2008zzb}, charm hadron spectra at LEP1 \cite{Barate:1999bg} and the bottom fragmentation function measured at LEP1 and SLD \cite{Abe:2002iq,Heister:2001jg,DELPHI:2011aa}.

In order to tune the evolution of the total charged particle multiplicity in $e^+e^-$ collisions as a function of energy the results of Refs. \cite{Derrick:1986jx,Aihara:1986mv,Berger:1980zb,Bartel:1983qp,Braunschweig:1989bp,Zheng:1990iq,Acton:1991aa,Abe:1996zi,Abreu:1996na,Abreu:2000nt,Abbiendi:2002vn,Heister:2003aj} spanning energies from 12 to 209 GeV were used.

The recoil-scheme tunes described in this subsection were performed before and independently of the Herwig~7.3.0 general tune described in \ref{\detokenize{review/appendix/tuning:herwig-7-3-general-tunes}}. The parameters \texttt{AlphaIn}, \texttt{pTmin}, \texttt{ClMaxLight}, \texttt{ClPowLight}, \texttt{PSplitLight}, \texttt{PwtSquark} and \texttt{PwtDIquark} were varied in both studies. The recoil-scheme study additionally varied the charm- and bottom-hadronization parameters listed below. The parameters \texttt{SngWt}, \texttt{DecWt}, \texttt{ProbabilityPowerFactor}, \texttt{ProbabilityShift} and \texttt{KinematicThresholdShift} were not part of the parameter set varied in this earlier study.

The full parameter set varied in this earlier study was:
\begin{enumerate}
\sphinxsetlistlabels{\arabic}{enumi}{enumii}{}{.}%
\item {}

the value of $\alpha_{s}$ at the $Z^0$ mass, ;

\item {}

the cut-off scale of the transverse momentum in the parton shower;

\item {}

the maximum mass above which clusters containing light quarks undergo cluster fission, see Eq. \eqref{equation:review/hadronization:eqn:clustersplit};

\item {}

the exponent controlling whether clusters containing light quarks undergo fission, see Eq. \eqref{equation:review/hadronization:eqn:clustersplit};

\item {}

the exponent controlling the masses of the daughter clusters for light quark clusters that undergo fission, see Eq. \eqref{equation:review/hadronization:eqn:clusterspect};

\item {}

the weight for producing a strange quark-antiquark pair in the hadronization;

\item {}

the weight for producing a diquark-antidiquark pair in the hadronization;

\item {}

the maximum mass above which clusters containing charm quarks undergo cluster fission, see Eq. \eqref{equation:review/hadronization:eqn:clustersplit};

\item {}

the exponent controlling whether clusters containing charm quarks undergo fission, see Eq. \eqref{equation:review/hadronization:eqn:clustersplit};

\item {}

the exponent controlling the masses of the daughter clusters for charm quark clusters that undergo fission, see Eq. \eqref{equation:review/hadronization:eqn:clusterspect};

\item {}

the parameter, which controls the smearing of the direction of hadrons containing perturbatively produced charm quarks, see Eq. \eqref{equation:review/hadronization:eqn:hadronsmear};

\item {}

the parameter, which controls the splitting of charm clusters to a single hadron above the threshold for producing two hadrons, see Eq. \eqref{equation:review/hadronization:eqn:singlehadron};

\item {}

the maximum mass above which clusters containing bottom quarks undergo cluster fission, see Eq. \eqref{equation:review/hadronization:eqn:clustersplit};

\item {}

the exponent controlling whether clusters containing bottom quarks undergo fission, see Eq. \eqref{equation:review/hadronization:eqn:clustersplit};

\item {}

the exponent controlling the masses of the daughter clusters for bottom quark clusters that undergo fission, see Eq. \eqref{equation:review/hadronization:eqn:clusterspect};

\item {}

the parameter, which controls the smearing of the direction of hadrons containing perturbatively produced bottom quarks, see Eq. \eqref{equation:review/hadronization:eqn:hadronsmear};

\item {}

the parameter, which controls the splitting of bottom clusters to a single hadron above the threshold for producing two hadrons, see Eq. \eqref{equation:review/hadronization:eqn:singlehadron};

\end{enumerate}

For the angular-ordered parton shower, tunes were performed for three choices of the quantity preserved in the kinematic reconstruction: the transverse momentum $p_\perp$, the parton virtuality $q^2$, and the dot product $q_i\cdot q_j$, as described in \hyperref[\detokenize{review/showers/qtilde:sub-shower-kinematics}]{Section \ref{\detokenize{review/showers/qtilde:sub-shower-kinematics}}}. For the dot-product-preserving scheme, results were obtained both with and without the veto described in \hyperref[\detokenize{review/showers/qtilde:sect-final-final-evolution}]{Section \ref{\detokenize{review/showers/qtilde:sect-final-final-evolution}}}, giving four configurations in total.

The values of the parameters obtained from the tuning are given in \hyperref[\detokenize{review/appendix/tuning:tab-ao-tunes}]{Table \ref{\detokenize{review/appendix/tuning:tab-ao-tunes}}},
while the values of $\chi^2$ are given in \hyperref[\detokenize{review/appendix/tuning:tab-ao-chisq}]{Table \ref{\detokenize{review/appendix/tuning:tab-ao-chisq}}}. For the phenomenological results of this earlier study, the dot-product-preserving tune with the veto was used as the reference configuration.

\begin{savenotes}\sphinxattablestart
\sphinxthistablewithglobalstyle
\centering
\sphinxcapstartof{table}
\sphinxthecaptionisattop
\sphinxcaption{The parameters obtained for different choices of the preserved quantity in the angular-ordered parton shower, using the Herwig setup of Ref.~\cite{Bewick:2019rbu}. For the dot product preserving scheme, we present the results obtained both with and without the veto described in \hyperref[\detokenize{review/showers/qtilde:sect-final-final-evolution}]{Section \ref{\detokenize{review/showers/qtilde:sect-final-final-evolution}}}.}\label{\detokenize{review/appendix/tuning:id43}}\label{\detokenize{review/appendix/tuning:tab-ao-tunes}}
\sphinxaftertopcaption
\begin{tabulary}{\linewidth}[t]{TTTTT}
\sphinxtoprule
\sphinxtableatstartofbodyhook

Preserved
&

$p_\perp$
&

$q^2$
&

$q_i\cdot q_j$ (no veto)
&

$q_i\cdot q_j$ (veto)
\\
\hline\sphinxstartmulticolumn{5}%
\begin{varwidth}[t]{\sphinxcolwidth{5}{5}}

Shower and Light quark hadronization parameters
\par
\vskip-\baselineskip\vbox{\hbox{\strut}}\end{varwidth}%
\sphinxstopmulticolumn
\\
\hline

\href{https://herwig.hepforge.org/doxygen/ShowerAlphaQCDInterfaces.html\#AlphaIn}{AlphaIn} ($\alpha^{\rm CMW}_S(M_Z)$)
&

0.1074
&

0.1244
&

0.1136
&

0.1186
\\
\hline

\texttt{pTmin}
&

0.900
&

1.136
&

0.924
&

0.958
\\
\hline

\texttt{ClMaxLight}
&

4.204
&

3.141
&

3.653
&

3.649
\\
\hline

\texttt{ClPowLight}
&

3.000
&

1.353
&

2.000
&

2.780
\\
\hline

\texttt{PSplitLight}
&

0.914
&

0.831
&

0.935
&

0.899
\\
\hline

\texttt{PwtSquark}
&

0.647
&

0.737
&

0.650
&

0.700
\\
\hline

\texttt{PwtDIquark}
&

0.236
&

0.383
&

0.386
&

0.298
\\
\hline\sphinxstartmulticolumn{5}%
\begin{varwidth}[t]{\sphinxcolwidth{5}{5}}

Charm quark hadronization parameters
\par
\vskip-\baselineskip\vbox{\hbox{\strut}}\end{varwidth}%
\sphinxstopmulticolumn
\\
\hline

\texttt{ClMaxCharm}
&

4.204
&

3.564
&

3.796
&

3.950
\\
\hline

\texttt{ClPowCharm}
&

3.000
&

2.089
&

2.235
&

2.559
\\
\hline

\texttt{PSplitCharm}
&

1.060
&

0.928
&

0.990
&

0.994
\\
\hline

\texttt{ClSmrCharm}
&

0.098
&

0.141
&

0.139
&

0.163
\\
\hline

\texttt{SingleHadronLimitCharm}
&

0.000
&

0.011
&

0.000
&

0.000
\\
\hline\sphinxstartmulticolumn{5}%
\begin{varwidth}[t]{\sphinxcolwidth{5}{5}}

Bottom quark hadronization parameters
\par
\vskip-\baselineskip\vbox{\hbox{\strut}}\end{varwidth}%
\sphinxstopmulticolumn
\\
\hline

\texttt{ClMaxBottom}
&

5.757
&

2.900
&

6.000
&

3.757
\\
\hline

\texttt{ClPowBottom}
&

0.672
&

0.518
&

0.680
&

0.537
\\
\hline

\texttt{PSplitBottom}
&

0.557
&

0.365
&

0.550
&

0.625
\\
\hline

\texttt{ClSmrBottom}
&

0.117
&

0.070
&

0.105
&

0.078
\\
\hline

\texttt{SingleHadronLimitBottom}
&

0.000
&

0.000
&

0.000
&

0.000
\\
\sphinxbottomrule
\end{tabulary}
\sphinxtableafterendhook\par
\sphinxattableend\end{savenotes}

\begin{savenotes}\sphinxattablestart
\sphinxthistablewithglobalstyle
\centering
\sphinxcapstartof{table}
\sphinxthecaptionisattop
\sphinxcaption{The values of \protect$\chi^2\protect$ per degree of freedom for the recoil-scheme fits of Ref.~\cite{Bewick:2019rbu}, using the same Herwig setup as Table~\ref{\detokenize{review/appendix/tuning:tab-ao-tunes}}. The values are
        \protect$\chi^{\prime2}\protect$ as described in the text for the tuning
        observables, normalised to the sum of the weights for the
        different bins.  The number of degrees of freedom for each set of
        observables is given. The \protect$\chi^2\protect$ corresponding to ATLAS
        jets, particle multi-plicities (mult), event shapes (event),
        identified-particle spectra (ident), quark jets (jet),gluon jets
        (gluon) and charged particle distributions (charged) are also
        shown.}\label{\detokenize{review/appendix/tuning:id44}}\label{\detokenize{review/appendix/tuning:tab-ao-chisq}}
\sphinxaftertopcaption
\begin{tabulary}{\linewidth}[t]{TTTTTT}
\sphinxtoprule
\sphinxtableatstartofbodyhook

Preserved
&

$p_\perp$
&

$q^2$
&

$q_i\cdot q_j$ (no veto)
&

$q_i\cdot q_j$ (veto)
&

Number of dof
\\
\hline\sphinxstartmulticolumn{6}%
\begin{varwidth}[t]{\sphinxcolwidth{6}{6}}

$\chi^2/ndf$ considering several sets of observables
\par
\vskip-\baselineskip\vbox{\hbox{\strut}}\end{varwidth}%
\sphinxstopmulticolumn
\\
\hline

Light jets
&

4.4
&

3.2
&

3.7
&

3.4
&

17746
\\
\hline

Charm  quarks
&

2.3
&

1.7
&

1.8
&

1.9
&

909
\\
\hline

Bottom quarks
&

6.0
&

6.5
&

5.1
&

4.1
&

1745
\\
\hline

ATLAS jets
&

0.16
&

0.41
&

0.19
&

0.54
&

22
\\
\hline\sphinxstartmulticolumn{6}%
\begin{varwidth}[t]{\sphinxcolwidth{6}{6}}

$\chi^2/ndf$ considering sub-samples of the “Light jets’’ set
\par
\vskip-\baselineskip\vbox{\hbox{\strut}}\end{varwidth}%
\sphinxstopmulticolumn
\\
\hline

mult
&

3.0
&

2.8
&

2.8
&

2.8
&

6780
\\
\hline

event
&

7.0
&

3.5
&

5.2
&

3.9
&

2689
\\
\hline

ident
&

10.7
&

10.0
&

9.8
&

10.1
&

953
\\
\hline

jet
&

4.6
&

3.2
&

4.1
&

3.6
&

3459
\\
\hline

gluon
&

1.1
&

1.2
&

1.2
&

1.2
&

1880
\\
\hline

charged
&

5.4
&

2.5
&

3.7
&

2.9
&

1850
\\
\sphinxbottomrule
\end{tabulary}
\sphinxtableafterendhook\par
\sphinxattableend\end{savenotes}

A similar philosophy was adopted to tune the parameters of the dipole shower, and the
hadronization model when used in conjunction with the dipole shower. As there
were no significant changes for the dipole parton shower
between Herwig 7.0 and Herwig 7.1 a slightly older
procedure, our default at the time Herwig 7.0 was released,
which only used data from the $Z^0$ pole and continuum near the
$\Upsilon(4S)$ resonance was used.

\subsection{Dipole shower}

A simple tune of the dipole shower was performed by adjusting only the shower cutoffs, while the hadronization parameters were not retuned. The intrinsic-transverse-momentum parameters \texttt{ValenceIntrinsicPtScale} and \texttt{SeaIntrinsicPtScale}, introduced in \hyperref[\detokenize{review/showers/intrinsic:shower-intrinsic}]{Section \ref{\detokenize{review/showers/intrinsic:shower-intrinsic}}}, were fitted separately for valence and sea quarks, giving 
\begin{align*}
    \texttt{ValenceIntrinsicPtScale} &= 1.26905\,{\rm GeV}\ ,\\ 
    \texttt{SeaIntrinsicPtScale} &=  1.1613\,{\rm GeV}\ .
\end{align*}

The infrared cutoff scales for the shower are set to $\texttt{IRCutoff} = 1.014259\,$GeV for \texttt{FFLightKinematics} and \texttt{FFMassiveKinematics}, while for all other dipoles it is set to 1.0\,GeV. These values are provided for this simple dipole-shower tune in the input files.

A more elaborate tune with the tool \texttt{Autotunes} \cite{Bellm:2019owc} has been obtained and is also provided as an extra snippet with our release. Here, also hadronization parameters have been adjusted.  

\begin{savenotes}\sphinxattablestart
\sphinxthistablewithglobalstyle
\centering
\sphinxcapstartof{table}
\sphinxthecaptionisattop
\sphinxcaption{The values of tuned parameters for the dipole shower, obtained with \texttt{Autotunes} and supplied in \texttt{Dipole\_AutoTunes\_gss.in} with Herwig~7.3.0.}\label{\detokenize{review/appendix/tuning:autotuned-parameters}}
\sphinxaftertopcaption
\begin{tabulary}{\linewidth}[t]{TTT}
\sphinxtoprule
\sphinxstyletheadfamily 

Tuned Parameter
&\sphinxstyletheadfamily 

Value
\\
\sphinxmidrule
\sphinxtableatstartofbodyhook
All input values for $\alpha_{s}$ 
&
0.118
\\
\hline

\texttt{g:ConstituentMass} {[}GeV{]}
&
0.95
\\

\texttt{b:NominalMass} {[}GeV{]}
&
4.700501
\\

\texttt{b:ConstituentMass} {[}GeV{]}
&
4.084889
\\
\hline

\texttt{ClSmrBottom}
&0.085964
\\

\texttt{ClSmrLight}
&0.698877
\\

\texttt{ClSmrCharm}
&0.246296
\\

\texttt{ClPowBottom}
&0.591646
\\

\texttt{ClPowLight}
&0.99945
\\

\texttt{ClPowCharm}
&3.386187
\\

\texttt{ClMaxBottom} {[}GeV{]}
&3.771649
\\

\texttt{ClMaxBottom} {[}GeV{]}
&3.771649
\\

\texttt{ClMaxCharm} {[}GeV{]}
&4.780456
\\

\texttt{ClMaxLight} {[}GeV{]}
&3.055256
\\

\texttt{PSplitLight}
&0.7779
\\

\texttt{PSplitCharm}
&0.628766
\\

\texttt{PSplitBottom}
&0.662911
\\

\texttt{SingleHadronLimitBottom}
&0.000446
\\

\texttt{SingleHadronLimitCharm}
&0.000508
\\

\texttt{SngWt}
&0.927141
\\

\texttt{DecWt}
&0.630787
\\
\sphinxbottomrule
\end{tabulary}
\sphinxtableafterendhook\par
\sphinxattableend\end{savenotes}

Furthermore, for all Kernels, the CMW scheme has to be switched on.

\subsection{Underlying Event, Colour Reconnection and MPI tunes}
\label{\detokenize{review/appendix/tuning:underlying-event-colour-reconnection-and-mpi-tunes}}

The tuning procedure of the parameters of Underlying Event (UE) and Colour Reconnection (CR) models
was evolving with the evolution of the models. However, one thing has remained unchanged,
this is the assumption that the tuning of Underlying Event and Colour Reconnection factorizes
from the tuning of the hadronization model, i.e.\ it does not affect significantly the LEP observables.
This assumption, for example, was explicitly checked when the first tunes of CR models were created
in \cite{Gieseke:2012ft}. Therefore, the UE and CR models were always tuned only to the
data from hadron-hadron collisions. 
More specifically, modern tunes involve the adjustment of the underlying event (UE) model parameters in \textsc{Herwig} using Minimum Bias and Underlying Event data collected at the LHC and Tevatron, spanning various center-of-mass energies, $\sqrt{s} = 0.9, 1.8, 7, 13$~TeV.
For each energy, a separate tune was initially obtained; these are then combined to produce a single, energy-extrapolated tune. Currently, we employ the extrapolation formula given in Eq.~\eqref{equation:review/ue:eq:ptmin} to perform a simultaneous fit across multiple energies, as discussed, for example, in \cite{Bellm:2019icn,Divisova:2025xqp}.

\section{Solution of the Sudakov equation}
\label{\detokenize{review/appendix/sudakov:solution-of-the-sudakov-equation}}\label{\detokenize{review/appendix/sudakov:sect-sudakov-solution}}\label{\detokenize{review/appendix/sudakov::doc}}

There are a range of methods which can be used to solve Eq. \eqref{equation:review/showers/general:eq:sudakov_master},
\begin{equation*}
\begin{split}\Delta\left(\kappa,\kappa_{h}\right)=\mathcal{R},\end{split}
\end{equation*}

which is repeated here for clarity.

In the FORTRAN HERWIG program this equation was solved by a brute force
numerical calculation, using an interpolation table for
$\Delta\left(\kappa,\kappa_{h}\right)$. In Herwig 7
alternative approaches are used in the angular-ordered and
dipole shower both based on the veto algorithm \cite{Sjostrand:2006za} which is described in
detail in Ref. \cite{Buckley:2011ms} Appendix A.3. This algorithm
determines the scale of the branchings without the need for any explicit integration of the Sudakov
form factor.

In the angular-ordered shower the veto algorithm is implemented
using simple functional forms to overestimate the kernels
while in the dipole shower the ExSample library
\cite{Platzer:2011dr} is used to adaptively compute overestimates of the splitting
kernels and to sample the distribution given by the Sudakov form factor.
Here we only discuss the procedure used in the angular-ordered shower in detail below.

This procedure is repeated to give a value of the evolution scale for
each possible type of branching, and the branching with the largest
value of evolution scale is selected, the ‘competition’ or ‘winner-takes-all’ algorithm.
This correctly generates both the type of branching, the evolution scale and any auxiliary
variables needed to describe the branching.

\subsection{Angular-Ordered shower}
\label{\detokenize{review/appendix/sudakov:angular-ordered-shower}}

The implementation of the veto algorithm used in the angular-ordered parton shower
involves generating
each branching according to a crude Sudakov form factor, based on an
\textit{overestimated} branching probability (Eq. \eqref{equation:review/showers/qtilde:eq:branchprob}),
simple enough that Eq. \eqref{equation:review/showers/qtilde:eq:sudakov_equals_random} can be solved
analytically. Each of these crudely determined branchings is subject to
a vetoing procedure based on a series of weights relating to the true
form factor. In this way the overestimated, crude emission rate and
emission distribution is reduced to the exact distribution.

The first ingredient we need in order to implement the algorithm is
therefore a crude approximation to the Sudakov form factor (Eq.
\eqref{equation:review/showers/qtilde:eq:sudakovmaster}), for each type of branching of a parent
parton $\widetilde{ij}$, $\widetilde{ij}\rightarrow i+j$. We
write these as
\begin{equation*}
\begin{split}\Delta_{\widetilde{ij}\to ij}^{{\rm over}}\left(\tilde{q},\tilde{q}_{h}\right) = \exp\left\{ -\int_{\tilde{q}}^{\tilde{q}_{h}}\mathrm{d}\mathcal{P}_{\widetilde{ij}\to ij}^{\mathrm{over,res.}}\right\},\end{split}
\end{equation*}

where
\begin{equation*}
\begin{split}\mathrm{d}\mathcal{P}_{\widetilde{ij}\to ij}^{\mathrm{over,res.}}=\frac{\mathrm{d}\tilde{q}^{2}}{\tilde{q}^{2}}\int_{z_{{\rm -}}^{\rm over}}^{z_{{\rm +}}^{\rm over}}\mathrm{d}z\,\frac{\alpha_{s}^{\mathrm{over}}}{2\pi}\, P_{\widetilde{ij}\to ij}^{\mathrm{over}}\left(z\right),\end{split}
\end{equation*}

is the overestimated probability that a resolvable branching occurs in
the interval
$\left[\tilde{q}^{2},\tilde{q}^{2}+\mathrm{d}\tilde{q}^{2}\right]$.
Overestimates of the splitting functions and the coupling constant are
denoted
$P_{\widetilde{ij}\to ij}^{{\rm over}}\left(z\right)\geq P_{\widetilde{ij}\to ij}\left(z,\tilde{q}\right)\ $
and
$\alpha_{s}^{{\rm over}}\geq\alpha_{s}\left(z,\tilde{q}\right)$,
while the limits $z_{\pm}^{\rm over}$ also denote overestimates of the
true $z$ integration region %
\begin{footnote}\sphinxAtStartFootnote
The overestimates of these limits were given in
Eqs. \eqref{equation:review/showers/qtilde:eq:zlimits2}, \eqref{equation:review/showers/qtilde:eq:zlimits3}.
\end{footnote} for all $\tilde{q}$. To
solve Eq. \eqref{equation:review/showers/qtilde:eq:sudakov_equals_random}
analytically we also require that
$P_{\widetilde{ij}\to ij}^{{\rm over}}\left(z\right)$ should be
analytically integrable and, in order to generate $z$ values, this
integral should be an invertible function of $z$.

Using this simplified Sudakov form factor we may analytically solve
$\Delta_{\widetilde{ij}\to ij}^{{\rm over}}\left(\tilde{q},\tilde{q}_{h}\right)=\mathcal{R}$ for
$\tilde{q}$ as
\begin{equation}\label{equation:review/appendix/sudakov:eq:crude_qtilde_solution}
\begin{split}\tilde{q}^{2}=\tilde{q}_{h}^{2}\,\mathcal{R}^{\frac{1}{r}},\end{split}
\end{equation}

where
\begin{equation*}
\begin{split}r=\frac{\mathrm{d}\mathcal{P}_{\widetilde{ij}\to ij}^{\mathrm{over,res.}}}{\mathrm{d}\ln\tilde{q}^{2}},\end{split}
\end{equation*}

which can be thought of as the number of emissions per unit of the
shower formation ‘time’ $\ln\tilde{q}^{2}$ (for the crude
distribution this is a constant). It is clear from Eq.
\eqref{equation:review/appendix/sudakov:eq:crude_qtilde_solution} how increasing this
rate $r$ causes the first branching to be generated ‘sooner’,
closer to $\tilde{q}_{h}$. When a value is generated for the
evolution scale of a branching, an associated $z$ value is then
generated according to
\begin{equation*}
\begin{split}z=I^{-1}\left[I\left(z_{-}^{\rm over}\right)+\mathcal{R}^{\prime}\left(I\left(z_{{\rm +}}^{\rm over}\right)-I\left(z_{{\rm -}}^{\rm over}\right)\right)\right],\end{split}
\end{equation*}

where $I\left(z\right)$ is the primitive integral of
$P_{\widetilde{ij}\to ij}^{{\rm over}}\left(z\right)$ over
$z$, $I^{-1}$ is the inverse of $I$ and
$\mathcal{R}^{\prime}$ is a uniformly distributed random number in
the interval $\left[0,1\right]$.

We now reject these values of $\tilde{q}$ and $z$ if:
\begin{itemize}
\item {} 

the value of $z$ lies outside the true phase-space limits,
\textit{i.e.} if $\mathbf{p}_{\perp}^{2}<0$;

\item {} 

$\frac{\alpha_{s}\left(z,\tilde{q}\right)}{\alpha_{s}^{{\rm over}}}<\mathcal{R}_{1}$;

\item {} 

$\frac{P_{\widetilde{ij}\to ij}\left(z,\tilde{q}\right)}{P_{\widetilde{ij}\to ij}^{{\rm over}}\left(z\right)}<\mathcal{R}_{2}$,

\end{itemize}

where $\mathcal{R}_{1,2}$ are random numbers uniformly distributed
between 0 and 1.

If we reject the value of $\tilde{q}$ we repeat the whole
procedure with $\tilde{q}_{h}=\tilde{q}$ until either we accept a
value of $\tilde{q}$, or the value drops below the minimum value
allowed due to the phase-space cutoffs, in which case there is no
radiation from the particle. As shown in Ref. \cite{Sjostrand:2006za} this
procedure, called the veto algorithm, exponentiates the rejection
factors and generates the values of $\tilde{q}$ and $z$
according to Eq. \eqref{equation:review/showers/qtilde:eq:sudakov_equals_random}
for one type of branching.

The backward evolution can be performed using the veto algorithm in the
same way as the forward evolution. We need to solve
\begin{equation*}
\begin{split}\Delta\left(x,\tilde{q},\tilde{q}_{h}\right)=\mathcal{R},\end{split}
\end{equation*}

to give the scale of the branching. We start by considering the backward
evolution of $i$ via a particular type of branching,
$\widetilde{ij}\rightarrow i+j$. We can obtain an overestimate of
the integrand in the Sudakov form factor
\begin{equation}\label{equation:review/appendix/sudakov:eq:sudakov_backward_approx}
\begin{split}\Delta_{\widetilde{ij}\to ij}^{{\rm over}}\left(x,\tilde{q},\tilde{q}_{h}\right)=\exp\left\{ -\int_{\tilde{q}}^{\tilde{q}_{h}}\frac{\mathrm{d}\tilde{q}^{\prime2}}{\tilde{q}^{\prime2}}\int_{x}^{z_{{\rm +}}^{\rm over}}{\rm d}z\mbox{ }\frac{\alpha_{s}^{{\rm over}}}{2\pi}\mbox{ }P_{\widetilde{ij}\to ij}^{{\rm over}}\left(z\right){\rm PDF}^{{\rm over}}\left(z\right)\right\},\end{split}
\end{equation}

where $P_{\widetilde{ij}\to ij}^{{\rm over}}\left(z\right)$,
$\alpha_{s}^{{\rm over}}$ and the overestimate of the limits must
have the same properties as for final-state branching. In addition
\begin{equation*}
\begin{split}{\rm PDF}^{{\rm over}}\left(z\right)\geq\frac{\frac{x}{z}f_{\widetilde{ij}}\left(\frac{x}{z},\tilde{q}\right)}{xf_{i}\left(x,\tilde{q}\right)}\,
\forall
\, z,\mbox{ }\tilde{q},\mbox{ }x.\end{split}
\end{equation*}

In this case the product
$P_{\widetilde{ij}\to ij}^{{\rm over}}\left(z\right){\rm PDF}^{{\rm over}}\left(z\right)$
must be integrable and the integral invertible. If we define
\begin{equation*}
\begin{split}r=\frac{\alpha_{s}^{{\rm over}}}{2\pi}\int_{x}^{z_{{\rm +}}^{\rm over}}\mathrm{d}z\mbox{ }P_{\widetilde{ij}\to ij}^{{\rm over}}\left(z\right){\rm PDF}^{{\rm over}}\left(z\right),\end{split}
\end{equation*}

then we can solve Eq. \eqref{equation:review/showers/qtilde:eq:sudakovbackward} using this
overestimated Sudakov giving
\begin{equation*}
\begin{split}\tilde{q}^{2}=\tilde{q}_{h}^{2}\,\mathcal{R}^{\frac{1}{r}}.\end{split}
\end{equation*}
The value of $z$ can then be generated according to
\begin{equation*}
\begin{split}z=I^{-1}\left[I\left(x\right)+\mathcal{R}^{\prime}\left(I\left(z_{{\rm +}}^{\rm over}\right)-I\left(x\right)\right)\right],\end{split}
\end{equation*}

where
$I\left(z\right)=\int\mathrm{d}z\, P_{\widetilde{ij}\to ij}^{{\rm over}}\left(z\right){\rm PDF}^{{\rm over}}\left(z\right)$,
$I^{-1}$ is the inverse of $I$ and
$\mathcal{R}^{\prime}$ is a random number uniformly distributed
between 0 and 1.

We now reject these values of $\tilde{q}$ and $z$ if:
\begin{itemize}
\item {} 

the value of $z$ lies outside the true phase-space limits,
\textit{i.e.} if $\mathbf{p}_{\perp}^{2}<0$;

\item {} 

$\frac{\alpha_{s}\left(z,\tilde{q}\right)}{\alpha_{s}^{{\rm over}}}<\mathcal{R}_{1}$;

\item {} 

$\frac{P_{\widetilde{ij}\to ij}\left(z,\tilde{q}\right)}{P_{\widetilde{ij}\to ij}^{{\rm over}}\left(z\right)}<\mathcal{R}_{2}$;

\item {} 

$\frac{\frac{\frac{x}{z}f_{a}\left(\frac{x}{z},\tilde{q}'\right)}{xf_{b}\left(x,\tilde{q}'\right)}}{{\rm PDF}^{{\rm over}}\left(z\right)}<\mathcal{R}_{3}$;

\end{itemize}

where $\mathcal{R}_{1,2,3}$ are random numbers uniformly
distributed between 0 and 1.

As with the final-state branching algorithm, if a set of values of
$\tilde{q}$ and $z$, generated according to the approximate
form factor in Eq.
\eqref{equation:review/appendix/sudakov:eq:sudakov_backward_approx} is rejected, a
further set is then generated by repeating the process with
$\tilde{q}_{h}=\tilde{q}$ in Eq.
\eqref{equation:review/appendix/sudakov:eq:sudakov_backward_approx}. This procedure is
repeated until either a generated set of branching variables passes all
four vetoes, or the generated value of $\tilde{q}$ falls below the
minimum allowed value, in which case the showering of the particle in
question ceases. To determine the species of partons involved, a trial
value of $\tilde{q}$ is generated for each possible type of
branching and the largest selected. By applying the four vetoing
criteria to each emission generated by the approximate, overestimated,
Sudakov form factor, the accepted values of $\tilde{q}$ and
$z$ are distributed according to the exact Sudakov form factor,
Eq. \eqref{equation:review/showers/qtilde:eq:sudakovbackward} \cite{Sjostrand:2006za}.

\section{Treatment of the running coupling}
\label{\detokenize{review/appendix/alphaS:treatment-of-the-running-coupling}}\label{\detokenize{review/appendix/alphaS:sub-the-running-coupling}}\label{\detokenize{review/appendix/alphaS::doc}}

The running coupling constant enters every dynamical aspect of the
parton shower, so a thorough treatment of it is mandatory for all parton
shower simulations. In general we work in an axial, or light-cone, gauge to derive the splitting functions.

Axial gauges have many special properties, most notable of these is that they
are ghost-free. Another related interesting feature of the light-cone gauge
is that, similar to QED where the Ward identities guarantee the equality of
the electron field and vertex renormalization constants, in light-cone gauge
QCD the Ward identities reveal that the 3-gluon vertex renormalization
constant $Z_{A^{3}}$, is equal to that of the transverse components
gluon field $Z_{A}^{1/2}$ \cite{Bassetto:1991ue}. This simplifies the
usual relation between the bare coupling $g_{S}^{\left(0\right)}$ and
renormalized coupling constant $g_{S}$ from
$g_{S}^{\left(0\right)}=Z_{A^{3}}Z_{A}^{-3/2}g_{S}$, to
$g_{S}^{\left(0\right)}=Z_{A}^{-1/2}g_{S}$, \textit{i.e.} in the light-cone
gauge, the running of the QCD coupling constant is due to the gluon
self-energy corrections alone. Thus, dimensionally regulated, one-loop
calculations of the gluon self-energy in this gauge possess an ultraviolet
divergence proportional to the usual QCD beta function
\cite{Bassetto:1991ue, Dalbosco:1986eb}.

In calculating higher order corrections to the splitting functions, we must
consider self-energy corrections to the emitted gluons and their associated
counter-terms. The self-energy corrections are equal to zero because the
gluons are on-shell and so the associated loop integrals have no scale, which
means they vanish in dimensional regularization.  This vanishing is due to a
complete cancellation of the ultraviolet and infrared parts of the
integrals. Therefore including the counter-terms cancels explicitly the
ultraviolet divergent parts of the loop integrals leaving behind infrared
divergent parts, which must have the same pole structure as the ultraviolet
parts \textit{i.e.} they must also be proportional to the beta function. The
residual virtual infrared divergence is cancelled by the associated real
emission corrections, in this case the two graphs where the emitted gluon
splits either to two on-shell gluons or to a quark-antiquark pair.  This
cancellation of infrared poles generates an associated logarithm of the
phase-space boundary divided by $\mu$ (the renormalization scale)
\cite{Bassetto:1984ik, Amati:1980ch} multiplied by the same coefficient as
the pole (the beta function). The phase-space boundary is equal to the
maximum possible virtuality of the daughter gluon, whose branchings
comprise the real emission corrections. For a time-like splitting,
$\widetilde{ij}\rightarrow i+j$ where $\widetilde{ij}$ is a
quark, $i$ is a daughter quark and $j$ is the daughter gluon, to
which we consider real and virtual corrections, a quick calculation in the
Sudakov basis Eq. \eqref{equation:review/showers/qtilde:eq:sudbasis} shows that the gluon virtuality
$q_{j}^{2}$ must satisfy the relation
\begin{equation*}
\begin{split}q_{j}^{2}\le\left(1-z\right)q_{\widetilde{ij}}^{2},\end{split}
\end{equation*}
where $1-z$ is the light-cone momentum carried by the emitted gluon,
and $q_{\widetilde{ij}}^{2}$ is the virtuality of the emitting quark.
The net effect of these real and virtual corrections is therefore to
simply correct the leading order $q\rightarrow qg$ splitting
function by a multiplicative factor
\begin{equation*}
\begin{split}1-\beta_{0} \, \alpha_{s}(\mu^{2}) \,\ln\left(\left(1-z\right)q_{\widetilde{ij}}^{2}/\mu^{2}\right)+\mathcal{O}\left(\alpha_{s}\right)\end{split}
\end{equation*}

where the omitted $\mathcal{O}\left(\alpha_{s}\right)$ terms are
non-logarithmic, non-kinematic, constant terms, $\beta_{0}$ is the
QCD beta function, and $\mu^{2}$ is the renormalization scale.

Two important points follow directly from this analysis. Firstly, for
soft configurations, $z\rightarrow1$, the effect of these loop
contributions can produce large, numerically significant, logarithms.
Secondly, plainly, by choosing the renormalization scale to be
$\left(1-z\right)q_{\widetilde{ij}}^{2}$, instead of the more
obvious $q_{\widetilde{ij}}^{2}$, the corrections vanish, or
rather, more correctly, they are absorbed in the coupling constant.

For $g\rightarrow gg$ splittings the same arguments hold but in
this case it is apparent that as well as large logarithms of
$1-z$, large logarithms of $z$ are also possible from soft
emission in the $z\rightarrow0$ region. We may simultaneously
include both types of correction by using
$z\left(1-z\right)q_{\widetilde{ij}}^{2}$, \textit{i.e.} the transverse momentum of the branching,  as the argument of the
running coupling, as
\begin{equation*}
\begin{split}z\left(1-z\right)q_{\widetilde{ij}}^{2} = z^2\left(1-z\right)^2
\tilde{q}_{\widetilde{ij} \to i,j}^{2} = p_T^2.\end{split}
\end{equation*}

From the point of view of the leading-logarithmic approximation, the choice of
scale is technically a higher order consideration, nevertheless, these
effects turn out to be highly phenomenologically significant,
particularly their effect on multiplicity distributions and cluster mass
spectra \cite{Amati:1979fg,Amati:1980ch}.

Thus, by default, the strong coupling is evaluated at $z^2(1-z)^2
\tilde{q}^2$.  The user has the possibility to selectively change the
argument from the default to $z(1-z) \, \tilde{q}^2$, that corresponds to
the virtuality $q^2$, for certain splitting processes only, using the
\href{https://herwig.hepforge.org/doxygen/SplittingFunctionInterfaces.html\#ScaleChoice}{ScaleChoice}
interface.

\subsection{The Monte Carlo scheme for \texorpdfstring{$\bm{\alpha_{s}}$}{αₛ}}
\label{\detokenize{review/appendix/alphaS:the-monte-carlo-scheme-for-alpha-s}}\label{\detokenize{review/appendix/alphaS:sub-the-monte-carlo}}

By choosing the scale of the running coupling as
advocated for above we have
\begin{equation}\label{equation:review/appendix/alphaS:eq:soft_one_loop_splitting_fn}
\begin{split}\lim_{z\rightarrow1} \alpha_{s} \left(\left(1-z\right)  q_{\widetilde{ij}}^{2}\right)P_{q\to qg}^{\left[1\right]}\left(z\right) =  \alpha_{s}\frac{2C_{F}}{1-z}\left(1-\alpha_{s}\beta_{0}\ln\left(1-z\right)\right) +\mathcal{O}(\alpha_{s}^3),\end{split}
\end{equation}

where we have momentarily abbreviated
$\alpha_{s}\left(q_{\widetilde{ij}}^{2}\right)$ by $\alpha_{s}$,
and we have used a superscript $\left[1\right]$ to denote that
$P_{q\to qg}^{\left[1\right]}$ is the \textit{one-loop} (\textit{i.e.} leading
order) $q\rightarrow qg$ splitting function. This is almost, but
not exactly equal to the soft $z\rightarrow1$ singular limit of
the \textit{two-loop} $q\rightarrow qg$ splitting function
$P_{q\to qg}^{\left[2\right]}$ with $\alpha_{s}$ evaluated
at $q_{\widetilde{ij}}^{2}$,
\begin{equation}\label{equation:review/appendix/alphaS:eq:soft_two_loop_splitting_fn}
\begin{split}\lim_{z\rightarrow1}\alpha_{s}\left(q_{\widetilde{ij}}^{2}\right) P_{q\to qg}^{\left[2\right]}\left(z\right)  = \alpha_{s}\frac{2C_{F}}{1-z} \left(1-\alpha_{s}\beta_{0}\ln\left(1-z\right)+\frac{\alpha_{s}}{2\pi}K_{g}\right) +\mathcal{O}(\alpha_{s}^3)\end{split}
\end{equation}
where %
\begin{footnote}\sphinxAtStartFootnote
In fact the constants $K_{g}$ are given by the finite remainder
of the real emission phase-space corrections due to the daughter
gluon splitting discussed in
\hyperref[\detokenize{review/appendix/alphaS:sub-the-running-coupling}]{\ref{\detokenize{review/appendix/alphaS:sub-the-running-coupling}}} (see \textit{e.g.} Eqs.~(5.28), (C.12) and (C.13) of
\cite{Catani:1996vz}).
\end{footnote}
\begin{equation*}
\begin{split}K_{g}=C_{A}\left(\frac{67}{18}-\frac{\pi^{2}}{6}\right)-T_{R}n_{f}\,\frac{10}{9}.\end{split}
\end{equation*}
On integrating over the phase-space of the two-loop splitting function the
$K_{g}$ term gives rise to terms proportional to
$\alpha_{s}^{2}\ln^{2}q_{\widetilde{ij}}^{2}$, \textit{i.e.} it gives
next-to-leading-logarithmic soft-collinear contributions to the Sudakov exponent of
the form $\alpha_{s}^{n}\ln^{n}q_{\widetilde{ij}}^{2}$ (as opposed to
leading-logarithmic contributions that are instead proportional to
$\alpha_{s}^{n}\ln^{n+1}q_{\widetilde{ij}}^{2}$). In a similar way
to that in which the higher order
$\beta_{0}\alpha_{s}\ln\left(1-z\right)$ term was included, we may
exploit the fact that the $z\rightarrow1$ dependence of the
$K_{g}$ term in $P_{q\to qg}^{\left[2\right]}\left(z\right)$ is
equal to that of $P_{q\to qg}^{\left[1\right]}\left(z\right)$, to
incorporate it in the running coupling as well.

This is done by replacing the usual
$\Lambda_{\overline{\mathrm{MS}}}$ QCD scale, from which the
coupling runs, to $\Lambda_{\mathrm{CMW}}$ \cite{Catani:1990rr}:
\begin{equation*}
\begin{split}\Lambda_{\mathrm{CMW}}=\Lambda_{\overline{\mathrm{MS}}}\exp\left(K_{g}/4\pi\beta_{0}\right),\end{split}
\end{equation*}

where $\mathrm{CMW}$ denotes the so-called Catani-Marchesini-Webber (CMW) \cite{Catani:1990rr} or \textit{Monte Carlo scheme}.
Expanding $\alpha_{s}P_{q\to qg}^{\left[1\right]}\left(z\right)$
again, as in Eq.~\eqref{equation:review/appendix/alphaS:eq:soft_one_loop_splitting_fn},
but with $\alpha_{s}$ evaluated at
$\left(1-z\right)q_{\widetilde{ij}}^{2}$ in the CMW scheme,
reproduces exactly the two-loop result in Eq.~\eqref{equation:review/appendix/alphaS:eq:soft_two_loop_splitting_fn}.
With this prescription the Sudakov form factor generally includes all
leading- and next-to-leading-logarithmic contributions, except for those due to
soft wide angle gluon emissions, however, for the case that the
underlying hard process comprises a single colour dipole, these are
also included (see \hyperref[\detokenize{review/showers/qtilde:sub-shower-dynamics}]{Section \ref{\detokenize{review/showers/qtilde:sub-shower-dynamics}}} and
\cite{Frixione:2007vw, Bonciani:2003nt}).

\subsection{Options for the treatment of 
  \texorpdfstring{$\bm{\alpha_{s}}$}{αₛ} 
  in the angular-ordered parton shower}
\label{\detokenize{review/appendix/alphaS:options-for-the-treatment-of-alpha-s-in-the-angular-ordered-parton-shower}}

Although we have made strong physical arguments restricting the argument
of the coupling constant and suggesting a more physical renormalisation
scheme, there is still some degree of freedom in how precisely
$\alpha_{s}$ is calculated. In what follows below we enumerate the
options associated with these in the program:
\begin{itemize}
\item {} 

The \href{https://herwig.hepforge.org/doxygen/ShowerAlphaQCDInterfaces.html\#InputOption}{InputOption}
switch selects the way in which initial conditions for running the
coupling constant are determined.
\begin{itemize}
\item {} 

The default setting, \texttt{AlphaIn}, uses our tuned value of $\alpha_{s}$ at the
$Z^{0}$ resonance to calculate a value of $\Lambda_{\mathrm{QCD}}$ from which the running coupling constant is evaluated.
This input can be reset from the default value %
\begin{footnote}\sphinxAtStartFootnote
The default value is tuned to $e^+e^-$ annihilation data as
described in \hyperref[\detokenize{review/appendix/tuning:sect-tuning}]{\ref{\detokenize{review/appendix/tuning:sect-tuning}}} and is typical of the
values one gets when fitting leading order QCD predictions to data.
\end{footnote} of 0.1186 using the
\texttt{AlphaIn} interface.

\item {} 

Alternatively one may select an option \href{https://herwig.hepforge.org/doxygen/ShowerAlphaQCDInterfaces.html\#LambdaQCD}{LambdaQCD},
which uses the input or default value of
$\Lambda_{\mathrm{QCD}}$ to calculate the coupling. The value may be set
using the \texttt{LambdaQCD} interface.

\end{itemize}

\item {} 

The \href{https://herwig.hepforge.org/doxygen/ShowerAlphaQCDInterfaces.html\#LambdaOption}{LambdaOption}
flag determines whether the value of
$\Lambda_{\mathrm{QCD}}$, which can be calculated from
$\alpha_{s}\left(m_{Z^{0}}\right)$ using the \texttt{AlphaIn}
interface or inputted directly via the \texttt{LambdaQCD} interface,
is already given in the $\mathrm{CMW}$ (Monte Carlo) scheme
of Ref.  \cite{Catani:1990rr} (option \href{https://herwig.hepforge.org/doxygen/ShowerAlphaQCDInterfaces.html\#LambdaOption}{Same}, this is the default behaviour),
described in \hyperref[\detokenize{review/appendix/alphaS:sub-the-monte-carlo}]{\ref{\detokenize{review/appendix/alphaS:sub-the-monte-carlo}}}, or it is in the
$\overline{\mathrm{MS}}$ scheme and thus needs to be
converted (option \href{https://herwig.hepforge.org/doxygen/ShowerAlphaQCDInterfaces.html\#LambdaOption}{Convert}).

\item {} 

The \href{https://herwig.hepforge.org/doxygen/ShowerAlphaQCDInterfaces.html\#NumberOfLoops}{NumberOfLoops}
parameter specifies the loop order of the beta function used to
calculate the running of $\alpha_{s}$. The default setting uses
the two-loop beta function.

\item {} 

The \href{https://herwig.hepforge.org/doxygen/ShowerAlphaQCDInterfaces.html\#ThresholdOption}{ThresholdOption}
option selects whether to use the \href{https://herwig.hepforge.org/doxygen/ShowerAlphaQCDInterfaces.html\#ThresholdOption}{Current}
or \href{https://herwig.hepforge.org/doxygen/ShowerAlphaQCDInterfaces.html\#ThresholdOption}{Constituent}
quark masses in determining the flavour thresholds in the evolution of
the coupling constant. The default is to use the
($\overline{\mathrm{MS}}$) current quark masses.

\item {} 

The \href{https://herwig.hepforge.org/doxygen/ShowerAlphaQCDInterfaces.html\#Qmin}{Qmin}
parameter specifies the value of $\mathrm{Qmin}$, the scale beneath which
non-perturbative effects are considered to render the usual
renormalization group running with a beta function determined at some
finite loop order invalid. Below this scale, which is currently tuned
to 0.935 GeV, a number of parameterizations of the scaling of the
coupling with energy may be selected according to the
\href{https://herwig.hepforge.org/doxygen/ShowerAlphaQCDInterfaces.html\#NPAlphaS}{NPAlphaS}
option described below.

\item {} 

The
\texttt{NPAlphaS}
option selects a parameterization of the scaling of the
running coupling with energy in what we regard as the non-perturbative
region, where the scale at which it is to be evaluated falls below the
value set by $\mathrm{Qmin}$. By setting
\href{https://herwig.hepforge.org/doxygen/ShowerAlphaQCDInterfaces.html\#NPAlphaS}{(NPAlphaS=Zero)} the coupling is simply
taken to be zero for scales $\mathrm{Q}<\mathrm{Qmin}$. For the
\href{https://herwig.hepforge.org/doxygen/ShowerAlphaQCDInterfaces.html\#NPAlphaS}{(NPAlphaS=Const)}
coupling \textit{freezes} \textit{out} at $\mathrm{Qmin}$, \textit{i.e.} it assumes the
constant value
$\tilde{\alpha}_{S}=\alpha_{s}\left(\mathrm{Qmin}\right)$ for all
scales below $\mathrm{Qmin}$. This is the default
parameterization. It is the same prescription used in early works on
resummation by Curci and Greco \cite{Curci:1979am, Curci:1981yr}. The options
\href{https://herwig.hepforge.org/doxygen/ShowerAlphaQCDInterfaces.html\#NPAlphaS}{(NPAlphaS=Linear)} and
\href{https://herwig.hepforge.org/doxygen/ShowerAlphaQCDInterfaces.html\#NPAlphaS}{(NPAlphaS=Quadratic)}
calculate the running coupling below $\mathrm{Qmin}$
according to $\tilde{\alpha}_{S}\mathrm{Q}/\mathrm{Qmin}$ and
$\tilde{\alpha}_{S}\left(\mathrm{Q}/\mathrm{Qmin}\right)^{2}$
respectively. Setting
\href{https://herwig.hepforge.org/doxygen/ShowerAlphaQCDInterfaces.html\#NPAlphaS}{(NPAlphaS=Exx1)}
assumes a quadratically decreasing running of the
coupling in the non-perturbative region from the value
\href{https://herwig.hepforge.org/doxygen/ShowerAlphaQCDInterfaces.html\#AlphaMaxNP}{AlphaMaxNP} down to
$\tilde{\alpha}_{S}$. Finally,
\href{https://herwig.hepforge.org/doxygen/ShowerAlphaQCDInterfaces.html\#NPAlphaS}{(NPAlphaS=Exx2)}
sets $\alpha_{s}$ equal to  \texttt{AlphaMaxNP}
for all input scales $\mathrm{Q}<\mathrm{Qmin}$, which amounts to
a minor variation of the default \textit{freeze-out} option.

\end{itemize}

\subsection{Options for the treatment of 
  \texorpdfstring{$\bm{\alpha}_{s}$}{αₛ} 
  in the dipole parton shower}
\label{\detokenize{review/appendix/alphaS:options-for-the-treatment-of-alpha-s-in-the-dipole-parton-shower}}\label{\detokenize{review/appendix/alphaS:dipole-shower-strong-coupling}}

The dipole parton shower, as well as the Matchbox module, in Herwig makes use of
distinct implementations of the strong coupling constant. One and two loop running
are implemented in \href{https://herwig.hepforge.org/doxygen/classmatchbox\_1\_1lo\_\_alpha\_\_s.html}{lo\_alpha\_s}
and \href{https://herwig.hepforge.org/doxygen/classmatchbox\_1\_1nlo\_\_alpha\_\_s.html}{nlo\_alpha\_s}, respectively.
\begin{itemize}
\item {} 

To set the input value and scale for the dipole shower $\alpha_{s}$
classes the \href{https://herwig.hepforge.org/doxygen/classmatchbox\_1\_1alpha\_\_s.html}{input\_alpha\_s}
and \href{https://herwig.hepforge.org/doxygen/classmatchbox\_1\_1alpha\_\_s.html}{input\_scale}
are to be used.

\item {} 

It is possible to set the allowed number of active flavours with the
\href{https://herwig.hepforge.org/doxygen/classmatchbox\_1\_1alpha\_\_s.html}{min\_active\_flavours} and
\href{https://herwig.hepforge.org/doxygen/classmatchbox\_1\_1alpha\_\_s.html}{max\_active\_flavours}
parameters, that effects the running as well as the threshold matching.

\item {} 

The threshold matching is performed at the beginning of each run
using gsl bisection to fix the running at the nominal quark masses.

\item {} 

Both the LO and NLO implementations treat the number of active flavours as constant
below a given input scale, the \href{https://herwig.hepforge.org/doxygen/classmatchbox\_1\_1lo\_\_alpha\_\_s.html}{freezing\_scale} parameter.

\item {} 

The NP behaviour in the dipole shower is screened by a so called screening scale.
The \href{https://herwig.hepforge.org/doxygen/classHerwig\_1\_1DipoleSplittingKernel.html}{ScreeningScale}
is an interface of each splitting kernel.
It corresponds to the  \texttt{(NPAlphaS=Const)} treatment in the angular-ordered parton shower.

\item {} 

The running at LO is performed with
\begin{equation*}
\begin{split}\alpha_{s}(q)=\frac{1}{\beta_0 L} \;\;\; \mathrm{with}\;\;\; \beta_0=\frac{33-2 n_F}{12\pi} \;\;\; \mathrm{and}\;\;\; L=\log\left(\frac{q^2}{\Lambda^2(n_F(q^2))}\right),\end{split}
\end{equation*}
while the two loop running makes use of
\begin{equation*}
\begin{split}\alpha_{s}(q)=\frac{1}{\beta_0 L} \left(1 - \frac{\beta_1}{\beta^2_0} \frac{\log(L)}{L} + LQT \right)\;\;\; \mathrm{with}\;\;\; \beta_1 = \frac{153-19 n_F}{24 \pi^2},\end{split}
\end{equation*}

where additional terms in the large $Q$ expansion LQT
\begin{equation*}
\begin{split}LQT= \frac{\beta^2_1}{\beta^4_0 L^2}  \left(\left(\log(L)-\frac{1}{2}\right)^2 - \frac{5}{4}\right),\end{split}
\end{equation*}

can be added using the \href{https://herwig.hepforge.org/doxygen/classmatchbox\_1\_1nlo\_\_alpha\_\_s.html}{two\_largeq\_terms} interface.
This is an approximation to the renormalization group running but the exact running
can be requested and calculated with the \href{https://herwig.hepforge.org/doxygen/classmatchbox\_1\_1nlo\_\_alpha\_\_s.html}{exact\_evaluation} switch.

\item {} 

The modifications needed for the Monte Carlo scheme are part of the Dipole Kernels options, called
\href{https://herwig.hepforge.org/doxygen/classHerwig\_1\_1DipoleSplittingKernel.html}{CMWScheme} .

\end{itemize}

\section{Massless dipoles}
\label{\detokenize{review/appendix/masslessDipole:massless-dipoles}}\label{\detokenize{review/appendix/masslessDipole:sect-massless-dipoles}}\label{\detokenize{review/appendix/masslessDipole::doc}}

This appendix contains the results for massless dipoles which are omitted
from the main text. See \hyperref[\detokenize{review/showers/dipole:dipole-shower-kinematics}]{Section \ref{\detokenize{review/showers/dipole:dipole-shower-kinematics}}} for more details.

\subsection{Final-Final dipole kinematics}
\label{\detokenize{review/appendix/masslessDipole:final-final-dipole-kinematics}}\label{\detokenize{review/appendix/masslessDipole:dipole-shower-kin-ff-massless}}

We require that the momenta prior to the
splitting can be written in terms of the momenta following the splitting
\begin{equation}\label{equation:review/appendix/masslessDipole:eqn:DS:FFKin:LightTildeKin}
\begin{split}\tilde{p}_k&= \frac{1}{1-y_{ij,k}} q_k \ ,
\\
\tilde{p}_{ij}&= q_i + q_j - \frac{y_{ij,k}}{1-y_{ij,k}} q_k \ .\end{split}
\end{equation}

With this requirement, the physical momenta following the splitting can be
written in terms of the splitting variables
\begin{equation}\label{equation:review/appendix/masslessDipole:eqn:DS:FFKin:LightNewMom}
\begin{split}q_i &= z_i\tilde{p}_{ij} + (1-z_i) y_{ij,k}\tilde{p}_k + k_\perp \ ,
\\
q_j &= (1-z_i) \tilde{p}_{ij} + z_i y_{ij,k}\tilde{p}_k - k_\perp \ ,
\\
q_k &= (1-y_{ij,k})\tilde{p}_k \ .\end{split}
\end{equation}

As we have a massless spectator, $z_i$ is identical to $z$.
We can write $y_{ij,k}$ in terms of $z \text{ and } p_\perp$
\begin{equation*}
\begin{split}y_{ij,k} = \frac{p_\perp^2}{z(1-z)Q^2} \ .\end{split}
\end{equation*}

and the limits on the splitting variables $z_i$
and $y_{ij,k}$ are simply
\begin{equation*}
\begin{split}z_{i,-} = 0 \ , \qquad z_{i,+} = 1 \ ,\\
y_- = 0 \ ,\qquad y_+ = 1 \ .\end{split}
\end{equation*}

Working in a frame where $Q = 0$, the kinematic upper
limit on the transverse momentum, $p_{\perp,\mathrm{max}}$, is simply the
magnitude of the emission 3-momentum in the limit that the spectator following
the splitting has zero momentum. The momentum conservation requirement in
Eq. \eqref{equation:review/showers/dipole:eqn:DS:FFKin:Q} can then be rearranged to give
\begin{equation*}
\begin{split}p_{\perp,\mathrm{max}}^2 = \frac{s}{4} \ .\end{split}
\end{equation*}

The limits on $z$ follow from the limit $y_{ij,k} < 1$,
\begin{equation*}
\begin{split}z_\pm = \frac{1}{2} \left[ 1 \pm \sqrt{1 - \frac{p_\perp^2}
{p_{\perp,\mathrm{max}}^2} } \right] \ .\end{split}
\end{equation*}

\subsection{Final-Initial dipole kinematics}
\label{\detokenize{review/appendix/masslessDipole:final-initial-dipole-kinematics}}\label{\detokenize{review/appendix/masslessDipole:dipole-shower-kinematics-fi-massless}}

Following the massless Sudakov parametrization, the physical momenta
following the splitting are written as,
\begin{equation*}
\begin{split}q_i & =  z\tilde{p}_{ij} + \frac{p_\perp^2}{ s_{ij,b} z} \tilde{p}_b + k_\perp \ , \\
q_j & =  (1-z)\tilde{p}_{ij} + \frac{p_\perp^2}{ s_{ij,b} (1-z)}\tilde{p}_b - k_\perp \ , \\
q_b & =  \frac{1}{x_{ij,b}}\tilde{p}_b.\\\end{split}
\end{equation*}

As in the massive case we have $z_i = z$.
We can simply insert the equations above into \eqref{equation:review/showers/dipole:eqn:DS:FIKin:x} to obtain
an explicit expression for $x_{ij,b}$ in terms of $z$ and
$p_\perp$,
\begin{equation*}
\begin{split}x_{ij,b} = \left[1 + \frac{p_\perp^2}{ s_{ij,b} z(1-z)} \right]^{-1} \ .\end{split}
\end{equation*}

The phase-space limits on the splitting variables
$z_i$ and $x_{ij,b}$ are,
\begin{equation*}
\begin{split}z_{i,-} = 0 \ , \qquad z_{i,+} = 1 \ ,
\\
x_- = x_s \ ,\qquad x_+ = 1 \ .\end{split}
\end{equation*}

Following from the result in \eqref{equation:review/showers/dipole:eqn:DS:FI:xlim} one can obtain the following
limits on the generated variables,
\begin{equation*}
\begin{split}p_{\perp,\mathrm{max}}^2 =
\frac{s_{ij,b}}{4} \left( \frac{1-x_s}{x_s} \right) \ ,\end{split}
\end{equation*}\begin{equation*}
\begin{split}z_\pm = \frac{1}{2}
\left[ 1 - \sqrt{1-\frac{p_\perp^2} {p_{\perp,\mathrm{max}}^2} } \right] \ .\end{split}
\end{equation*}

\subsection{Initial-Final dipole kinematics}
\label{\detokenize{review/appendix/masslessDipole:initial-final-dipole-kinematics}}\label{\detokenize{review/appendix/masslessDipole:dipole-shower-kinematics-if-massless}}

The physical momenta following the splitting can be written in terms of the
splitting variables as,
\begin{equation*}
\begin{split}q_a &=  \frac{1}{x_{jk,a}} \tilde{p}_{aj} \ ,
\\
q_j &= \left( \frac{1-x_{jk,a}}{x_{jk,a}} \right) (1-u_j) \tilde{p}_{aj}
+ u_j \tilde{p}_k - k_\perp \ ,
\\
q_k &= \left( \frac{1-x_{jk,a}}{x_{jk,a}} \right) u_j \tilde{p}_{aj}
+ (1-u_j) \tilde{p}_k + k_\perp \ ,\end{split}
\end{equation*}

and the splitting variables can be written in terms of the generated variables
$z$ and $p_\perp$ as,
\begin{equation*}
\begin{split}x_{jk,a} &= \frac{1-z+r}{2r} \left[ 1 - \sqrt{1 - \frac{4r z(1-z)}{(1-z+r)^2}} \right] \ ,
\\
u_j &= x_{jk,a} \left( \frac{r}{1-z} \right) \ ,\end{split}
\end{equation*}

where we have defined $r = \frac{p_\perp^2}{s_{aj,k}}$.

The limits on the splitting variables $u_j$ and
$x_{jk,a}$ are then simply,
\begin{equation*}
\begin{split}u_- = 0 \ , \qquad u_+ = 1 \ ,
\\
x_- = x_e \ ,\qquad x_+ = 1 \ .\end{split}
\end{equation*}

Following from the inequality in \eqref{equation:review/showers/dipole:eqn:IFKin:xLim} we obtain
the limits on the generated variables $z$ and $p_\perp$,
\begin{equation*}
\begin{split}p_{\perp,\mathrm{max}}^2 &= \frac{s_{aj,k}}{4} \left( \frac{1-x_e}{x_e} \right) \ ,
\\
z_{\pm} &= \frac{1}{2} \left[ (1+x_e) \pm (1-x_e) \sqrt{1-\frac{p_\perp^2}{ p_{\perp,\mathrm{max}}^2}} \right] \ .\end{split}
\end{equation*}

\subsection{Final-Final dipole kernels}
\label{\detokenize{review/appendix/masslessDipole:final-final-dipole-kernels}}\label{\detokenize{review/appendix/masslessDipole:dipole-shower-kernels-ff-masses}}\begin{equation*}
\begin{split}\langle V_{q_i g_j,k} \left( z_i, y_{ij,k} \right) \rangle
&= 8\pi \alpha_\text{S} C_\text{F}
\left[ \frac{2}{1-z_i(1-y_{ij,k})} - (1+z_i) \right ] \\
\langle V_{q_i \bar{q}_j,k} \left( z_i, y_{ij,k} \right) \rangle
&= 8\pi \alpha_\text{S} T_\text{R}
\left[ 1 - 2z_i(1-z_i) \right] \\
\langle V_{g_i g_j,k} \left( z_i, y_{ij,k} \right) \rangle
&= 16\pi \alpha_\text{S} C_\text{A}
\left[ \frac{1}{1-z_i(1-y_{ij,k})} +
\frac{1}{1-(1-z_i)(1-y_{ij,k})} - 2 + z_i(1-z_i) \right]\end{split}
\end{equation*}

\subsection{Final-Initial dipole kernels}
\label{\detokenize{review/appendix/masslessDipole:final-initial-dipole-kernels}}\label{\detokenize{review/appendix/masslessDipole:dipole-shower-kernels-fi-masses}}\begin{equation*}
\begin{split}\langle V_{q_i g_j}^b \left( z_i, x_{ij,b} \right) \rangle
&= 8\pi \alpha_\text{S} C_\text{F}
\left[ \frac{2}{1-z_i + (1-x_{ij,b})} - (1+z_i) \right ]\\
\langle V_{q_i \bar{q}_j}^b \left( z_i, x_{ij,b} \right) \rangle
&= 8\pi \alpha_\text{S} T_\text{R} \left[ 1 - 2z_i(1-z_i) \right] \\
\langle V_{g_i g_j}^b \left( z_i, x_{ij,b} \right) \rangle
&= 16\pi \alpha_\text{S} C_\text{A}
\left[ \frac{1}{1-z_i + (1-x_{ij,b})}
+ \frac{1}{z_i + (1-x_{ij,b})} - 2 + z_i(1-z_i) \right]\end{split}
\end{equation*}

\subsection{Initial-Final dipole kernels}
\label{\detokenize{review/appendix/masslessDipole:initial-final-dipole-kernels}}\label{\detokenize{review/appendix/masslessDipole:dipole-shower-kernels-if-massless}}\begin{equation*}
\begin{split}\langle V^{q_a g_j}_k \left( u_j, x_{jk,a} \right) \rangle
&= 8 \pi \alpha_\text{S} C_\text{F} \left[ \frac{2}{1-x_{jk,a}+u_j} - (1+x_{jk,a}) \right ] \\
\langle V^{g_a q_j}_k \left( u_j, x_{jk,a} \right) \rangle
&= 8 \pi \alpha_\text{S} T_\text{R} \left[ 1 - 2x_{jk,a}(1-x_{jk,a}) \right] \\
\langle V^{q_a q_j}_k \left( u_j, x_{jk,a} \right) \rangle
&= 8 \pi \alpha_\text{S} C_\text{F} \left[ x_{jk,a} + \frac{2(1-x_{jk,a})}{x_{jk,a}} \right] \\
\langle V^{g_a g_j}_k \left( u_j, x_{jk,a} \right) \rangle
&= 16 \pi \alpha_\text{S} C_\text{A} \left[ \frac{1}{1-x_{jk,a}+u_j} + \frac{1-x_{jk,a}}{x_{jk,a}} - 1 + x_{jk,a}(1-x_{jk,a}) \right]\end{split}
\end{equation*}

\clearpage

% hack to fix running header in bibliography section
\makeatletter
\if@runhead
  \markboth{References}{References}
  \gdef\@author{References}\gdef\@title{References}
\fi
\makeatother

\bibliographystyle{JHEP}
\bibliography{herwig}

\end{document}